%% file: LH_NewPhysics_Report.tex
\documentclass[hyperref,12pt]{cernyrep}
\usepackage{hyperref}
\usepackage{ifpdf}
\usepackage{url}
\usepackage{subfigure}
\usepackage[english]{babel}
\usepackage{color}
\hypersetup{
  colorlinks,
  citecolor=blue,
  linkcolor=red,
  urlcolor=blue}

\usepackage{tikz}
\usetikzlibrary{trees}
\usetikzlibrary{decorations.pathmorphing}
\usetikzlibrary{decorations.markings}

  \definecolor{jblue}  {RGB}{20,50,100}
  \definecolor{npurple}  {RGB} {153, 51, 204}
  \definecolor{wred}   {RGB}{217,0,56}
  \definecolor{white}   {RGB}{255,255,255}
  
  \definecolor{korange}   {RGB}{235, 80,  43}
  \definecolor{korange2}   {RGB}{245, 100,  63}
  \definecolor{kyelloworange}   {RGB}{255, 210,  110}
  \definecolor{kyelloworange2}   {RGB}{240, 170,  90}
  \definecolor{kred}   {RGB}{204,  102, 153}
  \definecolor{kpurple}   {RGB}{153,  61, 190}
  \definecolor{kpurplelight}   {RGB}{213,  161, 230}

	\tikzset{
	  photon/.style={decorate, decoration={snake}, draw=npurple,very thick},
	  boson/.style={decorate, decoration={snake}, draw=npurple,very thick},
	  electron/.style={draw=jblue,very thick, postaction={decorate},
	           decoration={markings,mark=at position .55 with {\arrow[draw=jblue]{>}}}
	  },
	  electron2/.style={draw=jblue,very thick, postaction={decorate},
	           decoration={markings,mark=at position .55 with {\arrow[draw=jblue]{<}}}
	  },
	  fermion/.style={draw=jblue,very thick, postaction={decorate},
	            decoration={markings,mark=at position .55 with {\arrow[draw=jblue]{}}}
	  },
	  gluon/.style={decorate, draw=korange,very thick, 
	    decoration={coil,amplitude=4pt, segment length=6pt}},
	  higgs/.style={draw=wred,very thick, postaction={decorate},
	           decoration={markings,mark=at position .55 with {\arrow[draw=wred]{>}}}
	  },
	  graviton/.style={draw=wred,very thick, postaction={decorate},
	           decoration={snake}
	  },
	  nothing/.style={draw=white,very thick}
	}

\input Commands.tex

\begin{document}

\setcounter{tocdepth}{0}
\thispagestyle{empty}

\vspace{1cm}

\begin{center}
{\Large {\bf LES HOUCHES 2013: PHYSICS AT TEV COLLIDERS \\[4mm]}}
{\Large {\bf NEW PHYSICS WORKING GROUP REPORT}}
\end{center}

\vspace{0.1cm}
\input authors.tex

 \vspace{1cm}
\begin{center}
{\large {\bf Abstract}}\\[.2cm]
\end{center}
We present the activities of the 
``New Physics'' working group for the ``Physics at TeV Colliders"
workshop (Les Houches, France, 3--21 June, 2013).  Our 
report includes new computational tool developments,
studies of the implications of the Higgs boson discovery on new physics,
important signatures for searches for natural new physics at the LHC,
new studies of flavour aspects of new physics,
and assessments of the interplay between direct dark matter searches and the LHC.
\vspace{1cm}
\begin{center}
{\bf Acknowledgements}\\[.2cm]
\end{center}
We would like to heartily thank the funding bodies, the organisers 
(G.~B\'elanger, F.~Boudjema, S.~Gascon, P.~Gras, D.~Guadagnoli, J.P.~Guillet, G.~Hamel de Monchenault, B.~Herrmann, S.~Kraml, G.~Moreau, E.~Pilon, P.~Slavich and D.~Zerwas), the staff and the other
participants of the Les Houches workshop for providing a stimulating and
lively environment in which to work.

\newpage

\vspace{1cm}

\thispagestyle{empty}
\setcounter{page}{2}

\begin{center}
\input addresses.tex

\end{center}

\newpage

\tableofcontentscern

\newpage

\input intro.tex
\addtocontents{toc}{\protect\contentsline{part}{\protect\numberline{} \hspace{-2cm}Introduction}{6}{}}
\AddToContent{G.~Brooijmans, R.~Contino, B.~Fuks, F.~Moortgat, P.~Richardson, S.~Sekmen, A.~Weiler}

\setcounter{figure}{0}
\setcounter{table}{0}
\setcounter{section}{0}
\setcounter{equation}{0}
\setcounter{footnote}{0}
\clearpage


\superpart{ Tools }

\input hdecay/hdecay.tex
\AddToContent{A.~Djouadi, J.~Kalinowski, M.~M\"uhlleitner, M.~Spira}
\renewcommand{\thesection}{\arabic{section}}

\input ehdecay/ehdecay.tex
\AddToContent{R.~Contino, M.~Ghezzi, C.~Grojean, M.~M\"uhlleitner, M.~Spira}
\renewcommand{\thesection}{\arabic{section}}

\input nmssmcalc/nmssmcalc.tex
\AddToContent{J.~Baglio, R.~Grober, M.~M\"uhlleitner,
  D.T.~Nhung, H.~Rzehak, M.~Spira, J.~Streicher, K.~Walz}
\renewcommand{\thesection}{\arabic{section}}

\renewcommand{\thesection}{\arabic{section}}
\renewcommand{\thesubsection}{\thesection.\arabic{subsection}}
\renewcommand{\thesubsubsection}{\thesubsection.\arabic{subsubsection}}
\renewcommand{\thefigure}{\arabic{figure}}
\renewcommand{\thetable}{\arabic{table}}
\renewcommand{\theequation}{\arabic{equation}}

\superpart{ The Higgs Boson }

\input htheo/HiggsFitTH.3.tex
\AddToContent{S.~Fichet, G.~Moreau}
\renewcommand{\thesection}{\arabic{section}}

\input hcouplings/hfit-LH.tex
\AddToContent{J.~Bernon, B.~Dumont, J.~F.~Gunion, S.~Kraml}
\renewcommand{\thesection}{\arabic{section}}

\input vbsh/wlwlLH.tex

\AddToContent{A.~Belyaev, E.~Boos, V.~Bunichev, 
 Y.~Maravin, A.~Pukhov, R.~Rosenfeld, M.~Thomas}
\renewcommand{\thesection}{\arabic{section}}

\input tth-lh-proc_final/tth.tex
\AddToContent{F.~Boudjema, R.~Godbole, D.~Guadagnoli, K.~Mohan}
\renewcommand{\thesection}{\arabic{section}}

\input hh/hh.tex
\AddToContent{A.~J.~Barr, M.~J.~Dolan, C.~Englert, M.~M.~M\"uhlleitner, M.~Spannowsky}
\renewcommand{\thesection}{\arabic{section}}

\input vbfhh/HH_VV_final.tex
\AddToContent{A.~Belyaev, O.~Bondu,  A.~Massironi, A.~Oliveira, R.~Rosenfeld, V.~Sanz}
\renewcommand{\thesection}{\arabic{section}}

\input vbfhh2/HHvbf.tex
\AddToContent{O.~Bondu, A.~Oliveira, R~.Contino, M.~Gouzevitch, A.~Massironi, J.~Rojo}
\renewcommand{\thesection}{\arabic{section}}

\input tgc/tgclep.tex
\AddToContent{A.~Falkowski, S.~Fichet, K.~Mohan, F.~Riva, V.~Sanz}
\renewcommand{\thesection}{\arabic{section}}

\renewcommand{\thesection}{\arabic{section}}
\renewcommand{\thesubsection}{\thesection.\arabic{subsection}}
\renewcommand{\thesubsubsection}{\thesubsection.\arabic{subsubsection}}
\renewcommand{\thefigure}{\arabic{figure}}
\renewcommand{\thetable}{\arabic{table}}
\renewcommand{\theequation}{\arabic{equation}}

\superpart{ Natural Models }

\input vlq/procVLQ.tex
\AddToContent{D.~Barducci, L.~Basso, A.~Belyaev, M.~Buchkremer, G.~Cacciapaglia, 
A.~Deandrea, T.~Flacke, J.H.~Kim, S.J.~Lee, S.H.~Lim, F.~Mahmoudi, 
L.~Panizzi, and J.~Ruiz-\'Alvarez}
\renewcommand{\thesection}{\arabic{section}}

\input compressed/compressed.tex
\AddToContent{B.~Fuks, P. Richardson, A.~Wilcock}
\renewcommand{\thesection}{\arabic{section}}

\input nsusy/nsusy_main.tex

\AddToContent{J.~Bernon, G.~Chalons, E.~Conte,
B.~Dumont, B.~Fuks, A.~Gaz, S.~Kraml, S.~Kulkarni, L.~Mitzka, S.~Pataraia, W.~Porod,
S.~Sekmen, D.~Sengupta, N.~Strobbe, W.~Waltenberger, F.~W\"urthwein, C.~Wymant}

\input lrmssm/lrmssm.tex
\AddToContent{A.~Alloul, L.~Basso, B.~Fuks, M. E.~Krauss, W.~Porod}
\renewcommand{\thesection}{\arabic{section}}

\input rho/draft_rho.tex
\AddToContent{R.~Contino, D.~Greco, C.~Grojean, D.~Liu, D.~Pappadopulo, A.~Thamm, R.~Torre,  A.~Wulzer}
\renewcommand{\thesection}{\arabic{section}}

\renewcommand{\thesection}{\arabic{section}}
\renewcommand{\thesubsection}{\thesection.\arabic{subsection}}
\renewcommand{\thesubsubsection}{\thesubsection.\arabic{subsubsection}}
\renewcommand{\thefigure}{\arabic{figure}}
\renewcommand{\thetable}{\arabic{table}}
\renewcommand{\theequation}{\arabic{equation}}

\superpart{ Flavour }

\input nmfvsusy/nmfvsusy.tex
\AddToContent{K.~De Causmaecker, B.~Fuks, B.~Herrmann, F.~Mahmoudi,
  B.~O'Leary, W.~Porod, S.~Sekmen, N.~Strobbe}
\renewcommand{\thesection}{\arabic{section}}

\input cpvmssm/cpvmssm.tex
\AddToContent{A.~Arbey, J.~Ellis, R.~M.~Godbole, F.~Mahmoudi}
\renewcommand{\thesection}{\arabic{section}}

\renewcommand{\thesection}{\arabic{section}}
\renewcommand{\thesubsection}{\thesection.\arabic{subsection}}
\renewcommand{\thesubsubsection}{\thesubsection.\arabic{subsubsection}}
\renewcommand{\thefigure}{\arabic{figure}}
\renewcommand{\thetable}{\arabic{table}}
\renewcommand{\theequation}{\arabic{equation}}

\superpart{ Dark Matter }

\input dmeft/DMeff.tex
\AddToContent{A. Arbey, M. Battaglia, G. B\'elanger, A. Goudelis, F. Mahmoudi, S. Pukhov}
\renewcommand{\thesection}{\arabic{section}}

\input nsusydm/NSUSYDM.tex
\AddToContent{D.~Barducci, S.~Belyaev, A.~Bharucha, W.~Porod and V.~Sanz}
\renewcommand{\thesection}{\arabic{section}}

\clearpage

\bibliography{LH_NewPhysics_Biblio}

\end{document}

%% file: Commands.tex
\addto\captionsenglish{}

\makeatletter
\def\myaddcontentsline@toc#1#2#3{%
   \addtocontents{#1}{\protect\thispagestyle{empty}}%
   \addtocontents{#1}{\protect \vspace{-.5cm}}
   \addtocontents{#1}{\protect\contentsline{#2}{#3}{}{}}
   \addtocontents{#1}{\protect \vspace{-.3cm}}}
\def\myaddcontentsline#1#2#3{%
  \expandafter\@ifundefined{addcontentsline@#1}%
  {\addtocontents{#1}{\protect \vspace{-.5cm}}
  \addtocontents{#1}{\protect\contentsline{#2}{#3}{}{}}
  \addtocontents{#1}{\protect \vspace{-.3cm}}}
  {\csname addcontentsline@#1\endcsname{#1}{#2}{#3}}{}}
\makeatother

\makeatletter
\def\mysuperaddcontentsline@toc#1#2#3{%
   \addtocontents{#1}{\protect\thispagestyle{empty}}%
   \addtocontents{#1}{\protect \vspace{.1cm}}
   \addtocontents{#1}{\protect\contentsline{#2}{#3}{\thepage}{}}
   \addtocontents{#1}{\protect \vspace{-.1cm}}}
\def\mysuperaddcontentsline#1#2#3{%
  \expandafter\@ifundefined{addcontentsline@#1}%
  {\addtocontents{#1}{\protect \vspace{.1cm}}
  \addtocontents{#1}{\protect\contentsline{#2}{#3}{\thepage}{}}
  \addtocontents{#1}{\protect \vspace{-.1cm}}}
  {\csname addcontentsline@#1\endcsname{#1}{#2}{#3}}{}}
\makeatother

\newcommand{\superpart}[1]{\clearpage \thispagestyle{empty}  \vphantom{blabla}\vspace{5cm} \centerline{\huge #1} 
\mysuperaddcontentsline{toc}{part}{\protect\numberline{} \hspace{-2cm}#1}%
\clearpage}
   
\newcommand{\AddToContent}[1]{\myaddcontentsline{toc}{chapter}{\protect  \textit{\hspace{1.5cm} \small#1}}}

\renewcommand{\thefigure}{\arabic{figure}}
\renewcommand{\thetable}{\arabic{table}}
\renewcommand{\theequation}{\arabic{equation}}

\catcode`\@=11
\font\manfnt=manfnt
\def\Watchout{\@ifnextchar [{\W@tchout}{\W@tchout[1]}}
\def\W@tchout[#1]{{\manfnt\@tempcnta#1\relax%
  \@whilenum\@tempcnta>\z@\do{%
    \char"7F\hskip 0.3em\advance\@tempcnta\m@ne}}}
\let\foo\W@tchout
\def\dubious{\@ifnextchar[{\@dubious}{\@dubious[1]}}

\def\@dubious[#1]{%
  \setbox\@tempboxa\hbox{\@W@tchout#1}
  \@tempdima\wd\@tempboxa
  \list{}{\leftmargin\@tempdima}\item[\hbox to 0pt{\hss\@W@tchout#1}]}
\def\@W@tchout#1{\W@tchout[#1]}
\catcode`\@=12

\newcommand{\leer}[1]{}

\newcommand{\mt}[1]{\ensuremath{m_{t'}}}
\newcommand{\mb}[1]{\ensuremath{m_{b'}}}
\newcommand{\me}[1]{\ensuremath{m_{\tau'}}}
\newcommand{\mv}[1]{\ensuremath{m_{\nu'}}}

%% file: authors.tex
\begin{center}
\textbf{G.~Brooijmans}$^{1}$, 
\textbf{R.~Contino}$^{2,3}$, %
\textbf{B.~Fuks}$^{3,4}$, %
\textbf{F.~Moortgat}$^{3}$, %
\textbf{P.~Richardson}$^{5}$, %
\textbf{S.~Sekmen}$^{3,6}$ %
\textbf{and}
\textbf{A.~Weiler}$^{3,7}$  %
\textbf{(convenors)}\\
A.~Alloul$^8$, 
A.~Arbey$^{3,9}$, 
J.~Baglio$^{10}$, 
D.~Barducci$^{11,12}$,
A.~J.~Barr$^{13}$, 
L.~Basso$^{4,14}$,
M.~Battaglia$^{3,15}$, 
G.~B\'elanger$^{16}$, 
A.~Belyaev$^{11,12}$, 
J.~Bernon$^{17}$, 
A.~Bharucha$^{18}$,
O.~Bondu$^{3}$,
F.~Boudjema$^{16}$, 
E.~Boos$^{19}$, 
M.~Buchkremer$^{20}$, 
V.~Bunichev$^{19}$, 
G.~Cacciapaglia$^{21}$, 
G.~Chalons$^{17}$,
E.~Conte$^{8}$,
M.~J.~ Dolan$^{22}$, 
A.~Deandrea$^{21}$, 
K.~De Causmaecker$^{3,23}$,
A.~Djouadi$^{24}$, 
B.~Dumont$^{17}$,
J.~Ellis$^{3,25}$,
C.~Englert$^{26}$, 
A.~Falkowski$^{24}$, 
S.~Fichet$^{27}$,
T.~Flacke$^{28}$, 
A.~Gaz$^{29}$,
M.~Ghezzi$^{30,31}$, 
R.~Godbole$^{32}$, 
A.~Goudelis$^{16}$,
M.~Gouzevitch$^{21}$, 
D.~Greco$^{2}$,
R.~Grober$^{10}$, 
C.~Grojean$^{33}$,
D.~Guadagnoli$^{16}$, 
J.~F.~Gunion$^{34}$,
B.~Herrmann$^{16}$,
J.~Kalinowski$^{35}$,
J.H.~Kim$^{28,36}$, 
S.~Kraml$^{17}$, 	  
M.~E.~Krauss$^{37}$, 
S.~Kulkarni$^{17}$, 
S.J.~Lee$^{28,38}$, 
S.H.~Lim$^{28,36}$,
D.~Liu$^{2,39}$, 
F.~Mahmoudi$^{3,40}$,
Y.~Maravin$^{41}$,
A.~Massironi$^{42}$, 
L.~Mitzka$^{37}$, 
K.~Mohan$^{32}$,
G.~Moreau$^{24}$,
M.~M.~M\"uhlleitner$^{10}$, 
D.T.~Nhung$^{10}$,
B.~O'Leary$^{37}$,
A.~Oliveira$^{21,43}$, 
L.~Panizzi$^{11,12}$,
D.~Pappadopulo$^{44}$, 
S.~Pataraia$^{45}$, 
W.~Porod$^{37}$,
A.~Pukhov$^{19}$,
F.~Riva$^{2}$, 
J.~Rojo$^{3}$,
R.~Rosenfeld$^{43}$,
J.~Ruiz-\'Alvarez$^{21}$,
H.~Rzehak$^{14}$, 
V.~Sanz$^{46}$,
D.~Sengupta$^{17,47}$, 
M.~Spannowsky$^5$,
M.~Spira$^{48}$, 
J.~Streicher$^{10}$,
N.~Strobbe$^{49}$, 
A.~Thamm$^{2}$, 
M.~Thomas$^{11}$,
R.~Torre$^{50,51}$,  
W.~Waltenberger$^{52}$, 
K.~Walz$^{10}$,
A.~Wilcock$^5$,
A.~Wulzer$^{50}$,
F.~W\"urthwein$^{53}$, 
C.~Wymant$^{16}$
\end{center}

%% file: addresses.tex
{\footnotesize
$^1$ Physics Department, Columbia University, New York, NY 10027, USA\\ 
$^{2}$ Institut de Th\'eorie des Ph\'enom\`enes Physiques, EPFL,1015 Lausanne, Switzerland \\
$^3$ Physics Department, CERN, CH-1211 Geneva 23, Switzerland\\
$^4$Institut Pluridisciplinaire Hubert Curien/D\'epartement
    Recherches Subatomiques,\\ Universit\'e de Strasbourg/CNRS-IN2P3,
    23 rue du Loess, F-67037 Strasbourg, France\\
$^5$ Institute for Particle Physics Phenomenology, University of Durham,
    Durham, DH1 3LE, United Kingdom\\
$^6$ Department of Physics, Florida State University, Tallahassee, Florida 32306, USA\\
$^7$ Deutsches Elektronen-Synchrotron DESY, D-22607 Hamburg, Germany\\
$^8$ Groupe de Recherche de Physique des Hautes \'Energies (GRPHE), 
    Universit\'e de Haute-Alsace, IUT Colmar, 34 rue du Grillenbreit BP 50568, 68008 Colmar Cedex, France\\
$^{9}$ Centre de Recherche Astrophysique de Lyon, Observatoire de Lyon,
Saint-Genis Laval Cedex, F-69561, France; CNRS, UMR 5574;\\
Ecole Normale Sup\'erieure de Lyon, France;\\
Universit\'e de Lyon, Universit\'e Lyon 1, F-69622 Villeurbanne Cedex, France\\
$^{10}$ Institut f\"ur Theoretische Physik, Karlsruher Institut f\"ur Technologie KIT, 76131 Karlsruhe, Germany\\
$^{11}$ School of Physics and Astronomy, University of Southampton, Highfield, Southampton SO17 1BJ, UK\\
$^{12}$ Particle Physics Department, Rutherford Appleton Laboratory, Chilton, Didcot, Oxon OX11 0QX, UK\\
$^{13}$ Denys Wilkinson Building, Department of Physics, Oxford, OX1 3RH, UK\\
$^{14}$ Physikalisches Institut, Albert-Ludwigs-Universit\"at Freiburg D-79104 Freiburg, Germany\\
$^{15}$ University of California at Santa Cruz,
Santa Cruz Institute of Particle Physics, CA 95064, USA\\
$^{16}$ LAPTh, Univ. de Savoie, CNRS, 9 Chemin de Bellevue, B.P. 110, Annecy-le-Vieux 74941, France\\
$^{17}$ LPSC, Universit\'e Grenoble-Alpes, CNRS/IN2P3, 53 Avenue des Martyrs, F-38026 Grenoble, France\\
$^{18}$ Physik Department T31, Technische Universit\"at M\"unchen, James-Franck-Stra\ss e~1, D-85748 Garching, Germany\\
$^{19}$ Skobeltsyn Institute of Nuclear Physics, Moscow State University, Moscow 119991, Russia \\
$^{20}$ Centre for Cosmology, Particle Physics and Phenomenology (CP3), Universit\'e catholique de Louvain, Chemin du Cyclotron, 2, B-1348, Louvain-la-Neuve, Belgium\\
$^{21}$ Universit\'e de Lyon, F-69622 Lyon, France; Universit\'e Lyon 1, Villeurbanne; CNRS/IN2P3, \\UMR5822, Institut de Physique Nucl\'eaire de Lyon, F-69622 Villeurbanne Cedex, France\\
$^{22}$ Theory Group, SLAC National Accelerator Laboratory, Menlo Park, CA 94025, USA\\
$^{23}$ Theoretische Natuurkunde, IIHE/ELEM and International Solvay Institutes, Vrije Universiteit Brussel,
  Pleinlaan 2, B-1050 Brussels, Belgium\\
$^{24}$  Laboratoire de Physique Th\'eorique, CNRS -- UMR 8627,  Universit\'e de Paris-Sud 11, F-91405 Orsay Cedex, France\\
$^{25}$ Theoretical Particle Physics and Cosmology Group, Department of Physics, King's College London, London WC2R 2LS, UK\\
$^{26}$ SUPA, School of Physics and Astronomy, University of Glasgow,
Glasgow G12 8QQ, UK\\
$^{27}$ International Institute of Physics, UFRN, 
Av. Odilon Gomes de Lima, 1722 - Capim~Macio - 59078-400 - Natal-RN, Brazil \\
$^{28}$ Department of Physics, Korea Advanced Institute of Science and Technology, 335 Gwahak-ro, Yuseong-gu, Daejeon 305-701, Korea \\
$^{29}$ University of Colorado,  Boulder, CO 80309-0390, USA\\
$^{30}$ Dipartimento di Fisica, Universit\`a di Torino and INFN, Torino, Italy\\
$^{31}$ Dipartimento di Fisica, Universita di Roma La Sapienza" and INFN, Sezione di Roma, Italy\\
$^{32}$ Centre for High Energy Physics, Indian Institute of Science, Bangalore, India\\
$^{33}$ ICREA at IFAE, Universitat Aut\`onoma de Barcelona, E-080193 Bellaterra, Spain\\
$^{34}$ Department of Physics, University of California, Davis, CA 95616, USA\\
$^{35}$ Instytut Fizyki Teoretycznej UW, Hoza 69, PL-00681 Warsaw, Poland\\
$^{36}$ Center for Theoretical Physics of the Universe, IBS, Daejeon, Korea \\
$^{37}$ Institut f\"ur Theoretische Physik und Astrophysik,
    Universit\"at  W\"urzburg, 97074  W\"urzburg, Germany\\
$^{38}$ School of Physics, Korea Institute for Advanced Study, Seoul 130-722, Korea\\
$^{39}$ State Key Laboratory of Theoretical Physics, Institute of Theoretical Physics, Chinese Academy of Sciences, Beijing, People's Republic of China \\
$^{40}$ Clermont Universit\'e, Universit\'e Blaise Pascal, CNRS/IN2P3, LPC, BP 10448, F-63000 Clermont-Ferrand, France\\
$^{41}$ Physics Department, Kansas State University, Manhattan, KS 66506, USA.\\
$^{42}$ Northeastern University, Boston, MA -  USA\\
$^{43}$ ICTP South American Institute for Fundamental Research \& 
Instituto de F\'{\i}sica Te\'orica \\
UNESP - Universidade Estadual Paulista - 
Rua Dr. B. T. Ferraz 271, 01140-070, S\~ao Paulo, SP, Brazil\\
$^{44}$ Department of Physics, University of California, Berkeley, USA and Theoretical Physics Group, Lawrence Berkeley National Laboratory, Berkeley, USA\\
$^{45}$ Bergische Universit\"at Wuppertal,  42119 Wuppertal, Germany\\
$^{46}$ Department of Physics and Astronomy, University of Sussex, Brighton BN1 9QH, UK \\
$^{47}$ Department of High Energy Physics, Tata Institute of Fundamental Research,
        1 Homi Bhabha Road,  Mumbai 400005, India. \\
$^{48}$ Paul Scherrer Institut, CH-5232 Villigen PSI, Switzerland\\
$^{49}$ Department of Physics and Astronomy, Ghent University, Belgium\\
$^{50}$ Dipartimento di Fisica e Astronomia, Universit\`a di Padova and INFN, Padova, Italy \\
$^{51}$ SISSA, Trieste, Italy\\
$^{52}$ HEPHY Vienna, \"OAW, Nikolsdorfer Gasse 18, 1050 Wien, Austria\\
$^{53}$ University of California, San Diego, 9500 Gilman Drive, La Jolla, CA 92093, USA
}

%% file: intro.tex
\noindent {\Large {\bf Introduction}}
\vspace{.5cm}

{\it G.~Brooijmans, R.~Contino, B.~Fuks, F.~Moortgat, P.~Richardson, S.~Sekmen, A.~Weiler}
\vspace{.5cm}

This document is the report of the New Physics
session of the 2013 Les Houches Workshop ``Physics at TeV Colliders''.  
The discovery of a Higgs boson in the first run of the Large Hadron Collider (LHC)
has led to a substantial refocusing of the searches for new physics.
As a consequence, the leading search areas now lie in the investigation
of deviations from the
Standard Model predictions in Higgs boson production and decay, as well as in the
probe of so-called ``natural'' models in which new
physics eliminates, or at the very least attenuates, the hierarchy problem introduced by the 
existence of a Higgs boson at the electroweak scale.  During the
workshop, which brings together theorists and experimenters, a substantial number of 
ideas around these topics were discussed, and for a number of these in-depth studies were initiated.  
This  report describes the results of those studies.

A first section presents progress specific to the software tools crucial in predicting 
Higgs boson properties in various models.  The first contribution details new ingredients
to {\sc HDecay}, a widely used program allowing one to calculate Higgs boson decay widths and branching ratios
both in and beyond the Standard Model.  The two other contributions in the ``Tools'' section of this document
introduce packages dedicated to the computation of Higgs boson properties in an effective Lagrangian approach for
the introduction of new physics effects, as well as in the specific case of the Next-to-Minimal Supersymmetric
Standard Model.

The second section of this report shows various examples of how to use the existence of the
Higgs boson to probe new physics.  Its first
contribution weighs the relative importance of experimental and theoretical uncertainties
in extracting the Higgs boson couplings to Standard Model particles.
The next three contributions exploit single-Higgs production: the
first of these addresses a way to assess the
custodial symmetry and CP properties primarily in the $HWW$ and $HZZ$ vertices, the second one
investigates a possible measurement
of the contributions of longitudinal and transverse vector boson polarizations
in Higgs boson production via vector boson fusion, and
the third one analyzes the $t\bar{t}H$ coupling. Next, three studies focus on Higgs-boson 
pair production: the first of these assumes a non-resonant scenario,
whereas the second and third ones postulate a resonant production channel in the vector boson fusion
mode. The last contribution in this section re-examines constraints on triple gauge
couplings obtained at the LEP collider in a form that can be readily combined with direct Higgs boson results.

In the third section, multiple ``natural'' models of new physics are studied.  Vector-like quarks are predicted 
in many new physics scenarios, and the first contribution in this section sets up three simplified models that
encapsulate all the relevant vector-like quark phenomenology for the LHC.  A second contribution studies the 
phenomenology of vector resonances in composite Higgs theories, converging on a simple benchmark model for 
searches at colliders.  The other three contributions examine specific supersymmetric
(SUSY) scenarios.
One investigates monotop signatures for cases where sparticle masses are close to the electroweak scale, but
have escaped detection because the SUSY spectrum is compressed. A second one
examines how existing searches for
stops and sbottoms can be used to form a coherent picture of constraints on third generation squarks, and a third
one takes as an example left-right SUSY to investigate how results of searches for new charged gauge
bosons should be presented
to extract constraints on various models.

Section four tackles flavour and CP violation.  One contribution evaluates the constraints on non-minimal
flavour violating effects from the Higgs boson discovery and several flavour and electroweak observables,
and a second one uses the data from the Higgs boson discovery, direct dark matter searches and 
electric dipole moment (EDM) constraints
in determining how CP-violating effects in the Minimal Supersymmetric Standard Model could be discovered.

Finally, the fifth section studies the interplay between direct dark matter searches and collider physics.
This includes a contribution on using dark matter effective field theory in interpreting dark matter 
searches at colliders, and a second one on the interplay between direct dark matter detection and the LHC in 
the specific case of the far focus point of natural SUSY.

The meeting in Les Houches fostered a large number of discussions between theorists and
experimenters, but, as mentioned above, in-depth studies could only be completed for a
number of the generated ideas on the required timescale. It is clear however that even those
that could not converge to a written contribution have paid off through the breadth of 
searches conducted by experimenters, their ways of presenting their results, and 
theorists' further understanding of the constraints imposed on experiments.
We expect that many more future results will 
benefit from the discussions held at the workshop.

%% file: hdecay/hdecay.tex

\chapter{Updates and Extensions of the Program HDECAY}

{\it A.~Djouadi, J.~Kalinowski, M.~M\"uhlleitner, M.~Spira}



\begin{abstract}
The program {\tt HDECAY} determines the decay widths and branching ratios of
the Higgs bosons within the Standard Model (with 3 and 4 generations)
and its minimal supersymmetric extension, including the dominant
higher-order corrections. New theoretical developments are briefly
discussed and the new ingredients incorporated in the program are
summarized.
\end{abstract}

\section{INTRODUCTION}
The search strategies for Higgs bosons searches at LEP, Tevatron, LHC and future
$e^+e^-$ linear colliders (LC) exploit various Higgs boson decay
channels. The strategies depend not only on the experimental setup
(hadron versus lepton colliders) but also on the theoretical scenarios:
the Standard Model (SM) or some of its extensions such as the Minimal
Supersymmetric Standard Model (MSSM) or variants as e.g.  including a
4th generation. It is of vital importance to have reliable predictions
for the branching ratios of the Higgs boson decays for these theoretical
models.

The current version of the program {\tt HDECAY}
\cite{Djouadi:1997yw,Djouadi:2006bz} can be used to calculate Higgs
boson partial decay widths and branching ratios within the SM with 3 and
4 generations, the MSSM and fermiophobic Higgs models and includes:
\\[0.3cm]
-- All decay channels that are kinematically allowed and which have
branching ratios larger than $10^{-4}$, i.e.~the loop mediated, the
three body decay modes and in the MSSM the cascade and the
supersymmetric decay channels
\cite{Spira:1997dg,Djouadi:2005gi,Djouadi:2005gj}. \\[0.3cm]
-- All relevant higher-order QCD corrections to the decays into quark
pairs and to the loop mediated decays into gluons are incorporated
\cite{Djouadi:1995gt}. \\[0.3cm]
-- Double off--shell decays of the CP--even Higgs bosons into massive
gauge bosons which then decay into four massless fermions, and all
important below--threshold three-- and four--body decays \cite{Djouadi:1995gv}.
\\[0.3cm]
-- In the MSSM, the complete radiative corrections in the effective
potential approach with full mixing in the stop/sbottom sectors; it uses
the renormalization group improved values of the Higgs masses and
couplings and the relevant next--to--leading--order corrections are
implemented
\cite{Carena:1995wu,Haber:1996fp,Carena:2000dp,Degrassi:2002fi}.
\\[0.3cm]
-- In the MSSM, all the decays into supersymmetric (SUSY) particles (neutralinos,
charginos, sleptons and squarks including mixing in the stop, sbottom
and stau sectors) when they are kinematically allowed
\cite{Djouadi:1992pu,Djouadi:1996mj,Djouadi:1996pj}. The SUSY particles
are also included in the loop mediated $\gamma \gamma$ and $gg$ decay
channels. \\[0.3cm]
The program, written in FORTRAN, provides a very flexible and convenient
use, fitting to all options of phenomenological relevance. The basic
input parameters, fermion and gauge boson masses and their total widths,
coupling constants and, in the MSSM, soft SUSY-breaking parameters can
be chosen from an input file. In this file several flags allow switching
on/off or changing some options [{\it e.g.} choosing a particular Higgs
boson, including/excluding the multi-body or SUSY decays, or
including/excluding specific higher-order QCD corrections].

\section{UPDATES}
Since the release of the original version of the program a number of
improvements and new theoretical calculations have been implemented. The
following points summarize the most important modifications of {\tt
HDECAY} after its release and beyond the updates summarized in
Ref.~\cite{Butterworth:2010ym}: \\[0.3cm]
-- Inclusion of the full mass effects to $H\to gg,\gamma\gamma$ at
   next-to-leading order in QCD within the Standard Model
   \cite{Spira:1995rr}.
   \\[0.3cm]
-- Inclusion of the leading electroweak corrections to all effective down-type
   fermion Yukawa couplings, i.e.~for the $\mu,\tau,s,b$ according to
   \cite{Hempfling:1993kv,Hall:1993gn,Carena:1994bv,Pierce:1996zz,
   Carena:1998gk}. In this context the sneutrino masses of the first
   two generations are allowed to be different from the third generation.
   \\[0.3cm]
-- Inclusion of the two-loop QCD corrections to the top decays
   \cite{Czarnecki:1998qc,Chetyrkin:1999ju,Blokland:2004ye,Blokland:2005vq,
   Czarnecki:2010gb,Gao:2012ja,Brucherseifer:2013iv}.
   \\[0.3cm]
-- Inclusion of the full CKM mixing effects in charged Higgs and top
   decays. This required the appropriate extension of the {\tt hdecay.in}
   input file.
   \\[0.3cm]
-- Inclusion of running mass effects and $\Delta_{b/s}$ corrections to
   the Yukawa couplings in charged Higgs decays into $b$ and $s$ quarks,
   where $\Delta_{b/s}$ denotes the leading SUSY-QCD and SUSY-electroweak
   corrections to the effective bottom/strange Yukawa couplings.
   \\[0.3cm]
-- Addition of the charged Higgs decays $H^+\to t\bar d/ t\bar s/ c\bar
   d$.
   \\[0.3cm]
-- Inclusion of charm loop contributions in the gluonic Higgs decays,
   $\phi\to gg$, for the SM and MSSM.
   \\[0.3cm]
-- Inclusion of bottom mass effects and double off-shell decays in
   $H\to hh/AA/AZ/H^+W^-/t\bar t$, $A\to hZ$ and $H^+\to t\bar b$.
   \\[0.3cm]
-- Extension of {\tt HDECAY} to the general Two Higgs Doublet model (2HDM)
   \cite{Harlander:2013qxa}. This required the extension of the
   {\tt hdecay.in} input file and the inclusion of several new decay
   modes that are not possible within the MSSM. The input file allows
   to work with two different set-ups for the 2HDM.
   \\[0.3cm]
-- Inclusion of rescaled Higgs couplings to SM particles according to
the effective interaction Lagrangian
\begin{eqnarray}
{\cal L}_{int} & = & -\sum_\psi c_\psi m_\psi \bar\psi \psi \frac{H}{v}
+ 2 c_W m_W^2 W^{+\mu} W^-_\mu \frac{H}{v} + c_Z m_Z^2 Z^{\mu} Z_\mu
\frac{H}{v} \nonumber \\
& + & \left\{ \frac{\alpha_s}{8\pi} c_{gg} G^{a\mu\nu}G^a_{\mu\nu}
+ \frac{\alpha}{8\pi} c_{\gamma\gamma} F^{\mu\nu}F_{\mu\nu}
+ \frac{\sqrt{\alpha \alpha_2}}{4\pi} c_{Z\gamma}
F^{\mu\nu}Z_{\mu\nu}\right\} \frac{H}{v}
\end{eqnarray}
where $G^{a\mu\nu}$, $F^{\mu\nu}$ and $Z^{\mu\nu}$ are the field
strength tensors of the gluon, photon and $Z$-boson fields. The
couplings $\alpha$, $\alpha_2$ and $\alpha_s$ are the electromagnetic
(in the Thompson limit), isospin ($g^2 = 4 \pi \alpha_2$) and strong
couplings, respectively, $v$ is the Higgs vacuum expectation value and
$H$ the Higgs boson field. Note that we added novel point-like
couplings of the Higgs boson to gluons, photons and $Z$ bosons
affecting the Higgs decays $H\to gg/\gamma\gamma/Z\gamma$. Electroweak
corrections are only kept in the SM part of the individual decay
amplitudes, i.e.~the parts for $c_\psi=c_W=c_Z=1$ and
$c_{gg}=c_{\gamma\gamma}=c_{Z\gamma}=0$, while QCD corrections have been
included in all parts of the decays widths, since the dominant parts
factorize. This approach deviates from the general addition of dimension-six
operators as pursued in Ref.~\cite{Contino:2013kra} where additional
tensor structures have been added at the dimension-six level.

The above rescaling of the Higgs couplings modifies e.g.~the Higgs decay
widths into quarks as
\begin{equation}
\Gamma(H\to q\bar q) = \frac{3G_FM_H}{4\sqrt{2}\pi}
\overline{m}_q^2(M_H) c_b \left\{ c_b + \delta_{elw} \right\} \left\{
1+\delta_{QCD} + \frac{c_t}{c_b} \delta_t \right\}
\end{equation}
where $\delta_{elw}$ denotes the electroweak corrections, $\delta_{QCD}$ the
pure QCD corrections and $\delta_t$ the top-quark induced QCD
corrections with the latter involving the top Yukawa coupling instead of
the bottom one. The coefficient is expressed in terms of the Fermi
constant $G_F$, the Higgs mass $M_H$ and the running $\overline{\rm MS}$
bottom mass $\overline{m}_q$ at the scale of the Higgs mass.

The gluonic Higgs decay, taken as an example of a case with a novel tensor
structure involving the point-like coupling factor $c_{gg}$, is given by
\begin{eqnarray}
\Gamma(H\to gg)  &=& \frac{G_F\alpha_s^2 M_H^3}{36\sqrt{2}\pi^3}
\Bigg[ 
 \bigg| \sum_{Q=t,b,c} c_Q \, A_Q\left(\tau_Q\right)
 \bigg|^2  c_{eff}^2 \, \kappa_{soft}  \nonumber \\
&& + \delta_{elw} \left( \sum_{Q,Q'=t,b,c} c_Q \, A_Q\left(\tau_Q\right)
A_Q^*\left(\tau_{Q'}\right) \right)
 c_{eff}^2 \, \kappa_{soft}  \nonumber \\
&& + 2\, \mathrm{Re}\!\left( \sum_{Q=t,b,c} c_Q \,
A^*_Q\left(\tau_Q\right)
\frac{3}{2} c_{gg} \right) c_{eff} \, \kappa_{soft}  
    + \left|\frac{3}{2} c_{gg}\right|^2
    \kappa_{soft}  \\
&& + \sum_{Q,Q'=t,b} c_Q\, A^*_Q\left(\tau_Q\right)
c_{Q'}\, A_Q\left(\tau_{Q'}\right) \kappa^{NLO} (\tau_Q,\tau_{Q'})
\Bigg] \, , \nonumber 
\end{eqnarray}
where $\tau_Q = 4 m_Q^2 /M_H^2$ and $\delta_{elw}$ denotes the
electroweak corrections \cite{Aglietti:2004nj,Aglietti:2004ki,
Degrassi:2004mx,Actis:2008ug,Actis:2008ts}. The loop function
$A_Q(\tau_Q)$ is normalized to unity for large quark masses and can be
found in Ref.~\cite{Spira:1995rr}. The contributions $c_{eff}$ and
$\kappa_{soft}$ denote the QCD corrections originating from the
effective Lagrangian in the heavy top quark limit,
\begin{equation}
{\cal L}_{eff} = c_{eff}~\frac{\alpha_s}{12\pi} G^{a\mu\nu}G^a_{\mu\nu}
\frac{H}{v}
\end{equation}
and the residual corrections due to diagrams involving gluon exchange
and light-quark contributions, respectively. They are included up to
the next-to-next-to-next-to-leading order (NNNLO) \cite{Inami:1982xt,Djouadi:1991tka,Chetyrkin:1997iv,
Chetyrkin:1997un,Kramer:1996iq,Schroder:2005hy,Chetyrkin:2005ia,
Baikov:2006ch}. At the next-to-leading order (NLO), they are given by
\cite{Inami:1982xt,Djouadi:1991tka},
\begin{equation}
c_{eff} = 1 + \frac{11}{4}\, \frac{\alpha_s}{\pi}\; , \qquad \qquad
\kappa_{soft} = 1 + \left( \frac{73}{4} - \frac{7}{6} N_F \right)
\frac{\alpha_s}{\pi}
\end{equation}
with $N_F=5$ light quark flavours.  Finally $\kappa^{NLO}$ represents
the finite top and bottom mass effects at NLO beyond the limit of heavy
quarks, i.e.~beyond the terms contained in $c_{eff}$ and $\kappa_{soft}$
\cite{Spira:1995rr}.

All other Higgs decay modes are treated analogously in the case of
rescaled Higgs couplings.
\\[0cm]

The logbook of all modifications and the most recent version of the
program can be found on the web page
\url{http://people.web.psi.ch/spira/proglist.html}. \\



%% file: ehdecay/ehdecay.tex

\chapter{eHDECAY - a Fortran Code for the Computation of Higgs Decays in
the Effective Lagrangian Approach}

{\it R.~Contino, M.~Ghezzi, C.~Grojean, M.~M\"uhlleitner, M.~Spira}




\begin{abstract}
We present the Fortran code {\tt eHDECAY}. It is based on a
modification of the program {\tt HDECAY}~\cite{Djouadi:1997yw,Djouadi:2006bz}, in which the full list of 
leading bosonic operators of the Higgs effective Lagrangian has been
implemented. This has been done for a linear and a non-linear
realization of the electroweak symmetry and for two benchmark composite Higgs
models. In the decay widths all the relevant QCD corrections have been
included. The electroweak corrections on the other hand can only be
implemented  in a consistent way for the linear realization in the
vicinity of the Standard Model (SM).
\end{abstract}

\section{INTRODUCTION}
After the discovery of a new boson by the ATLAS~\cite{Aad:2012tfa} and 
CMS~\cite{Chatrchyan:2012ufa}  collaborations,
any hint of the existence of new additional  particles is still lacking. 
The approach of the Higgs sector in terms of an effective
Lagrangian allows us to parametrize our ignorance of New Physics (NP) beyond
the SM (BSM), and thereby to describe the properties of the new 
boson and to investigate its nature. In Ref.~\cite{Contino:2013kra} we have
reviewed in detail the 
low-energy effective Lagrangian for a light Higgs-like boson. In order
to investigate the effective Lagrangian beyond tree-level a multiple
expansion has been performed in the SM coupling parameter $\alpha/\pi$
and in powers of $E/M$, with $E$ being the energy of the process and
$M$ the NP scale, where new massive states appear. If the
Higgs-like boson is part of a weak doublet there is an additional
expansion parameter $v/f \ll 1$ with $f\equiv M/g_\star$ and $g_\star$
the typical NP coupling. The weak scale is defined in terms of the
Fermi constant $G_F$ by $v\equiv 1/(\sqrt{2} G_F)^{1/2} \approx
246$\,GeV. In Ref.~\cite{Contino:2014xyz} the relation between the 
non-linear and the linear effective Lagrangian approach has been
discussed. Furthermore, the implementation in the Fortran code {\tt
  eHDECAY} has been presented in detail. In this contribution,
using the example of the Higgs boson decay into two gluons, the
importance of the higher order QCD corrections and of the mass effects
in the corrections shall be discussed.  
The program can be downloaded from the url: 
\url{http://www.itp.kit.edu/~maggie/eHDECAY/}.  

\section{HIGHER ORDER CORRECTIONS AND MASS EFFECTS}
As the leading part of the QCD corrections in general factorizes with respect to the
expansion in the number of fields and derivatives of the effective
Lagrangian, they can be included by taking over the results from the
SM. The electroweak (EW) corrections, on the contrary, require
dedicated computations that are 
not available at present. They can only be implemented in the
framework of the Strongly Interacting Light Higgs Lagrangian (SILH)~\cite{Giudice:2007fh},
 in which the coupling deviations from the SM are small, 
and up to orders $v/f$. In the following the higher order QCD
corrections and the mass effects shall be discussed for the non-linear
implementation. We denote by $c_\psi$ the modification of the Higgs
couplings to fermions in terms of the SM coupling and by $c_{gg}$ the
effective Higgs coupling to gluons. Hence, with $h$ denoting the
scalar field, the related effective Lagrangian reads
\begin{eqnarray}
{\cal L}_{\psi,G} = - \sum_{\psi=u,d,l} c_\psi \bar{\psi} \psi \frac{h}{v} +
\frac{c_{gg}}{2} G^a_{\mu\nu} G^{a\mu\nu} \frac{h}{v}\;.
\end{eqnarray}
Here $G^a_{\mu\nu}$ denotes the field strength tensor for gluons,
\begin{eqnarray}
G_{\mu\nu}^a = \partial_\mu G^a_\nu - \partial_\nu G^a_\mu + g_S
f^{abc} G_\mu^b G_\nu^c \;, \qquad a,b,c=1,...,8 \;,
\end{eqnarray}
with the strong coupling constant $g_S$ and the gluon field
$G_\mu$. The decay rate into gluons implemented in {\tt eHDECAY} in
the framework of the non-linear Lagrangian is then given by
\begin{eqnarray}
\Gamma(gg)\big|_{NL}  &=& \frac{G_F\alpha_s^2 m_h^3}{4\sqrt{2}\pi^3} \Bigg[ 
 \bigg| \sum_{q=t,b,c} \frac{c_q}{3} \, A_{1/2}\left(\tau_q\right)
 \bigg|^2  c_\textit{\scriptsize eff}^2 \, \kappa_\textit{\scriptsize soft}  \nonumber \\
&& + 2\, \mathrm{Re}\!\left( \sum_{q=t,b,c} \frac{c_q}{3} \, A^*_{1/2}\left(\tau_q\right) \frac{2 \pi c_{gg}}{\alpha_s} \right) c_\textit{\scriptsize eff} \, \kappa_\textit{\scriptsize soft}  
    + \left|\frac{2\pi c_{gg}}{\alpha_s}\right|^2
    \kappa_\textit{\scriptsize soft}  \label{ehdecay_label1} \\
&& + \frac{1}{9} \sum_{q,q'=t,b} c_q\, A^*_{1/2}\left(\tau_q\right)
c_{q'}\, A_{1/2}\left(\tau_{q'}\right) \kappa^\textit{\scriptsize NLO} (\tau_q,\tau_{q'})
\Bigg] \, , \nonumber 
\end{eqnarray}
where $\tau_q = 4 m_q^2 /m_h^2$ and the loop function
\begin{eqnarray}
A_{1/2} (\tau) = \frac{3}{2} \tau [1+(1-\tau) f(\tau)] \;,
\end{eqnarray}
which is normalized to 1 in the limit of large quark masses. In the
decay width we use the pole masses for the top, bottom and charm
quarks, $m_t=172.5$\,GeV, $m_b=4.75$\,GeV and $m_c=1.42$\,GeV, and
$\alpha_s$ is computed up to the next-to-next-to-next-to-leading order (N$^3$LO)
at the scale $m_h$ for $N_F=5$
active flavours, $\alpha_s=0.114$. The function $f(\tau)$ is given by 
\begin{equation}
 f\left(\tau\right)=\left\lbrace \begin{array}{ll}
\displaystyle \arcsin^2\frac{1}{\sqrt{\tau}} & \quad \tau\geq1 \\[0.5cm]
\displaystyle -\frac{1}{4}\left[\ln\frac{1+\sqrt{1-\tau}}{1-\sqrt{1-\tau}}-i\pi\right]^2 & \quad \tau<1\, . \end{array} \right. 
\end{equation}
The QCD corrections have been taken into account up to
N$^3$LO QCD in the limit of heavy
loop-particle masses. The effect from soft gluon radiation, given by
the coefficient $\kappa_\textit{\scriptsize soft}$, factorizes in this limit. The
corrections from hard gluon and hard quark exchange with virtuality
$q^2 \gg m_t^2$ are encoded in the coefficient $c_\textit{\scriptsize eff}$. Namely, for
$m_h \ll 2m_t$  the top quark can be integrated out, leading to the
effective five-flavour Lagrangian 
\begin{eqnarray}
{\cal L}_{\mathrm{eff}} = -2^{1/4} G_F^{1/2} C_1 G^0_{a\mu\nu}
G^{0\mu\nu}_a h \;.
\end{eqnarray}
The superscript 0 denotes the bare fields. The dependence on the top
quark mass $m_t$ is included in the coefficient function $C_1$. We
then have for $\kappa_\textit{\scriptsize soft}$ and $c_\textit{\scriptsize eff}$,
\begin{eqnarray}
\kappa_\textit{\scriptsize soft}   &=&   \frac{\pi}{2 m_h^4} \, \mathrm{Im}\,\Pi^{GG} (q^2
= m_h^2) \\[0.2cm] 
c_\textit{\scriptsize eff} &=& -\frac{12 \pi \, C_1}{\alpha_s^{(5)} (m_h)} \, ,
\end{eqnarray}
with the vacuum polarization $\Pi^{GG} (q^2)$ induced by the gluon
  operator. The N$^3$LO expressions for $C_1$~\cite{Chetyrkin:1997un,Kramer:1996iq,Schroder:2005hy,Chetyrkin:2005ia}
  in the on-shell 
  scheme and for $\mathrm{Im}\,\Pi^{GG}$ have been given in Ref.~\cite{Baikov:2006ch}. The next-to-leading order (NLO) expressions for
  $\kappa_\textit{\scriptsize soft}$ and $c_\textit{\scriptsize eff}$ read~\cite{Inami:1982xt,Djouadi:1991tka,Chetyrkin:1997iv} 
\begin{eqnarray}
\kappa_\textit{\scriptsize soft}^\textit{\scriptsize NLO} = 1 + \frac{\alpha_s}{\pi} \left(\frac{73}{4} - \frac{7}{6} N_F \right)\, , \qquad
c_\textit{\scriptsize eff}^\textit{\scriptsize NLO} = 1 + \frac{\alpha_s}{\pi} \frac{11}{4} \, ,
\end{eqnarray}
in agreement with the low-energy theorem~\cite{Ellis:1975ap,Shifman:1979eb,Kniehl:1995tn}. 
Here $\alpha_s$ is evaluated at the scale $m_h$ and computed for
$N_F=5$ active flavours. In {\tt eHDECAY} it is consistently computed
up to N$^3$LO.  
The additional mass effects at NLO~\cite{Spira:1995rr} in
the top and bottom loops are taken into account by the function
$\kappa^\textit{\scriptsize NLO} (\tau_q, \tau_{q^\prime})$, in the last line of
Eq.~(\ref{ehdecay_label1}). This function quantifies the difference
between the NLO QCD corrections for the top (bottom) contribution taking into
account finite mass effects in the loop, and the result for the top
(bottom) contribution in the limit of a large loop particle mass. These mass
effects shall be discussed in the following. 

In order to investigate the mass effects in the NLO QCD corrections we
choose a scenario for $m_h=125$\,GeV with SM bottom- and
charm-couplings, {\it i.e.} $c_b=c_c=1$ and a modified top quark
coupling $c_t=0.85$. Note, that at the LHC with
$300$~fb$^{-1}$ the top quark coupling can be
determined with a precision of about 15\%
\cite{CMS:2013xfa,ATLAS:2013hta,Dawson:2013bba}. 
Furthermore, we set the effective Higgs-gluon-gluon
coupling to zero, $c_{gg}=0$. In Table~\ref{ehdecay_label2} we show at
leading order (LO) and at N$^3$LO the total Higgs decay width into
gluons, as well as the individual contributions from the top- and
bottom-loops, and the top-bottom interference. Note, that the total
gluonic decay width includes the charm loop contribution as well. 
\begin{table}[h]
\renewcommand{\arraystretch}{1.2}
\begin{center}
\begin{tabular}{|c|c|c|c||c|}
\hline 
     & {\small $\Gamma_{tt}$ [GeV]} & {\small $\Gamma_{bb}$ [GeV]} &
       {\small $\Gamma_{tb}$ [GeV]} &
      {\small $\Gamma^{gg}_{\mbox{\scriptsize tot}}$ [GeV]}\\ \hline
{\small LO} & 1.433 $\times 10^{-4}$ & 2.174 $\times 10^{-6}$ & -2.028
$\times 10^{-5}$ & 1.217 $\times 10^{-4}$ \\ \hline
{\small N$^3$LO w/o} & 2.674 $\times 10^{-4}$ & 4.056 $\times 10^{-6}$ &
-3.785 $\times 10^{-5}$ & 2.271 $\times 10^{-4}$ \\ \hline
{\small N$^3$LO w/} & 2.682 $\times 10^{-4}$ & 3.768 $\times 10^{-6}$ &
-3.509 $\times 10^{-5}$ & 2.304 $\times 10^{-4}$
\\
\hline \hline
{\small $K$, w/o} & 1.87 & 1.87 & 1.87 & 1.87 \\ \hline
{\small $K$, w/} & 1.87 & 1.73 & 1.73 & 1.89 \\ \hline
\end{tabular}
\caption{\label{ehdecay_label2} The total partial decay width for the Higgs
  decay into gluons $\Gamma^{gg}_{\mbox{\scriptsize tot}}$ and the
  individual contributions from the top and bottom loops with
  the corresponding interferences. The scenario is given by
  $c_t=0.85$, $c_b=c_c=1$ and $c_{gg}=0$. At LO (1st line), at N$^3$LO
  without inclusion of the mass effects in the top and bottom
  loops at NLO QCD (2nd line) and with the inclusion of the mass effects (3rd
  line). The 4th and 5th line are the $K$-factors without and with
  inclusion of the NLO top and bottom mass effects.}
\end{center}
\renewcommand{\arraystretch}{1.0}
\end{table}
The N$^3$LO QCD corrections are computed in the limit of large loop
particle masses (2nd line in the Table) and taking into account finite
quark mass effects at NLO in the top and bottom loops (3rd line). 
This means, that in the first case we set $\kappa^\textit{\scriptsize NLO}=0$ in
Eq.~(\ref{ehdecay_label1}). As expected, the higher order corrections
are large, with the N$^3$LO QCD corrections increasing the 
total width by almost up to 90\%. The results in
Table~\ref{ehdecay_label2} furthermore show, that the mass effects at
NLO QCD are relevant for the bottom loop where they amount to 8\%, while they
are negligible for the top loop. 

The $K$-factor is defined as the ratio between the decay width at N$^3$LO
and at LO. It is shown for the individual contributions and the total
gluonic decay width, without the NLO bottom and top mass effects (4th
line) and including them through the function $\kappa^\textit{\scriptsize NLO}$ (5th
line). Taking into account finite masses in the
NLO QCD loops has an 8\% effect on the
$K$-factor for the bottom loop contribution, while for the $K$-factor
of the total width it is only 1\%.

Setting also $c_t=1$, hence taking the SM-limit, the LO width and the 
N$^3$LO QCD width, including mass effects at NLO, amount to 
\begin{eqnarray}
\Gamma^{\mbox{\scriptsize LO}} (h^{SM} \to gg) &=& 1.724 \times
10^{-4} \mbox{\,GeV} \nonumber \\
\Gamma^{\mbox{\scriptsize N$^3$LO QCD}} (h^{SM} \to gg) &=& 3.259
\times 10^{-4} \mbox{\,GeV} \label{ehdecay_label4} \;.
\end{eqnarray}
The 15\% decrease in the top-Yukawa couplings hence decreases the total
SM width (without EW corrections) by 30\%. 
Note, that at a future $e^+e^-$ collider the Higgs decay rate into
gluons will be accessible with a precision of a few percent through a
measurement of the branching ratio~\cite{Baer:2013cma,Gomez-Ceballos:2013zzn}.
To be consistent with the non-linear approach, in
Eq.~(\ref{ehdecay_label4}) no electroweak corrections have been 
included. Including electroweak corrections, the SM width is changed by
$\sim 5$\%, 
\begin{eqnarray}
\Gamma^{\mbox{\scriptsize N$^3$LO QCD}}_{\mbox{\scriptsize EW}}
(h^{SM} \to gg) &=& 3.424 \times 10^{-4} \mbox{\,GeV} \; . 
\end{eqnarray}

We now investigate a scenario with a non-vanishing effective Higgs coupling to
gluons, $c_{gg}=0.001$, and SM-like Yukawa couplings,
$c_t=c_b=c_c=1$. In a composite Higgs scenario with fully composite
right-handed top quark, for example, we expect $c_{gg}$ to be of
the order of $\alpha_s/(4 \pi) \, m_t^2/m_*^2$, where $m_*$
denotes the mass of the top partners~\cite{Contino:2013kra}. This leads
to a $c_{gg}$ of order 0.001 for a top partner mass in the TeV
range. A value of $c_{gg}=0.001$ furthermore corresponds to the
contribution of a top squark with a mass value of about the top quark mass
in a {\it natural} supersymmetric scenario\footnote{Applying the Higgs
low-energy theorem, for the light stop contribution in the
decoupling limit we have 
$c_{gg}=\frac{\alpha_s}{24\pi}(m_t^2/m_{\tilde
t_1}^2+m_t^2/m_{\tilde t_2}^2-\sin^2 (2 \theta_{\tilde{t}})
\delta m^4/(4 m_{\tilde t_1}^2 m_{\tilde
t_2}^2))$~\cite{Haber:1984zu,Muhlleitner:2006wx}, with
$m_{\tilde{t}_{1,2}}$ 
denoting the masses of the two stops, $\theta_{\tilde{t}}$ the
stop mixing angle and $\delta m^2$ the squared mass difference of
the two stops.}.
The total partial decay width for
$m_h=125$\,GeV into a gluon pair is given in Table~\ref{ehdecay_label3}
and, separately, the quark loop and the effective Higgs-$gg$
contributions with their interference term. The quark loop
contribution stems from the sum of the top-, bottom- and charm-quark
contributions. Shown are the contributions
at LO and at N$^3$LO without and with NLO bottom and top quark loop
mass effects. 
\begin{table}[h]
\renewcommand{\arraystretch}{1.2}
\begin{center}
\begin{tabular}{|c|c|c|c||c|}
\hline 
     & $\Gamma_{qq}$ [GeV] & $\Gamma_{gg}$ [GeV] & $\Gamma_{qq-gg}$ [GeV]
     & $\Gamma^{gg}_{\mbox{\scriptsize tot}}$ [GeV]\\ \hline
LO & 1.72 $\times 10^{-4}$ & 5.13 $\times 10^{-6}$ & 5.91 $\times
10^{-5}$ & 2.37 $\times 10^{-4}$\\ \hline 
N$^3$LO w/o & 3.22 $\times 10^{-4}$ & 7.84 $\times 10^{-6}$ & 9.99
$\times 10^{-5}$ &
4.29 $\times 10^{-4}$ \\ \hline 
N$^3$LO w/ & 3.26 $\times 10^{-4}$ & 7.84 $\times 10^{-6}$ & 9.99
$\times 10^{-5}$ & 4.33 $\times 10^{-4}$
\\
\hline \hline
$K$-factor w/o & 1.87 & 1.53 & 1.69 & 1.81 \\ \hline
$K$-factor w/ & 1.89 & 1.53 & 1.69 & 1.83 \\ \hline
\end{tabular}
\caption{\label{ehdecay_label3} The total partial decay width
  $\Gamma^{gg}_{\mbox{\scriptsize tot}}$ for the Higgs decay into gluons and the
  individual contributions from the sum over the quark loops, the
  effective Higgs-$gg$ coupling and the interference of these two contributions.
  The scenario is given by
  $c_t=c_c=c_b=1$ and $c_{gg}=0.001$. At LO (1st line), at N$^3$LO
  without inclusion of the mass effects in the top and bottom
  loops at NLO QCD (2nd line) and with the inclusion of the mass effects (3rd
  line). The 4th and 5th line are the $K$-factors without and with
  inclusion of the NLO top and bottom mass effects.}
\end{center}
\renewcommand{\arraystretch}{1.0}
\end{table}
As can be inferred from Table~\ref{ehdecay_label3} and
Eq.~(\ref{ehdecay_label4}) a non-vanishing $c_{gg}=0.001$ increases
the total N$^3$LO SM width by 33\%. The $K$-factor for the individual
contribution from the effective Higgs-$gg$ coupling is 1.53 and 
roughly $20$\% smaller than the one for the quark loops. This can be
traced back to the factor $c_\textit{\scriptsize eff}^2 = 1.22$ in the N$^3$LO
corrections to the quark loop contributions, {\it
  cf.}\,Eq.~(\ref{ehdecay_label1}). In fact, the $K$-factors without
mass effects for the quark loops in Tables~\ref{ehdecay_label2}
and~\ref{ehdecay_label3} are given by $c_\textit{\scriptsize eff}^2
\kappa_\textit{\scriptsize \scriptsize soft}=1.87$, and the $K$-factor for the effective
Higgs-$gg$- contribution is given by $\kappa_\textit{\scriptsize \scriptsize soft}=1.53$. The
$K$-factor 1.81 for the total width $\Gamma^{gg}_{\mbox{\scriptsize
    tot}}$ is due to the relative weight of the long distance and
short distance contributions. The inclusion of
finite mass effects in the NLO QCD corrections, finally, again has a
negligible effect on the total N$^3$LO $K$-factor, being ${\cal O}(1\%)$. 

In summary, the inclusion of finite top and bottom quark masses in the
N$^3$LO QCD corrections, has an effect of 8\% on the individual bottom
quark contribution, while the over-all $K$-factor is changed by 1\%
only in scenarios close to the SM. A non-vanishing effective
Higgs-gluon-gluon coupling can significantly change the total width,
also for small coupling values. This is to be expected, as the decay
at leading order is already loop-mediated. 

\section*{CONCLUSIONS}
We have presented the program {\tt eHDECAY} which is based on an
extension of the Fortran code {\tt HDECAY} to include the full list of
leading bosonic operators of the Higgs effective Lagrangian. The
higher-order QCD corrections have been consistently included both in
the linear and the non-linear approach. Electroweak corrections have
been implemented only for the linear realization of the electroweak
symmetry, in the vicinity of the SM, as their inclusion in the general
case would require dedicated calculations not 
available at present. Taking the example of the Higgs decay into
gluons we have discussed mass effects in the NLO QCD corrections
as well as the inclusion of an effective Higgs coupling to gluons. 

\section*{ACKNOWLEDGEMENTS}
RC, MM and MS would like to thank the organisers of Les Houches for the
very nice atmosphere of the workshop.


%% file: nmssmcalc/nmssmcalc.tex

\chapter{NMSSMCALC - a Fortran Package for Higher Order Higgs Boson Masses
  and Higgs Decay Widths in the Real and the Complex NMSSM}

{\it J.~Baglio, R.~Grober, M.~Muhlleitner,
  D.T.~Nhung, H.~Rzehak, M.~Spira, J.~Streicher, K.~Walz}



\begin{abstract}
We present the Fortran package NMSSMCALC. It provides the loop-corrected Higgs
boson masses and calculates their decay widths including the dominant
higher order corrections as well as off-shell effects in the
CP-con\-ser\-ving and the CP-violating Next-to-Minimal Supersymmetric
Extension of the Standard Model (NMSSM). Special emphasis is put on
the inclusion of the supersymmetric-QCD and supersymmetric-electroweak corrections to the
decay widths into quark and lepton pair final states, which have been
evaluated for the first time for the NMSSM during the development of this tool.
\end{abstract}

\section{INTRODUCTION}
The discovery of the Higgs boson by the LHC experiments ATLAS
\cite{Aad:2012tfa} and CMS \cite{Chatrchyan:2012ufa}
in 2012 has been followed by an intense research program in order to 
determine its properties and to pin down its true nature. While the
boson looks very Standard Model (SM)-like, there is still room for
interpretations within models beyond the SM (BSM), 
in particular supersymmetric (SUSY) extensions have been intensely
studied. In order to catch up with the increasing amount of data
and precision of the experimental measurements, the theory predictions
have to become more refined, necessarily including higher order
corrections. This is also essential for the proper distinction between
different models. The program package NMSSMCALC
\cite{Baglio:2013vya,Baglio:2013iia} complies with these
requirements by providing the loop-corrected masses for the neutral
Higgs sector of the NMSSM \cite{Fayet:1974pd,Barbieri:1982eh,Dine:1981rt,Nilles:1982dy,Frere:1983ag,Derendinger:1983bz,Ellis:1988er,Drees:1988fc,Ellwanger:1993xa,Ellwanger:1995ru,Ellwanger:1996gw,Elliott:1994ht,King:1995vk,Franke:1995tc,Maniatis:2009re,Ellwanger:2009dp}. It furthermore computes the decay widths of the neutral and
charged NMSSM Higgs bosons including the dominant higher order
corrections as well as possibly important off-shell decays. It is the
first package which includes at this level of precision not only the
CP-conserving but also the CP-violating case. The program package can
be downloaded from the url: \url{http://www.itp.kit.edu/~maggie/NMSSMCALC/}. 

\section{PROGRAM DESCRIPTION}
\underline{\it The Model:}
In the NMSSM an additional singlet superfield $\hat{S}$ is introduced
compared to the minimal supersymmetric extension of the SM (MSSM). The
$\mu$ parameter is generated dynamically when the scalar component of
$\hat{S}$, which couples to the two Higgs doublets $\hat{H}_u$ and
$\hat{H}_d$ via $\lambda \hat{S} \hat{H}_u \hat{H}_d$, acquires a
vacuum expectation value (VEV) $v_s$. A massless axion is prevented by
breaking the Peccei-Quinn symmetry \cite{Peccei:1977hh,Peccei:1977ur}
explicitly through the introduction of a cubic coupling for $\hat{S}$, 
$\kappa \hat{S}^3/3$, 
in the scale-invariant superpotential. The MSSM $\mu$-term as well as
the tadpole and bilinear terms of the singlet superfield are assumed
to be zero in the superpotential. Mixing between the sfermion 
generations is neglected and the soft SUSY breaking terms linear and
quadratic in the singlet field are set to zero.
We denote by $\tan\beta$ the ratio of the absolute values of the two
vacuum expectation values, $v_u$ and $v_d$, of the scalar components
of the two Higgs doublets $\hat{H}_u$ and $\hat{H}_d$, which can be
complex in the CP-violating NMSSM. At tree-level the
Higgs sector of the NMSSM can be described by $\tan\beta$, by the two NMSSM 
specific couplings $\lambda$ and $\kappa$, by the soft SUSY breaking
trilinear couplings $A_\lambda$\footnote{The trilinear coupling
  $A_\lambda$ can be traded for the charged Higgs boson mass
  $M_{H^\pm}$.} and $A_\kappa$ and by
the vacuum expectation value $v_s$ of the singlet field. In case of
CP-violation the parameters $\lambda$ and 
$\kappa$ and, as mentioned above, the VEVs can be
complex, while they are real in the
CP-conserving case. All soft-SUSY breaking mass parameters of the
scalar fields are real, however, the soft SUSY breaking gaugino mass
parameters and trilinear couplings can be complex, if CP-violation is
assumed.  The phases of the Yukawa couplings, which are complex in
general, are reabsorbed by a redefinition of the quark
fields. After electroweak (EW) symmetry breaking three degrees of
freedom of the Higgs 
doublet fields are absorbed to give masses to the massive gauge
bosons, so that we are left with seven Higgs bosons. In the
CP-conserving case, there are three CP-even Higgs bosons $H_i$
($i=1,2,3$) and two CP-odd Higgs bosons $A_j$ ($j=1,2$) as well as two
charged Higgs states $H^\pm$. They are ordered by ascending mass with
$M_{H_1} \le M_{H_2} \le M_{H_3}$ and $M_{A_1} \le M_{A_2}$. In the
CP-violating NMSSM there are no mass eigenstates with definite CP quantum
numbers. The five neutral mass eigenstates are denoted by $H_i$
($i=1,...,5$) and ordered by ascending mass.

The input and output files of the program package feature the SUSY Les
Houches Accord (SLHA) \cite{Skands:2003cj,Allanach:2008qq}. The input
file is named {\tt inp.dat} and must contain the specification of the
model (CP-conserving or violating) and the SLHA SM input parameters,
which have been extended by the $W$ boson pole mass which is used in
the mass and decay width calculations.
The values for $\tan\beta$, for the NMSSM specific parameters and
for the soft SUSY breaking masses and trilinear couplings have to be
provided in the SLHA block {\tt EXTPAR}.
In the CP-violating case the input has to be supplemented by the
block {\tt IMEXTPAR} with the imaginary parts of the corresponding real
parameters, and by a newly introduced block with the phase
$\varphi_u$. The phases $\varphi_u$ and $\varphi_s$ describe the phase
differences between the three vacuum expectation values
$\langle H^0_u \rangle$, $\langle H^0_d \rangle$ and $\langle S \rangle$
of the neutral components of the Higgs doublet and singlet fields. In a
second input file, {\tt bhdecay.in}, the CKM parameters and several flags for the
calculation of the decay widths are set. 

The package consists of a wrap file {\tt nmssmcalc.f} and three main
files: {\it (i)} {\tt CalcMasses.F} for the calculation of the one-loop
corrected NMSSM Higgs boson masses in the real and the complex
NMSSM. They are obtained in the Feynman-diagrammatic approach in two
different renormalisation schemes both in the real and in the complex
NMSSM. Details can be found in
Refs.~\cite{Baglio:2013iia,Ender:2011qh,Graf:2012hh,Nhung:2013lpa}. 
{\it (ii)} {\tt bhdecay.f}, which computes the NMSSM Higgs boson decay widths
and branching ratios in the real NMSSM. {\it (iii)} {\tt bhdecay\_c.f},
providing the decay widths and branching ratios of CP-violating NMSSM
Higgs bosons. 

The wrap file reads in the input parameters for the Higgs mass
calculation from {\tt inp.dat} and calls {\tt CalcMasses.F} to which it
then passes on the input values. The latter calculates the
loop-corrected neutral Higgs boson masses, which are written
out together with the mixing angles in an SLHA output
file {\tt slha.in}. Subsequently {\tt nmssmcalc.f} calls {\tt
  bhdecay.f} (in the CP-violating case {\tt bhdecay\_c.f}), which
reads in {\tt slha.in} and computes all NMSSM Higgs decay widths
and branching ratios. They are written out in an SLHA output file {\tt
slha\_decay.out}. Sample input and output files can be found
on the program webpage. Note, that the user can also specify the names
of the input file and of the output files, provided by the mass and
decay routines, in the command line when running the program. 

The decay widths have been implemented including the dominant higher order
QCD corrections. In addition, the higher order SUSY-QCD corrections and
the approximate SUSY-electroweak corrections at one-loop level
are included in the Higgs boson decays into fermion pairs.  The decays
into stop and sbottom pairs contain the SUSY-QCD corrections in the case
of the CP-conserving NMSSM. Off-shell decays are taken into account in
the decays into massive gauge boson pairs, into gauge and Higgs boson
final states, into Higgs pairs and into heavy quark pairs. In this context
bottom quark mass effects in the off-shell decays into top quark final
states $tt^*$ for the neutral Higgs bosons and into the top-bottom
final state $t^*b$ for the charged Higgs boson have been calculated
and implemented. Note, that EW corrections beyond the approximate
SUSY-EW contributions are not included in general, as this would
require further missing calculations, due to the additional
NMSSM singlet field compared to the MSSM.  

\section{SUSY CORRECTIONS TO FERMIONIC DECAYS}
In the decays into fermion final states the SUSY-QCD and the SUSY-EW
corrections become important in parts of the parameter space. The
dominant contributions can be resummed and included in effective
Yukawa couplings \cite{Hempfling:1993kv,Hall:1993gn,Carena:1994bv,Pierce:1996zz,Carena:1998gk,Carena:1999py,Carena:2002bb,Guasch:2003cv,Noth:2008tw,Noth:2010jy,Mihaila:2010mp}, by deriving them from an effective
Lagrangian. In the NMSSM care has to be taken to properly
take into account the singlet contribution. For the bottom-Yukawa
part {\it e.g.} the corresponding effective Lagrangian is given
by (see \cite{Carena:1999py,Guasch:2003cv} in the case of the MSSM)
\begin{eqnarray}
{\cal L}_{\mathrm{eff}} = - y_b  \overline{{b}}_R\left[(1+\Delta_{1})
 H_d + \frac{\lambda^* e^{i \varphi_u} (1+\Delta_1)
   \Delta_b}{\mu_{\scriptsize  
\mathrm{eff}}^*\tan\beta} S^* H_u^{*} \right] b_L   + h.c. \;,
\end{eqnarray}  
where $H_d$, $H_u$ and $S$ are the scalar components of $\hat{H}_d$,
$\hat{H}_u$ and $\hat{S}$, respectively. 
The Yukawa coupling is denoted by $y_b$, and the indices $L,R$ denote the
left- and right-chiral projections of the bottom quarks. In the 
case of the CP-conserving 
NMSSM $\varphi_u=0$ and also $\lambda$ and $\mu_{\scriptsize  
\mathrm{eff}}$ are real. The corrections
$\Delta_b$ and $\Delta_1$ include the SUSY-QCD and SUSY-EW corrections
inducing a modification of the relation between the bottom quark mass
$m_b$ and the Yukawa coupling $y_b$. The explicit expressions for the
CP-conserving and CP-violating case can be found in
\cite{Baglio:2013iia}. Also the corrections for the neutral Higgs
boson decays into strange quark and lepton pair final states,
respectively, as well as for the charged Higgs decays into
up/charm/top and bottom quark pair and into up/charm/top and strange
quark pair can be found there. 

\section{HIGGS BOSON PHENOMENOLOGY}
In this section we discuss some interesting features in the NMSSM Higgs
boson phenomenology which can be investigated by  applying our program
package {\tt NMSSMCALC}. 

\begin{figure}[ht]
\begin{center}
\includegraphics[width=0.45\textwidth]{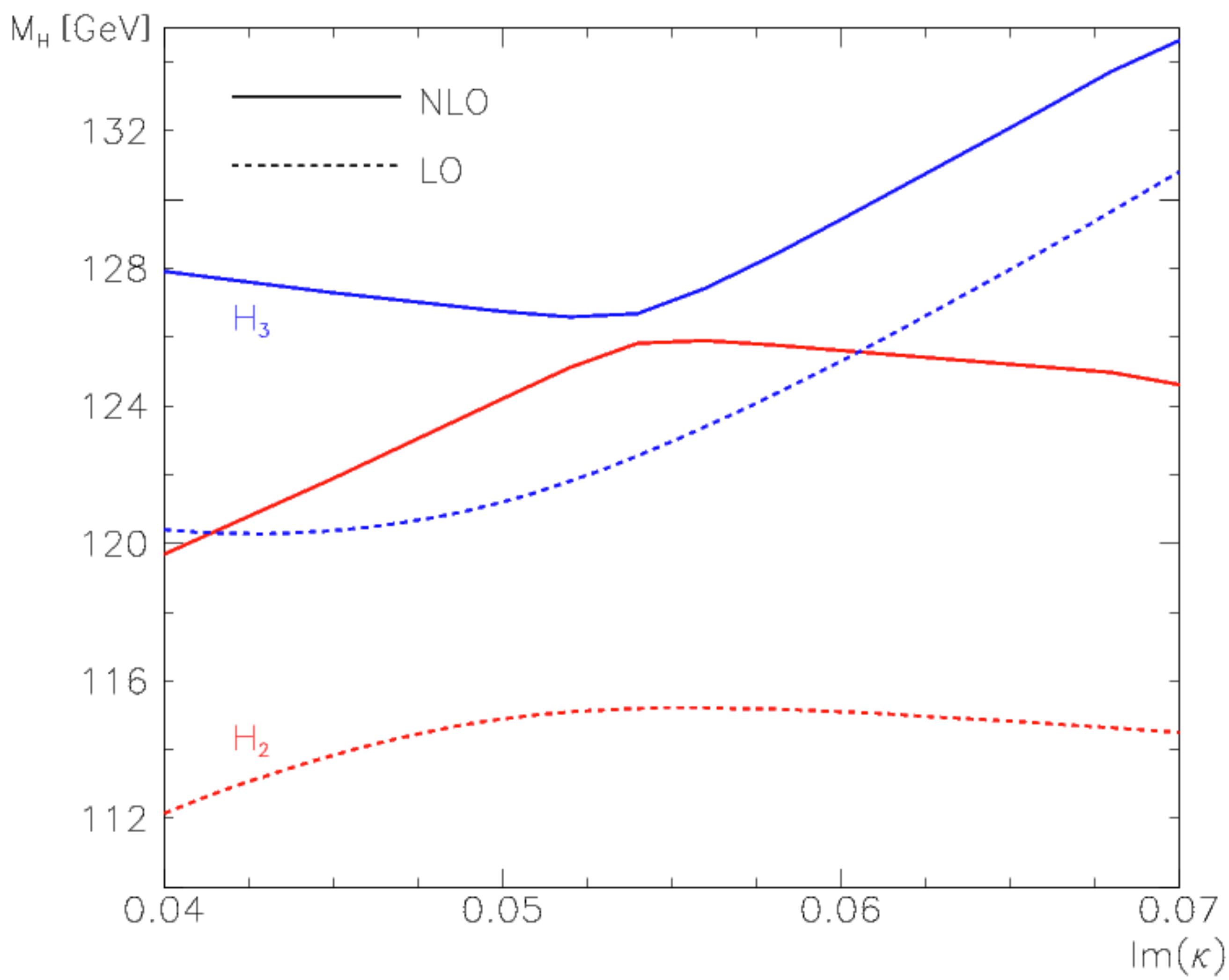}
\includegraphics[width=0.45\textwidth]{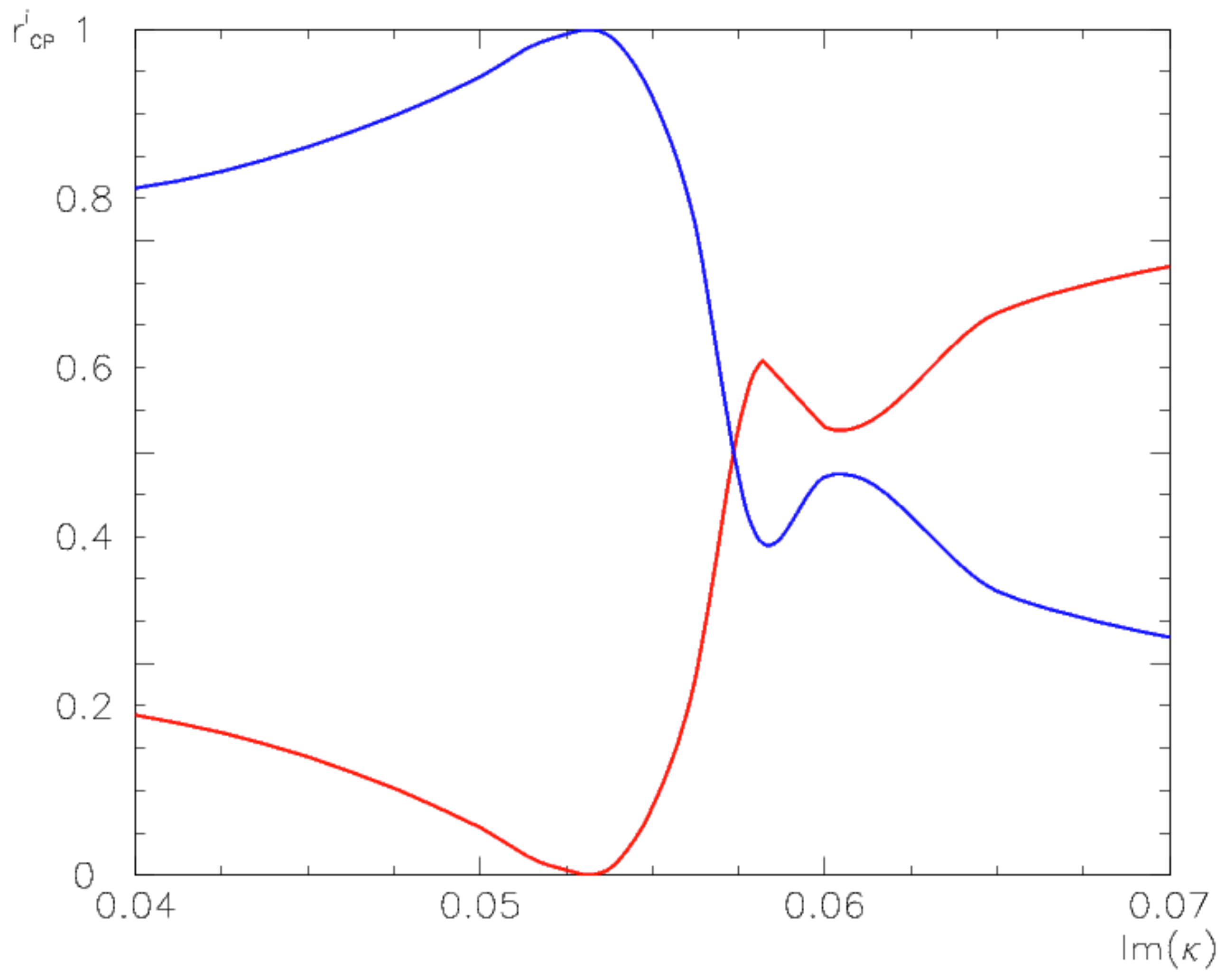}
 \caption{\label{nmssmcalc_label1} Left: Higgs boson masses $M_{H_2}$ (red/light
   grey) and $M_{H_3}$ (blue/dark grey) as function of $\mathrm{Im}(\kappa)$
   at LO (dashed) and NLO (full). Right: The amount of CP-violation
   $r^i_{\mathrm{CP}}$ for $H_2$ (red/light grey) and $H_3$ (blue/dark
   grey) as function of $\mathrm{Im} (\kappa)$.
}
\end{center}
\end{figure}

\noindent
\underline{\it Higgs boson masses in the complex NMSSM:}
We first analyse the impact of non-vanishing complex pha\-ses, {\it
  i.e.}~CP-violation, on Higgs boson phenomenology, in particular the
Higgs boson masses. Contrary to the MSSM, in the NMSSM CP-violation
already appears at tree-level, described by a non-vanishing phase
$\varphi_y$ given by 
\begin{eqnarray}
\varphi_y = \varphi_\kappa - \varphi_\lambda + 2\varphi_s - \varphi_u \;,
\end{eqnarray}
where $\varphi_\kappa$, $\varphi_\lambda$ and $\varphi_s$ denote the
phases of $\kappa$, $\lambda$ and the VEV of the singlet field $S$,
respectively. In Fig.~\ref{nmssmcalc_label1} (left) we show the
tree-level masses and the loop-corrected masses of the neutral Higgs
bosons $H_2$ and $H_3$. The input parameters are given by
\begin{eqnarray}
\begin{array}{lllllllll}
\tan\beta &=& 2\; , & \quad \mathrm{Re}(\lambda) &=& 0.65 \; , & \quad
\mathrm{Re}(\kappa) &=& 0.085 \; , \\
A_\kappa &=& -95 \mbox{ GeV} \; , & \quad M_{H^\pm} &=&
500 \mbox{ 
  GeV} \; , & \mathrm{Re}(\mu_{\mathrm{eff}}) &=& 200 \mbox{ GeV} \; ,
\end{array}
\end{eqnarray}
with the effective $\mu$ parameter $\mu_{\mathrm{eff}} =
\lambda v_s \exp (i\varphi_s) /\sqrt{2}$. The real parts of the
parameters are denoted by $\mathrm{Re}$, and we have set $\varphi_s=0$. The
soft SUSY breaking gaugino mass parameters 
$M_{1,2,3}$ as well as the left- and right-handed soft SUSY breaking
mass mass parameters of the lepton sector 
and correspondingly of the quark sector for the three different generations 
have been chosen as
\begin{eqnarray}
M_1 &=& 150 \mbox{ GeV}, \qquad M_2 = 340 \mbox{ GeV}, \qquad M_3 = 1
\mbox{ TeV}, \nonumber \\
M_{\tilde{E}_{R1,2}} &=& M_{\tilde{L}_{1,2}} =
M_{\tilde{D}_{R1,2,3}} = M_{\tilde{Q}_{1,2}} = 1.5 \mbox{ TeV},
\\
M_{\tilde{U}_{R1,2}} &=& 1 \mbox{ TeV}, \qquad M_{\tilde{U}_{R3}} =
M_{\tilde{Q}_3} = 700 \mbox{ GeV}, \qquad
M_{\tilde{E}_{R3}} = M_{\tilde{L}_3} = 250 \mbox{ GeV} \;. \nonumber
\end{eqnarray} 
The soft SUSY breaking trilinear couplings are set to, 
($U\equiv u,c,t$, $D \equiv d,s,b$, $L\equiv e,\mu,\tau$)
\begin{eqnarray}
\mathrm{Re}(A_U) = 300 \mbox{ GeV}, \qquad A_D =
A_L = 1.5 \mbox{ TeV} \;.
\end{eqnarray}
We furthermore include CP-violation at tree-level by choosing a
non-vanishing phase $\varphi_u$ and non-zero imaginary parts for
$\lambda$ and $\kappa$, respectively. 
Additionally, we choose an imaginary part for $A_U$ which enters
through loop-corrections in the masses (only $A_t$ induces a
non-negligible effect, though). Hence,
\begin{eqnarray}
\varphi_u = 0.1, \qquad \mathrm{Im}(\lambda) = 0.1, \qquad
\mathrm{Im}(\kappa) \in [0.04,...,0.07], \qquad 
\mathrm{Im}(A_U) = -50 \mbox{ GeV} \;.
\end{eqnarray}
Keeping $\varphi_u$ and $\mathrm{Im}(\lambda)$ constant, we vary
$\mathrm{Im}(\kappa)$ and hence $\varphi_y$. From the dashed lines in 
Fig.~\ref{nmssmcalc_label1} (left) we see that CP-violation at 
tree-level clearly has an impact on the mass values. With increasing
$\mathrm{Im}(\kappa)$ the masses of the next-to-lightest $H_2$ and of
the third lightest $H_3$ approach each other, before
departing from each other again. The one-loop corrections increase the
masses by up to 10\%, as can be inferred from the full lines. Also
here the strong dependence 
on the complex phase persists. Interestingly, for $\mathrm{Im}(\kappa)
\approx 0.054$ the two masses are almost degenerate and both close to
126~GeV. In this case the two Higgs bosons would build up the signal
observed at the LHC. 
At this value of $\mathrm{Im}(\kappa)$ in fact, $H_2$ and $H_3$
interchange their roles, as explained below when discussing
Fig.~\ref{nmssmcalc_label1} (right).

In Fig.~\ref{nmssmcalc_label1} (right) we show the amount of CP-violation
$r^{i=2,3}_{\mbox{CP}}$ of $H_2$ and $H_3$, respectively. For the boson
$H_i$ it is quantified by the loop-corrected mixing matrix elements
${\cal R}_{ij}$ as 
\begin{eqnarray}
r^i_{\mathrm{CP}} \equiv ({\cal R}_{i1})^2 + ({\cal R}_{i2})^2 + ({\cal
  R}_{i3})^2  \; , \quad i=1,...,5 \;.
\end{eqnarray}
The $5\times 5$ matrix ${\cal R}$ rotates the
interaction to the loop-corrected mass eigenstates $H_i$ where the
ordering is such that the first three (last two) columns correspond to the CP-even
(CP-odd) components of $H_u, H_d$ and $S$. A pure CP-even (CP-odd) state
corresponds to $r^{i}_{\mathrm{CP}}=1$ $(0)$. 

From Fig.~\ref{nmssmcalc_label1} (right) it can be inferred, that the initially
CP-even-like $H_3$ becomes more and more CP-even-like with increasing $\mathrm{Im}
(\kappa)$, reaching its maximum value of almost one at the point where
$H_2$ and $H_3$ are nearly mass degenerate. Subsequently, it develops
a CP-odd component which increases with $\mathrm{Im} (\kappa)$. The
next-to-lightest $H_2$ shows the opposite behaviour. Being initially
CP-odd-like, this component increases until $H_2$ is almost purely CP-odd-like
at $\mathrm{Im} (\kappa) \approx 0.054$. Beyond this point $H_2$ and
$H_3$ interchange their roles, with $H_2$ being more and more
CP-even-like. Finally, we 
note that in the limit of CP-conservation we have $M_{H_2}=115.8$~GeV
and $M_{H_3}=125.5$~GeV, with a CP-odd $H_2$ and a CP-even
$H_3$. Hence $H_3$ plays the role of the 125 GeV resonance, discovered
at CERN, while for the CP-violating case with $\mathrm{Im} (\kappa)=0.07$ this
role is taken over by $H_2$. 

This discussion shows that non-vanishing CP-violating phases can have
an important impact on Higgs boson phenomenology, in particular if
CP-violation is already present a tree-level. 

\noindent
\underline{\it SUSY corrections to decays into fermions:} In order to
discuss the impact of the SUSY-QCD and SUSY-EW corrections on the
Higgs decays into fermions we take a toy scenario with large values 
of $\tan\beta$ and $\mu_{\mathrm{eff}}$ for which these corrections
become sizeable. We keep the same parameter set as in the previous
scenario with the exception that we set all imaginary parts and phases
to zero, so that we are in the CP-conserving case. Furthermore we
change $\tan\beta$, $\lambda$, $M_{H^\pm}$ and $\mu_{\mathrm{eff}}$ to 
\begin{eqnarray}
\tan\beta= 30 \; , \qquad \lambda = 0.1 \; , \qquad M_{H^\pm} = 150
\mbox{ GeV} \;, \qquad \mu_{\mathrm{eff}} = 500 \mbox{ GeV} \;.
\end{eqnarray}
This leads to a mass spectrum with the second lightest Higgs boson
having a mass of 125~GeV. As expected, due to the large value of $\tan\beta$ and
$\mu_{\mathrm{eff}}$ SUSY corrections in the decays into fermions
become important. In the $H_2$ decays they can increase the
partial decay width by up to $\sim 15$\% in the $\tau$ lepton final
state and reduce the decay width into bottom quarks by up to 8\%. In
the $H^\pm$ decays the corrections become even more important,
reaching up to $\sim 15$\% in the $\tau \nu_\tau$ final state and
$\sim 18$\% in the decays into bottom charm or bottom up. 

\section*{CONCLUSIONS}
We have presented the program package {\tt NMSSMCALC} for the computation of
the higher order corrected NMSSM Higgs boson masses and decay widths
in the CP-conserving and the CP-violating cases. For the computation
of the decay widths it is at present the most-up-to-date program
tool, which includes besides the most important QCD corrections also
SUSY-QCD and SUSY-EW corrections for the decays into fermion pairs,
not only for the real but also the complex NMSSM. In application of our
package we have discussed the impact of complex phases on Higgs boson
phenomenology, which can be considerable. In a toy example the
importance of the SUSY-QCD and SUSY-EW corrections for the decays into
fermionic final states has been demonstrated.

\section*{ACKNOWLEDGEMENTS}
MM and MS would like to thank the organisers of Les Houches for the
great and fruitful atmosphere of the workshop.



%% file: htheo/HiggsFitTH.3.tex

\chapter{The Theoretical Uncertainties in Higgs Signal Strength Fits}

{\it S.~Fichet, G.~Moreau}



\begin{abstract}

We carry out fits of a simple anomalous Lagrangian to  latest Higgs data  using various statistical treatments
of theoretical uncertainties.
We discuss the discrepancies obtained from frequentist and Bayesian statistics, and from employing either Gaussian or uniform priors.
Slight discrepancies appear between fits with  Gaussian and uniform priors. One also observes differences 
between frequentist and Bayesian both because of the regions definition and because of the treatment of nuisance parameters. 
These various discrepancies do not have a significant impact on the conclusions of the fits.  We point out that as the LHC will collect more data, 
the frequentist and Bayesian treatments of nuisance parameters will converge and the dependence on theoretical uncertainty priors 
will increase.

\end{abstract}

\section{INTRODUCTION}

In addition to the recent discovery of a resonance around $125.5$~GeV~\cite{Aad:2012tfa,Chatrchyan:2012ufa} that is most probably the Brout-Englert-Higgs boson, the ATLAS and CMS 
collaborations have provided a long list of production and decay rate measurements. 
This precious list of data constitutes a new source of information on physics beyond the Standard Model (SM). 
Indeed, deviations of the observed Higgs boson rates with respect to their SM predictions may reveal the presence of underlying new physics (NP), while the absence of such deviations  translates as  constraints on NP models.
A large amount of Higgs coupling studies has already been realized, with so far no significant signs from unknown physics.  However, given the far-reaching 
implications of such results, it is mandatory to take carefully into account all the possible sources of uncertainty. 
In this contribution we carry out a global fit to the available data, taking into account the potentially significant uncertainties present on the theoretical side. 
In particular we implement the possibility of a flat prior distribution for the theoretical error and carry out both frequentist and Bayesian treatments.

\section{Likelihood function and theoretical uncertainties}  
\label{HiggsfitTH_se:likes}

The results from Higgs searches are given in terms of signal strengths $\mu(X,Y)$, the ratio of the observed rate for some process $X\to h \to Y$ relative to the prediction for the SM Higgs,
\begin{equation}
 \mu(X,Y)=\frac{N_{ obs}}{\left[ \sigma(X\rightarrow h) \, \mathcal{B}(h\rightarrow Y ) \, \varepsilon^{XY} {\cal L}\right]_{SM}}\,. \label{HiggsfitTH_mu_exp}
 \end{equation}
 We use the full dataset collected so far with luminosities of ${\cal L} \sim 5$~fb$^{-1}$ at the center of mass energy $\sqrt s=7$~TeV and ${\cal L} 
 \sim 20$~fb$^{-1}$ at $\sqrt s=8$~TeV~\cite{ATLAS-CONF-2013-014,ATLAS-CONF-2013-034,CMS-PAS-HIG-12-045}. 
 Here $\mathcal B$ is the branching ratio of the decay and the coefficient $\varepsilon^{XY}\in[0,1]$ characterizes the efficiency of event selection for a given subcategory.
   An experimental channel is defined by its final state ($\gamma\gamma$, $ZZ$, $WW$, $b\bar{b}$, $\tau\tau$) and is often divided into subchannels having different sensitivity to the various production processes. The accessible production mechanisms at the LHC are {\it i)} gluon-gluon Fusion (ggF), {\it ii)} Vector Boson
Fusion (VBF), {\it iii)} associated production with an electroweak gauge
boson $V=W,Z$ (Vh), and {\it iv)} associated production with a $t\bar{t}$ pair (tth).
The theoretical signal strengths for Higgs searches can be expressed as 
(see~\cite{Moreau:2012da,Djouadi:2013qya,Dumont:2013wma} for more details)
\begin{equation}
\bar{\mu}(X,Y)=\frac {\left[ \sigma(X\rightarrow h) \, \mathcal{B}(h\rightarrow Y ) \, \varepsilon^{XY}\right]_{NP }}  {\left[ \sigma(X\rightarrow h) \, \mathcal{B}(h\rightarrow Y ) \, \varepsilon^{XY}\right]_{SM}}\,. \label{HiggsfitTH_mu_BSM}
 \end{equation}
 In all generality, efficiencies in the SM with and without higher order operators 
 are not necessarily the same, \textit{i.e.}~$\varepsilon_{NP} \neq \varepsilon_{SM}$, because kinematic distributions can be modified in a non-trivial way by new physics effects.

The model of new physics we adopt consists of a modification of the Standard Model tree-level couplings, 
\begin{equation}
\mathcal{L}_h^{tree}=c_V \frac{h}{v}2m_W^2W_\mu^+W_\mu^-+c_V \frac{h}{v}m_Z^2(Z_\mu)^2
-c_f\sum_f  \frac{h}{v}m_f \bar{\Psi}_{fL}\Psi_{fR}+h.c.
\,,
 \end{equation}
by  the parameters $(c_V,c_f) $. The theoretical signal strengths are thus function of these anomalous couplings, \textit{i.e.}~$\bar{\mu}(c_V,c_f)$. 
Note these deviations to the SM are not the most general ones. 
Indeed, assuming the existence of new physics at a mass scale somewhat higher than the EW scale, the effects of new physics are captured in a low-energy effective Lagrangian, 
which in turn induces a peculiar pattern of anomalous couplings~\cite{Dumont:2013wma}. 
However we limit ourselves to the two dimensional subset $(c_V,c_f)$ because our focus is on the statistical methods rather than the effective Lagrangian. The efficiencies are not modified in this simple case.

Given the released experimental information, the likelihood we are able to reconstruct is a product of Gaussians. 
Although experimental systematics are in principle included in these data, the non-trivial correlations among subchannels they induce are most of the time not available. They will not be discussed further in this work. 
 Labeling by $I$ the various independent subsets of channels, the likelihood reads 
\begin{equation}
L_0\propto\prod_I e^{-(\bar{\mu}_I-\mu_{I})^2/2(\Delta^{\rm ex}_{I})^{2}}\,, \label{HiggsfitTH_eq:L0}
 \end{equation}
where $\Delta^{\rm ex}_{I}$ denotes the experimental uncertainties.

Let us  turn to the picture of theoretical uncertainties. The leading theoretical uncertainties come from the QCD prediction of the Higgs production cross sections. At the proton level, they come from parton distribution function (PDF) errors. 
At the parton level they originate  from the lack of knowledge of the higher order contributions in the perturbative expansion, and can be equivalently recast into the dependence on the QCD renormalization scale. Besides cross sections, this same source of uncertainty also affects the branching ratios.
As they  affect the predicted quantities, it turns out that these theoretical uncertainties approximately cancel in the predicted signal strengths $\bar \mu$, because they appear both in the numerator and the denominator. Instead  the experimental signal strength $\mu$ gets fully affected as no cancellation occurs.   
 Here we will  parametrize the theoretical errors on experimental signal  strengths under the form
\begin{equation}
\mu_I \rightarrow \mu_I\times (1+\delta_I \frac{\Delta_{I}^{\rm th}}{\mu_I} )\,,
\end{equation}
where the $\delta_I$ are nuisance parameters, and the $\Delta_{I}^{\rm th}$ set the magnitudes of the error in a given subchannel.

We adopt two different statistical frameworks, the one  of Bayesian statistics and the  one of (so-called) hybrid frequentist statistics. In both of these arguably well-defined frameworks,  a probability density function ({\em pdf}) is associated with all input parameters. From now we denote it as the ``prior'' {\em pdf}. The prior {\em pdf} multiplies the likelihood, such that one has to deal with the distribution 
\begin{equation}
L_0(c_V,c_f, \delta_I) p(c_V,c_f,\delta_I)\,,
\end{equation}
called ``posterior'' in Bayesian statistics. 
  The parameters of interest $c_V$, $c_f$ will be given a uniform prior. Our focus is rather on the nuisance parameters which are modeling the theoretical uncertainties. 
 We give the same probability distributions to all of the $\delta_I$'s.  It is clear that theoretical uncertainties may induce new correlations among the various subchannels. However we take the study of such correlations to be beyond the scope of the present contribution, and leave it for future work~\cite{toappear}. Here we rather focus on the various discrepancies related to the statistical framework. 
 
Apart from conceptual differences, in practice there are two differences between frequentist and Bayesian statistics that lead to potentially different results for Higgs fits. First, nuisance parameters are integrated over in the Bayesian framework, while the likelihood is instead maximized with respect to them in the frequentist framework. Second, the frequentist confidence regions and Bayesian credible regions have different definitions ({\it c.f.}~\cite{Nakamura:2010zzi}). It is therefore necessary to discuss the impact of these differences on Higgs fits.

A widespread practice in the community  is to combine theoretical uncertainties  in quadrature with the experimental uncertainties, that is
make the following replacement in Eq.(\ref{HiggsfitTH_eq:L0}), 
\begin{equation}
(\Delta_{I}^{\rm ex})^2 \to \Delta_I^2 \equiv (\Delta_{I}^{\rm ex})^2+(\Delta_{I}^{\rm th})^2\,.
 \end{equation}
We stress that this approach actually amounts to assume a Gaussian prior for the $\delta_I$'s, under the form
  \begin{equation}
p(\delta_I)\propto e^{-\delta_I^2/2 }\,.
 \end{equation}
  This is true in both frequentist and Bayesian frameworks. 
  We emphasize that a Gaussian prior is however not particularly motivated in order to model the uncertainties. A uniform prior over an appropriate interval  may for example seem more objective. In any case, the prior dependence of the fit needs to be controlled.  
In addition, in case of  a prior other than Gaussian, the frequentist and Bayesian approaches also differ at the level of marginalization itself, \textit{i.e.} the marginalized likelihoods are no longer identical.

All these aspects constitute motivations to compare both frequentist with Bayesian fits and Gaussian with uniform priors.
Before turning to numerical results, it is instructive to consider the behavior of these fits in various  limits. 
First, in the limit   $\Delta_{I}^{\rm th}\rightarrow0$ at fixed $\Delta^{\rm ex}_{I}$, the interval of $\delta_I$ shrinks to zero, \textit{i.e.} $p(\delta_I)$ tends to a Dirac peak, \textit{i.e.} $\delta_I$ is  set to zero. In that limit, the difference between fits based on the various prior shapes vanishes, and the only remaining  difference between frequentist and Bayesian fits comes from the definition of the best-fit regions. 
Second, in the ``high statistics limit'' $\Delta^{\rm ex}_{I}\rightarrow 0$ at fixed $\Delta^{\rm th}_{I}$, the $L_0$ likelihood function of Eq.(\ref{HiggsfitTH_eq:L0}) tends to a Dirac peak, 
so that there is  no more difference between frequentist and Bayesian treatments of nuisance parameters. Indeed, in this limit, the likelihoods take the form 
  \begin{equation}
L\propto\prod_I p(\delta_I)|_{\delta_I=({\bar \mu_I}/\mu_I-1)/(\Delta_{I}^{\rm th}/\mu_I)}\,.
 \end{equation}
The marginalized likelihoods therefore depend maximally on the prior for $\delta$, 
such  that the issue of theoretical uncertainties will become dominantly important as the LHC  accumulates more data.

\section{Numerical results}

\begin{figure}
\includegraphics[width=7.5cm]{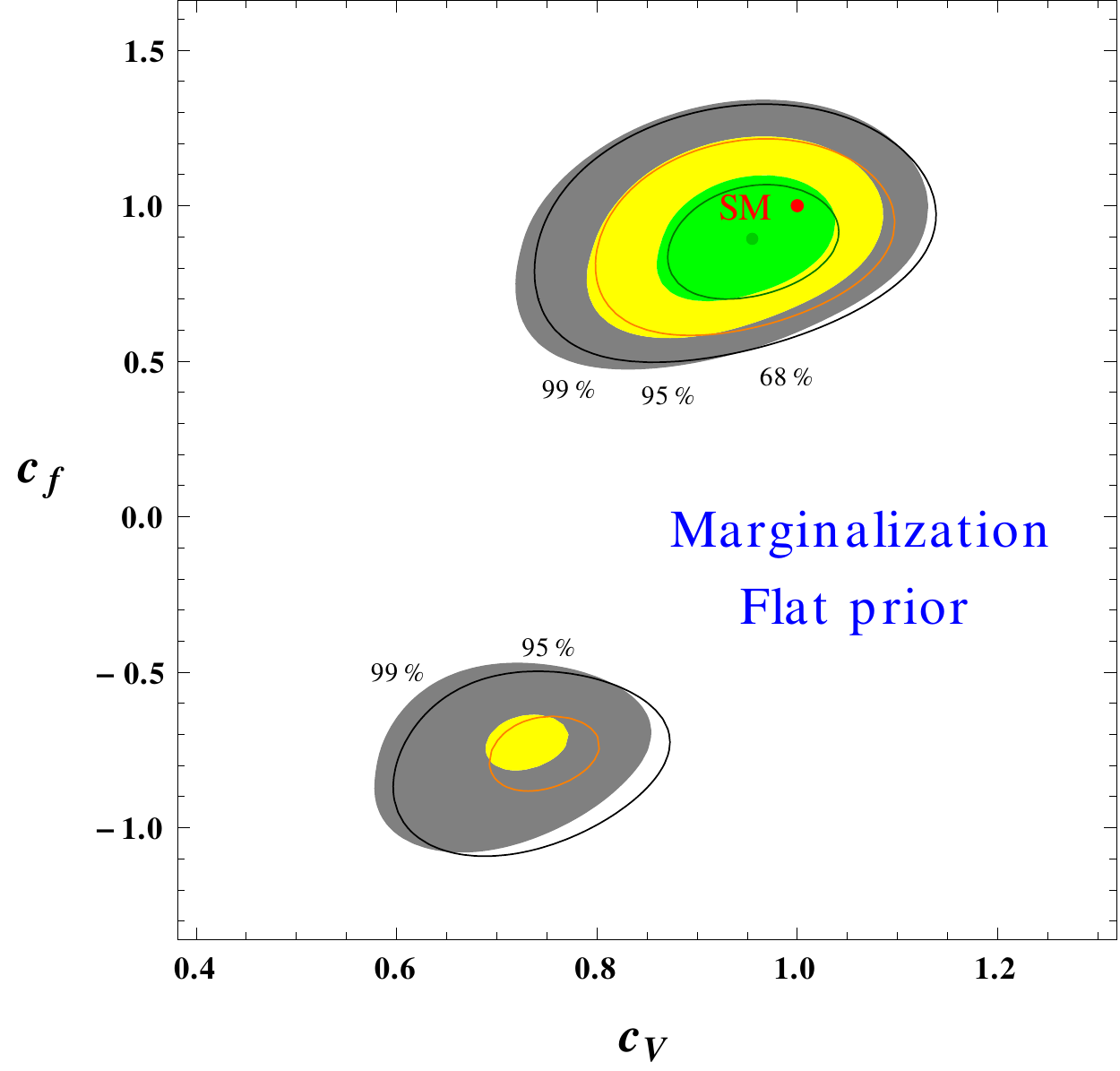}
\includegraphics[width=7.5cm]{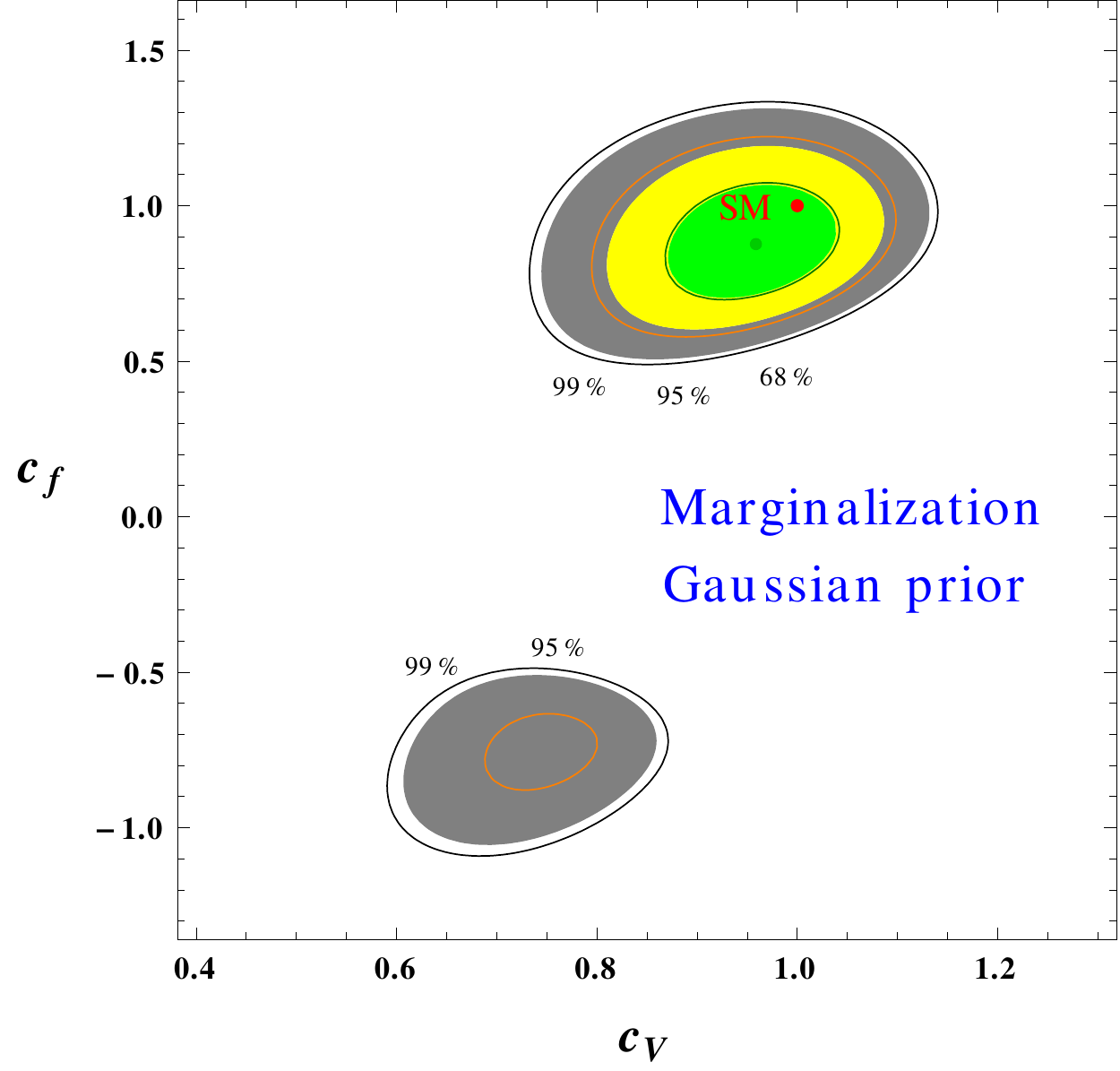}
\caption{
The $68.27\%$, $95.45\%$, $99.73\%$ confidence level (CL) (from marginalized frequentist likelihoods) and Bayesian credible (from posteriors) regions are shown
respectively as colored regions and contour levels in the $(c_V,c_f)$ plane. 
The green point is the frequentist best-fit location. Left and right panels respectively correspond to the flat and Gaussian prior case.
The SM prediction point is also displayed [in red].
}\label{HiggsfitTH_fig:CombMarg1D}
\end{figure}

The marginalized frequentist likelihood and Bayesian posteriors with either flat or Gaussian prior are shown in Fig.~\ref{HiggsfitTH_fig:CombMarg1D}. 
First of all, we can observe that the frequentist and Bayesian approaches give very similar outcome. This shows explicitly the fact that the two approaches 
have reached a certain degree of convergence since the priors on $c_V,c_f$  are already dominated by the data\footnote{Notice however that this statement depends on the model under consideration. In particular with the present data, a global fit of the dimension 6 effective Lagrangian still has some weakly constrained directions in the operator space~\cite{Dumont:2013wma}.}. 

For the Gaussian case, even though the likelihood functions are exactly the same, 
we observe that the $95.45\%$CL region is still present in the Bayesian fit, while it disappears in the frequentist case -- this difference being only 
induced by the definition of the best-fit regions. It turns out from the right figure that the Bayesian regions are more conservative. 
For the flat prior case, one can see that the  difference between frequentist and Bayesian treatments of nuisance parameters leads to slightly different 
positions and shapes for the best-fit regions.

For a given statistical approach (either frequentist or Bayesian), small differences are observed between the flat and Gaussian priors. As discussed in the previous Section, the smallness of these discrepancies is related to the fact that we are currently in the regime where experimental uncertainties are large with respect to theoretical uncertainties, \textit{i.e.} $\Delta^{\rm ex}_{I}\gg\Delta^{\rm th}_{I} $. This situation will change with future and more accurate LHC data.


\section*{CONCLUSIONS}

In this contribution we discuss the  discrepancies induced on Higgs fits from various  statistical treatments of theoretical uncertainties, using the most recent data available.   
We present and compare results obtained from frequentist and Bayesian statistics, and from employing either Gaussian or uniform priors.
Discrepancies between frequentist and Bayesian are observed in the Gaussian case because of the regions definition.  
In the flat prior case, further discrepancies appear because of the treatment of nuisance parameters.
These discrepancies do not have a significant impact on the conclusions of the fits. 
It is worth pointing out that in the limit of high statistics, the frequentist and Bayesian treatments 
of nuisance parameters converge, while the theoretical uncertainty prior dependence increases. 
The study of the impact from correlations induced by theoretical uncertainties is left for further work~\cite{toappear}.

\section*{ACKNOWLEDGEMENTS}
S.~Fichet thanks the University of Orsay/France for hospitality offered
as some of the work contained in these proceedings was performed there.
The authors also acknowledge the organizers of the ``Les Houches'' Workshop for the perfect organization
and friendly atmosphere during the Workshop.



%% file: hcouplings/hfit-LH.tex

\chapter{Testing Custodial Symmetry and CP Properties of the 125~GeV Higgs Boson}

{\it J.~Bernon, B.~Dumont, J.~F.~Gunion, S.~Kraml}


\begin{abstract}
Performing a fit to all publicly available data, we analyze the extent
to which the latest results from the LHC and Tevatron constrain the
couplings of the 125~GeV Higgs boson.
In particular, we test custodial symmetry through $HWW$ and $HZZ$ coupling modifications and study possible CP-violating contributions.
Moreover, we consider consequences for the $Z\gamma$ channel.
\end{abstract}

\section{INTRODUCTION}

That the mass of the Higgs boson is about 125~GeV is a very fortunate circumstance in that we can  
detect it in many different production and decay channels~\cite{Aad:2012tfa,Chatrchyan:2012ufa}.
Indeed, many distinct signal strengths, defined as production$\times$decay rates relative to Standard Model (SM) expectations, $\mu_i\equiv (\sigma\times {\rm BR})_i/(\sigma\times {\rm BR})_i^{\rm SM}$, 
have been measured with unforeseeable 
precision already with the 7--8~TeV LHC run~\cite{ATLAS-CONF-2013-034,CMS-PAS-HIG-13-005}. 
From these signal strengths one can obtain information about the couplings of 
the Higgs boson to electroweak gauge bosons and fermions (of the third generation), and 
loop-induced couplings to photons and gluons.

Fits to various combinations of reduced Higgs couplings ({\it i.e.}\ Higgs couplings to fermions and gauge bosons relative to their SM values) have been performed by the experimental collaborations themselves 
as well as in a large number of theoretical papers, see {\it e.g.} Ref.~\cite{Belanger:2013xza} and references therein.  
In Ref.~\cite{Belanger:2013xza} some of us combined the information
provided by ATLAS, CMS and the Tevatron experiments on the 
$\gamma\gamma$, $ZZ^{(*)}$, $WW^{(*)}$, $b\bar{b}$ and $\tau\tau$  final states 
including the error correlations among various production modes.
The five theoretically ``pure'' production modes which are accessible are gluon--gluon fusion (ggF),   
vector boson fusion (VBF), associated production with a $W$ or $Z$ boson (WH and ZH, commonly denoted as VH), 
and associated production with a top-quark pair (ttH). 
The scheme conveniently adopted by the experimental collaborations is to group these five 
modes into just two effective modes, ggF + ttH and VBF + VH,  and present 
contours of constant likelihood for particular final states in the 
$\mu({\rm ggF + ttH})$ versus $\mu({\rm VBF + VH})$ plane.
Using this information, we obtained ``combined likelihood ellipses'', which can be used in a simple, generic way to 
constrain non-standard Higgs sectors and new contributions to the loop-induced processes, provided they 
have the same Lagrangian structure as the SM.  

In particular, these likelihoods can be used to derive constraints on a
model-dependent choice of generalized Higgs couplings, the implications of which we studied in Ref.~\cite{Belanger:2013xza} 
for several well-motivated models. In this contribution, we go a step further and study to which extent 
the current global coupling fit can test 
i) custodial symmetry and ii) possible CP-violating contributions.

\section{TESTING CUSTODIAL SYMMETRY}

We fit the latest Higgs data using the following parametrization of the Higgs couplings: 
\begin{equation}
\mathcal{L}=g\left[ C_W m_W W^\mu W_\mu + C_Z \frac{m_Z}{\cos \theta_W} Z^\mu Z_\mu - C_U \frac{m_t}{2m_W} \bar{t}t - C_D \frac{m_b}{2 m_W} \bar{b}b - C_D \frac{m_\tau}{2 m_W} \bar{\tau}\tau \right]H. 
\label{eq:hfit-HL}
\end{equation}
We set $C_W, C_Z > 0$ by convention and use ${\rm sgn}(C_U)={\rm sgn}(C_D) = 1$ unless explicitly mentionned. For convenience, we define $C_{WZ}$ as the ratio of the $HWW$ coupling to the $HZZ$ coupling,
\begin{equation}
C_{WZ}\equiv\frac{C_W}{C_Z}.
\end{equation}
We define $C_g$ and $C_\gamma$ to be the ratio of the $Hgg$ and $H\gamma\gamma$ couplings to their SM values. In this paper
we do not consider potential BSM loop contributions $\Delta C_g$ and $\Delta C_\gamma$ from new particles in the $gg\to H$ and $H\to \gamma \gamma$ loops and, therefore, $C_g$ and $C_\gamma$ are simply computed in terms of the parameters appearing in Eq.~(\ref{eq:hfit-HL}).

Our aim is to obtain confidence level (CL) intervals on $C_{WZ}$. For that, the signal strengths $\mu$ in the various observed channels are used to compute a $\chi^2$:
\begin{equation}
\chi^2=\sum_k \frac{(\bar{\mu}_k - \mu_k)^2}{\Delta \mu_k^2},
\end{equation}
where $k$ runs over all available production$\times$decay channels, $\mu_k$ is the experimental signal strength for the channel $k$, and $\bar{\mu}_k$ is the predicted signal strength from the set of reduced couplings $C_i$.
Whenever the 2D information $(\mu_{\rm ggF+ttH}, \mu_{\rm VBF+VH})$ is available, a Gaussian fit is performed and the correlation between the two signal strengths is obtained and taken into account.
Details on this procedure are explained in Ref.~\cite{Belanger:2013xza} (see also the discussion in Ref.~\cite{Boudjema:2013qla}).

When the experimental result is only given for VH production and not for WH and ZH separately, we combine the calculated signal strengths for a given decay mode as:
\begin{equation}
\bar{\mu}_{\rm VH} = \frac{\sigma(pp\to ZH)}{\sigma(pp\to VH)} \ \bar{\mu}_{\rm ZH} + \frac{\sigma(pp\to WH)}{\sigma(pp\to VH)} \ \bar{\mu}_{\rm WH} \; .
\end{equation}
For the VBF production mode, we compute a reduced $C_{\rm VBF}$ coupling as
\begin{equation}
C_{\rm VBF}^2=\frac{C_Z^2 \ \sigma({\rm ZBF}) + C_W^2 \ \sigma({\rm WBF}) + C_Z C_W \ \sigma({\rm interference})}{\sigma({\rm ZBF})+\sigma({\rm WBF})+\sigma({\rm interference})},
\end{equation}
where the cross sections are obtained with VBFNLO \cite{Arnold:2011wj}.

\begin{figure}[b!]
\begin{center}
\includegraphics[scale=0.3]{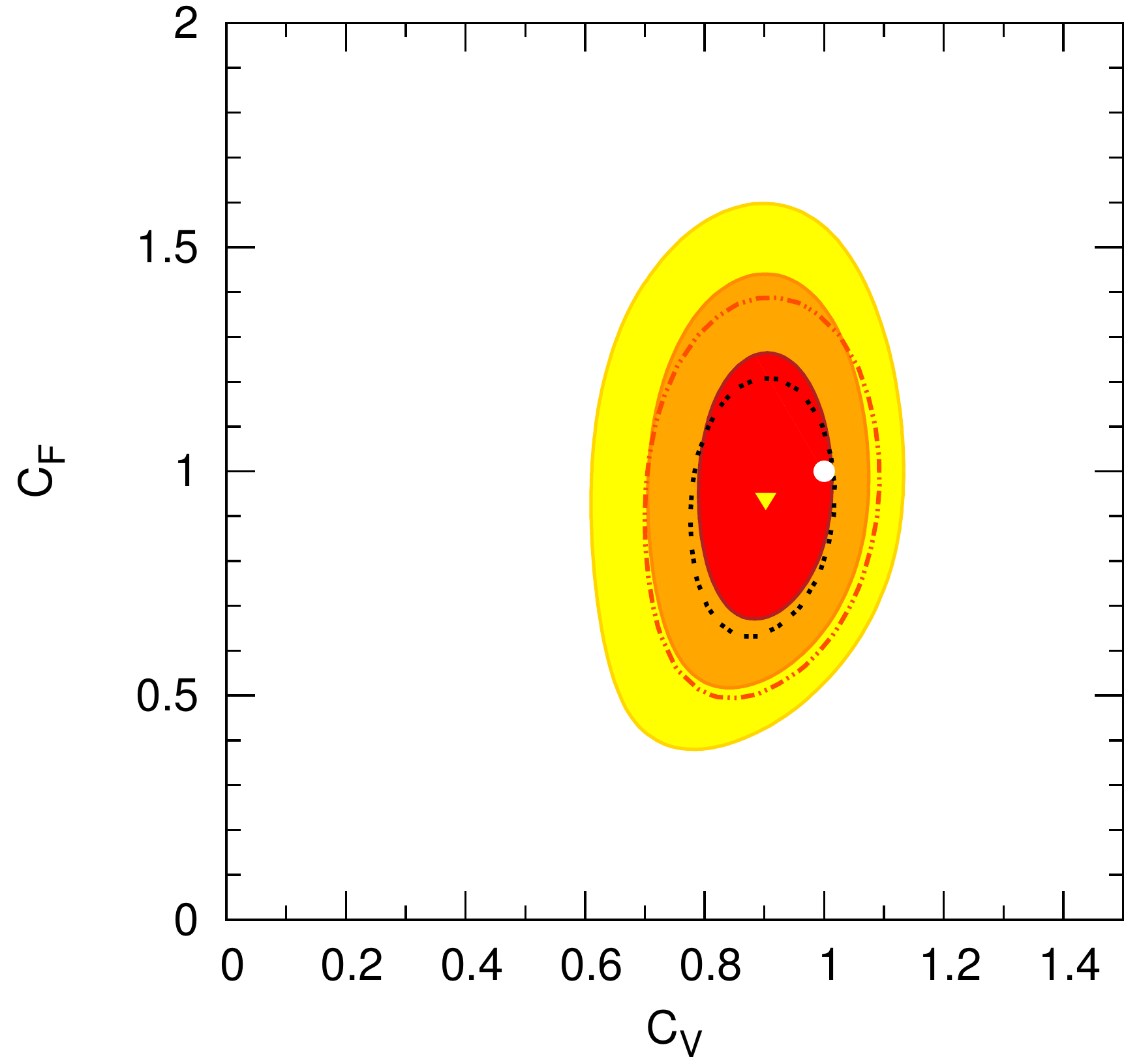}
\includegraphics[scale=0.3]{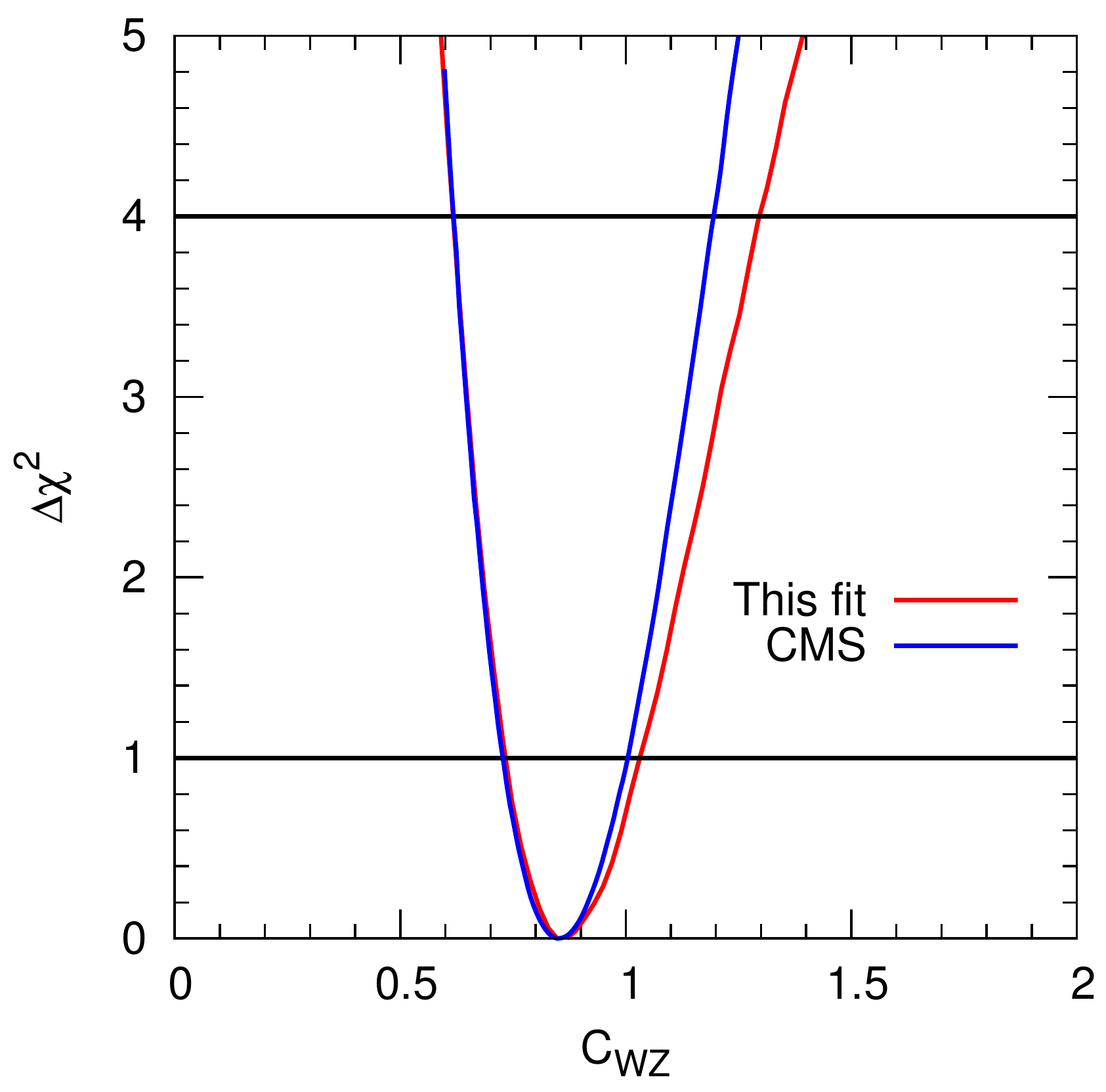}
\includegraphics[scale=0.3]{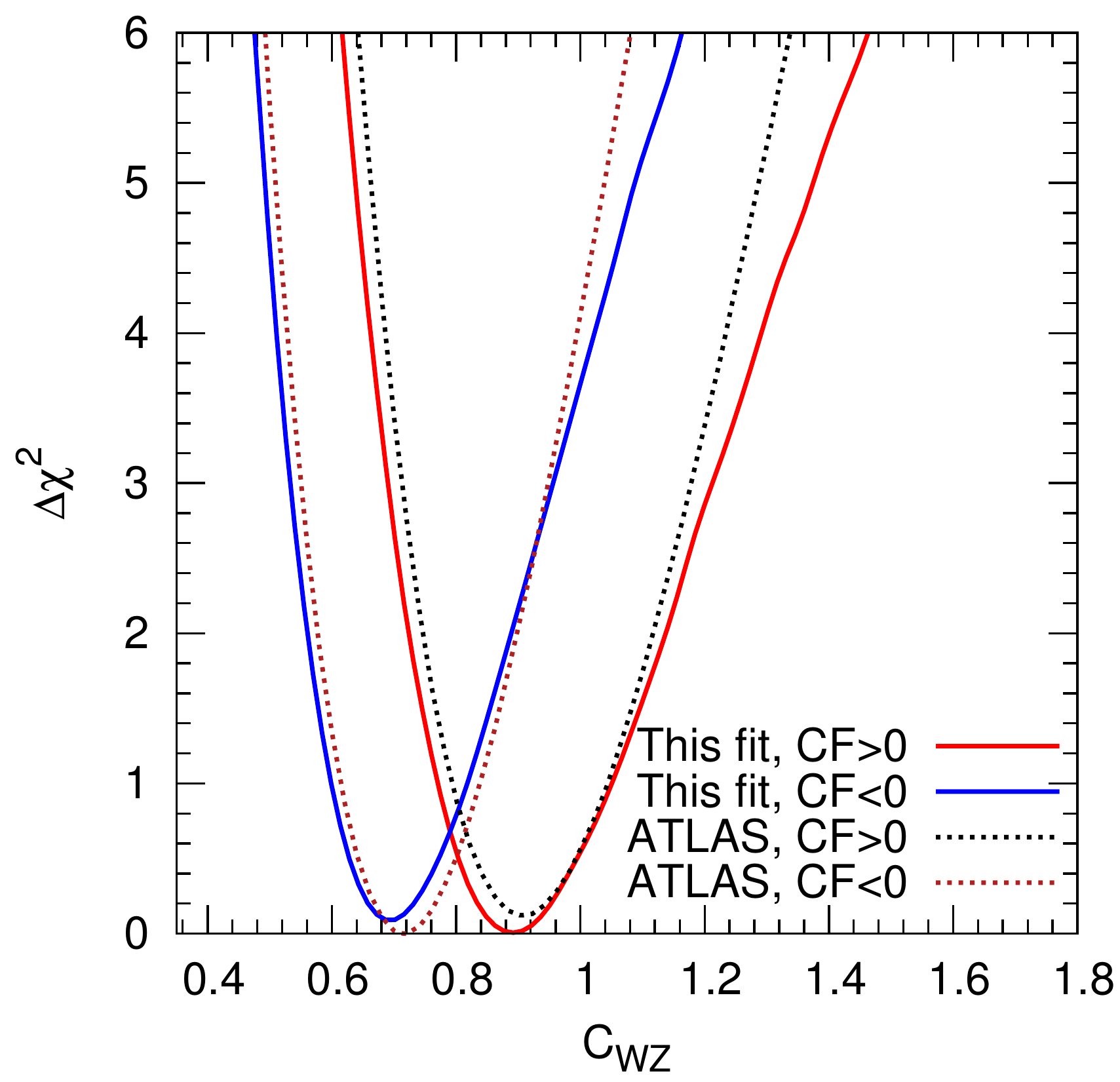}
\vspace{-0.2cm}
\caption{Left plot: Fit of $C_V$ and $C_F$. The dashed lines are the 68\% and 95.4\% CL contours as reported by the CMS collaboration in Fig.~6 of Ref.~\cite{CMS-PAS-HIG-13-005}, while the red, orange and yellow regions are the 
68\%, 95.4\% and 99.7\% CL regions of our fit. Middle plot: Fit of $C_{WZ}$ while $C_Z$ and $C_F>0$ are profiled over with comparison to the CMS fit, Fig.~7 of Ref.~\cite{CMS-PAS-HIG-13-005}. Right plot: Fit of $C_{WZ}$ while $C_Z$ and $C_F$ are profiled over with comparison to the ATLAS fit, Fig.~11 of Ref.~\cite{Aad:2013wqa}.}
\label{fig:hfit-compCMS}
\end{center}
\end{figure}

For validation, we first compare our fits using only CMS or ATLAS data to the results of fits performed by the CMS~\cite{CMS-PAS-HIG-13-005} and ATLAS~\cite{Aad:2013wqa} collaborations themselves. For CMS, two cases are considered. First, custodial symmetry and universal fermion couplings to the Higgs boson are assumed, \textit{i.e.} $C_V\equiv C_Z=C_W$ and $C_F\equiv C_U=C_D$. Second, a universal fermion coupling $C_F$ is assumed while $C_{WZ}$ and $C_Z$ are scanned over. For ATLAS, the scan has been performed over $C_F$, $C_{WZ}$ and $C_Z$, for different signs of $C_F$. 
The results are displayed in Fig.~\ref{fig:hfit-compCMS}. 
Since we do not have access to the full experimental information, our fits naturally somewhat differ from the ones done by the collaborations. Nonetheless, as one can see, our fits provide a very good approximation.

\begin{figure}[!t]
\begin{center}
\includegraphics[scale=0.30]{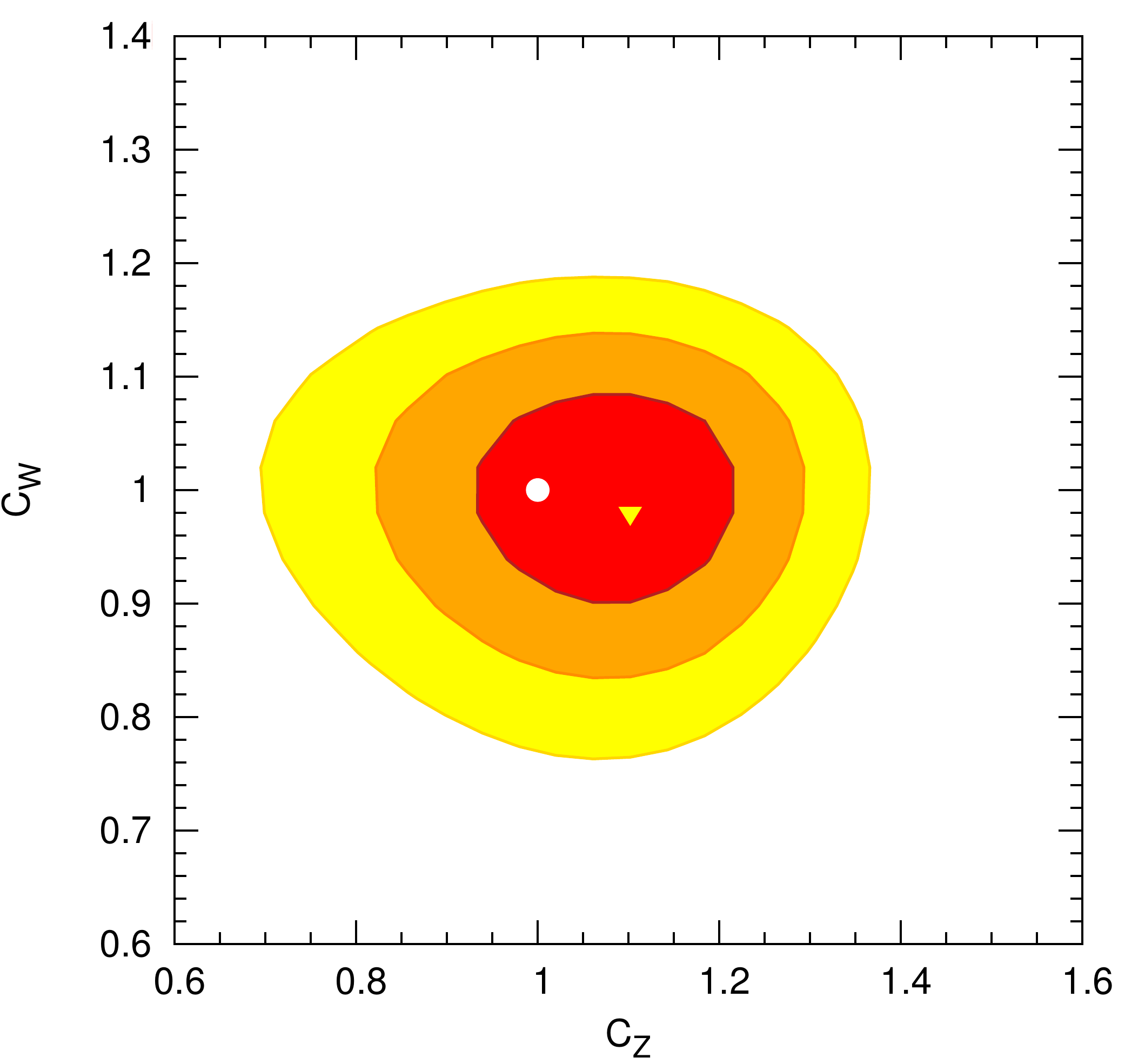}
\includegraphics[scale=0.30]{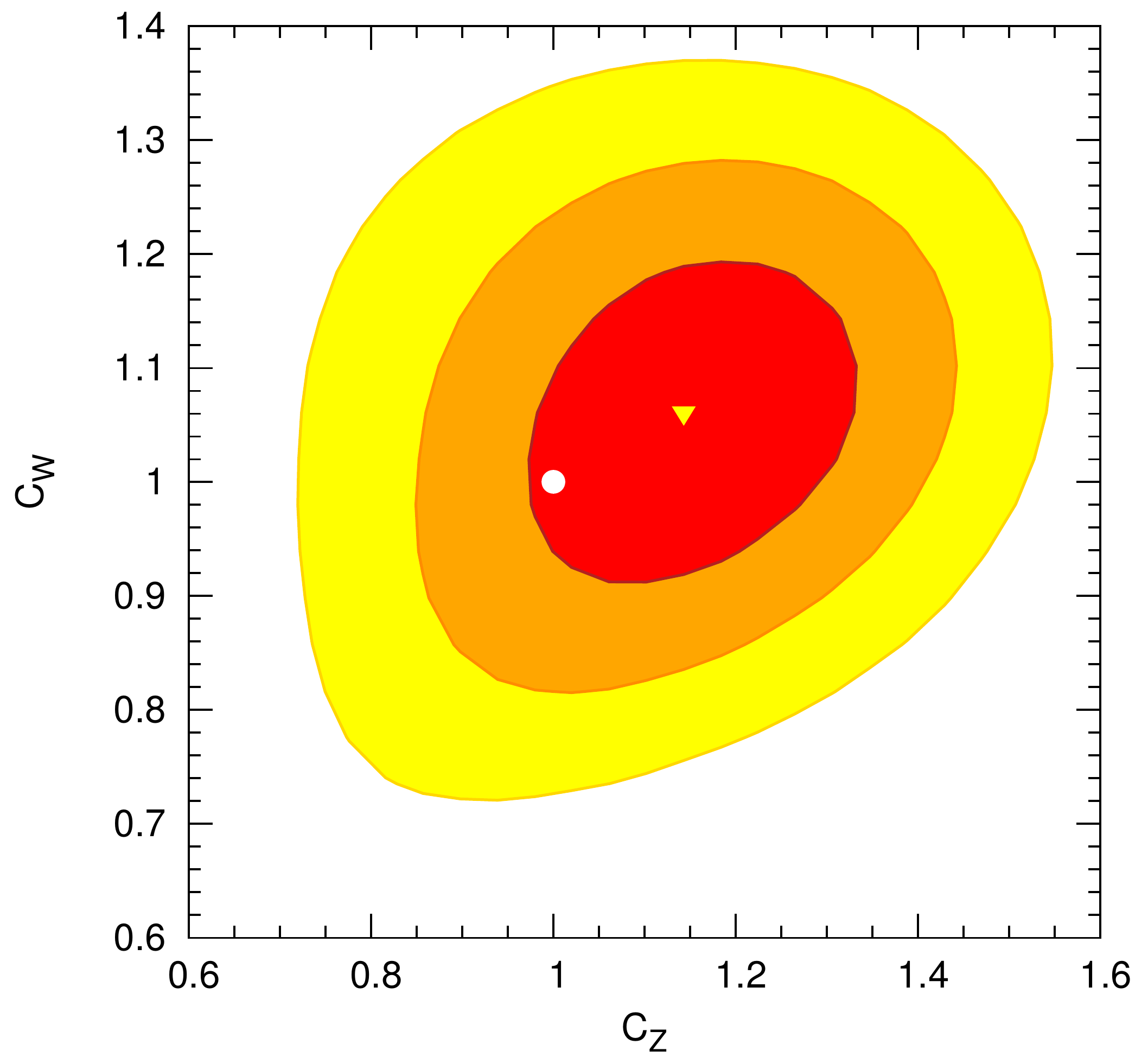}
\vspace{-0.2cm}
\caption{$C_W, C_Z$ fit assuming $C_U=C_D=1$ (left) and varying $C_U, C_D>0$ (right). The red, orange and yellow regions are respectively the 68\%, 95.4\%, and 99.7\% CL regions. The white point shows the SM expectation while the yellow triangle shows the best-fit point: ($C_W=0.980,\, C_Z=1.102$) (left) and ($C_W=1.061,\, C_Z=1.143$) (right).}
\label{fig:hfit-firstfit}
\end{center}
\end{figure}

Next, we fit the Higgs couplings to $W$ and $Z$ bosons using the full data sets available from the LHC and Tevatron, {\it i.e.}\ combining all available results. We consider the cases that a)~the fermionic couplings $C_U$ and $C_D$ are as in the SM, \textit{i.e.}\ $C_U=C_D=1$ and b)~the fermionic couplings are allowed to vary independently. The results are reported in the 2D plane of $C_W$ versus $C_Z$ plane in Fig.~\ref{fig:hfit-firstfit}. 
When the fermionic couplings are set to their SM values, no correlation is seen between $C_W$ and $C_Z$. The situation is different when $C_U$ and $C_D$ are varied independently. Indeed, when $C_W$ takes values larger (smaller) than 1, the signal strengths $\mu(H\to WW)$ and $\mu(H\to \gamma\gamma)$ become large (small) and thus deviate considerably from the experimental values. When $C_D$ is allowed to deviate from one, the total width of the observed state can be made larger (smaller) in order to decrease (increase) the signal strengths $\mu(H\to WW)$, $\mu(H\to\gamma\gamma)$ and $\mu(H\to ZZ)$ while keeping an acceptable rate for $H \to b\bar b$ and $H \to \tau\tau$. Finally, to accommodate themeasured $H\to ZZ$ value, $C_Z$ is preferably larger than 1. This leads to the observed correlation between $C_W$ and $C_Z$.
Note that in case a) the best fit for $C_W$ is slightly below unity while that for $C_Z$ it is slightly above unity. 
In case b) the best fits for values $C_W$ and $C_Z$ are both slightly above unity. 
The corresponding 68\% and 95.4\% CL intervals for $C_W$ and $C_Z$ are given in Table~\ref{tab:hfit-CL_CWCZ}.

\begin{table}[!htb]
\begin{center}
\begin{tabular}{|c||c|c||c|c|} 
  \hline
   & $C_W$: 1$\sigma$ range & $C_Z$: 1$\sigma$ range & $C_W$: 2$\sigma$ range & $C_Z$: 2$\sigma$ range \\   
   \hline
  $C_U=C_D=1$ & [0.94, 1.06] & [0.99, 1.17] & [0.87, 1.11] & [0.89, 1.26] \\  
  \hline
  $C_U,C_D>0$ & [0.96, 1.15] & [1.04, 1.27] & [0.87, 1.24] & [0.91, 1.39] \\ 
  \hline
\end{tabular}
\end{center}
\vspace{-0.1cm}
\caption{68\% and 95.4\% CL intervals of $C_{W}$ and $C_Z$ in two different scenarios.}
\label{tab:hfit-CL_CWCZ}
\end{table}

\begin{figure}[t!]
\begin{center}
\includegraphics[scale=0.30]{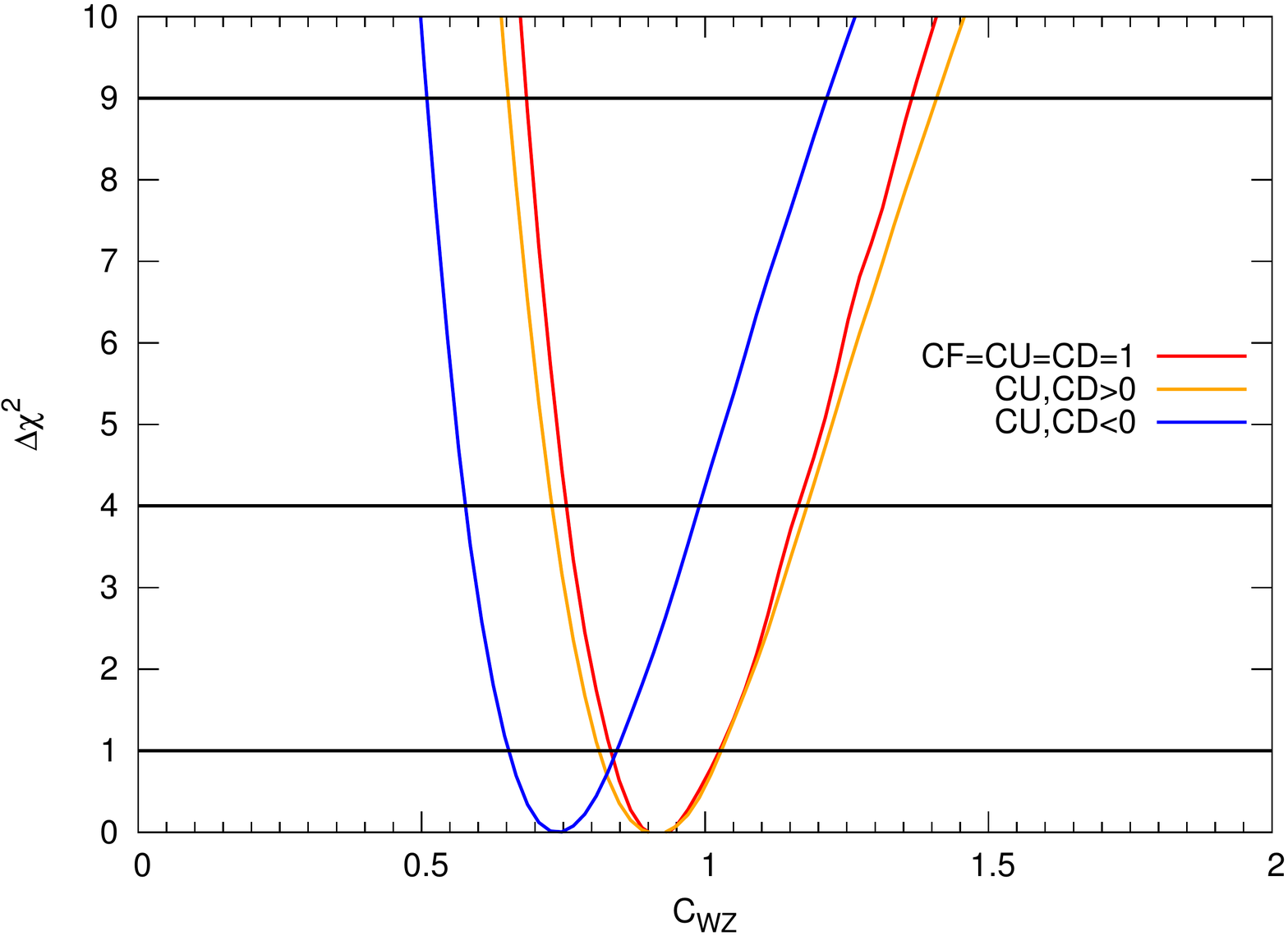}
\vspace{-0.1cm}
\caption{1-dimensional fit of $C_{WZ}$ for three different choices of fermionic couplings; when free, $C_U$ and $C_D$ are profiled over.}
\label{fig:hfit-cwz}
\end{center}
\end{figure}

The $\Delta \chi^2$ profile of $C_{WZ}$ is shown in Fig.~\ref{fig:hfit-cwz}, for the cases $C_U=C_D=1$ (red line), varying $C_U, C_D>0$ (orange line) and varying $C_U, C_D<0$ (blue line).  The corresponding 68\% and 95.4\% CL intervals for $C_{WZ}$ and the $\chi^2$ at the best-fit points are listed in Table~\ref{tab:hfit-CL_CWZCZ}.
The first two scenarios are compatible at the 1$\sigma$ level with the SM. The best-fit case is obtained when $C_U$ and $C_D$ are positive and varied independently. The best-fit point for this scenario is: $C_U = 0.90$, $C_D = 1.02$, $C_Z = 1.18$, $C_{WZ} = 0.90$, $C_W = 1.06$ with a $\chi^2_{\rm min}$ of $15.70$ for $20$ degrees of freedom.

\begin{table}[!htb]
\begin{center}
\begin{tabular}{|c|c|c|c|}
  \hline
   & 1$\sigma$ range & 2$\sigma$ range & $\chi^2_{\rm min} / {\rm ndof}$\\
  \hline
  $C_U=C_D=1$ & [0.83, 1.02] & [0.76, 1.16] & $16.87/22=0.77$ \\  
  \hline
  $C_U,C_D>0$ & [0.81, 1.03] & [0.73, 1.18] & $15.70/ 20 = 0.79$\\  
  \hline
  $C_U,C_D<0$ & [0.65, 0.84] & [0.58, 0.99] & $18.84 /20= 0.94$ \\  
  \hline
\end{tabular}
\end{center}
\vspace{-0.1cm}
\caption{68\% and 95.4\% CL intervals of $C_{WZ}$ and $\chi^2_{\rm min}/{\rm ndof}$ in three different scenarios.}
\label{tab:hfit-CL_CWZCZ}
\end{table}

Even though the SM unity values of $C_{W}$ and $C_Z$ are compatible with the experimental data already at the $1 \sigma$ level, this fit cannot exclude values greater than unity. Of course, values above unity can only be realized if the $H$ of Eq.~(\ref{eq:hfit-HL}) has a component coming from triplet or higher Higgs representations. Differences between $C_W$ and $C_Z$ are  of course severely constrained from the precise measurement of the Peskin--Takeuchi $T$ parameter at LEP. It is however still possible to obtain $C_W \neq C_Z$ in an effective approach at the price of a fine-tuned cancellation between operators~\cite{Farina:2012ea}. Another possibility would be to generate an apparent splitting between $C_W$ and $C_Z$ from the tensor couplings $H (Z_{\mu\nu})^2$ and $H W^+_{\mu\nu}W^-_{\mu\nu}$~\cite{Dumont:2013wma}.

\section{TESTING CP-VIOLATING ADMIXTURES}

The Higgs coupling to vector bosons has the  general form 
\begin{eqnarray}
  VVH : \quad C_V\; \frac{gM_V^2}{m_W}\; g^{\mu \nu} \; ,
\end{eqnarray}
where as above $C_V$ measures the departure from the SM: $C_V=1$ for a pure scalar  (CP-even)
state with SM-like couplings and $C_V=0$ for a pure pseudoscalar (CP-odd) state. Above, we found that $C_V$ is always well compatible with 1. However, this does not mean that CP is conserved in the Higgs sector. 
Likewise, $C_V\not=1$ would not automatically be an indication of CP violation (although $C_V>1$ would require higher Higgs representations). Instead, it is possible that 
two or more states of an enlarged Higgs sector share the couplings to $W$ and $Z$ bosons between them. 
(In this case,  the squared couplings of each state $H_i$ to gauge bosons should sum to unity, $\sum_i (C_V^i)^2=1$.) 
For testing CP mixing, one would need to exploit angular distributions, see Refs.~\cite{Godbole:2004xe,Accomando:2006ga} for reviews. 

A more decisive test of a possible CP-odd admixture comes from the fermion sector because of the general scalar and pseudo--scalar structure of the Higgs coupling to fermions~\cite{Godbole:2004xe,Accomando:2006ga}. Concretely, we have 
\begin{eqnarray}
   f\bar f H : \quad   -\bar f (v_f +i a_f  \gamma_5 ) f \, \frac{gm_f}{2m_W} \,,
\label{eq:hfit-CP-Hff}
\end{eqnarray}
where in the SM one has $v_f=1$ and $a_f=0$, while a purely CP-odd Higgs would have $v_f=0$ and $a_f=1$. 
In the notation of Eq.~(\ref{eq:hfit-HL}), $v_f={\rm Re}(C_F)$ and $a_f={\rm Im}(C_F)$, $F=U,D$, and 
the normalisation of the coupling, ${\rm Re}(C_F)^2 + {\rm Im}(C_F)^2=  |C_F|^2$,
should be taken arbitrary as in the previous section.  Effects of CP mixing will show up at loop level, in particular in the $gg\to H$ and $H\to \gamma\gamma$ rates.
A test of the CP properties of the observed Higgs boson from a global fit to the signal strengths was first  
presented in Ref.~\cite{Djouadi:2013qya}. 
Following Ref.~\cite{Djouadi:2013qya}, at leading order the Higgs rates normalized to 
the SM expectations can be written as
\begin{eqnarray}
\frac{\Gamma( H \to \gamma\gamma)}{\Gamma( H \to \gamma\gamma)\vert_{\rm SM}} 
&  \simeq & 
\frac{\big \vert \frac{1}{4} C_W A^+_1[m_W] + (\frac{2}{3})^2\, {\rm Re}(C_U)  \big \vert^2
+ \vert (\frac{2}{3})^2 \frac{3}{2} {\rm Im}(C_U) \vert^2} 
{\big \vert \frac{1}{4} A^+_1[m_W] + (\frac{2}{3})^2  \big \vert^2 } \;, 
\nonumber \\ 
\frac{\sigma( gg \to H)}{\sigma( gg \to H)\vert_{\rm SM}} & = & 
\frac{\Gamma( H \to gg)}{\Gamma( H \to gg)\vert_{\rm SM}}  \simeq 
\big \vert {\rm Re}(C_U)  \big \vert^2 + \vert \frac{3}{2} {\rm Im}(C_U)  
\big \vert^2 \;, 
\label{eq:hfit-widthsCP} 
\end{eqnarray}
with $A^+_1[m_W] \simeq -8.34$ for $M_H = 125.5$ GeV. For convenience, the contribution of the $b$ 
quark has been omitted in the above equations but is taken into account in the numerical analyses.
Note that a pure pseudoscalar state not only implies Re$(C_U)=0$,  
but also $C_W=C_Z=0$; this is clearly excluded, as VBF production 
as well as $H\to ZZ$ and $WW$ decays ($4\ell$ and $2\ell 2\nu$ signals) have been observed;  
besides, there would be no $W$ boson contribution to the $H\to \gamma \gamma$ rate. 

A CP-odd admixture to the observed state at 125.5~GeV is however still an interesting possibility and would necessarily indicate an enlarged Higgs sector. To test this possibility, we include Re$(C_U)$ and Im$(C_U)$ as independent parameters and perform the following fits:\\
\begin{description}
	\item{i)} \ \ free ${\rm Im}(C_U)$ and $C_V$ while $|C_U|^2=C_D=1$,
	\item{ii)} \ free ${\rm Re}(C_U)$, ${\rm Im}(C_U)$ and $C_V$ while $C_D=1$, 
	\item{iii)} free ${\rm Re}(C_U)$, ${\rm Im}(C_U)$, $C_D$, $C_W$ and $C_Z$.\\
\end{description}

\begin{figure}[t!]
\begin{center}
\includegraphics[scale=0.26]{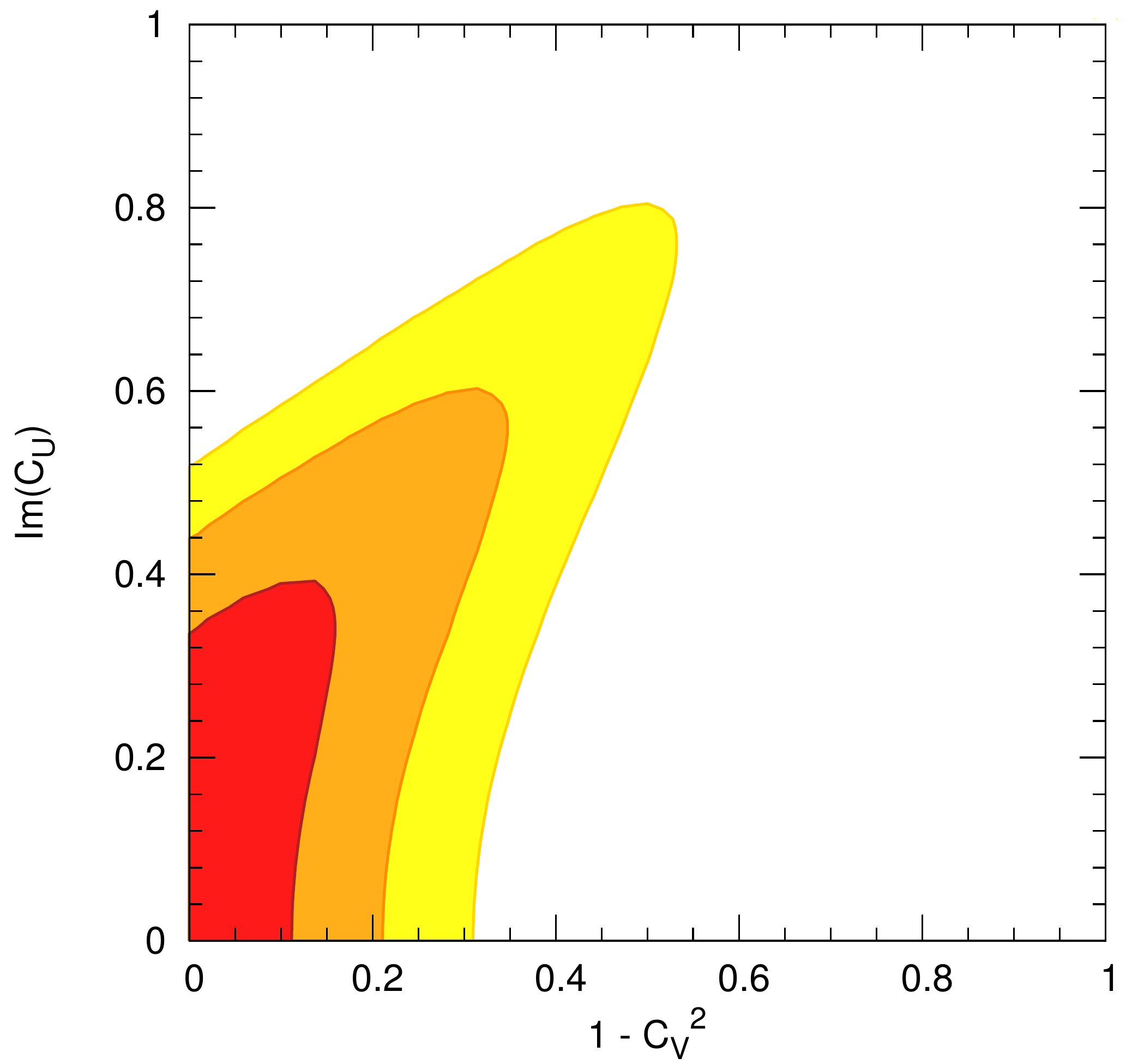}
\includegraphics[scale=0.26]{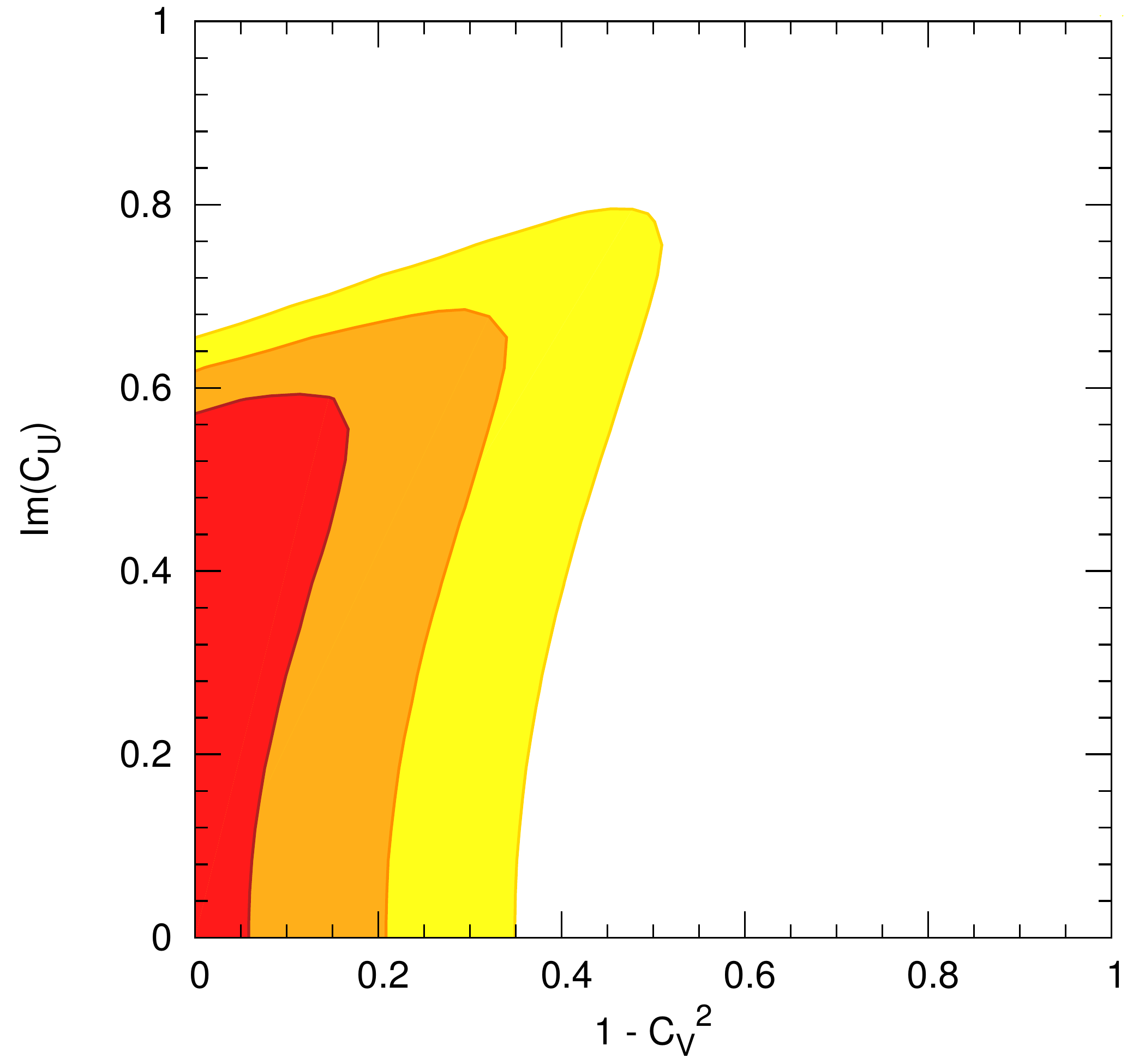}
\vspace{-0.2cm}
\caption{68\% (red), 95.4\% (orange) and 99.7\% (yellow) CL best-fit regions in the $\big(1-C_V^2, \ {\rm Im}(C_U) \big)$ plane, on the left for fit i), on the right for fit ii). In both cases, the best-fit point lies in the $1-C_V^2 < 0$ region.}
\label{fig:hfit-ImCU-CV}
\end{center}
\end{figure}

\begin{figure}[t!]
\begin{center}
\includegraphics[scale=0.26]{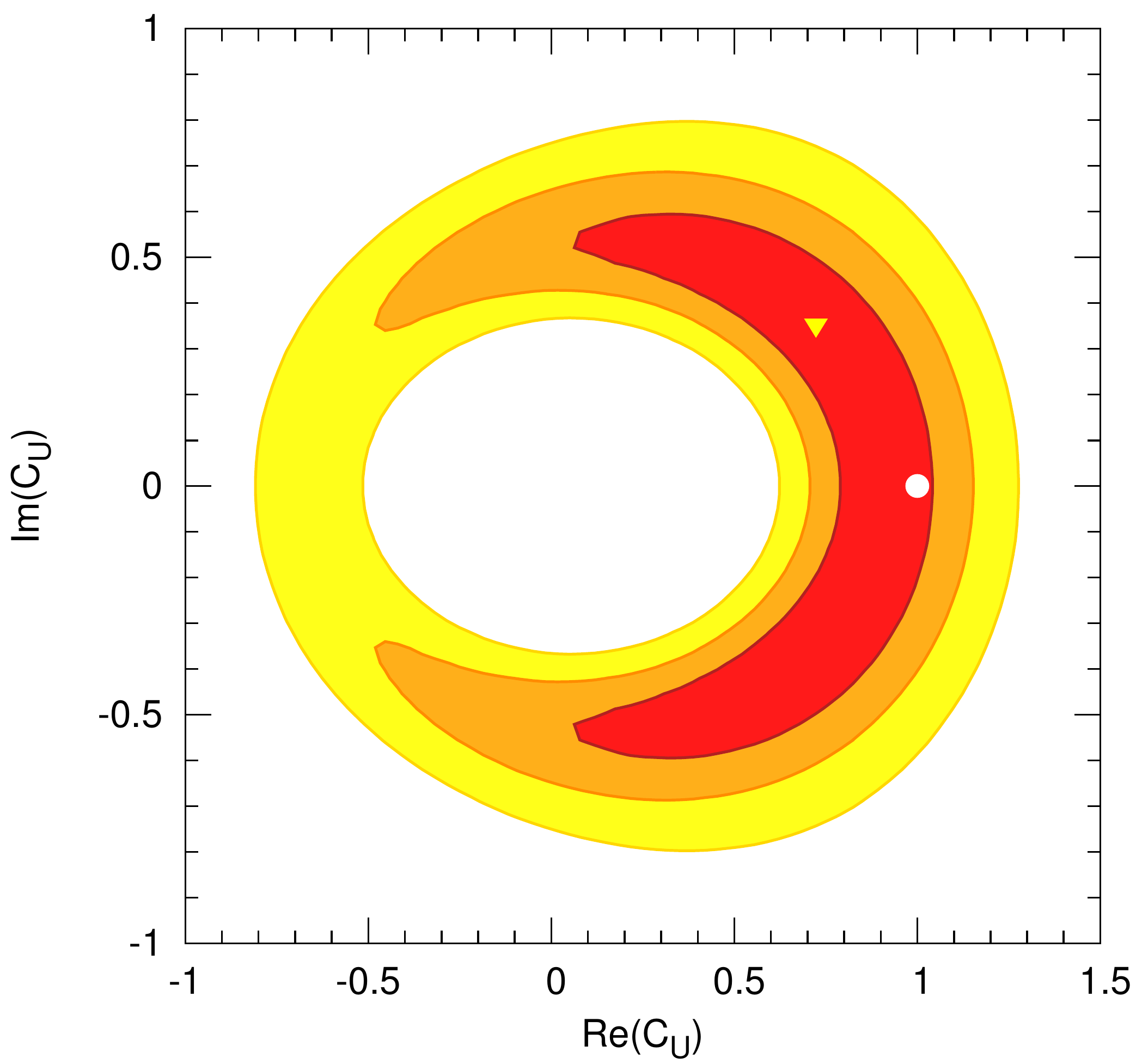}
\includegraphics[scale=0.26]{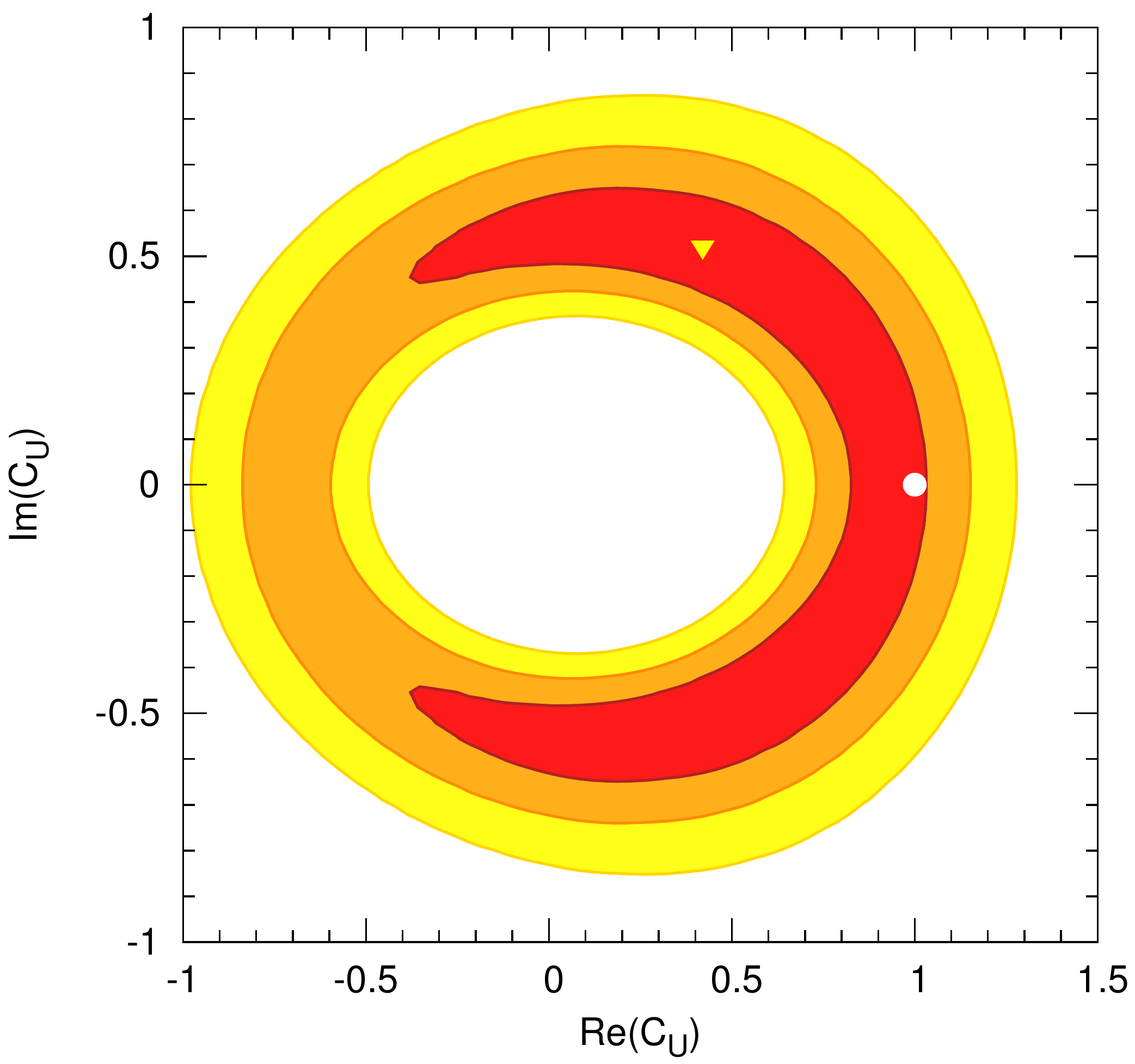}
\vspace{-0.2cm}
\caption{68\% (red), 95.4\% (orange) and 99.7\% (yellow) CL best-fit regions in the $\big( {\rm Re}(C_U), \ {\rm Im}(C_U)\big)$ plane, on the left for fit ii) with $C_V\equiv C_W=C_Z$, on the right for fit iii) with free $C_W$ and $C_Z$. The white point shows the SM expectation while the yellow triangle shows the best-fit point.}
\label{fig:hfit-ImCU-ReCU}
\end{center}
\end{figure}

The results are shown in Figs.~\ref{fig:hfit-ImCU-CV} and \ref{fig:hfit-ImCU-ReCU}.\footnote{Our analysis extends that of Ref.~\cite{Djouadi:2013qya} in that it 1)~includes the latest data and 2)~uses the 2D information including correlations in the $(\mu_{\rm ggF+ttH}, \mu_{\rm VBF+VH})$ plane, instead of signal strengths per cut categories.} 
The best-fit points and the corresponding $\chi^2_{\rm min}/{\rm ndof}$ are given in Table~\ref{tab:hfit-BestFit}.
Upper limits on a given parameter are computed from the 1-dimensional $\Delta \chi^2$ distribution obtained by profiling over all other parameters of the fit. 

In the $\big(1-C_V^2, {\rm Im}(C_U) \big)$ plane, Fig.~\ref{fig:hfit-ImCU-CV}, we see that relaxing the $|C_U|^2=1$ constraint does not affect much the $1-C_V^2$ parameter but enhances considerably the maximal possible value of ${\rm Im}(C_U)$ at 68\% and 95\% CL, as also seen in Table~\ref{tab:hfit-ImBound}. 
In fit i) the best fit point actually has ${\rm Im}(C_U)=0$ and $C_V=1.03$, which corresponds to a pure scalar.  
In fit ii) the best fit point has ${\rm Re}(C_U)=0.72$ and ${\rm Im}(C_U)=0.35$, which would indicate a CP-violating contribution with an overall reduced $|C_U|=0.8$, while $C_V$ remains very SM-like.  
(Note however the SM values ${\rm Im}(C_U)=1-C_V^2=0$ are perfectly compatible with data at the $1\sigma$ level.)
Moreover, fit ii) gives a 95.4\% CL upper bound of $1-C_V^2<0.26$ and ${\rm Im}(C_U)<0.64$. 
This is a direct bound on a possible CP-odd admixture to the observed 125.5~GeV state.

Without custodial symmetry, in fit iii),  the best-fit point corresponds to a non-vanishing value, ${\rm Im}(C_U)=0.52$,
which is even larger than the CP-even component,  ${\rm Re}(C_U)=0.42$. The reduced value $|C_U|=0.67$ is compensated by an enhanced $C_Z=1.15$ (which with custodial symmetry this does not work because $C_W$ is more constrained then $C_Z$). Overall, there is still an upper bound of ${\rm Im}(C_U)<0.69$ (0.80) at 95.4\% (99.7\%) CL. 

The correlation between the possible CP-even and CP-odd components of $C_U$, without constraining $|C_U|$ to unity, is illustrated in Fig.~\ref{fig:hfit-ImCU-ReCU}. Note in particular that a pure CP-odd coupling to the top quark, 
\textit{i.e.}\ ${\rm Re}(C_U)=0$,  is compatible with the experimental data at the 1.6$\sigma$ level in fit ii) and 
at the 1$\sigma$ level in fit iii). 
The SM values ${\rm Re}(C_U)=1$ and ${\rm Im}(C_U)=0$ are nonetheless still within the 68\% CL region. 
Note finally that the quality of the fits, {\it i.e.}\ the $\chi^2_{\rm min}/{\rm ndof}$, 
is roughly the same in all fits considered here.

\begin{table}[!htb]
\begin{center}
\begin{tabular}{|c|c|c|c|c|c|c|}
  \hline
   Fit & ${\rm Re}(C_U)$ & ${\rm Im}(C_U)$ & $C_D$ & $C_W$ & $C_Z$ & $\chi^2_{\rm min}/{\rm ndof}$\\
  \hline
  i) & 1.00 & 0.00 & 1.00 & \multicolumn{2}{c|}{1.03} & $17.43 / 22=0.79$ \\  
  \hline
  ii) & 0.72 & 0.35 & 1.00 & \multicolumn{2}{c|}{1.04} & $16.14 / 21=0.77$ \\  
  \hline
  iii) & 0.42 & 0.52 & 1.05 & 0.98 & 1.15 & $14.67 / 19=0.77$ \\  
  \hline
\end{tabular}
\end{center}
\vspace{-0.4cm}
\caption{Best-fit points and $\chi^2_{\rm min}/{\rm ndof}$ in the three different fits with CP-violating contributions.}
\label{tab:hfit-BestFit}
\end{table}

\begin{table}[!htb]
\begin{center}
\begin{tabular}{|c|c|c|c|}
  \hline
   Fit &  $1\sigma$ & $2\sigma$ & $3\sigma$ \\
  \hline
  i) & 0.28 & 0.50 & 0.72 \\  
  \hline
  ii) & 0.55 & 0.64 & 0.75  \\  
  \hline
  iii) & 0.61 & 0.69 & 0.80  \\  
  \hline
\end{tabular}
\end{center}
\vspace{-0.4cm}
\caption{Upper bounds on $|{\rm Im}(C_U)|$ at the 1, 2 and 3$\sigma$ levels 
when profiling over all other parameters in fits i)--iii).}
\label{tab:hfit-ImBound}
\end{table}

\section{PREDICTIONS FOR $H \to Z\gamma$}

\begin{figure}[t!]
\begin{center}
\includegraphics[scale=0.6]{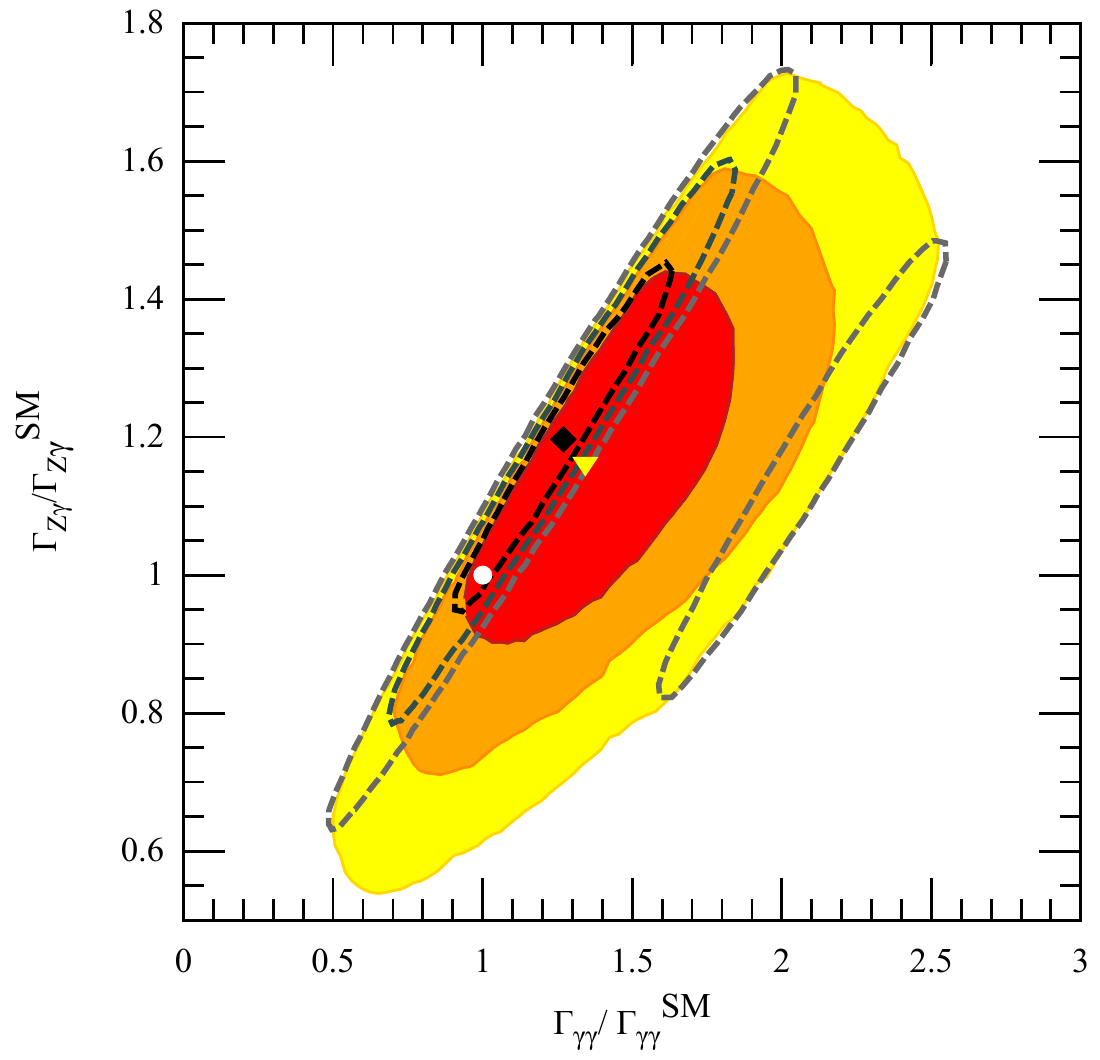}
\vspace{-0.2cm}
\caption{Fit of ${\rm Re}(C_U)$, ${\rm Im}(C_U)$ and $C_V$ shown in the $\left(\Gamma_{ \gamma\gamma}/\Gamma_{\gamma\gamma}^{\rm SM}, \ \Gamma_{Z\gamma}/\Gamma_{Z\gamma}^{\rm SM} \right)$ plane. 
The red, orange and yellow regions are the 68\%, 95.4\% and 99.7\% CL regions, respectively. The yellow triangle marks the best fit point, while the white point is the SM expectation. Overlaid are for comparison the  
68\% (dashed black), 95.4\% (dashed dark gray) and 99.7\% (dashed light gray) CL contours for the CP-conserving case with ${\rm Im}(C_U)=0$.}
\label{fig:hfit-HZg}
\end{center}
\end{figure}

Analogously to Eq.~(\ref{eq:hfit-widthsCP}), we can write the partial width of the $H\to Z\gamma$ decay, 
normalized to its SM expectation, as 
\begin{equation}
\frac{\Gamma( H \to Z\gamma)}{\Gamma( H \to Z\gamma)\vert_{\rm SM}}
 \simeq 
\frac{\big \vert 2 \frac{v_t}{c_W} {\rm Re}(C_U) B^+_{1/2}[m_t] + C_W B^+_1[m_W] \big \vert^2 + 4\big \vert 2 \frac{v_t}{c_W} {\rm Im} (C_U) B^-_{1/2}[m_t] \vert^2}
{\big \vert 2 \frac{v_t}{c_W} B^+_{1/2}[m_t] + B^+_1[m_W] \big \vert^2} \;,
\end{equation}
with $v_t=2I^3_t-4Q_t s_W^2$, $B^+_{1/2}[m_t] \simeq -0.35$, $B^+_{1}[m_W] \simeq 5.78$, and $B^-_{1/2}[m_t] \simeq 0.54$ for $m_H = 125.5$~GeV, see Refs.~\cite{Djouadi:2005gi,Djouadi:2005gj}. 
This allows us to make a prediction for $H\to Z\gamma$ based on the fits presented above. 
As an example, Fig.~\ref{fig:hfit-HZg} illustrates how the strong correlation between $H \to \gamma\gamma$ and $H \to Z\gamma$ is relaxed when allowing for a non-vanishing ${\rm Im}(C_U)$, which affects the $\gamma\gamma$ mode more strongly than the $Z\gamma$ mode.

\section{CONCLUSIONS AND OUTLOOK}

Using all publicly available results from the LHC and Tevatron experiments,
we investigated the extent to which custodial symmetry and CP violation can be tested 
in a global Higgs coupling fit. We found that the equality between the reduced 
$C_W$ and $C_Z$ couplings and their compatibility 
with the SM value of unity hold at 68\% CL.
However, this analysis  does not exclude values of $C_W, C_Z$ greater than 1 
and the possibility that the observed $H$ boson contains a component coming from Higgs representations higher than doublets or singlets thus still remains.
Moreover, we observed that the present experimental data allows for a sizeable CP-odd contribution, ${\rm Im}(C_U)$,  in the coupling to the top quark, although the SM still remains compatible with the data at 68\% CL. A limit of $1-C_V^2<0.26$ at 95.4\% CL has been derived for the CP-odd admixture to the 125.5~GeV observed state. Alternatively, this means that a pure CP-odd component acting as the observed state is excluded at more than $4\sigma$. From the fermionic sector, however, we found that a pure CP-odd component is compatible at the $1\sigma$ level with the experimental data while the SM CP-even component lies in the same region as well.
In addition, when allowing for a non-vanishing ${\rm Im}(C_U)$ we found that the correlation between $\gamma\gamma$ and $Z\gamma$ decay modes is significantly relaxed.

It is important to recognize that the kinds of fits that we have performed on the observed $H$ boson data do not take into account constraints that may be present in the context of particular models.  For example, in the case of two Higgs doublet models (2HDMs) plus singlets, the CP-violating component of the $Htt$ coupling, $a_t$ (Eq.~(\ref{eq:hfit-CP-Hff})), must be zero if $C_V=1$.  More generally, sum rules \cite{Grzadkowski:1999ye,Grzadkowski:1999wj} imply that the size of $a_t$ is limited by the extent to which $C_V$ deviates from unity. Thus, it could happen that more precise data will yield fits to the properties of the $H$ boson that violate the 2HDM+singlets sum rules, thereby requiring higher Higgs representations, even if $C_V=1$ with high precision.  In any case, in order to discuss fits within a given model context, it is necessary to employ a parametrization of the various $C_i$ of Eq.~(\ref{eq:hfit-HL}) appropriate to that model, which will automatically guarantee that all sum rules and related constraints are enforced. For 2HDMs, this was done in Ref.~\cite{Cheung:2013rva}. 

Last but not least we note that two of us are currently developing a new version of our code for fitting the Higgs likelihood, which is entirely written in {\tt Python} and is intended for public release in Spring 2014~\cite{hlike}.
This new code is modular and has all the experimental results stored in a flexible {\tt XML} database that includes the full likelihoods in the 2D plane $(\mu_{\rm ggF+ttH}, \mu_{\rm VBF+VH})$ when available \cite{ATLAS-data-Hgamgam,ATLAS-data-HZZ,ATLAS-data-HWW} instead of a Gaussian approximation. The new code moreover provides a rather general and user-friendly way of specifying the model input in terms of reduced couplings, or cross sections and branching fractions, or signal strengths directly. In all cases, there is the option to include the effects of invisible or undetected decay modes in the fit.

\section*{ACKNOWLEDGEMENTS}

This work was supported in part by the PEPS-PTI project ``LHC-itools'', by IN2P3 under contract PICS FR--USA No.~5872 and by the US DOE grant DE-SC-000999.  J.B.\ is supported by the ``Investissements d'avenir, Labex ENIGMASS''.



%% file: vbsh/wlwlLH.tex



\chapter{Probing Higgs Physics with Vector-Boson Scattering}

{\it A.~Belyaev, E.~Boos, V.~Bunichev, 
 Y.~Maravin, A.~Pukhov, R.~Rosenfeld, M.~Thomas}


%


\begin{abstract}
  We suggest the combination of two main observables
which provides a unique sensitivity to  the ratio of the longitudinal versus 
transverse polarizations of the $W$ and $Z$ bosons in the vector-boson scattering processes.
Therefore, the analysis we present also provides the sensitivity to the Higgs boson 
couplings to the gauge bosons and consequently to the theory underlying the Higgs sector.
We conclude that the analysis of vector boson fusion provides a model independent and robust
method to study the  Higgs boson couplings to the gauge bosons.

\end{abstract}


\section{Introduction} 

The historical discovery of a Higgs-like particle at the LHC \cite{Aad:2012tfa,Chatrchyan:2012ufa}  ushered a new era in 
the determination of the properties of the electroweak symmetry breaking (EWSB) sector.
Since the longitudinal polarizations of the electroweak gauge bosons ($V_L$'s, $V= W^{\pm}, Z$) have their origin in the EWSB sector, 
determining their interactions, especially in the study of $V_LV_L$ scattering, is of fundamental importance to unravel the mechanism of EWSB.
In particular, we now know that $V_LV_L$ scattering is at least partially unitarized by the Higgs-like particle.

This importance has been known for many years. The first calculations of $V_L V_L$ scattering were performed in the context of the so-called 
Effective W Approximation (EWA) \cite{Dawson:1984gx,Chanowitz:1985hj,Borel:2012by}  with the use of the Equivalence Theorem (ET) \cite{Cornwall:1974km}, 
that states that at high energies the amplitudes for $V_L V_L$ scattering
can be calculated using the corresponding degrees of freedom in the EWSB sector.
The first realistic study of  $V_L V_L$ scattering in a strongly coupled EWSB sector but assuming ET and EWA and adopting 
several unitarization prescriptions were performed in the 1990's \cite{Bagger:1993zf,Bagger:1995mk} (see also \cite{Butterworth:2002tt}). 
Basic techniques such as as forward jet tagging, central jet vetoing, and cuts on the transverse momenta were introduced
to select processes with vector boson fusion (VBF).
The first studies that went beyond the EWA performing a complete calculation of $WW$ scattering  were \cite{Belyaev:1998ih,Ballestrero:2008gf}.

One of the most difficult issues in extracting the physics of EWSB from $VV$ scattering is the so-called transverse pollution, that is, 
the irreducible background coming from the transversely polarized gauge bosons. Much work has been done to devise cuts that can
reduce the transverse pollution and this is the subject of this contribution. Below we start with a brief review of the most recent developments
in these efforts.

In the gauge boson rest frame the distribution of transverse and longitudinal polarizations are given by:
\begin{equation}
P_{\pm} (\cos \theta^\ast) = \frac{3}{8} (1 \pm \cos \theta^\ast) ^2, \;\;\;\;\;
P_{L} (\cos \theta^\ast) = \frac{3}{4} (1 - \cos^2 \theta^\ast), 
\end{equation}
where  $\theta^\ast$ is the angle between the fermion in the decay products and the gauge boson boost to its rest frame.
It is worth clarifying the meaning of the the $P_\pm$ 
symbol in connection to the transverse polarisation of the vector boson 
and the polarisation of the respective fermion from its decay
which defines the $\theta^\ast$ angle.
The $P_+$ distribution takes place when the polarisation of the vector boson $V$ and the 
respective fermion $\psi$  {\it coincide}, i.e. $P_+$ corresponds to
the ($V_{left},\psi_{left}$)  or ($V_{right},\psi_{right}$) case.
On the other hand, the $P_-$ distribution  takes place when the polarisation of the vector boson $V$ and the 
respective fermion $\psi$  are {\it opposite}, i.e. $P_-$ corresponds to
the  ($V_{left},\psi_{right}$) or ($V_{right},\psi_{left}$) case.

Han {\it et al.} \cite{Han:2009em} proposed to directly reconstruct the 4-momenta of the decay products of the gauge boson and hence measure the
$\cos \theta^\ast$ distribution and fit it to
\begin{equation}
P(\cos \theta^\ast) = f_L P_L(\cos \theta^\ast) + f_+P_+(\cos \theta^\ast) + f_- P_-(\cos \theta^\ast)
\label{eq:fit}
\end{equation} 
with $f_L + f_- + f_+ = 1$. They show that the fit is robust against full hadronization.
Doroba {\it et al.} \cite{Doroba:2012pd} proposed a new variable to isolate $W_L W_L$ scattering in same-sign $WW$ production.
This variable is related to the observation that $W_L$'s tend to be emitted at smaller angles with respect to the initial quarks
and hence the final quarks are more forward. 
Therefore, they require a small transverse momenta of the forward jets in order to improve $W_T$ rejection.
The jet substructure techniques used by Han {\it et al.} to reconstruct hadronically decaying gauge bosons were recently further
improved in \cite{Cui:2013spa}, where a multivariate $W$ jet tagging method is employed. They also used the cuts suggested by 
Doroba {\it et al.} \cite{Doroba:2012pd}.
Freitas and Gainer \cite{Freitas:2012uk} showed that the significance of the VBF signal can be increased by using the matrix element
method but further investigation including showering and detector simulation is still required to quantify their findings.
More recently, Chang {\it et al.} \cite{Chang:2013aya} used $WW$ scattering to study the sensitivity to additional Higgs bosons in a complete
calculation without relying on EWA, employing the usual selection cuts to maximize the VBF contribution.
\begin{figure}[htb]
\includegraphics[width=0.5\textwidth]{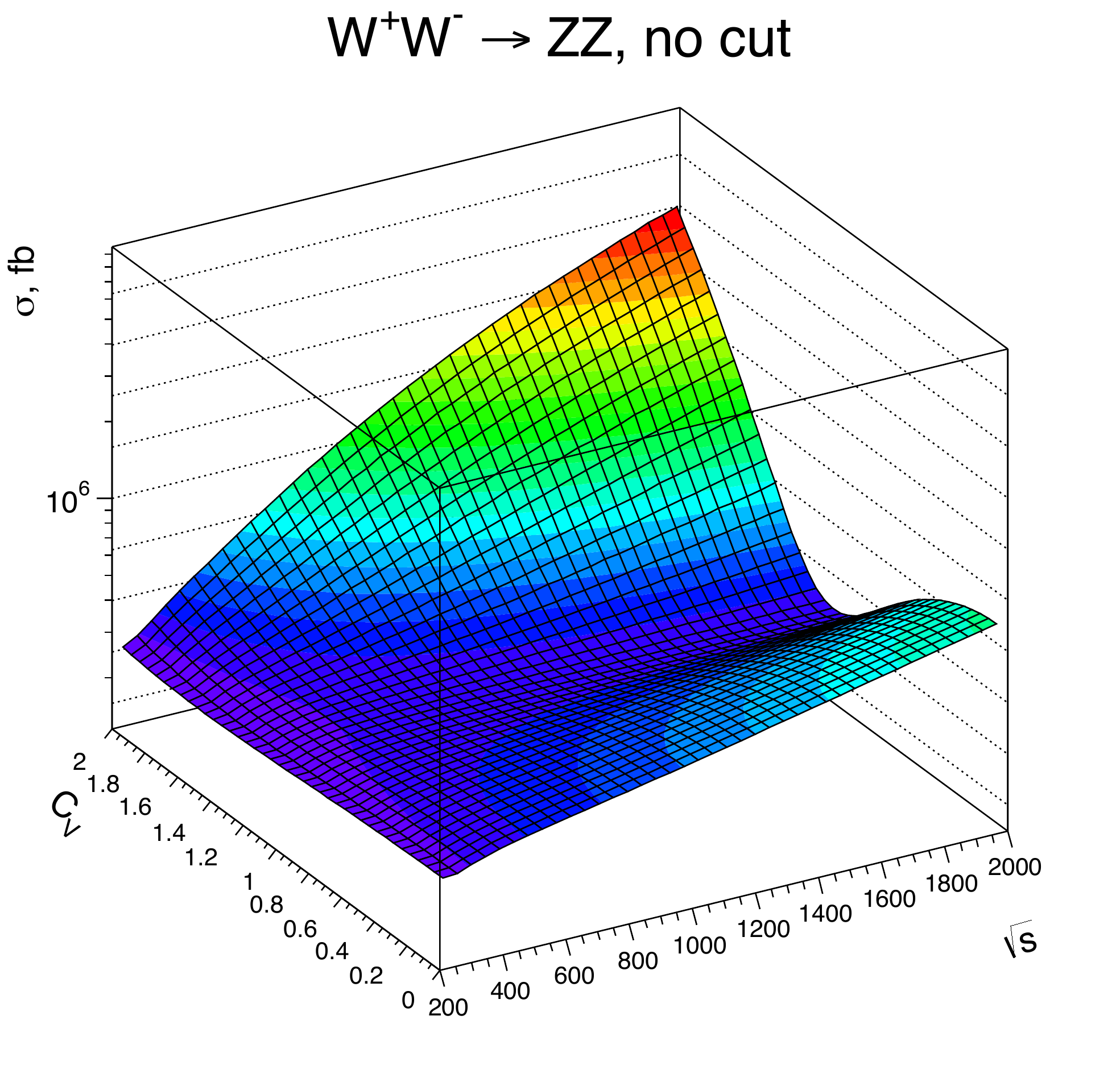}%
\includegraphics[width=0.5\textwidth]{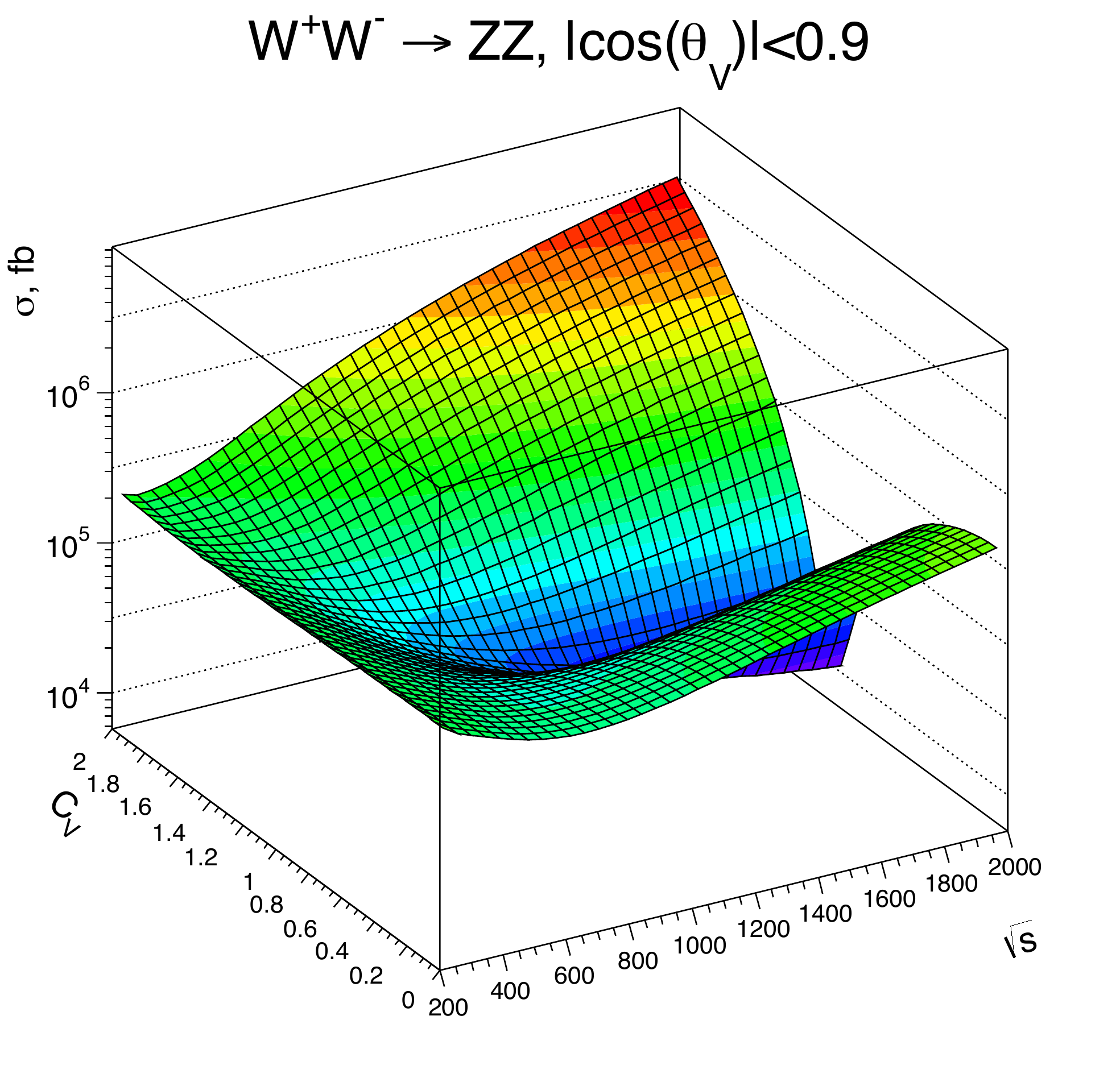}\\
\caption{Total $W^+ W^- \to ZZ$ without cuts (left plot) and with  $|\cos(\theta_Z)|<0.9$ (right plot) as a function of center-of-mass energy $\sqrt{s}$ and
anomalous coupling $c_V$.
\label{fig:WW_ZZ_cv}}
\end{figure}

In this Les Houches contribution, we study the sensitivity of $VV$ scattering to new physics. We focus on possible deviations
of the Higgs coupling to electroweak gauge bosons from its SM value, parameterized by an anomalous coefficient $c_V$, normalised such that $c_V = 1$ in the SM.
This corresponds to adding the contribution of higher dimensional effective operators to the SM lagrangian, such as 
$\partial_\mu (H^\dagger H) \partial^\mu (H^\dagger H)$ (see, {\it e.g.}, \cite{Giudice:2007fh}). 
The extreme case of a Higgsless case corresponds to $c_V=0$.
In particular, our goal is to devise optimal cuts capable of selecting the contribution from the longitudinally polarized gauge bosons and hence
increasing the sensitivity to $c_V$.

One important  point to stress is that the current way to measure $c_V$ is from direct Higgs production in gluon fusion through the decay $H \to V V^\ast$, which is somewhat model dependent because of the loop-induced gluon-gluon-Higgs coupling. In contrast, the method proposed here relies only on $VV$ scattering 
and therefore the measurement of $c_V$ in this case is more model-independent.  

In order to motivate our work, we show in the left panel of Fig.~\ref{fig:WW_ZZ_cv}
the total cross section for  $W^+ W^- \to ZZ$ as a function of center-of-mass energy $\sqrt{s}$ and
anomalous coupling $c_V$ without any cuts. One can see an increase for $c_V \neq 1$ due to the
non-cancellation of the growth with energy of $V_LV_L$scattering. This behaviour is enhanced when 
a simple angular cut such as $|\cos(\theta_Z)|<0.9$ is applied, since it tends to select the longitudinal
polarizations, as demonstrated in  the right panel of the figure.

This contribution is organized as follows. In the next Section we discuss in a parton level analysis 
the selection criteria we propose to implement in order to enhance the contribution from
the longitudinally polarized gauge bosons. In Section 3 a preliminary analysis is performed to understand
whether the efficiency of the proposed criteria survive at the full $2 \to 6$ level at the LHC. Finally, 
we conclude the study in Section 4.

\section{Analysis at the $VV \to VV$ level\label{sec:wlwl-parton}}

Let us consider the properties of  vector boson scattering 
in the  $VV \to VV$ process, where $V=W^\pm, Z$.
\begin{figure}[htb]
\centering
\includegraphics[width=0.5\textwidth]{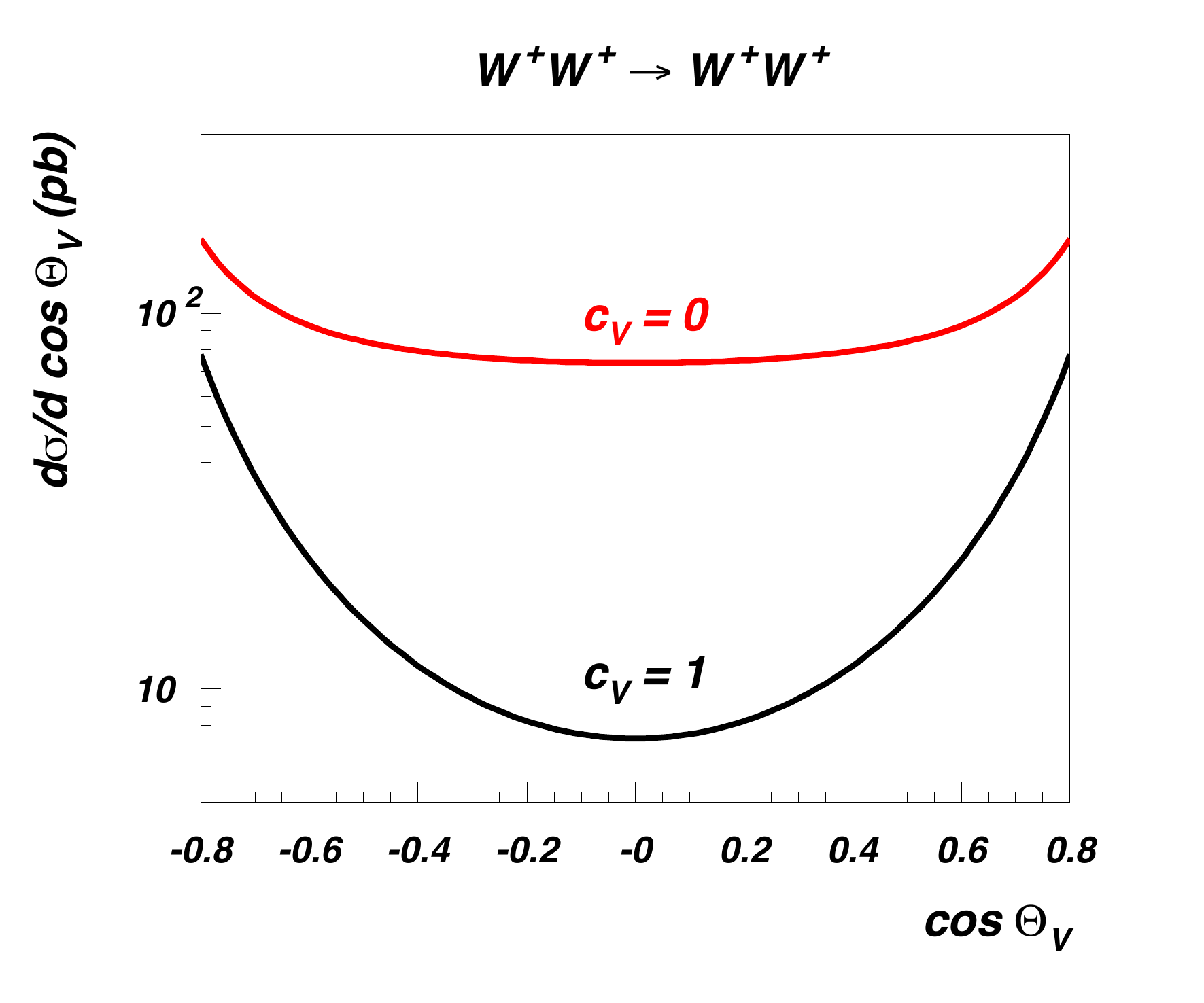}%
\includegraphics[width=0.5\textwidth]{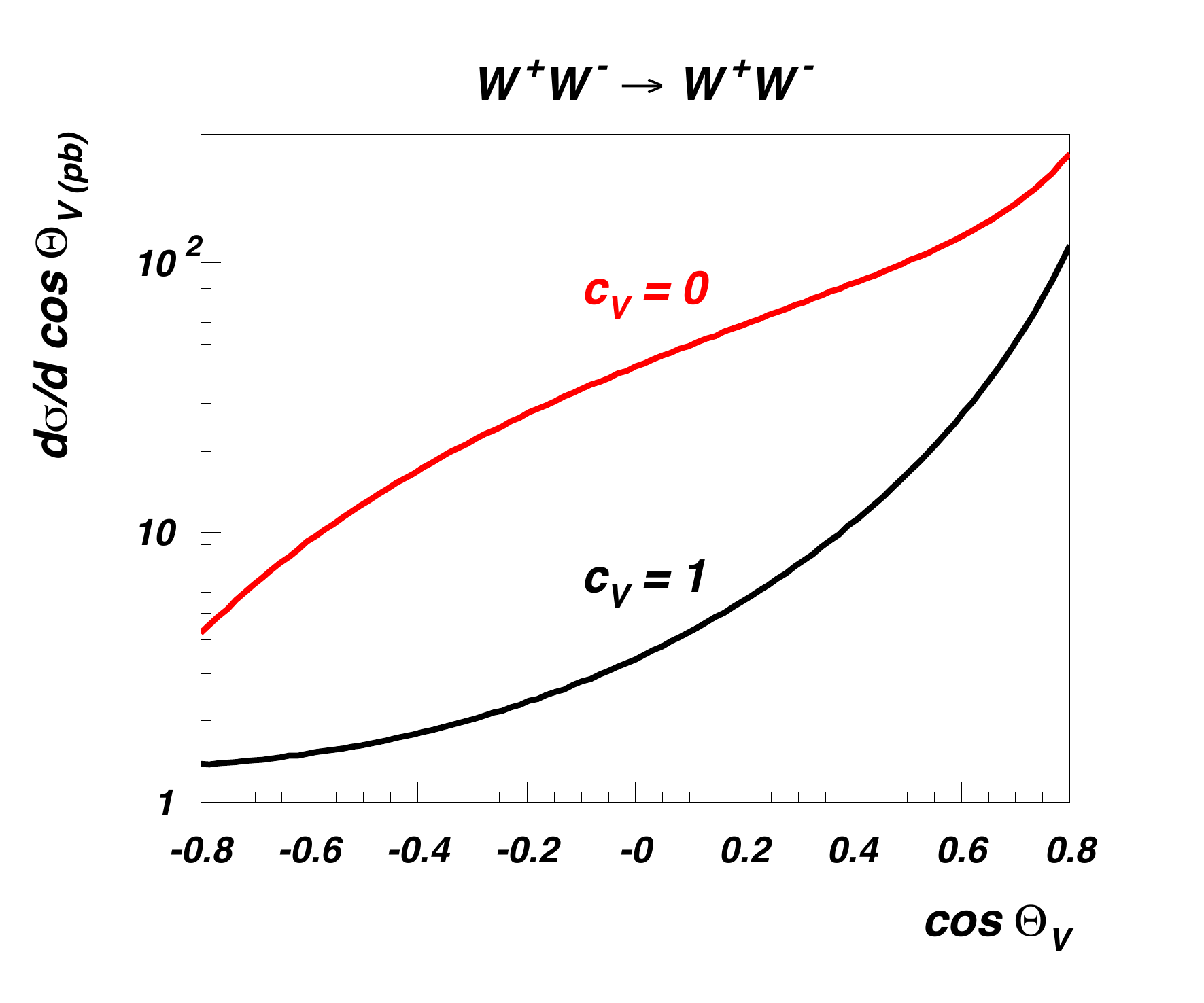}\\
\vskip -1cm
\includegraphics[width=0.5\textwidth]{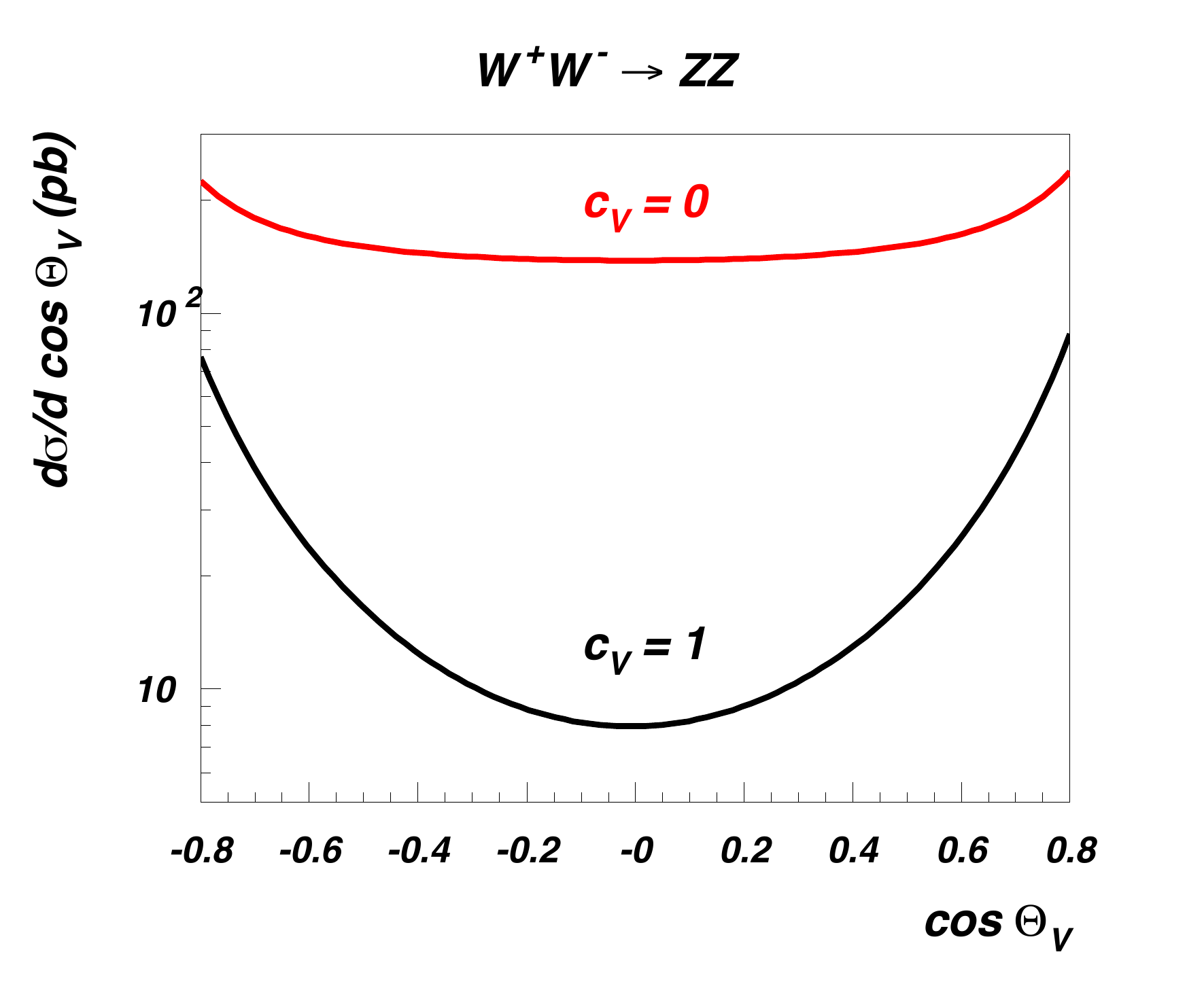}%
\includegraphics[width=0.5\textwidth]{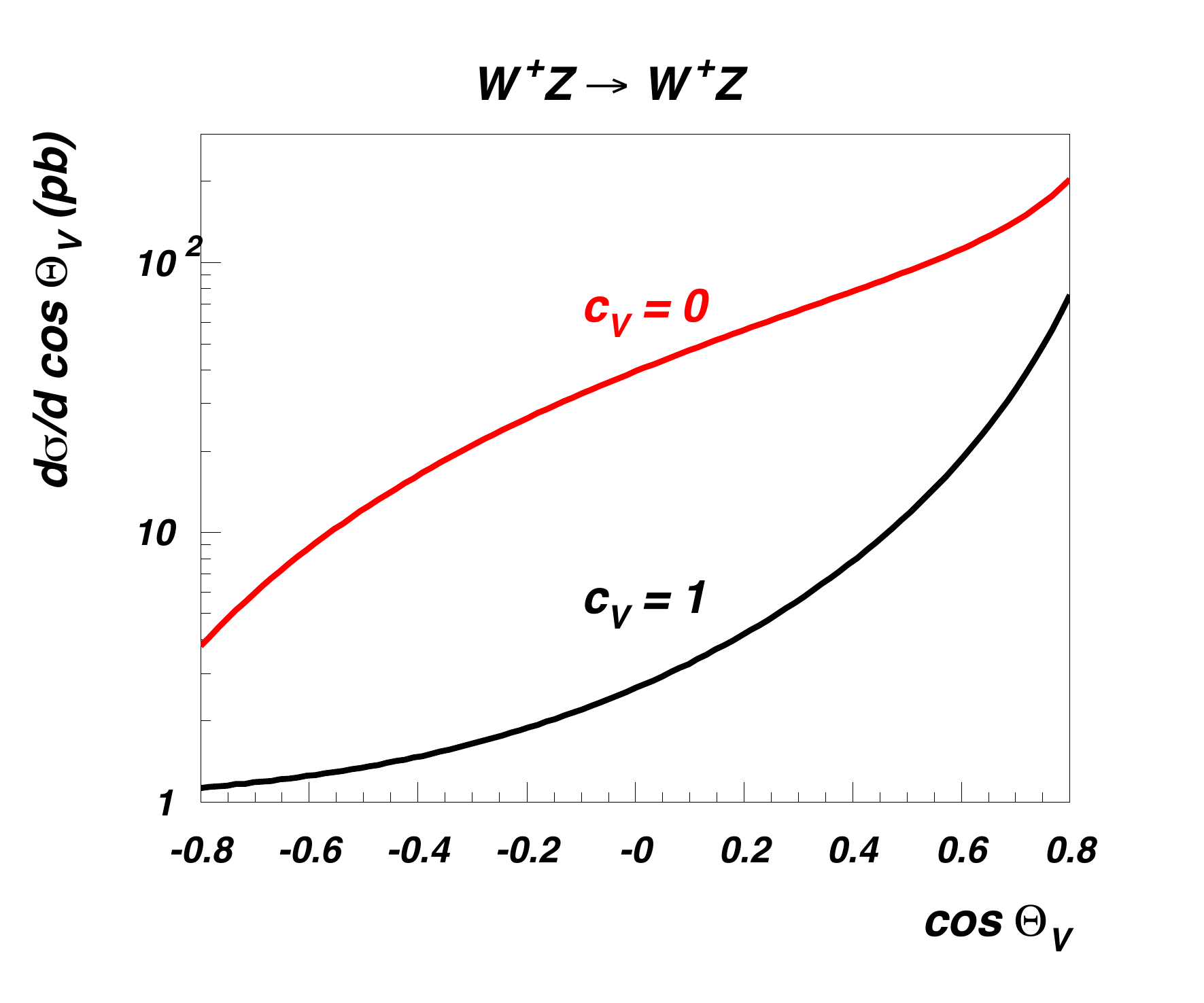}
\caption{$\cos\theta_V$ angular distributions for $VV\to VV$ process for $\sqrt{s}=1$~TeV
with (black  curves, $c_V=1$, SM case) and without Higgs boson (red curves, $c_V=0$).
\label{fig:VV_VV_C}}
\end{figure}
In Fig.~\ref{fig:VV_VV_C}
we present differential cross section for  $VV \to VV$ process
with respect to the scattering angle of the gauge boson $V$
for $\sqrt{s}=1$ TeV.
\begin{figure}[htb]
\centering
\includegraphics[width=0.65\textwidth]{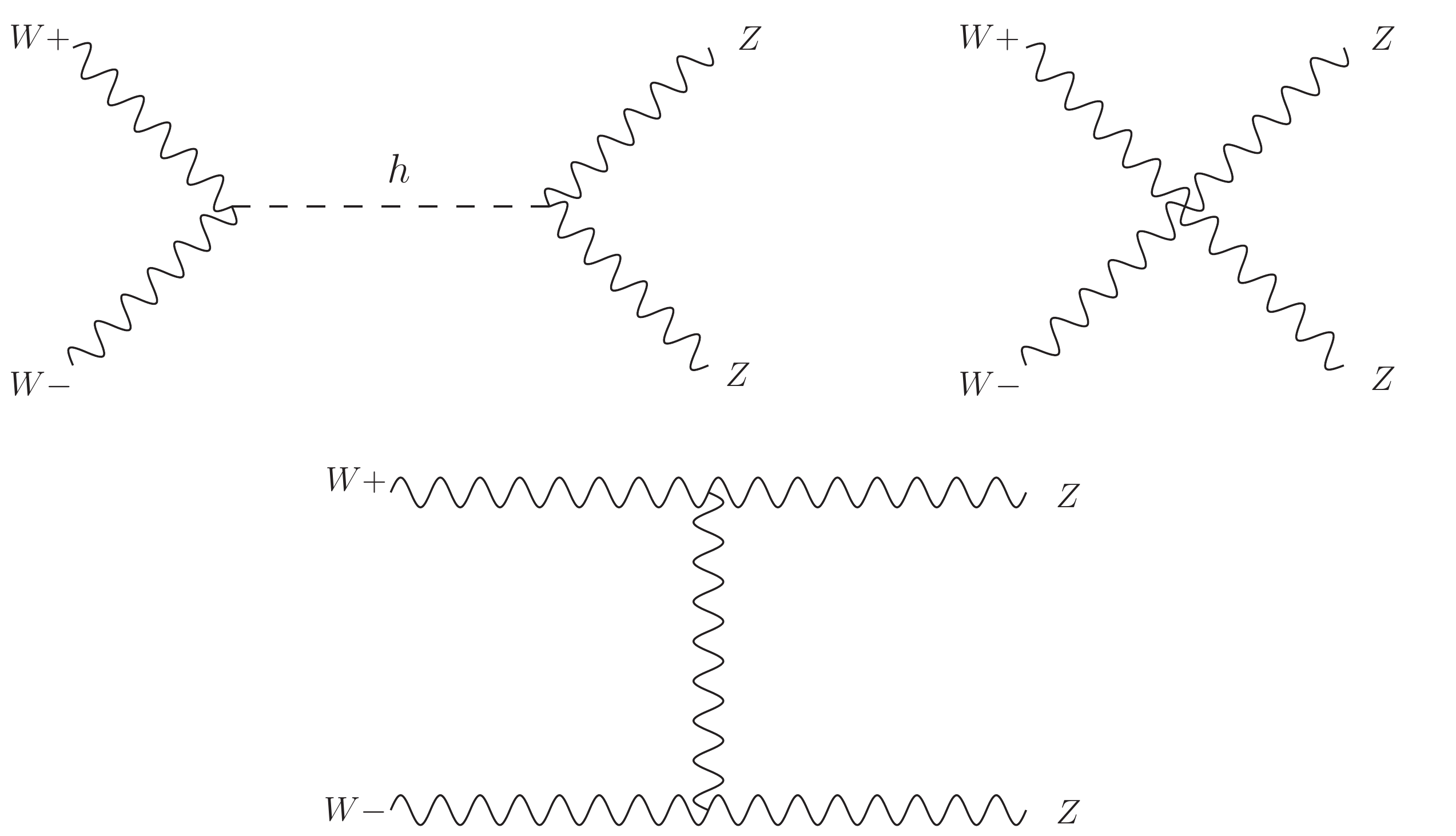}%
\caption{Diagrams contributing to the $W^+, W^- \rightarrow Z, Z$ process.
\label{fig:diags}}
\end{figure}
Due to the t- and u-channels corresponding to the exchange of an electroweak gauge boson 
(as exemplified in Fig.~\ref{fig:diags}) or a Higgs boson,
the angular distributions are peaked in the forward-backward directions, as seen in Fig.~\ref{fig:VV_VV_C}. 
The processes on the left plots
have symmetric final state and hence the distributions are symmetric, while this is not the case for the right plots.
The incoming $W^+$ boson tends eventually to scatter in the forward direction, as shown in the right plots, 
and the incoming $W^-$ or $Z$ bosons tend to scatter in the backward direction
because of the presence of the $t$-channel diagrams.
\begin{figure}[htb]
\includegraphics[width=0.5\textwidth]{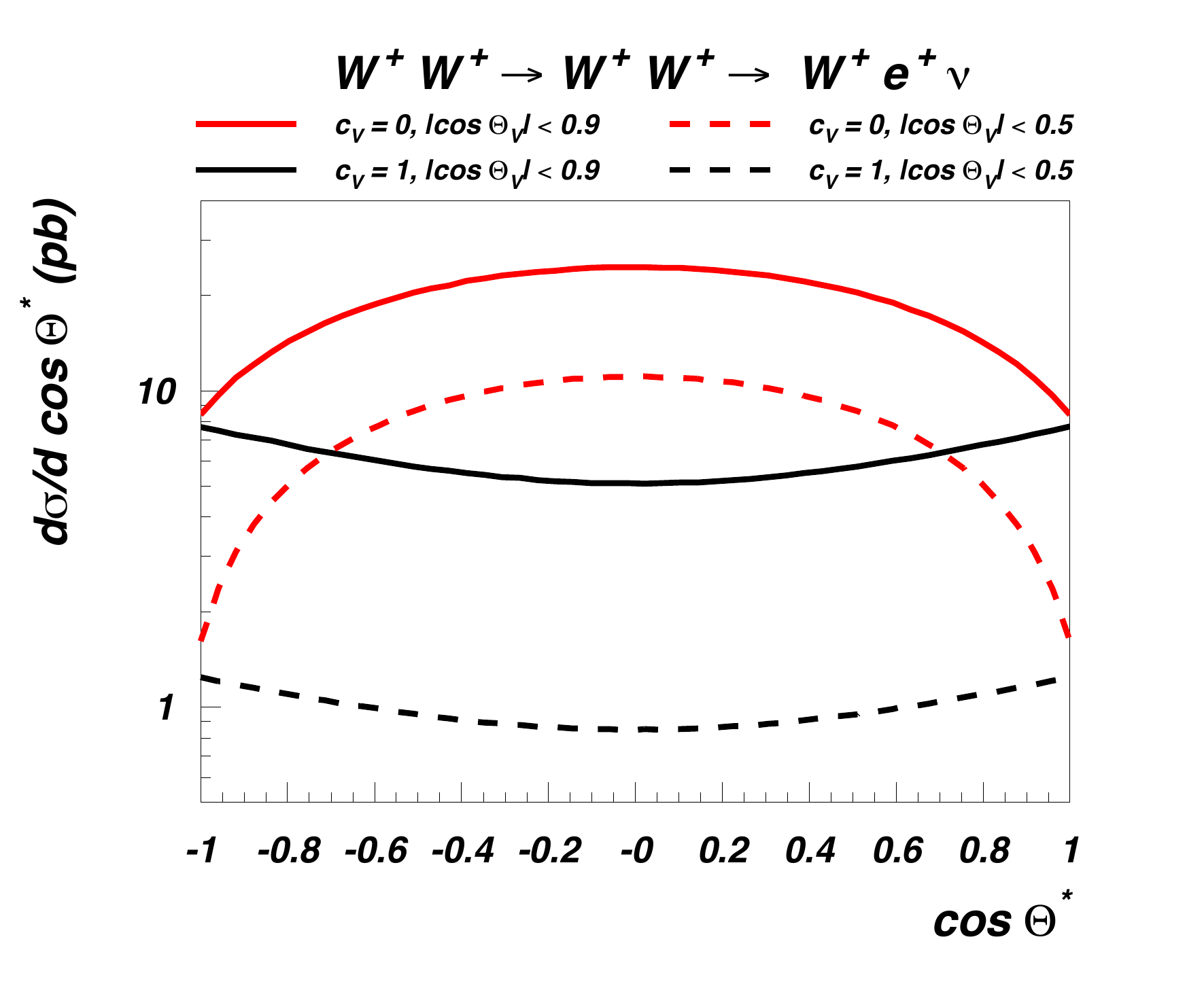}%
\includegraphics[width=0.5\textwidth]{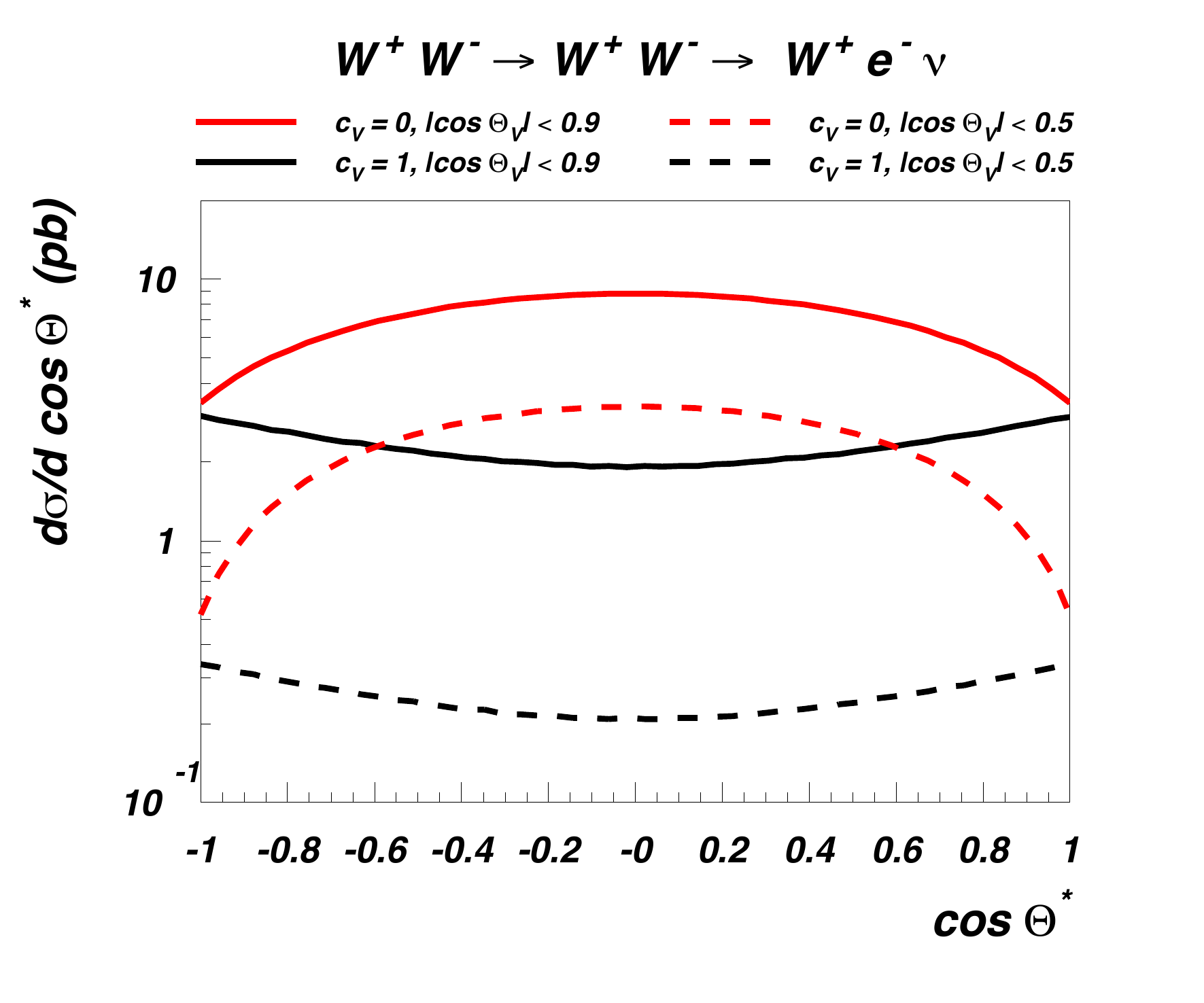}\\
\vskip -1cm
\includegraphics[width=0.5\textwidth]{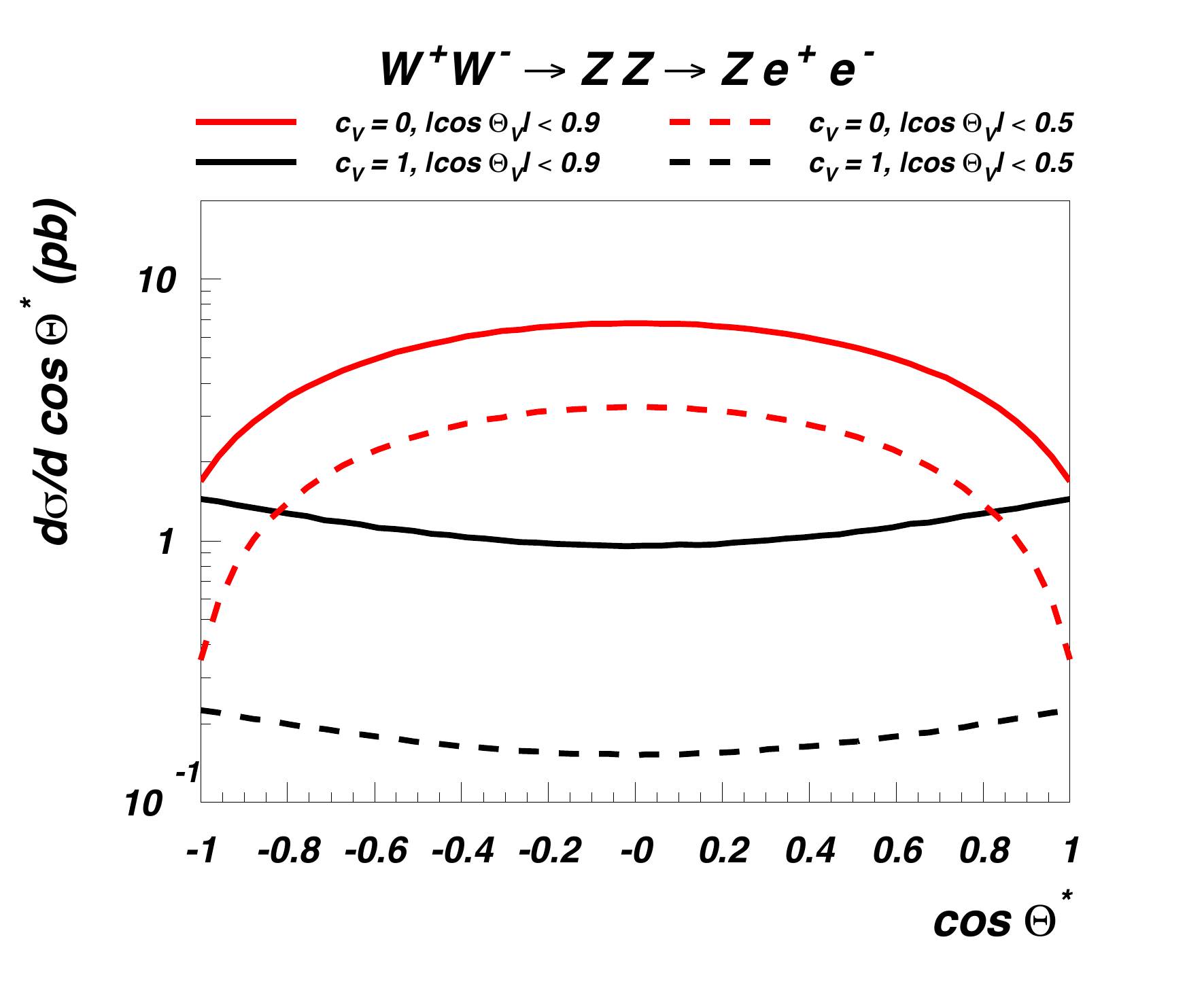}%
\includegraphics[width=0.5\textwidth]{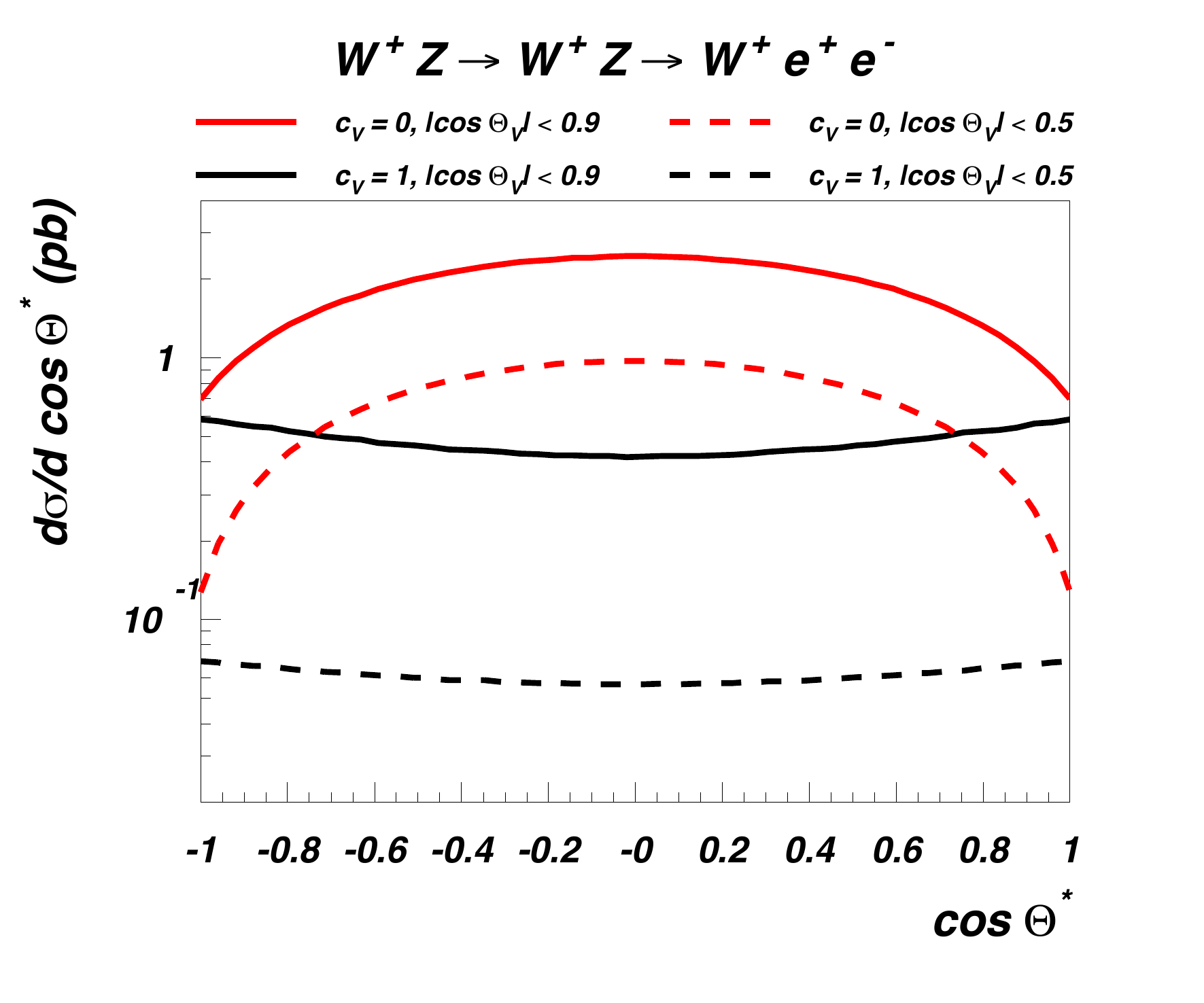}
\caption{$\cos\theta^\ast$ angular distributions for $VV\to VV \to V ll(Vl\nu)$ process for $\sqrt{s}=1$~TeV
with (black  curves, $c_V=1$, SM case) and without Higgs boson (red curves, $c_V=0$).
}
\label{fig:VV_VV_P}
\end{figure}
In the absence of a Higgs boson ($c_V=0$),
the relative contribution from the longitudinally polarized gauge bosons increases resulting in a larger cross section and an angular 
distribution less depleted in the central region, as shown by the red curves in Fig.~\ref{fig:VV_VV_C}.
This provides the first important observation:
the forward-backward regions in $VV \to VV$ scattering are mainly related to the transversely polarized gauge bosons
while the differences in the longitudinally polarized bosons are mainly pronounced in the central region.

In order to proceed further we decay one of the final state EW bosons and analyse the angular distribution in $\cos\theta^\ast$
of the charged fermion, which is sensitive to the degree of polarization of the parent EW gauge boson.
The results for two different angular cuts in $\cos\theta_V$ are shown in Fig.~\ref{fig:VV_VV_P}.
The difference between the SM and a higgsless model is more dramatic, with a change in the shape of the distributions, and it
increases as the angular cut becomes tighter, since we are selecting more longitudinally polarized gauge bosons in this case.
This observation makes this variable a very important one to discriminate between longitudinal and transversely polarized
gauge bosons. 
\begin{figure}[htb]
\includegraphics[width=0.5\textwidth]{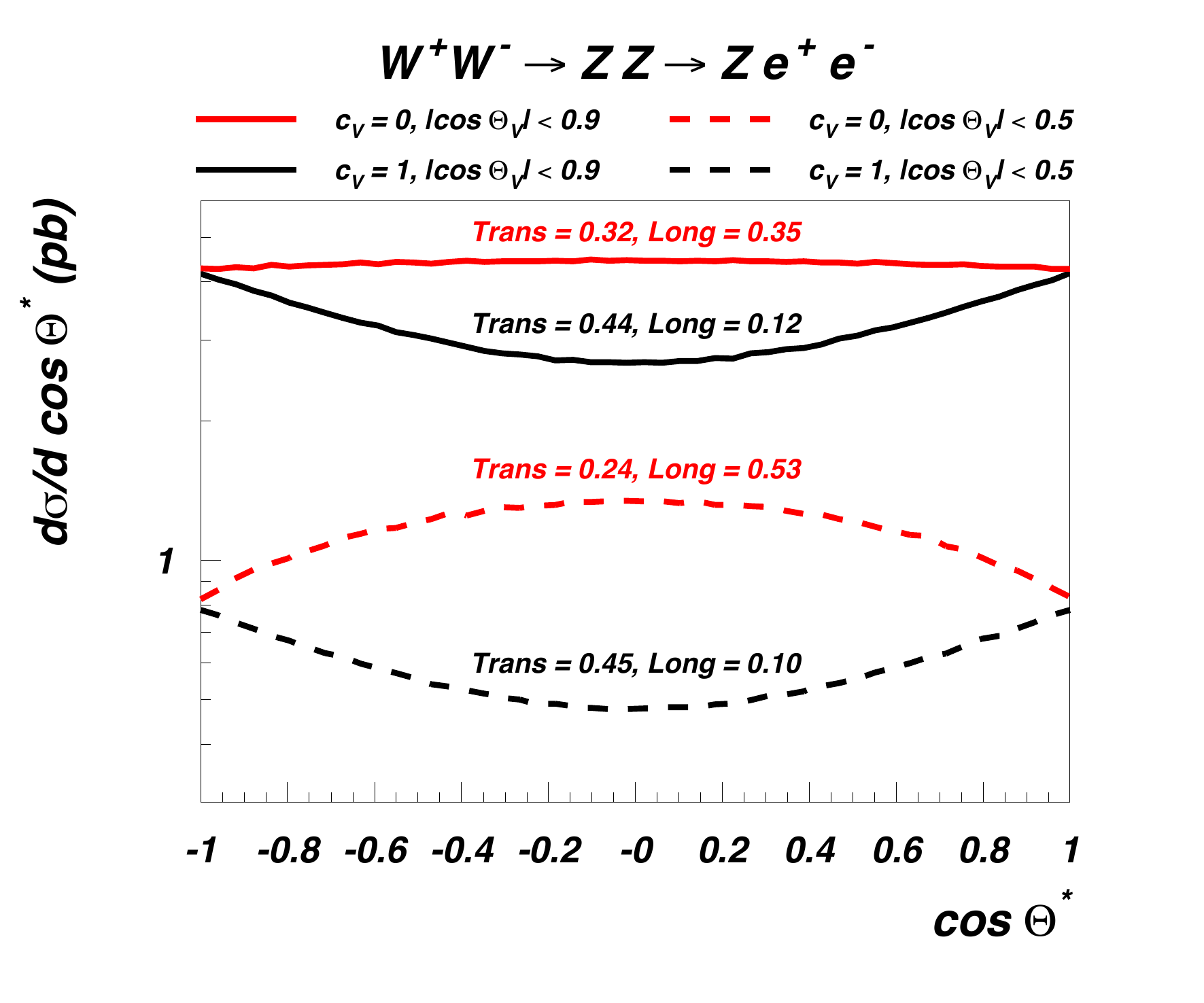}%
\includegraphics[width=0.5\textwidth]{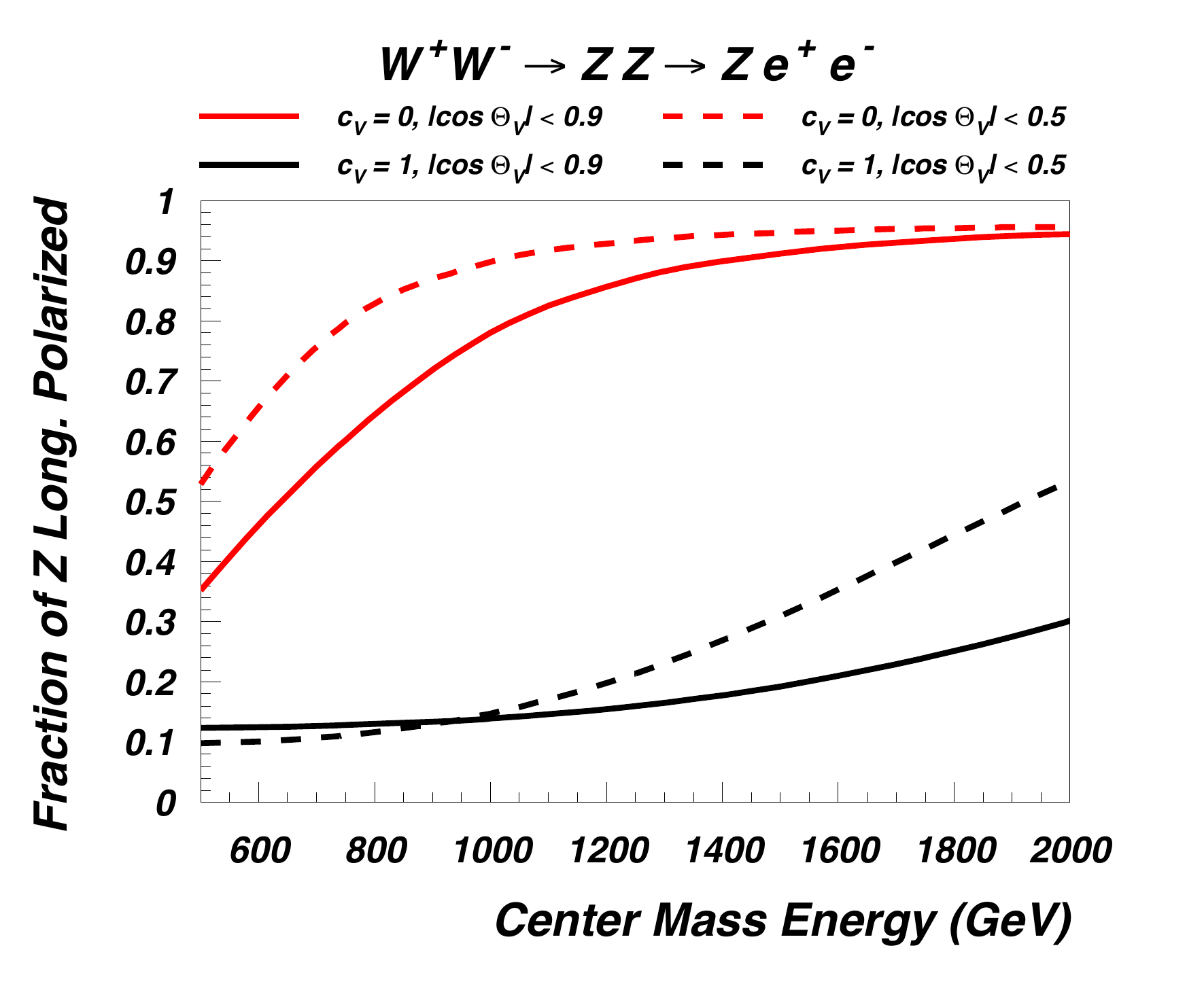}
\caption{Left: Fitting parameter results of the $Z$ boson polarization at $\sqrt{s}=0.5$~TeV (Left and Right polarizations were fitted separately but results were identical to accuracy given in plot.) Right: Fraction of the $Z$ bosons which are longitudinal as a function of energy for $c_V = 0,1$ with $|\cos\theta_V|<0.9$ and $0.5$.
\label{fig:VV_VV_fit}}
\end{figure}
In  Fig.~\ref{fig:VV_VV_fit} we recover the degree of polarization of the parent EW gauge boson by performing
a fit using Eq.~\ref{eq:fit}. In the left plot we show that in the higgsless case, one can enrich the fraction
of longitudinally polarized gauge bosons by performing cuts on $\cos\theta_V$, at the expense of 
reducing the number of events of course. In the right plot we show
how the fraction of longitudinally polarized gauge boson increases with the center-of-mass energy, reaching
almost 100\% in the $c_V=0$ case.


\section{LHC sensitivity to longitudinal vector boson scattering and Higgs boson couplings to gauge bosons}

In this section  we study the LHC sensitivity to probe the 
fraction of longitudinal polarisation of the gauge bosons
produced in the vector boson fusion (VBF), which will provide in  turn the sensitivity
to the Higgs boson couplings to gauge bosons.
In this contribution and in this section 
in particular we present our study
on the process
$pp\to jj Z Z \to e^+ e^- \mu^+ \mu^- j j$
($p=u,\bar{u},d,\bar{d}$, $j=u,\bar{u},d,\bar{d}$)
and the respective LHC sensitivity to
$HVV$ coupling.

The matrix element  for the  complete set of diagrams for this process
was evaluated using MADGRAPH package~\cite{Alwall:2011uj}, which was also used for the event  simulation.
For this parton level study we have used the following initial kinematic cuts:
\begin{eqnarray}
\mbox{Acceptance cuts:} & p_T^j>30~\mbox{GeV},\  |\eta_j|<4.5  & \nonumber\\
                      & p_T^e>20~\mbox{GeV},\  |\eta_e|<2.5   & \nonumber\\
                      & p_T^\mu>20~\mbox{GeV},\  |\eta_e|<2.5 &  	\\
\mbox{VBF cuts:\cite{He:2007ge}} & \Delta\eta_{jj}>4,\ E_j>300~\mbox{GeV}  & \\
\mbox{Z boson ID  cuts:} & |M_{ee,\mu\mu}-M_Z|\le 10~\mbox{GeV}   & 
\end{eqnarray}
In our calculations we have been using 
CTEQ6L1 PDF parameterisation and QCD scale 
fixed to $M_Z$.
The cross section for this process was found to be 0.0298 fb 
including Higgs boson exchange and  0.0362 fb 
when Higgs boson contribution was removed.
The cross section of this process is quite low since it includes leptonic decay for both
$Z$-bosons. For the case of semi-leptonic decays of $Z$-bosons, the cross section is about 40 
times larger. Moreover one can estimate that including $WW$ and $WZ$ processes with semi-leptonic
decay would lead to the event rate which is about a factor of 250 higher than the one 
mentioned above for $pp\to jj Z Z \to e^+ e^- \mu^+ \mu^- j j$.
We will keep this in mind when estimating prospect of our results.

\begin{figure}[htb]
\includegraphics[width=0.5\textwidth]{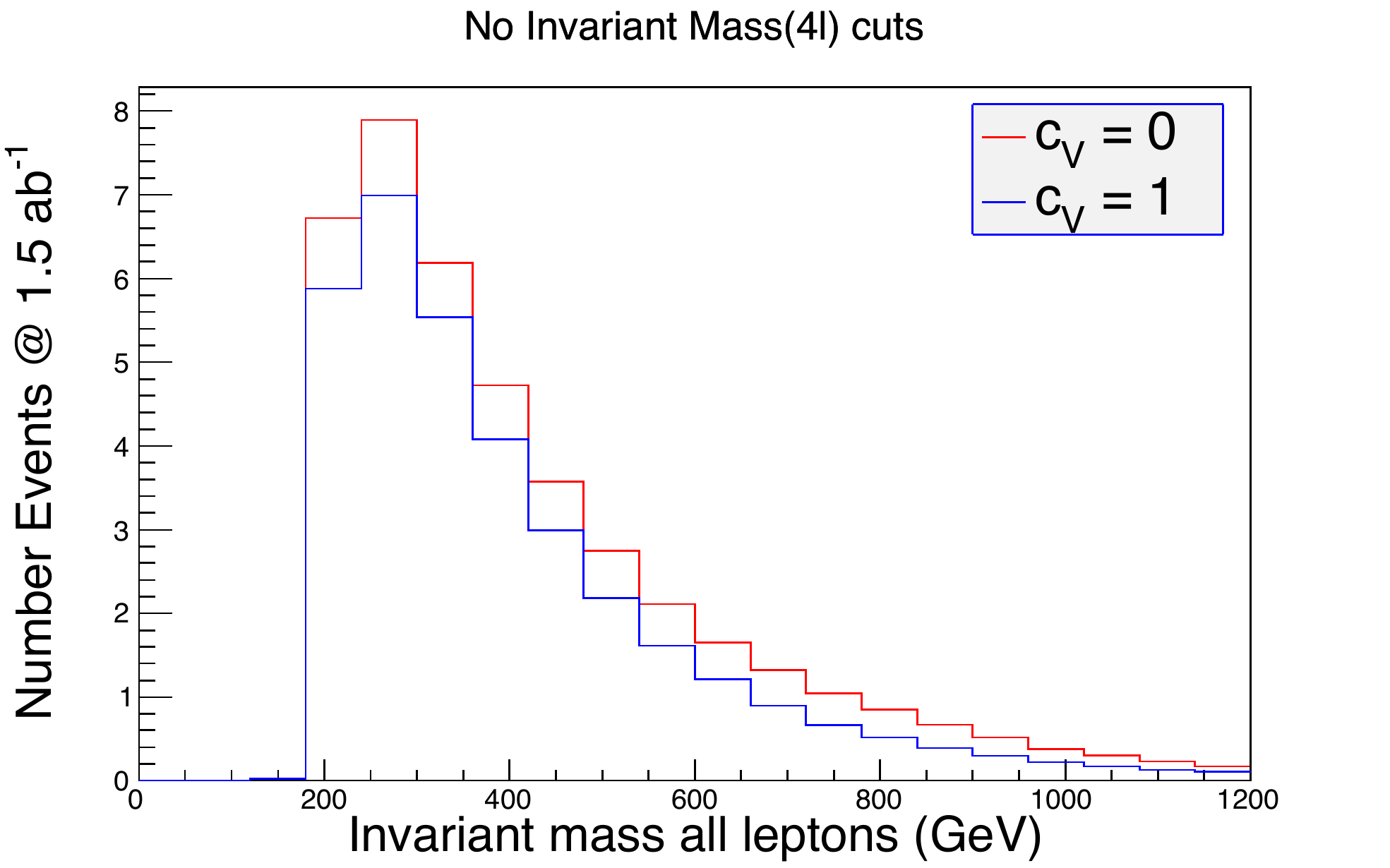}%
\includegraphics[width=0.5\textwidth]{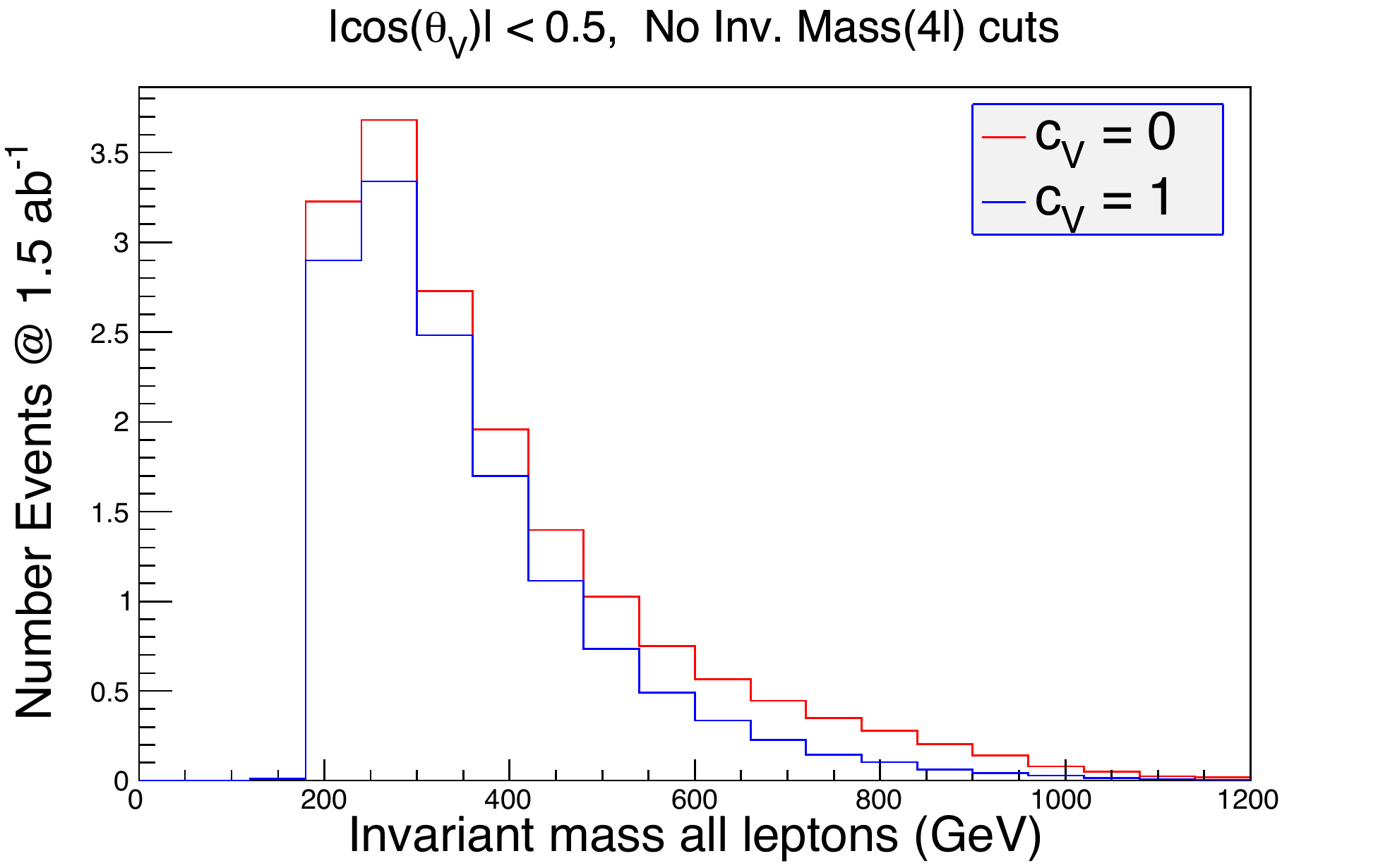}\\
\caption{Four-lepton invariant mass distribution, $M_{4l}$ for 1.5 ab$^{-1}$ @LHC13TEV
representing the invariant mass of the vector-boson scattering
in $pp\to jj Z Z \to e^+ e^- \mu^+ \mu^- j j$ process:
left (right) frames present the $M_{4l}$ distribution  for $|\cos\theta_V|<0.9 (0.5)$ cuts
respectively. The red histogram is for $c_V=0$, the blue one represents $c_V=0$ (SM) case.}
\label{fig:MVV}
\end{figure}

For the $pp\to jj Z Z \to e^+ e^- \mu^+ \mu^- j j$ 
processes let us assume an integrated luminosity of 1.5 ab$^{-1}$,
which is being discuss as one of the high luminosity benchmarks at the future LHC13TeV,
for which we will have about 50 events from this process for analysis.
We start our analysis with a presentation of the invariant mass distribution of the $ZZ$ pair
which is derived from the four-lepton invariant mass, $M_{4l}$, presented in Fig.~\ref{fig:MVV}.
The left (right) figure presents the $M_{4l}$ distribution for $|\cos\theta_V|<0.9 (0.5)$ cuts
respectively. One can see that the $c_V=0$ distribution presented by the red histogram
is visibly above the the $c_V=1$ distribution presented by the blue histogram. 
One can also see the slight effect of increasing the $|\cos\theta_V|$ cut from 0.9 to 0.5:
the difference between $c_V=0$ and $c_V=1$ become bigger especially in the region
of higher  $M_{4l}$ where the $c_V=0$ case is enhanced by the longitudinal vector boson 
scattering process. At the same time the difference between the $c_V=0$ and $c_V=1$ 
cases become slightly smaller in the lower   $M_{4l}$ region where the cross section
is dominated by the transverse vector boson scattering.

It is also worth clarifying the details of how the angle $\theta_V$ of vector 
boson scattering in the $VV$ mass fame was  was deduced for the $pp\to jj Z Z \to e^+ e^- \mu^+ \mu^- j j$ 
process.
First of all we find the momenta $p_1$ and $p_2$ of the initial quarks $q_1, q_2$
in the $ q_1 q_2 \to q_3 q_4 Z Z$ process from a) total invariant mass of the final state particles and
b) from the total momentum of the final state particle along the z-axis.
Then we find two  pairs of the  final and initial quarks, say, $(q_1,q_3)$ and $(q_2,q_4)$
with the minimal angle between them in the centre-of-mass frame.
This will give us access to the four momentum of each virtual vector bosons, $p^V_1, p^V_2$ in the initial state:
$p^V_1=q_3-q_1$ and $p^V_2=q_4-q_2$ which allows us to calculate the $\theta_V$ angle in the 
centre-of-mass frame of the $VV \to VV$ scattering.

\begin{figure}[htb]
\includegraphics[width=0.5\textwidth]{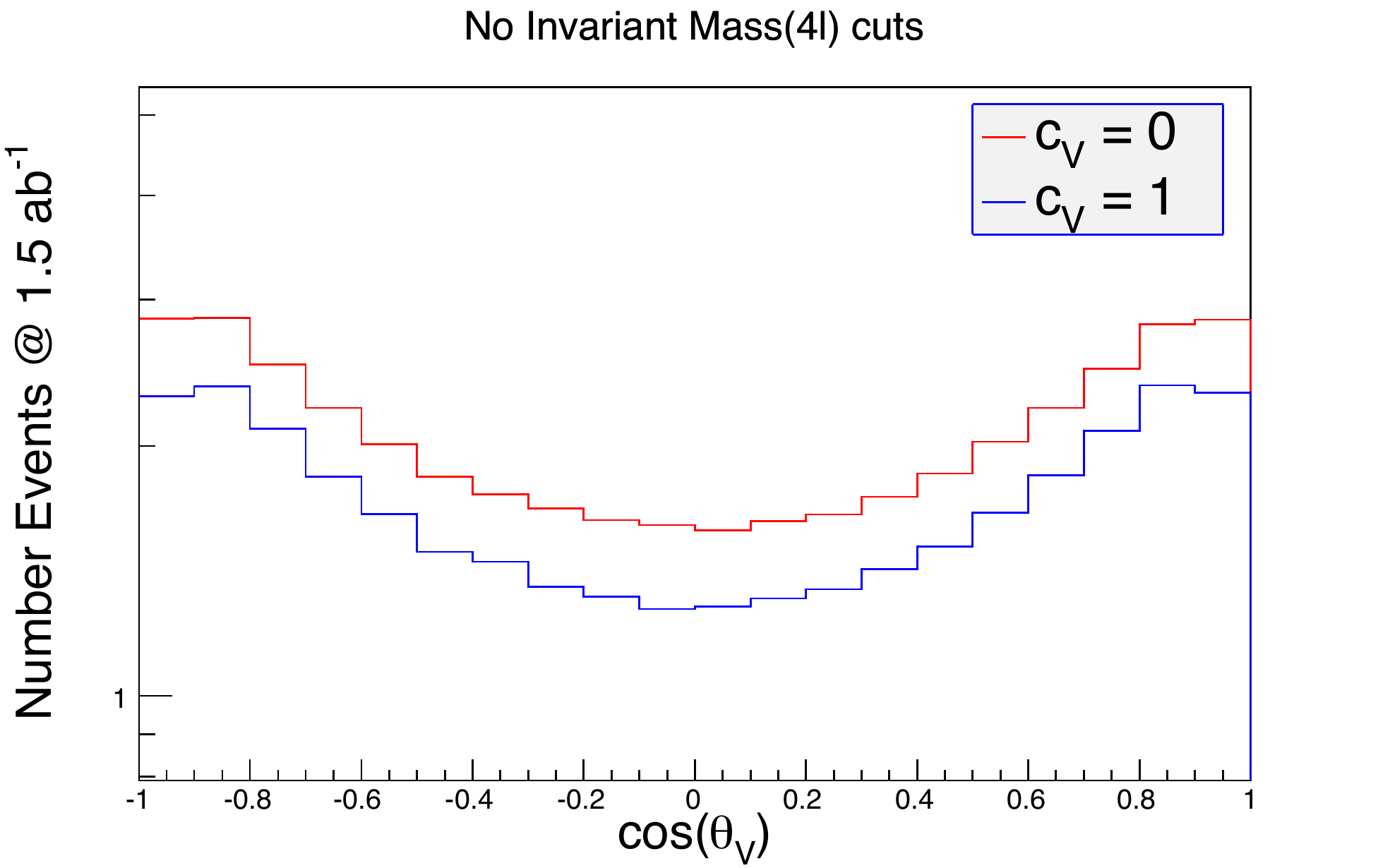}%
\includegraphics[width=0.5\textwidth]{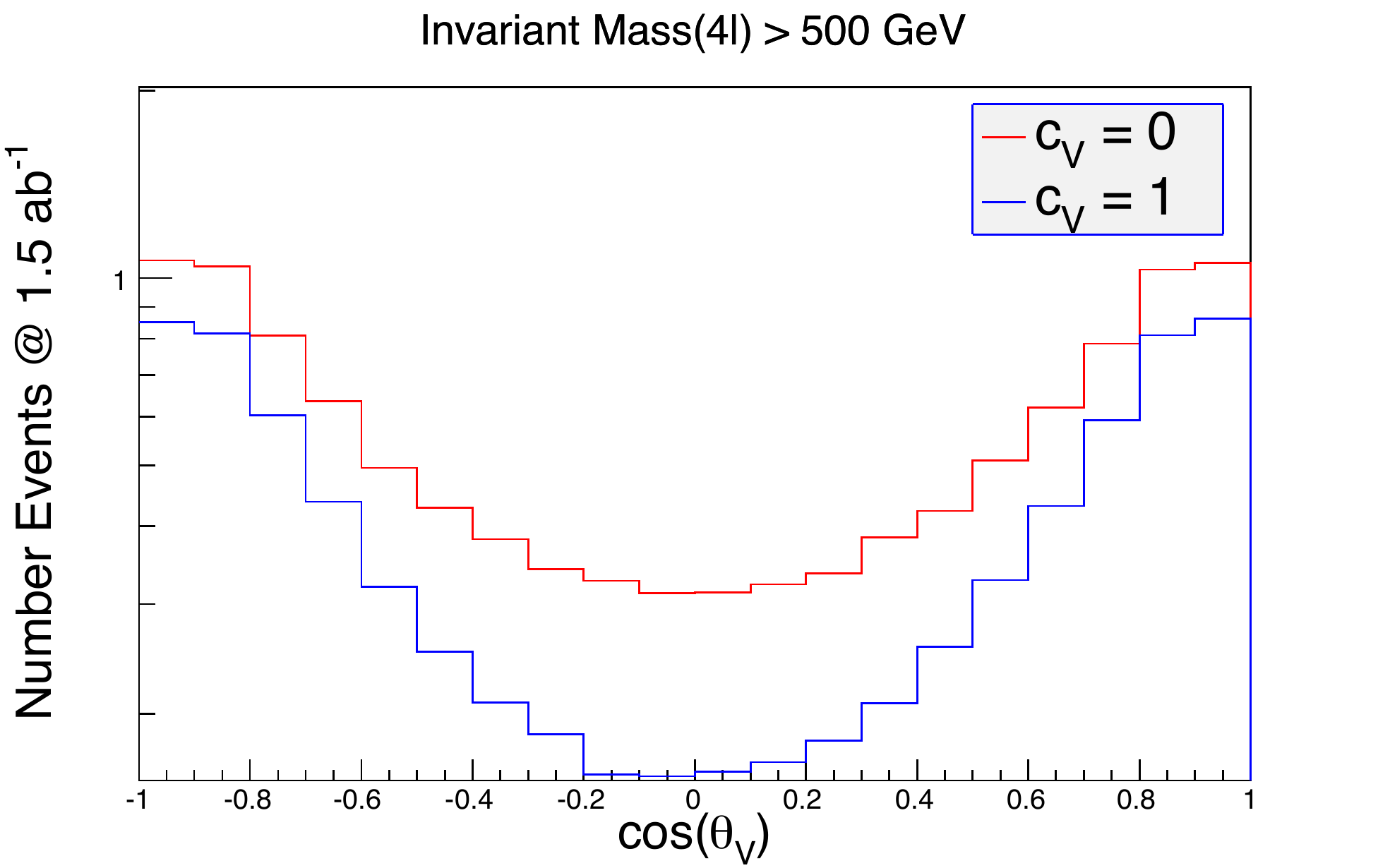}\\
\caption{ The $\cos(\theta_V)$ distribution for 1.5 ab$^{-1}$ @LHC13TEV
for  $pp\to jj Z Z \to e^+ e^- \mu^+ \mu^- j j$ process.
Left: no  $M_{4l}$ cut applied, Right:
the distributions after  $M_{4l}>500$~GeV cut.}
\label{fig:thetav}
\end{figure}

At the next step we study the ability of a cut on $M_{4l}$
to increase the sensitivity to longitudinal $VV$ scattering and the consequent sensitivity to the 
$HVV$ coupling. In Fig.~\ref{fig:thetav} we present the effect of the  $M_{4l}$ cut
on the $\cos(\theta_V)$ distribution by comparing the case when no  $M_{4l}$ cut is applied (left)
with the distributions after cutting for $M_{4l}>500$~GeV.
One can see the dramatic effect of the $M_{4l}$ cut which aimed to reject events with the
transversely polarised scattering which is background in the case of our study.
Indeed, after applying the $M_{4l}>500$~GeV cut one can see that a difference in the middle 
of the $\cos(\theta_V)$ distribution -- i.e. in the central region of the $VV$ scattering 
between $c_V=0$ and $c_V=1$ cases, becomes very pronounced which reproduces the results at the
level of $VV\to VV$ scattering which we found in Section~\ref{sec:wlwl-parton}

\begin{figure}[htb]
\includegraphics[width=0.5\textwidth]{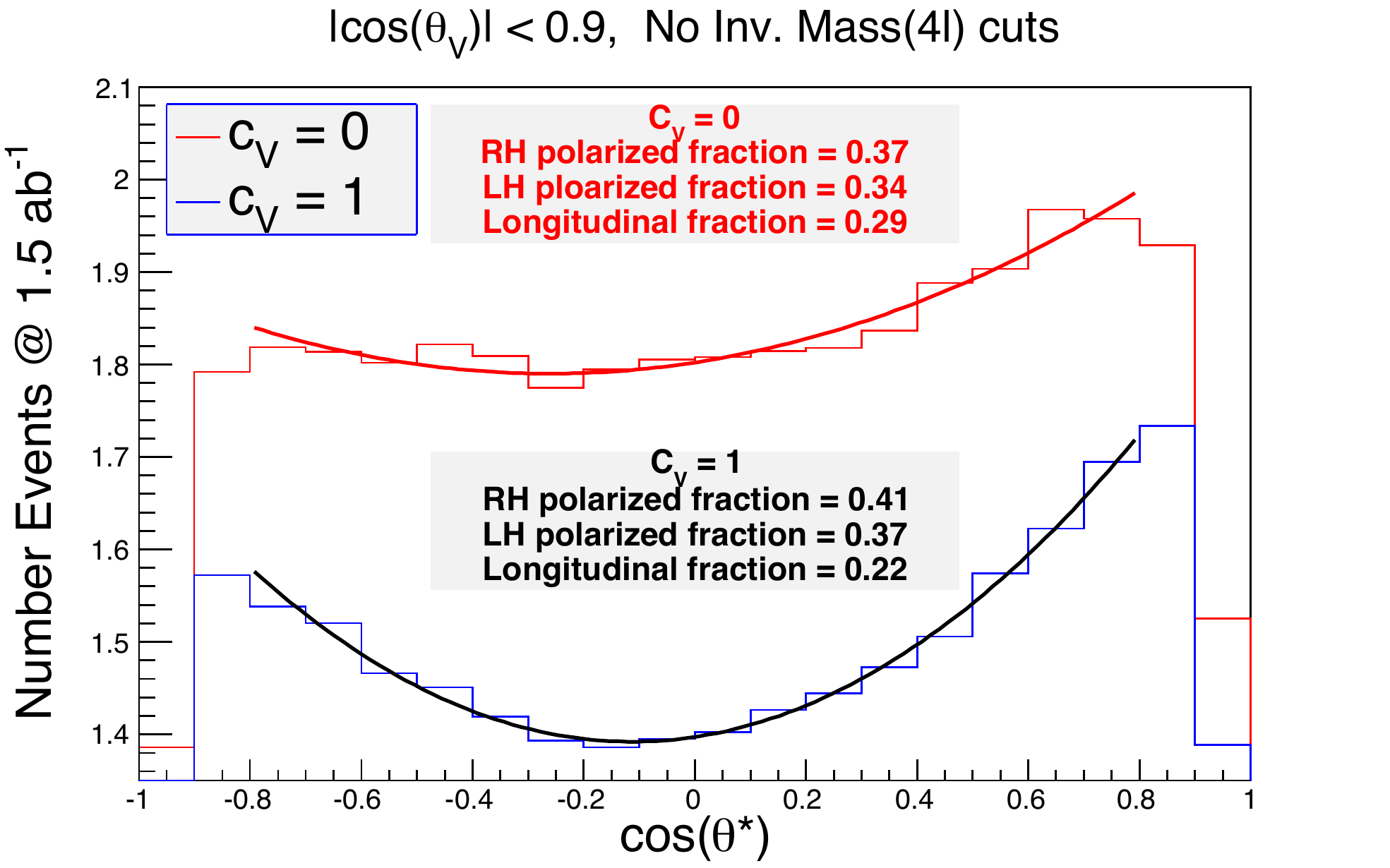}%
\includegraphics[width=0.5\textwidth]{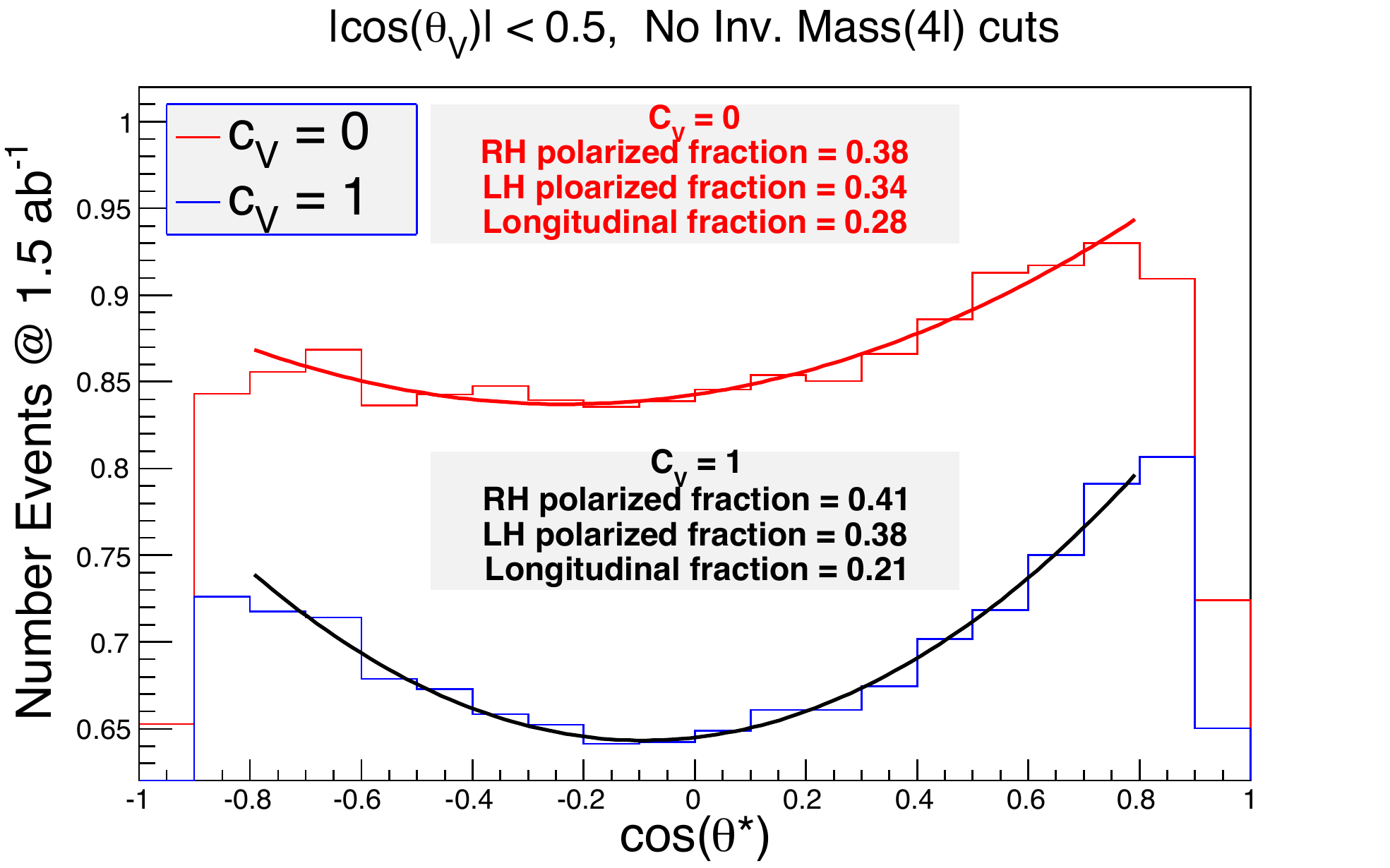}\\
\includegraphics[width=0.5\textwidth]{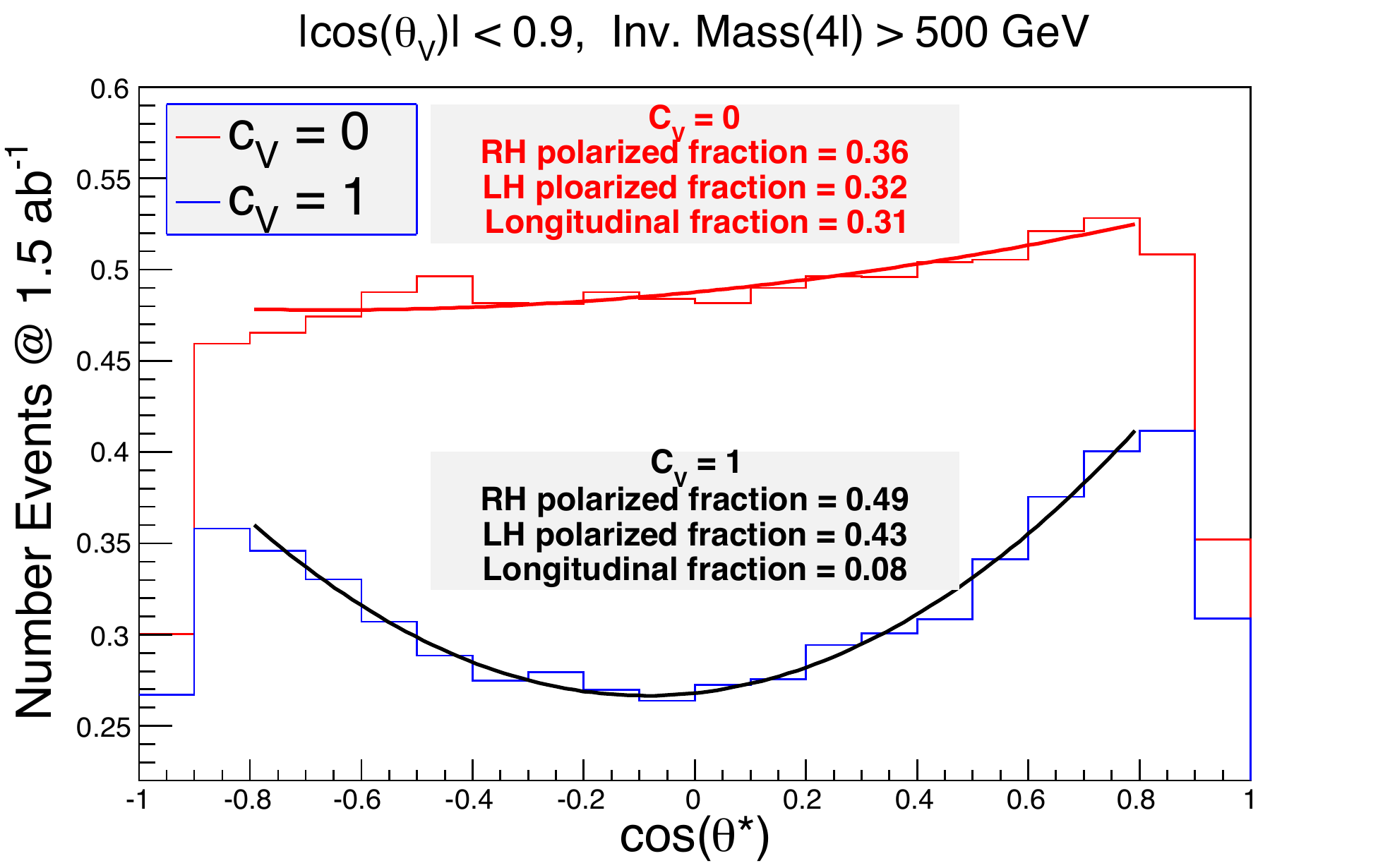}%
\includegraphics[width=0.5\textwidth]{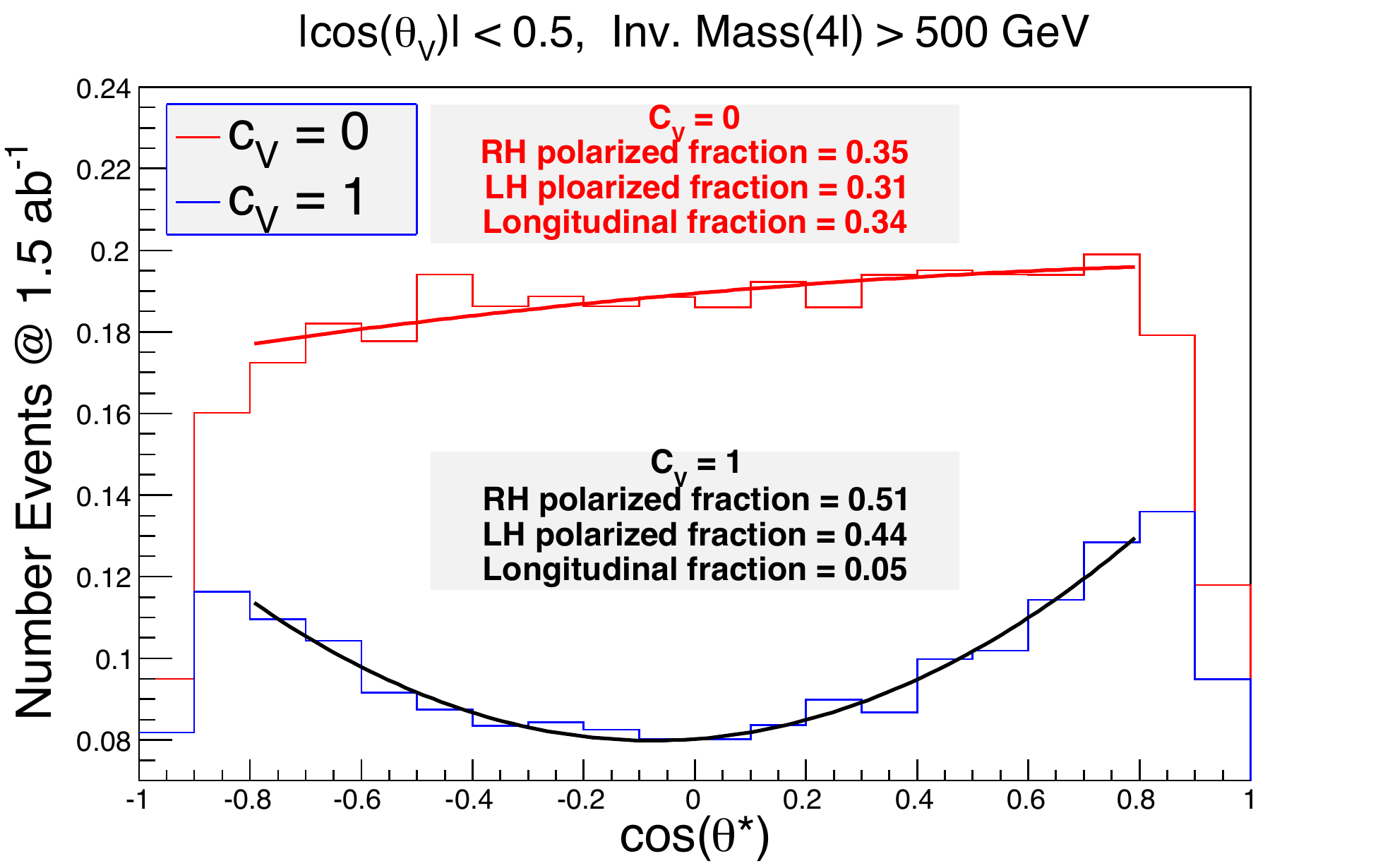}\\
\caption{The $\cos(\theta^*)$ distribution 
for  $pp\to jj Z Z \to e^+ e^- \mu^+ \mu^- j j$ process for 1.5 ab$^{-1}$ @LHC13TEV
for four different sets of cuts:
a) (top-left)$|\cos(\theta_V)|<0.9$; b) (top-right)$|\cos(\theta_V)|<0.5$; c)(bottom-left) $|\cos(\theta_V)|<0.9$ and $M_{4l}>500$~GeV
d)(bottom-right) $|\cos(\theta_V)|<0.5$ and $M_{4l}>500$~GeV.$pp\to jj Z Z \to e^+ e^- \mu^+ \mu^- j j$} \label{fig:thetastar}
\end{figure}

Finally, we perform an analysis of the $\cos(\theta^*)$ distribution and present the respective
results in Fig.~\ref{fig:thetastar}. The $\cos(\theta^*)$ is defined with respect to the electron
in the centre-of-mass of the $e^+e^-$ system and the direction of the boost to this system.
There are four frames in Fig.~\ref{fig:thetastar}
presenting $\cos(\theta^*)$ distributions for four different cases of kinematic cuts,
aimed to consequently increase the fraction of the longitudinal polarisation
of the $Z$-bosons in the $c_V=0$ case and enhance the difference between the 
$c_V=0$ and $c_V=1$ cases:
a) $|\cos(\theta_V)|<0.9$;~~b) $|\cos(\theta_V)|<0.5$;~~c) $|\cos(\theta_V)|<0.9$ and $M_{4l}>500$~GeV;~~d) $|\cos(\theta_V)|<0.5$ and $M_{4l}>500$~GeV.
For each case we fit the $\cos(\theta^*)$ distribution using the standard fitting routines of ROOT package~\cite{Brun:1997pa},
and for each of $c_V=0$ and $c_V=1$ we find the fraction of the longitudinal and transverse Z-bosons.
One can see that the $\cos(\theta_V)$ cut has a much smaller effect than the $M_{4l}$ cut
on enhancing the sensitivity to the difference between the $c_V=0$ and $c_V=1$ cases.
Among the four cuts, the biggest difference between the fraction of the longitudinally polarised (LP) Z-bosons
for $c_V=0$ and $c_V=1$ is for the cut d): the fit gives LP=0.34 for $c_V=0$ 
and LP=0.05 for $c_V=1$ (SM).
At the same time due to the lack of statistics, the fit error (which we do not presented in plots),
is quite large: it varies from about 100\% for cut a) statistics
to around 300\% in case of cuts d) statistics where the events rate is reduced by factor of 9.
At the same time we should stress that including semi-leptonic channels as well as a
complete set of processes --- $ZZ,WW,WZ$ --- will increase the statistics by a factor of about 250,
which will decrease the statistical fit error down to about 2.5\% level
for 1.5 ab$^{-1}$ integrated luminosity and to about 10\% even for 100 fb${^-1}$
integrated luminosity. The sensitivity to $c_V$ parameter is eventually expected to be similar.
Moreover the optimisation of the $M_{4l}$ and  $\cos(\theta_V)$ cuts as well as involving the total cross section as another variable for discrimination of different $c_V$ scenarios
will be able to improve the accuracy of the $c_V$ measurement even further.

\section{Conclusions}

In this contribution we present our preliminary results
of the LHC sensitivity to the $HVV$ couplings via the vector boson 
fusion process.
This sensitivity is {\it independent} of that which can be deduced from
direct Higgs searches at the LHC. Moreover,
the measurement of the  $HVV$ coupling ($V=Z,W^\pm$)
is not quite trivial from direct searches: it is derived
from several production processes and 
it depends on all other Higgs couplings, including couplings to new particles.
However only one Higgs coupling, $HVV$, enters the VBF
process, so it could be measured in a much more model-independent way from VBF
process.

The VBF process provides a unique sensitivity to the $HVV$ coupling since the Higgs boson 
provides a unitarisation of the $VV\to VV$ amplitudes, so 
any deviation from $c_V=1$ will lead to an enhancement of the
cross section of the longitudinal $VV$ scattering.
We have found and important correlation of the longitudinally polarised vector boson fraction (FL)
and the 
the vector boson scattering angle $\theta_V$ in the $VV$ centre-of-mass frame
and well as a FL correlation with the invariant mass of the $VV$ system, $M_{VV}$.
Using a combination of $\theta_V$, $M_{VV}$ and $\theta^*$
observables for the $pp\to jj Z Z \to e^+ e^- \mu^+ \mu^- j j$
process, chosen as an example for this preliminary study,
we have performed a fit of 
the $\theta^*$ distribution to find the
FL value, and have demonstrated an important sensitivity
of the VBF to the $HVV$ coupling given that statistics will be high enough.

For a combination of semi-leptonic channels as well as
complete set of processes --- $ZZ,WW,WZ$ --- which will increase statistics by a factor about 250,
one could expect VBF sensitivity to the $c_V$ coupling at a level of 10\% or better
even with 100$fb^{-1}$ @LHC13TEV.

One should also note that VBF provides not only an independent and precise way to measure the $HVV$
coupling, but it is also robust against systematic errors since it relies not only on the absolute 
cross section measurement but also on the {\it shape} of the $\cos(\theta^*)$ distribution.

\section*{Acknowledgements}
We thank the Les Houches workshop organisers  for an unprecedentedly creative
atmosphere which allowed us to generate ideas for this contribution.
We  acknowledge the use of IRIDIS HPC Facility at the University of Southampton for this study.
AB is supported in part by the NExT Institute and MCT is supported by an STFC studentship grant.
RR is supported in part by grants from CNPq and Fapesp.
EB, VB, and AP are partially supported by RFBR grant 12-02-93108 and NSh grant 3042.2014.2.


%% file: tth-lh-proc_final/tth.tex
\def \tth{t\bar{t}h}
\def \ttb{t\bar{t}}
\def \gtth{g_{tth}}
\def \gamgam{\gamma \gamma}
\def \Rg[#1]{\frac{\Gamma(h\to #1)}{\Gamma(h \to #1)^{SM} }}
\def \sigh[#1]{\frac{\sigma( #1)}{\sigma( #1)^{SM} }}
\def \nn{\nonumber}



\chapter{Probing the Structure of Top-Higgs Interactions at the LHC}

{\it F.~Boudjema, R.~Godbole, D.~Guadagnoli, K.~Mohan}


\begin{abstract}
We investigate the methods to explore the nature of the $\tth$ coupling at the LHC.
To that end we focus on the associated production of the Higgs boson with a $\ttb$ pair. 
We analyze the feasibility of this process for determining the CP properties of the 
coupling of the Higgs boson to the top quark. We first show the constraints implied by the 
Higgs rates from the currently available LHC data. We then focus on specific kinematic 
observables that can be used to determine the coupling itself.
\end{abstract}

\section{Introduction}

The 7-8 TeV runs of the LHC have led to the discovery of a 125 GeV boson 
\cite{Chatrchyan:2012ufa,Aad:2012tfa,CMS-PAS-HIG-13-005,ATLAS-CONF-2013-034}. The properties measured so 
far show very good consistency with those expected for the Standard-Model (SM) Higgs boson. Despite the 
fact that these runs have not revealed any sign of physics beyond the SM (in particular, production of 
new particles), 
the fact remains that the SM cannot address a few pressing questions, such as the baryon asymmetry of the 
universe, nor does the SM provide a candidate for Dark Matter. These issues call for New Physics.
Furthermore, the observation of a 125 GeV boson as well as the absence so far of New Physics at the TeV 
scale form a puzzle from the point of view of Naturalness.
Probing the Higgs sector is therefore of utmost importance, and crucial is in particular a precise 
determination of its couplings to other particles. One of the most important couplings of the Higgs boson is 
to the top quark, the heaviest SM particle.
Not only is this coupling responsible for the main production channel of the SM Higgs boson at the LHC but the 
interaction with the top quark also has important consequences on spontaneous symmetry breaking 
within the SM -- notably, vacuum stability arguments -- as well as beyond the SM -- supersymmetry, for
instance, where the top quark drives electroweak symmetry breaking in some scenarios. Yet a {\em direct} 
measurement of the Higgs-top coupling is lacking.  

In this note, we focus our attention on the possibility of a direct determination of the Higgs-top coupling.
The Higgs-top coupling can generally be parametrized in the form
\begin{equation}
\mathcal{L}_{tth}=\gtth\bar{t}(a+ib\gamma_{5})\phi t~,
\label{eqn:tth}
\end{equation}
where $\gtth =m_{t}/v$ normalizes the coupling to the SM strength. The coefficients $a$ and $b$ are 
assumed to be real. In the SM, where the Higgs boson is a scalar, $a=1$ and $b=0$. For a pure pseudoscalar $a=0$ 
and $b \neq 0$. 
A Higgs boson with mixed CP properties is realized if both $a\neq0$ and $b\neq0$. The exact values 
of these coefficients will depend on the specific model. Here we are interested in a model-independent 
approach to determine the nature of the interaction from data.

Note that the electron and neutron dipole moments \cite{Brod:2013cka} as well as the decay and production 
rates of the Higgs boson measured at the LHC \cite{Djouadi:2013qya,Cheung:2013kla} provide important constraints 
on the strength of the $a, b$ coefficients in Eq.~(\ref{eqn:tth}). However, an unambiguous 
determination of them is only possible by measuring directly Higgs production in association with a top and anti-top quark ($\tth$ production). 
In order to illustrate this point, in the following section we determine the 
constraints that Higgs rates can place on $a$ and $b$ using currently available data. We then proceed to 
analyze $\tth$ production at LHC and construct some observables that could be used to determine the nature 
of the $\tth$ interaction itself.

\section{Indirect probes of an anomalous $\tth$ coupling}

Within the SM, and with Higgs and top masses as measured, there are four main production modes of the 
Higgs boson at the LHC: gluon fusion, vector-boson fusion (VBF), Higgs production in association with a $W$/$Z$ 
boson (VH) and Higgs production in association with a $\ttb$ pair.
The gluon-fusion production mode has the largest cross-section at the LHC, and the dominant contribution 
to this process comes from a top loop. The Higgs decay to two photons has also a contribution due to a top 
loop, although the dominant one comes from a $W$-boson loop.
ATLAS and CMS have already put indirect constraints on the value of $a$, assuming that there are no other 
sources contributing to the effective couplings $g g \to h$ or $h \to \gamma \gamma$. At $95\%$ 
confidence level these constraints read~\cite{ATLAS-CONF-2013-034,CMS-PAS-HIG-12-020}
\begin{eqnarray}
a\in [-1.2,-0.6]\cup[0.6,1.3]&&\qquad 
{\rm ATLAS,}\nonumber\\
\label{indconstr}
a\in [0.3,1.0]\qquad\qquad\qquad\; &&\qquad 
{\rm CMS}.\nonumber
\end{eqnarray}
In the rest of this section we extend this analysis by allowing in the fit both $a$ and $b$ couplings in 
Eq.~(\ref{eqn:tth}).
As customary, the signal strength measured in a particular channel 
$i$ at the LHC is defined as
\begin{equation}
\hat{\mu}_i = \frac{n_{exp}^i}{(n_S^i)^{SM}}~,
\label{eqn:mui_exp}
\end{equation}
where $n_{exp}^i$ is the number of events observed in the channel $i$ and $(n_S^i)^{SM}$ is the expected 
number of events as predicted in the SM. In order to contrast specific model predictions with the 
experimentally derived $\hat \mu_i$ we define (as usual)
\begin{equation}
\mu_i = \frac{n_S^i}{(n_S^i)^{SM}}=\frac{\Sigma_p \sigma_p \epsilon_p^i}{\Sigma_p \sigma_p^{SM} \epsilon_p^i}
\times \frac{BR_i}{BR_i^{SM}}~.
\label{eqn:mui}
\end{equation}
Here $n_S^i$ corresponds to the expected number of events predicted in the hypothesized model under 
consideration; $\sigma_p$ corresponds to the cross-section in the $p^{th}$ production mode, 
i.e.~the cross-section for Higgs production in one of the four production modes listed earlier; 
$BR_i$ is the branching ratio of the Higgs boson in the 
$i^{th}$ channel; $\epsilon^i_p$ is the efficiency of the $p^{th}$ production mode to the selection cuts 
imposed in the $i^{th}$ channel. Note that the efficiencies in the numerator and denominator of 
Eq.~(\ref{eqn:mui}) are taken to be the same. This is true at leading order for the gluon fusion process. 

Besides direct $\tth$ production, which will be discussed separately in the next section, the
$a$ and $b$ couplings contribute to two main quantities, namely Higgs production from a gluon-gluon pair
and Higgs decay into two photons. We discuss these quantities in turn.
The ratio of the decay width of the Higgs boson to two photons to the SM decay width, at leading order and 
neglecting the small contribution from lighter fermions, can be written in the form \cite{Djouadi:2005gj}
\begin{equation}
\label{eqn:hyy}
\Rg[\gamgam]=\frac{|\kappa_W A^a_W(\tau_W) + a\frac{4}{3}A^a_t(\tau_t)|^2 + |b\frac{4}{3}A^b_t(\tau_t)|^2}
{| A^a_W(\tau_W) + \frac{4}{3}A^a_t(\tau_t)|^2}~.
\end{equation}
Here $A^i_j$ corresponds to the form factors defined below
\begin{eqnarray}
A^a_t(\tau)&=&\frac{2}{\tau^2}(\tau + (\tau -1)f(\tau)), \\ \nn
A^a_W(\tau)&=&-\frac{1}{\tau^2}(2\tau^2 +3\tau +3(2\tau -1)f(\tau)), \\ \nn
A^b_t(\tau)&=&\frac{2}{\tau}f(\tau)
\end{eqnarray}
with $\tau_i=\frac{m_h^2}{4m_i^2}$ and
\begin{eqnarray}
f (\tau) = \left\{ \begin{array}{lc}
\mbox{arcsin}^2 \sqrt{\tau}  & \mbox{for }\tau \leq 1 \\
- \frac{1}{4} \left[ \log \frac{1+\sqrt{1-\tau^{-1}}}{1-\sqrt{1-\tau^{-1}}} - i \pi \right]^2 & 
\mbox{for }\tau > 1 \end{array} \right.\,.
\end{eqnarray}
\noindent $\kappa_W$ is a multiplicative factor that rescales the SM $H W^\mu W_\mu$ coupling 
(for the SM $\kappa_W=1$). 
Similarly, for Higgs boson production through gluon fusion, neglecting the small $b$-quark contribution
and at leading order, one could write
\begin{equation}
\label{eqn:ggh}
\sigh[gg\to h]=\Rg[gg]=a^2+ b^2 \frac{|A^b_t(\tau_t)|^2}{|A^a_t(\tau_t)|^2}~.
\end{equation}

\noindent 
Combining both ATLAS and CMS data~\cite{ATLAS-CONF-2013-034,ATLAS-CONF-2013-012,CMS-PAS-HIG-13-001,%
CMS-PAS-HIG-13-002,CMS-PAS-HIG-13-003,CMS-PAS-HIG-13-004} we determine the best-fit values of $a$ and $b$ 
by minimizing the $\chi^2$ defined as 
\begin{equation}
\chi^2=\sum\limits_{i=(p,d)}^{}\frac{(\mu_i - \hat{\mu}_i)^2}{(\sigma^{exp}_i)^2},
\end{equation}
where the sum is over each of the production (ggF,VBF,Vh, tth) and decay ($h\to$ $ZZ^*$, $WW^*$, $\gamgam$,
$b\bar{b}$, $\tau \tau$) modes and $\sigma^{exp}_i$ is the experimental error. 
We adopt this Gaussian approximation rather than a full-fledged likelihood analysis, since our objective 
is to obtain approximate allowed ranges for non-SM couplings. Note that our fits for $a$ alone are in 
reasonable agreement with ATLAS and CMS fits, and that we have not taken into account recent data 
on $\tth$ production\cite{CMS-PAS-HIG-13-020,CMS-PAS-HIG-13-015,CMS-PAS-HIG-12-025,CMS-PAS-HIG-13-019,%
ATLAS-CONF-2013-080} and we do plan on doing so in the future. However the large errors these 
measurements come with imply that they will not strongly affect the fit.
\begin{figure}[h]
\centering
\includegraphics[scale=0.6]{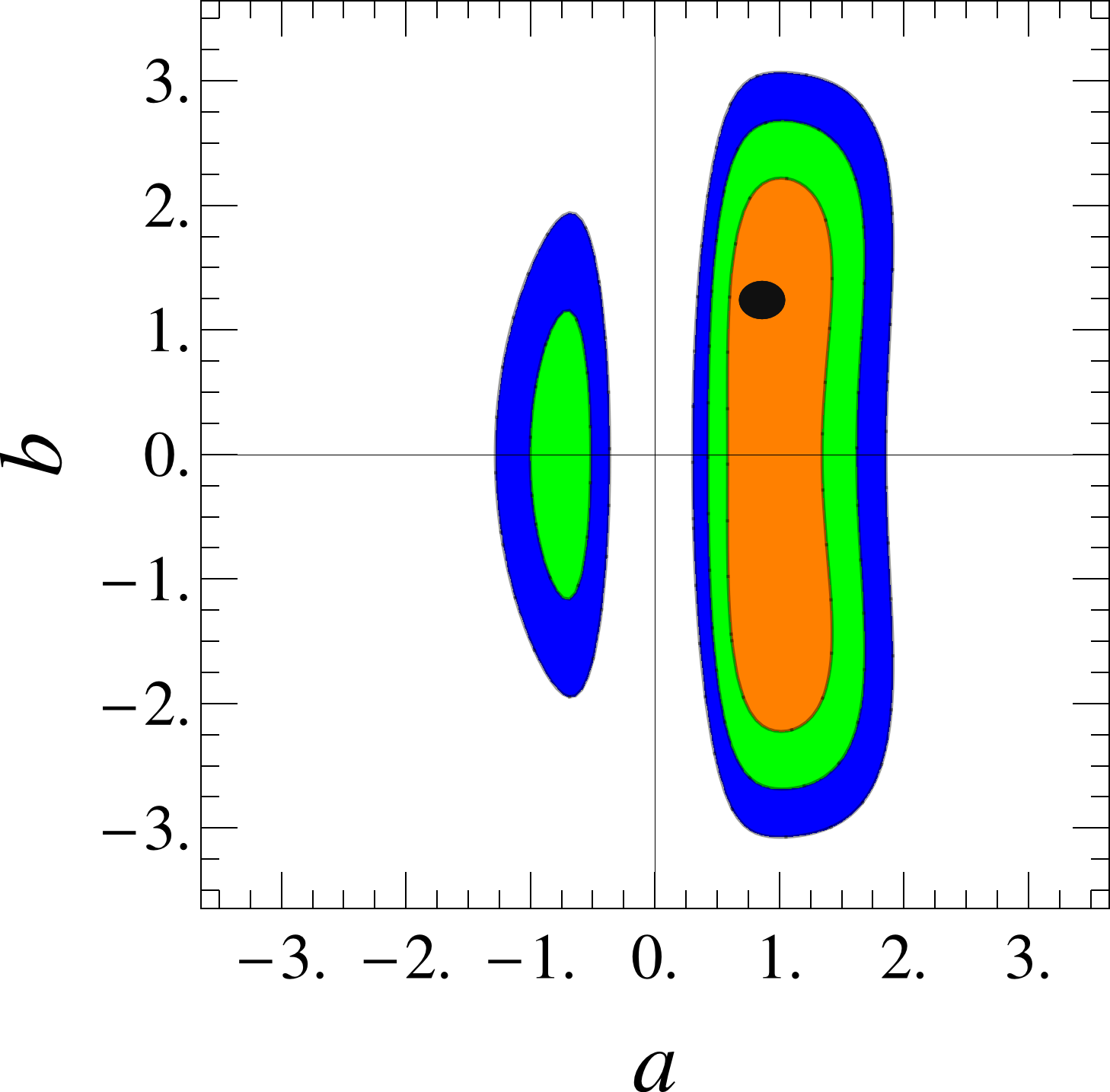}
\caption{Two-dimensional fit to the Higgs-top couplings $a$ and $b$ in Eq.~(\ref{eqn:tth}) using Higgs 
decay and production rates from ATLAS and CMS data.}
\label{fig:fit}
\end{figure}
The fit results are displayed in Fig.~\ref{fig:fit}, with the $1$, $2$ and $3 \sigma$ regions being shown in 
orange, green, and blue respectively. The best-fit point is given by the black dot. While the $a$ coupling 
is constrained fairly well, already within this limited amount of data, it is immediately evident that the 
constraint to the pseudoscalar coupling to fermions, $b$, is still fairly loose. Values of $|b|$ even above 
$2$ are still allowed.

These results can be easily understood by looking at Eqs.~(\ref{eqn:hyy}) -- (\ref{eqn:ggh}). 
In $gg \rightarrow h$ the $a$ and $b$ coefficients enter quadratically, weighed by the loop functions 
$A_t^a$ and $A_t^b$ respectively. Therefore, while $gg \rightarrow h$ production is useful to constrain 
the overall $|a|^2$ and $|b|^2$ magnitudes, it is unable to distinguish $a$ versus $b$ effects.
Including the $h \rightarrow \gamma \gamma$ decay channel substantially improves the discriminating power. 
The important point is that, in this decay channel, the scalar-coupling contribution, contrary to the 
pseudoscalar contribution, interferes with the $W$ boson contribution. In particular, for $a >0$, as in the SM, this 
interference is destructive. On the other hand, for $a$ negative, the branching ratio gets enhanced with 
respect to the SM by both the scalar and the pseudoscalar contributions, thus making $\Gamma(h 
\rightarrow \gamma \gamma)$ too large.
This is the reason why the $a>0$ and $a<0$ allowed regions are completely separated in Fig.~\ref{fig:fit}, 
and why the $a< 0$ allowed region is smaller. Specifically, $a=0$ does not fit the data either because in 
this case the $W$ loop is too large and cannot obviously be compensated by the $b$ contribution, 
irrespective of the value of $b$. In conclusion, from these data alone the constraints on $b$ are expected
to be (and remain) much looser than those on $a$, as confirmed by our Fig.~\ref{fig:fit}.

One must therefore resort to a more direct approach to determine the Higgs-top coupling, also to exclude
the presence of additional contributions not accounted for by $a$ and $b$. This will be discussed in the
following section.

\section{Associated production of the Higgs with a $\ttb$ pair}

Of the four production modes ($gg \to h; Vh; VV \to h; tt h$, with $V=W^\pm, Z$) of the Higgs boson at the LHC, 
$\tth$ production has the smallest cross-section. The complicated final state of the process, with the top 
quark decaying to a bottom quark and a $W$ boson, which in turn may decay either hadronically or 
leptonically, and the large backgrounds to the process make this a difficult channel to study at the LHC. 
However, as accentuated in the discussion in the previous section, $\tth$ production is necessary, among 
the other reasons, in order to unambiguously determine the parity of the Higgs coupling to the top quark. 

\begin{figure}
\includegraphics[scale=0.4]{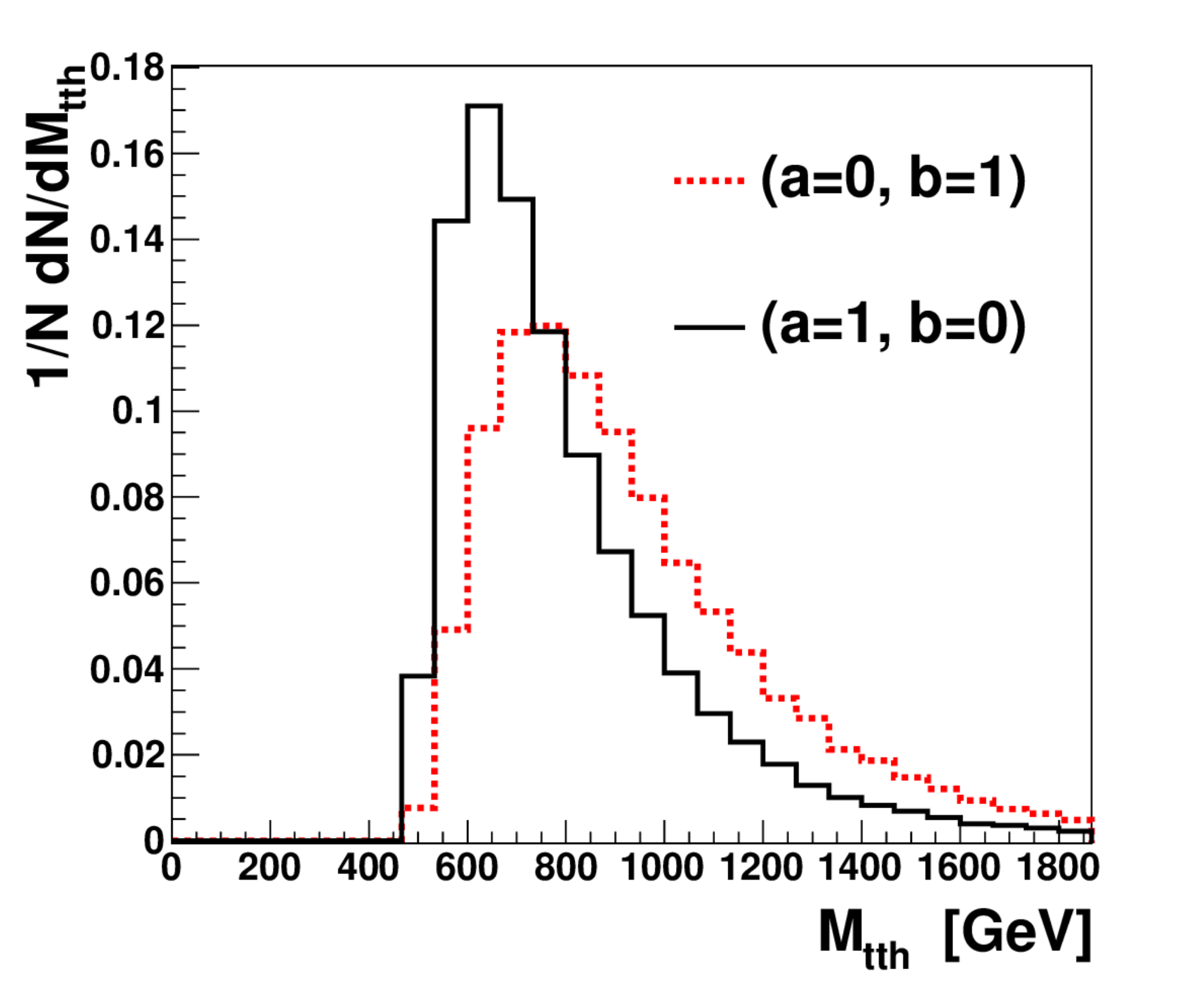}
\includegraphics[scale=0.4]{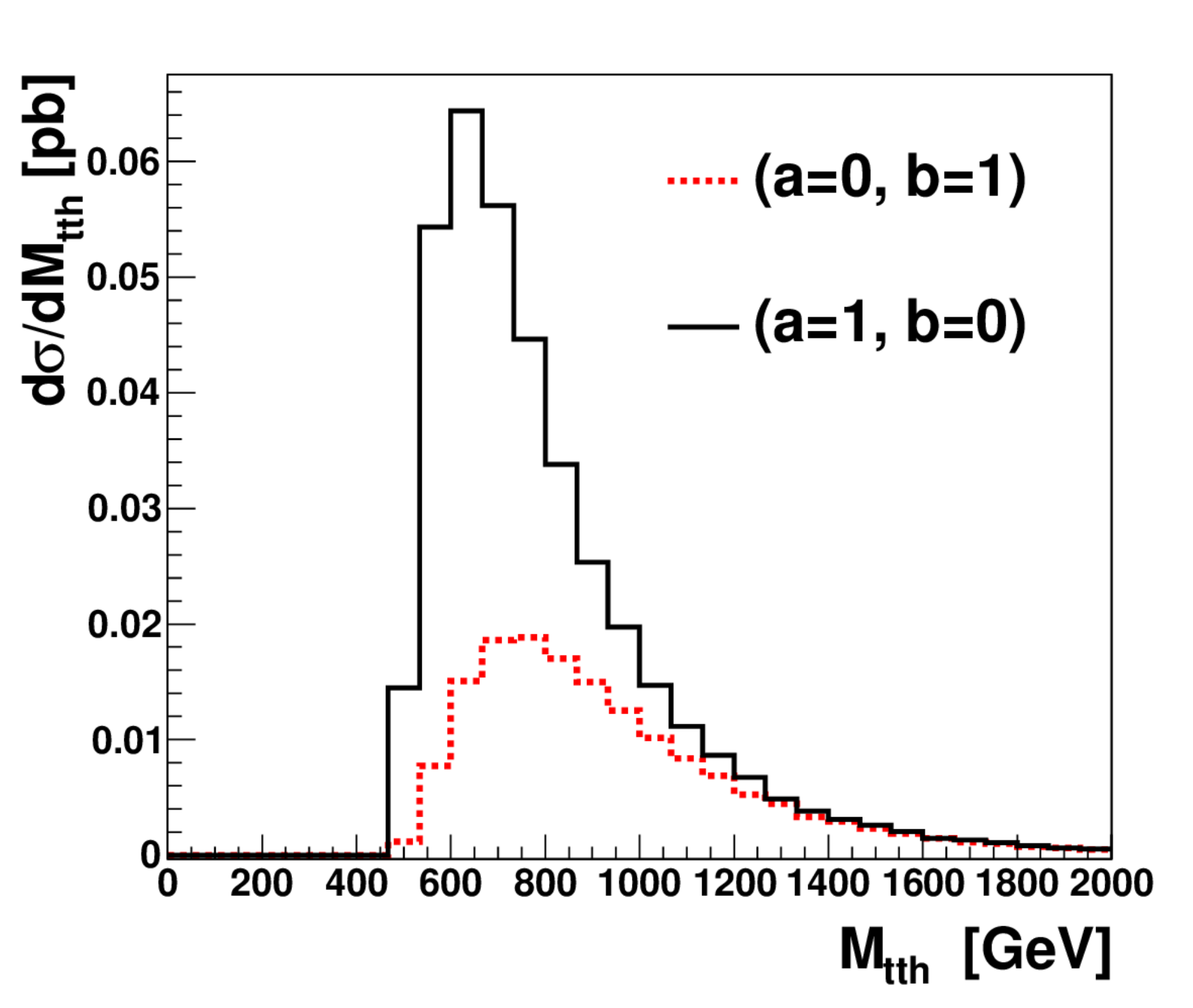}
\caption{
(Left panel) The invariant-mass distribution of the $\tth$ system, normalized to unity.
(Right panel) The differential cross-section with respect to the invariant mass distribution of the 
$\tth$ system.
}
\label{fig:set1}
\end{figure}

In this section we wish to point out the major differences that a scalar versus a 
pseudoscalar coupling case entails for $\tth$ production at the LHC. As a first step we consider 
observables suitable for the dileptonic decay mode, i.e.~with both the top and anti-top quarks decaying 
leptonically. We will comment on analogous observables that can be constructed for other decay 
modes of the top.

The first distribution we consider is the production cross-section near threshold. It has been pointed out
that the threshold behavior of the cross-section for a scalar versus a pseudoscalar Higgs boson is very
different at an $e^+ e^-$ collider \cite{Godbole:2011hw,Godbole:2007uz,Bhupal-Dev:2007is}. More
specifically, the {\em rate} of increase of the cross-section with the centre of mass energy of the 
collision is suppressed in the case of the pseudoscalar Higgs coupling by a factor of $\rho$, where 
$\rho = (\sqrt{s} - 2m_t - m_h)/\sqrt{s}$ parametrizes the proximity to the production threshold. This 
factor can be easily understood from arguments of parity and angular-momentum conservation 
\cite{Bhupal-Dev:2007is}. The same behavior is observed in the quark-initiated process of a $pp$ collision,
which is a spin-1, $s$-channel mediated process, but this contribution is negligible at the LHC. This being 
said, the dominant $gg$-initiated Higgs-production cross-section does also exhibit a faster rise at 
threshold for the scalar than for the pseudoscalar case, even if this rise is not as pronounced as in the 
case of $s$-channel spin-1 production. Similar considerations on parity and orbital angular momentum hold 
here, the difference being that more partial waves contribute in the $gg$-initiated process.

In the left panel of Fig.~\ref{fig:set1} we show the normalized invariant mass distributions of the $\tth$
system for the pseudoscalar $(a=0,b=1)$ and scalar $(a=1,b=0)$ cases.
We see that the rate of increase of the cross-section with the invariant mass of the 
$\tth$ system is much more rapid for the scalar than for the pseudoscalar case. This is an important 
distinguishing feature and could be used to probe the nature of the Higgs-top quark coupling.
The right panel of Fig.~\ref{fig:set1} shows the same distribution, but normalized to the total 
cross-section (i.e.~$d\sigma/dM_{tth}$). We observe that for the same magnitude of the coupling strength,
the cross-section for the pseudoscalar case is suppressed in comparison to scalar $\tth$ production.

\begin{figure}
\centering
\includegraphics[scale=0.4]{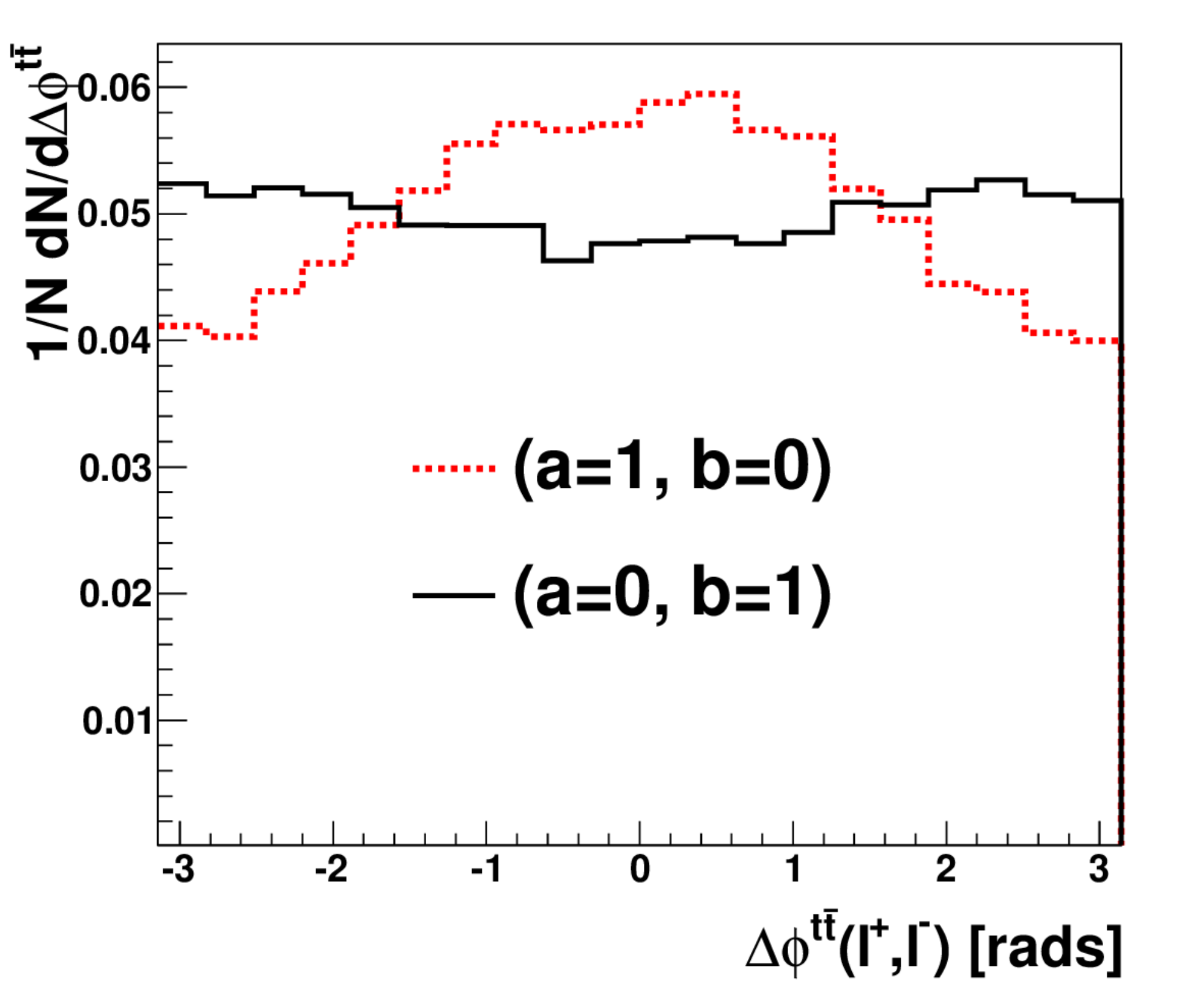}
\caption{
Difference in the azimuthal angles of the lepton momenta, with the $\ell^+$ momentum 
evaluated in the rest frame of the top quark and the $\ell^-$ momentum evaluated in the rest frame of the 
anti-top quark.
}
\label{fig:delphill_tt}
\end{figure}

Looking at the form of the Lagrangian in Eq.~(\ref{eqn:tth}) one expects that the nature 
of the Higgs-top coupling should also affect spin correlations between the top and anti-top quarks,
which in turn are passed on to the kinematic distributions of their decay products. 
In  
Fig.~\ref{fig:delphill_tt} we show $\Delta\phi^{\ttb}(\ell^+,\ell^-)$, defined as the difference between the 
azimuthal angle of the $\ell^+$ momentum in the rest frame of the top quark and the azimuthal angle of the 
$\ell^-$ momentum evaluated in the rest frame of the anti-top quark~\cite{Heinemeyer:2013tqa}\footnote{%
In constructing the $\ell^\pm$ momenta as described, we keep fixed for all events the choice of the $x$ 
and $y$ axes, and the $z$ axis is chosen, as customary, to lie along the beam direction. While individually
the azimuthal angles for the $\ell^+$ and $\ell^-$ momenta do depend on the choice of the $x$ and $y$
axes, their difference, as in $\Delta \phi$, does not. $\Delta \phi$ depends only on the choice of the 
beam axis. One can construct $\Delta \phi$ from the following formula
\begin{equation}
\label{eq:cosdphi}
\cos(\Delta \phi^{\ttb}(\ell^+,\ell^-)) =
\frac{(\hat z \times {\vec p}^{~\bar t}_{\ell^-}) \cdot (\hat z \times {\vec p}^{~t}_{\ell^+})}
{|{\vec p}^{~\bar t}_{\ell^-}||{\vec p}^{~t}_{\ell^+}|}~,
\end{equation}
that shows dependence only on the $\hat z$ direction. In this formula, the superscripts $t$ ($\bar t$) 
indicate that the given momentum is calculated in the rest frame of the $t$ ($\bar t$). Analogous 
relations apply for the other $\Delta \phi$ definitions to follow.
}.  It is evident 
that the information about the nature of the coupling is indeed encoded in the $t$ and $\bar t$ 
polarizations and hence in the spin correlations between them. Such an angle is however hard to 
reconstruct, in particular because it requires reconstruction of the top and anti-top momenta 
in presence of two escaping particles (the $\nu$'s). We therefore explore a similar observable in
more convenient frames.

\begin{figure}[t]
\includegraphics[scale=0.4]{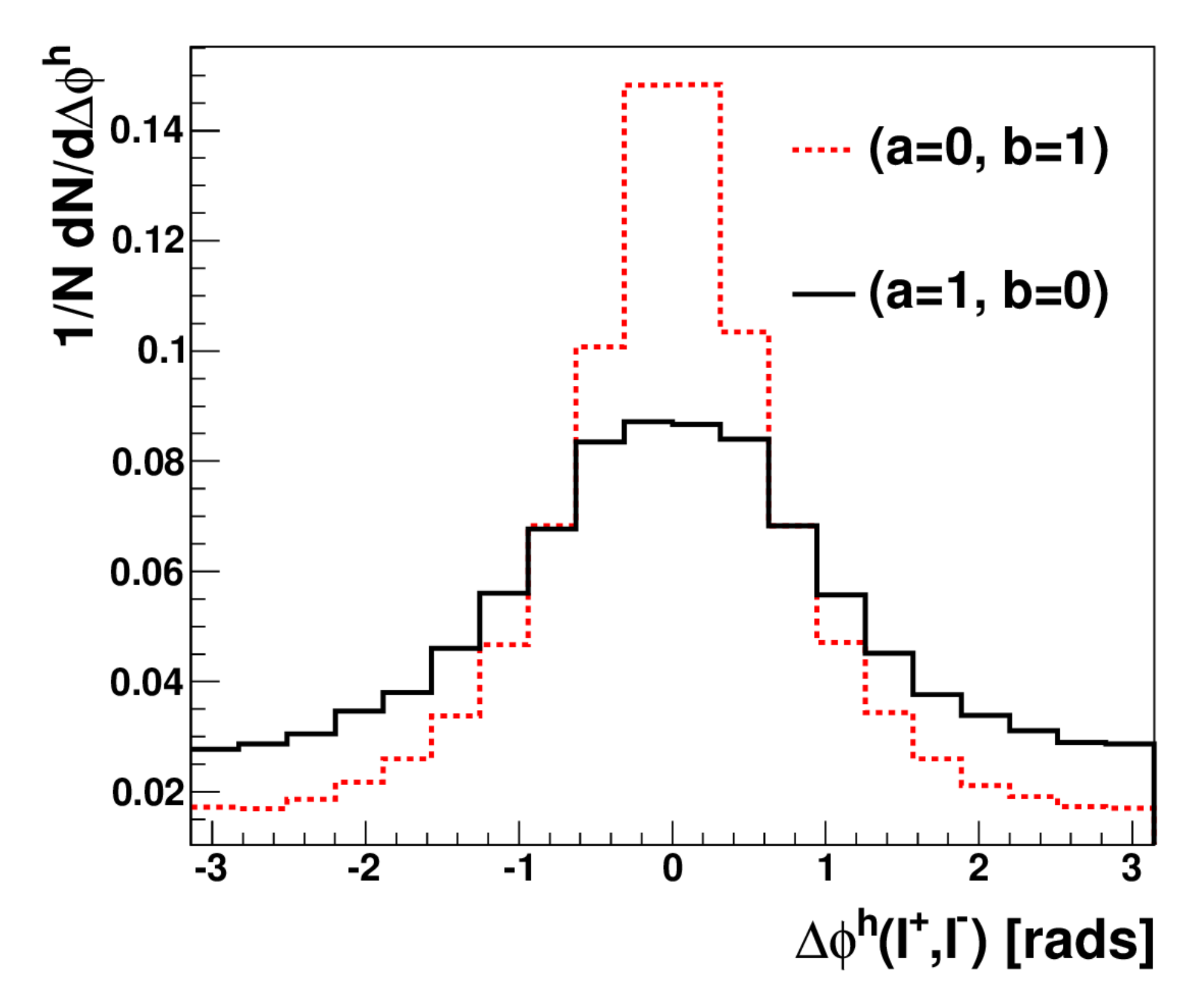}
\includegraphics[scale=0.4]{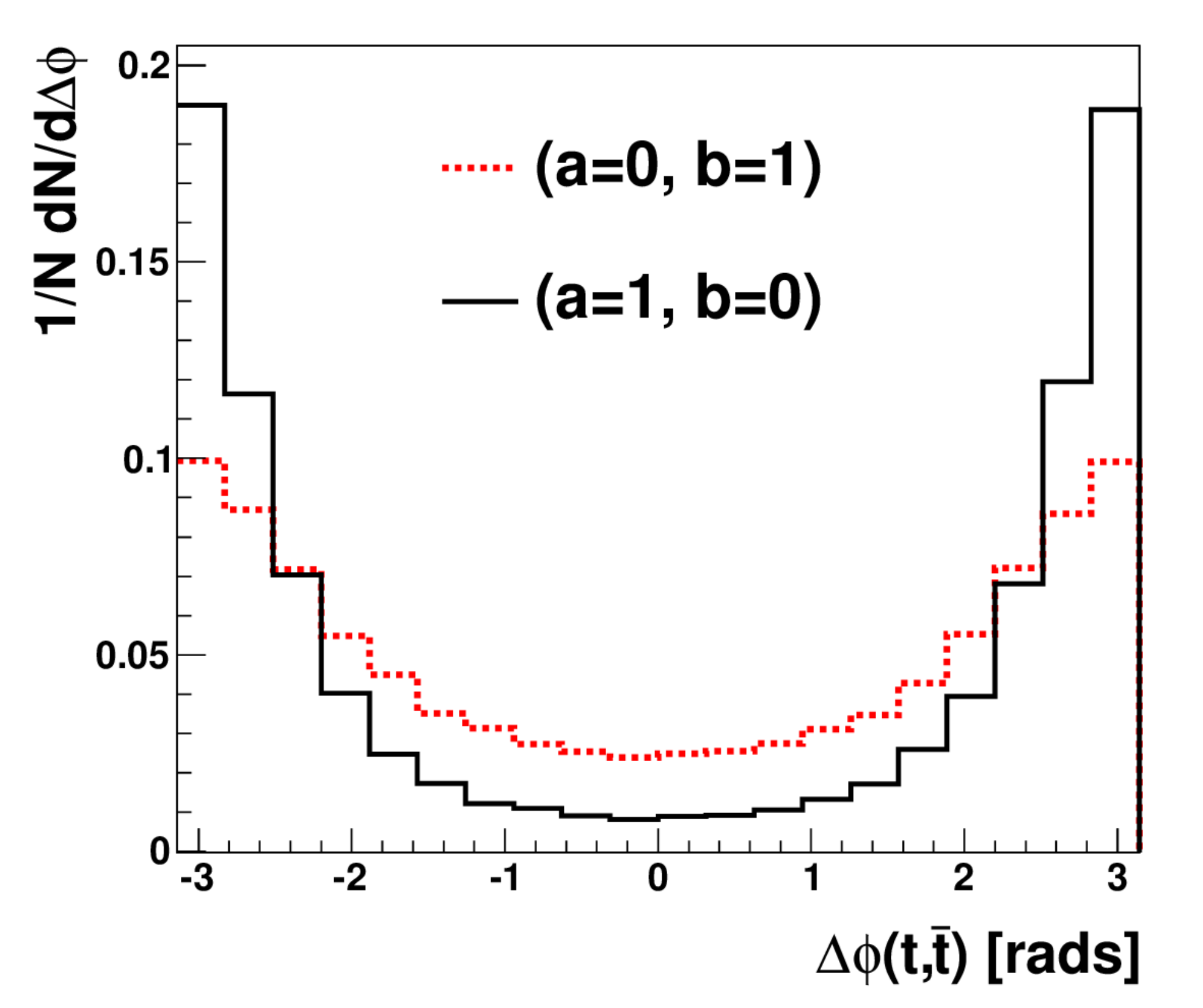}
\caption{(Left panel): Difference in the azimuthal angles of the lepton momenta, with both momenta
evaluated in the rest frame of the Higgs boson. (Right panel): Azimuthal angle difference between the top and anti-top quarks in the lab frame.}
\label{fig:set2}
\end{figure}
Results are presented in Fig.~\ref{fig:set2}. 
The left panel shows the azimuthal angle difference between the $\ell^+$ and $\ell^-$ with the lepton momenta evaluated in the rest frame of the Higgs boson.
The right panel shows the azimuthal angle difference between the top and anti-top $(\Delta \phi (t,\bar{t}) )$ quarks evaluated in the lab frame.
These observables are more straightforward to reconstruct at the LHC.
Furthermore, it is easy to find proxies of these distributions suitable for hadronic and semi-leptonic 
decays of the top quark. In particular, in semileptonic or fully hadronic decays one may replace one or respectively both leptons with the parent $W$ boson, and define azimuthal-angle observables accordingly, 
namely as $\Delta \phi(W^\pm, \ell^\mp)$ for the semileptonic case, and $\Delta \phi(W^\pm, W^\mp)$ for
the hadronic case, with momenta evaluated in the rest frame of the Higgs boson. Although not shown here, we 
have checked that these distributions are similar to the dilepton distribution shown in 
Fig.~\ref{fig:set2}.

\section*{Conclusions}

We have considered the general Higgs-top quark coupling and have explored the possibility to probe 
this coupling in a model-independent framework. We find that the information provided by the Higgs rates 
does not suffice to provide conclusive evidence about the nature (scalar versus pseudoscalar) of the coupling. 
One must therefore resort to a more direct method of probing the coupling. We investigated some of the 
possible kinematic observables that could be used to this end. We find that the information about the 
nature of the coupling is encoded in the threshold behaviour of the cross-section as well as in kinematic 
distributions that reflect the $t \bar t$ spin correlations, which are affected by the parity of the 
$\tth$ coupling. We also note that it is possible to extract information about this coupling by using 
these distributions to construct asymmetries.

\section*{Acknowledgements}

KM acknowledges the financial support from CSIR India, the French CMIRA and ENIGMASS Labex and LAPTh, 
Annecy-le-Vieux for hospitality while part of this work was done. RG wishes to thank the Department of 
Science and Technology, Government of India, for support under grant no. SR/S2/JCB-64/2007.



%% file: hh/hh.tex

\chapter{A Realistic Analysis of Non-Resonant BSM Higgs Pair Production}

{\it A.~J.~Barr, M.~J.~Dolan, C.~Englert, M.~M.~M\"uhlleitner, M.~Spannowsky}



\begin{abstract}
  After the Higgs boson discovery the question of the relation of the
  symmetry breaking potential to the TeV scale is more pressing than
  ever. Models that invoke the notion of TeV-scale compositeness are
  potential solutions that interpret the Higgs boson as a
  pseudo-Nambu-Goldstone mode of a strongly interacting sector that
  allows to postpone the hierarchy problem to a higher scale. In these
  scenarios the Higgs potential, induced by a Coleman-Weinberg type
  mechanism, can be dramatically different from the Standard Model
  (SM). Multi-Higgs production provides the only direct avenue to
  analyze the Higgs potential directly. We therefore perform a
  realistic hadron-level analysis of the di-Higgs final state in the
  $b\bar b \tau^+\tau^-$ channel for the representative MCHM4
  benchmark model, that makes new physics contributions to di-Higgs
  production most transparent.
\end{abstract}

\section{Introduction}

The end of Run I of the LHC has left a changed landscape in particle
physics, which theorists are still coming to terms with. The discovery
of the Higgs boson with mass 125~GeV
\cite{Aad:2012tfa,Chatrchyan:2012ufa} has provided final confirmation
of the realisation of the Higgs mechanism, in the process obliterating
any number of Higgsless theories constructed by theorists. At the same
time, and despite a multitude of searches performed by the ATLAS and
CMS experiments, no evidence of Beyond-the-Standard-Model (BSM)
physics has yet emerged from the LHC. However, according to the vague
dictates of the naturalness criterion signs of BSM physics should
emerge at the higher energies available in Run II: new states are
expected around the TeV scale in order to solve the hierarchy
problem. Along with softly-broken supersymmetry, new strong
interactions are the only other construction which can solve the
hierarchy problem in a way which leads to testable predictions. One of
the most prominent among these scenarios is the interpretation of the
observed Higgs boson as a Pseudo-Nambu Goldstone of a
strongly-interacting sector
\cite{Kaplan:1983fs,Kaplan:1983sm,Dimopoulos:1981xc,Georgi:1984ef,Georgi:1984af,Dugan:1984hq}.

A generic prediction of such models is the existence of new resonances
which are partners of the Standard Model top quark. These can
drastically alter Higgs phenomenology. In the first instance, the top
partners can propagate in the one-loop triangle diagram associated
with Higgs production from gluon fusion. However, composite Higgs
models also predict the existence of higher dimensional $HHt \bar t$
interactions whose effects can only be accessed in di-Higgs
production. Multi-Higgs production thus emerges as one of the most
attractive means of probing the composite Higgs scenario. In this
contribution to the Les Houches proceedings we study the di-Higgs
phenomenology and LHC14 prospects of one particular realisation of the
Pseudo-Nambu Goldstone-ism (pNG) paradigm, the so-called MCHM4 model
which we now briefly discuss.

\section{The Model}

Composite Higgs models are based upon the idea that the Higgs boson
could be a bound state of a strongly interacting sector rather than a
fundamental scalar. Such a state can be naturally lighter than the
other states associated with the strongly interacting sector if the
Higgs is the pNG of a large global symmetry $\mathcal{G}$ associated
with the strong dynamics which is then dynamically broken to a smaller
subgroup $\mathcal{H}$ at a scale $f$. The SM gauge group
$SU(2)_L\times U(1)_Y$ is then gauged into $\mathcal{H}$, and the
coset $\mathcal{H}/\mathcal{G}$ is required to contain an $SU(2)_L$
doublet, which can be associated with the SM Higgs boson.  The Higgs
potential is generated at loop-level and breaks electroweak
symmetry. Deviations from SM Higgs phenomenology are parametrised by
the dimensionless quantity $\xi=v/f$, with the SM-like behaviour being
realised in the limit $\xi \to 0$ (and infinitely heavy top
partners). A recent review of the status of composite Higgs models can
be found in Ref.~\cite{Bellazzini:2014yua}.

The simplest such model is given by a strong sector with the gauge
symmetry $\mathcal{G}=SO(5)\times U(1)_X$ which is broken down to
$\mathcal{H}=SO(4)\times U(1)_X$. Since $SO(4)$ is isomorphic to
$SU(2)_L \times SU(2)_R$ the SM gauge group $SU(2)_L \times U(1)_Y$
can easily be embedded. In the minimal composite Higgs model
MCHM4~\cite{Agashe:2004rs} the Standard Model fermions transform as
spinorial representations of $SO(5)$.  This leads to
\textit{universal} rescalings of all the Higgs couplings and of the
trilinear Higgs coupling by $\sqrt{1-\xi}$.  Note, that this is not
the case in other composite Higgs scenarios. Furthermore, there are
also direct four-point couplings between two Higgs bosons and two
fermions, $HHf\bar f$, whose coefficients are proportional to $\xi
m_f/v^2$. They therefore vanish in the SM limit of $\xi\to 0$. While
the MCHM4 and associated models have a rich and varied phenomenology,
these facts are sufficient for our study of the di-Higgs
phenomenology.

\begin{figure}[!t]
  \centering
  \includegraphics[width=0.75\textwidth]{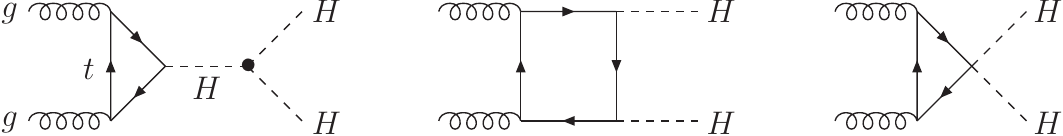}
  \caption{\label{fig:diag} Representative Feynman diagrams entering
    double Higgs production in MCHM4.}
\end{figure}

\begin{figure}[!t]
  \centering
  \includegraphics[width=0.48\textwidth]{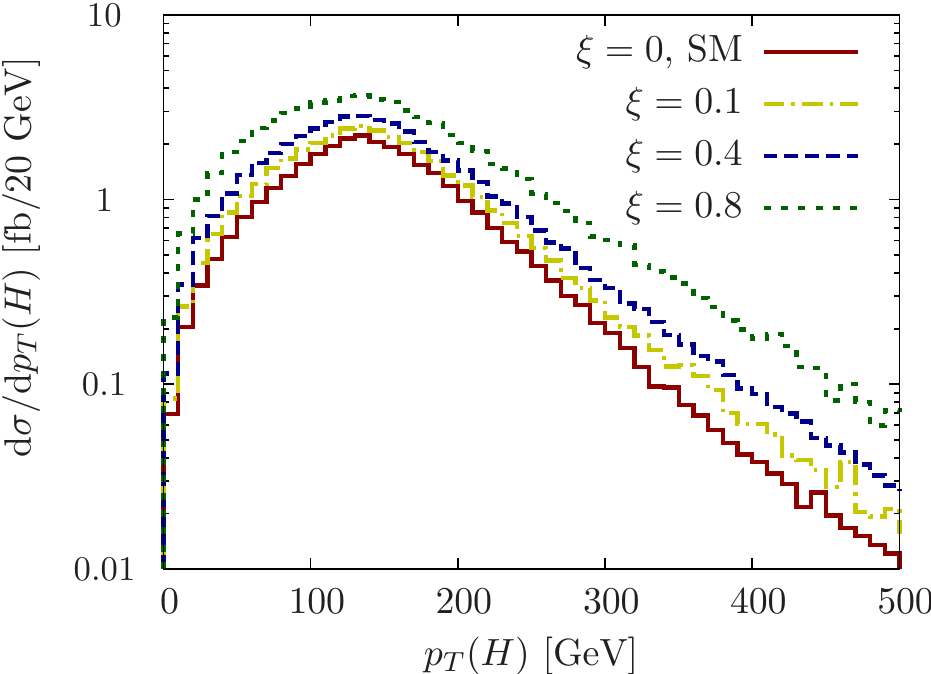}
  \hfill
  \includegraphics[width=0.48\textwidth]{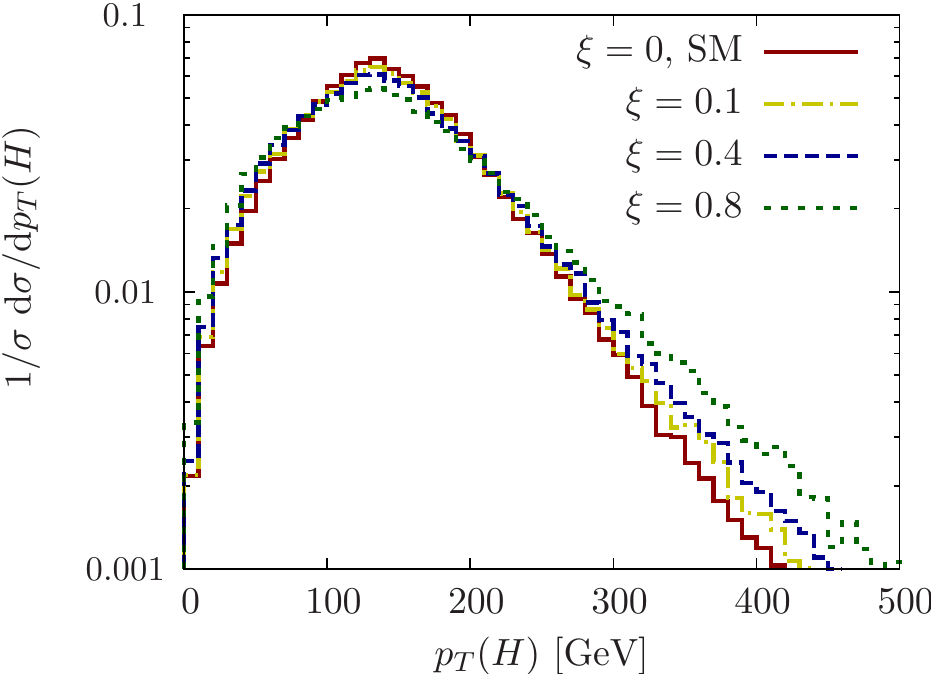}
  \caption{\label{fig:pt} Comparison of the inclusive Higgs transverse
    momentum spectrum for different values of $\xi$, a total
    $K$ factor estimate is included~\cite{Baglio:2012np}. We also show
    a shape-comparison in the right panel. It demonstrates the
    standard lore of composite scenarios: there are large deviations
    when the characteristic new physics scale starts to get resolved.}
\end{figure}

\subsection{Simulation results}
Our analysis is a re-evaluation of Ref.~\cite{Barr:2013tda} adapted to
the MCHM4 model as introduced in the previous section. The
phenomenology of the Higgs decays in MCHM4 remains unmodified with
respect to the SM, as a consequence of the global re-scalings of the
SM-like couplings, i.e. all the partial decay widths and hence also
the total Higgs width decrease by $1-\xi^2$ so that the Higgs
branchings ratios remain identical to SM case. Therefore there is a
strict discrimination between single Higgs phenomenology and double
Higgs phenomenology in MCHM4, especially because of the novel $HHt
\bar t$ couplings introduced in the previous section. This is the
reason why we choose to work in MCHM4, even if this model suffers from
electroweak precision constraints due to breaking of custodial isospin
symmetry.\footnote{These constraints can be weakened somewhat by new
  heavy fermions
  \cite{Lodone:2008yy,Gillioz:2008hs,Anastasiou:2009rv,Gillioz:2013pba}.}
We consider only gluon-induced $HH$ production \cite{Glover:1987nx} in
the following as this is the dominant contribution to $HH$ production
at the 14 TeV LHC run; representative Feynman diagrams that contribute
to this process in MCHM4 are depicted in Fig.~\ref{fig:diag}.

Events are generated using an MCHM4-adapted framework of
Refs.~\cite{Barr:2013tda,Dolan:2012rv} in the Les Houches event
standard and subsequently showered and hadronized with {\sc{Herwig++}}
\cite{Bahr:2008pv}. The cross-sections then agree with those obtained
in Ref.~\cite{Grober:2010yv}. We report results directly in relation
to the SM limit, $\xi=0$. Since QCD corrections to di-Higgs production
are driven by soft gluon emission
\cite{Frederix:2014hta,deFlorian:2013jea,Grigo:2013rya} we expect this
ratio to be robust against modified higher order QCD effects which
could in principle modify the total $K$ factor away from $K\simeq 2$
\cite{Baglio:2012np,Dawson:1996xz}, especially when heavy top partners
are included to the picture \cite{Dawson:2012mk}.

The simulation incorporates detector resolution effects based on the
ATLAS ``Krak\' ow'' parameterisation \cite{ATL-PHYS-PUB-2013-004},
which implements a conservative estimate of the expected detector
performance during the high-luminosity LHC run with a target of
$3/{\rm{ab}}$. We account for the effects of pile-up (with a mean
number of $pp$ collisions $\mu=80$), and include $\sum E_T$-dependent
resolutions for jets and missing transverse momentum.

We include both leptonic and hadronic $\tau$ candidates and assume
that events pass the trigger in the presence of two $\tau$s with
$p_T>40~{\rm{GeV}}$ or one $\tau$ with $p_T>60~{\rm{GeV}}$. The $\tau$
efficiencies and fake rates are based on
Ref.~\cite{ATL-PHYS-PUB-2013-004}, jets are reconstructed using the
anti-$k_t$ algorithm \cite{Cacciari:2011ma,Cacciari:2008gp} with
radial parameter $0.6$. We require exactly two reconstructed $\tau$
leptons as well as exactly two jets which need to pass $b$
tags. Furthermore, the $b$-tagged jets have to reconstruct the Higgs
mass within a 25 GeV window. Separating $H\to \tau^+\tau^-$ from $Z\to
\tau^+\tau^-$ is one of the main obstacles of this analysis. To
reflect the corresponding systematics, we discuss results for an
optimistic and a pessimistic estimate on the reconstruction of the
invariant $\tau$-pair mass. The former is performed without
$\slashed{p}_T$ resolution, demanding $100~{\rm{GeV}} < m_{\tau\tau} <
150~{\rm{GeV}}$ while the latter includes a conservative treatment of
the $\slashed{p}_T$ resolution and requires $80~{\rm{GeV}} <
m_{\tau\tau} < 130~{\rm{GeV}}$ thus including a dominant\footnote{It
  is shown in Ref~\cite{Barr:2013tda} that the background from
  $t\bar{t}\rightarrow bb\tau\tau\nu\nu$ can be greatly reduced using
  a selection on the kinematic variable $m_{T2}$} background
contribution from $Z\to \tau^+\tau^-$.  Further details can be found
in \cite{Barr:2013tda}.

Performing the analysis we find that trigger criteria and the event
selection do not induce a notable bias: the event selection
efficiencies are almost independent of $\xi$. The dominant
discriminating power, hence, results from the modified cross section,
that results from an enhancement over the entire relevant Higgs $p_T$
range, see Fig.~\ref{fig:pt}. Additional sensitivity to $\xi$ comes from
the relative increase of harder events for larger values of
$\xi$.

\begin{figure}[!t]
  \centering
  \includegraphics[width=0.48\textwidth]{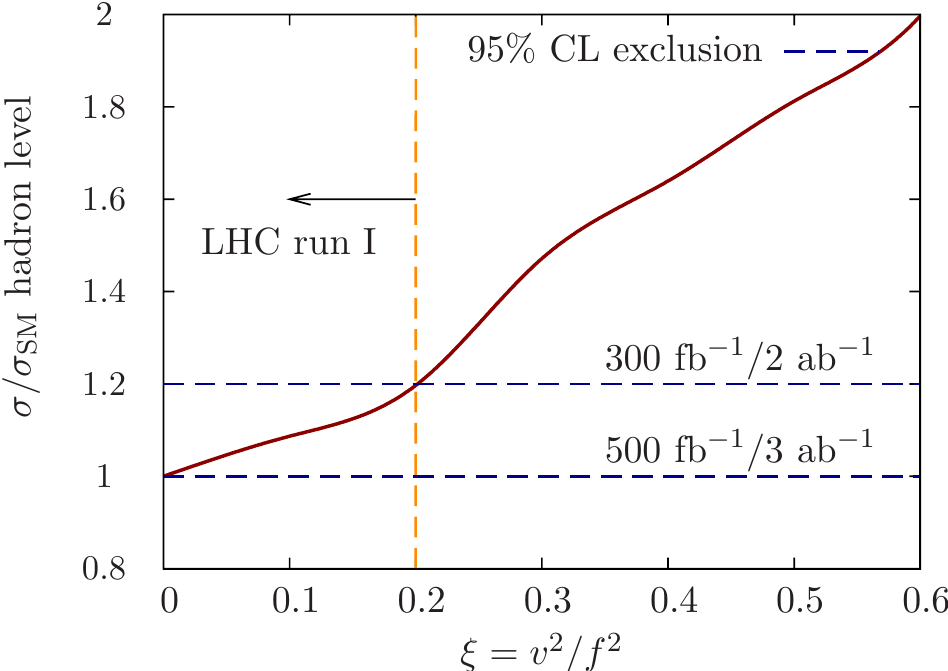}
  \caption{\label{fig:ptf} Double Higgs production cross section of
    MCHM4 in the full hadron-level analysis of $HH\to b\bar b
    \tau^+\tau^-$. For the two luminosities at each dashed line, see
    text.}
\end{figure}

The combined result of the described analysis is shown in
Fig.~\ref{fig:ptf}, where we show the 95\% CL constraints that are
expected for a measurement in the $HH\to b\bar b \tau^+\tau^-$
channel. The different values of the integrated luminosity reflect our
two choices of including the systematics of the $\tau$ pair
reconstruction. We also include the current constraint on these
model classes from Higgs data, $\xi \lesssim 0.2$.
Fig.~\ref{fig:ptf} excellently demonstrates that di-Higgs production
is an excellent tool to constrain the parameter region of composite
Higgs models with the LHC target luminosity of $3/{\rm{ab}}$. We
expect quantitatively improved result for, e.g.,
MCHM5~\cite{Contino:2006qr}, which exhibits a strong dependence on
$\xi$, in particular when additional dynamical top partners are taken
into account~\cite{Gillioz:2012se,Dolan:2012ac}.

\section*{Conclusions}
Di-Higgs production in composite Higgs scenarios is significantly
enhanced over the SM. In this article, we have adapted a hadron-level
analysis of $pp\to HH\to b\bar b \tau^+ \tau^-$ to the MCHM4 model. It
is expected that sensitivity to a
SM-like cross section can be obtained at the LHC 14 TeV with high
luminosity. As a consequence, the LHC will become sensitive to deviations 
that are apparent when the Higgs boson emerges as part of a strongly
interacting sector with a parametrically suppressed mass. In the
absence of a signal, i.e. an enhancement of the di-Higgs cross
section, the majority of the parameter space of the MCHM4 can be
excluded. Since it is known that other composite models often exhibit an even larger sensitivity in the di-Higgs final state, this result is rather general.

\section*{Acknowledgements}
We thank the organizers of the 2013 Les Houches workshop for keeping a well stocked bar for the duration of our stay.



%% file: vbfhh/HH_VV_final.tex
\graphicspath{{vbfhh/}}

\chapter{Resonant Higgs Pair Production in Vector Boson Fusion at the LHC}

{\it A.~Belyaev, O.~Bondu,  A.~Massironi, A.~Oliveira, R.~Rosenfeld, V.~Sanz}


%


\begin{abstract}
We examine resonant Higgs pair production at
the LHC in vector boson fusion. This channel directly tests
the couplings of the longitudinally polarized gauge bosons to new physics
and therefore it is important as a test to any new model. 
In particular, we use as benchmark a model 
of warped extra dimensions where a KK-graviton can be produced
on-shell through vector boson fusion and subsequently decays into
a pair of Higgs bosons.
We concentrate on the final state with 4 b quarks and 2 jets and devise cuts
to decrease the irreducible QCD and EW backgrounds.
We did not include effects of showering and hadronization in this preliminary analysis.
For this reason we do not impose any cuts that
relies on the reconstruction of either the resonance or the Higgs masses, cuts which would significantly reduce the SM backgrounds.
Although our results are over-pessimistic at this point, we believe
that there is great potential in this channel once a more realistic analysis with appropriate cuts is performed.
\end{abstract}

\section{INTRODUCTION}
In the Standard Model (SM), the production of a pair of Higgs bosons (H) is the only way to directly examine the Higgs self-coupling $\lambda$. 
Production of such pairs can proceed either via gluon fusion (GF) or vector boson fusion (VBF).

In the case of the SM, this coupling is predicted once the Higgs mass is known. 
Following the LHC discovery \cite{Chatrchyan:2012ufa,Chatrchyan:2013lba,Aad:2012tfa}, the self coupling is determined to be:
\begin{equation}
\lambda_{SM} = \frac{m_h^2}{2 v^2} \approx 0.13.
\end{equation}

This in turn predicts a rather small Higgs pair production cross section for a centre of mass energy of $\sqrt{s} = 14$~TeV :
$\sigma(pp \rightarrow gg \rightarrow hh +X) \approx 30$~fb and $\sigma(pp \rightarrow (WW,ZZ) \rightarrow hh +X) \approx 3$~fb \cite{Plehn:1996wb,Djouadi:1999rca}.

There exist several models Beyond the Standard Model (BSM) which predict different values for this coupling and that can lead to enhanced cross sections for Higgs pair production, 
both for the case of GF \cite{Dib:2005re,Gillioz:2012se} and VBF \cite{Contino:2010mh} processes.

However, even larger enhancements arise if there are new particles with masses within LHC reach that can decay on-shell into a Higgs boson pair, leading to the possibility of resonant Higgs boson pair production. Such a production process was studied in the context of the MSSM\cite{Djouadi:1999rca}, NMSSM\cite{Ellwanger:2013ova}, and
2-Higgs doublet models\cite{Moretti:2007ca}, where the resonances are identified with heavy Higgses. Models with Warped Extra Dimensions (WED) of the Randall-Sundrum type \cite{Randall:1999vf,Randall:1999ee} also contain particles that can contribute to resonant Higgs pair production, namely the radion and the KK-graviton.

A recent study was performed to demonstrate the feasibility of detection of resonant Higgs pair production in GF in the final state with 4 bottom quarks. This study featured a new tagging algorithm, that was developed in order to deal with the different boosting regimes of the decaying Higgses \cite{Gouzevitch:2013qca}. 
The benchmark model used in this study was a Randall-Sundrum (RS) model where the radion or the KK-graviton can decay into a Higgs boson pair.

The present contribution uses the same benchmark model, focusing on the KK-graviton contribution, to study resonant double Higgs production via vector boson fusion. 
The VBF signature, with the handle of two extra high-$p_T$ jets could make this channel competitive with the GF searches.
The couplings of the longitudinally polarized gauge bosons to the resonance play a crucial role and hence can be sensitive to details of the model.

\section{THE MODEL}
We write the couplings of the radion ($r$) and the KK-graviton ($G_{\mu \nu}$, later denoted by $G$) at the linear level with matter as (see, e.g. \cite{Lee:2013bua} and references therein):
\begin{equation}
{\cal L} = -\frac{c_i}{\Lambda} G_{\mu \nu} T^{\mu \nu}_i - \frac{d_i}{\sqrt{6}\Lambda} r T_i
\end{equation}
where $\Lambda$ is the compactification scale, $ T^{\mu \nu}_i$ is the energy-momentum tensor of species $i$, $T_i$ is its trace, and the coefficients
$c_i$ and $d_i$ are related to the overlap of the 5D profiles of the corresponding particles.
The compactification scale can be related to the curvature $\kappa$ and the size $L$ of the compact extra dimension as:
\begin{equation}
\Lambda = e^{-\kappa L} \bar{M}_{Pl},
\end{equation}  
where $\bar{M}_{Pl} = 2.43 \times 10^{18}$ GeV  is the reduced Planck mass. In order to solve the hierarchy problem 
one must have $\kappa L$ about 35. An important quantity that will be used to characterize the model is 
$\tilde{\kappa}$ defined as
\begin{equation}
\tilde{\kappa} \equiv \frac{\kappa}{\bar{M}_{Pl}} 
\end{equation}
The mass of the first KK-graviton resonance $M_G$ is given by
\begin{equation}
M_G = x_1 \tilde{\kappa} \Lambda,
\end{equation}
where $x_1 = 3.83$.

Since in models of warped extra dimensions the profiles of both the radion and the KK-graviton are peaked towards the infrared (IR) brane
we assume that the dominant couplings are with particles that are also mostly localized in the IR, namely the Higgs boson (H), 
the longitudinal components of the electroweak gauge bosons ($W_L$ and $Z_L$) and the right-handed top quark ($t_R$) 
\cite{Agashe:2007zd,Fitzpatrick:2007qr}, in which case we take $c=d=1$ for these particles. The coupling to the transverse
polarizations of the gauge bosons (which we denote $c_g$) is suppressed due to the fact that their wave function is flat in the extra dimension and
for definiteness we use $c_g = 0.0137$ (this value corresponds to  $\kappa L = 35$  and  has a very mild dependence on $M_G$).
This fixes the couplings that are relevant
for resonant Higgs pair production in VBF. In the following we concentrate only on the KK-graviton $G$ contribution.
The model was implemented in MadGraph 5~\cite{Alwall:2011uj} 
using a modified version of the FeynRules RS implementation in \cite{aquino}
 and checked with CalcHEP \cite{Belyaev:2012qa}.

\begin{figure*}[h]\begin{center}
\includegraphics[width=0.5\textwidth]{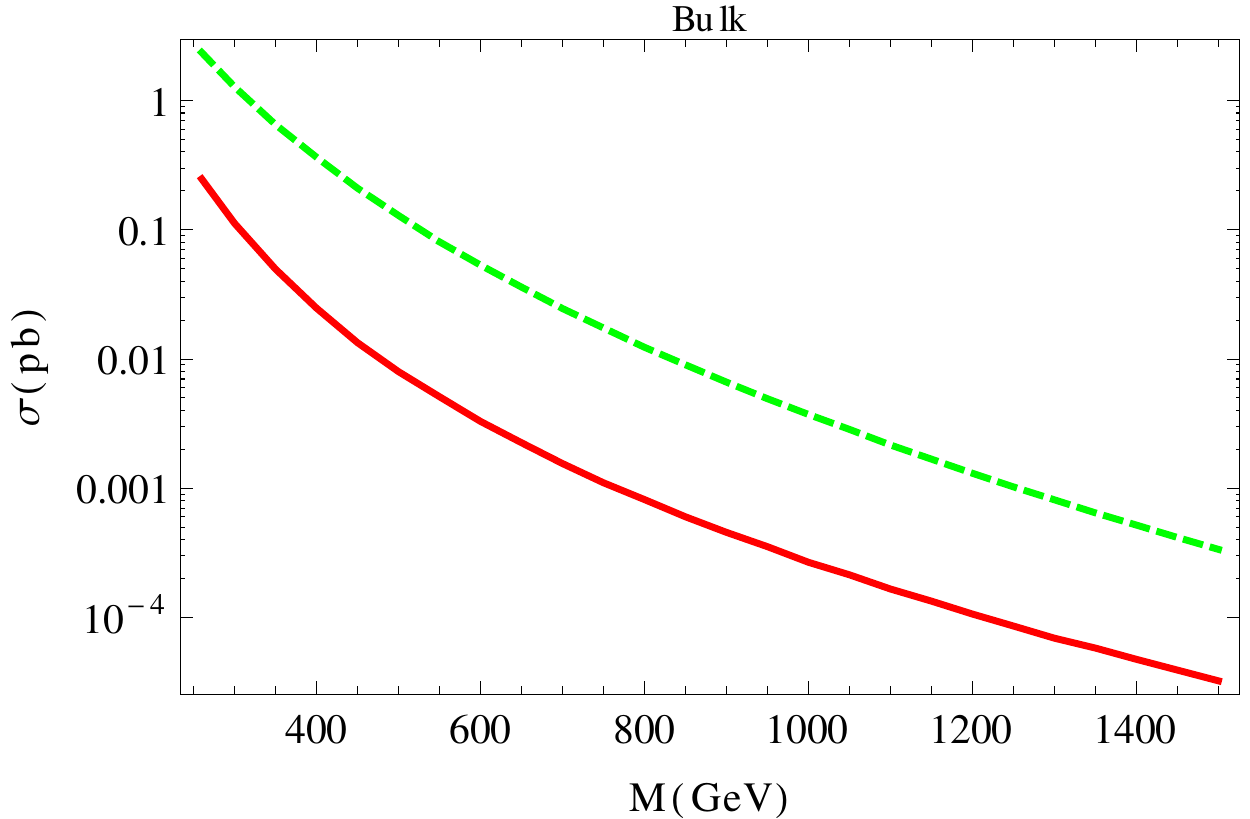}
\caption{Production cross section of KK-graviton (in pb) as a function of its mass with at the LHC13 with $\tilde{\kappa} = 0.1$ for the GF (dashed green line) and 
VBF (solid red line) processes. }
\label{fig:crosssection}\end{center}\end{figure*}

Fig.~\ref{fig:crosssection} shows the total cross section for the GF and VBF production of a KK-graviton at the LHC13 as a function of
its mass for fixed $\tilde{\kappa} = 0.1$. Typically the GF cross section is one order of magnitude larger than the corresponding VBF one.

\begin{figure}[h]
\centering
\includegraphics[width=0.7\textwidth]{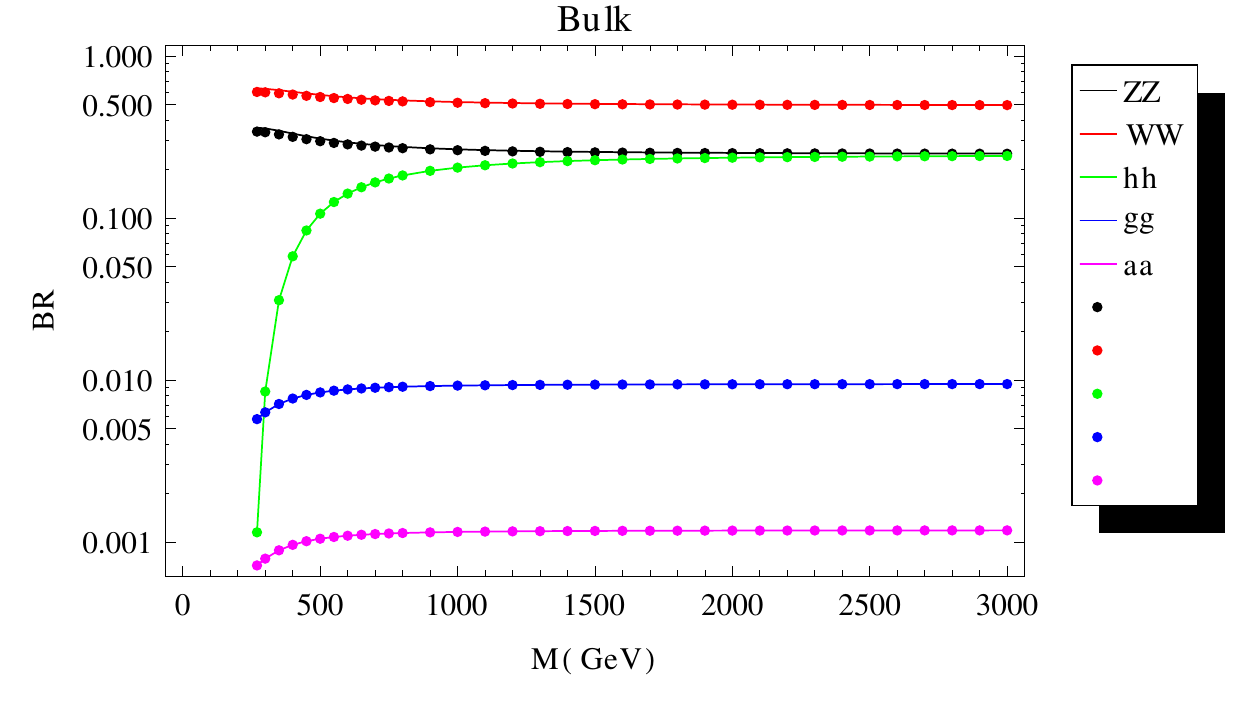}
\caption{
Branching ratios for bulk KK-graviton assuming no coupling with top quark. 
The solid lines represent analytical calculations and the dots are the numerical results from the model implementation. \label{fig:width}}
\end{figure}

We show in Fig.~\ref{fig:width} the branching ratios of the KK-graviton as a function of its mass, compared with the analytical predictions of \cite{Lee:2013bua}.
The branching ratios are insensitive to the parameter $\tilde{\kappa}$.
The total width for KK-graviton in all the suitable parameter space is less than 5 GeV \cite{Giudice:2000av}. 
Since the cross section results are not sensitive to the KK-graviton width, we keep
it fixed at 1 GeV in what follows.  

\begin{figure}[h]
\centering
\input{vbfhh/higgsDiagramTikzVBFhh}
\input{vbfhh/higgsDiagramTikzWhh}
\caption{Examples of Feynman diagrams for Higgs pair production via a bulk KK-graviton resonance in the VBF (left) and associated production (right) cases.}
\label{fig:resonantVBFhh}
\end{figure}
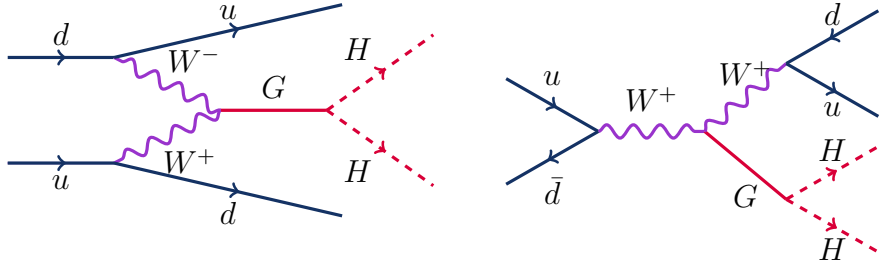

We will focus in the rest of this contribution on the process $pp\to G jj$, where the two jets come from vector boson fusion (see Fig.~\ref{fig:resonantVBFhh} (left)). It should be noted that KK-graviton production in association with a $W$ or $Z$ boson (see Fig.~\ref{fig:resonantVBFhh} (right)) can in principle lead to the same final state. The cross section of the former is expected to be larger than the latter, and the kinematics of the two processes are expected to differ, leading to possibly different final selections. 
In principle these two contributions can not be separated but imposing specific cuts can favour one over the other. 
For instance, requiring $M_{jj} > 100$ GeV reduces the contribution from the associated production process.

\section{SIGNAL GENERATION AND VBF CUTS}

We will now focus on studying the VBF production process $pp \to G jj$ following the model described in the previous section, for the prospect of the HL-LHC, assuming proton-proton collisions at a centre of mass energy of $\sqrt{s} = 13$~TeV for an integrated luminosity of $ L = 3\,{\rm ab}^{-1}$.

We generated 20k signal events in the model setting $\tilde{\kappa} = 0.1$ and varying the KK-graviton mass $M_G$ from $260$ to $1450$ GeV. 
Only light quarks $u$ and $d$ in the proton and as final state jets have been considered. 
The KK-graviton was required to be produced on-shell and further decayed into a pair of Higgs bosons in Madgraph, with a mass of $m_H = 125$~GeV.

The following cuts were used for signal generation:
\begin{equation}
p_T^j > 10 \; \mbox{GeV}, \;\;\;  M_{jj} > 100 \;\mbox{GeV}, \;\;\; |\eta_j|<5 ,  \;\;\; \Delta R(jj) > 0.01 
\label{eq:basic}
\end{equation}
\begin{equation}
p_T^b > 20 \; \mbox{GeV}, \;\;\; |\eta_b|< 2.5 \label{eq:basicb}
\end{equation}

We will further concentrate on the decay of both Higgs bosons into b-quarks, leading to a $b\bar{b}b\bar{b}jj$ final state. 
Assuming SM branching fractions, this decay final state would represent roughly $33$\% of the total amount of Higgs boson pairs produced.

\section{BACKGROUNDS FOR $b\bar{b} b\bar{b} jj$ FINAL STATE}

To study potential detection of this channel at the LHC13, several processes leading to the same final state have to be considered, and will constitute an irreducible background. 
Three types of irreducible background have been considered :
\begin{itemize}
\item QCD production of four $b$--quarks and two additional jets. This is the background with the largest cross section; however it does not exhibit the same kinematics and can 
be kept under control using appropriate cuts. These events were generated with Alpgen \cite{Mangano:2002ea}.
\item The Standard Model VBF production of Higgs boson pairs, with the Higgs boson decaying into $b$--quarks. Even though this background is the most similar to our signal in terms of kinematics, the di-Higgs pair production cross section is small and it does not have a resonant structure. These events were generated with Madgraph.
\item Electroweak background producing $Z$ bosons pairs decaying to $b$-quarks in association with jets. These events were generated with Madgraph.
\end{itemize}

Table~\ref{tab:samples} summarizes the different samples used, along with the expected cross-sections and number of generated events.

\begin{table}[h]
\centering
\begin{tabular}{l|c|c|c}
Process & Generator & $\sigma$ (pb) & Number of events \\
\hline
QCD $b\bar{b}b\bar{b}jj$ & Alpgen & 148  & 500k\\
EWK $Z(b\bar{b})b\bar{b} jj$ & Madgraph & 1.46 & 200k \\
EWK $Z(b\bar{b})Z(b\bar{b}) jj$ & Madgraph & 0.06 & 210k\\
VBF SM $H(b\bar{b})H(b\bar{b}) jj$ & Madgraph & $2.71 \times 10^{-4}$  & 50k\\ \hline
\end{tabular}
\caption{\label{tab:samples} Background samples and the corresponding cross sections used for the analysis.}
\end{table}

The cuts performed at the generator level for the QCD background were:
\begin{eqnarray}
&p_T^j& > 20\; \mbox{GeV}, \;\; |\eta_j| < 5, \;\;  \Delta R(jj) > 0.3, \;\; \Delta R(bj) > 0.3, \;\; p_T^b > 20 \; \mbox{GeV}, \;\; |\eta_b| < 2.5, \nonumber \\
&\Delta R(bb)& > 0.1 
\end{eqnarray}
and for the electroweak background were:
\begin{eqnarray}
&p_T^j& > 20\; \mbox{GeV}, \;\; |\eta_j| < 5, \;\;  \Delta R(jj) > 0.3, \;\; \Delta R(bj) > 0.4, \;\; p_T^b > 20 \; \mbox{GeV}, \;\; |\eta_b| < 2.5,  \nonumber \\
&\Delta R(bb)& > 0.1,  \;\; M_{jj} > 100 \; \mbox{GeV}. 
\end{eqnarray}
The $ \Delta R(bb)$ is relaxed with respect to the default value in order to take into account the case of a boosted Higgs. 
The $M_{jj}$ cut, design to favour the VBF topology, was implemented  only in the Madgraph generated background since it is not straightforward to do so in Alpgen.

\section{RECONSTRUCTION OF THE PROCESS $pp \to G \to HH jj \to b\bar{b} b\bar{b}jj$}

We cluster the ``jets" with the anti-$k_T 5$  algorithm using {\tt FastJet3} \cite{Cacciari:2011ma}. This procedure makes sense 
even at a parton level preliminary analysis, since two partons that are close enough can cluster on a jet.

Following \cite{Gouzevitch:2013qca} we base the reconstruction algorithm for event selection on 
the number of jets per event. For the presence of jets with non-trivial substructure and high invariant mass, 
we use the mass drop definition of a fat jet  \cite{Butterworth:2008iy}.
We describe below the analysis flow we adopted. 




We require at least two jets in the event after $b$ and fat jet--tagging. 
The VBF jets are chosen as the pair that holds the largest invariant mass among the non-tagged jets. 
In this preliminary study we are considering only events with four b's and assuming perfect $b$--tag performance. 
In deriving our results we use more realistic cuts on the VBF jets:
\begin{equation}
p_T^j > 30 \; \mbox{GeV}, \;\;\;  M_{jj} > 400 \;\mbox{GeV}, \;\;\; |\eta_j|<4.7 
\label{eq:qual}
\end{equation}

After this first selection we reconstruct the Higgses from the remaining jets.
We also take advantage of the fact that the di-Higgs system is boosted by using the following cuts:
\begin{equation}
M_{HH} > 250 \;\mbox{GeV}, \;\;\;  p_T^{HH} > 60 \;\mbox{GeV}, \;\;\; \Delta \eta(HH) < 2 \label{eq:qualHH}
\end{equation}

We follow closely \cite{Gouzevitch:2013qca} and classify events by the number of mass-drop tagged jets (fat jets). We define mass drop by the tagger parameters:

\begin{equation}
\mu = 0.67 , \;\;\;  y_{cut} = 0.09, \;\;\;  M_{\mbox{\small{fat jet}}} > 120 \;\mbox{GeV}  \label{eq:tagger}
\end{equation}

We detail below event selection and classification of our tagging algorithm.

\begin{itemize}
  \item If the two hardest jets in the event are found fat and satisfy the di-Higgs quality requirements of equation (\ref{eq:qualHH}) the event is assigned to the {\bf 2-tag}.
  \item If one of the two fat jets do not pass quality requirements or  only one of the jets among the two hardest is fat, then events with fewer than three jets after cuts are discarded.
 We then apply eqs.~\ref{eq:basic} and \ref{eq:basicb} and if the event passes those cuts it is assigned to {\bf 1-tag} sample. 
  \item If the event is not assigned as {\bf 2-tag} or {\bf 1-tag} sample, we check the event can be classified as being two resolved Higgses. \\
Events with fewer than six jets passing the basic cuts are discarded.
To select the Higgs candidates we pair the four non-VBF jets $i,j,k,l$ minimizing the mass difference $|M_{ij}-M_{kl}|$. If the two reconstructed Higgs candidates satisfy the cuts (\ref{eq:qualHH}) the event is classified as belonging to the {\bf 0-tag} sample.  
\end{itemize}

We are now ready to discuss the preliminary parton level results.

\section{RESULTS}

\begin{figure*}[hbtp]\begin{center}
\includegraphics[width=.49\textwidth]{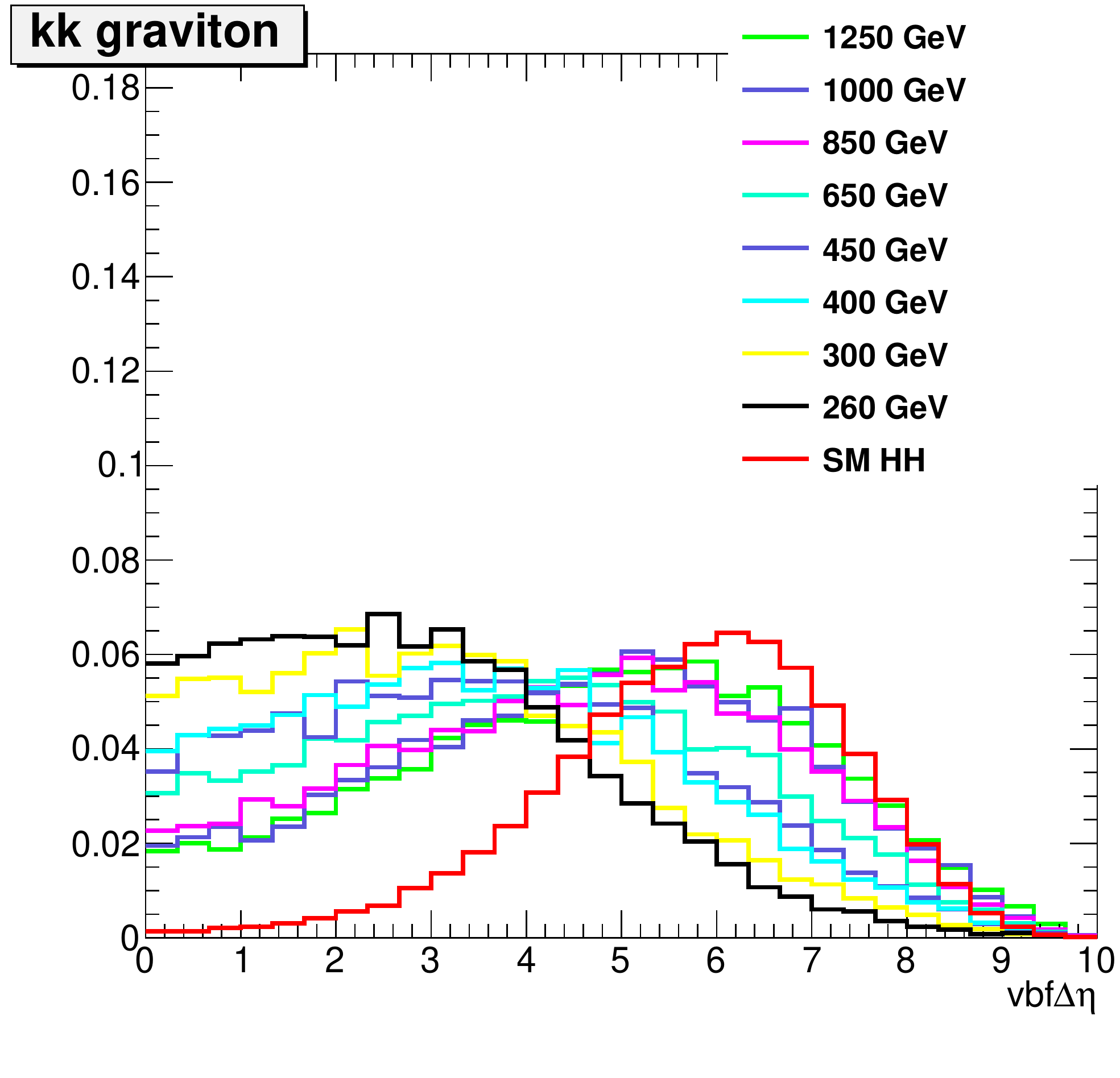}
\caption{ Separation in pseudo rapidity between the two VBF jets $\Delta\eta (jj)$.  No additional cuts applied, the different contributions have been normalized to the same area.
 \label{fig:njets}}
\end{center}\end{figure*}

We first discuss the general properties of the bulk graviton signal hypothesis. 
In figure (\ref{fig:njets}) we show the pseudo-rapidity difference among the two VBF jets. 
The kinematics of the VBF jets in the signal differ from the case of the Standard Model VBF Higgs production, which tend to be more central. 
This is due to the fact that the KK-graviton vertices with massive vector bosons have a particular $p_T$ dependence, and from the fact that the 
graviton is a spin 2 particle while the SM Higgs is a scalar. This effect has been already explored e.g. in reference \cite{Djouadi:2013yb}.


\begin{figure*}[hbtp]\begin{center}
\includegraphics[width=0.5\textwidth]{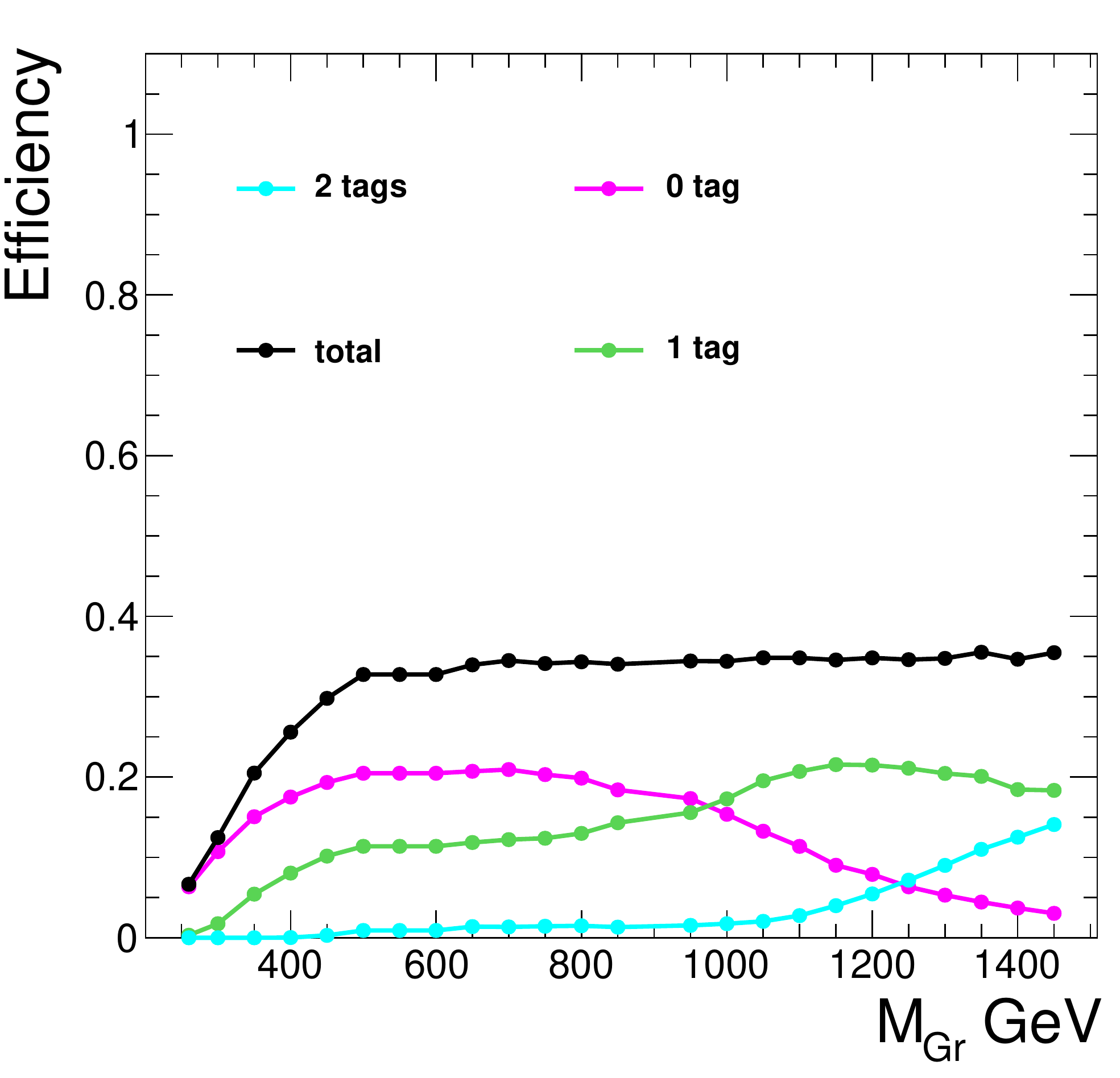}
\caption{Parton level number separation on number of mass drops after analysis flow. 
 \label{fig:njets2}}
\end{center}\end{figure*}

We present the cut flow of our analysis in Table~ \ref{table:cut_flow}, where we show the impact of the cuts on the VBF jets  system (\ref{eq:qual}) and on di-Higgs system (\ref{eq:qualHH}).
To visually complement the cut flow we also display the parton level signal reconstructed efficiency separated by the number of mass drop tags in Fig.~\ref{fig:njets2}. 
In this figure we see that for KK-graviton mass hypothesis larger than 1 TeV there is a systematic increase of events with two tags, 
meaning the two $b$--quarks products of each one of the Higgses start to merge in one jet. 
This change of behaviour  at 1 TeV is expected from calculating the distance radius between two partons giving the boost of their mother particle \cite{Gouzevitch:2013qca}. 

\begin{table}[h,t,b]
\centering
\resizebox{\textwidth}{!}{
\begin{tabular}{|c|c|c|c|c|c|c|}
\hline
Sample & basic cuts  & jet merging &  &   &  &  \\
 &  (eqs. \ref{eq:basicb},\ref{eq:basic}) &  (akt5) & $M_{jj} > 400$ GeV & $M_{HH}>250$ GeV & $p_T^{HH} > 60$ GeV & $\Delta\eta_{HH}<$ 2 \\ \hline
1450  GeV & 0.53 & 0.49 & 0.44 & 0.44 & 0.41 & 0.35 \\
1250  GeV & 0.52 & 0.47 & 0.43 & 0.43 & 0.40 & 0.35 \\
1050  GeV & 0.52 & 0.48 & 0.44 & 0.44 & 0.40 & 0.35 \\
850  GeV & 0.51 & 0.47 & 0.42 & 0.42 & 0.39 & 0.34 \\
650  GeV & 0.51 & 0.43 & 0.39 & 0.39 & 0.36 & 0.33 \\
450  GeV & 0.54 & 0.38 & 0.34 & 0.34 & 0.32 & 0.30 \\
400  GeV & 0.55 & 0.33 & 0.29 & 0.29 & 0.28 & 0.26 \\
300  GeV & 0.58 & 0.17 & 0.14 & 0.14 & 0.13 & 0.12 \\
260  GeV & 0.60 & 0.09 & 0.07 & 0.07 & 0.07 & 0.07 \\ \hline
SM	$H(b\bar{b})H(b\bar{b})$ jj & 0.41 & 0.38 & 0.36 & 0.36 & 0.30 & 0.12 \\
$Z(b\bar{b})$ $b\bar{b}$ jj & 0.50 & 0.36 & 0.14 & 0.10 & 7.91E-02 & 4.55E-02 \\
$Z(b\bar{b})Z(b\bar{b})$ jj & 0.62 & 0.51 & 0.17 & 0.12 & 9.66E-02 & 6.61E-02 \\
$b\bar{b}$ $b\bar{b}$ jj & 0.70 & 0.20 & 0.11 & 6.73E-02 & 5.49E-02 & 4.55E-02 \\ \hline
\end{tabular}
}
\caption{\small Cut flow on signal and BKG. The displayed numbers are efficiencies. 
\label{table:cut_flow}}
\end{table}  

At very low mass ($M_{G} <$ 400 GeV) both Higgses are not very boosted, so there is larger probability 
of jets to superpose. We loose a large portion of the signal due this effect, due to for example the intrinsic mis-tag of events caused by an accidental merging of 
the b's of the two different Higgses.

The most powerful variables for background rejection in this channel are related with its resonant character and rely on reconstructing the resonance.
However, since we performed only a parton-level study without showering we restrain ourselves of using such variables and our results are possibly over-pessimistic.
We present our results for the significance of the signal in Table~\ref{table:nevents}.

\begin{table}[h]
\centering
\begin{tabular}{|c|c|c|c|c|}
\hline
 & $pp \to G(HH) jj$ & $\sigma \times$eff. (pb) & Nevents (3000/fb) & $S/\sqrt{B}$ \\
 & $\sigma$ (pb)   & &  &  \\
 \hline				
1450 GeV & 6.91E-06 & 8.17E-07 & 2.45E+00 & 7.74E-06 \\
1250 GeV & 1.86E-05 & 2.14E-06 & 6.41E+00 & 2.03E-05 \\
1050 GeV & 7.87E-05 & 9.13E-06 & 2.74E+01 & 8.66E-05 \\
850 GeV & 2.63E-04 & 3.00E-05 & 9.01E+01 & 2.85E-04 \\
650 GeV & 1.28E-03 & 1.40E-04 & 4.20E+02 & 1.33E-03 \\
450 GeV & 8.87E-03 & 8.80E-04 & 2.64E+03 & 8.34E-03 \\
400 GeV & 1.41E-02 & 1.20E-03 & 3.60E+03 & 1.14E-02 \\
300 GeV & 1.55E-02 & 6.42E-04 & 1.93E+03 & 6.09E-03 \\
260 GeV & 5.72E-03 & 1.27E-04 & 3.80E+02 & 1.20E-03 \\\hline
\end{tabular}
\caption{\small Cross section and events for HL-LHC,  after all selections of Table \ref{table:cut_flow}. The parameters used are $\tilde{\kappa} = 0.1$ and $\kappa L=35$.
The $S/\sqrt{B}$ ratio is calculated only with respect to the EW backgrounds and without the cut on resolution the mass of the Higgses and KK-graviton candidates. 
\label{table:nevents}}
\end{table}  

\begin{figure*}[hbtp]\begin{center}
\includegraphics[width=0.45\textwidth]{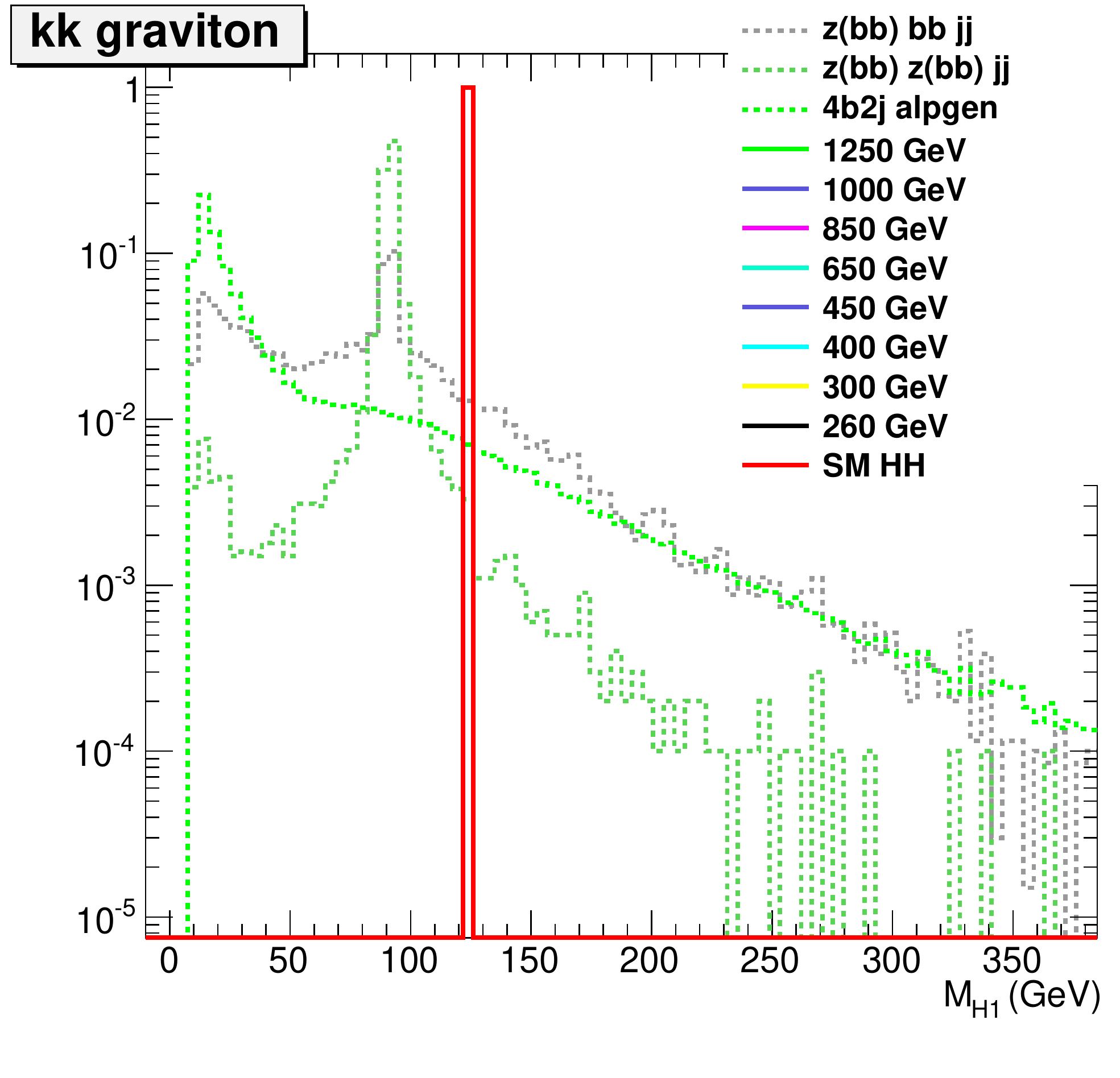}
\includegraphics[width=0.45\textwidth]{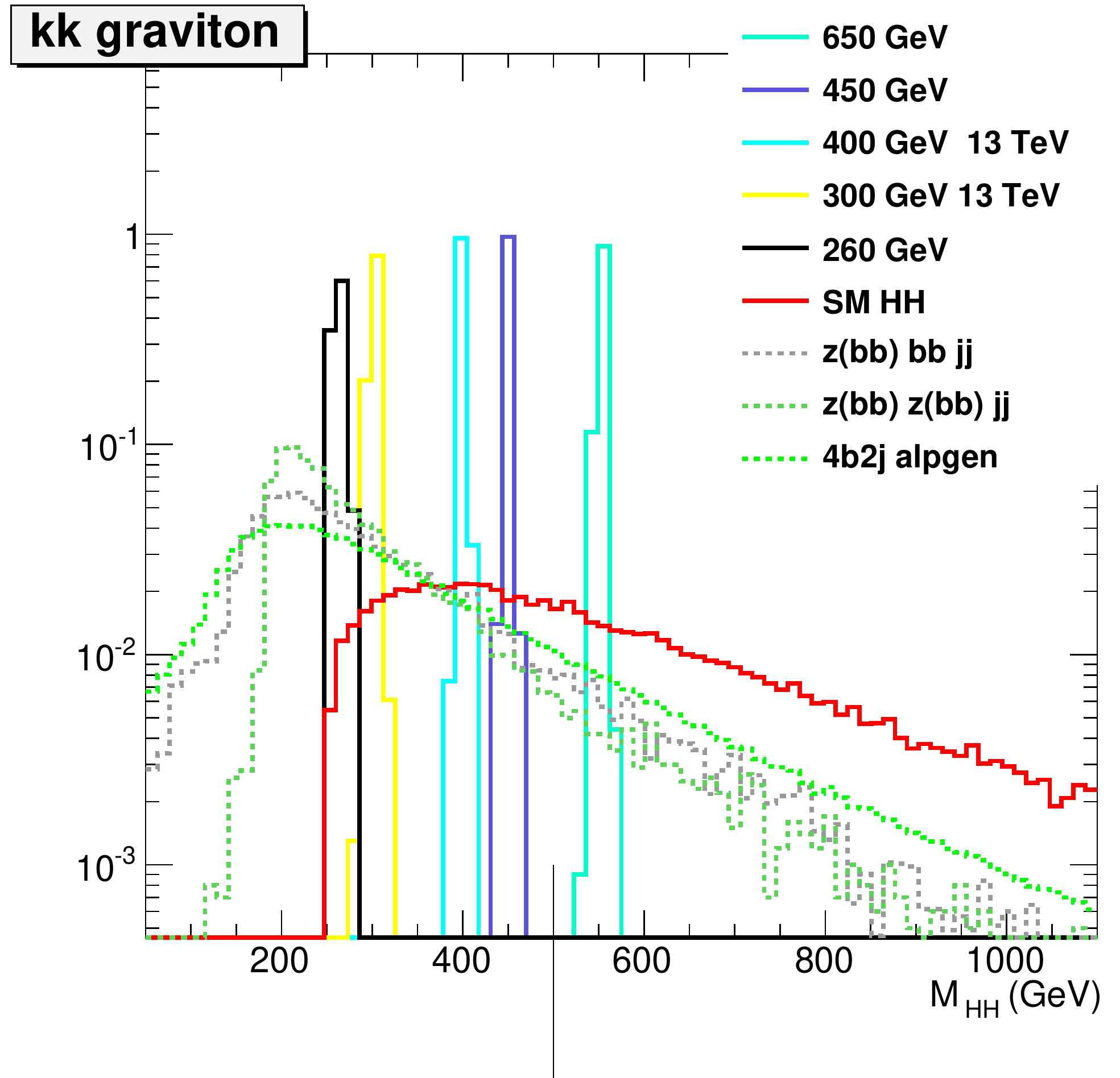}
\caption{Parton level distributions of reconstructed objects with no cuts applied. Distributions are shape normalized. {\bf Left:} Leading reconstructed Higgs. {\bf Right:} Reconstructed bulk 
KK-graviton. Those distributions are only for illustration.
 \label{fig:njets3}}
\end{center}\end{figure*}

To demonstrate the discovery potential of the channel despite our preliminary sensitivity results 
and push the need of a higher simulation level study we also display the parton level distributions 
of the mass of the leading reconstruct Higgs and the reconstructed di-Higgs resonance in Fig.~\ref{fig:njets3}.

\section{CONCLUSIONS}

The VBF production of a Higgs pair is sensitive to new physics that couple to the longitudinally polarized gauge bosons.
Hence it is complementary to Higgs pair production via gluon fusion. 
We performed a first preliminary parton level study on the search for a bulk KK-graviton resonance at a HL-LHC using the di-Higgs channel with a VBF topology.
We prospect the HL-LHC significance of such signal. After basic event quality cuts we do not find a large significance. However, one should note that our results 
are obtained using the signal and background efficiencies without applying cuts on 
resonance resolution (on both Higgs and KK-Graviton). 
We believe that applying such cuts at the parton level study would produce over-optimistic results. In this sense the numbers in Table~\ref{table:nevents} are a
gross underestimate of the significance and a complete analysis based on more a realistic simulation that allows us to make more accurate predictions 
needs to be performed to rightly assess the detectability of a VBF resonant Higgs pair production.

\section*{ACKNOWLEDGEMENTS}
We thank M.~Gouzevich and J.~Rojo for valuable discussions on the QCD backgrounds. 
We acknowledge the use of IRIDIS HPC Facility at the University of Southampton for this study.
AB is supported in part by the NExT Institute and MCT is supported by an STFC studentship grant.
RR is supported in part by grants from CNPq and Fapesp (2011/11973-4).




%% file: vbfhh/higgsDiagramTikzVBFhh.tex
		\begin{tikzpicture}[
				thick,
				level/.style={level distance=1.4cm, line width=0.3mm},
		  		level 2/.style={sibling distance=1.4cm},
		  		level 3/.style={sibling distance=1.4cm}
	]
			\coordinate
			child[grow=right] 
			{
				child [] { 
		      		child[] {
			         child []{
							edge from parent [nothing]
			         }	
			         child [sibling distance=2cm]{
			         		edge from parent [higgs, dashed]
		            		node[below] {$H\;\;\;\;\; $}
			         }
			         child [sibling distance=2cm]{
			         		edge from parent [higgs, dashed]
		            		node[above] {$H\;\;\;\;\; $}
			         }
			         child[grow=left] { 
				         	child[] {
				         		edge from parent [boson, solid]
							}
				         	child[] {	
								child[grow=left, solid] {
									child[grow=right] { 
										child [level distance=3cm]{
			            					edge from parent [electron]	
			              				node[below] {$d$}
										}
										child [grow=north east, level distance=0pt]{
										}
				            			edge from parent [electron]	
				              		node[below] {$u$}
									}
			            		}
			            		edge from parent [boson, solid]	
			            		node[below] {$\;\;\;\;\;  W^+$}
							}
			         }
				 		edge from parent [graviton, solid]		
		         		node[above] {$G$}
		       	}
					edge from parent [nothing]
		         node[above] {$\;\;\;\;\;\; W^-$}
				}
				child[above, level distance=3cm] {
					edge from parent [electron]
					node [above] {$u$}
				}
				edge from parent [electron]
				node [above] {$d$}
			};
		\end{tikzpicture}
\hspace{6 mm}

%% file: vbfhh/higgsDiagramTikzWhh.tex
		\begin{tikzpicture}[
					thick,
			level/.style={level distance=1.4cm, line width=0.4mm},
			level 2/.style={sibling distance=1.4cm},
			level 3/.style={sibling distance=1.4cm},
			level 4/.style={sibling distance=1.4cm}
	]
			\coordinate
			child[grow=-30]{
				child[grow=east] { 
					child[grow=-40] {    		
							child[grow=-30] {
								edge from parent [higgs,dashed]
		              		node[below] {$H$}
							}
							child[grow=30]{
								edge from parent [higgs,dashed]    
		              		node[above] {$H$}
							}						    
						edge from parent [graviton]
						node[below=3pt] {$G$}
					}
					child[grow=40] {
						child[grow=-30] {
							edge from parent [electron]
		              	node[below] {$u$}
						}
						child[grow=30]{
							edge from parent [electron2]    
		              	node[above] {$\bar{d}$}
						}
						edge from parent [boson]
						node [above] {$W^{+}$}
					}    
		  			edge from parent [boson]
		   			node[above=3pt] {$W^{+}$}           
		 		}
				child[grow=210] {
		   			edge from parent [electron]
		   			node[below=3pt] {$\bar{d}$}
		 		}
		 		edge from parent [electron] node [above=3pt] {$u$}
			};
		\end{tikzpicture}

%% file: vbfhh2/HHvbf.tex
 




\def\smallfrac#1#2{\hbox{${{#1}\over {#2}}$}}
\newcommand{\be}{\begin{equation}}
\newcommand{\ee}{\end{equation}}
\newcommand{\bea}{\begin{eqnarray}}
\newcommand{\eea}{\end{eqnarray}}
\newcommand{\bi}{\begin{itemize}}
\newcommand{\ei}{\end{itemize}}
\newcommand{\ben}{\begin{enumerate}}
\newcommand{\een}{\end{enumerate}}
\newcommand{\la}{\left\langle}
\newcommand{\ra}{\right\rangle}
\newcommand{\lc}{\left[}
\newcommand{\rc}{\right]}
\newcommand{\lp}{\left(}
\newcommand{\rp}{\right)}
\newcommand{\as}{\alpha_s}
\newcommand{\aq}{\alpha_s\left( Q^2 \right)}
\newcommand{\amz}{\alpha_s\left( M_Z^2 \right)}
\newcommand{\aqq}{\alpha_s \left( Q^2_0 \right)}
\newcommand{\aqz}{\alpha_s \left( Q^2_0 \right)}
\def\toinf#1{\mathrel{\mathop{\sim}\limits_{\scriptscriptstyle
{#1\rightarrow\infty }}}}
\def\tozero#1{\mathrel{\mathop{\sim}\limits_{\scriptscriptstyle
{#1\rightarrow0 }}}}
\def\toone#1{\mathrel{\mathop{\sim}\limits_{\scriptscriptstyle
{#1\rightarrow1 }}}}
\def\frac#1#2{{{#1}\over {#2}}}
\def\gsim{\mathrel{\rlap{\lower4pt\hbox{\hskip1pt$\sim$}}
    \raise1pt\hbox{$>$}}}         
\def\lsim{\mathrel{\rlap{\lower4pt\hbox{\hskip1pt$\sim$}}
    \raise1pt\hbox{$<$}}}         
\newcommand{\mrexp}{\mathrm{exp}}
\newcommand{\dat}{\mathrm{dat}}
\newcommand{\one}{\mathrm{(1)}}
\newcommand{\two}{\mathrm{(2)}}
\newcommand{\art}{\mathrm{art}} 
\newcommand{\rep}{\mathrm{rep}}
\newcommand{\net}{\mathrm{net}}
\newcommand{\stopp}{\mathrm{stop}}
\newcommand{\sys}{\mathrm{sys}}
\newcommand{\stat}{\mathrm{stat}}
\newcommand{\diag}{\mathrm{diag}}
\newcommand{\pdf}{\mathrm{pdf}}
\newcommand{\tot}{\mathrm{tot}}
\newcommand{\minn}{\mathrm{min}}
\newcommand{\mut}{\mathrm{mut}}
\newcommand{\partt}{\mathrm{part}}
\newcommand{\dof}{\mathrm{dof}}
\newcommand{\NS}{\mathrm{NS}}
\newcommand{\cov}{\mathrm{cov}}
\newcommand{\gen}{\mathrm{gen}}
\newcommand{\cut}{\mathrm{cut}}
\newcommand{\parr}{\mathrm{par}}
\newcommand{\val}{\mathrm{val}}
\newcommand{\tr}{\mathrm{tr}}
\newcommand{\checkk}{\mathrm{check}}
\newcommand{\reff}{\mathrm{ref}}
\newcommand{\extra}{\mathrm{extra}}
\newcommand{\draft}[1]{}
\def\beq{\begin{equation}}  
\def\eeq{\end{equation}}  

\def\bgamma{\boldsymbol{\gamma}}
\def\nn{\nonumber}
\def \so{\sigma_I^{DIS}(x_I,Q^2_I)}
\def \sh{\frac{d\sigma^{hh}}{dX}}
\def\sdy{\frac{d\sigma^{\mathrm{DY}}}{dQ_I^2dY_I}}
\def \npdf{N_{\mathrm{pdf}}}
\def \gtilda{\tilde\Gamma_J^{\mathrm{OBS}}}
\def \n0{N_j^{(0)}}
\def \a{\alpha}
\def \b{\beta}
\def \g{\gamma}
\def \c{\xi}
\def \z{\zeta}
\def\lapprox{\lower .7ex\hbox{$\;\stackrel{\textstyle <}{\sim}\;$}}
\def\gapprox{\lower .7ex\hbox{$\;\stackrel{\textstyle >}{\sim}\;$}}
\def\half{\smallfrac{1}{2}}
\def\GeV{{\rm GeV}}
\def\TeV{{\rm TeV}}
\def\ap{{a'}}
\def\vp{{v'}}
\def\e{\epsilon}
\def\d{{\rm d}}
\def\calN{{\cal N}}
\def\shat{\hat{s}}
\def\barq{\bar{q}}
\def\qq{q \bar q}
\def\uu{u \bar u}
\def\dd{d \bar d}
\def\pp{p \bar p}
\def\xa{x_{1}}
\def\xb{x_{2}}
\def\xaa{x_{1}^{0}}
\def\xbb{x_{2}^{0}}
\def\smx{\stackrel{x\to 0}{\longrightarrow}}
\def\Li{{\rm Li}}

\newcommand{\tmop}[1]{\ensuremath{\operatorname{#1}}}
\newcommand{\tmtextit}[1]{{\itshape{#1}}}
\newcommand{\tmtextrm}[1]{{\rmfamily{#1}}}
\newcommand{\tmtexttt}[1]{{\ttfamily{#1}}}

\chapter{Strong Double Higgs Production at the LHC in the $4b$ and $2b2W$ Final States}

{\it O.~Bondu, A.~Oliveira, R~.Contino, M.~Gouzevitch, A.~Massironi, J.~Rojo}



\begin{abstract}
The measurement of Higgs pair production will be one of the cornerstones
of the LHC physics program in the next years.
Double Higgs production via vector boson fusion, in particular, 
is sensitive to the strong interactions of a composite Higgs boson, and
allows a direct extraction of the $hhVV$ quartic interaction.
In this contribution we study the feasibility of probing strong double Higgs production via vector boson fusion
at the $13\,$TeV LHC in the  $4b$ and $2b2W$ final states.
By performing a simple parton-level analysis, 
we find that, although experimentally challenging, these final states lead to a good sensitivity on New Physics.
Our results are encouraging and motivate  more realistic analyses to 
include parton shower, jet reconstruction and $b$-tagging efficiencies.
\end{abstract}


\section{Introduction}

The measurement of double Higgs production will be one of the main
physics goals of the LHC program in its upcoming high-energy and high-luminosity phase.
Double Higgs production is directly sensitive to the Higgs trilinear coupling, and thus 
provides information on the scalar potential responsible for electroweak symmetry breaking.
It is also sensitive to the underlying strength of the Higgs interactions
at high energies, and can test the composite nature of the Higgs boson~\cite{Giudice:2007fh,Contino:2010mh}.

In the Standard Model (SM), the dominant mechanism for the production of two Higgs bosons at the LHC is 
gluon fusion (see Ref.~\cite{Baglio:2012np} and references therein), analogously to single Higgs production.
For a center-of-mass energy $\sqrt{s} = 14\,$TeV, the recently computed next-to-next to leading order (NNLO)
total cross section is approximately $40\,$fb~\cite{deFlorian:2013jea}.
Feasibility studies in the case of a light Higgs boson have been performed for several different final states, including
$b\bar b\gamma\gamma$~\cite{Baur:2003gp,Barger:2013jfa},
$b\bar{b}\tau^+\tau^-$~\cite{Baur:2003gpa,Barr:2013tda,Dolan:2012rv,Dolan:2013rja},
$b\bar{b}W^+W^-$~\cite{Dolan:2012rv,Papaefstathiou:2012qe} and
$b\bar{b}b\bar{b}$~\cite{Baur:2003gpa,Dolan:2012rv,Gouzevitch:2013qca,Cooper:2013kia}.
Another relevant production mode for Higgs boson pairs at the LHC is vector boson fusion (VBF),
where a soft emission of two vector bosons from the incoming protons is followed by 
the hard $VV \to hh$ scattering ($V=W,Z$). 
In the SM, the leading order (LO) rate of this channel is small, around $1\,$fb at $\sqrt{s} = 14\,$TeV.
The  full NNLO calculation recently performed in Ref.~\cite{Liu-Sheng:2014gxa}
has shown that the total cross section is perturbatively very stable, 
with a NNLO/LO K-factor deviating from 1 by only~$\sim$10\%.
The production rate can be however strongly enhanced in theories where the Higgs
is a composite pseudo Nambu-Goldstone boson (pNGB) of new strong dynamics at the TeV scale~\cite{Kaplan:1983fs}.
In those theories the Higgs anomalous couplings imply a growth of the $VV\to hh$ cross section with the
partonic center-of-mass energy, $\sigma \propto \hat s/f^4$, where $f$ is the pNGB decay constant.
The enhanced sensitivity to the underlying strength of the Higgs interactions thus makes double Higgs production
via VBF  a key process to test the nature of the electroweak symmetry breaking dynamics.

A first detailed study of double Higgs production via VBF at the LHC 
was performed in Ref.~\cite{Contino:2010mh} for a Higgs mass $m_{h}=180\,$GeV by
focussing on the $4W$ final state.
In this work we want to revisit the feasibility of observing this process at the LHC 
for $m_{h}=125\,$GeV.
Although a realistic final state simulation of all Higgs pair production
channels can be now achieved up to next-to-leading order (NLO) matched 
to parton showers in the {\tt MadGraph5\_aMCatNLO} framework~\cite{Frederix:2014hta},
we will keep our analysis as simple as possible and work at the parton level.
We will consider the two final states with the largest branching fractions:
$hh\to 4b$ and $hh \to 2b2W$.
We will compute the relevant backgrounds for these final states and devise 
simple kinematic cuts to isolate the signal at the LHC with $300\,\text{fb}^{-1}$ as well as at its 
future High-Luminosity upgrade to $3\,\text{ab}^{-1}$.
In each case, we will study the sensitivity to New Physics (NP) in
a way as model independent as possible.
As we will show, a High-Luminosity upgrade of the LHC (HL-LHC) is essential
in order to make a detailed study of double Higgs production,  
although already at $300\,\text{fb}^{-1}$ important information can be obtained on the Higgs anomalous couplings
in the most favorable NP scenarios.

The outline of this paper is as follows.
In Sect.~\ref{sec:theomod} we present our 
modeling of signal and background.
Sections~\ref{sec:4b2j} and~\ref{sec:2w2b2j} contain our analysis of Higgs pair production via VBF respectively
for the $4b$  and $2W2b$ final states.
In Sect.~\ref{sec:con} we conclude and discuss future directions
for these studies.

\section{Theoretical modeling of signal and background}
\label{sec:theomod}

In this section we discuss the theoretical modeling of
signal and background events for Higgs pair production
via vector boson fusion at the LHC for the two final states under consideration
\begin{align}
& pp \to hh jj \to 4b \, jj \\[0.2cm]
& pp \to hh jj \to 2W 2b\, jj \to l^+ l^- \!\not\!\! E_T \,2b\, jj \, ,
\end{align}
where $j$, $b$ and $l=e,\mu$ stand respectively for jets (from light quarks and gluons),
$b$-tagged jets and light charged leptons (electrons and muons).
For simplicity, in the following we will refer to the $ l^+ l^- \!\not\!\! E_T \,2b\, jj$ final state as $2W2bjj$,
always assuming the leptonic decay of the $W$ bosons. 
We set $m_h=125\,$GeV and assume a collider center-of-mass energy $\sqrt{s}=13\,$TeV.
All events are generated at the parton level, without any shower or hadronization effect.
No detector simulation or inclusion of underlying event and pileup corrections is attempted: in this first feasibility study we
want to explore the most optimistic scenario. For this reason, we will also assume perfect $b$-jet tagging and omit the related 
identification efficiencies in our estimates.
As our starting point,  we impose the following selection cuts to all signal and background samples
\begin{gather}
\label{eq:acceptancecuts}
\begin{gathered}
p_{Tj} \ge 25~{\rm GeV} \, , \quad p_{Tb} \ge 25~{\rm GeV} \, , \quad
p_{Tl_1} \ge 20~{\rm GeV} \, ,  \quad p_{Tl_2} \ge 10~{\rm GeV}  \\[0.15cm]
|\eta_j|\le 4.5 \, , \quad |\eta_{b}|\le 2.5\, ,  \quad |\eta_{l}|\le 2.5  \\[0.15cm]
\Delta R_{jb} \ge 0.4 \, , \quad\Delta R_{bb} \ge 0.2 \, , \quad
\Delta R_{jl} \ge 0.4 \, , \quad\Delta R_{bl} \ge 0.4 \, , \quad\Delta R_{ll} \ge 0.4 \, , 
\end{gathered}
\\[0.6cm]
\label{eq:vbfcuts}
m_{jj} \ge 500~{\rm GeV} \, , \qquad \Delta R_{jj}\ge 4.0 \, .
\end{gather}
The cuts of  Eq.~(\ref{eq:acceptancecuts}) are simple acceptance requirements, while those of Eq.~(\ref{eq:vbfcuts}) are specifically
designed to isolate the VBF signal. The $hhjj$ process also follows from gluon-gluon fusion at NNLO, but we expect this contribution
to be subdominant compared to the VBF process after the cuts of Eq.~(\ref{eq:vbfcuts}).
In practice, in a realistic analysis one will have to classify events by jet topology and isolate the VBF and gluon-fusion initiated
contributions.
The value of $\Delta R_{jb}$ has been chosen so as to reproduce the jet reconstruction of a cone algorithm with 
minimum cone size $R=0.4$. 
The looser cut on $\Delta R_{bb}$ is motivated by the recent
development of jet substructure techniques that provide the possibility
to discriminate the $h\to b\bar{b}$ decay over the background even when
the two prongs of the Higgs decay end up in the same jet~\cite{Butterworth:2008iy}.

The signal event samples have been generated with  {\tt MadGraph5}~\cite{Alwall:2011uj}, including the decays 
of the Higgs bosons into the relevant final states.
We have used the NNPDF2.1LO PDF set~\cite{Ball:2011uy} with 
dynamically generated renormalization and factorization scales  (default choice in {\tt MadGraph5}).
Anomalous Higgs couplings have been parametrized according to the  effective Lagrangian 
for non-linearly realized $SU(2)_L \times U(1)_Y$. Specifically, we have rescaled the  couplings of the Higgs
to vector bosons and the Higgs trilinear coupling as follows~\cite{Contino:2010mh}
\be
c_{V}\, g^{SM}_{hVV} \, hVV \, , \qquad c_{2V}\, g^{SM}_{hhVV} \, hhVV \, ,
\qquad c_{3}\, g^{SM}_{hhh}\, hhh \label{Eq:BSM} \, ,
\ee
where $g^{SM}_i$ are the corresponding SM couplings, so that the Standard Model limit is recovered by setting $c_V=c_{2V}=c_3=1$.
In the case of composite Higgs theories where the electroweak symmetry is linearly realized  at high energies and the Higgs
boson is part of an $SU(2)_L$ doublet, the coefficients $c_V$, $c_{2V}$ and $c_3$ deviate from 1 at order $(v/f)^2$.
In general, one expects $O(g_*^2 v^2/m_*^2)$ corrections, where  $g_*$ is the Higgs coupling
strength to new states and $m_*$ is their mass.
For simplicity we have neglected the direct contribution of new states, such as vector and scalar resonances, to the scattering amplitude
(see for example Ref.~\cite{Contino:2011np} for an analysis in the case of composite Higgs theories).
Eq.~(\ref{Eq:BSM}) has been implemented by a suitable modification of
the Standard Model {\tt UFO}~\cite{Degrande:2011ua} file in {\tt Madgraph5}.\footnote{We are grateful
to Benjamin Fuks for assistance with this implementation.}
We have chosen to explore the following range of Higgs couplings:
\begin{equation}
\begin{split}
0.5 &\le  c_V \le 1.5  \\
0.0 &\le  c_{2V} \le 2.0  \\
0.0 &\le  c_3 \le 2.0 \, .
\end{split}
\end{equation}
While this is a reasonable choice, useful for the illustrative purposes of this work, one should keep in mind that
current Higgs searches at the LHC  typically set slightly more stringent bounds on $\Delta c_V = 1 -c_V$ at the level of $0.10-0.20$,
depending on the assumptions made (see for example Refs.~\cite{Espinosa:2012im,Ellis:2013lra,Pomarol:2013zra}). 
Tighter limits on $\Delta c_V$ come from electroweak precision tests in absence of additional NP contributions
to the electroweak observables, see for example Ref.~\cite{Ciuchini:2013pca}.
The couplings $c_{2V}$ and $c_3$ have instead no direct experimental determination so far.
Double Higgs production via VBF is the only process at the LHC that can give access to the coupling $c_{2V}$,
while $c_3$ can in principle be extracted also from di-Higgs production via gluon-fusion.

In Table~\ref{tab:xsec} we collect the  cross sections of the signal in the Standard Model
and in various BSM scenarios after the selections cuts of Eqs.~(\ref{eq:acceptancecuts}) and~(\ref{eq:vbfcuts}).
Branching fractions to the two final states under consideration are included assuming that they are not modified compared to their SM value:
${\rm BR}(hh\to b\bar b b\bar b) =0.333$, and 
${\rm BR}(hh\to b\bar b WW \to b\bar b\, l^+ \nu \,l^- \bar \nu) = 1.17 \cdot 10^{-2}$~\cite{Dittmaier:2011ti,Dittmaier:2012vm,Heinemeyer:2013tqa}.
In the SM case, we also report the value of the $pp \to hhjj$ cross section with undecayed Higgs bosons both
before cuts and after applying the cuts of Eqs.~(\ref{eq:acceptancecuts}),~(\ref{eq:vbfcuts})  to the two jets.
%
\begin{table}[tbp]
\centering
\begin{tabular}{l|c|c|c}
\hline
Model & Final state & Cross section [fb] & $N_{\rm ev} \; ({\cal L}=3\,\text{ab}^{-1})$\\
\hline \hline
& & & \\[-0.3cm]
SM (no cut)  &   $hhjj$  &  0.83 & 2500 \\
SM               &   $hhjj$  & 0.12  & 360  \\[0.1cm]
\hline
 & & & \\[-0.3cm]
SM  &   \multirow{7}{*}{$hh jj \to 4b jj$}  & 0.049   & 150 \\
$c_V=0.5$  &     & 0.54   &  1600  \\
$c_V=1.5$  &     & 2.72    &  8100  \\
$c_{2V}=0$  &     & 1.23    &  3700   \\
$c_{2V}=2$  &     &  0.78   &  2300  \\
$c_{3}=0$  &      & 0.14    & 420   \\
$c_{3}=2$  &      &  0.042  & 130   \\[0.1cm]
\hline
& & & \\[-0.3cm]
SM  & \multirow{7}{*}{ $hh jj \to  l^+ l^- \!\not\!\! E_T \,2b\, jj$} &  $8.6\cdot 10^{-4}$   & 2.6   \\
$c_V=0.5$  &     & $2.0\cdot 10^{-3}$   &  6    \\
$c_V=1.5$  &     &  $9.8\cdot 10^{-2}$   & 290  \\
$c_{2V}=0$  &     &  $1.9\cdot 10^{-2}$   &  54  \\
$c_{2V}=2$  &     &  $1.1\cdot 10^{-2}$   &  33  \\
$c_{3}=0$  &      &  $2.4\cdot 10^{-3}$   &  7 \\
$c_{3}=2$  &      &  $7.4\cdot 10^{-4}$   &   2.2 \\[0.1cm]
\hline
\end{tabular}
\caption{\label{tab:xsec}
\small Cross sections for Higgs pair production in the
vector-boson fusion channel at the LHC with $\sqrt{s} = 13\,$TeV.
All numbers have been obtained after applying the basic selection cuts
of Eqs.~(\ref{eq:acceptancecuts}),~(\ref{eq:vbfcuts}) except for the first row, where no cut is applied.
Branching fractions to the two final states under consideration are included assuming that they are not modified compared to their SM value,
except for the first two rows where the cross section for undecayed Higgs bosons is reported.
The last column indicates the corresponding number of  events for an integrated luminosity
$\mathcal{L}=3\,$ab$^{-1}$.
The final cross sections and number of signal and background events expected after all
the analysis cuts 
is summarized in Tables~\ref{tab:4b2j} and~\ref{tab:2b2w2j}.
}
\end{table}
%
One can see that in the latter case (selection cuts applied only to the VBF jets)
the SM signal cross section is  0.12 fb before  Higgs decays,
which corresponds to about 36 Higgs boson pairs produced in
the VBF channel with $\mathcal{L}=$ $300\,\text{fb}^{-1}$ and to about 
360 Higgs pairs with $\mathcal{L}=$ $3\,\text{ab}^{-1}$.
These integrated luminosities correspond to the expectations for the
future LHC runs and the possible High-Luminosity upgrade respectively.~\footnote{Most likely, a High-Luminosity LHC will run at the
higher center-of-mass energy $\sqrt{s} = 14\,$TeV. The impact of going from 13 to $14\,$TeV is however less important
than that of a factor 10 increase in luminosity.}
Our LO {\tt MadGraph5} computation
is in good agreement with the recent calculation of
Ref.~\cite{Liu-Sheng:2014gxa} within scale uncertainties.
One can also see from Table~\ref{tab:xsec} that the event yields
are much larger in the $4bjj$ final state than in the $2b2Wjj$ one,
the latter being suppressed by the leptonic branching fraction of the~$W$ bosons.
In the $4bjj$ final state, we expect 
about 150 events in the SM with $\mathcal{L}=3\,\text{ab}^{-1}$ after basic selection cuts, with
event rates that can be higher by up to a factor $\sim 50$ in the most
optimistic BSM scenarios.

For the $4bjj$ final state, we have considered three types of backgrounds: 
the QCD (non-resonant) $pp \to 4bjj$ production, and the resonant $pp \to Z(bb) bbjj$ and $pp \to Z(bb)Z(bb)jj$ processes 
with respectively one and two $Z$ bosons decaying to $b\bar b$. 
To generate the QCD $4bjj$ events, we have used the {\tt ALPGEN}
parton level event generator~\cite{Mangano:2002ea}, while
$Z(bb) bbjj$ and $Z(bb)Z(bb) jj$ events have been  generated with {\tt MadGraph5}\footnote{These 
samples were also used in the contribution ``Resonant Higgs Pair Production in 
Vector Boson Fusion at the LHC" by A.~Belyaev, O.~Bondu, A.~Massironi, A.~Oliveira,
R.~Rosenfeld and V.~Sanz, appearing in these proceedings.}.
The $Z(bb)Z(bb)jj$  background turns out to be much smaller than the other two
after applying our analysis cuts and will be then neglected in the following.
The  cross sections of $4bjj$ and $Z(bb) bbjj$ after the selection cuts of Eqs.~(\ref{eq:acceptancecuts}),~(\ref{eq:vbfcuts}) 
are reported in Table~\ref{tab:4b2j} in the $\sigma_I$ column.

In the case of the $2W2bjj$ final state, the dominant background comes from the SM process
$pp \to WW b\bar{b}jj$, with the subsequent leptonic decay of the $W$ bosons.
The bulk of its cross section (about $3/4$ of the total) is dominated by the $pp \to t\bar{t}jj$ process,
which is much easier to generate due to the smaller final state multiplicity.
We have simulated both $WW b\bar{b}jj$ and  $t\bar{t}jj$ events with {\tt MadGraph5}.
The corresponding cross sections after the selection cuts of Eqs.~(\ref{eq:acceptancecuts}),~(\ref{eq:vbfcuts}) are reported in Table~\ref{tab:2b2w2j} in the 
$\sigma_I$ column.

As a final remark, we stress that all the above backgrounds have been generated at the parton level.
Once shower and detector effects are included, additional backgrounds will have to be included
which can be potentially important, like for example  $p p \to 4j 2b$ where two light jets are mistagged as $b$-jets. 
Their estimate is beyond the scope of this work and is left for a future analysis.

\section{Double Higgs production via VBF in the $4bjj$ final state}
\label{sec:4b2j}

In this section we discuss our analysis of  Higgs pair production via VBF in the $4bjj$ final state.

As shown in Table~\ref{tab:4b2j}, after the basic selection cuts of Eqs.~(\ref{eq:acceptancecuts}),~(\ref{eq:vbfcuts})
the QCD $4bjj$ process is by far the largest background. 
As a first way to reduce it, we can exploit the peculiar kinematics of the signal, where
two pairs of $b$-jets come from the decay of the Higgs bosons.
We construct the two Higgs candidates by identifying the two $b$-jet pairs $(bb)_1$ and $(bb)_2$ 
that minimize the dijet mass difference~\cite{Gouzevitch:2013qca}.
Their invariant masses are then required to lie in a window within 15\% of the nominal Higgs mass ($m_h=125$ GeV in our study)
\be
\label{eq:higgsmass}
|m_{(bb)_i}-m_h| \le 0.15\, m_h\, , \qquad (i=1,2)\, ,
\ee
which roughly corresponds to the hadronic jet mass resolution in ATLAS and CMS. 
The cross sections of signal and  background after this cut are reported
in Table~\ref{tab:4b2j} in the $\sigma_{II}$ column.
%
\begin{table}[tbp]
\begin{center}
\begin{tabular}{l||c|c|c|c||c}
\hline
Sample & $\sigma_I$  [fb] & $\sigma_{II}$ [fb] & $\sigma_{III}$ [fb] & $\sigma_{IV}$ [fb] & $N_{\rm ev} = \sigma_{IV} \cdot {\cal L}$ \\
\hline
\hline
& & & & & \\[-0.35cm]
SM & 0.04895 & 0.04894 & 0.04254 & 0.002263 & 6.788 \\
$c_{V}$ = 0.5 & 0.5424 & 0.5424 & 0.4869 & 0.1958 & 587.5 \\
$c_{V}$ = 1.5 & 2.717 & 2.717 & 2.393 & 0.6798 & 2039 \\
$c_{2V}$ = 0 & 1.237 & 1.237 & 1.101 & 0.3865 & 1160 \\
$c_{2V}$ = 2 & 0.7863 & 0.7861 & 0.713 & 0.3275 & 982.5 \\
$c_{3}$ = 0 & 0.14 & 0.14 & 0.1186 & 0.003405 & 10.21 \\
$c_{3}$ = 2 & 0.04156 & 0.04155 & 0.03687 & 0.001795 & 5.385 \\[0.1cm]
\hline
\hline
& & & & & \\[-0.35cm]
$4bjj$ & 7138 & 133.8 & 59.81 & 0.1185 & 355.5 \\
$Zbbjj \to 4bjj$ & 126.8 & 1.918 & 1.039 & $< 0.0024$ & $< 7$ \\[0.1cm]
\hline
\end{tabular}
\caption{\small 
\label{tab:4b2j}
Cut-flow for the analysis in the $4b2j$ final state.
Cross sections are reported for the signal and backgrounds after sequentially imposing:
the selection cuts of Eqs.~(\ref{eq:acceptancecuts}),~(\ref{eq:vbfcuts}) ($\sigma_I$);
the Higgs reconstruction cut of Eq.~(\ref{eq:higgsmass})~($\sigma_{II}$); 
the VBF cut of Eq.~(\ref{eq:strongervbf})~($\sigma_{III}$); the cut of 
Eq.~(\ref{eq:m4bcut}) ($\sigma_{IV}$).
At each stage all the previous cuts are also imposed.
The values of the signal cross sections $\sigma_I$ are the same as  those reported in Table~\ref{tab:xsec}.
The last column shows the expected number of events after all cuts, $N_{ev} = \sigma_{IV} \cdot {\cal L}$, 
for an integrated luminosity ${\cal L}= 3\,\text{ab}^{-1}$.
Notice that $b$-tagging efficiencies have not been included.
}
\end{center}
\end{table}
%

\begin{figure}[tp]
\centering
\includegraphics[width=0.496\textwidth]{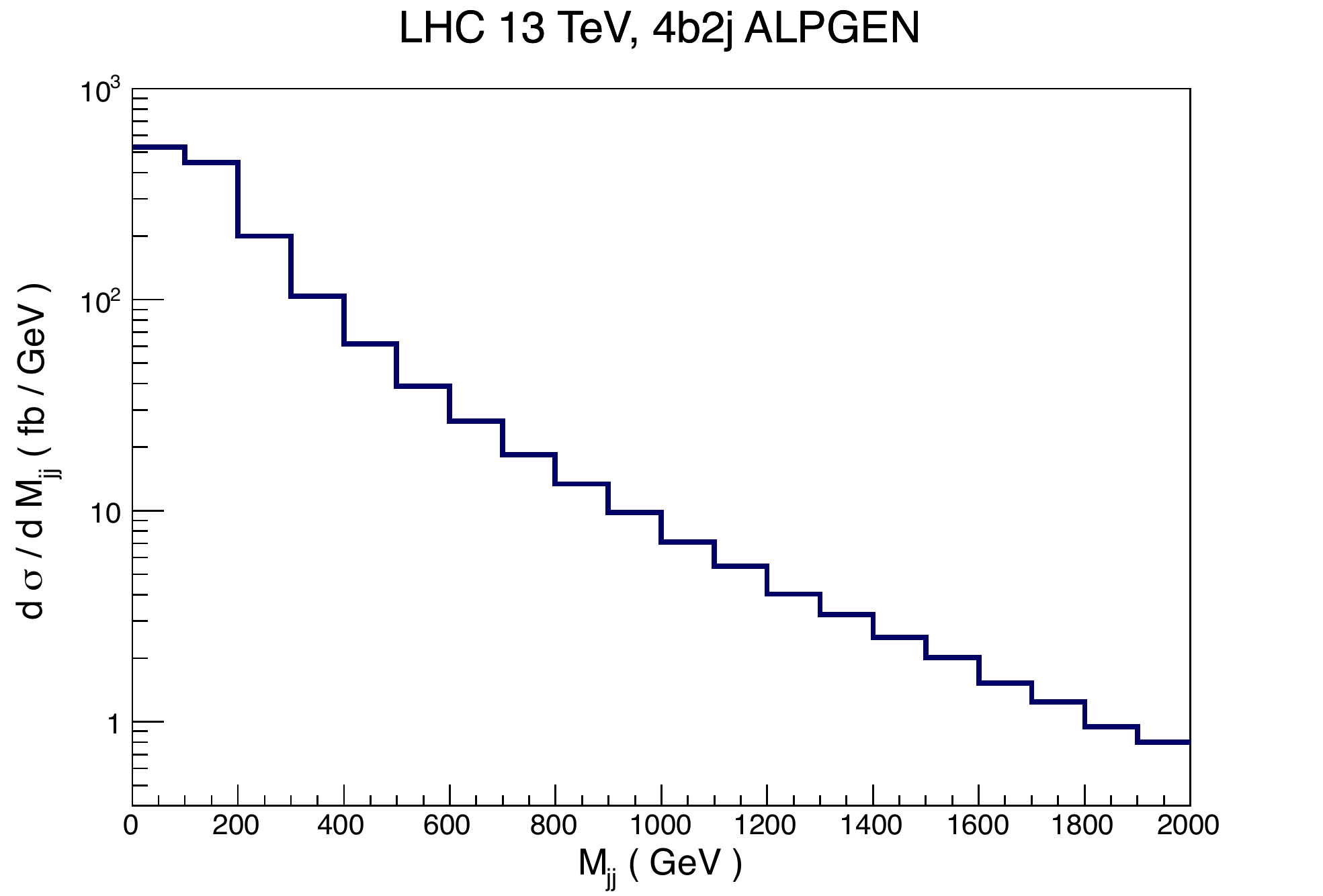}
\includegraphics[width=0.49\textwidth]{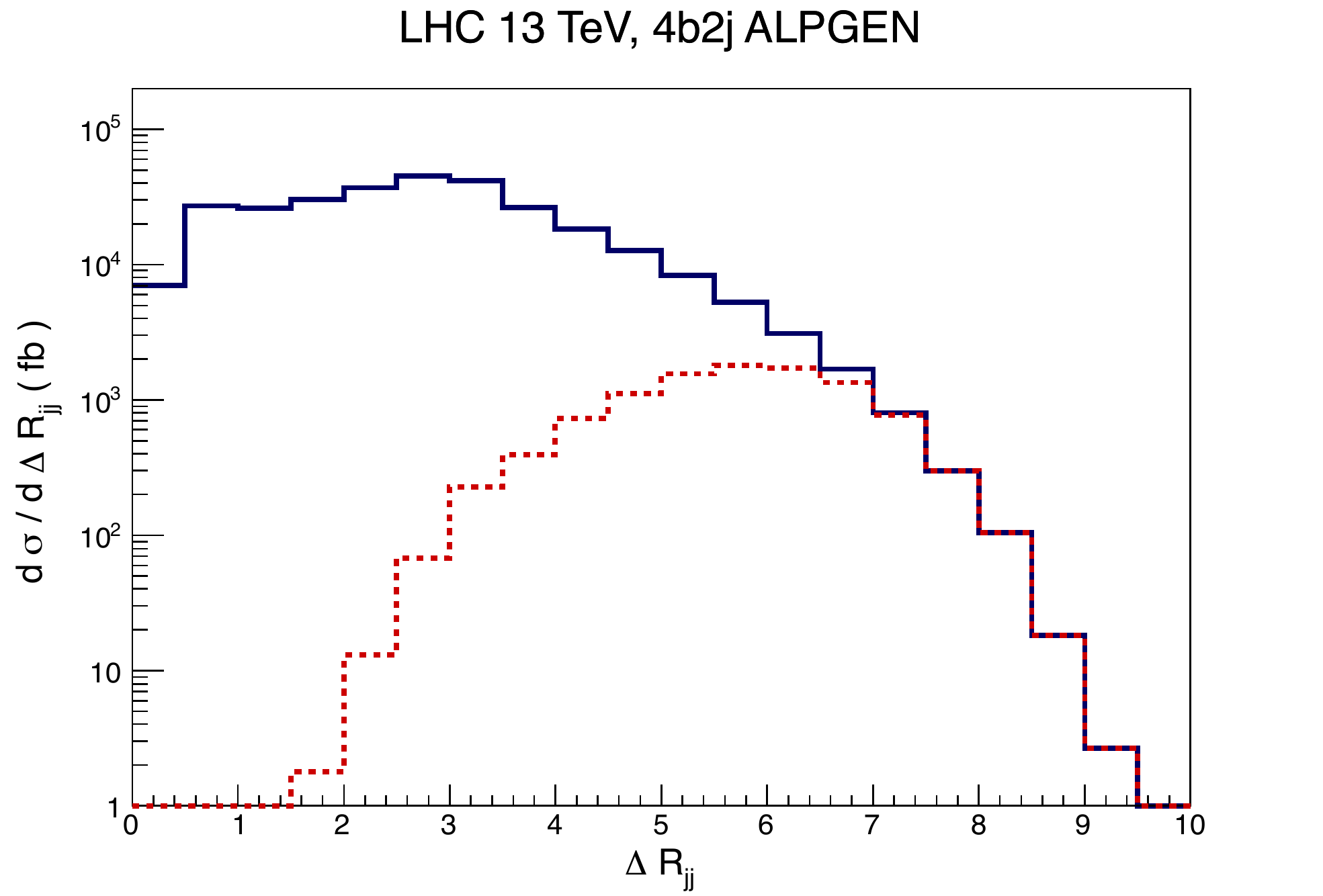}
 \caption{\small \label{fig:mjjdRjj} 
Distribution of the invariant mass $m_{jj}$ (left plot) and separation $\Delta R_{jj}$ (right plot) of the two light jets in the QCD background $4bjj$.
The blue solid curves are obtained after the selection cuts of Eqs.~(\ref{eq:acceptancecuts}),~(\ref{eq:vbfcuts}); the dashed red curve in the right plot
is obtained after imposing the additional VBF cut of Eq.~(\ref{eq:strongervbf}).
The $4bjj$ events have been simulated with the \texttt{ALPGEN} parton-level generator.
 }
\end{figure}

A further reduction of the background can be obtained by imposing a harder cut on the invariant mass of the VBF jets
\begin{equation}
\label{eq:strongervbf}
m_{jj} \ge 800\,\text{GeV}\, ,
\end{equation}
which has a much steeper distribution for the $4bjj$ background than for the signal, see the left plot of Fig.~\ref{fig:mjjdRjj}.
%
%
Requiring an additional stronger cut on the $\Delta R$ separation of the two VBF jets does not help, since its distribution is highly correlated
with that of $m_{jj}$, as illustrated in the right plot of Fig.~\ref{fig:mjjdRjj}.~\footnote{We thank Mauro Moretti for an important discussion on this point.}
The cross sections of signal and  background after the additional VBF cut of Eq.~(\ref{eq:strongervbf}) 
are summarized  in Table~\ref{tab:4b2j} in the $\sigma_{III}$ column.

\begin{figure}[tbp]
\centering
\includegraphics[width=0.6\textwidth]{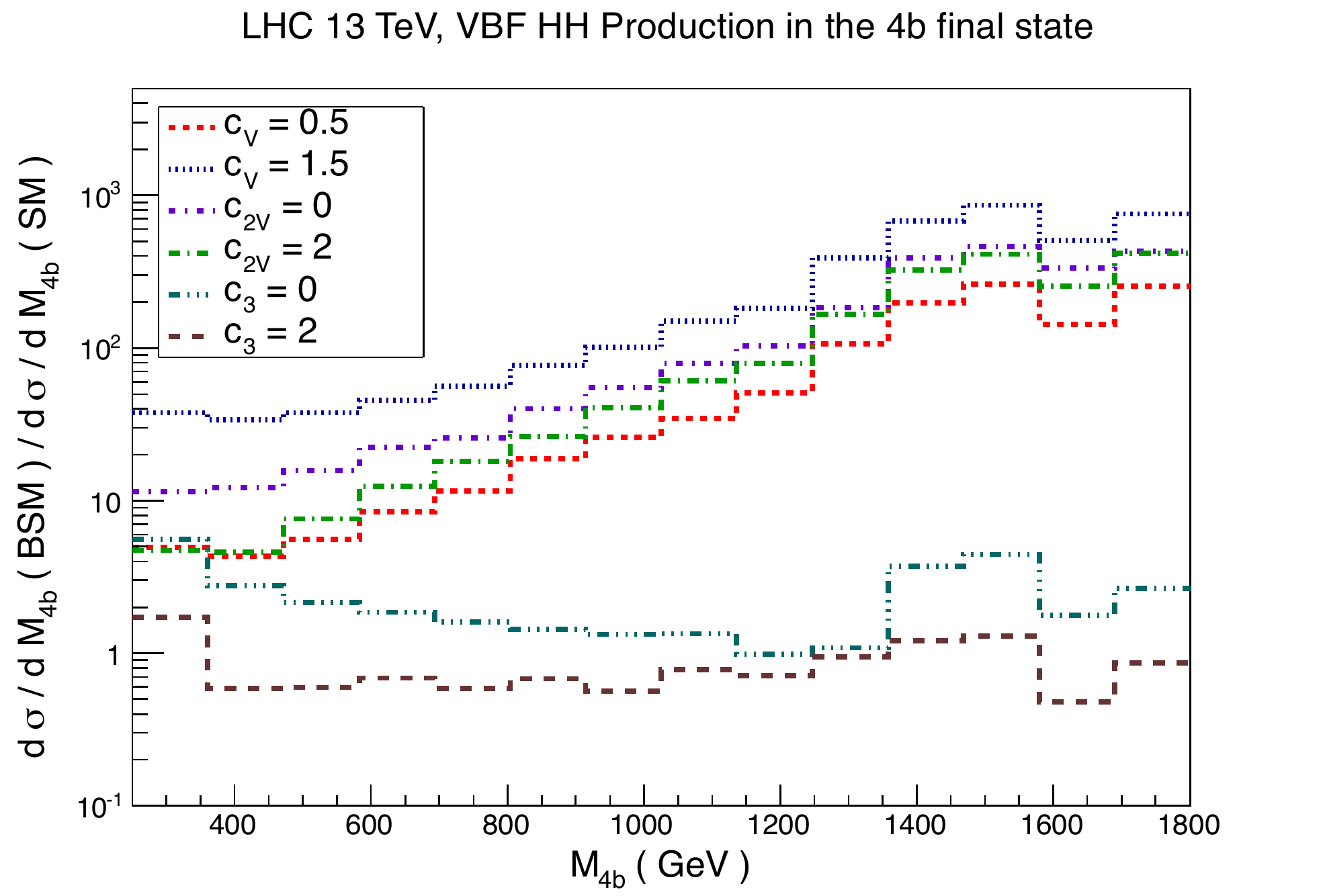}
 \caption{\small \label{fig:mHHhisto_rat} 
Invariant mass distribution of the $4b$ system 
after the basic selection cuts of Eqs.~(\ref{eq:acceptancecuts}),~(\ref{eq:vbfcuts})
for the signal in various  BSM scenarios. 
All curves are normalized to the SM signal.
 }
\end{figure}

\begin{figure}[tbp]
\centering
 \includegraphics[width=0.60\textwidth]{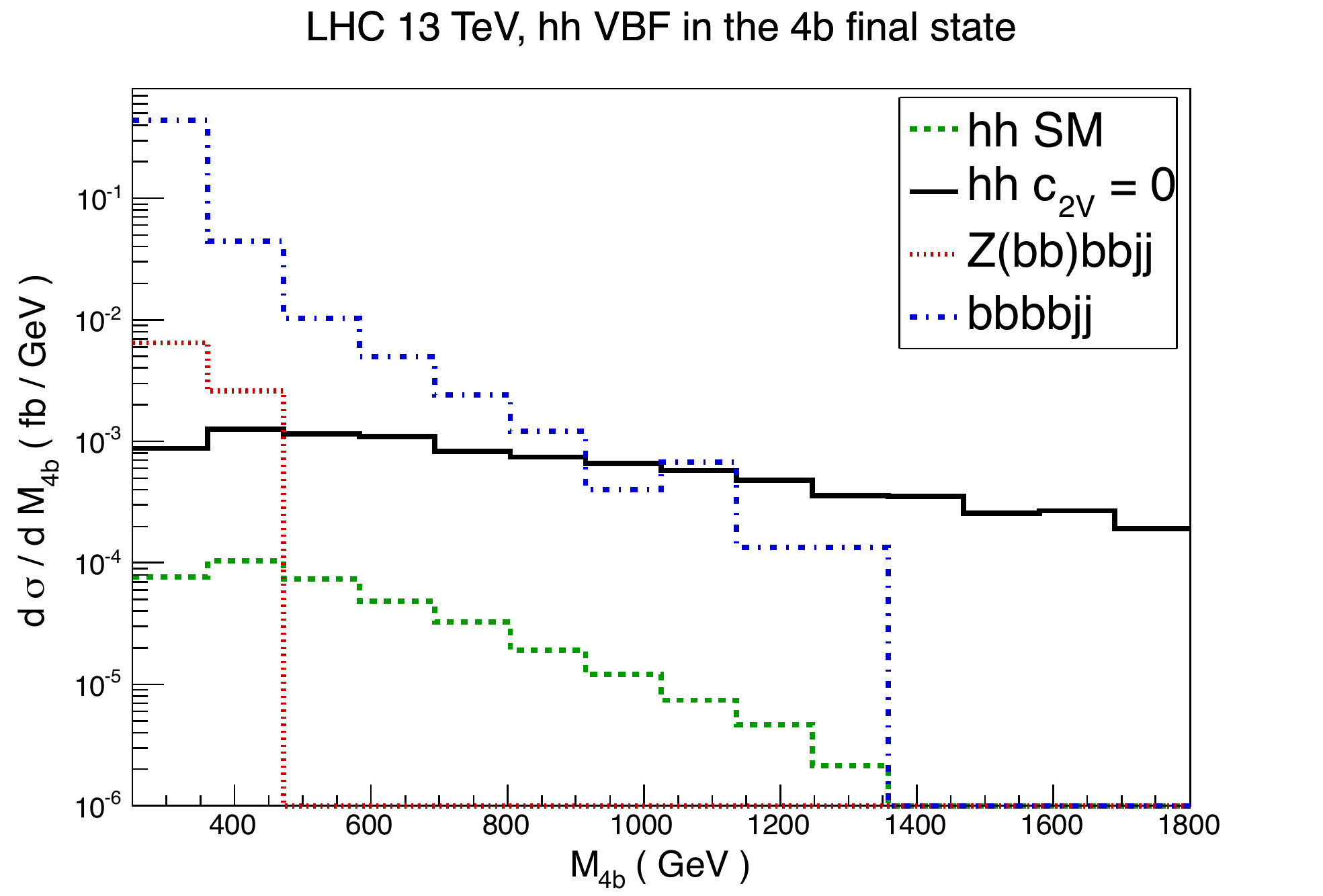}
 \caption{\small \label{fig:mHHhisto} 
Invariant mass distribution of the $4b$ system 
after the basic selection cuts of Eq.~(\ref{eq:acceptancecuts}),~(\ref{eq:vbfcuts}) and the 
additional cuts of Eqs.~(\ref{eq:higgsmass}),~(\ref{eq:strongervbf}),
for the signal and the background.
The abrupt fall-off of the dotted red  and dot-dashed blue curves around their end points
is due to the limited statistics of our background event samples. 
 }
\end{figure}

Although Eqs.~(\ref{eq:higgsmass}) and (\ref{eq:strongervbf}) are very efficient in suppressing the $4bjj$ background 
without reducing the signal, at this stage the signal-over-background ratio ($S/B$) is still quite small. 
This indicates that an analysis of  Higgs pair production via VBF in the  decay channel $hh\ to 4b$ is rather challenging,
in agreement with the estimate of Ref.~\cite{Contino:2010mh}.
There is however one more feature of the signal that can be exploited to isolate it from the background.
In the case of anomalous Higgs couplings to the vector bosons, more specifically for $(c_V^2-c_{2V}) \not = 0$, 
the cross section of the underlying $VV\to hh$ scattering grows  with the partonic energy. 
This is indeed the distinctive prediction of theories with strong EWSB dynamics~\cite{Giudice:2007fh}.
After convoluting over the luminosities of the two initial vector bosons, such energy growth leads to an invariant mass
distribution of the di-Higgs system which is much flatter than for the SM case.
This is clearly shown in Fig.~\ref{fig:mHHhisto_rat}, where  the distribution of the invariant
mass of the four $b$-jet system, $m_{4b}$, is plotted for various BSM scenarios by normalizing it to the SM case.
%
%
As expected, in the case of modified Higgs trilinear coupling,  the $m_{4b}$ distribution is affected mostly at threshold,
and does not significantly differ from the SM case away from it.
In the case of the background, on the other hand, the cross section steeply falls at large $m_{4b}$, see Fig.~\ref{fig:mHHhisto}.~\footnote{Although
we have generated more than 10 million $4b jj$ events with \texttt{ALPGEN}, at very large $m_{4b}$ invariant masses we are limited by statistics, 
and the corresponding distribution has a steep fall-off near its end point. This however occurs in a region where the BSM signal is much larger than the 
QCD background, so our estimates are not affected.
}
%
%
%
It is then clear that the invariant mass $m_{4b}$ can be used as a way to enhance the sensitivity on the signal
in the case of  anomalous $c_V$ and $c_{2V}$ couplings.
Although a realistic analysis will make use of the whole differential distribution to maximize the signal significance and
extract the Higgs couplings, here we apply a simple cut on $m_{4b}$  as a way to quantify 
the fraction of  events with large di-Higgs invariant mass 
and estimate the significance of the signal.
We require:
\be
\label{eq:m4bcut}
m_{4b}\ge 1000~{\rm GeV} \ .
\ee
The signal and background cross sections obtained after this cut are reported in Table~\ref{tab:4b2j} in the
$\sigma_{IV}$ column. The corresponding number of events expected for an integrated luminosity ${\cal L}=3\,\text{ab}^{-1}$
is shown in the last column of the same Table.

Our results indicate that a simple cut-and-count analysis of the $4bjj$ final state is inadequate to isolate the SM signal from the overwhelming QCD background.
The possibilities of extracting the Higgs trilinear coupling are also quite limited in this channel, even in a High-Luminosity phase of the LHC.
On the other hand, the  invariant mass distribution of the di-Higgs system can be used to uncover the signal in the case of strong double Higgs production,
implying in particular that a potentially interesting reach on the coupling $c_{2V}$ can be obtained with sufficiently large integrated luminosity.

\section{Double Higgs production via VBF in the $2b2Wjj$ final state}
\label{sec:2w2b2j}

Next we discuss our analysis for the $2b2Wjj$ final state.

\begin{figure}[tbp]
\centering
\includegraphics[width=0.60\textwidth]{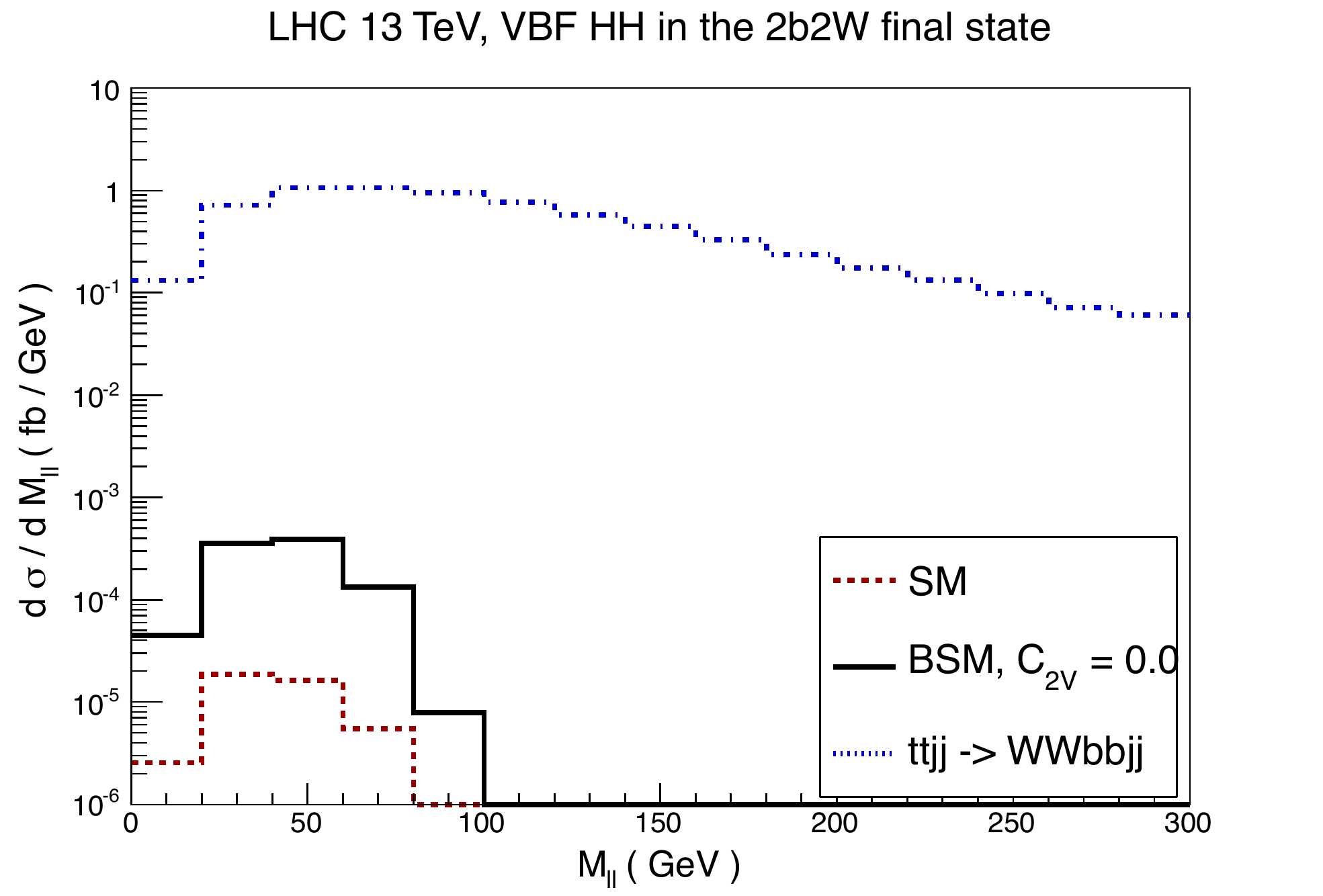}
 \caption{\small \label{fig:mll} 
Invariant mass distribution of the $ll$ system for the signal and the background
after the selection cuts of Eqs.~(\ref{eq:acceptancecuts}),~(\ref{eq:vbfcuts}), and the cut of Eq.~(\ref{eq:higgsrecobb}).
 }
\end{figure}

As for the study of the $4bjj$ channel, it is useful to impose additional cuts, besides those of Eqs.~(\ref{eq:acceptancecuts}),~(\ref{eq:vbfcuts}), that exploit the
kinematic of the signal, where two Higgs bosons are produced on-shell and then decay.
We require the following set of Higgs-reconstruction cuts:
\begin{align}
\label{eq:higgsrecobb}
|m_{bb}-m_h| & \le 0.15 \, m_h \\[0.5cm]
\label{eq:higgsrecoWWll}
\begin{split}
m_{ll} & \le 70\,{\rm GeV} \\[0.2cm]
m_T(WW) & \le 125\,\text{GeV}\, .
\end{split}
\end{align}
The transverse mass $m_T(WW)$ is defined as
\begin{equation}
m_T(WW)  \equiv  \left( \left(\sqrt{ m_{ll}^2 + |\vec p_{Tll}|^2 } + \sqrt{ m_{ll}^2 + |\vec p_{Tmiss}|^2 }\right)^2 - | \vec p_{T ll} +  \vec p_{Tmiss} |^2 \right)^{1/2}\, ,
\end{equation}
where  $\vec p_{Tll}$ and $m_{ll}$ are respectively the transverse momentum and the invariant mass
of the $ll$ system, and $\vec p_{Tmiss}$ is the missing transverse momentum.
The cross sections of signal and background after imposing Eqs.~(\ref{eq:higgsrecobb}) and~(\ref{eq:higgsrecoWWll}) 
are shown in Table~\ref{tab:2b2w2j} in the column $\sigma_{II}$.
%
\begin{table}[tp]
\begin{center}
\begin{tabular}{l||c|c|c||c}
\hline
Sample & $\sigma_{I}$ [fb] & $\sigma_{II}$ [fb] & $\sigma_{III}$ [fb] &  $N_{\rm ev} = \sigma_{III} \cdot {\cal L}$  \\
\hline
\hline
& & & & \\[-0.35cm]
SM & 0.0008607 & 0.0005184 & $4\cdot 10^{-5}$ & 0.1 \\
$c_{V}$ = 0.5 & 0.0020 & 0.0012 & $4.5\cdot 10^{-4}$ & 1.4 \\
$c_{V}$ = 1.5 & 0.098 & 0.059 & 0.01508 & 45.2 \\
$c_{2V}$ = 0.0 & 0.019 & 0.011 & 0.003421 & 10.3 \\
$c_{2V}$ = 2.0 & 0.011 & 0.0063 & 0.002688 & 8.1 \\
$c_{3}$ = 0.0 & 0.0024 & 0.0014 & $6.0\cdot 10^{-5}$ & 0.2 \\
$c_{3}$ = 2.0 & 0.00074 & 0.00044 & $2.4\cdot 10^{-5}$ & 0.1 \\
\hline
\hline
& & & & \\[-0.35cm]
$WWbbjj$ & 788.1 & 14.86 & --  & --  \\
$t\bar{t}jj \to  WWbbjj$ & 779.7 & 14.52 & $< 0.0020$  & $< 6.2$ \\[0.1cm]

\hline
\end{tabular}
\caption{\small \label{tab:2b2w2j}
Cut-flow for the analysis in the $2b2Wjj$ final state.
Cross sections are reported for the signal and backgrounds after sequentially imposing:
the selection cuts of Eqs.~(\ref{eq:acceptancecuts}),~(\ref{eq:vbfcuts}) ($\sigma_I$);
the Higgs reconstruction cuts of Eqs.~(\ref{eq:higgsrecobb}),~(\ref{eq:higgsrecoWWll})~($\sigma_{II}$); 
the cuts of  Eq.~(\ref{eq:m2b2lcut}) ($\sigma_{III}$).
At each stage all the previous cuts are also imposed.
The values of the signal cross sections $\sigma_I$ are the same as  those reported in Table~\ref{tab:xsec}.
The $t\bar tjj$ background appearing in the last line is included as a resonant subprocess of the $WWbbjj$ background.
The last column shows the expected number of events after all cuts, $N_{ev} = \sigma_{III} \cdot {\cal L}$, 
for an integrated luminosity ${\cal L}= 3\,\text{ab}^{-1}$.
Notice that $b$-tagging efficiencies have not been included.
}
\end{center}
\end{table}
%
The cuts on $m_{ll}$ and on the transverse mass $m_{T}(WW)$ exploit the fact that in the signal the lepton pair and the missing
energy come from the decay of the second Higgs (see for example Refs.~\cite{Contino:2010mh,Chatrchyan:2013iaa,ATLAS-CONF-2013-030}). 
The invariant mass $m_{ll}$ tends to be small due to the spin correlation of the two leptons
implied by the decay of a spin-0 particle, see Fig.~\ref{fig:mll}. The $m_{T}(WW)$ distribution has instead a sharp kinematic edge at $m_h$.
%
%
A further refinement of the Higgs reconstruction could be obtained by imposing a lower cut on the transverse momentum of the
$ll$ system, which might be useful to eliminate additional soft backgrounds not considered in this analysis. A lower cut on
the transverse missing energy would also help.
Here we try to  keep our strategy as simple  as possible,  and do not attempt any further optimization of the Higgs reconstruction.

\begin{figure}[tbp]
\centering
\includegraphics[width=0.60\textwidth]{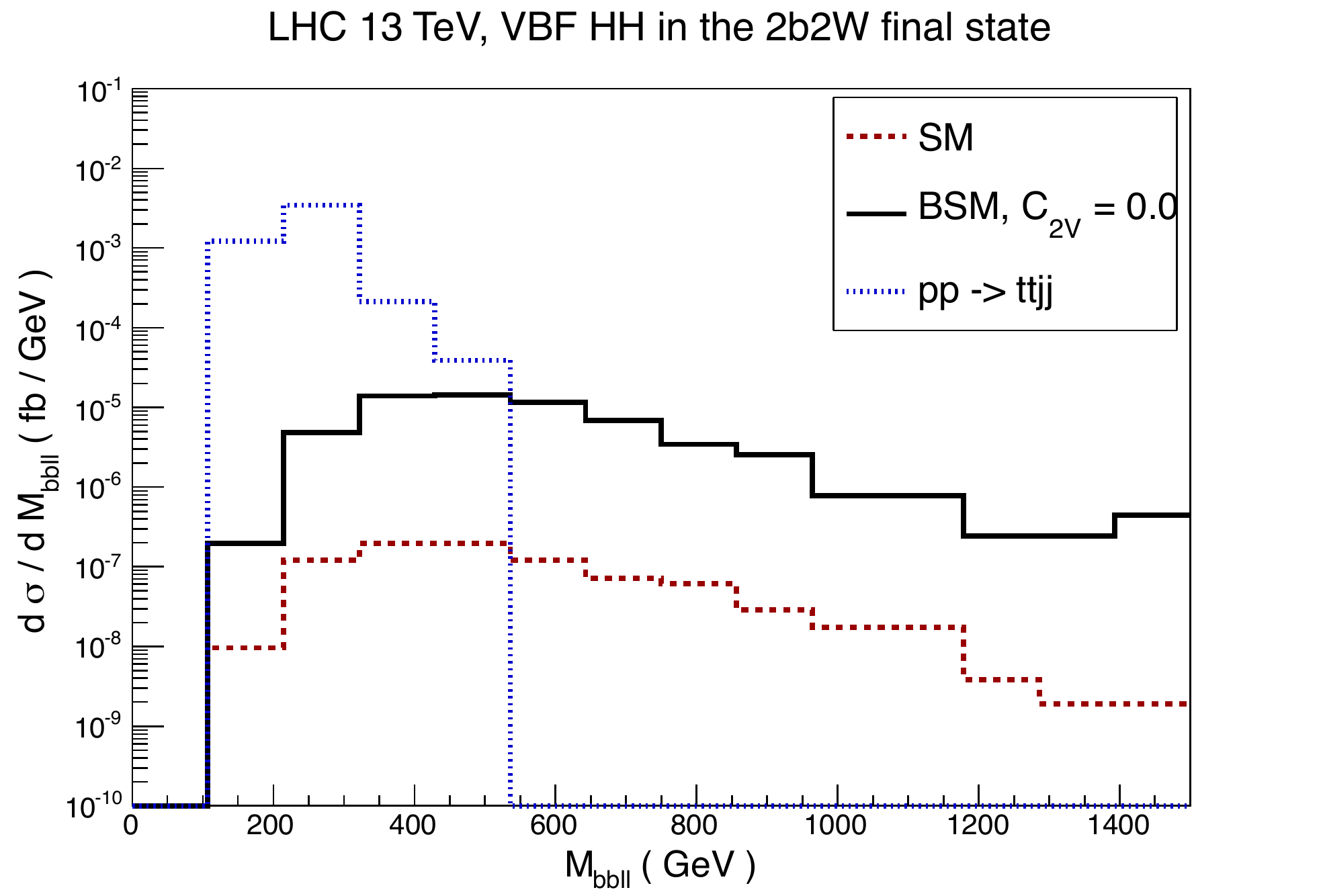}
 \caption{\small \label{fig:mllbb} 
Invariant mass distribution of the $bbll$ system for the signal and the background
after the cuts of Eqs.~(\ref{eq:acceptancecuts}),~(\ref{eq:vbfcuts}),~(\ref{eq:higgsrecobb}),~(\ref{eq:higgsrecoWWll}) and~(\ref{eq:pTbbcut}),
The abrupt fall-off of the dotted blue curve around its end point
is due to the limited statistics of our $t\bar tjj$ event sample. 
 }
\end{figure}

At this stage the background is still overwhelming the signal, and its largest component is $t\bar tjj$.
A large reduction can be however obtained by selecting events where the $llbb$ system is very energetic.
In particular, we impose:
\begin{align}
\label{eq:m2b2lcut}
m_{2b2l} & \ge 500~{\rm GeV}  \\[0.2cm]
\label{eq:pTbbcut}
p_{Tbb}  & \ge 200\,\text{GeV}\, .
\end{align}
The cut on the invariant mass of the $bbll$ system, $m_{2b2l}$, is quite efficient to suppress the background at the cost of a moderate
reduction of the signal, see Fig.~\ref{fig:mllbb}. 
%
%
The cut on the transverse momentum of the $bb$ pair, $p_{Tbb}$, is strongly correlated with $m_{2b2l}$ in the case of the signal and
helps reducing the $t\bar tjj$ background.

After imposing Eqs.~(\ref{eq:m2b2lcut}),~(\ref{eq:pTbbcut}) the cross sections are those reported in the column $\sigma_{III}$ of Table~\ref{tab:2b2w2j}.
Although the number of generated events is too small in the case of the $WWbbjj$ background
to allow us to get an precise calculation of its cross section after all cuts, a good estimate comes from the $t\bar t jj \to WWbbjj$ subprocess.
It is reasonable to expect that the latter will account for the bulk of the $WWbbjj$ rate even after imposing Eqs.~(\ref{eq:m2b2lcut}),~(\ref{eq:pTbbcut}).
Under this assumption, Table~\ref{tab:2b2w2j}  shows that despite the small number of events,
in the case of strong double Higgs production the $2b2W$ channel offers a good sensitivity on $c_V$ and $c_{2V}$.
Isolating the SM signal or  extracting the Higgs trilinear coupling, on the other hand, seems extremely challenging, even at a High-Luminosity 
phase of the LHC.

\section{Conclusions and outlook}
\label{sec:con}

We have explored the feasibility of measuring Higgs
pair production in the vector boson fusion channel at the LHC.
While Higgs pair production via gluon fusion is key to pin
down the Higgs self-coupling,~$c_3$, the VBF channel
is especially sensitive to anomalous Higgs couplings to vector bosons, $c_V$ and $c_{2V}$.
Modifications of these couplings from their SM values imply a growth with the energy of the $VV\to hh$ partonic cross section,
which in turn leads to potentially large enhancements of the production
rate at the LHC.
A previous study assumed a Higgs mass of $m_h=180$ GeV and
focused on the $hh\to 4W$ decay mode, concluding that the vector
boson fusion channel is extremely challenging due to the large
background and the small signal rates, even in the most optimistic BSM scenarios~\cite{Contino:2010mh}.

In this work we have revisited this process for a $m_h=125$ GeV
Higgs boson and concentrated on the $hh\to 4b$ and $hh\to 2b2W$ final states, which are those with the largest branching fractions.
We  performed a  parton level analysis and followed a simple cut-and-count strategy.
Our main results are summarized in Tables~\ref{tab:4b2j} and~\ref{tab:2b2w2j}.
We find that isolating the SM signal over the background is extremely challenging, if not impossible, in both channels.
In the case of anomalous $c_V$ and $c_{2V}$ couplings, on the other hand, the signal can be more easily uncovered
due to the larger rate and to a much harder spectrum of di-Higgs invariant mass that follows from the 
energy growth of the underlying partonic scattering.
Our analysis suggests that a good sensitivity on $c_{2V}$ can be obtained at the High-Luminosity LHC with $3\,\text{ab}^{-1}$.
 The sensitivity on $c_3$, on the other hand, is extremely limited, and a measurement of this couplings seems rather challenging
in the $4b$ and $2W2b$ decay modes even with high luminosity.
In general, the $4b$ channel has a larger signal rate but it is plagued by a sizable QCD background.
The $2b2W$ final state is cleaner and can lead to a large $S/B$ in the
most optimistic BSM scenarios,  but it is characterized by smaller rates due
to the leptonic branching ratios of the $W$ bosons. 

While our analysis was performed at the parton level and was intended to give a first assessment on the feasibility
of observing di-Higgs production via VBF, it is clear that a more detailed estimate will require a more
sophisticated simulation with hadronization and jet reconstruction.
A  realistic $b$-tagging algorithm is also required, since the finite light-jet mistag rates could substantially
enhance the QCD backgrounds.
In any case, jet substructure techniques and
methods such as scale-invariant tagging that allow one to merge the
boosted and resolved regimes, have the potential to help
the $S/B$ discrimination even in this more realistic scenario.\footnote{In addition,
a suitable optimization of the jet radius $R$ can be used to improve
the performance of the kinematic reconstructions of the Higgs mass in the
$b\bar{b}$ final state minimizing the contamination from QCD radiation,
underlying event and pile-up~\cite{Cacciari:2008gd}.}
We plan to present this work in a future publication, where we will
determine the exact ranges of the $c_V$ and $c_{2V}$ parameters than can be explored at the HL-LHC.

In any case, is clear that even at the HL-LHC, the measurement of Higgs
pair production in the vector boson fusion channel is very challenging.
In this sense, a future circular collider at 100 TeV would substantially
increase the production rates and allow one to access a wider variety
of final states like $2b2\gamma$ or $2b2\tau$, with complementary
BSM sensitivity.
As an additional goal of our future publication, 
we plan to study this process at
$\sqrt{s}=100$ TeV, and assess what is the reach in terms
of BSM physics.

Let us finally mention that in this work we have only considered the
non-resonant production of Higgs boson pairs in the VFB channel.
Resonant production via the exchange of a new massive  states could also
increase substantially the event rates, and lead to a similar boosted final state
topologies as the ones we find in the effective field theory approach.

\section*{Acknowledgments}
We are grateful to 
Olivier Mattelaer for assistance with the {\tt Madgraph5} simulations and to Mauro Moretti and Fulvio Piccinini for 
assistance with the {\tt ALPGEN} simulations.
The work of R.C. was partly supported by the ERC Advanced Grant No. 267985 
\textit{Electroweak Symmetry Breaking, Flavour and Dark Matter: One Solution for Three Mysteries (DaMeSyFla)}.




%% file: tgc/tgclep.tex


\newcommand{\tmpkm}[1]{{\bf \boldmath \textcolor{red}{[KM: #1]}}}



\chapter{Triple Gauge Couplings at LEP revisited}

{\it A.~Falkowski, S.~Fichet, K.~Mohan,  F.~Riva, V.~Sanz}


\begin{abstract}
We study the constraints on anomalous triple couplings of electroweak gauge bosons imposed by WW cross section measurements at LEP.
We give the bounds on all eleven CP-even and CP-odd anomalous couplings when only one coupling is switched on at a time. 
Then we move to the effective theory approach with the Standard Model supplemented by dimension-6 operators. 
For the three coefficients of CP-even dimension-6 operators  that modify the triple gauge couplings we provide a simultaneous  fit including correlations.  

\end{abstract}

\section{Introduction}

Triple gauge couplings (TGCs) of electroweak (EW) gauge bosons can be probed in collisions where a pair of gauge bosons is produced. 
This program started in the second phase of the LEP experiment (once the center of mass energy was sufficient for  $W^+ W^-$ production), and has continued at the Tevatron and the LHC. 
The early motivation for this measurement was to establish the  $SU(2)_L\times U(1)_Y$ symmetry structure of EW self-interactions.  
From the modern perspective, the study of TGCs is yet another precision test of the Standard Model (SM)  that could potentially reveal the existence of new physics. 

In the SM, the self-interactions of EW gauge bosons arise from the non-abelian structure of the field-strength tensor in the gauge kinetic terms:
\begin{equation}
{\cal L}^{\rm SM} \supset 
 - {1\over 4} W_{\mu \nu}^i W_{\mu \nu}^i, 
\qquad
W_{\mu \nu}^i = \partial_\mu W_\nu^i - \partial_\nu W_\mu^i + g_L \epsilon^{ijk} W_\mu^j W_\nu^k, 
\end{equation}
where $W_\mu^i$, $B_\mu$  are the gauge fields  $SU(2)_L \times U(1)_Y$, and the corresponding gauge couplings are denoted $g_L$, $g_Y$.  
It follows that the TGCs in the SM are given by ${\cal L}_{\rm TGC}^{\rm SM}  = -g_L \epsilon^{i j k} \partial_\mu W_\nu^i W_\mu^j W_\nu^k$. 
After EW symmetry breaking, this leads to interactions between two charged and one neutral boson of the SM:  
\begin{equation} 
{\cal L}_{\rm TGC}^{\rm SM} =  i \sum_{V=\gamma,Z}  g_{WWV}  \left [   \left ( W_{\mu \nu}^+ W_\mu^-  -  W_{\mu \nu}^- W_\mu^+ \right ) V_\nu  +   V_{\mu\nu}\,W_\mu^+W_\nu^- \right ], 
\end{equation} 
whereas interactions between 3 neutral gauge bosons are absent in  the SM Lagrangian.  
The coupling strength is uniquely fixed by the gauge couplings:  $g_{WW\gamma} = e$, and $g_{WWZ} = e g_L/g_Y$, where $e \equiv  g_L g_Y/\sqrt{g_L^2 + g_Y^2}$ is the electric charge. 
Going beyond the SM, TGCs  between two charged and one neutral boson  can be generally parametrized as 
\begin{equation} 
\label{tgclep_tgcpar}
\begin{split}
{\cal L}_{\rm TGC} = & \sum_{V=\gamma,Z} g_{WW\gamma}   \left ( {\cal L}_{\rm TGC}^+   + {\cal L}_{\rm TGC}^- \right),
\\  
 {\cal L}_{\rm TGC}^+= & 
 i  (1 + \delta g_{1}^V)\,  \left ( W_{\mu \nu}^+ W_\mu^-  -  W_{\mu \nu}^- W_\mu^+ \right ) V_\nu 
 + i  (1 + \delta \kappa_V )\, V_{\mu\nu}\,W_\mu^+W_\nu^-
 \\ &
+    i   \frac{ \lambda_V}{m_W^2} W_{\mu \nu}^+W_{\nu \rho}^- V_{\rho \mu}
- g_5^V   \epsilon_{\mu\nu\rho\sigma} \left ( W_\mu^+ \partial_\rho W_\nu^- - \partial_\rho W_\mu^+  W_\nu^- \right ) V_\sigma ,
\\  
 {\cal L}_{\rm TGC}^-= & 
  i \tilde \kappa_V  \tilde V_{\mu \nu } W_\mu^+ W_\nu^- 
  +   i   \frac{ \tilde \lambda_V}{m_W^2} W_{\mu \nu}^+W_{\nu \rho}^- \tilde V_{\rho \mu}
 -  \tilde g_4^V W_\mu^+ W_\nu^- (\partial_\mu V_\nu + \partial_\nu V_\mu),
\end{split}
\end{equation}
where $V_{\mu\nu} = \partial_\mu V_\nu - \partial_\nu V_\mu$, $\tilde V_{\mu\nu} =   \epsilon_{\mu\nu\rho\sigma} \partial_\rho V_\sigma$. 
The unbroken electromagnetic gauge invariance implies $\delta g_{1}^\gamma = g_4^\gamma = g_5^\gamma = 0$. 
We split these interactions  into the CP-even (+) and CP-odd (-) parts.\footnote{C acts as $W^\pm \to - W^\mp$, $V \to -V$. P acts as $V_0 \to V_0$, $V_i \to -V_i$, $\partial_i \to -\partial_i$.}
The first five terms in  ${\cal L}_{\rm TGC}^+$ are C- and P-even, and follow the standard parametrization of Ref.~\cite{Hagiwara:1986vm,Hagiwara:1993ck}.\footnote{
The couplings \cite{Hagiwara:1993ck} are related to the ones in  Eq.~(\ref{tgclep_tgcpar})  by $\kappa_V = 1 + \delta \kappa_V$, $g_{1}^Z = 1 + \delta g_{1}^Z$.   
Note that we take the sign of $g_{WWV}$ opposite to that  of Ref.~\cite{Hagiwara:1993ck} because we use a different convention to define covariant derivatives:  $D = \partial - i g V$, as opposed to   $D = \partial + i g V$ in Ref.~\cite{Hagiwara:1986vm}. This  corresponds to flipping the sign of all gauge fields.}
The last term proportional to $g_5^V$ is C- and P-odd.
The first two terms in ${\cal L}_{\rm TGC}^-$ are C-even and P-odd, while the last one is C-odd and P-even. 
At the end of the day,   we have  6 CP-even and 5 CP-odd couplings characterizing anomalous TGCs. 
The CP-even operators interfere with the SM amplitudes, therefore their effects show up at the linear order in the coupling, while the CP-odd couplings show up only at the quadratic order.    
In the SM limit all these couplings are zero at tree level~\footnote{SM loops contribute to TGCs~\cite{Argyres:1995ib}.}. 

Constraints on anomalous TGCs can be translated into constraints on masses and couplings of new physics models. 
In some scenarios these constraints are complementary to those provided by direct searches and other precision studies. 
Basically, TGCs can be generated by loops of any particles with EW charges. 
For example, in the MSSM one can generate TGCs via loops of sfermions, gauginos and Higgses \cite{Argyres:1995ib,Fichet:2013ola}.  
Light supersymmetric particles could evade direct constraints due to lack of missing energy as in RPV SUSY, or small cross sections, as in electroweakino production, yet they could be constrained by precise measurements of  TGCs. 
See \cite{Fichet:2013ola} for TGCs in warped extra dimensions, composite  Higgs, and their interplay with  Higgs, EW precision physics and direct searches. 

The effect of new physics is very transparent in the effective theory approach to physics beyond the SM (BSM). 
At the leading order in the effective theory expansion there are three operators associated with CP-even deformations of EW TGCs, and two more with CP-odd ones. 
Of these five operators, two affect {\em only} the self-couplings of EW gauge bosons, and can be probed {\em only} via precision measurements of TGCs. 
Further three operators affect both the self-couplings and the couplings to the Higgs boson, leading to important synergy between the Higgs and TGC studies.  
Finally, two of these operators contribute to the Peskin-Takeuchi S-parameter \cite{Peskin:1991sw}  leading to an interplay between the TGC and precision constraints from the Z-pole observables.

In this article we revisit the  bounds on TGCs from the measurements of the total and differential $W^+ W^-$ production at LEP-2.  
We use the combined results from all four LEP collaborations \cite{Schael:2013ita}.  
We extract bounds on anomalous TGCs for all the 11 couplings  in Eq.~(\ref{tgclep_tgcpar}).    
In this case, due to a  large number of parameters and large degeneracies, we restrict to a one-by-one fit where only one anomalous coupling is switched on at a time. 
We then move to the parametrization of the TGCs in terms of the dimension-6 operators beyond the SM. 
We present constraints on the coefficients of the three CP-even  operators contributing to anomalous TGCs, including their correlation. 
Previously, equivalent fits were performed by LEP's DELPHI collaboration \cite{Abdallah:2010zj},  but no combined LEP constraints  were available in this form.  
Our result is given in the form that can be readily combined with the Higgs results and other precision experiments.

\section{One-by-one constraints on anomalous TGCs}
\label{tgclep_sec1b1}

In this section we derive bounds on TGCs  in the case when only one anomalous coupling is switched on at a time.  
In Ref.~\cite{Schael:2013ita} the LEP collaborations report combined measurements of the  $e^- e^+ \to  W^- W^+$ cross sections at different center-of-mass energies $\sqrt{s}$. 
We use the  total cross sections for eight  values of $\sqrt{s}$ from Table~5.3 of Ref.~\cite{Schael:2013ita}, and the corresponding correlation matrix $\rho$ from Table~E.3 of Ref.~\cite{Schael:2013ita}.  
We  also use the differential cross section as function of the  scattering angle between $W^-$ and $e^-$  for four values of averaged $\sqrt{s}$ listed in Table~5.6 of Ref.~\cite{Schael:2013ita}. 
For the differential cross sections the correlation matrix is not given; we assume these measurements are uncorrelated with each other and with the total cross section measurement.\footnote{This is clearly not a realistic assumption. Note however that the differential cross section measurements use a smaller subset of the data than the total cross section measurements, therefore it is not trivial to estimate the correlation between these observables.} 
We  collectively denote the measured central values of these observables as  $\sigma_{\rm exp}^I$, and the experimental  errors as   $\Delta \sigma_{\rm exp}^I$. 
To compare these results with theoretical predictions, we first compute analytically these cross sections at tree level as function of the anomalous  couplings  in Eq.~(\ref{tgclep_tgcpar}). 
The result can be split as  $\sigma_{\rm LO}^I= \sigma_{\rm LO, \rm SM}^I + \delta \sigma_{\rm LO}^I$, where $\sigma_{\rm LO,SM}^I$ is the tree-level SM cross-section, and $\delta \sigma_{\rm LO}^I$ is the shift of the cross section due to anomalous TGCs that depends linearly (interference terms) and quadratically (new physics squared terms) on these couplings.
In principle, the $e^- e^+ \to  W^- W^+$ cross section could be affected by other deformations of the SM, for example modifications of the Z-boson mass and couplings to fermions, or the mixing between the $U(1)_Y$ and $SU(2)_L$ gauge bosons.   
These are however strongly constrained by other precision measurements, especially by the Z-pole constraints from LEP-1 and SLC \cite{ALEPH:2005ab}. 
Taking into  account these constraints, these deformations have negligible effects on the $WW$ production at LEP.
To evaluate $ \delta \sigma_{\rm LO}^I$ we take the SM gauge couplings  $g_L = 0.650$,  $g_Y =0.358$, and the masses  $m_Z = 91.1876$~GeV, $m_W  = 80.385$~GeV.  
Then the theoretical prediction is taken    $\sigma_{\rm th}^I = \sigma_{\rm NLO,SM}^I +  \delta \sigma_{\rm LO}^I$.

For the SM NLO predictions   $\sigma_{\rm SM, NLO}^I$ we use the results from the RACOONWW code  \cite{Denner:1999gp} that we borrow from  from Table~E.4 (total cross section) and Fig.~5.4 (differential cross section) of Ref.~\cite{Schael:2013ita}. 
Finally, we construct  the $\chi^2$ function: 
\begin{equation}
\label{tgclep_chi2}
\chi^2 = \sum_{IJ}  
{ \sigma_{\rm NLO,SM}^I +  \delta \sigma_{\rm LO}^I  - \sigma_{\rm exp}^I \over \Delta \sigma_{\rm exp}^I } \rho_{I J}
{\sigma_{\rm NLO,SM}^J +  \delta \sigma_{\rm LO}^J  - \sigma_{\rm exp}^J \over \Delta \sigma_{\rm exp}^J}. 
\end{equation}
Minimizing this $\chi^2$ we obtain the central values, $1~\sigma$ error and the 95\% CL intervals   for the anomalous TGCs when only one such coupling is present at a time. 
For the CP-even ones we find   
\begin{eqnarray} 
\delta g_{1}^Z  &=& -0.10^{+0.05}_{-0.05},
\qquad    -0.19 < \delta g_{1}^Z < 0.00, 
\nonumber \\ 
\delta \kappa_\gamma &=& -0.03^{+0.04}_{-0.04},
\qquad     -0.10 < \delta \kappa_\gamma<  0.04, 
\nonumber \\  
\delta \kappa_Z &=&  -0.07^{+0.04}_{-0.04},
\qquad    -0.14  <\delta \kappa_Z <  0.01,  
\nonumber \\  
\lambda_\gamma &=&  -0.05^{+0.04}_{-0.04}, 
\qquad    -0.12 < \lambda_\gamma < 0.04, 
\nonumber \\ 
 \lambda_Z &=& -0.08^{+0.04}_{-0.04},
 \qquad   -0.15   < \lambda_Z < 0.00, 
\nonumber \\ 
g_{5}^Z &=&  -0.10^{+0.09}_{-0.09},
\qquad     -0.28 < g_{5}^Z < 0.08 \ .    
\end{eqnarray}
The fit to the LEP WW data can be improved by $\Delta \chi^2 \approx 4$  with respect to the SM if  one of the anomalous couplings  $\delta g_{1,Z}$, $\delta \kappa_Z$, or $\lambda_Z$ is present. 
Note however that generic models beyond the SM will induce several anomalous couplings rather than just one. 
In the one-by-one fit the results depend very weakly on whether the quadratic corrections in anomalous couplings  to the WW observables are kept or neglected.  

For the CP-odd anomalous couplings we find 
\begin{eqnarray} 
\tilde \kappa_\gamma &=& 0.00^{+0.15}_{-0.15} ,
\qquad   | \tilde \kappa_\gamma|  <  0.24 ,
\nonumber \\ 
\tilde \kappa_Z &=& 0.00^{+0.11}_{-0.11},
\qquad   |\tilde \kappa_Z | < 0.17  ,
\nonumber \\ 
\tilde \lambda_\gamma &=& 0.00^{+0.12}_{-0.12},
\qquad  |\tilde \lambda_\gamma| < 0.18 ,
\nonumber \\   
\tilde \lambda_Z &=& 0.00^{+0.09}_{-0.09},
\qquad  |\tilde \lambda_Z| < 0.14 ,
\nonumber \\ 
\tilde g_{4}^Z &=&  0.00^{+0.20}_{-0.20},
\qquad  |\tilde g_{4}^Z | < 0.32 \ .
\end{eqnarray}
In this case  $ \delta \sigma_{\rm LO}^I$ depends only quadratically on the  anomalous couplings, which is the reason why the constraints are weaker than for the CP-even ones.  
Apparently, the fit to the WW LEP data cannot be improved at all if only CP-odd anomalous couplings are present.

\section{Constraints on dimension-6 operators from TGCs}

\subsection{Effective Lagrangian} \label{EFT}

With the LHC discovery of the Higgs scalar and the study of its properties, experiments have finally addressed all aspects of the SM and more information on the EW sector has been gained. In particular, a linearly-realized EW symmetry breaking (EWSB) sector, described by an $SU(2)_L$ doublet $H$ that obtains VEV, is clearly favored. 
Given the experimental data it is fair to assume that $H$ is the only source of EW symmetry breaking. 
Moreover, no other states beyond those predicted by the SM  have been found in the first LHC run. 
Thus, it is reasonable to assume that fundamental interactions at the weak scale can be described by the SM Lagrangian supplemented by higher-dimensional operators representing the effects of heavy new physics states.

We  consider the effective Lagrangian of the form 
\begin{equation}
{\cal L}_{\rm eff}  = {\cal L}_{\rm SM} + {\cal L}^{D=5} + {\cal L}^{D=6}  + \dots 
\end{equation} 
The first term is the SM Lagrangian.  
The following ones contain $SU(3)_C \times SU(2)_L \times U(1)_Y$ invariant operators of dimension $D>4$ constructed out of the SM gauge, fermion and Higgs fields that modify the predictions of the SM.
In the effective Lagrangian philosophy, one assumes that the lowest dimension operators have the largest impact on observables.    
The only operators at $D=5$ that one can construct are of the form  $(LH)^2$; these give masses to neutrinos and are irrelevant for the present discussion.   
At $D=6$ there are 3 CP-even and 2 CP-odd operators that affect the self-couplings of EW gauge bosons.
In a convenient basis\footnote{%
Several other bases are commonly used in the literature. 
In the {\em GIMR basis}~\cite{Buchmuller:1985jz,Grzadkowski:2010es}, instead of the operator ${\cal O}_{W}$  in Eq.~(\ref{tgclep_dim6}), one has a combination of ${\cal O}_D\equiv J^W_h.J^W_f $, ${\cal O}_{4f}\equiv J_f^W.J^W_f $, ${\cal O}'_{D^2}\equiv |H^\dagger D^\mu H |^2 $  (the $J^W_i$ are $SU(2)_L$ currents) that contribute to TGCs~\cite{Dumont:2013wma}.
In the {\em SILH basis}  \cite{Giudice:2007fh, Contino:2013kra}, instead of the operator ${\cal O}_{WB}$  in Eq.~(\ref{tgclep_dim6}), one has a combination of ${\cal O}_{HW}\equiv D_\mu H^\dagger \sigma^i (D_\nu H)W_{\mu \nu}^i$,  ${\cal O}_{HB}\equiv D_\mu H^\dagger D_\nu H B_{\mu \nu}$, ${\cal O}_{W}$  and the operators ${\cal O}_{BB}\equiv |H|^2 B_{\mu \nu}  B_{\mu \nu}$,  ${\cal O}_{WW}\equiv |H|^2 W_{\mu \nu}^i  W_{\mu \nu}^i$ and ${\cal O}_{B}\equiv H^\dagger \overleftrightarrow D_\mu H  D_\nu B_{\mu \nu}$. Of these, only  two combinations of ${\cal O}_{HW},{\cal O}_{HB},{\cal O}_{W}$  contribute to TGCs~\cite{Giudice:2007fh,Elias-Miro:2013mua,Pomarol:2013zra}. 
All these bases are equivalent up to total derivatives, and only two independent combinations contribute to TGCs,  whatever the basis.}
 they can be written as~\footnote{For a translation between Higgs operators in the SILH basis and TGCs see Ref.~\cite{Alloul:2013naa}.}  
\begin{eqnarray}  
\label{tgclep_dim6}
{\cal L}_{\rm TGC}^{D=6} &=& 
{c_{WB}  g_L g_Y   \over m_W^2}  B_{\mu\nu}  W_{\mu\nu}^i H^\dagger \sigma^i H 
+\frac{i c_W g_L}{2m_W^2}\left( H^\dagger  \sigma^i \overleftrightarrow {D^\mu} H \right )( D^\nu  W_{\mu \nu})^i 
+ \frac{c_{3W} g_L^3}{m_W^2} \epsilon^{ijk} W_{\mu \nu}^{i} W_{\nu \rho}^{j} W_{\rho \mu}^{k}    
\nonumber \\ &+& 
  \tilde c_{WB} {g_L g_Y \over m_W^2}\tilde  B_{\mu\nu}  W_{\mu\nu}^i H^\dagger \sigma^i H 
+ \frac{\tilde c_{3W}\,  g_L^3}{m_W^2}\, \epsilon^{ijk} W_{\mu \nu}^{i} W_{\nu \rho}^{j} \tilde W_{\rho\mu}^{k} .    
\end{eqnarray}  
The coefficients $c_i$ are formally of order $m_W^2/M^2$,  where $M$ is the mass scale of new physics,  and need to satisfy $c_i \ll 1$ for the effective theory approach to be valid.
The anomalous  TGCs in Eq.(~\ref{tgclep_tgcpar}) are expressed by the coefficients of these 5 operators as  
\begin{eqnarray}
\delta \kappa_\gamma &=& 4  c_{WB} ,  
\nonumber \\
\delta \kappa_Z &=& - 4 {g_Y^2 \over g_L^2} c_{WB}   -  {g_L^2 + g_Y^2 \over 2 g_L^2}  c_W ,
\nonumber \\
\delta g_Z &=&  -  {g_L^2 + g_Y^2 \over 2 g_L^2}  c_W  ,
\nonumber \\
\lambda_\gamma = \lambda_Z &=&  - 6 g_L^2  c_{3W} ,
\nonumber \\
\tilde \kappa_\gamma &=& 4  \tilde c_{WB} ,
\nonumber \\
 \tilde \kappa_Z &=& - 4 {g_Y^2 \over g_L^2}  \tilde c_{WB} ,
\nonumber \\
\tilde \lambda_\gamma = \tilde \lambda_Z &=&  - 6 g_L^2 \tilde c_{3W}, 
\end{eqnarray}
while  $g_4^Z$ and $g_5^Z$ are not generated by dimension-6 operators. 
Notice that in minimally coupled theories only ${\cal O}_W$ can be generated perturbatively at loop-level,  the other operators being  
 necessarily loop-induced~\cite{Giudice:2007fh,Dumont:2013wma,Einhorn:2013kja}.

Three of the operators in Eq.(~\ref{tgclep_dim6}) contribute not only to TGCs but also affect the Higgs boson couplings.  
One obtains 
\begin{eqnarray} 
\Delta {\cal L}_{\rm Higgs} &=&   
\label{tgclep_dlhiggs}
  c_{WB}  {h \over v}{4 g_Y^2 \over g_L^2 +g_Y^2}   \left [ 
2  Z_{\mu \nu} Z_{\mu \nu}     
- {g_L^2 - g_Y^2 \over g_L g_Y }   \gamma_{\mu\nu} Z_{\mu\nu} -   \gamma_{\mu\nu} \gamma_{\mu\nu}    \right ] 
\nonumber \\  &+& 
 c_{W}   {h \over v} \left [ 
  (  \partial_\nu W_{\mu \nu}^+  W_\mu^- + {\rm h.c.} )
+  \partial_\nu Z_{\mu \nu} Z_\mu 
+  { g_Y \over g_L}  \partial_\nu \gamma_{\mu \nu} Z_\mu  
 \right ]
\nonumber \\  &+& 
  \tilde c_{WB}  {h \over v}{4 g_Y^2 \over g_L^2 +g_Y^2}   \left [ 
2  Z_{\mu \nu} \tilde Z_{\mu \nu}     
- {g_L^2 - g_Y^2 \over g_L g_Y }   \gamma_{\mu\nu} \tilde Z_{\mu\nu}  -   \gamma_{\mu\nu} \tilde \gamma_{\mu\nu}   \right ] . 
\end{eqnarray}
However,
 since other dimension-6 operators also contribute to the same Higgs couplings, processes like $h\to \gamma\gamma$, $h\to Z\gamma$ and the total width of $h\to VV^*$ ($V=W,Z$) cannot constrain the parameters of Eq.~(\ref{tgclep_dlhiggs}) in a model independent way \cite{Dumont:2013wma,Elias-Miro:2013mua,Pomarol:2013zra}. For example,  the operators $-{c_{BB} g_Y^2 \over  m_W^2} |H|^2 B_{\mu \nu}  B_{\mu \nu}$ and $- {c_{WW}  g_L^2 \over m_W^2} |H|^2 W_{\mu \nu}^i  W_{\mu \nu}^i$ in the dimension-6 Lagrangian also contribute to the CP-even Higgs coupling  to photons, and therefore the LHC  measurement of the $h \to \gamma \gamma$ rate constrains the combination  
$c_{WB} + c_{WW} + c_{BB}$, and not $c_{WB}$ alone. 
Model independent constraints on the operators of Eq.(~\ref{tgclep_dlhiggs}) could come from measurements of the differential distribution $h\to V \bar f f$ or $V^*\to Vh$ \cite{Ellis:2013ywa,Isidori:2013cga}, but these observables do not provide at present any meaningful constraints.
One concludes that the TGC constraints are {\em complementary} to the Higgs constraints~\cite{Dumont:2013wma, Pomarol:2013zra}. 

Finally, we note that the ${\cal O}_{WB}$ and ${\cal O}_{W}$ operators contribute to the  S parameter, 
$\Delta S = {32 \pi \over g_L^2} (c_W + 8 c_{WB})$.  
However, $\Delta S$  depends also on another dimension-6 operator  $i {c_{B} g_Y \over 2 m_W^2} H^\dagger \overleftrightarrow D_\mu H  D_\nu B_{\mu \nu}$. 
So, again, there is no model independent bound on the magnitude of anomalous TGCs from EW precision tests.  

\subsection{Fit to LEP WW data} 

We now present the simultaneous fit of the coefficients $c_i$ of the dimension-6 operators in Eq.~(\ref{tgclep_dim6}) to the LEP WW data. 
We follow a similar procedure as described in Section~\ref{tgclep_sec1b1},  with one important modification. 
In the effective theory approach it is not consistent to retain the quadratic correction on $c_i$ to the observables, unless dimension-8 operators are also included in the analysis. 
Indeed, the coefficients $c_i$ are formally ${\cal O}(m_W^2/M^2)$ where $M$ is the mass scale of new physics. 
Thus,  $c_i^2$ is ${\cal O}(m_W^4/M^4)$, just like the coefficients of dimension-8 operators.
 For this reason, we  only include linear terms in $c_i$ in our expression for $\delta \sigma_{\rm LO}^I$ in   Eq.~(\ref{tgclep_chi2}). 
As before, we ignore other deformations of the SM that are more strongly constrained by EW precision tests on the Z-pole. 
 In particular, we assume that the new physics contributions to the kinetic mixing between the $U(1)_Y$ and $SU(2)_L$ gauge bosons, 
which  in our parametrization is proportional to $c_W + c_B + 8 c_{WB}$,  is negligible as required by Z-pole constraints on the  S parameter. 
With this procedure we obtain 
\begin{equation}
c_{WB} = -0.01 \pm 0.03,  
\qquad 
c_{W} = 1.18 \pm 0.56, 
\qquad 
c_{3W} = - 0.30 \pm 0.16,   
\end{equation}
with the correlation matrix 
\begin{equation}
\rho = \left ( \begin{array}{ccc}
1 & -0.72 &  -0.78 
\\ 
 -0.72 & 1 &  0.99 
\\
-0.78 & 0.99 & 1 
 \end{array} \right ). 
\end{equation}
The huge errors for $c_{W}$ and $c_{3W}$ are  due to a near-degeneracy along the direction $c_{3W} \approx -0.3 c_{W} + 0.05$ with $c_{WB} \approx 0$.
Note that because of that  the central values of $c_{W}$ and $c_{3W}$ do not have any physical meaning: for $c_i \sim 1$ the contribution of higher-than-6 dimensional operators cannot be neglected.
 In fact, including the corrections quadratic  in $c_i$ to $\delta \sigma_{\rm LO}^I$ into the fit would completely change the central values and the errors. 
This degeneracy arises because the new physics contributions to the $e e \to W W$ cross section mostly affect just one class of polarization cross sections where left-handed electrons  produce one helicity-1 and one helicity-0 W  boson. 
These polarization  cross sections  happen to depend  the combination $c_{3W} + 0.3 c_{W}$, with a weak dependence on $\sqrt{s}$,  and the flat direction occurs where the new physics deformation of these cross sections is minimized. 
Actually, there is  also a large contribution to the polarization cross sections where two helicity-0 W  bosons are produced. 
These depend on $c_{W}$ but not on $c_{3W}$, which naively should break the degeneracy. 
However, in this case the linear effects of $c_W$ on $e_L e_L \to WW$ and $e_R e_R \to WW$ cross sections are of opposite sign and   approximately cancel between each other.
The flat direction persists also in the 2D fit  when $c_{WB}$ is constrained to vanish.
Nevertheless, the fit we provide can be useful for constraining new physics that predicts some relations between the parameters of the effective theory that do not coincide with the near-degenerate direction in the $c_{W}$--$c_{3W}$ subspace. 
As an example, in the case when  $c_{3W} =0$  we obtain 
\begin{equation}
c_{WB} = 0.03 \pm 0.02, \qquad c_{W} = 0.18 \pm 0.07, \qquad  \rho = \left ( \begin{array}{cc}
1 & 0.79  
\\ 
 0.79 & 1 
 \end{array} \right ).  
\end{equation}
This 2D fit only weakly depends on whether or not we include the corrections quadratic  in $c_i$ to $\delta \sigma_{\rm LO}^I$.

\section*{Conclusions}

We presented a new analysis of TGCs using LEP data on WW production cross section. 
Our  results go beyond the previous fits in two ways: {\it i}) in the one-by-one constraints on TGCs, we derived limits on C- and P-violating couplings, and  {\it ii}) we performed a general  analysis of  TGCs in the framework of the effective theory with dimension-6 operators beyond the SM.  
For the latter, we presented simultaneous constraints on the three CP-conserving dimension-6 operators contributing to TGCs, including their correlations.  
This is relevant as  typical new physics models  generate more than one operator. 
We also identified a flat direction in the effective theory fit, due to an accidental approximate degeneracy of the dimension-6 operator contribution to the  $ee \to WW$ cross section. 
This flat direction should be lifted by including the constraints from single W production at LEP, and di-boson production at the LHC.  
This will be attempted in future publications. 

\section*{Acknowledgments}
AF was supported by the ERC advanced grant Higgs@LHC.
FR acknowledges support from the Swiss National Science Foundation, under the Ambizione grant PZ00P2 136932.



%% file: vlq/procVLQ.tex

\chapter{Model Independent Analyses of Vector-Like Quarks}

{\it D.~Barducci, L.~Basso, A.~Belyaev, M.~Buchkremer, G.~Cacciapaglia, 
A.~Deandrea, T.~Flacke, J.H.~Kim, S.J.~Lee, S.H.~Lim, F.~Mahmoudi, 
L.~Panizzi, and J.~Ruiz-\'Alvarez}


\begin{abstract}
We propose three simplified models based on a singlet and two doublet Vector-like Quarks, which describe all the relevant phenomenology for the LHC, single and pair production and decays into third generation and light quarks, in terms of 3 independent parameters.
Such models can be used to reinterpret present searches, define new searches that can potentially close in unexplored parameter regions, and also perform systematic studies of the flavor bounds.
\end{abstract}

\section{INTRODUCTION: SIMPLIFIED MODELS}

Heavy partners of the top quark are predicted by many New Physics scenarios, including Little Higgs Models, Extra Dimensions, and Composite Higgs Models~\cite{Vignaroli:2012nf,Contino:2008hi,Contino:2006qr,Contino:2006nn,Anastasiou:2009rv,Matsedonskyi:2012ym}. Such new partners can be scalars as in supersymmetric models or vector-like fermions, and are generally introduced in order to cancel the contributions of the quadratic divergences to the Higgs mass renormalization. The observation of new heavy quarks thus plays an important role in the investigation of the Higgs sector, and in understanding the generation of quark masses.

Although vector-like quarks are usually assumed to mix with the third generation only following hierarchy or naturalness arguments (for recent works, see for instance Refs.~\cite{Berger:2012ec,DeSimone:2012fs,Aguilar-Saavedra:2013wba,Aguilar-Saavedra:2013qpa}), the top partners can mix in a sizable way with lighter quarks while remaining compatible with the current experimental constraints~\cite{Cacciapaglia:2011fx,Buchkremer:2013bha}. Recent New Physics scenarios now take into account this possibility and must be considered with attention. Indeed, the top partners interactions with the electroweak and Higgs bosons are generically allowed through arbitrary Yukawa couplings, implying that the branching ratios into light quarks can be possibly competitive with the top quark. Another possibility that has been recently investigated is to couple the vector-like quarks to heavier states, like for instance a composite vector, thus decaying directly into three quarks~\cite{Redi:2013eaa}.

In these proceedings, we will limit ourselves to the latter case of mixing via Yukawa couplings: in this case, the new quarks can only decay into a standard quark plus a boson, $W^\pm$, $Z$ or Higgs. The relative branching ratios are purely determined by the weak quantum numbers of the multiplet the new quark belongs to~\cite{delAguila:2000rc}. Experimental collaborations are now considering the interplay between the various decays in their searches: a combination of the various channels can be found on the public result pages of both ATLAS~\cite{ATLAS-CONF-2013-051,ATLAS-CONF-2013-056,ATLAS-CONF-2013-018,ATLAS-CONF-2013-060} and CMS~\cite{Chatrchyan:2013uxa,CMS-PAS-B2G-12-019,CMS-PAS-B2G-13-003,CMS-PAS-B2G-12-021} collaborations, with final states in the third generation.
Couplings to light generation have also been considered in single production~\cite{Aad:2011yn}, and in some few cases in the decays (see for instance Ref.~\cite{ATLAS-CONF-2013-051} for the case of a b-partner).
Here we want to provide a simple and model independent framework where all the experimental effort can be reinterpreted, and also provide new channels to analyze.
To do so, we adopt the simplified model approach, and propose to study a single representation that couples to standard quarks via Yukawa interactions.
A simple Lagrangian can be written down, following Refs.~\cite{Atre:2011ae,Atre:2013ap,Buchkremer:2013bha}. The lagrangian can be largely simplified by the observation that the mixing to the light quarks is dominantly chiral: the new vector-like quarks couple to left-handed quarks for singlets and triplets of SU(2), and to right-handed quarks for doublets.
During the Les Houches workshop, it was decided to focus on 3 benchmark scenarios: one with a singlet top partner $T$, one with a doublet with standard hypercharge $(T,B)$, and one with exotic hypercharge $Y_{\rm doublet} = 7/6$, consisting of $(X,T)$ where $X$ has the exotic charge $+5/3$.
These three cases well approximate the phenomenology of realistic models, even in the case where more states are present with near-degenerate masses (see Appendix of Ref.~\cite{Buchkremer:2013bha}).
Also, 3 parameters are enough to encode all the model dependence relevant for the LHC:
\begin{itemize}
\item[-] $M$, the vector-like mass of the multiplet, which is equal to the mass of the new quarks up to small corrections due to the mixing to the standard quarks;
\item[-] $g_\ast$, which represent a common coupling strength to light quarks in units of standard couplings, and which is only relevant when discussing single production;
\item[-] $R_L$, which describes the rate of decays to light quarks with respect to the third generation, so that $R_L = 0$ corresponds to coupling to top and bottom only, while $R_L = \infty$ coupling to light quarks only.
\end{itemize}
As an example, the Lagrangian for the singlet case $T$ is given by
\begin{eqnarray}
& \mathcal{L}_{\rm T} = g_\ast \left\{ \sqrt{\frac{R_L}{1+R_L}} \frac{g}{\sqrt{2}} [ \bar{T}_L W^+_\mu \gamma^\mu d_L ] + \sqrt{\frac{1}{1+R_L}} \frac{g}{\sqrt{2}} [ \bar{T}_L W^+_\mu \gamma^\mu b_L ] + \right. & \nonumber\\
& \sqrt{\frac{R_L}{1+R_L}} \frac{g}{2 \cos \theta_W} [ \bar{T}_L Z_\mu \gamma^\mu u_L ] + \sqrt{\frac{1}{1+R_L}} \frac{g}{2 \cos \theta_W} [ \bar{T}_L Z_\mu \gamma^\mu t_L ] + &\\
&\left. -  \sqrt{\frac{R_L}{1+R_L}} \frac{g M}{2 m_W} [ \bar{T}_R H u_L ] - \sqrt{\frac{1}{1+R_L}} \frac{g M}{2 m_W} [ \bar{T}_R H t_L + \frac{m_t}{M}\;  \bar{T}_L H t_R ]  \right\} + h.c. & \nonumber
\end{eqnarray}
where the subscripts $L$ and $R$ label the chiralities of the fermions. The last term proportional to the top mass $m_t$ encodes a correction to the Higgs coupling which is suppressed by the top mass over the vector-like mass $M$, and it is only relevant for low masses.
Analogously, we write down the Lagrangian for the doublet $(X,T)$:
\begin{eqnarray}
& \mathcal{L}_{\rm (X,T)} = g_\ast \left\{ \sqrt{\frac{R_L}{1+R_L}} \frac{g}{\sqrt{2}} [ \bar{X}_R W^+_\mu \gamma^\mu u_R ] + \sqrt{\frac{1}{1+R_L}} \frac{g}{\sqrt{2}} [ \bar{X}_R W^+_\mu \gamma^\mu t_R ] + \right. & \nonumber\\
& \sqrt{\frac{R_L}{1+R_L}} \frac{g}{2 \cos \theta_W} [ \bar{T}_R Z_\mu \gamma^\mu u_R ] + \sqrt{\frac{1}{1+R_L}} \frac{g}{2 \cos \theta_W} [ \bar{T}_R Z_\mu \gamma^\mu t_R] + &\\
&\left. -  \sqrt{\frac{R_L}{1+R_L}} \frac{g M}{2 m_W} [ \bar{T}_L H u_R ] - \sqrt{\frac{1}{1+R_L}} \frac{g M}{2 m_W} [ \bar{T}_L H t_R + \frac{m_t}{M}\;  \bar{T}_R H t_L ]  \right\} + h.c. & \nonumber
\end{eqnarray}
In the latter note the absence of the coupling of $T$ with the $W$ because the mixing is dominantly right-handed (a left-handed term exists, however it is suppressed by the light quark mass).
The Lagrangian for the doublet $(T,B)$ can be constructed as above.
Note that in the Lagrangians above, we coupled the new quarks to the first generation: the reason for this choice is that a coupling to both first and second generation would lead to strong flavor bounds, thus we chose the first generation to enhance the single-production cross sections.

The Lagrangians for the three simplified models have been implemented in FeynRules~\cite{Christensen:2008py,Alloul:2013bka}, and the validated model files can be found on the official website~\footnote{http://feynrules.irmp.ucl.ac.be/wiki/VLQ}. The model files also include gauge couplings between two vector-like quarks and electroweak gauge bosons, which may be relevant for some electroweak pair production processes. The three free parameters are set to the benchmark values: $g_\ast = 0.1$, which is in rough agreement with bounds on first generation couplings, $M = 1$ TeV and $R_L = 0.5$.

The main goal of this working group on vector-like quarks is to use these simplified models as test scenarios at the LHC for both present searches and new ones.
As the case of couplings to third generation have been already extensively studied by the experimental collaborations, we will focus on single production channels and the couplings to light quarks.
Single production, in fact, is the only channel that allows a direct test of the coupling strength $g_\ast$, i.e. of the level of mixing between the vector-like quarks and the standard model ones.
During the workshop at Les Houches, samples of data have been produced and tested using MadGraph version 5~\cite{Alwall:2011uj}, and for the LHC at 8 TeV of center of mass energy.

One interesting point that has been explored is the relevance of final states with 3 or more leptons.
In fact, the increase in the lepton multiplicity 
when choosing the signature from the vector-like quark will eventually improve signal to background ratio,
even though it would also lead to a decrease in the signal rates.
The question is which lepton multiplicity is optimal for the best signal significance 
against the respective backgrounds.
It was shown in Ref.~\cite{Belyaev:2012ai} that in case of Minimal Universal Extra Dimensions (MUED)
the tri-lepton signature plays a leading role for the LHC sensitivity
to the MUED scale. At the same time, for a generic vector-like quark model, it is not obvious
since the lepton multiplicity drops faster than in the MUED case
due to real $Z$ and $W$-bosons  which appear in the decay chains. Eventually their 
decay branching ratios are dominated by hadronic channels.
A preliminary analysis performed under the assumption of decays to third generation only~\footnote{Work in progress by A.~Belyaev, D.~Barducci and L.~Panizzi.}, i.e.~$R_L = 0$, confirms that tri-lepton signatures play the leading role for the LHC sensitivity to pair production.
This statement is also confirmed by recent CMS studies~\cite{Chatrchyan:2013uxa}
on inclusive searches for a vector-like $T$ quark 
at 8 TeV LHC. The current combined limit on the $T$ mass 
varies between 687 and 782 GeV depending on the different channels contribution, and it
is primarily driven by the tri-lepton signature
which is far better than the single-lepton channel
and visibly better than di-lepton channel.
Therefore the  tri-lepton channel is expected to play also 
the leading role for 13 TeV LHC projection.
We also studied the impact of tri- and four-lepton signatures to the case of significant mixing to light quarks, i.e. $R_L > 0$, as reported in Section~\ref{tri-pair}.

Multi lepton signatures can also play a role in single production: in this case, however, due to the smaller number of gauge bosons in the final state, the cross sections for the leptonic signatures are rather small.
Thus, fully hadronic signatures have the advantage of larger effective cross sections and a careful reconstruction of the final state kinematics may allow to beat the large backgrounds, as described in Section~\ref{had-single}
On the other hand, from the results in Section~\ref{tri-single}, tri-lepton final states may have a better chance at higher energies.
Finally, the presence of the Higgs in the final state coming from decays of the vector-like quarks~\cite{Azatov:2012rj,Atre:2013ap} offers new interesting search channels, as studied in Section~\ref{Higgs}.

The simple parametrisation we propose here can also be used for a systematic study of the flavor bounds on the mixing to light and heavy generations. This step is in fact crucial in order to have an independent estimate of the maximum allowed mixing, i.e. maximum possible $g_\ast$, and therefore the impact of single production to the future search strategies.
In fact, single production decreases with increasing vector-like mass in a slower fashion than pair production, therefore at higher masses single production may become more important than QCD pair production.
Efforts in the direction of a systematic study of flavor bounds have been initiated after this Les Houches workshop.

\section{PAIR PRODUCTION: TRI- AND FOUR-LEPTON SIGNATURES\protect\footnote{Contributing authors: G.~Cacciapaglia, M.~Buchkremer}} \label{tri-pair}

Selecting events with one charged lepton, same-sign and opposite-sign
di-leptons have been shown to provide efficient channels for vector-like
quark searches (see, e.g., the CMS searches~\cite{Chatrchyan:2013wfa,Chatrchyan:2013uxa}). Despite a
reduced production rate, three and four lepton signatures may also be used
to uncover such new heavy states. Events with multileptons, jets and missing
energy have been considered in the framework of pair-produced heavy quarks
coupling to the top quark, but not yet to the lighter generations. In this
section, we report on tentative new strategies for future top partner
searches at the TeV scale, with three and four charged leptons in the final
state, and considering general assumptions on their mixings with all three
Standard Model quark families.

\begin{figure}[htb]
\begin{center}
\includegraphics[width=.95\textwidth]{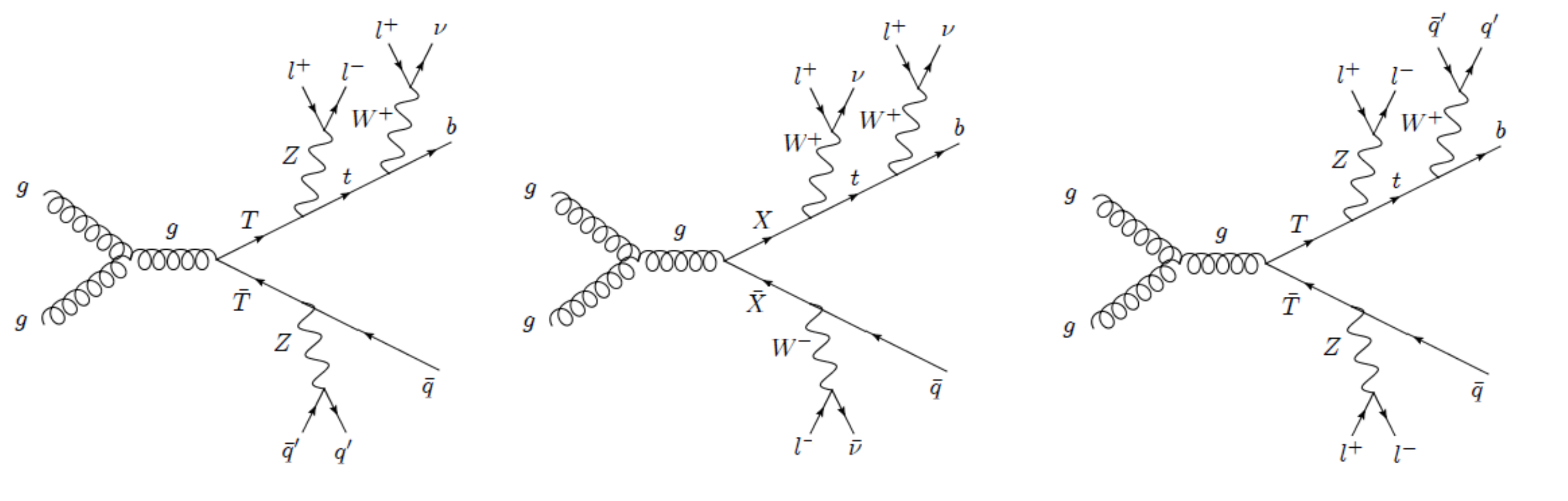}
\end{center}
\caption{Representative $T\bar{T}$ and $X\bar{X}$\ pair-production channels to three-lepton (left, middle) and four-lepton (right) final states in non-exclusive mixing scenarios (e.g., with $R_{L}>0$%
).} 
\label{fig:Feynman}
\end{figure}

For convenience, we exemplify our analysis with the $T$ singlet and $(X,T)$
doublet scenarios.\ As a representative parameter point for our analysis, we
considered $g^{\ast }=0.1$ and $M=500$ GeV\ for the mass of the heavy
quarks. The $R_{L}$ parameter is set to 0.5, allowing for generic mixings
with the first, second and third SM\ quark families. The vector-like quarks can therefore decay both into a third generation quark, top or bottom, and a light jet.\ This benchmark scenario allows for
non-exclusive decay modes, so that channels with mixed decays, where one heavy quark decays to a top and the other to a light jet, contribute to
several signal regions. Interestingly, the possibility for top partners
coupling simultaneously to the light and the third quark generations opens
the production modes%
\begin{equation}
pp\rightarrow Q\bar{Q}\rightarrow Vt+V^{\prime }j\mbox{ with }Q=X,T\mbox{
and }V^{(^{\prime })}=W,Z,H,
\end{equation}%
which are not allowed for the exclusive mixing scenarios $R_{L}\rightarrow 0$
and $R_{L}\rightarrow \infty $, and lead to novel signatures with boosted
objects, one top quark and at least two jets with large transverse momentum,
one being a $b-$quark.\ Such topologies are always present for $R_{L}>0$,
and allow for final states with three and four leptons.\ As an illustration,
we depict in Fig.~\ref{fig:Feynman} the multilepton signatures%
\begin{eqnarray}
pp &\rightarrow &T\bar{T}\rightarrow ZtZ\bar{q}\rightarrow
l^{+}l^{-}l^{+}\nu b\bar{q}q^{\prime }\bar{q}^{\prime }, \\
pp &\rightarrow &X\bar{X}\rightarrow W^{+}tW^{-}\bar{q}\rightarrow l^{+}\nu
l^{-}\nu l^{-}\nu b\bar{q}, \\
pp &\rightarrow &T\bar{T}\rightarrow ZtZ\bar{q}\rightarrow
l^{+}l^{-}bq^{\prime }\bar{q}^{\prime }l^{+}l^{-}\bar{q},
\end{eqnarray}%
as representative decay channels for pair-produced top partners. For the
simplified scenarios considered here, one of the two decay channels $ZtZj$
and $WtWj$ is always present.\ Assuming non-exclusive coupling to the first
or the third generation, i.e., considering $R_{L}=0.5$, the total
contribution of the $top+jets$ final states in $pp\rightarrow Q\bar{Q}$
pair-production sums up to a total branching ratio of $%
2R_{L}/(1+R_{L})^{2}=44.4\%$ with respect to the full signal.

The final states of interest for our analysis are%
\begin{eqnarray}
pp &\rightarrow &l^{\pm }l^{\pm }l^{\mp }+n\mbox{ }jets+\mbox{MET}, \\
pp &\rightarrow &l^{+}l^{-}l^{+}l^{-}+n\mbox{ }jets+\mbox{MET}\mbox{,}
\end{eqnarray}%
where the number of jets depends on the underlying process. Multilepton
signatures are well known to avail themselves of low background
contamination.\ The Standard Model \ processes leading to possibly
significant background to trilepton events involve $t\bar{t}$ production in
association with $W$ and $Z$ bosons, diboson, and triboson decays. As for
four lepton candidate events, the backgrounds involving two pairs of leptons
can be produced in hard interactions, including $ZZ$ and $Zt\bar{t}$
production in association with jets. The prompt decays to three (four)
charged leptons from $WZ$ ($ZZ$) production with (without) transverse
missing energy may also generate a significant source of misidentification
if both bosons decay leptonically.\ Given its large production cross
section, $t\bar{t}+jets$ provides another potential background to trilepton
signals if the two underlying $W$ bosons decay leptonically and one bottom
quark gives a third isolated lepton. Yet, the latter (reducible) background
has not been included in our analysis as we cannot estimate it at the
parton-level. Similarly, possible contributions from non-prompt or fake
lepton candidates have been neglected.\ These background sources are
expected to be severely suppressed when imposing appropriate cuts on the
number of jets, total hadronic transverse energy $H_{T}$ and missing
transverse energy MET. 

Minimal requirements on relevant variables such as, e.g., the missing
transverse momentum of the signal or the number of jets, have been set so
to avoid the main background contributions.\ The most problematic process
for multilepton signals with three and four lepton candidates consists in
the irreducible $Zt\bar{t}+jets$, whose events can involve one $Z$ boson,
two $W$ bosons, two bottom quarks and large transverse energy. In Table \ref{tab:XSections}, we summarise\ the production cross-sections for
the signals and for the various backgrounds in the multilepton channels. All
the background samples (i.e., $t\bar{t}$, $ZZ$, $WZ$, $Wt\bar{t}$ and $Zt%
\bar{t}$) used in this report have been generated at leading order with
MadGraph5~\cite{Alwall:2011uj}, using the CTEQ6L1 parton distribution functions.
Although we did not include them in the present analysis, the K-factors for
pair-produced VLQ are known to lie in the range $1.5-1.8$ for Vector-Like
quark masses between 500\ GeV\ and 1 TeV~\cite{Cacciari:2011hy}. Parton showering and
hadronization have been carried out with PYTHIA6~\cite{Sjostrand:2006za}.\ Event analysis has been performed with MadAnalysis5~\cite{Conte:2012fm}. Detector simulation is left for a more realistic analysis.

\begin{table}[tb]
\begin{center}
\begin{tabular}{|c||c|c|}
\hline
Process & $\sigma \times BR(l^{\pm }l^{\pm }l^{\mp })$ $($fb$)$ & $\sigma
\times BR(l^{+}l^{-}l^{+}l^{-})$ $($fb$)$ \\ \hline\hline
$T\ $singlet signal ($M=500$ GeV) & 31.85 (55.10) & 4.26 (7.36) \\ \hline
$(X,T)$ doublet signal ($M=500$ GeV) & 112.06 (195.50) & 11.89 (18.33) \\ 
\hline\hline
$ZZ$ background & 25.62 & 57.97 \\ \hline
$WZ$ background & 1092 & $-$ \\ \hline
$Wt\bar{t}$ background & 17.93 & $-$ \\ \hline
$Zt\bar{t}$ background & 19.70 & 3.434 \\ \hline
\end{tabular}
\caption{Leading order cross-sections computed
with MadGraph5 at $\sqrt{s}=14$ TeV for the three- and the four-lepton
signal and background event samples, with no added jets. The NLO+NLLL
predictions for the signal are given in parentheses~\cite{Cacciari:2011hy}.} \label{tab:XSections}
\end{center}
\end{table}

The MadGraph default cuts are used at the generation level, and we select
events with electrons and muons having $p_{T}^{l}\geq 10$ GeV, within $|\eta
|_{l}<2.5$ ($l=e,$ $\mu $). All jets are required to have transverse
momentum larger than 20\ GeV\ and pseudorapidity $|\eta |<5$. The standard
identification criteria for jet-lepton separation are applied. It is
required that all final particles are isolated within a cone of $\Delta R=0.4
$. As for isolating the multilepton signal, we imposed the following
requirements, in order:

\begin{itemize}
\item \textit{Cut 1}: a minimum of three charged leptons must be identified
with each $p_{T}>10$ GeV, including at least one with $p_{T}>20$ GeV.

\item \textit{Cut 2}: the signal events are selected so that they contain
at least two jet constituents, including one $b-$tagged jet.\ 

\item \textit{Cut 3}: all the events with total hadronic transverse energy $%
H_{T}$ smaller than 300\ GeV are rejected.
\end{itemize}

\begin{figure}[htb]
\begin{center}
\includegraphics[width=.45\textwidth]{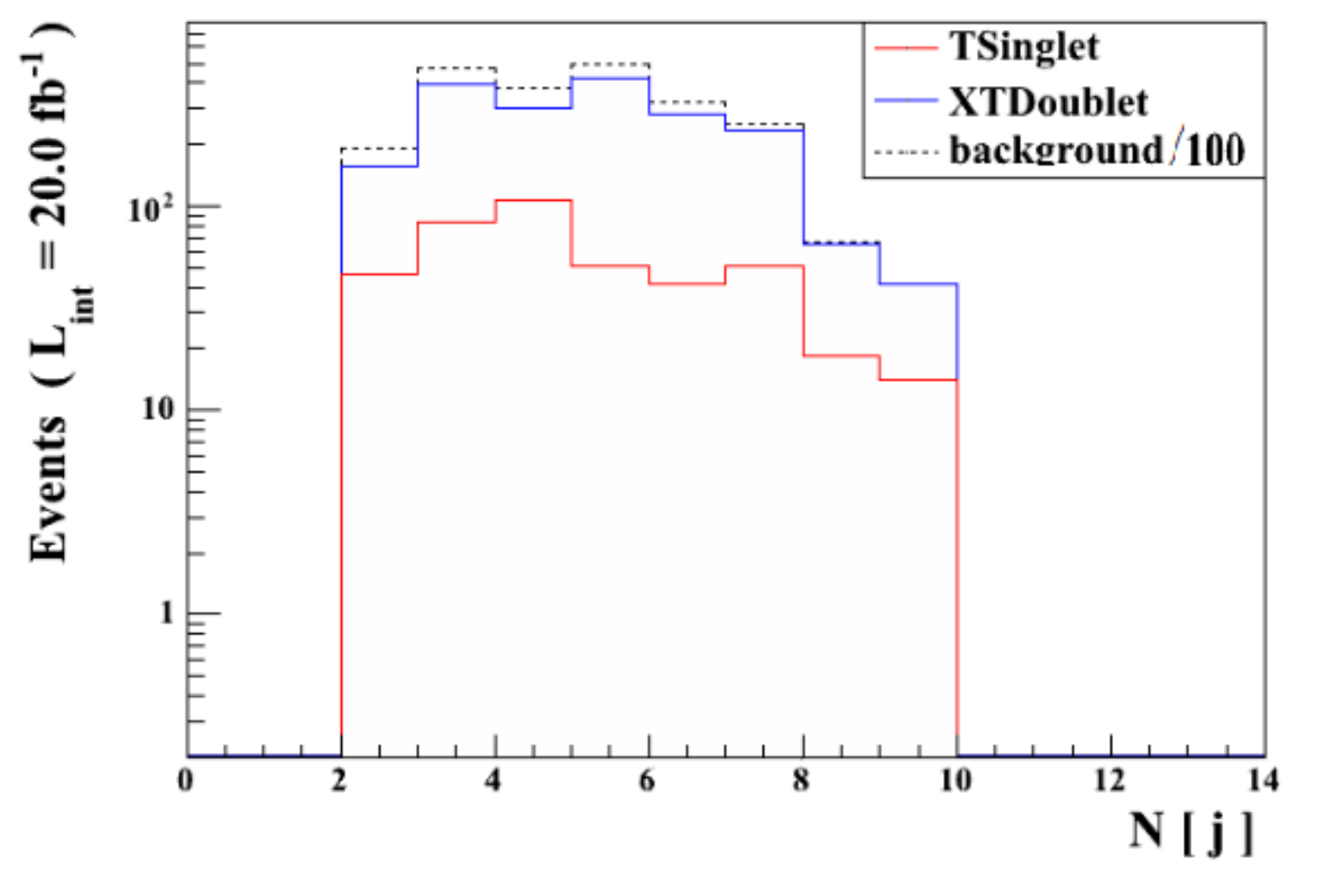}
\includegraphics[width=.45\textwidth]{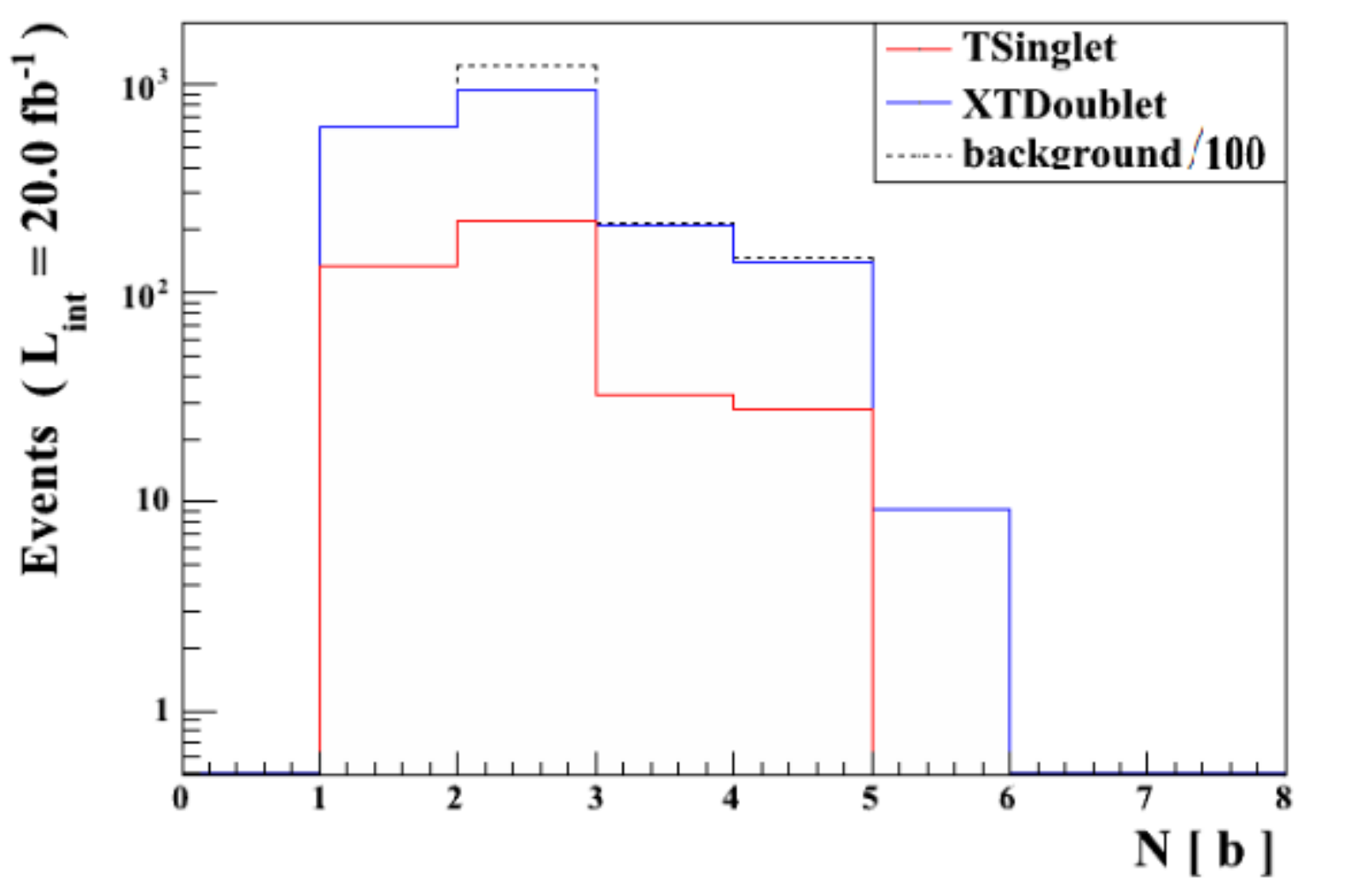}
\includegraphics[width=.45\textwidth]{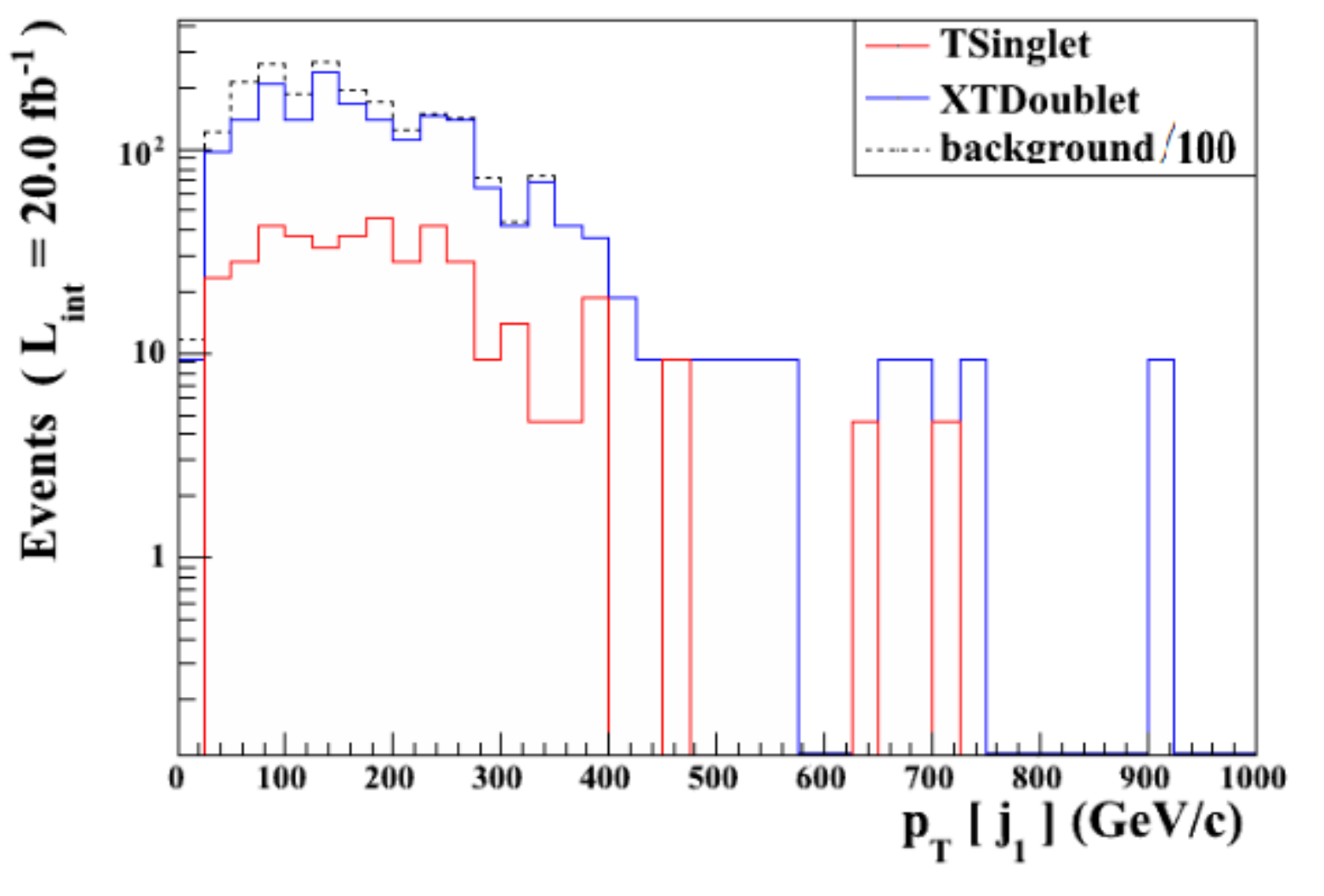}
\includegraphics[width=.45\textwidth]{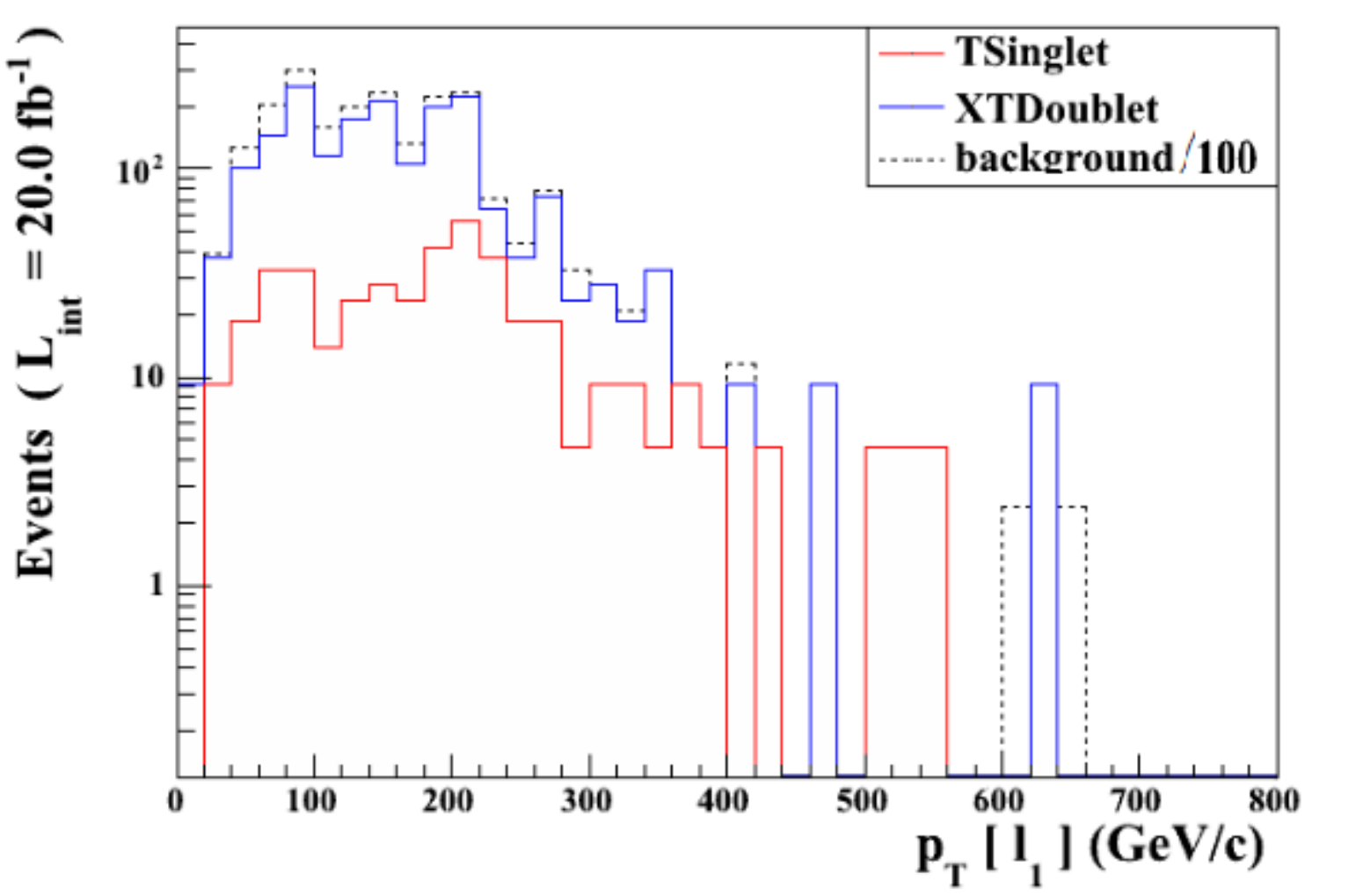}
\caption{Leading order kinematical
distributions for the light jets multiplicity (top left), the light jets
transverse momentum (bottom left), the $b$-jets multiplicity (top right) and
the $b$-jets transverse momentum (bottom right), in the three- and
four-lepton channels, for an integrated luminosity of 20 fb$^{-1}$\ at the
LHC\ 14\ TeV. The total expected numbers of event is plotted for the $T$
singlet (red) and $(X,T)$ doublet (blue) signal models, versus the
backgrounds (black), after imposing the standard cuts.} 
\label{fig:Distributions}
\end{center}
\end{figure}

The optimisation of the cuts, so as to maximise the signal-over-background
significance, is left for future work.\ Asking further for missing
transverse energy $\mbox{MET}\geq 20$ GeV\ allows for a significant
suppression of the QCD\ multi-jet background. Altogether, the above basic
requirements significantly reduce the\ diboson backgrounds, while the
contributions originating from $Zt\bar{t}+jets$, $WZ+jets$ and $ZZ+jets$
remain significant. We observe that the distributions for the signal and the
background jet multiplicities, shown in Fig.~\ref{fig:Distributions},
display similar shapes and are both peaked at $N_{j}=5$ and $N_{b}=2$.
The $p_{T}$ and $\eta $ distributions of the leading jets
and leptons for the multilepton event samples are also displayed against backgrounds.

Although a more aggressive cut on the total transverse energy, $H_{T}>500$
GeV,\textbf{\ }significantly reduces the background for the considered
benchmark, applying a more stringent selection on the transverse momenta of
either the leading jets or the leptons also enhances sizably the signal
significance. We also checked that requiring a forward-jet tag with a
transverse momentum as large as 80\ GeV\ suppresses most of the background
arising from $Wt\bar{t}+jets$, while additionally asking $p_{T}>70$ GeV\ for
the second hardest jet removes most of the $Zt\bar{t}+jets$ contribution. A\
similar selection efficiency can be obtained by imposing $p_{T}>100$ GeV\
and $p_{T}>40$ GeV for the first and second hardest leptons, respectively.
Alternatively, requiring the presence of \textit{exactly} one $b$ quark with 
$p_{T}>20$ GeV\ and $|\eta |<2.4$ allows to extract most of the signal,
provided that no other $b-$jet is tagged.\ Setting cuts on $p_{T}^{b}$ only
mildly improves the selection efficiency. Other directions for refinement
could involve mass resconstruction in specific decay modes, or using boosted
techniques to increase the sensitivity.\ 

Despite the fact that the total number of tri- and four-lepton events is expected to
decrease significantly for larger masses, these signatures provide
encouraging alternatives to search for pair-produced Vector-Like quarks in
the forthcoming searches. This is true, in particular, for top partners
having sizable mixings with both the light and the third SM\ quark
generations, that are allowed to decay simultaneously to top quarks and
light jets with high transverse momenta.

\section{SINGLE PRODUCTION: FULLY HADRONIC SIGNATURE\protect\footnote{Contributing authors: G.~Cacciapaglia, A.~Deandrea, J.L.~Ruiz-\'Alvarez}} \label{had-single}

Single production of vector-like quarks can give information about the size of the mixing with the standard quarks, but also on the nature of the new particle itself. As an example, in this contribution we will consider the case of the doublet $(X,T)$, even though similar considerations can be done for the standard doublet $(T,B)$.
Due to the suppression of the coupling to the $W$, the top partner $T$ can be produced via coupling to the third generation only in association with a top quark, due to the absence of tops in the colliding protons.
If an even small coupling to the first generation is allowed, then a new channel is open
\begin{equation}
p p \to T j \to H t j \to b \bar{b}\, b j j\, j\,.
\end{equation}
The production here takes place via the coupling with a $Z$ boson and an up quark.
The decay, on the other hand, can be dominated by the third generation: here we will focus in particular on the decay to a Higgs, which fares about 50\%, because we are interested in fully hadronic final states.
Final states with leptons can be obtained when the $T$ decays to a $Z$ boson, however the event rate pays for the small leptonic $Z$ fraction (the leptonic channel will be studied in the next section).
The aim of our preliminary study is to show that it is in principle possible to reconstruct the $T$ in a fully hadronic final state, and distinguish it from the dominant backgrounds.

\begin{table}[tb]
\centering
\begin{tabular}{|c||c|c|}
  \hline
  Process & XS ($pb$) & Ex. Events \\ \hline
  QCD (bbjjj) & 500 & 10000000 \\\hline
  W+jets & 37509 & 750180000 \\ \hline
  Z+jets & 3503.71 & 70074200 \\ \hline
  $t\; \bar{t}$ & 234 & 4680000 \\ \hline
  $t$ & 114.85 & 2297000 \\ \hline
  Diboson & 96.82 & 1936400 \\
  \hline
\end{tabular}
\caption{Cross sections and number of events for backgrounds.} \label{tab:bkgrd}
\end{table} 

The final state we are interested in consists of 5 jets with high $p_T$, three of which being b-jets, coming from the $T$ decays, and a more forward jet at production. In this analysis we considered the backgrounds listed in Table~\ref{tab:bkgrd}, where cross sections and number of events at 8 TeV for 20 fb$^{-1}$ are also listed.
All the events have been produced using MadGraph5~\cite{Alwall:2011uj}, and Pythia 6~\cite{Sjostrand:2006za} for the hadronization of the samples at parton-level. Proper matching between the hard radiation generated by MadGraph and the soft ISR/FSR radiation added by Pythia have been implemented. For the QCD sample, jets were produced with a $p_{T}>30$ GeV and within $|\eta|<5$. All the other background samples have jets with $p_{T}>10$ GeV, while no pseudo-rapidity cut have been imposed. In samples containing at least one $Z$ (di-boson processes $ZZ$ and $WZ$, and $Z$+jets) the mass of the di-lepton pair was required to be $M_{ll}>50$ GeV.
Finally, the signal sample has been produced with $p_{T}>10$ GeV  on jets. For the signal with $M_{T}=734$ GeV around 700 events were expected at 8 TeV with 20 fb$^{-1}$, where the couplings have been chosen to reproduce the benchmark point in Ref.~\cite{Buchkremer:2013bha}. For this mass point the signal has a cross section around 200 fb.

An important feature of the fully hadronic channel is the possibility of having a bump-hunt strategy: in fact, a full reconstruction of all the decay products coming from the top partner is possible. Our goal is therefore to find a bump in the invariant mass distribution for the five jets coming from $T$. The same distribution on the backgrounds should be flat-falling, while the cuts may affect this expected shape. The main difficulty with this strategy is the dependence on the good identification of the jets that reconstruct the $T$ decay. 

\begin{figure}[tb]
\begin{center}
\includegraphics[scale=0.3, viewport=0 0 650 650, clip=true]{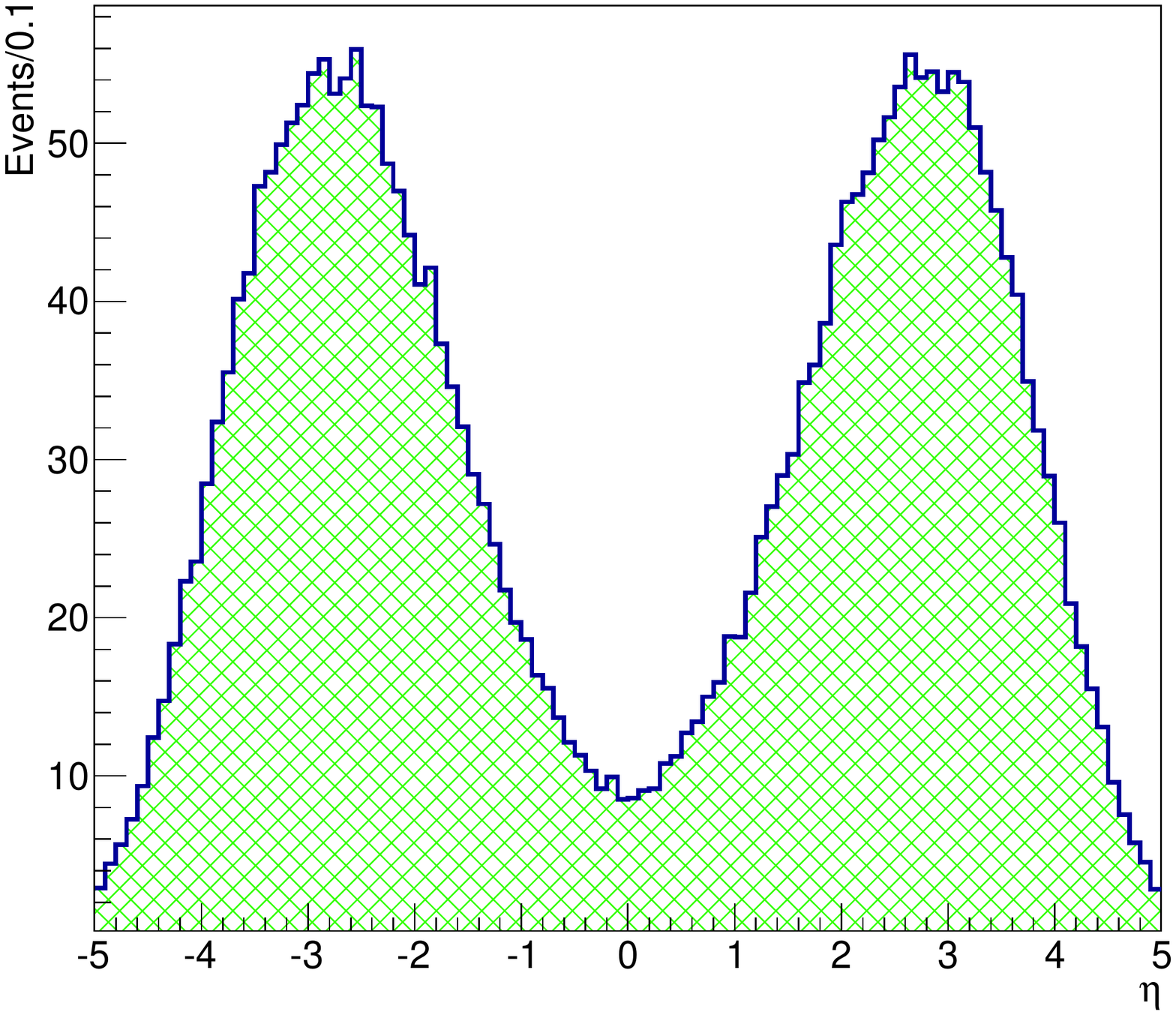}
\includegraphics[scale=0.3, viewport=0 0 650 650, clip=true]{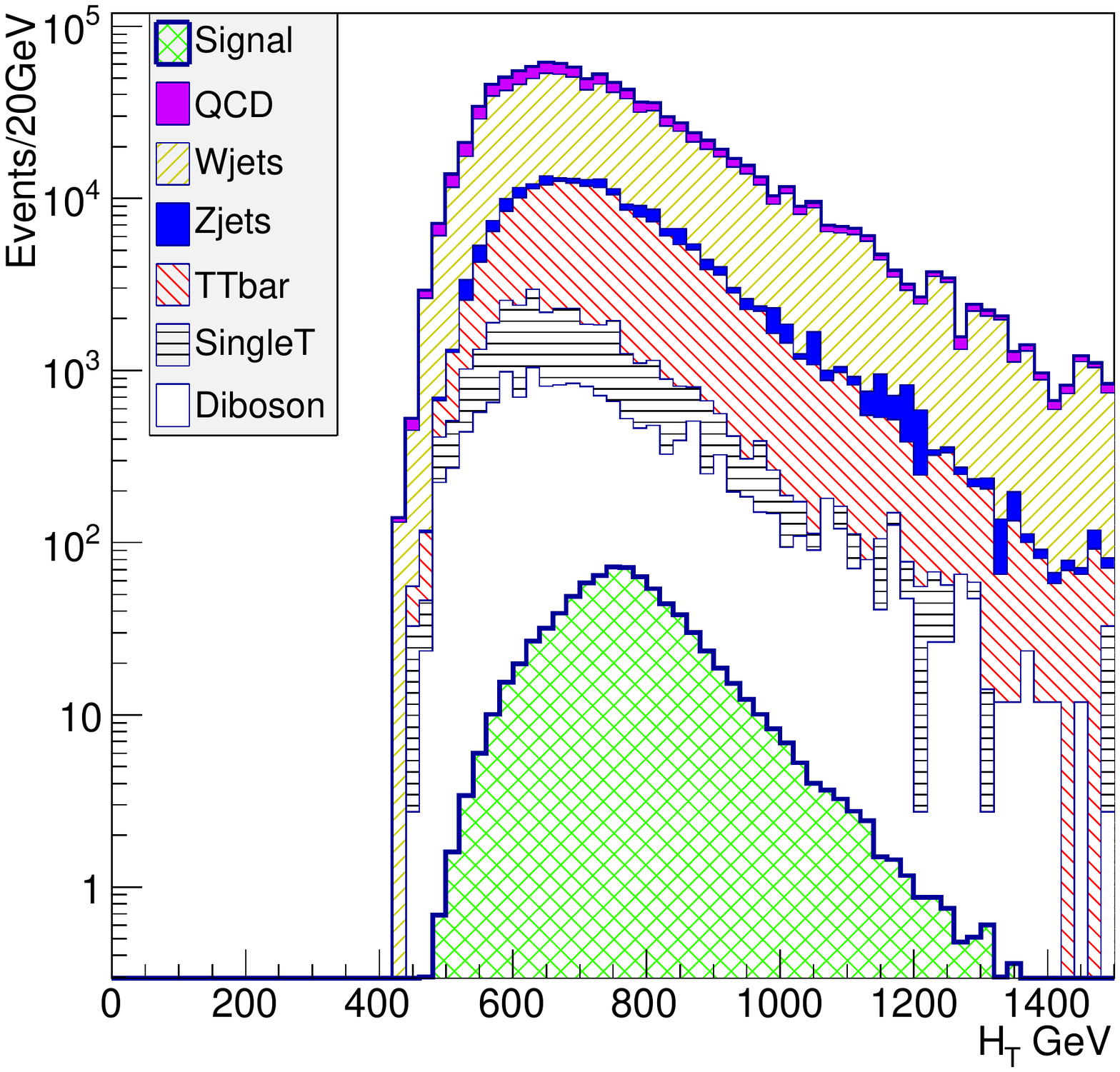}
\caption{\small \it  LEFT: pseudorapidity $\eta$ of the associated light jet at parton level. RIGHT: total hadronic energy for backgrounds (stacked) and signal (overimposed).
\label{fig:Eta6thjet}}
\end{center}
\end{figure}

As a positive point, various characteristics of the signal can be exploited in order to differentiate it from the backgrounds. We propose in this short study two variables of interest. First, the production process of the top partner generates also a light quark that adds an extra jet in the event. In Figure~\ref{fig:Eta6thjet} the pseudorapidity $\eta$ distribution for this additional jet is shown at parton level, showing that it is mainly a forward jet as expected. It should be useful therefore to require that each event has at least one jet with $|\eta|>2.5$. Second, as the top partner has a large mass, signal events should have a substantial deposit of energy from all the decay products in the detector. This can be studied with the total hadronic energy of the event, defined as $H_{T}=\sum |p_{T}^{j}|$. In Figure~\ref{fig:Eta6thjet} the distribution for this variable is shown for the signal and different backgrounds. From the $H_{T}$ plot, it is possible to say that cutting around 600 GeV will help to extract the signal.

\begin{figure}[h!]
\begin{center}
\includegraphics[scale=0.3, viewport=0 0 700 650, clip=true]{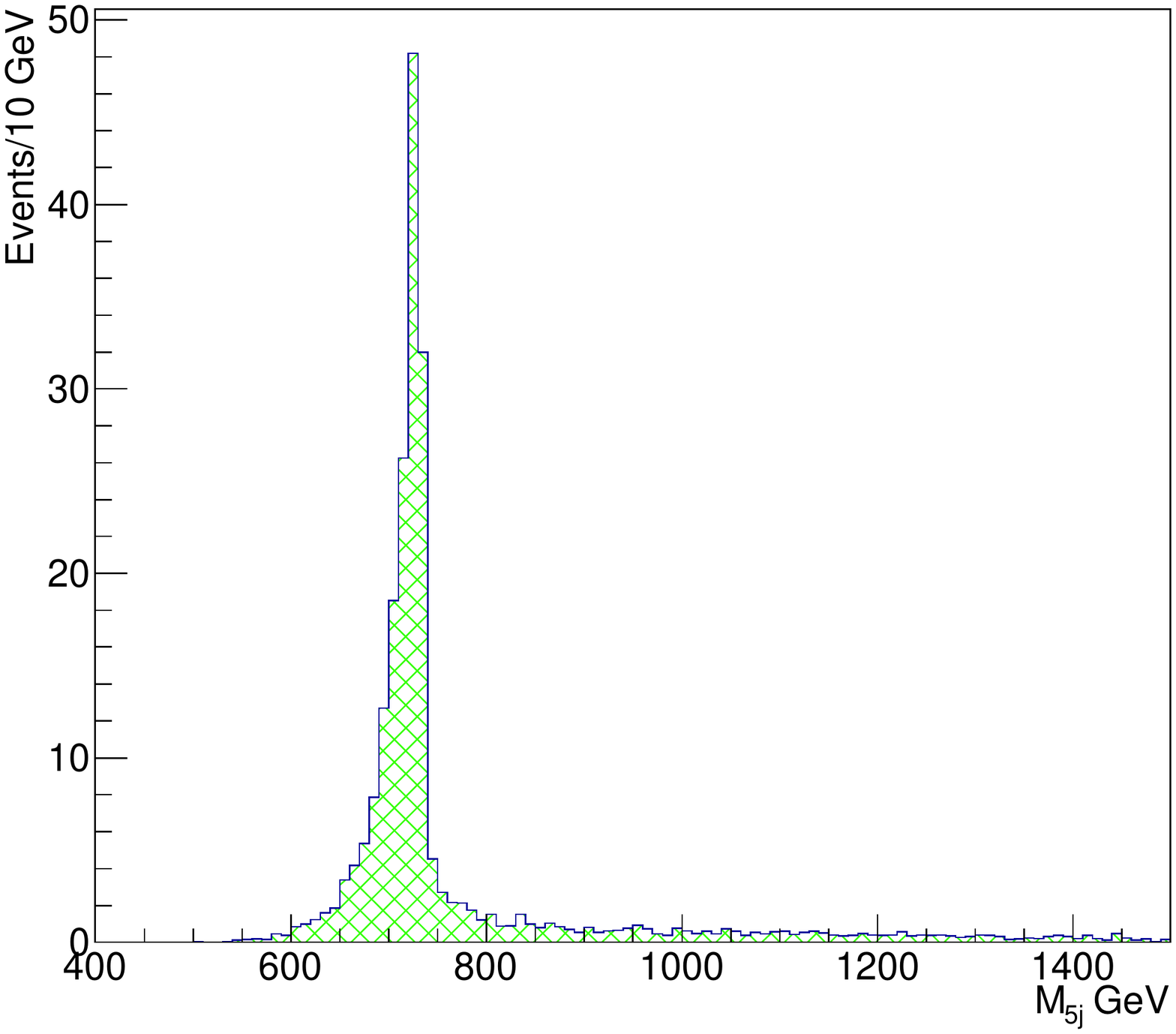}
\caption{\small \it  Reconstructed $T$ mass for the signal sample. 
\label{fig:MassPeak}}
\end{center}
\end{figure}

Finally, in order to reduce the combinatorics when reconstructing the $T$ it should be useful to require two b tags, either for the full identification of the Higgs and for the identification of the top and one of the jets coming from the Higgs. Even though requiring $H\rightarrow b \bar{b}$ closes up some decay channels of the Higgs, this choice will surely lead to a better reconstruction of the mass peak, and additionally will lead to kill a lot of the background coming from QCD processes.
In Figure~\ref{fig:MassPeak} the mass peak of the $T$ is shown using two b tags and selecting other three jets to be compatible with the top and Higgs as decay products coming form the top partner.

\section{SINGLE PRODUCTION: TRI-LEPTON SIGNATURE\protect\footnote{Contributing author: L.~Basso}} \label{tri-single}

As previously mentioned, an interesting scenario is the case of the singlet vector-like top partner coupling both to first and third generation quarks. In the case of single production, the $T'$ is produced in association with a light jet, and it can decay into a $Z$ boson and a top quark.
A very rich final state can hence be studied. In this subsection, we concentrated on the largest lepton multiplicity, namely a trilepton signature, as follows:
\begin{equation}\label{VLQ_eq_trilep}
pp \to j T' \to j Z t \to j (\ell^\pm \ell^\mp) b\ell^\pm \nu \, .
\end{equation}
Backgrounds to this signature are mainly the reducible $WZjj$, $t\overline{t}$, and $t\overline{t}\ell \nu$, and the irreducible $tZj$. The difference between the two backgrounds encompassing top quarks arise from the source of the third lepton: in the $t\overline{t}$ case, a third lepton comes from a semi-leptonic $B$-hadron decay in the b-jet, whilst in the $t\overline{t}\ell \nu$ case, a prompt lepton is considered (from 
$t\overline{t}W$). Regarding $t\overline{t}$, the main suppression comes from requiring that the leptons are isolated, as the one stemming from the $b-$jet passes this selection only in about few percent of the cases. Further, it is strongly reduced by the reconstruction of the leptonic vector bosons ($W$ and $Z$) as in the signal (despite leptonic $W$ bosons being present, the combinatorics in reconstructing first the $Z$ boson heavily affect it). We therefore expect it to be comparable to all others, despite the large $t\overline{t}$ production cross section of $25$ pb (when both top quarks decay semi-leptonically).

\begin{figure}[tb]
\begin{center}
  \includegraphics[angle=0,width=0.8\textwidth ]{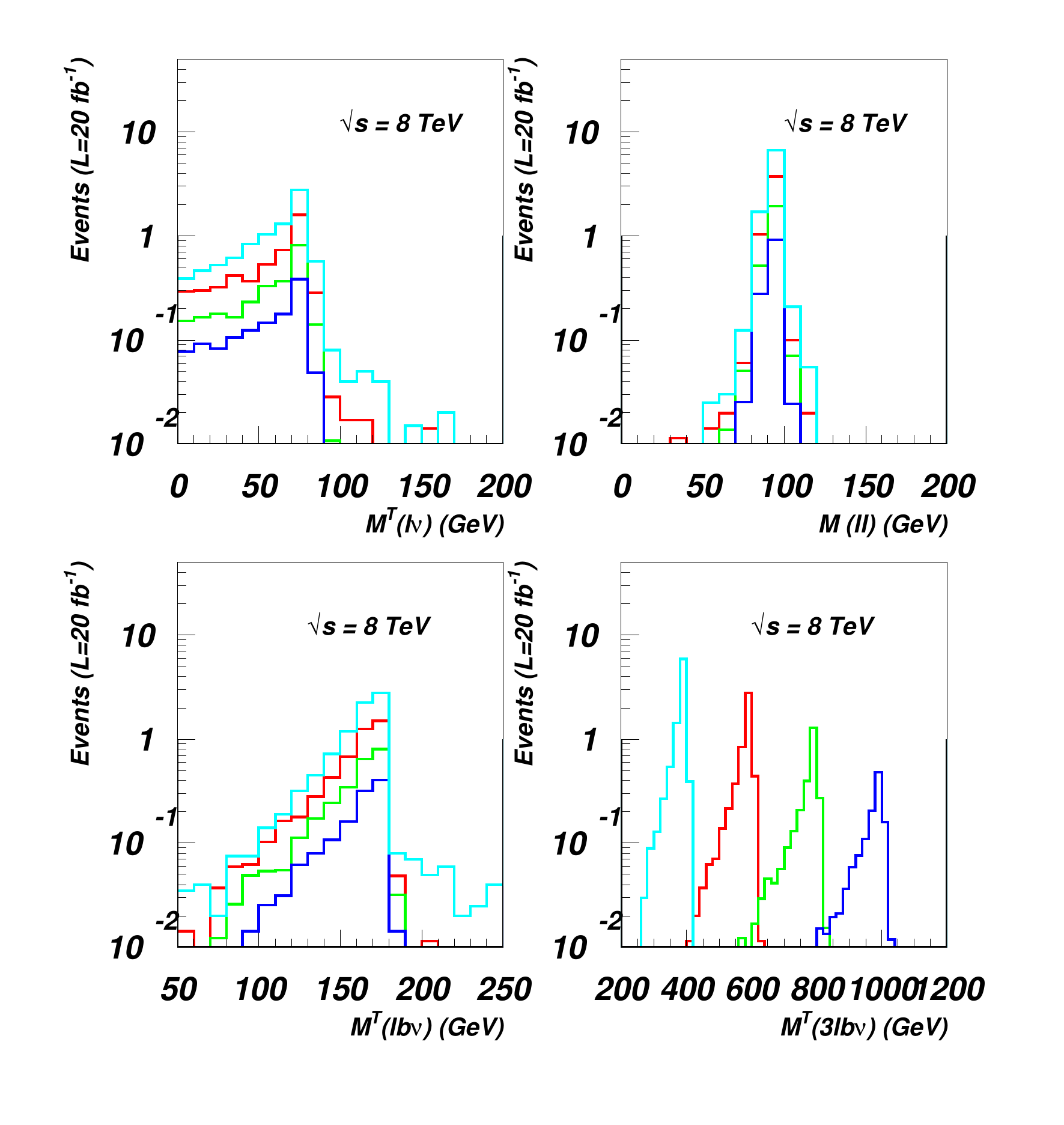}
  \caption{Plots for signal: (top-left) $W$-boson reconstruction: transverse mass of 1 lepton; (top-right) $Z$-boson reconstruction: invariant mass of pair of lepton closest in value to $M_Z$; (bottom-left) top reconstruction: transverse mass of $W$-boson and $b$-quark; (bottom-right) $T'$ reconstruction: transverse mass of $b-$quark and all 3 leptons. \label{VLQ_fig_3l}}
\end{center} \end{figure}

We start by pinning down the properties of the signal at parton level. 
The cut flow for this analysis is as follows: first we apply typical detector acceptances as follows: $p^{\ell}_T > 20$ GeV, $p^j_T > 40$ GeV, $|\eta_{j(\ell)}| < 2.5$, where $\ell = e,\,\mu$ and $\eta_j$ is restricted only to the tracker to allow for $b$-tagging. Further, all final state particles are required to be isolated in a cone of $\Delta R_{jj} = \Delta R_{\ell j} = 0.5$ and $\Delta R_{\ell \ell} = 0.2$, for jets and leptons, respectively. The analysis is carried out in the MadAnalysis package~\cite{Conte:2012fm}.
Then, the pair of leptons that best reconstructs the $Z$ boson is chosen, with the third lepton assigned to the $W$ boson, and further combined with the $b-$quark to reconstruct the top quark. We see in Figure~\ref{VLQ_fig_3l} that this algorithm works very well. A window around the $Z$ mass, $|M(\ell \ell)-91.8$ GeV$| \leq 20$ GeV,is selected. Regarding the $W$ boson, an asymmetric window  $20$ GeV $\leq M_T(\ell \nu) \leq 100$ GeV is selected. Finally, around the top quark, another asymmetric window  $100$ GeV $\leq M_T(b \ell \nu) \leq 190$ GeV is considered. The optimisation of these windows in order to maximise the signal-over-background sensitivity is left for future work.

Cross sections for the process of eq.~(\ref{VLQ_eq_trilep}) for $\sqrt{s}=8$ TeV are too small to leave one with a sufficient amount of events: with the same benchmark~\cite{Buchkremer:2013bha} considered in the previous section, we obtain $\sigma(400~\rm{GeV})=2.5$ fb and $\sigma(600~\rm{GeV})=1.4$ fb.
 We therefore study the sensitivity at $14$ TeV and $\mathcal{L}=20$ fb$^{-1}$ including also the backgrounds described earlier on. At $14$ TeV, cross sections for $pp\to tZj \to 3\ell + X$ ($\ell = e,\,\mu$) are as follows:
\begin{eqnarray*}
T'~(400 \rm{~GeV}): \qquad \sigma = 4.0 \rm{~fb}, &\qquad\qquad& tZj: \qquad \sigma = 15.5 \rm{~fb,}\\
T'~(600 \rm{~GeV}): \qquad \sigma = 2.6 \rm{~fb}, &\qquad\qquad& WZjj: \hspace{0.3cm} \sigma = 246.7\rm{~fb~,}\; \rm{~(with~generation~cuts)}\\
T'~(800 \rm{~GeV}): \qquad \sigma = 1.6 \rm{~fb}, &\qquad\qquad& t\bar{t}\ell\nu:  \qquad  \sigma = 5.7 \rm{~fb,} \\
&\qquad\qquad& t\bar{t}:  \hspace{1.2cm}   \sigma = 25.0 \rm{~pb,} \; (\sigma_{2\ell})
\end{eqnarray*}
For a more realistic analysis, we performed our study at hadron level, after showering (but without account for detector simulation, hence in a ``perfect detector'' scenario). All signal and background samples have been produced with \rm{CalcHEP 3.4}~\cite{Belyaev:2012qa} (a part for the $t\overline{t}$ one done with \rm{MadGraph 5}~\cite{Alwall:2011uj}), and then showered with \rm{Pythia 6.4}~\cite{Sjostrand:2006za}. Jet reconstruction has been performed with FastJet~\cite{Cacciari:2011ma}, employing the anti-$k_T$ algorithm with radius $R=0.5$. Further, $b$-tagging efficiency is set to $70\%$ and light flavour mistagging to $10\%$. In our event selection, we require exactly 3 leptons, at least 2 jets of which exactly one is a $b-$jet, to further reduce $t\overline{t}$ backgrounds and the irreducible $tZj$, where $j$ is mainly a $b-$jet.
Numbers of events for $\mathcal{L}=20$~fb$^{-1}$ and relative efficiencies are given in Table~\ref{VLQ_tab_3leff-14}, for our backgrounds and 3 benchmark masses for the signal.

\begin{table}[h]
\begin{center}
\begin{tabular}{|c||c|c||c|c||c|c|} \hline
$\#$ & $400$ GeV & $\varepsilon (\%)$ & $600$ GeV & $\varepsilon (\%)$ & $800$ GeV & $\varepsilon (\%)$ \\ \hline
$0$ & 79.4 & $-$  & 51.4 & $-$  & 31.5 & $-$  \\
$1$ & 48.4 & 61.0 & 34.6 & 67.4 & 20.8 & 66.0 \\
$2$ & 6.7  & 13.7 &  4.3 & 15.4 &  3.3 & 15.8 \\
$3$ & 5.2  & 78.4 &  4.0 & 74.1 &  2.3 & 70.2 \\ \hline
$4$ & 4.5  & 86.1 &  3.5 & 89.3 &  2.1 & 89.5 \\
\hline
\end{tabular}
\begin{tabular}{|c||c|c||c|c||c|c||c|c|}  \hline
$\#$ & $tZj$ & $\varepsilon (\%)$ & $WZjj$ & $\varepsilon (\%)$ & $t\bar{t}\ell\nu$ & $\varepsilon (\%)$ & $t\bar{t}$ & $\varepsilon (\%)$ \\ \hline
$0$ & 309.2 & $-$  & 4934.0 & $-$  & 114.3 & $-$  & 5$\cdot 10^{5}$ &  $-$ \\
$1$ & 194.5 & 62.9 & 2078.8 & 42.1 &  35.9 & 31.4 &     937.7  &  0.2 \\
$2$ &  23.4 & 12.0 &  57.3  &  2.8 &   4.6 & 12.8 &      69.4  &  7.4 \\
$3$ &  19.9 & 84.8 &  47.5  & 83.0 &   1.2 & 25.8 &      35.9  & 51.7 \\ \hline
$4$ &  17.3 & 87.2 &  14.1  & 29.6 &   0.5 & 40.8 &      19.2  & 53.3 \\
\hline
\end{tabular}
\end{center}
\caption{Events and efficiencies for $20$ fb$^{-1}$ after the application of cuts (efficiency always with respect to previous item) for (top) signal and (bottom) backgrounds. $\#0$: no cuts, $\#1$: only leptonic cuts, $\#2$: only hadronic cuts and $b-$tagging, $\#3$: $W$ and $Z$ reconstruction, $\#4$: top reconstruction. \label{VLQ_tab_3leff-14}}
\end{table}

\begin{figure}[tb] \begin{center}
  \includegraphics[angle=-90,width=0.6\textwidth ]{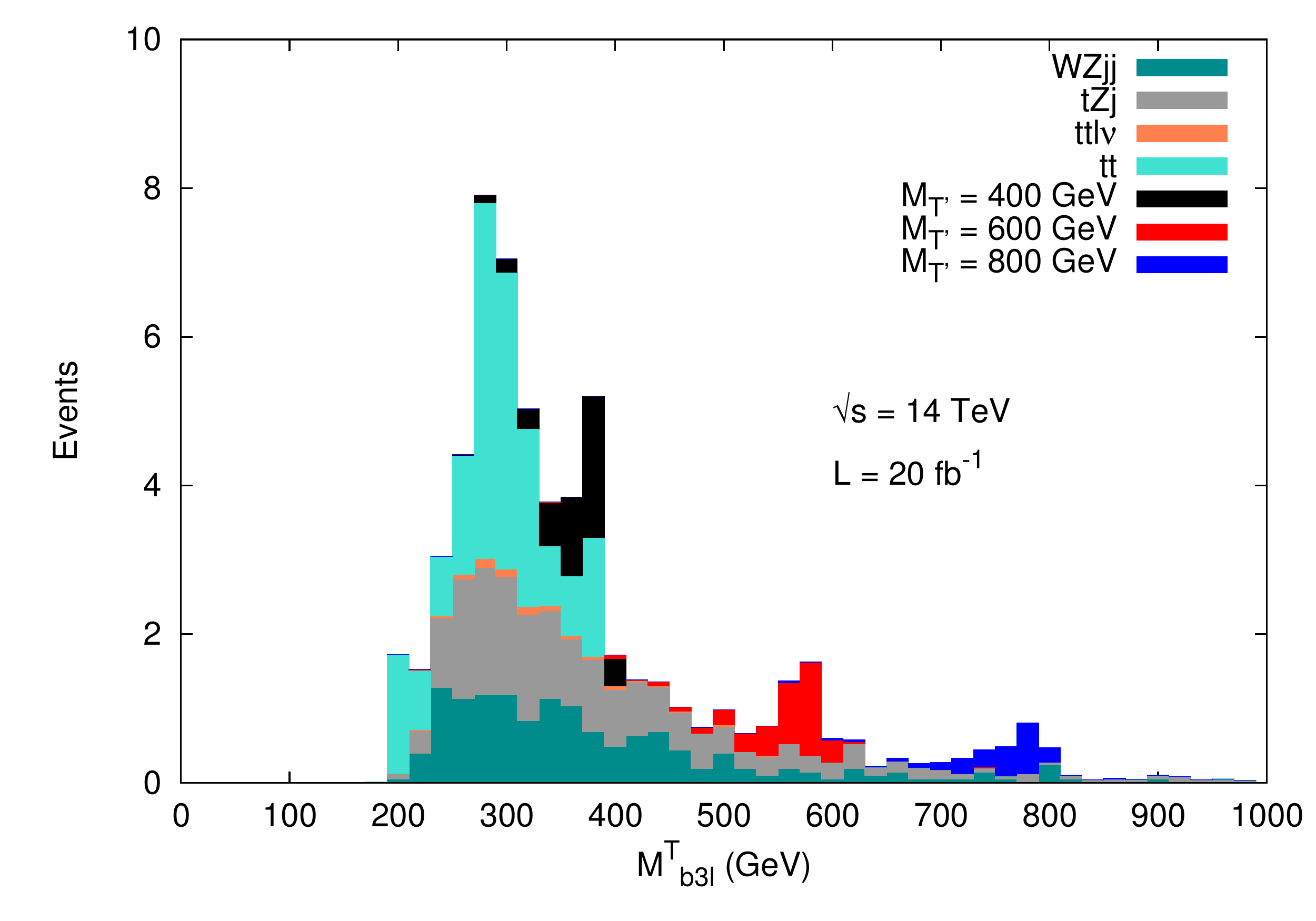}
  \caption{$T'$ reconstruction: transverse mass of $b-$quark and all 3 leptons.  \label{VLQ_fig_tp}}
  \end{center}
\end{figure}

After the last cut (i.e., the top reconstruction), the signal is clearly visible in the transverse mass of the $b-$jet and the 3 leptons, see Figure~\ref{VLQ_fig_tp}. It is clear that the largest background is the irreducible SM $tZj$, while the larger $WZjj$ is effectively suppressed by the $b$-tagging Also $t\overline{t}$ is effectively reduced, and it has not a tail as the others. Notice that they all peak at roughly $300$ GeV. $20$ fb$^{-1}$ seems to be a sufficient integrated luminosity to probe this channel for $T'$ masses up to $\mathcal{O}(600)$ GeV, although a more realistic detector simulation is required to confirm its feasibility.

\section{HIGGS SEARCHES FOR VECTOR-LIKE QUARK PARTNERS\protect\footnote{Contributing authors: T.~Flacke, J.H.~Kim, S.J.~Lee and S.H.~Lim}} \label{Higgs}

Vector-like (VL) quark partner searches in Higgs final states have been performed in the context of top partner searches~\cite{Chatrchyan:2013uxa},
which assumes that dominant production channel for the top partner is through QCD pair production. The LHC bound for a  top partner with 100$\%$ branching fraction (BR) into top plus Higgs is found to be $M>706$ GeV~\cite{Chatrchyan:2013uxa}. On the contrary, a similar search for a VL bottom quark partner~\cite{CMS-PAS-B2G-12-019} does not yield bounds if  100$\%$ BR into bottom plus Higgs is assumed, and bounds for light quark family partners have mostly been obtained for partner decays into gauge bosons and jets~\cite{Delaunay:2013pwa}.\footnote{For prospects of VL quark searches in Higgs channels {\it cf. e.g.} Refs.~\cite{Vignaroli:2012sf,Vignaroli:2012nf,Atre:2013ap}.} Only Ref.~\cite{Flacke:2013fya} established bounds on VL light quark partners in the Higgs channel from current LHC data, where the standard model Higgs searches from $H\rightarrow\gamma\gamma$ channel have been used. In this analysis, we generalize the study of Ref.~\cite{Flacke:2013fya}, which was performed in the context of minimal composite Higgs models~\cite{Agashe:2004rs}, into  simplified models~\cite{Buchkremer:2013bha}.

\subsection{Vector-Like quark searches in Higgs final states -- Models}
In the following, we discuss searches for vector-like quark partners in Higgs final states.
As we want to be more general than the simplified scenarios described in the Introduction, we will use in this study the Lagrangian proposed in Ref.~\cite{Buchkremer:2013bha}, which describes single production and decay of a vector-like up-type quark partner \footnote{We work in unitary gauge. And we focus on up-type partners here. Concerning the discussion of Higgs interactions in this section, down-type partners can be treated completely analogously.}: 
\begin{eqnarray}
\mathcal{L}_{T single} &=&  \kappa_W V_{L/R}^{4i} \frac{g}{\sqrt{2}}\; [\bar{T}_{L/R} W_\mu^+ \gamma^\mu d^i_{L/R} ]  + \kappa_Z V_{L/R}^{4i} \frac{g}{2 c_W} \; [\bar{T}_{L/R} Z_\mu \gamma^\mu u^i_{L/R} ]  \nonumber \\
&-& \kappa_H V_{L/R}^{4i} \frac{M}{v}\; [\bar{T}_{R/L} h u^i_{L/R} ] + h.c. \, , \label{eq:topL}
\end{eqnarray}
where $i$ is a family index. For simplicity, we focus on partners of purely the up-, charm-, or top-quark in what follows, {\it i.e.} we choose to keep partners of only one family in the following analysis.
The Lagrangian in Eq.~(\ref{eq:topL}) can be identified with $\mathcal{L}_T$ in the introduction by renaming $V_{L}^{4u} = \sqrt{\frac{R_L}{1+R_L}}$ and $V_{L}^{4t} = \sqrt{\frac{1}{1+R_L}}$, and setting $V_R^{4i} = 0$ and $\kappa_W = \kappa_Z = \kappa_H = g_\ast$ (thus neglecting the $m_t/M$ term): large differences compared to this case can be achieved in models with large mixing between VL quarks. 
The ``mixing'' couplings described by the Lagrangian~(\ref{eq:topL}) arise when expressing the gauge and Higgs interactions of the SM-like quarks and the up-type partner in the mass eigenbasis. Mixing of SM-like quarks and the quark partner in the left-handed quark sector is strongly constrained from electroweak precision tests ({\it cf.} {\it e.g.}  Refs.~\cite{Delaunay:2010dw,Redi:2011zi}), while in the right-handed sector substantial mixing is possible~\cite{Agashe:2006at}. To capture phenomenologically viable models, we study two sample scenarios. 

\bigskip

In Model I, we assume $V_{R}^{4i}\gg V_{L}^{4i}$ and $\kappa_H\gg \kappa_{W,Z}$.\footnote{An explicit realization of this parameter choice has been discussed in Ref.~\cite{Flacke:2013fya} in terms of a partially composite right-handed quark model.} The full effective Lagrangian of the model is given by
\begin{equation}
\mathcal{L}_{\rm eff}=\mathcal{L}_{\rm SM}+\bar{T}\left(i\partial_\mu+ e \frac{2}{3} A_\mu -g\frac{2}{3}\frac{s^2_w}{c_w} Z_\mu+ g_3 G_\mu  \right)\gamma^\mu T - M\bar{T}T
-\left[\lambda^{\rm eff, i}_{\rm mix}h\bar{T}_{L}u^i_{R}+\mbox{h.c.}\right]. \label{Lpceff}
\end{equation}
Matching the last term to the simplified model parameters in Eq.~(\ref{eq:topL}) yields $\lambda^{\rm eff, i}_{\rm mix}=\kappa_H V_{R}^{4i} M/v$. The effective coupling $\lambda^{\rm eff, i}_{\rm mix}$ can be sizable without being in conflict with electroweak precision constraints. This has important consequences for the $T$ production channels discussed in the next section.  Concerning the decay, $T\rightarrow h u^i$ is the only allowed decay channel, such that this sample model can only be tested in Higgs final states at the LHC.

\bigskip

For Model II,  we assume $V_{R}^{4i}\simeq V_{L}^{4i}$.\footnote{This model setup arises for example in the fully composite quark model discussed in Ref.~\cite{Flacke:2013fya}.} The effective Lagrangian of this model is given by
\begin{eqnarray}
\mathcal{L}_{\rm eff}&=&\mathcal{L}_{\rm SM}+\bar{T}\left(i\partial_\mu+ e \frac{2}{3} A_\mu -g\frac{2}{3}\frac{s^2_w}{c_w}Z_\mu+ g_3 G_\mu  \right)\gamma^\mu T - M\bar{T} T \nonumber\\
&&-\left[\lambda^{\rm eff, i}_{\rm mix}\frac{m_Z}{M_1}\bar{T}_{L}Z_\mu \gamma^\mu u^i_{L}+\sqrt{2}\lambda^{\rm eff, i}_{\rm mix}\frac{m_W}{M_1}\bar{T}_{L}W_\mu \gamma^\mu d^i_{L}+\lambda^{\rm eff}_{\rm mix}h\bar{T}_{L}u^i_{R} +\mbox{h.c.}\right], \label{Lfceff}
\end{eqnarray}
where matching with the simplified model Lagrangian Eq.~(\ref{eq:topL}) yields $\lambda^{\rm eff, i}_{\rm mix}=\kappa_H V_{R}^{4i} M/v=\kappa_W V_{L}^{4i} M/v=\kappa_Z V_{L}^{4i} M/v $. In this sample model, electroweak precision constraints require the couplings $\lambda^{\rm eff, i}_{\rm mix}$ to be small. Therefore, single-production of the $T$ is suppressed. The branching fractions of the $T$ decays into $W,Z$, or Higgs and a SM-like quark are approximately $50\%$, $25\%$ and $25\%$~\cite{Flacke:2013fya}. 

\subsection{BSM Production channels with Higgs final states}

\begin{figure}[t]
\begin{center}
\includegraphics[scale=.5]{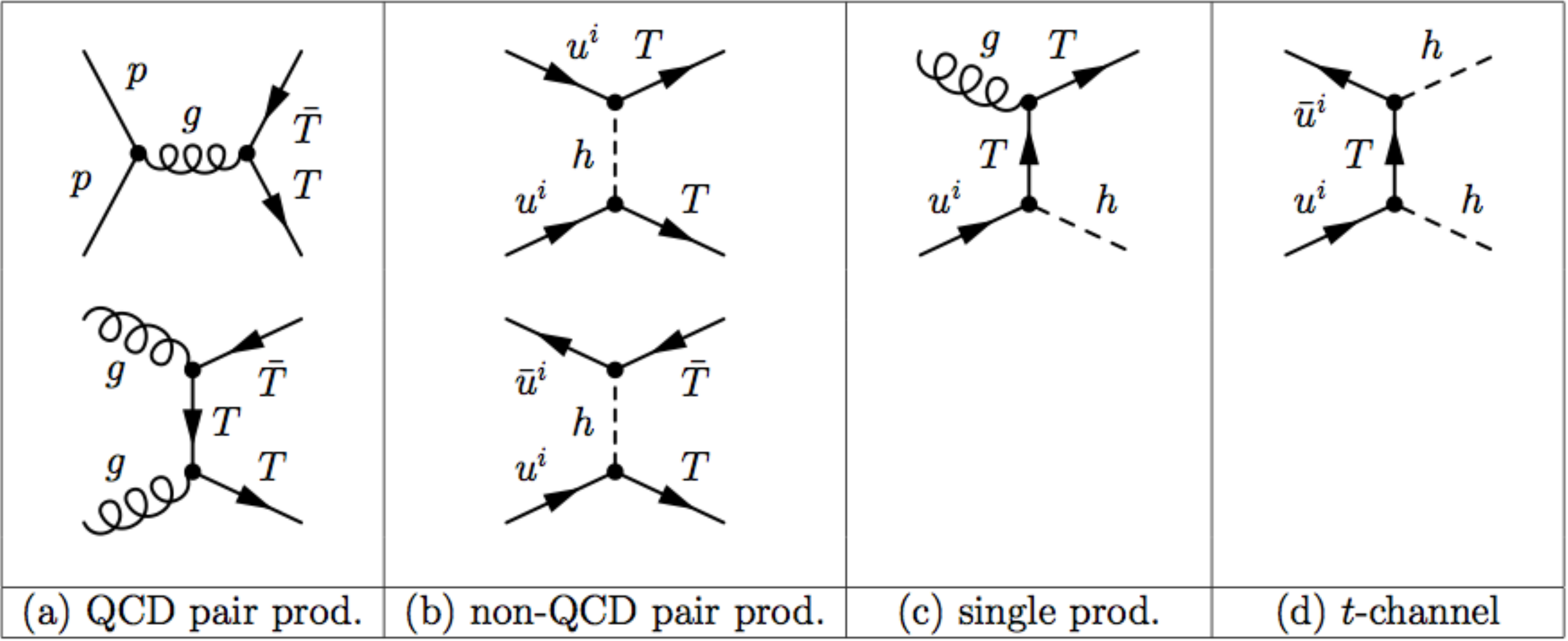} 
\end{center}
\caption{Main channels leading to BSM Higgs final states from Model I and II. The label $p$ in the first diagram in (a) stands for any parton of the proton: $g$, $q$ and $\bar{q}$.}
\label{fig:prod}
\end{figure}
 
\begin{figure}
\begin{center}
\begin{tabular}{cc}
\includegraphics[scale=0.8]{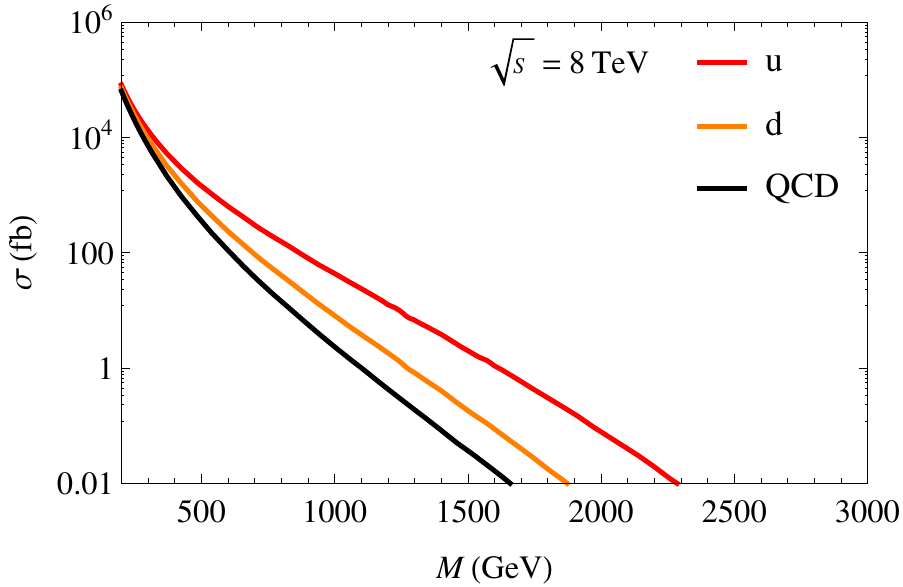} &
\includegraphics[scale=0.8]{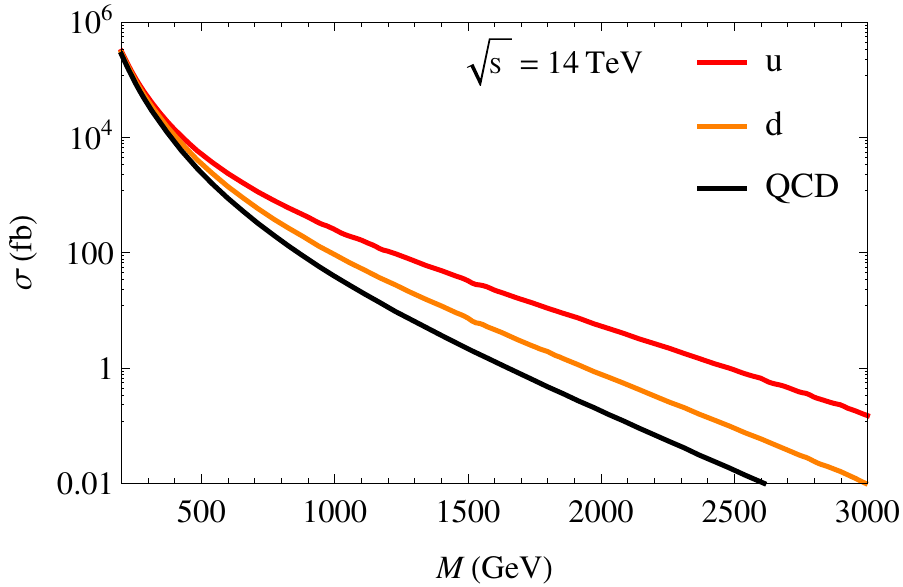}
\end{tabular}
\end{center}
\caption{Production cross section for a pair of vector-like quark partners in Model I  as a function of the partners' mass $M$ for LHC at $8 \mathrm{ TeV}$ (left) and $14 \mathrm{ TeV}$ (right). 
The first two lines from the top correspond to the pair production cross section with $\kappa_H V^{4i}_R=v/M$ for partners of the up (red) and down (orange),
while the third line (black) denotes the QCD pair production cross section. The non-QCD pair production cross sections for partners of the $s,c$ and $b$ quarks are PDF suppressed. Thus, the pair production cross section for these quark partners is to a good approximation given by the QCD pair production cross section.
}
\label{fig:prodXsec}
\end{figure}

\begin{figure}
\begin{center}
\begin{tabular}{cc}
\includegraphics[scale=0.8]{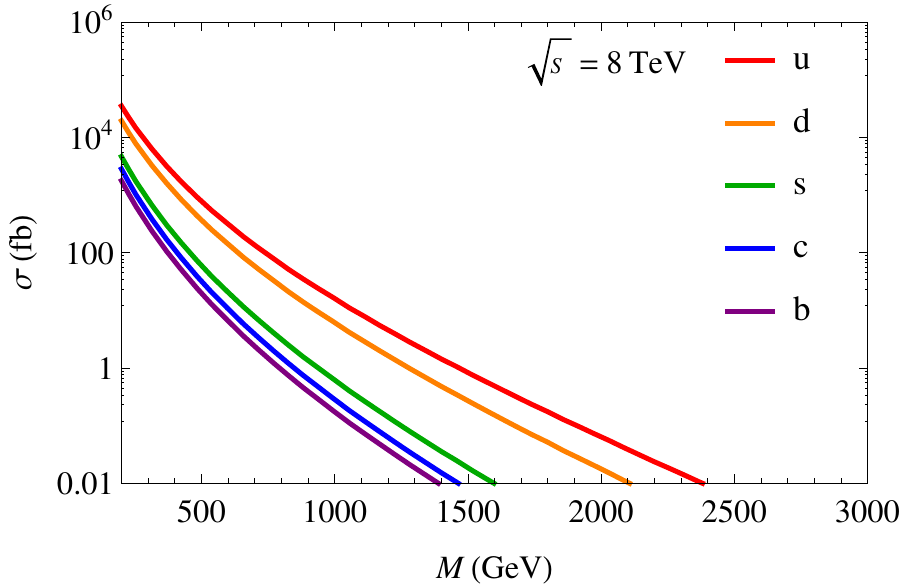} &
\includegraphics[scale=0.8]{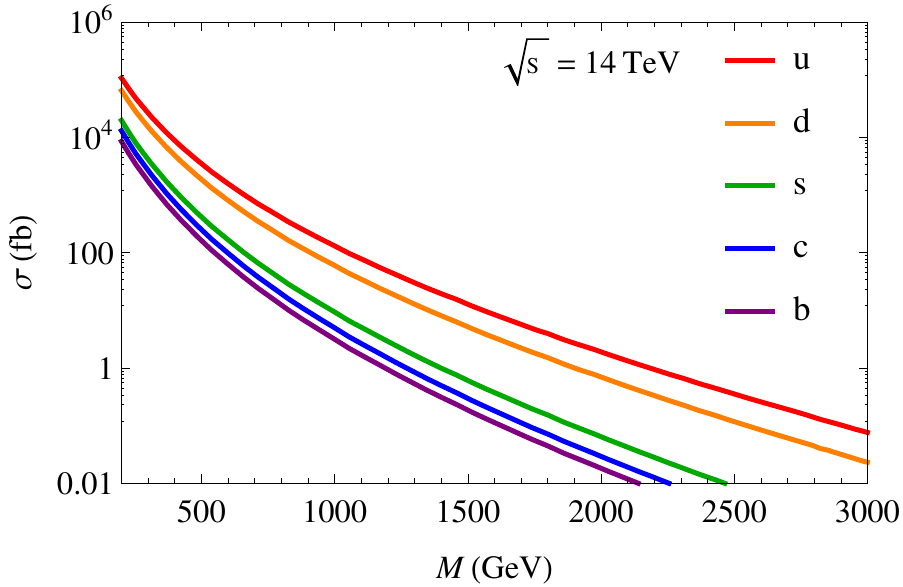}
\end{tabular}
\end{center}
\caption{Production cross section for a single $T$  as a function of its mass $M$ for LHC at $8 \mathrm{TeV}$ (left) and $14 \mathrm{TeV}$ (right).  Lines denote (from right to left): Single production cross section with $\kappa_H V^{4i}_R=v/M$ for partners of the $u,d,s,c,b$ quark.}
\label{fig:prodXsecsing}
\end{figure}

\begin{figure}
\begin{center}
\begin{tabular}{cc}
\includegraphics[scale=0.8]{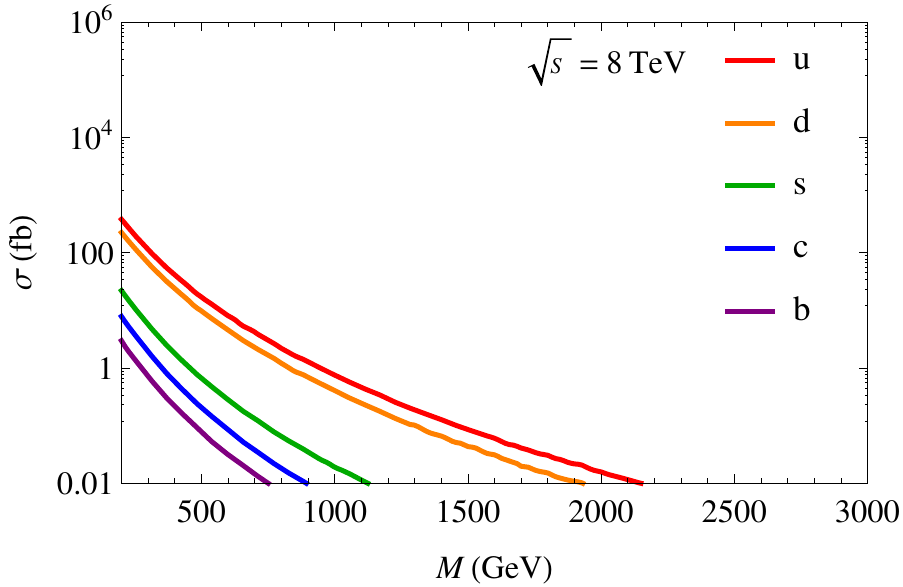} &
\includegraphics[scale=0.8]{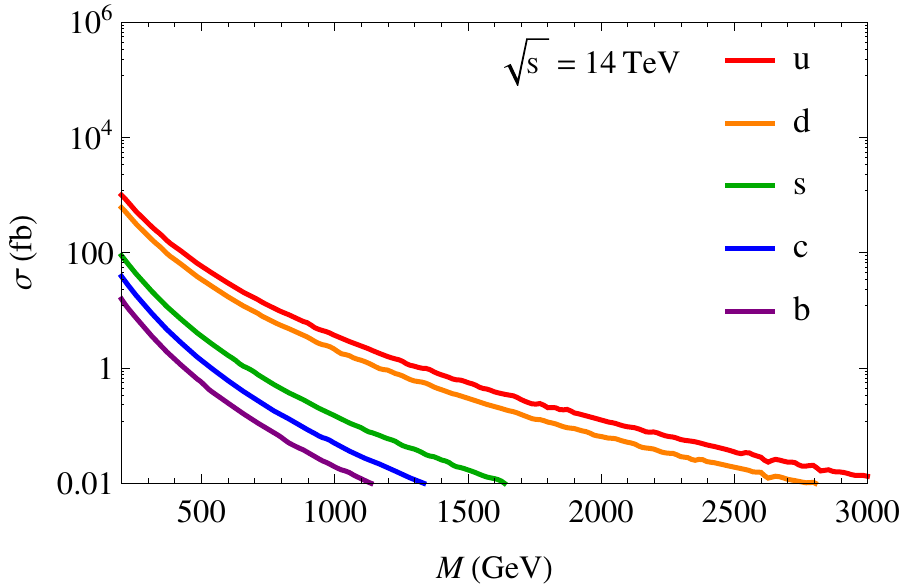}
\end{tabular}
\end{center}
\caption{Production cross section for two Higgses via $t$-channel exchange of a $T$  quark for LHC at $8 \mathrm{TeV}$ (left) and $14\mathrm{TeV}$ (right).  Lines denote (from right to left): Cross section with $\kappa_H V^{4i}_R=v/M$ for the exchange of $u,d,s,c,b$ quark partners.}
\label{fig:prodXseczero}
\end{figure}

The dominant  channels which lead to BSM final state Higgses from our sample models I and II are shown in Fig.~\ref{fig:prod}. For small $\lambda^{\rm eff, i}_{\rm mix}$ (which is required in Model II but not in Model I), BSM production of Higgses arises mainly through QCD production of a $T$ pair shown in panel (a) and their decay into a Higgs and an SM-like quark. The remaining channels (non-QCD $T$ pair production in panel (b), single $T$ production in panel (c), and di-Higgs production in panel (d)) only play a role when $\lambda^{\rm eff, i}_{\rm mix}$ is not suppressed. For vector-quark partners which mainly interact with the up-quark, these channels considerably enhance the $T$ pair production cross section and also yield sizable single $T$ production and BSM di-Higgs production, while for second or third family quark partners, these channels are PDF suppressed. Figures~\ref{fig:prodXsec}-\ref{fig:prodXseczero} show the respective cross-sections for $T$ pair-, and single production and direct BSM di-Higgs production as a function of the partner quark mass $M$ for LHC at $\sqrt{s} = 8 \mbox{ TeV}$ and $14 \mbox{ TeV}$ for a reference value of $\lambda^{\rm eff, i}_{\rm mix}=1$, which corresponds to $\kappa_H V^{4i}_R=v/M$ in terms of the simplified model parameters. We also included the production cross-sections for down-type partner quarks which can be implemented analogously to the up-type partners.

The processes shown in Fig.~\ref{fig:prod} yield $T$ pairs (for both, Model I and II), as well as $Th$, or a Higgs pair for Model I. In the case of Model I, the $T$ decays into $h u^i$. The main BSM signature of this model is therefore a di-Higgs with zero, one, or two high-$p_T$ $u^i$. For Model II, QCD $T$ pair production dominates, and the $T$ decays into $W$, $Z$ or $h$ with an associated quark (top/bottom or jet, depending on which $V^{4i}_R$ dominates). Therefore for Model II, final states with gauge bosons are available\footnote{For a top-partner, the strongest bound  from such signatures is $M > 696 \mbox{ GeV}$~\cite{Chatrchyan:2013uxa}. For a bottom-partner the strongest constraint is $M > 700 \mbox{ GeV}$~\cite{CMS-PAS-B2G-12-019}. For light quark partners, no bounds are available yet.} while final states with a Higgsboson are reduced due the BRs.

\subsection{Results for vector-like partners of light quarks} 

The di-Higgs signature provides a very promising search channel for Model I and (to a lesser extend) for Model II, but ATLAS and CMS did not release results in these channels yet. In the remainder of this section we therefore focus on obtaining bounds on Models I and II from the currently available Standard Model Higgs searches.
To distinguish Higgses from SM and the above discussed BSM production channels, notice that Higgses arising from heavy $T$ decays result in a high $p_T$ Higgs and a high $p_T$ quark which are untypical for SM Higgs production. Measurements of $p_T$ distributions in Higgs decays can therefore yield a good discriminator.

In Ref.~\cite{ATLAS-CONF-2013-072}, the ATLAS collaboration presented results on differential cross sections of the Higgs in the $h\rightarrow\gamma\gamma$ channel. In particular, Ref.~\cite{ATLAS-CONF-2013-072} studies the $p^{\gamma\gamma}_T$, $N_{\mathrm{jets}}$, and the highest $p^{\mathrm{jet}}_T$ distributions which are in good agreement with the SM predictions.
We simulate the BSM contribution to these distributions which arise in Model I and II from the processes shown in Fig.~\ref{fig:prod}. Bounds on the vector-quark partner parameter space are obtained by comparing the BSM contributions in the respective $p^{\gamma\gamma}_T$, $N_{\mathrm{jets}}$, and highest $p^{\mathrm{jet}}_T$ bins to the measured bounds determined in Ref.~\cite{ATLAS-CONF-2013-072}. The details of the simulation and data evaluation can be found in Ref.~\cite{Flacke:2013fya}.

\begin{figure}[t]
\begin{center}
\begin{tabular}{cc}
\includegraphics[scale=0.75]{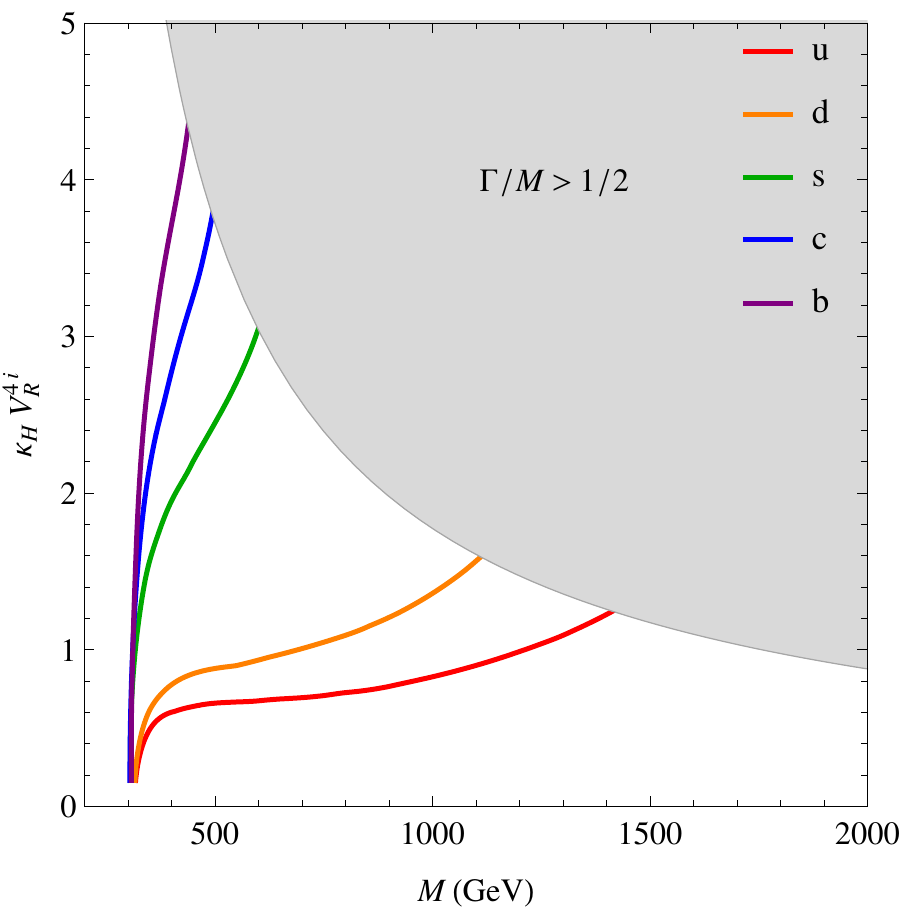} &
\includegraphics[scale=0.75]{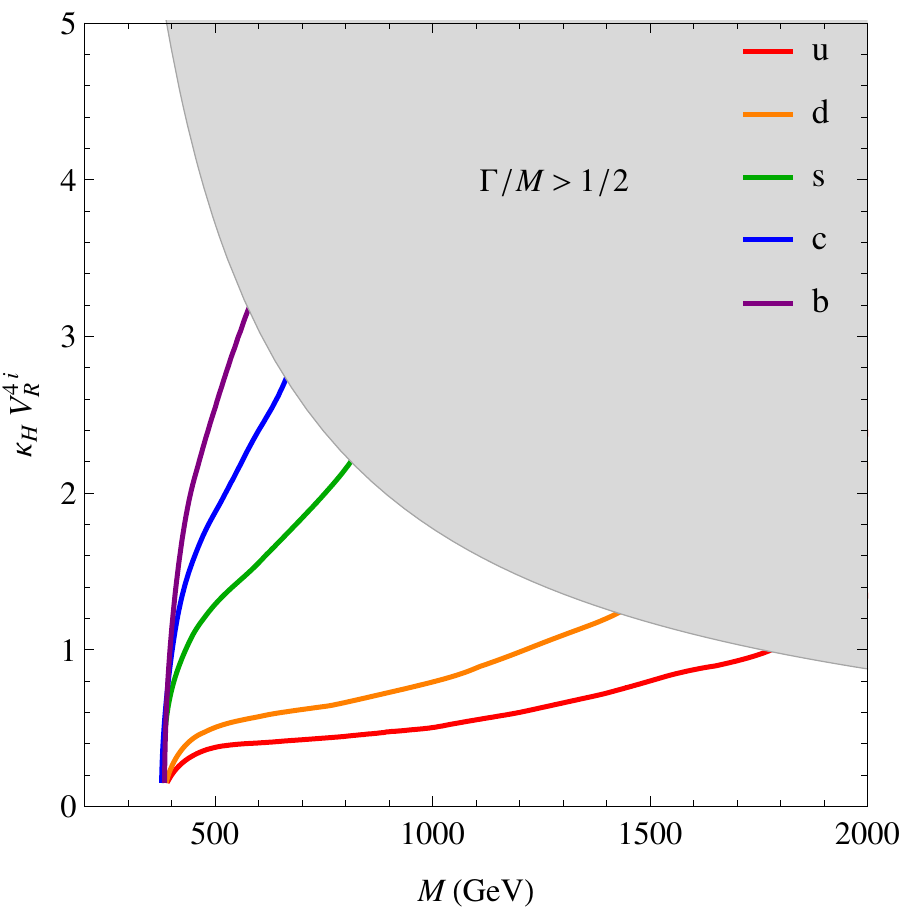}
\end{tabular}
\end{center}
\caption{The exclusion plots for partners of the $u,d,s,c$ and $b$ quark (curves from right to left) within sample Model I.
Left: 95\% CL exclusion limits including the signal bins of Ref.~\cite{ATLAS-CONF-2013-072}. Right: Would-be 95\% CL exclusion limits when both signal and overflow bin data of Ref.~\cite{ATLAS-CONF-2013-072} is included in the analysis. 
The parameter region to the top-left region from the respective curve is excluded at 95\% CL. For reference, the gray region in each of the plots shows $\Gamma / M > 1/2$ and the coupling above which the narrow-width-approximation does not apply anymore. Ref.~\cite{Flacke:2013fya} did not search through this region.}
\label{fig:pcresult}
\end{figure}

Figure~\ref{fig:pcresult} shows the resulting bounds on the vector-quark partner of a $u,b,s,c$ and $b$ quark. On the left, we show the $95\%$ CL bounds which arise when only including the signal bins of Ref.~\cite{ATLAS-CONF-2013-072}.\footnote{Ref.~\cite{ATLAS-CONF-2013-072} chose as their highest signal bins for their $p_T$ distributions $100 \mbox{ GeV}<p_T^{\gamma\gamma}<200 \mbox{ GeV}$ and $100 \mbox{ GeV}<p_T^{j1}<140 \mbox{ GeV}$. Events with higher $p_T$ are collected in $p_T^{\gamma\gamma}$ and $p_T^{j1}$ overflow bins.} On the right, we show the bounds which would be obtained when including the overflow bins into the analysis. The signal and overflow bins are treated on a different footing in Ref.~\cite{ATLAS-CONF-2013-072}, and the left figure therefore represents the (conservative) constraints obtained from our analysis. The right plot indicates how much these bounds could be increased by a dedicated search for high $p_T^{\gamma\gamma}$ (which yields the dominant improvement) and  high $p_T^{j1}$ events. For reference, in both plots, the grey-shaded region corresponds to the parameter space in which the $T$ width is larger than $M/2$ such that the narrow width approximation does not apply. Our analysis is not relying on the narrow width approximation. This information is mainly provided because it is relevant for future searches which aim to reconstruct the invariant mass of the $T$.

From Fig.~\ref{fig:pcresult} we obtain a quark flavor and $\kappa_H$- independent bound of $M > 310$ GeV at 95\% CL from the signal bin analysis ($M > 385$ GeV when including the overflow bins). These bounds arise solely from QCD pair produced $T$'s ({\it c.f. } Fig.~\ref{fig:prodXsec}) which subsequently decay into a Higgs and a quark. The flavor and $\kappa_H$ dependent increase of the bound arises due to the increase of the $T$ pair-, and single production ({\it c.f. } Figs.~\ref{fig:prodXsec} and \ref{fig:prodXsecsing}) while the increase of the direct di-Higgs production ({\it c.f. } Fig.~\ref{fig:prodXseczero}) plays a minor role. For Model I, in which the vector-like quark partner only decays into a Higgs and a quark, the constraints presented above are the only currently available bounds for light quark partners. Only for top quark partners a stronger bound of $M> 706 \mbox{ GeV} $is available from a dedicated CMS search for top partners in Ref.\cite{Chatrchyan:2013uxa}. 

\bigskip

The analogous analysis for Model II yields a flavor and $\kappa_H$ independent bound of $M>212$ GeV at 95\% CL from the signal bin analysis ($M > 240$ GeV when including the overflow bins). The constraint is weaker than for Model I because only $\sim 25\%$ of the $T$ decays yield a final state Higgs. Furthermore, there is no $\kappa_H$ or flavor dependent increased bound because in Model II QCD production of $T$ pairs dominates. For top- and bottom partners within Model II, stronger bounds from searches for pair produced quark partners with an electroweak gauge boson and a $b$ quark in the final state exist ($M> 696 \mbox{ GeV}$ for a $t$-partner~\cite{Chatrchyan:2013uxa} and $M > 700 \mbox{ GeV}$ for a $b$-partner~\cite{CMS-PAS-B2G-12-019}), while for $u,d,s$ and $c$ partners, the constraint presented above represents the strongest bound within the Model II setup.

\section*{CONCLUSIONS}

The search for vector-like partners of the top quark is entering a crucial era, as the bounds from the searches at the LHC are approaching the TeV scale.
For this reason, it is useful to take a model-independent look at the searches and allow for all possible channels and couplings.
In this report, we propose a simple parametrisation of the lagrangian with 3 parameters, which can approximate with good accuracy the phenomenology of many realistic models.
A crucial novelty is the fact that we allow for couplings to light quarks, and not only third generation.
The coupling to light quarks generates new signatures, some of which can contribute to searches designed for decays into the third generation.
An example of this is given by final states with three or four leptons.
Sizable couplings to the light quarks also allows for larger single-production processes in association with a light jet, and interesting final states can be obtained when the vector-like quark further decays into a top.
Two exploratory studies of this case have been presented: in the channel $p p \to T j \to H t j$, the final state can be reconstructed in the fully hadronic decays of the Higgs and top and a bump-hunt strategy in the 5-jet invariant mass may allow for a distinction of the signal over backgrounds; another promising channel is generated by $p p \to T j \to Z t j$, where both the $Z$ and top decays leptonically.
In the latter case, the suppression due to the leptonic branching ratios requires the larger energy of the LHC in order to have a detectable number of events.
Finally, we presented a complete overview of the final states with a Higgs boson, both in the case of a generic model with couplings to light quarks, and in a specific scenario inspired by composite Higgs models.
The studies presented in this report finally show that, although the massive effort by the experimental collaborations ATLAS and CMS have provided us with plenty of channels to constraint models with vector-like quarks, the most general scenario still offers new unexplored and promising channels at the LHC.

\section*{ACKNOWLEDGEMENTS}

LB and JRA would like to thank Eric Conte and Jeremy Andrea for their support with Madanalysis and insightful comments.
LB has received partial support from the Theorie-LHC France initiative of the CNRS/IN2P3 and by the French ANR 12 JS05 002 01 BATS@LHC, and from the Deutsche Forschungsgemeinschaft through the Research Training Group grant GRK\,1102 \textit{Physics at Hadron Accelerators} and by the Bundesministerium f\"ur Bildung und Forschung within the F\"orderschwerpunkt.
GC and AD acknowledge the Labex-LIO (Lyon Institute of Origins) under grant ANR-10-LABX-66 and the Theorie-LHC France initiative of the CNRS/IN2P3 for partial support. AD is partially supported by Institut Universitaire de France. 




%% file: compressed/compressed.tex
\def\be{\begin{equation}}
\def\ee{\end{equation}}
\newcommand{\Herwig}{\small \textsf{Herwig++} \normalsize}
\newcommand{\fastjet}{\textsf{FastJet}}
\renewcommand{\d}{\mathrm{d}}
\newcommand{\specialcell}[2][c]{%
  \small \begin{tabular}[#1]{@{}c@{}}#2\end{tabular} \normalsize}

\newcommand{\todo}[1]{\textbf{(todo: #1)}\marginpar{\rule{10mm}{3mm}}}


\chapter{Cornering Compressed Supersymmetric Spectra with Monotops}

{\it B.~Fuks, P. Richardson, A.~Wilcock}


\begin{abstract}
We investigate the sensitivity of the Large Hadron Collider to compressed 
supersymmetric scenarios in which one of the possible experimental signals 
consists of a single top quark produced in association with missing transverse 
energy, also referred to as monotop. We perform our study using Monte Carlo 
simulations of $10~\mathrm{fb}^{-1}$ of collisions expected to occur at a 
center-of-mass energy of 14~TeV.  Focusing on leptonically decaying monotop 
states, we present an analysis strategy sensitive to regions of supersymmetric 
parameter space featuring small sparticle mass splittings and illustrate
its strengths in the context of a particular set of benchmark points.

\end{abstract}

\section{INTRODUCTION}
While the Standard Model (SM) of particle physics has proven enormously successful 
in predicting most high-energy physics data, it exhibits a set of conceptual problems 
and is therefore believed to be the low-energy limit of a more fundamental theory. 
This has led to the development of a plethora of new physics models, the most popular 
and well studied of which being weak scale supersymmetry (SUSY)~\cite{Nilles:1983ge,
Haber:1984rc}.  The absence of any new physics hints at the Large Hadron Collider 
(LHC) means that bounds on the superpartner masses are being pushed to higher scales. Most 
of these constraints can however be evaded for specific benchmark scenarios where the 
low-energy part of the SUSY mass spectrum is compressed.  In compressed scenarios, 
superparticles decay into missing energy carried by the lightest sparticle and soft 
leptons and jets.  Consequently, the transverse momenta of the decay products fall below the typical 
trigger thresholds of the LHC experiments and the smaller expected amount of missing 
transverse energy makes kinematical quantities traditionally employed to reduce the SM 
background less effective. These observations have motivated several studies trying to 
constrain compressed models by non-standard means, such as benefiting from monojet or 
monophoton signatures~\cite{Alves:2010za,LeCompte:2011cn,Dreiner:2012gx,
Bhattacherjee:2012mz,Dreiner:2012sh,Dutta:2013gga}.

In our approach, we focus on the production of monotop systems that are defined as 
states comprised of a single top quark and missing transverse energy and that are 
expected to be easily observable at the LHC for a large range of new physics masses 
and coupling strengths~\cite{Andrea:2011ws,Wang:2011uxa,Alvarez:2013jqa,Agram:2013wda}. 
In the framework of simplified compressed SUSY scenarios where the electroweak 
superpartners are neglected (with the exception of the lightest neutralino), a 
monotop signature is expected to arise from the production of the three-body final 
state of a gluino, top squark and top quark. In this case, both superpartners give 
rise to a small amount of missing energy produced together with soft objects. 
Restricting ourselves to the case of a leptonically decaying top quark, we illustrate 
how to benefit from the presence of the latter to get sensitivity to the initially 
produced sparticles, and adopting a typical monotop selection strategy, we show that
10~fb$^{-1}$ of LHC collisions at a center-of-mass energy of 14~TeV should be 
sufficient for observing hints of new physics in the context of compressed SUSY scenarios.

The outline of this contribution is as follows: In 
Section~\ref{sec:compressed_simulation}, we describe the technical setup for the 
Monte Carlo simulations of both the new physics signal and the relevant sources of 
SM background.  Our analysis strategy allowing for extracting a monotop signal from 
the background is detailed in Section~\ref{sec:compressed_selection} and the results 
are presented in Section~\ref{sec:compressed_results}.  Our conclusions are given 
in Section~\ref{sec:compressed_conclusions}.

\section{SIMULATION} \label{sec:compressed_simulation}

\renewcommand{\arraystretch}{1.2}
\begin{table}[t]
\centering
\begin{tabular}{c||c||c|c}
  Process & Simulation details & $\sigma^{\rm total} [\mathrm{pb}]$ & $N^{\rm surviving}_{\mathrm{event}}$ \\ \hline \hline
  $W ( \rightarrow l \nu ) $ + light-jets &\specialcell{$W$-boson production simulated at NLO and\\ matched to LO production of the $W$-boson\\ with 1 or 2 extra jets using {\sc Sherpa}~2.0~\cite{Gleisberg:2003xi}.}& $67453 $ &  $ \approx 0 $ \\ \hline
  $\gamma^* / Z ( \rightarrow l \bar{l} / \nu \bar{\nu} ) $ + jets  & \specialcell{As above and requiring $m_{ll}, m_{\nu \nu} > 10 \GeV$.} & $38990  $ & $ \approx 0  $ \\ \hline
  $t\bar{t}$ &\specialcell{Parton-level hard events simulated at NLO using\\ the {\sc PowhegBox}~\cite{Frixione:2007nw} and matched to \Herwig \\ for parton showering and hadronization.}  & $781 $ &  $  1387$\\ \hline
  Single top [$t$-channel] & \specialcell{As above~\cite{Alioli:2009je}.}& $244.3  $  &  $ 1.2  $\\ \hline
  Single top [$s$-channel] & \specialcell{As above~\cite{Re:2010bp}.}& $10.4  $  &  $ 0.1 $\\ \hline
  $tW$ production & \specialcell{As above, suppressing doubly-resonant\\ diagrams following the prescription of Ref.~\cite{Frixione:2008yi}}& $77.1  $  &  $ 73 $ \\ \hline
  $Wb\bar{b}$ with $W \rightarrow l \nu$  & \specialcell{Parton-level hard events, with $W \rightarrow l \nu $, simulated\\ at LO with {\sc MadGraph}~5  and matched to \\\Herwig \small for parton showering and hadronization. \normalsize} & $122.1$  &  $ \approx 0$\\ \hline
  Diboson  &\specialcell{Simulated at NLO using the built-in\\ {\sc Powheg} implementation in \Herwig~\cite{Hamilton:2010mb}.} & $157.7$ &  $ \approx 0$  \\  \hline
  $(m_{\tilde{t}_1},m_{\tilde{\chi}_1^0}) = (200, 190) \GeV $ &\specialcell{Details given in text.} & $ 0.8 $ & $272 $ \\ 
\end{tabular}
\caption{Cross sections and simulation details for the background processes and 
an example signal scenario. In all cases, $l=e, \mu, \tau$ and $\nu$ denotes any 
neutrino. Also shown are the number of events, $N_{\mathrm{events}}$, surviving all 
selection criteria described in Section~\ref{sec:compressed_selection}. Results 
correspond to 10~fb$^{-1}$ of LHC collisions at a center-of-mass energy of 14~TeV. }
\label{tab:compressed_cross}
\end{table}
\renewcommand{\arraystretch}{1.0}

Event generation for the hard scattering signal process, including the subsequent 
decay of the top quark $t \rightarrow b\, W \rightarrow b \, l \nu_l$ where 
$l=e, \mu, \tau$ has been simulated using the {\sc MadGraph}~5~\cite{Alwall:2011uj} 
program with the parton density function (PDF) set CTEQ6L1~\cite{Pumplin:2002vw}. 
Additionally, the decays of the $\tilde{t}_1$ and $\tilde{g}$ through the modes 
$\tilde{t}_1 \rightarrow c\, \tilde{\chi}_1^0 $ and 
$\tilde{g} \rightarrow q \, \bar{q} \, \tilde{\chi}_1^0$ have been simulated with 
\Herwig 2.7~\cite{Bahr:2008pv,Bellm:2013lba}.  Finally, parton-level hard event 
samples have been matched with the parton shower and hadronization infrastructure 
provided by \Herwig.  In this study we consider scenarios with maximal stop mixing 
and with purely bino $\tilde{\chi}_1^0$.  The corresponding 
signal cross section for an example compressed spectra scenario with 
$m_{\tilde{t}_1}=m_{\tilde{g}}=200 \GeV$ and $m_{\tilde{\chi}_1^0}=190 \GeV$ can be found in
Table~\ref{tab:compressed_cross}.

The signature of a leptonically decaying monotop state consists of a hard lepton, 
a jet originating from the fragmentation of a $b$-quark and missing transverse energy. 
As such, the main sources of background arise from $t \bar{t}$ events where one of the 
two top quarks decays leptonically and from events related to the production of a 
single-top quark in association with a $W$-boson, where either the top quark or the 
$W$-boson decays leptonically.  Other contributing background processes consist of 
the other single-top production modes, $W$-boson plus jets production, 
$\gamma^*/Z$-boson plus jets production and diboson production. The respective total 
cross sections of these background processes are presented in 
Table~\ref{tab:compressed_cross} along with details of the Monte Carlo programs used 
in their simulation.  In all cases, the CTEQ6L1~\cite{Pumplin:2002vw} and 
CTEQ6M~\cite{Pumplin:2002vw} PDF sets were used in the simulation of processes 
generated at leading order (LO) and next-to-leading order (NLO) respectively.  While 
QCD multijet production processes should also be taken into account, a correct 
estimation of their contribution relies more on data-driven approaches than Monte 
Carlo simulations. Therefore, we have chosen to neglect it and ensure good control of 
this source of background through appropriate event selection criteria, as detailed 
in Section~\ref{sec:compressed_selection}. Moreover, we neglect all possible sources 
of instrumental background, this task going beyond the scope of this work.

\section{EVENT SELECTION} \label{sec:compressed_selection}

The object reconstruction used in this study is based on the typical approach of 
the ATLAS experiment in studies of single-top production, see for example 
Ref.~\cite{Aad:2012ux}, while the event selection criteria has been chosen to 
reflect the particle content of final states originating from the signal process. 
As such, we demand the presence of exactly one lepton candidate in each event, with 
electron (muon) candidates required to have a transverse momentum $p_T^l > 20$~GeV 
and a pseudorapidity $|\eta|<2.47$ (2.5). Lepton candidates are constrained to be 
isolated by restricting the sum of the transverse momenta of all charged particles 
in a cone of $\Delta R < 0.2$\footnote{We define 
$\Delta R = \sqrt{\Delta \phi^2 + \Delta \eta^2}$  where $\Delta \phi$ and 
$\Delta \eta$ denote the differences in the azimuthal angle with respect to the 
beam direction and in pseudorapidity of the charged particle and the lepton, 
respectively.} around the particle to be less than 10\% of its transverse momentum.

Jets have been reconstructed from all visible final-state particles fulfilling 
$|\eta| < 4.9$ using the {\sc FastJet}~\cite{Cacciari:2011ma} implementation of the 
anti-$k_T$ algorithm with a radius parameter of $R = 0.4$~\cite{Cacciari:2008gp}. 
Moreover, the transverse momentum of the reconstructed jet candidates has been 
required to satisfy $p_{T} >10$~GeV and jets which overlap with candidate electrons 
within a distance of $\Delta R < 0.2 $ have been discarded. Having done so, only jets 
with $p_T > 20 \GeV $ and $|\eta| < 2.5$ are retained and any lepton candidate within 
a distance $\Delta R < 0.4 $ of the closest remaining jet has been discarded.  We 
further identify jets as originating from a $b$-quark if they lie within 
$\Delta R < 0.3$ of a $b$-hadron and impose a $p_T$-dependent $b$-tagging probability 
as described in Ref.~\cite{ATLAS-CONF-2011-089}, which corresponds to an average efficiency 
of 70\% for $t\bar{t}$ events. We then select events by requiring the presence of 
exactly one $b$-jet with $p_T^b > 30$~GeV and $|\eta| < 2.5$. Finally, to reflect 
the expectation that the decay products of the $\tilde{t}_1$ and $\tilde{g}$ are 
largely invisible, any event containing an extra jet with transverse momentum 
$p_T > \mathrm{min}(p_T^b, 40 \GeV)$ has been discarded.

The following selection criteria have then been imposed in order to increase the 
sensitivity, $s$, of the analysis to the signal process, where we define 
$s = S/ \sqrt{S+B}$ in which $S$ and $B $ are the number of signal and background 
events passing all selection criteria, respectively.  The missing transverse momentum 
in the event, $p_T^{\mathrm{miss}}$, has been determined from the vector sum of the 
transverse momenta of all visible final-state particles. The magnitude of this 
quantity, $E_T^{\mathrm{miss}}$, is required to satisfy $E_T^{\rm{miss}} > 150 \GeV$. 
The choice of such a loose $E_T^{\mathrm{miss}}$ selection criterion is motivated by the 
observation that while a higher requirement would improve background rejection it 
would also significantly reduce the signal acceptance, leading to an overall reduction 
in the sensitivity. This low $E_T^{\mathrm{miss}}$ limit is moreover feasible in the 
context of event triggers for the LHC experiments by requiring the presence of a 
relatively hard, isolated lepton in the selected events, which opens up the 
possibility of using leptonic triggers.  We have verified that the transverse 
momentum requirement on the candidate lepton could be increased to $p_T^l>30 \GeV$, 
which allows for an improvement of the trigger efficiencies, without a significant 
degradation of the signal.

While the selection criterion on $E_T^{\mathrm{miss}}$ must be low to avoid the rejection 
of signal events, the sensitivity of the analysis can be further improved by 
constraining the orientation of $p_T^{\mathrm{miss}}$ with respect to the identified 
lepton. As such, we impose a minimum value of the $W$-boson transverse mass, defined by
\be\label{eq:compressed_mtw}
  m_T^W = \sqrt{2 p_T^l E_T^{\rm{miss}} (1 -
    \cos(\Delta \phi_{l,p_T^{\mathrm{miss}}}))} \ ,
\ee
where $\Delta \phi_{l,p_T^{\mathrm{miss}}}$ is the difference in azimuthal angle between 
the lepton and $p_T^{\mathrm{miss}}$.  When the missing transverse momentum in the event 
originates solely from the leptonic decay of a $W$-boson, the distribution is peaked 
at a lower value of $m_T^W$ than when its source is both a $W$-boson decay and a pair 
of invisible particle as in the signal case. Therefore, we require events to satisfy 
\mbox{$m_T^W > 120 \GeV$}, which also ensures that the non-simulated QCD multijet 
background contribution is under control~\cite{Aad:2012twa,Chatrchyan:2012bd}.

In order to reduce the number of background events in which the identified lepton 
and $b$-jet do not originate from a single top quark, the restriction on the invariant 
mass of the lepton plus $b$-jet system, $m_{bl} <150 \GeV$, has been imposed.  
Finally, a minimum value of the invariant mass of the lepton, $b$-jet and missing 
transverse momentum system, $m \left(E_T^{\mathrm{miss}},l,b \right)$, has been enforced, 
\mbox{$m \left(E_T^{\mathrm{miss}},l,b \right) > 1.0 \TeV$}. This serves to further reduce 
the number of $t\bar{t}$ and $tW$ events passing all the selection criteria.

\section{RESULTS} \label{sec:compressed_results}
\begin{figure}[t]
\begin{subfigure}
    \centering
    \includegraphics[trim=0mm 0mm 0mm 3mm, clip, height=5.5cm]{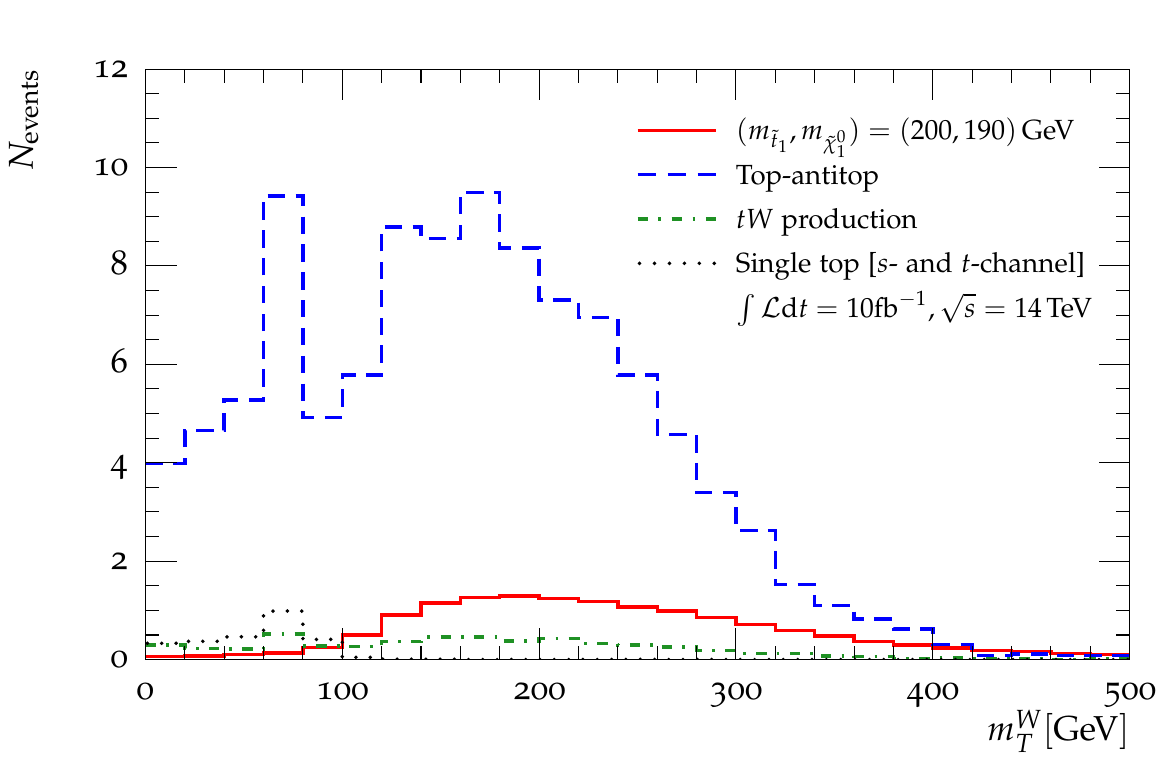} 
  \end{subfigure}
  \begin{subfigure}
    \centering
   \includegraphics[trim=0mm 3mm 0mm 0mm, clip, height=5.5cm]{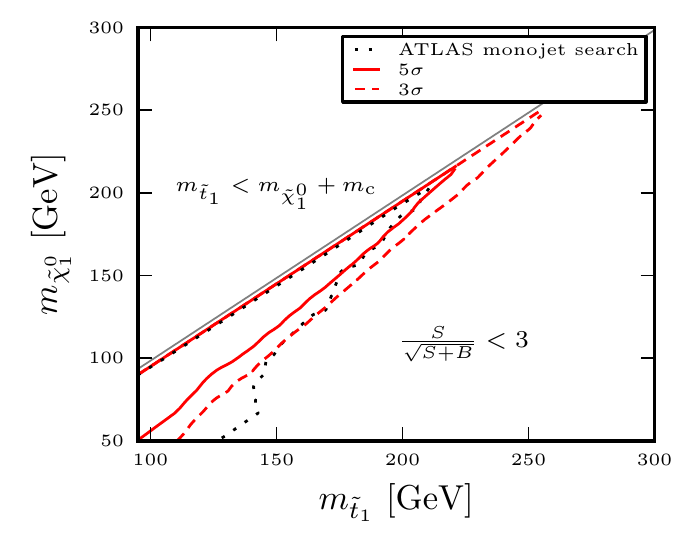}
  \end{subfigure}
  \caption{Left: $W$-boson transverse mass distribution $m_T^W$, as defined in 
Eq.~\eqref{eq:compressed_mtw}, for surviving background contributions and the 
example signal scenario of Section~\ref{sec:compressed_simulation} after all 
selection criteria have been applied, excepted $m_T^W > 120 \GeV$. Right: LHC 
sensitivity to SUSY monotop signals in the $(m_{\tilde{t}_1},m_{\tilde{\chi}_1^0})$ plane 
with $m_{\tilde{t}_1}=m_{\tilde{g}}$. We superimpose a previous exclusion bound set by an 
ATLAS analysis searching for stops decaying into a charm quark and the lightest 
neutralino using monojet events. }
  \label{fig:compressed_scan}
\end{figure}

The numbers of events surviving all selection criteria described in 
Section~\ref{sec:compressed_selection} are listed separately in
Table~\ref{tab:compressed_cross} for the different background contributions and 
the example compressed spectra scenario of Section~\ref{sec:compressed_simulation}. The 
effect of our analysis strategy is further illustrated in 
Figure~\ref{fig:compressed_scan} which shows the $m_T^W$ distributions after all 
selection requirements, excepted $m_T^W> 120 \GeV$, for the remaining background 
sources ($t\bar{t}$ and single-top) and the considered signal scenario. In the case 
of the background, a peak, related to events in which both the lepton and missing 
transverse momentum originate from the decay of a single $W$-boson, is visible in the 
region \mbox{$m_T^W \simeq 80 \GeV$}.  In contrast, the signal distribution exhibits 
a suppression in the region $m_T^W \lesssim 120 \GeV$, which motivates our choice of 
selection criterion for this observable. Relatively large background contributions 
from $t\bar{t}$ and, to a lesser extent, $tW$ production are however still foreseen 
after the requirement $m_T^W>120 \GeV$ has been imposed.  Despite this, the sensitivity 
of the LHC to this point reaches $s =6.1$ after all selections, although a more precise 
quantitative statement requires further studies involving detector effects or the 
instrumental background. This however goes beyond the scope of this prospective work 
which only aims to demonstrate the feasibility of searching for monotop SUSY signals 
at the LHC in the case of compressed superparticle mass spectra.

To more extensively study the LHC sensitivity to different compressed spectra SUSY 
scenarios, a scan of the $(m_{\tilde{t}_1},m_{\tilde{\chi}_1^0})$ plane with 
$m_{\tilde{t}_1} = m_{\tilde{g}}$ has been performed. We have derived in this way contours 
corresponding to $5\sigma$ and $3\sigma$ observation boundaries, that we respectively 
show by solid and dashed red lines in Figure~\ref{fig:compressed_scan}. A more careful 
optimization of the event selection criteria could however possibly extend the reach 
of our search strategy.  We superimpose on our results exclusion bounds at the 95\% 
confidence level, indicated by a black dotted line, set by the ATLAS collaboration 
on the basis of 20.3~fb$^{-1}$ of collision data at a center-of-mass energy of 
8~TeV~\cite{ATLAS-CONF-2013-068}. They arise from a monojet search for the 
pair production of top squarks that each decay into a charm quark and the lightest 
neutralino.\footnote{The analysis of Ref.~\cite{ATLAS-CONF-2013-068} also 
presents an exclusion boundary derived using charm-flavour identification techniques. 
This extends the limits away from the $m_{\tilde{t}_1}= m_{\tilde{\chi}_1^0}+m_c$ 
boundary into a region of the parameter space not probed by our analysis. For this 
reason, it has not been shown on Figure~\ref{fig:compressed_scan}.} It is likely that 
searches for monotop compressed SUSY signals could provide an independent check of the 
exclusion bounds obtained from a more standard monojet analysis. Moreover, as both 
channels are statistically uncorrelated, an improvement of the bounds could result from 
combining the two results.

\section{CONCLUSION}\label{sec:compressed_conclusions}
In this study, we have investigated the feasibility of using monotop searches to 
get a handle on SUSY scenarios featuring a compressed spectrum. We have considered 
the process $pp \rightarrow  \tilde{g} \, \tilde{t}_1 \, \bar{t}$, where the top 
quark decays leptonically and the sparticles decay via 
$\tilde{t}_1 \rightarrow c\, \tilde{\chi}_1^0$ and 
$\tilde{g} \rightarrow q\,\bar{q}\, \tilde{\chi}_1^0$. We consider small mass 
splittings between the $\tilde{t}_1$, $\tilde{g}$ and $\tilde{\chi}_1^0$, meaning 
jets produced in the superparticle decays are soft and the corresponding observable 
signal is a single top quark with missing transverse momentum, referred to as 
monotop. Using Monte Carlo simulations of 10~fb$^{-1}$ of LHC collisions at a 
center-of-mass energy of 14~TeV, we have shown that signals of this type will in
principle be reachable at the LHC for mass spectra scenarios close to the 
$m_{\tilde{t}_1}= m_c + m_{\tilde{\chi}_1^0}$ boundary in the case where 
$m_{\tilde{t}_1}=m_{\tilde{g}}$. This result motivates further investigation of SUSY
monotop signals, both into the optimization of the event selection criteria and the 
case of hadronic top quark decay, as well as consideration of detector effects and 
instrumental background or alternative compressed spectrum setups.

\section*{ACKNOWLEDGEMENTS} 
The authors would like to thank the Les Houches school for physics for their 
hospitality during the workshop where some of the work was performed, as well as 
the workshop organizers. AW and PR acknowledge the support of the European Union via 
MCNet, PITN-GA-2012-315877.  AW also acknowledges support from the Science and Technology Facilities 
Council.  The work of BF has been partially supported by the Theory-LHC France 
initiative of the CNRS/IN2P3 and by the French ANR 12 JS05 002 01 BATS@LHC.



%% file: nsusy/nsusy_main.tex

\chapter{Constraining Natural Supersymmetry from the LHC Stop and Sbottom Search Results at 8 TeV}

{\it J.~Bernon,
G.~Chalons,
E.~Conte,
B.~Dumont,
B.~Fuks,
A.~Gaz,
S.~Kraml,
S.~Kulkarni,
L.~Mitzka,
S.~Pataraia,
W.~Porod,
S.~Sekmen,
D.~Sengupta,
N.~Strobbe,
W.~Waltenberger,
F.~W\"urthwein,
C.~Wymant}



\begin{abstract}
Both the ATLAS and CMS collaborations have been searching for light stops and sbottoms in a variety of different channels. 
The mass limits published by the experimental collaborations however typically assume 100\% branching ratio for a given decay mode. A coherent picture of the status of light third-generation squarks, in particular in combination with light higgsinos (aka natural supersymmetry) is however still missing.  
We report on the progress in developing such a coherent picture by means of a scan over physical parameters, 
namely stop and sbottom masses and mixing angles, and higgsino masses.  
We present results obtained with {\sc SModelS} in the Simplified Models approach and  
describe the status of the implementation and validation of  various stop and sbottom analyses 
from ATLAS and CMS in the {\sc MadAnalysis}\,5 framework. 
Finally, we describe the extensions done in {\sc MadAnalysis}\,5 to adapt it for this project. 
\end{abstract}

\section{INTRODUCTION}

In order to solve the gauge hierarchy problem of the Standard Model (SM), 
supersymmetric (SUSY) particles that have sizable couplings to the Higgs sector should 
have masses not too far above the electroweak (EW) scale. This concerns in particular the squarks 
of the third generation, stops and sbottoms, which should be lighter than about a TeV in order not to
create a severe naturalness problem. Moreover, the higgsino mass parameter $\mu$ should be small, 
because its intimate relation with the EW scale: $-M_Z^2/2=|\mu|^2+m_{H_u}^2$. 
Since the standard searches for gluinos and light-flavor squarks at the LHC have so far produced only null results and  precision measurements in flavor physics are frustratingly consistent with SM predictions, the scenario 
with light stops and sbottoms and light higgsinos, but multi-TeV first and second generation squarks -- commonly dubbed ``natural supersymmetry'' (NSUSY) -- is increasingly becoming the new paradigm of SUSY phenomenology.  

The ATLAS and CMS collaborations have been searching for light stops and sbottoms in a variety of channels~\cite{atlassusy,cmssusy}. This has led to mass limits of the order of 500--700~GeV. These limits however depend on various assumptions. In particular, 100\% branching fraction for a given decay mode is typically assumed. In realistic scenarios, stops and sbottoms can however have a variety of decays, in particular into charginos and neutralinos, with the branching ratios depending not only on the mass pattern (kinematics) but also on the mixing angles in the stop/sbottom and the chargino/neutralino sectors. It is thus interesting to assess in detail how the current ATLAS and CMS searches, performed during the first phase of LHC running, constrain NSUSY in general. This assessment is the aim of this project.

\section{NATURAL SUSY PARAMETER SCAN} \label{sec:nsusy-scan_parameters}

In order to achieve a good survey of the relevant parameter space, we choose to work with physical stop and sbottom masses and mixing angles ($m_{\tilde t_1}$, $m_{\tilde t_2}$, $m_{\tilde b_1}$, $m_{\tilde b_2}$, $\theta_{\tilde t}$, $\theta_{\tilde b}$). The reason for this is the strong dependence on the $A_t$ parameter and the large radiative corrections in the stop/sbottom sector, which make it difficult to reach all corners of physical parameter space when working with soft-breaking terms. In the chargino/neutralino sector it is more convenient to compute the masses and mixings from the Lagrangian parameters $M_1$, $M_2$, $\mu$ and $\tan\beta$. 

For our first scan, we fix slepton and first and second generation squark masses at 5~TeV, the gluino mass at 2~TeV, and $M_1=M_2=1$~TeV. 
We then vary 
\begin{eqnarray}
  m_{\tilde t_1} &=& 150,\, 200,\, 300,\, 400,\, \ldots ,\, 1000~{\rm GeV}\,; \nonumber \\
  m_{\tilde b_1} &=& 150,\, 200,\, 300,\, 400,\, \ldots ,\, 1000~{\rm GeV}\,; \nonumber \\
  \mu &=& 100,\, 200,\, 300,\, 400,\, 500~{\rm GeV}\,; \nonumber \\
  \theta_{\tilde t} &=& 0,\, 45,\, 90~{\rm deg}\,; \nonumber \\
  \theta_{\tilde b} &=& 0,\, 45,\, 90~{\rm deg}\,; \nonumber\\
  \tan\beta &=& 10,\, 50\,. \label{eq:nsusy-ranges}
\end{eqnarray}
Further scans also consider $\theta_{\tilde t,\tilde b}=135^\circ$ and $180^\circ$, and lower gluino masses of 750 and 1000 GeV.
The computation of the chargino and neutralino masses and of all decay branching ratios is done with {\sc SPheno}~\cite{Porod:2003um,Porod:2011nf}. The $\tilde t_2$ is assumed to be heavy and irrelevant for LHC phenomenology at 8~TeV. We do not worry about the value of the light Higgs mass because $m_h\simeq 126$~GeV can always be achieved by adjusting $m_{\tilde t_2}$ and $A_t$.  Note that in order to have a neutralino as the lightest SUSY particle (LSP), the ranges in Eq.~(\ref{eq:nsusy-ranges}) are effectively limited to $m_{\tilde t_1},m_{\tilde b_1}>\mu$. 

The stop and sbottom masses and mixing angles are however not completely independent: because of the $SU(2)_L$ symmetry the $\tilde t_L$ and $\tilde b_L$ masses depend on the  same mass parameter $M_{\tilde Q_3}$. This implies the sum rule 
\begin{equation}
  m_W^2\cos 2\beta \, = \, m_{\tilde t_1}^2\cos^2\theta_{\tilde t} + m_{\tilde t_2}^2\sin^2\theta_{\tilde t}
  - m_{\tilde b_1}^2\cos^2\theta_{\tilde b} - m_{\tilde b_2}^2\sin^2\theta_{\tilde b} - m_t^2 + m_b^2
\label{eq:nsusy-sqmassrelation}
\end{equation}
at tree level. This means we have to be careful when $\theta_{\tilde t}=0$ or $90^\circ$ in our scan. 
In case of $\theta_{\tilde t}=0$, one has a pure $\tilde t_L$ in the game and thus also a nearby $\tilde b_L$. We compute the mass of $\tilde b_L$ using the tree-level mass relation above. In case that this mass is below the $m_{\tilde b_1}$ chosen for this scan point, we flip the ordering of the sbottom masses and change the mixing matrix accordingly.
Moreover, when $\theta_{\tilde t} =90^\circ$, we consider only $\theta_{\tilde b} =90^\circ$.
In total the first scan set, Eq.~(\ref{eq:nsusy-ranges}), thus contains 3696 points. 

The choice of small $\mu$ together with large $M_{1,2}$ leads to a higgsino LSP.  Moreover, the other higgsinos, 
$\tilde\chi^0_2$ and $\tilde\chi^\pm_1$, are almost mass-degenerate with the $\tilde\chi^0_1$. For $\mu\ge 200$~GeV, the $\tilde\chi^0_2$--$\tilde\chi^0_1$ mass difference is around 10~GeV while the $\tilde\chi^\pm_1$--$\tilde\chi^0_1$ mass difference is about 3--4~GeV. The masses for \mbox{$\tan\beta=10$} are given in Table~\ref{tab:nsusy-higgsinomasses}. For $\tan\beta=50$ one obtains very similar numbers, differing only by about 1~GeV. 

Since the $\tilde\chi^0_1$, $\tilde\chi^0_2$ and $\tilde\chi^\pm_1$ are basically pure higgsinos, leaving aside kinematic effects, the stop (sbottom) branching ratios depend only on the stop  (sbottom) mixing angle and $\tan\beta$. Illustrative examples are given 
in Tables~\ref{tab:nsusy-branchings-tb10}~and~\ref{tab:nsusy-branchings-tb50}. 
It is worth noting that the $\tilde t_1\to b\tilde\chi^+_1$ decay is never completely dominant unless $m_{\tilde t_1}-m_{\tilde\chi^0_1}<m_t$. Likewise, $\tilde b_1\to b\tilde\chi^0_{1,2}$ is never  completely dominant unless the decay into $t\tilde\chi^-_1$ is kinematically suppressed.  
Note, however, that the numbers in Tables~\ref{tab:nsusy-branchings-tb10} and \ref{tab:nsusy-branchings-tb50} can change drastically if $\tilde t_1\to \tilde b_1 W$ or $\tilde b_1\to \tilde t_1 W$ decays are kinematically allowed.  

\begin{table}
\caption{Lightest neutralino and chargino masses (in GeV) as a function of $\mu$ (in GeV), for $\tan\beta=10$. }
\label{tab:nsusy-higgsinomasses}
\begin{center}\begin{tabular}{c|c|c|c|c|c}
  $\mu$ & 100 & 200 & 300 & 400 & 500 \\
  \hline
  $m_{\tilde\chi^0_1}$ & 94.9 & 194.2 & 293.4 & 392.3 & 490.8\\
  $m_{\tilde\chi^0_2}$ & 98.2 & 202.6 & 302.4 & 402.3 & 502.1 \\
  $m_{\tilde\chi^\pm_1}$ & 102.9 & 197.6 & 296.8 & 395.8 & 494.6 
\end{tabular}\end{center}
\end{table}

\renewcommand{\arraystretch}{1.3}
\begin{table}
\caption{Stop and sbottom branching ratios for $m_{\tilde t_1,\tilde b_1}=600$~GeV, $\mu=200$~GeV and $\tan\beta=10$.}
\label{tab:nsusy-branchings-tb10}
\begin{center}\begin{tabular}{c|c|c|c||c|c|c|c}
  $\theta_{\tilde t_1}$ [deg] & 0 & 45 & 90 & $\theta_{\tilde b_1}$  [deg] & 0 & 45 & 90 \\
  \hline
  BR$(\tilde t_1\to t\tilde\chi^0_{1,2})$ &  0.96 & 0.59 & 0.44 & BR$(\tilde b_1\to b\tilde\chi^0_{1,2})$ & 0.04 & 0.08 & 0.56 \\
  BR$(\tilde t_1\to b\tilde\chi^+_1)$\hphantom{.}  &  0.04 & 0.41 & 0.56 & BR$(\tilde b_1\to t\tilde\chi^-_1)$\hphantom{0} & 0.96 & 0.92 & 0.44
\end{tabular}\end{center}
\caption{Stop and sbottom branching ratios for $m_{\tilde t_1,\tilde b_1}=600$~GeV, $\mu=200$~GeV and $\tan\beta=50$.}
\label{tab:nsusy-branchings-tb50}
\begin{center}\begin{tabular}{c|c|c|c||c|c|c|c}
  $\theta_{\tilde t_1}$  [deg]  & 0 & 45 & 90 & $\theta_{\tilde b_1}$  [deg] & 0 & 45 & 90 \\
  \hline
  BR$(\tilde t_1\to t\tilde\chi^0_{1,2})$ &  0.57 & 0.49 & 0.44 & BR$(\tilde b_1\to b\tilde\chi^0_{1,2})$ & 0.43 & 0.55 & 0.56 \\
  BR$(\tilde t_1\to b\tilde\chi^+_1)$\hphantom{.}  &  0.43 & 0.51 & 0.56 & BR$(\tilde b_1\to t\tilde\chi^-_1)$\hphantom{0} & 0.57 & 0.45 & 0.44
\end{tabular}\end{center}
\end{table}
\renewcommand{\arraystretch}{1.}

\section{SIMPLIFIED MODELS RESULTS}

In order to get a first overview of how the current searches perform in case of a higgsino LSP, we pass the scenarios created in our scan (for the 2 TeV gluino case) through {\sc SModelS}~\cite{Kraml:2013mwa}.  
{\sc SModelS} decomposes each scenario into its Simplified Model Spectra (SMS) topologies and 
compares the $\sigma\times{\rm BR}$ of each topology to the 95\% confidence level (CL) upper limits 
given by the experimental collaborations. The  procedure and the analyses considered are explained in detail in Ref.~\cite{Kraml:2013mwa}.  Note that because of the small mass difference of $\Delta m < 5$~GeV, the $\tilde\chi^\pm_1$ decays into the LSP are treated as invisible in {\sc SModelS};  the $\tilde\chi^0_2$ decays on the other hand are assumed to be visible ($\Delta m > 5$~GeV). The exception is the $\mu=100$~GeV case, where the $\tilde\chi^{\pm}_1$--$\tilde\chi^0_2$ mass pattern is inverted.

\begin{figure}
\centering
\includegraphics[width=0.48\textwidth]{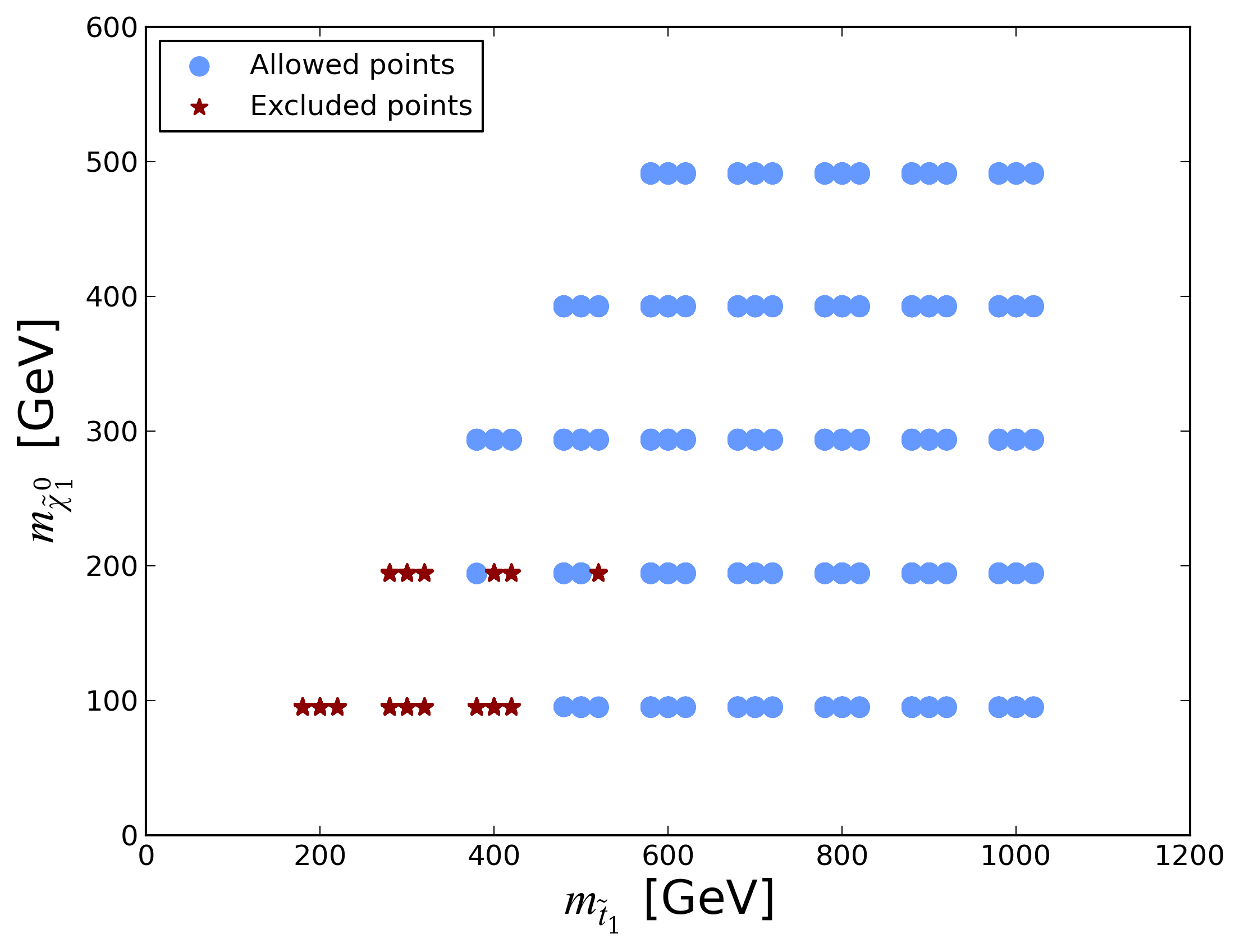}\quad
\includegraphics[width=0.48\textwidth]{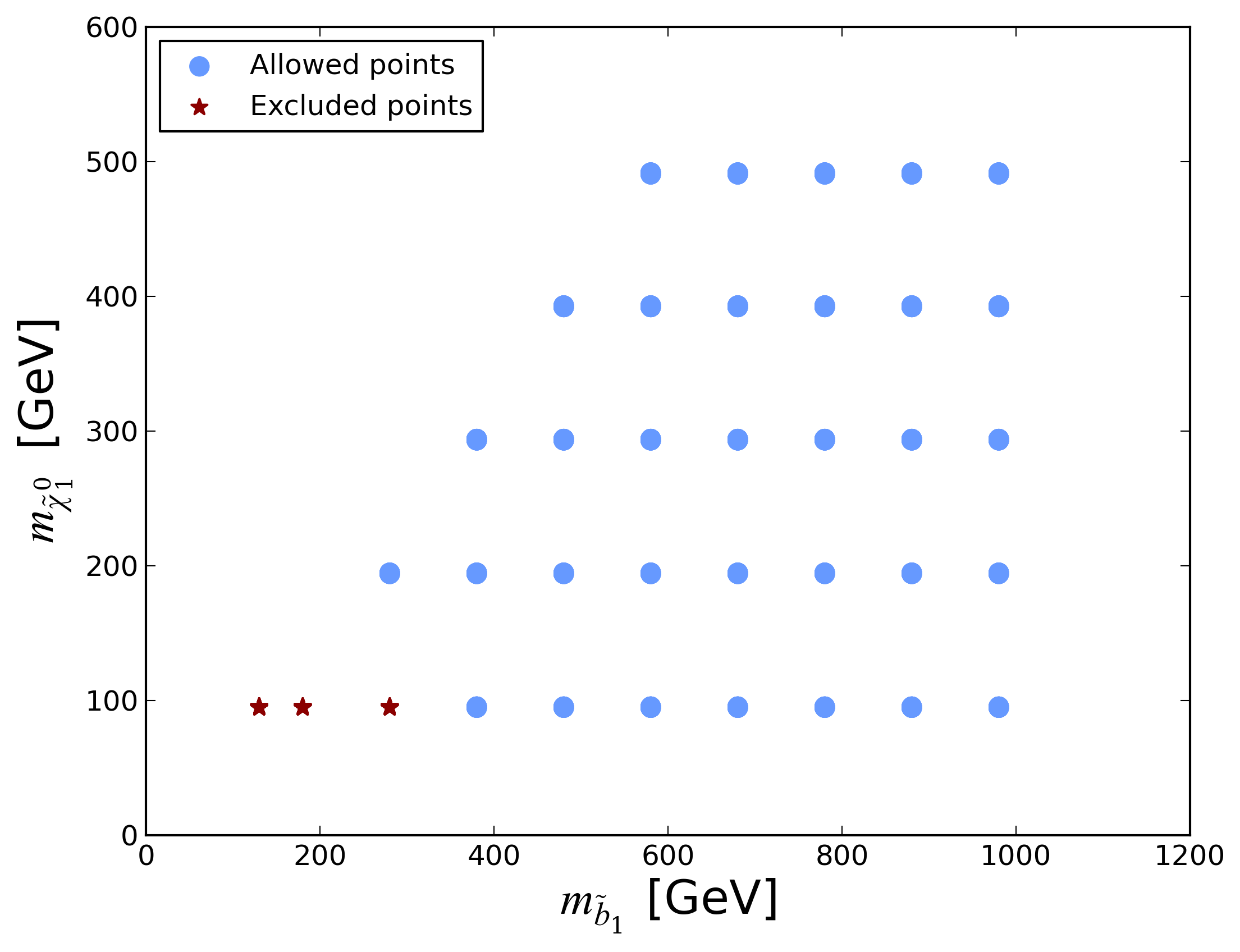}
\caption{Simplified Models results, obtained with {\sc SModelS}, for the scan with $m_{\tilde g}=2$~TeV described in Section~\ref{sec:nsusy-scan_parameters}. Excluded mass combinations are marked by a red star, while combinations that are not excluded for at least one combination of the scanned parameters are marked by a blue point. 
In the $\tilde\chi^0_1$ versus $\tilde t_1$ mass plane (left panel), the results are presented for 3 stop mixing angles,  with a small offset introduced in the stop mass to make the points visible. For each ($\tilde t_1$,\,$\tilde\chi^0_1$) mass combination, the connected points thus represent, from left to right, a pure $\tilde t_L$, a mixed $\tilde t_{LR}$ and a pure $\tilde t_R$ ($\theta_{\tilde t} = 0, 45, 90$\,deg). The results in the  $\tilde b_1$ versus $\tilde\chi^0_1$ mass plane (right panel) depend less on the bottom mixing angle, thus $\theta_{\tilde b}$ is marginalized over.} 
\label{fig:nsusy-stop_sbot_lsp_combined}
\end{figure}

Figure~\ref{fig:nsusy-stop_sbot_lsp_combined} shows the excluded and not excluded points in the 
$\tilde\chi^0_1$ versus $\tilde t_1$ and $\tilde b_1$ mass planes (left and right panels, respectively). 
In the $\tilde\chi^0_1$ versus $\tilde t_1$ case, the results are separated for different mixing angles by introducing a small artificial offset in the stop mass in the figure. We see that for $m_{\tilde\chi^0_1}\simeq 100$~GeV, stop masses 
up to 400~GeV are excluded, irrespective of the mixing angle, bottom mass, or $\tan\beta$. For 500~GeV stop quarks,  
there are already some parameter combinations such that the point cannot be strictly excluded. 
At $m_{\tilde\chi^0_1}\simeq 200$~GeV, right-handed and mixed stop quarks of 400~GeV mass are excluded, 
while a pure $\tilde t_L$ is not. Moving on to a stop mass of 500~GeV,  only the $\tilde t_R$ is strictly excluded, while a $\tilde t_L$ or a mixed stop quark evades the limits. 

This can be understood as follows. For the scan considered, the only relevant SMS topologies for which 
experimental limits are available consist in the production of a top-antitop
and bottom-antibottom pair of quarks in association with missing energy ($t\bar t+{\rm MET}$ and $b\bar b+{\rm MET}$).
At $\mu=100$~GeV, these topologies arise from 
$\tilde t_1\to t\tilde\chi^0_{1,2}$ and $\tilde b_1\to b\tilde\chi^0_{1,2}$ decays, respectively. 
At $\mu\ge 200$~GeV, $t\bar t+{\rm MET}$ arises from  
$\tilde t_1\to t\tilde\chi^0_{1}$ and $\tilde b_1\to t\tilde\chi^-_1$ decays, while  $b\bar b+{\rm MET}$ arises from
$\tilde t_1\to b\tilde\chi^+_1$ and $\tilde b_1\to b\tilde\chi^0_{1}$ decays (\textit{cf.}~the higgsino mass splittings shown in Table~\ref{tab:nsusy-higgsinomasses})\footnote{Note however that these are subject to  theoretical uncertainties of the order of 1\% in the calculation of the mass spectrum.}.
Let us now take the scenario $\mu=200$~GeV and $m_{\tilde t_1}=500$~GeV as a concrete example. 
At $\theta_{\tilde t}=90$\,deg ($\tilde t_1=\tilde t_R$), the point is [marginally] excluded by the 
ATLAS search in the $b\bar b+{\rm MET}$ channel~\cite{Aad:2013ija}. 
For a mixed stop, the BR$(\tilde t_1\to b\tilde\chi^-_1)$ is somewhat reduced and the scenario escapes this exclusion. 
At $\theta_{\tilde t}=0$\,deg ($\tilde t_1=\tilde t_L$), the point is excluded by the 
CMS leptonic stop search ($t\bar t+{\rm MET}$ topology)~\cite{Chatrchyan:2013xna} for $\tan\beta=10$, 
but not for $\tan\beta=50$. The reason is that at high $\tan\beta$ the branching ratio into 
$t + \tilde\chi^0_1$ is reduced; moreover, only the decay into the $\tilde\chi^0_1$ is accounted for in the $t\bar t+{\rm MET}$ signature. Concretely, we have BR$(\tilde t_1\to t\tilde\chi^0_1)\simeq 26\%$, so only $6.6\%$ of the 
$\tilde t_1\tilde t_1^*$ events can lead to $t\bar t+{\rm MET}$, which is not enough for an exclusion. 
At the same time BR$(\tilde t_1\to b\tilde\chi^+_1)\simeq 40\%$ is also insufficient for an exclusion through the $b\bar b+{\rm MET}$ topology. 

For a better sensitivity, it would be necessary to consider mixed events with one $\tilde t_1$ decaying into $t\tilde\chi^0_1$ and the other one into $b\tilde\chi^+_1$, and/or combine the results from different channels.  
Results for $tb+{\rm MET}$ topologies are however not available, and for a combination of results we would need the likelihood of the exclusion limits instead of just the the 95\%~CL values (and even then the statistically correct treatment is not obvious). 

\begin{figure}
\centering
\includegraphics[width=0.54\textwidth]{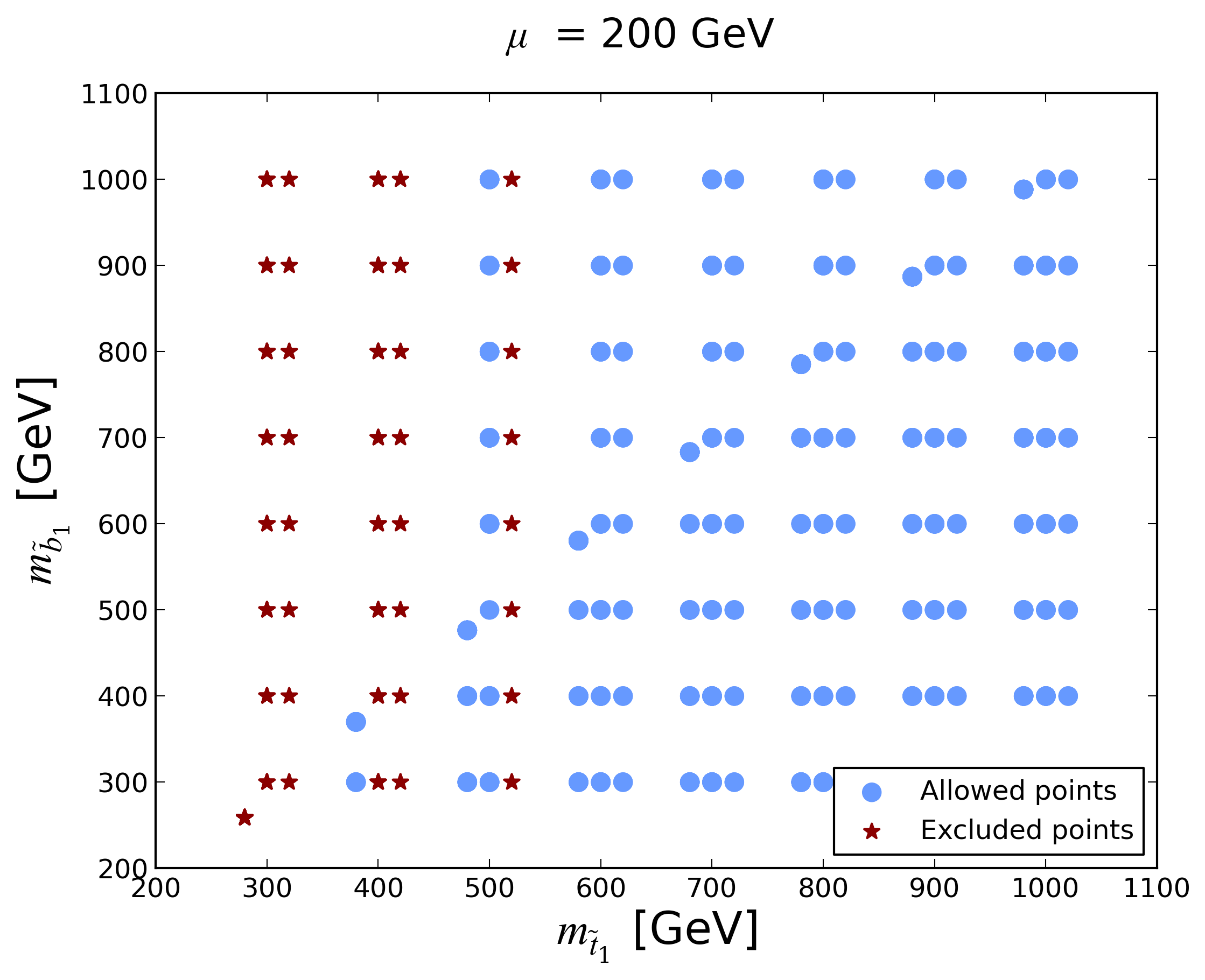}
\vspace*{-2mm}
\caption{Same as Figure~\ref{fig:nsusy-stop_sbot_lsp_combined} (left) but in the $\tilde t_1$ versus $\tilde b_1$ mass plane for fixed $\mu=200$~GeV. See text for details.}
\label{fig:nsusy-stop_sbot_combined}
\end{figure}

In Figure~\ref{fig:nsusy-stop_sbot_combined} the same results are presented in the $\tilde t_1$ versus $\tilde b_1$ mass plane for a fixed neutralino mass of about 200~GeV. Again, three different points for a given ($\tilde t_1,\,\tilde b_1$) mass combination  represent three different stop mixing angles ($\theta_{\tilde t} = 0, 45, 90$~deg). 
Note that for $\theta_{\tilde t} = 0$~deg, the sbottom mass cannot be chosen freely but is related to the stop mass as explained in Section~\ref{sec:nsusy-scan_parameters}, Eq.~\eqref{eq:nsusy-sqmassrelation}. This restricts the $\theta_{\tilde t} = 0$ case to $m_{\tilde b_1} \leq m_{\tilde t_1}$ and also causes the small offset of $m_{\tilde b_1}$ for the left-hand points in the figure. The figure confirms the conclusions from Figure~\ref{fig:nsusy-stop_sbot_lsp_combined}.  
We see that stop masses around 300~GeV are always excluded, while 
stop masses of $600$~GeV and above are not excluded for any sbottom mass or stop mixing angle. 
In the best case scenario $\theta_{\tilde t} = 90$~deg stop masses up to 500 GeV can be excluded irrespective of the sbottom mass. The reach is  less good for a mixed stop ($\theta_{\tilde t} = 45$~deg), again irrespective of the sbottom mass. 
Interestingly, the  case $\theta_{\tilde t} = 0$, where $m_{\tilde b_1} \leq m_{\tilde t_1}$ and hence both stop and sbottom 
events should contribute, is less well constrained than the other scenarios. The reason is to large extent the strong $\tan\beta$ dependence in this case.


\section{ANALYSIS IMPLEMENTATION IN {\sc MadAnalysis\,5}} \label{sec:nsusy-recast}

In order to go beyond Simplified Models, we decided to implement the relevant ATLAS and CMS searches 
in {\sc MadAnalysis}\,5~\cite{Conte:2012fm,Conte:2013mea,ma5update}, interfaced with {\sc Delphes}\,3. Within this framework, 
we produce and read the event files, implement all experimental selection requirements (commonly dubbed \textit{cuts}), and produce histograms and cut flows. 
The aim is to produce a database of re-implemented and validated analyses that can readily be used for 
re-interpretation studies. 
Table~\ref{tab:nsusy-anllist} shows a list of relevant analyses which should be implemented within this project.

\renewcommand{\arraystretch}{1.2}
\begin{table}
\caption{Analyses to be implemented within this project.}
\begin{center}
\begin{tabular}{|l|l|}
\hline
Analysis & Reference \\
\hline
\multicolumn{2}{|l|}{0 lepton} \\
\hline
$E_T^{\rm miss}$, $2$ $b$ jets & ATLAS-SUSY-2013-05~\cite{Aad:2013ija} \\
$E_T^{\rm miss}$, $\ge 3$ $b$ jets & ATLAS-CONF-2013-061~\cite{ATLAS-CONF-2013-061}  \\
$H_T$, $E_T^{\rm miss}$, $\ge 1$ $b$ jets & CMS-SUS-12-024~\cite{Chatrchyan:2013wxa}  \\ 
$E_T^{\rm miss}$, n $b$ jets with $\alpha_T$ & CMS-SUS-12-028~\cite{Chatrchyan:2013lya}  \\
$H_T$, $H_T^{\rm miss}$ & CMS-SUS-13-012~\cite{Chatrchyan:2014lfa}  \\
\hline
\multicolumn{2}{|l|}{1 lepton} \\
\hline
$\tilde{t}$ search with $E_T^{\rm miss}$, $\ge 2$ b jets & ATLAS-CONF-2013-037~\cite{ATLAS-CONF-2013-037}  \\
$E_T^{\rm miss}$, $\ge 3$ $b$ jets & ATLAS-CONF-2013-061~\cite{ATLAS-CONF-2013-061}  \\
$E_T^{\rm miss}$, $\ge 2$ $b$ jets & CMS-SUS-13-007~\cite{Chatrchyan:2013iqa}  \\
$\tilde{t}$ search with $E_T^{\rm miss}$, $\ge 2$ b jets & CMS-SUS-13-011~\cite{Chatrchyan:2013xna}   \\
\hline
\multicolumn{2}{|l|}{2 leptons} \\
\hline
$\tilde{t}$ search with $E_T^{\rm miss}$, jets & ATLAS-CONF-2013-048~\cite{ATLAS-CONF-2013-048}   \\
SS 2$\ell$ & CMS-SUS-13-013~\cite{Chatrchyan:2013fea}  \\
OS 2$\ell$ & CMS-SUS-13-016~\cite{CMS-PAS-SUS-13-016}  \\
\hline
\end{tabular}
\end{center}
\label{tab:nsusy-anllist}
\end{table}
\renewcommand{\arraystretch}{1.}

\noindent
Direct searches for electroweak-inos may also be relevant, but will  be considered only at a later stage.
The work of  implementing one-by-one the analyses listed in  Table~\ref{tab:nsusy-anllist} 
started only recently,  because first several technical developments were necessary in 
{\sc MadAnalysis}\,5, as will be described in Section~\ref{sec:nsusy-MA5new}. 
Before reporting in detail on the implementation and validation of the experimental searches, 
a few comments are in order.

For each analysis, validation of the implementation done within {\sc MadAnalysis}\,5 is a crucial step. 
For that purpose, we consider the same benchmark points as in the experimental paper or in the analysis note and we generate associated samples of events in a manner as close as possible to what has been done by ATLAS or CMS. We then try to reproduce all publicly available results for these benchmark points: histograms of kinematic variables, number of events after given cuts defining a signal region, and, if available, cut flows.

Unfortunately, the information publicly available in the experimental papers and in the analyses notes often  
does not suffice for an unambiguous validation. First of all, the definition of the benchmark points, in particular 
when they  correspond to Simplified Model scenarios, is often incomplete because the mixing angles in the 
relevant squark and electroweak-ino sectors are omitted. In some cases, the relevant branching ratios are also 
not clearly defined. A simple way to overcome such problems 
would be that the experimental collaborations follow more closely the recommendations of Ref.~\cite{oai:arXiv.org:1203.2489} and systematically provide the employed supersymmetric spectra under the Supersymmetry Les Houches Accord (SLHA) format~\cite{Allanach:2008qq,Skands:2003cj}, {\it e.g.}\ on the Twiki pages or on HepData~\cite{Whalley:1989mt} 
or INSPIRE~\cite{inspire}.
ATLAS currently provides SLHA files on HepData for a number of SUSY analyses, which is extremely useful; 
we hope that this will be generalized to all ATLAS and CMS SUSY analyses in the future.

Generating events in the same way as done by ATLAS or CMS is an additional source of uncertainty, since the matching procedure between the parton shower and matrix element description is not publicly available. 
Furthermore, rather softs objects are almost always difficult to treat correctly, given the [lack of]  publicly available 
information on trigger and identification efficiencies.

Finally, the description of the analyses ({\it i.e.} the description of the selection criteria, event cleaning, signal isolation cuts, efficiencies, {\it etc.}) is sometimes not clear and/or incomplete and makes it necessary to contact the authors of the analysis for clarification of crucial elements. This can make the implementation and validation of some analyses a tedious process, which is only successful if enough inside information  can be obtained from the experimental collaboration.

\subsection{CMS-SUS-13-011: search for stop quarks in the single-lepton final state} \label{sec:nsusy-TopSquark1lepton}

The CMS search for stops in the single lepton and missing energy, $\ell + E^{\rm miss}_T$, final state with full luminosity at a center-of-mass energy
$\sqrt{s} = 8$~TeV~\cite{Chatrchyan:2013xna}, has been taken as a ``template analysis'' to develop a common language and framework for the analysis implementation. It also allowed us to test the new developments in {\sc MadAnalysis}\,5 (see Section~\ref{sec:nsusy-MA5new}) which were necessary for carrying out this project.

The analysis targets two possible decay modes of the stop quark: $\tilde{t}_1 \to t \tilde{\chi}^{0}_1$ and $\tilde{t}_1 \to b \tilde{\chi}^{+}_1$, where one of the $W$-bosons produced in the decay of the top quark or the chargino decays leptonically and the other one hadronically. 
In the cut-based version of the analysis\footnote{The search also contains an analysis based on boosted decision tree multivariate discriminants; such analyses generically cannot be externally reproduced and are therefore ignored in this work.}, two sets of signal regions (SRs) with different cuts, dedicated to one of the two decay modes, are defined. These two sets are further divided into ``low $\Delta M$'' and ``high $\Delta M$'' categories, targeting small and large mass differences with the LSP, respectively. Finally, each of these four categories are further sub-divided using four different $E^{\rm miss}_T$ requirements. In total, 16 different, 
potentially overlapping SRs are defined. Two cuts are based on rather complex and specific kinematic variables designed to reduce the dilepton $t\bar{t}$ background: a $\chi^2$ resulting from the full reconstruction of the hadronic top quark and $M^W_{T2}$ -- a variant of the $M_{T2}$ observable. The implementation of the $\chi^2$ quantity in our code was straightforward thanks to the {\sc C++} {\sc Root} code provided on the CMS Twiki page, whilst as shown in Section~\ref{sec:nsusy-kinem}, the $M_{T2}^W$ variable has been implemented from the algorithm presented in Ref.~\cite{Bai:2012gs}.
Overall, this analysis is very well documented. 
A missing piece of information, however, is details on the lepton and $b$-jet efficiencies.

The validation of the reimplementation of the analysis can be done using the eleven benchmark points 
presented in the experimental paper: four for the ``T2tt'' simplified model (in which the stop always decays as $\tilde{t}_1 \to t \tilde{\chi}^{0}_1$), and seven for the ``T2bW'' simplified model (in which the stop always decays as $\tilde{t}_1 \to b \tilde{\chi}^{+}_1$), with different assumptions on the various masses. The distributions of the kinematic variables used in the analysis are given in Figure~2 of Ref.~\cite{Chatrchyan:2013xna} after the preselection cuts, with at least one benchmark point for illustration. Also provided are the corresponding histograms after the \mbox{$M_T > 120$~GeV} cut, as supplementary material on the CMS Twiki page. We use this information, together with the final number of events in the individual SRs ({\it i.e.} after all selection cuts) for given benchmark points provided in Tables~4 and 6 of Ref.~\cite{Chatrchyan:2013xna} for the validation of our reimplementation within {\sc MadAnalysis}\,5.

The validation material both before and after cuts defining the SRs is truly valuable information since one can separately check on the one hand the implementation of the kinematic variables and the preselection/cleaning cuts, 
and on  the other hand the series of cuts defining the SRs. Furthermore, the large number of benchmark points allows us to check in detail the quality of the reimplementation in complementary regions of phase space.

The validation process was based on (partonic) event samples, in LHE format, provided by the CMS collaboration. The provision of such event files greatly reduced the uncertainties in the first
stage of validation since it avoided possible differences in the configuration of the used
Monte Carlo tools. In the case of the CMS-SUS-13-011 analysis, the setup of {\sc MadGraph}~5~\cite{Alwall:2011uj},
the event generator employed for generating the hard scattering matrix elements, is crucial, in particular
with respect to the merging of samples with different (parton-level) jet multiplicities as performed by the CMS collaboration.
The LHE files were passed through {\sc Pythia}~6~\cite{Sjostrand:2006za} for parton showering and hadronization, then processed by our modified version of {\sc Delphes}3 (see Section~\ref{sec:nsusy-moddelphes3}) for simulation of the detector effects. The number of events after cuts and histograms produced by {\sc MadAnalysis}\,5 were then normalized to the correct luminosity after including cross sections at next-to-leading
order and next-to-leading logarithmic (NLO+NLL) accuracy, as tabulated by the LHC SUSY Cross Section Working Group~\cite{8tevxs_susy}.  Section~\ref{sec:MA5SRM} contains some snippets of our implementation of this search, to illustrate the new {\sc MadAnalysis}\,5 syntax for signal regions and cuts.

\begin{figure}
\centering
\includegraphics[width=0.4\textwidth]{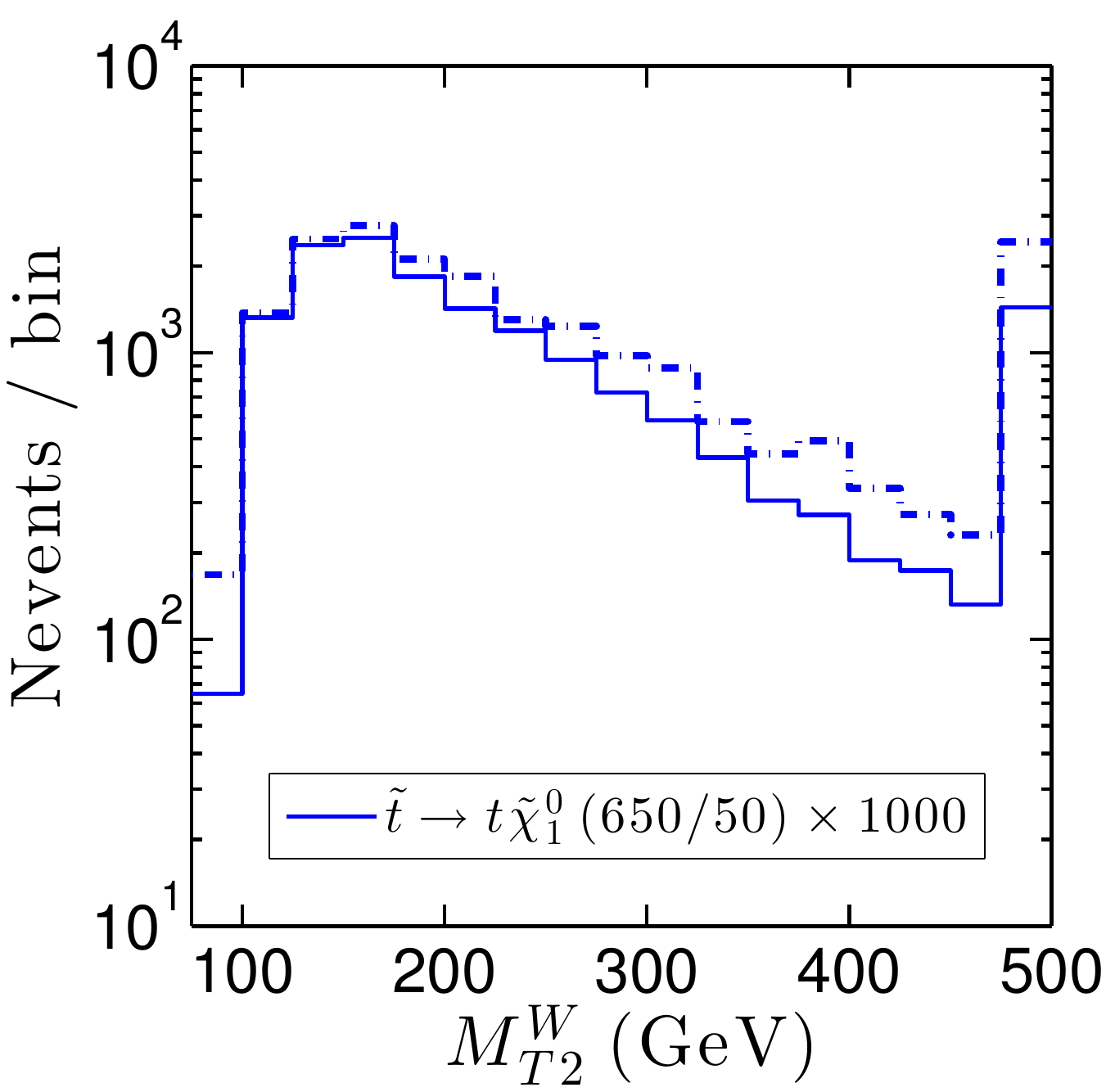}\quad
\includegraphics[width=0.4\textwidth]{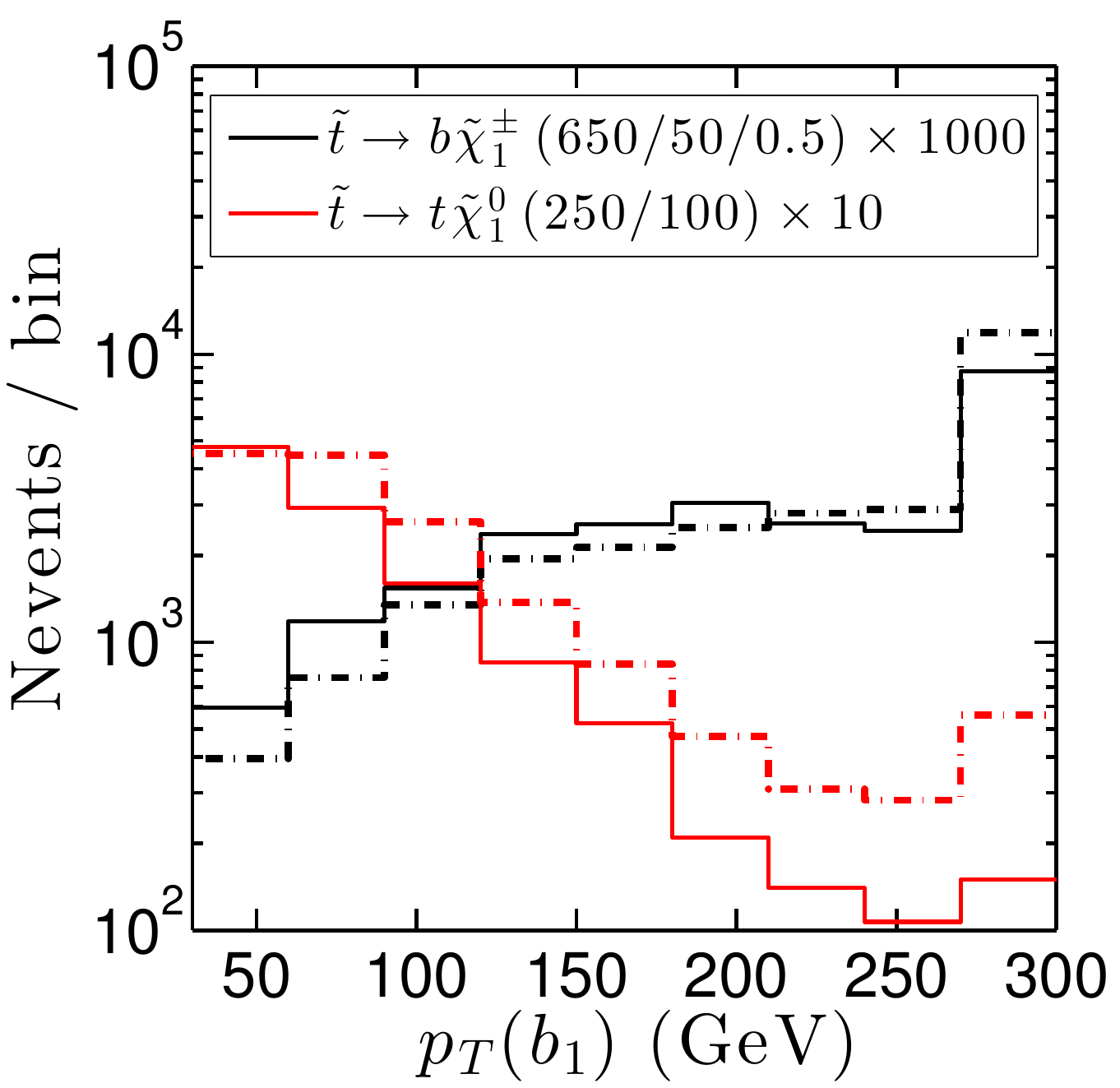}
\vspace*{-2mm}
\caption{Distributions of the kinematic variable $M^W_{T2}$ and the $p_T$ of the leading $b$-tagged jet after preselection cuts of the analysis CMS-SUS-13-011. The solid lines correspond to the CMS results, given in Figure~2 of Ref.~\cite{Chatrchyan:2013xna}, while the dashed lines are obtained from our re-interpretation within {\sc MadAnalysis}\,5.}
\label{fig:nsusy-kinvarsus13011}
\end{figure}

Some examples of histograms reproduced for the validation are shown in Figure~\ref{fig:nsusy-kinvarsus13011}. The shapes of the distributions shown -- as well as all other distributions that we obtained but do not show here -- follow closely the ones from CMS, which indicates the correct implementation of the analysis and all the kinematic variables. (Note that discrepancies in bins where the number of events is relatively small, as seen on a logarithmic scale, suffers from larger statistical uncertainties and hence should not be over-interpreted.)
The expected yields for several benchmark points in two example SRs are given in Tables~\ref{tab:nsusy-NevStopbcharginoLowDeltaMMET2501} and~\ref{tab:nsusy-NevStopTneutralinoLowDeltaMMET2502}. While the agreement is good for benchmark points with $m_{\tilde{t}_1} \ge 600$~GeV, large discrepancies appear when the stop is lighter and when the mass splittings are reduced. We are currently investigating the origin of these discrepancies. Two possible effects are under scrutiny: first, after generating the samples of events with {\sc MadGraph}~5, CMS is applying a correction because of the incorrect Monte Carlo modeling of the initial-state radiation as explained in Appendix B of Ref.~\cite{Chatrchyan:2013xna}. The effect can be as large as 20\%, depending on the $p_{T}$ of the system recoiling against the initial-state radiation jets. Second, the selection and isolation efficiencies we are currently using for the leptons may be inaccurate, as not enough information is provided by CMS\footnote{Efficiencies as function of $p_T$ are provided for other CMS analyses, like for instance in Ref.~\cite{Chatrchyan:2013fea}. 
However, as details in lepton selection differ between these analyses, it is not possible to use the information provided in one CMS note
to validate another analysis. A more consistent approach in providing such detailed information for all analyses would be
desirable.}.

\begin{table}
\caption{Final number of events for the SR $\tilde{t} \rightarrow b \tilde{\chi}^{\pm}_1, \rm{low\,} \Delta M, E^{\rm miss}_T > 250~{\rm GeV}$ of the analysis CMS-SUS-13-011. For each benchmark point, the first number indicates the stop mass, the second the LSP mass and the third one is $x$, defined as $m_{\tilde{\chi}^+_1} = x \cdot m_{\tilde{t}_1} + (1 - x) m_{\tilde{\chi}^0_1}$.
\label{tab:nsusy-NevStopbcharginoLowDeltaMMET2501}}
\renewcommand\arraystretch{1.2}
\begin{center}
\begin{tabular}{l|c|c}
\hline
\multicolumn{3}{c}{$\tilde{t} \rightarrow b \tilde{\chi}^{\pm}_1, \rm{low\,} \Delta M, E^{\rm miss}_T > 250~{\rm GeV}$} \\ 
benchmark point & CMS result & {\sc MadAnalysis}\,5 result \\ 
$\tilde{t} \rightarrow b \tilde{\chi}^{\pm}_1 \  (650/50/0.5)$ & $8.4 \pm 0.4$ & $8.0$ \\ 
$\tilde{t} \rightarrow b \tilde{\chi}^{\pm}_1 \  (650/50/0.75)$ & $11 \pm 0.5$ & $10.4$ \\ 
$\tilde{t} \rightarrow b \tilde{\chi}^{\pm}_1 \  (600/100/0.5)$ & $7.9 \pm 0.5$ & $7.8$ \\ 
$\tilde{t} \rightarrow b \tilde{\chi}^{\pm}_1 \  (450/50/0.25)$ & $8.7 \pm 1.4$ & $15.2$ \\ 
$\tilde{t} \rightarrow b \tilde{\chi}^{\pm}_1 \  (250/50/0.5)$ & $21 \pm 3.4$ & $34.2$ \\ 
$\tilde{t} \rightarrow b \tilde{\chi}^{\pm}_1 \  (250/50/0.75)$ & $56 \pm 6.4$ & $87.1$ \\ 
\hline 
\end{tabular}
\end{center}
\end{table}

\begin{table}
\caption{Final number of events for the SR $\tilde{t} \rightarrow t \tilde{\chi}^0_1, \rm{low\,} \Delta M, E^{\rm miss}_T > 250~{\rm GeV}$ of the analysis CMS-SUS-13-011. For each benchmark point, the first number indicates the stop mass, the second the LSP mass.
\label{tab:nsusy-NevStopTneutralinoLowDeltaMMET2502}}
\renewcommand\arraystretch{1.2}
\begin{center}
\begin{tabular}{l|c|c}
\hline
\multicolumn{3}{c}{$\tilde{t} \rightarrow t \tilde{\chi}^0_1, \rm{low\,} \Delta M, E^{\rm miss}_T > 250~{\rm GeV}$} \\ 
benchmark point & CMS result & {\sc MadAnalysis}\,5 result \\ 
$\tilde{t} \rightarrow t \tilde{\chi}^0_1 \  (650/50)$ & $6.2 \pm 0.1$ & $7.2$ \\ 
$\tilde{t} \rightarrow t \tilde{\chi}^0_1 \  (250/50)$ & $12 \pm 1.2$ & $16.9$ \\ 
\hline 
\end{tabular}
\end{center}
\end{table}

\subsection{ATLAS-CONF-2013-048: search for stop quarks in the dilepton final state}

The ATLAS search for stops in the dilepton and missing transverse energy, $2 \ell + E^{\rm miss}_T$, final state with full luminosity at $\sqrt{s} = 8$~TeV~\cite{ATLAS-CONF-2013-048} has also been implemented within {\sc MadAnalysis}\,5. Four SRs are defined with increasing requirements on $M_{T2}$ and 
the hadronic activity in the final state. All SRs target the pair production of stop quarks, followed by a $\tilde{t}_1 \to b \tilde{\chi}^{+}_1$ decay. The different cuts on $M_{T2}$ and on the hadronic activity are designed to provide sensitivity to both small and large mass differences between the stop quark and the chargino, as well as between the chargino and the LSP.
The four SRs are then divided according to the flavor of the two leptons: Same Flavor (SF, $ee + \mu\mu$) and Different Flavor (DF, $e\mu$).

Two benchmark points have been employed in this analysis, \mbox{$(m_{\tilde{t}_1}, m_{\tilde{\chi}^+_1}, m_{\tilde{\chi}^0_1})$} $=$ \mbox{$(150, 120, 1)$~GeV} and \mbox{$(400, 250, 1)$~GeV}.
The gauge-eigenstate composition ({\it i.e.} the mixing) of the superpartners, which affects the kinematics, is not specified in the paper. We take the stop to be left-handed, the lightest chargino to be wino-like, and the LSP to be bino-like.
The validation material consists of histograms showing the transverse hadronic activity in the event, $p_{Tb}^{\ell\ell}$, 
and the kinematic variable $M_{T2}$ in the different SRs (Figures~1--5 of Ref.~\cite{ATLAS-CONF-2013-048}). 
Apart from $M_{T2}$, the distributions are all presented after all cuts. 
Moreover, Appendix A of the conference note provides a cut flow for the benchmark point $(m_{\tilde{t}_1}, m_{\tilde{\chi}^+_1}, m_{\tilde{\chi}^0_1}) = (400, 250, 1)$~GeV. This cut flow is indeed valuable 
for the validation process, which is however still in an early stage.

\subsection{ATLAS-SUSY-2013-05: Search for direct third-generation squark pair production in final states with missing transverse momentum and two $b$-jets}

In this ATLAS analysis \cite{Aad:2013ija}, $\tilde{b}_1$ and $\tilde{t}_1$ quarks are searched for in final states with 
large missing transverse momentum and two jets identified as $b$-jets. The results are presented for
an integrated luminosity of $20.1$~fb$^{-1}$ of data and
two possible sets of SUSY mass spectra are
investigated. In the first one, the lightest sbottom, $\tilde{b}_1$,  is the only 
colored sparticle being produced and is assumed to
decay exclusively via the $\tilde{b}_1 \rightarrow b \tilde{\chi}_1^0$ decay mode. In the
second set of spectra, the lightest stop quark, $\tilde{t}_1$, is this time the only colored sparticle
that can be produced and it decays via the $\tilde{t}_1 \rightarrow b
\tilde{\chi}_1^\pm$ channel, with undetectable products of the subsequent decay of the
chargino $\tilde{\chi}_1^\pm$ due to the small mass splitting between the
$\tilde{\chi}_1^\pm$ and the $\tilde{\chi}_1^0$ states.
Two sets of SRs, denoted by SRA and SRB, are defined to provide sensitivity to the kinematic topologies
associated with the two sets of SUSY mass spectra. SRA targets
signal events with large mass splittings between the squark and the neutralino, 
while SRB is designed to enhance the sensitivity when the
squark--neutralino mass difference is small; for the benchmarks points considered
here, $\Delta m = m_{\tilde{\chi}_1^\pm} - m_{\tilde{\chi}_1^0}= 5$ GeV. 

Let us first describe the selection cuts common to both regions. Events are selected by requiring
a large amount of missing transverse energy,
\mbox{$E^{\rm miss}_T > 150$~GeV}, and any event containing an identified muon or electron is vetoed. For the SR
selections, all jets with a pseudorapidity $|\eta| < 2.8$ are ordered according to their $p_T$, and two out of the
$n$ identified jets are required to be $b$-tagged. The following variables are then defined:
\begin{itemize}
\item[--] $\Delta \phi_{\rm min}$ is defined as the minimum azimuthal distance,
$\Delta \phi$, between any of the three leading jets and the $\mathbf{p}_T^{\rm
miss}$ vector. Multijet background events are typically characterized by small
values of $\Delta \phi_{\rm min}$;

\item[--] $m_{\rm eff}$ is defined as the scalar sum of the $p_T$ of the $k$
leading jets and the $E^{\rm miss}_T$, $k=2$ for SRA and $k=3$ for SRB;

\item[--] $H_{T,3}$ is defined as the scalar sum of the $p_T$ of the $n$ jets,
without including the three leading jets;

\item[--] $m_{bb}$ is defined as the invariant mass of the system of the two $b$-tagged jets;

\item[--] $m_{\rm CT}$ is the contransverse mass \cite{Tovey:2008ui}, a
kinematic variable that can be used to measure the masses of
pair-produced semi-invisibly decaying heavy particles. In this analysis, the
two $b$-jets originating from the squark decays are the visible particles and the invisible
particles are either the two $\tilde{\chi}_1^0$ particles or the decay products of the charginos,
depending on the considered benchmark scenario. A correction to $m_{\rm CT}$  for the
transverse boost due to initial state radiation is also applied~\cite{Polesello:2009rn}. The $m_{\rm
CT}$ variable was implemented in our analysis using the publicly available library 
provided by the authors of Ref.~\cite{Polesello:2009rn}\footnote{The
library can be downloaded from \texttt{http://projects.hepforge.org/mctlib}.}.
\end{itemize}\noindent

In the SRA, the first two leading jets must be $b$-tagged and are
identified as the sbottom or stop decay products. The event is vetoed if any
additional central jet ($|\eta| < 2.8$) with $p_T > 50$ GeV is found. To reject
the multijet background large $\Delta \phi_{\rm min}$ and $E^{\rm
miss}_T/m_{\rm eff}$ are required. To reduce the SM backgrounds a cut on $m_{bb}
> 200$ GeV is applied. As a final selection, five different thresholds on $m_{\rm CT
}$ ranging from 150 GeV to 350 GeV are demanded. 

In SRB, the sensitivity to small squark-neutralino mass difference is increased
by selecting events whose leading jet have a very large $p_T$, which is likely
to have been produced from initial state radiation, recoiling against the
squark-pair system. High thresholds on the leading jet and on the missing transverse
momentum, which are required to be almost back-to-back in $\phi$, are imposed.
The leading jet is required to be non-$b$-tagged and two additional jets are
required to be $b$-tagged. Just like for SRA, large values of $\Delta \phi_{\rm
min}$ and $E^{\rm miss}_T/m_{\rm eff}$ are required, thereby suppressing the
multijet background. The selection for SRB is finally completed by demanding
that the additional hadronic activity is bounded from above, $H_{T,3} < 50$ GeV.

As validation material, several kinematic variable spectra are available in
Ref.~\cite{Aad:2013ija}. 
For SRA, both the $m_{\rm CT}$ and $m_{bb}$ distributions are presented
for two benchmark points,
$(m_{\tilde{b}_1},m_{\tilde{\chi}_1^0})=(500,1)$~GeV and
$(m_{\tilde{t}_1},m_{\tilde{\chi}_1^0})=(500,100)$~GeV with $\Delta m = 5$ GeV.
After the SRA selection with $m_{\rm CT} > 250$ GeV, 3\% of the simulated events
are retained for the signal point corresponding to
$(m_{\tilde{b}_1},m_{\tilde{\chi}_1^0})=(500,1)$~GeV. 

For SRB, the missing transverse-momentum distribution
is shown with all
other selection criteria applied. Moreover, the distribution of $H_{T,3}$ is provided with all cuts but
the $H_{T,3}$ requirement. The results are again shown for two benchmark points:
$(m_{\tilde{b}_1},m_{\tilde{\chi}_1^0})=(300,200)$ GeV and
$(m_{\tilde{t}_1},m_{\tilde{\chi}_1^0})=(250,150)$ GeV with $\Delta m = 5$ GeV.

To validate our implementation of this analysis we simulated $10^5$ events for
each benchmark point using \textsc{MadGraph}~5 interfaced to
\textsc{Pythia}~6, with the CTEQ6L1~\cite{Pumplin:2002vw} set of parton distribution functions. 
The signal cross
sections were normalized to the 8 TeV predictions computed at the NLO+NLL accuracy in
the strong coupling constant \cite{8tevxs_susy}. 
At the time of writing, the validation is still in progress. This task is made difficult by
the fact that only the final distributions, once almost all the cuts have been
applied, are provided; since no cut-flows are available, tracking the
validation procedure step-by-step is not possible. 
A complete description of the configuration employed in the event simulation 
together with a cut-flow would make the validation considerably easier.



\section{NEW DEVELOPMENTS IN {\sc MadAnalysis\,5} and {\sc Delphes}}\label{sec:nsusy-MA5new}

\subsection{Region selection manager: dealing with multiple signal regions} \label{sec:MA5SRM}

In many experimental analyses performed at the LHC, including all of the new physics searches considered
in this work, a branching set of selection criteria (``cuts'')
is used to define several different sub-analyses (``regions'') within the same analysis.
Versions of {\sc MadAnalysis}\,5  older than v.1.1.10, however, were not expected
to deal with this situation: the program was mainly dedicated to the design of prospective
analyses, for which more than one sub-analysis is typically unnecessary.
In other coding frameworks, multiple regions can be implemented but only with a nesting of conditions checking the cuts, which grows exponentially more complicated with the number of cuts.
The scope of this project therefore motivated us to develop the {\sc MadAnalysis}\,5 package
to facilitate the handling of analyses with multiple regions.

Analyses in the {\sc MadAnalysis}\,5 framework are
divided into three functions:
\begin{itemize}
  \item \texttt{Initialize}, which is dedicated to the initialization of the signal regions, histograms
    and cuts;
  \item \texttt{Execute}, which consists of the analysis to be applied to each event; and
  \item \texttt{Finalize}, which controls the production of histograms and cut-flow charts,
    or in other words, the results of the analysis.
\end{itemize}
The function \texttt{Finalize} has been internally automated within {\sc MadAnalysis}\,5,
so that the user can skip its implementation if desired; the two other methods however should be
written by the user to suit his/her physics needs.

For the rest of this section we will illustrate the handling of multiple regions in the new
{\sc MadAnalysis}\,5 framework by showing extracts of
our implementation of the search CMS-SUS-13-011 (see Section \ref{sec:nsusy-TopSquark1lepton}).
This search defines 16 different SRs, and hence our \texttt{Initialize} function
contains the initialization/declaration of 16 regions, all using the same syntax. 
For illustration, two of the 16 SRs are declared as 
\begin{verbatim}
Manager()->AddRegionSelection("Stop->t+neutralino,LowDeltaM,MET>150");
Manager()->AddRegionSelection("Stop->t+neutralino,LowDeltaM,MET>200");
\end{verbatim}
The declaration of each region relies on the \texttt{AddRegionSelection} method
of the analysis manager class, of which \texttt{Manager()} is an instance.
It takes as its argument
a string which uniquely defines the SR under consideration.

Selection cuts must also be declared in \texttt{Initialize}. They fall into
two categories: those which are
common to all regions, and those which are not (\textit{i.e.} those which apply only
to some of the regions, thus serving to define the different regions).
Two examples of the declaration of common cuts are:
\begin{verbatim}
  Manager()->AddCut("1+ candidate lepton");
  Manager()->AddCut("1 signal lepton");
\end{verbatim}
The \texttt{AddCut} method of the analysis manager class has been used, which takes
as argument a string which uniquely defines the cut. The declaration of
cuts shared by some (but not all) regions makes use of the
same \texttt{AddCut} function but requires a second argument: either a string or
an array of strings, consisting of the names of all the regions to which
the cut under consideration applies. For instance, we have
\begin{verbatim}
  string SRForMet150Cut[] = {"Stop->b+chargino,LowDeltaM,MET>150",
    "Stop->b+chargino,HighDeltaM,MET>150",
    "Stop->t+neutralino,LowDeltaM,MET>150",
    "Stop->t+neutralino,HighDeltaM,MET>150"};
  Manager()->AddCut("MET > 150 GeV",SRForMet150Cut);
\end{verbatim}
for a cut on the missing transverse energy that applies to four of the SRs of the
CMS-SUS-13-011 analysis.

Histograms are initialized using the \texttt{AddHisto} method.
As for the \texttt{AddCut} method, a string argument is required
to act as a unique identifier for the object (histogram or cut), and a further optional argument consisting
of a string or array of strings can be used to associate the object to desired regions.

We now move to the description of the
\texttt{Execute} function that contains the analysis itself. It mainly relies on
standard methods described in the manual of
{\sc MadAnalysis}\,5~\cite{Conte:2012fm}, which are used
to declare particle objects and compute observables related either to specific particles
or to the event as a whole. We therefore restrict ourselves here to
a description of the new manner in which cuts are applied and histograms filled,
using the analysis manager class developed in this
work. Having declared and filled a vector
\texttt{SignalLeptons} with objects satisfying the signal lepton
definition used in CMS-SUS-13-011,
we impose a selection cut demanding exactly one signal lepton
by including, in the \texttt{Execute} method of the analysis, the lines
\begin{verbatim}
  unsigned int nsl = SignalLeptons.size();
  if( !Manager()->ApplyCut((nsl==1), "1 signal lepton"))
    return;
\end{verbatim}
Calling the \texttt{ApplyCut} using the syntax \mbox{\texttt{if(!Manager()->ApplyCut($\ldots$)) {return;}}} ensures that we stop analysing any given event as early as possible if all regions fail the cuts.

In the new framework, histogramming becomes as easy as
applying a cut. For example in this search we are interested in
the transverse-momentum spectrum of the leading lepton, and thus our code contains
\begin{verbatim}
  Manager()->FillHisto("pT(l)", SignalLeptons[0]->momentum().Pt());
\end{verbatim}
This fills the histogram which was previously declared with the name \texttt{"pT(l)"} in
the \texttt{Initialize} method.
The filling or not of histograms, according to whether cuts have been passed, is handled
automatically.

For both selection cuts and histograms, all the
kinematical observables described in Ref.~\cite{Conte:2012fm} can be
employed, as well as the new \texttt{isolCones()} method attached to the \texttt{RecLeptonFormat}
class. This returns a vector of \texttt{IsolationConeType}
objects describing the transverse activity
in a cone of radius $\Delta R$ centered on the lepton. The properties of such objects can
be accessed through
\begin{itemize}
 \item \texttt{deltaR()}: returns the size of the cone,
 \item \texttt{sumPT()}: returns the scalar sum of the transverse momenta of the tracks lying
   in a cone of radius $\Delta R$ centered on the lepton,
 \item \texttt{sumET()}: returns the scalar sum of the transverse energy deposits
   in a cone of radius $\Delta R$ centered on the lepton.
\end{itemize}
These features can be used together with the modifications of {\sc Delphes}~3
described in Section~\ref{sec:nsusy-moddelphes3}.

\subsection{Special kinematic variables}\label{sec:nsusy-kinem}

It is helpful if special kinematic variables, which appear repeatedly in the experimental analyses, 
are directly available for the user. For this reason we have added two such variables, 
namely $M_{T2}$~\cite{Lester:1999tx} and $M_{T2}^W$~\cite{Bai:2012gs}, 
in the \texttt{PHYSICS} class of {\sc MadAnalysis}\,5. 

The $M_{T2}$ variable has been implemented following the algorithm of
Ref.~\cite{Cheng:2008hk}. It can be computed with the function 
\begin{verbatim}
  PHYSICS->Transverse->MT2(p1,p2,met,mass) 
\end{verbatim}
in the \texttt{Execute} function of a given analysis.
In this notation we assume the pair production of a new state decaying into
an invisible particle, whose mass is represented by the variable \texttt{mass},
and a visible particle. The final state thus contains two visible particles 
referred to by the \texttt{p1} and \texttt{p2} objects and some
missing transverse energy represented by the object \texttt{met}.

The $M_{T2}^W$ variable has been implemented following the algorithm presented
in Ref.~\cite{Bai:2012gs}. The observable can be computed using
\begin{verbatim}
  PHYSICS->Transverse->MT2W(jets,lep,met)
\end{verbatim}
in the \texttt{Execute} function of an analysis. 
Here, the variable \texttt{jets} is a vector containing all jets of the event under consideration, 
\texttt{lep} a single lepton candidate and \texttt{met} the missing transverse energy of the event. 
Only the three leading jets are considered
for the computation of the $M_{T2}^W$ variable and, if available,
the $b$-tagging information is used.

Further kinematical variables will be added as the need arises.

\subsection{Modifications in {\sc Delphes3}} \label{sec:nsusy-moddelphes3}

\textsc{Delphes}~\cite{deFavereau:2013fsa} is a {\sc C++} framework which allows one to simulate a generic detector used in collider experiments. \textsc{Delphes} does not simulate fully the particle-matter interactions but uses detector response parameterizations and reconstructs the main physics objects. The speed of the simulation is therefore enhanced and the accuracy level is suitable for phenomenological investigations.\\
From the computing side, \textsc{Delphes} is a very modular framework where developers are invited to tune the default parameterization and add their own contributions. This modularity is based on a simulation process split into modules (derived from the \texttt{TTask} \textsc{Root} class). The configuration file using the {\sc Tcl} script language allows to easily add or to remove some modules. The content of the produced \textsc{Root} files can be configured in the same way.
According to the goals of the present physics project, a tuning of the \textsc{Delphes3} release has been performed in order to supply all required information for recasting. The main changes are the following:
\begin{itemize}
\item In the initial \textsc{Delphes}3 simulation processing, an isolation criterion is applied to both leptons and photons. Only particles satisfying the criterion are saved in the \textsc{Root} file. In our case, this isolation criterion is a part of the analysis selection. That is why a new \textsc{Delphes} module called \texttt{CalculationIsolation} has been implemented to compute some isolation variables: the scalar sum of the track transverse momenta, the scalar sum of the calorimeter tower transverse energies and the number of tracks in the isolation cone. These variables are calculated for different isolation cone sizes, $\Delta R=0.5,\, 0.4,\, 0.3$ and $0.2$. The initial \textsc{Delphes} module filtering the lepton and photon candidates according to their isolation is skipped in order to keep all the candidates in the \textsc{Root} files. The isolation selection cut is thus postponed to the analysis step.
This development has been achieved together with the \texttt{isolCones()} method of the \texttt{RecLeptonFormat} class
of {\sc MadAnalysis}~5 (Section~\ref{sec:MA5SRM}).

\item As isolated leptons and isolated photons are not defined during the \textsc{Delphes} simulation, the module \texttt{UniqueObjectFinder} which gives a unique identification to reconstructed objects is by-passed.

\item Adding isolation variables to the data format increases the size of produced \textsc{Root} files. A cleaning of the collections is in order to reduce the file size. Collections such as calorimeter towers and particle-flow objects are not stored. The remaining heavy collection is the collection of generated particles at the
hard scattering process level and after parton showering and hadronization. This information is however useful to perform some cross-checks by matching reconstructed objects with generated particles. A skim is applied in order to keep only particles produced
at the hard-scattering process level, final state leptons and $b$-quarks after showering. The size of the produced \textsc{Root} file is in this way divided by ten with respect to that obtained with the default configuration file.

\item The parameterization related to the $b$-jet tagging efficiency and the rate of $b$-jet misidentification has been tuned in order to better mimic the performance of the CMS detector~\cite{CMS-PAS-BTV-09-001,CMS-PAS-BTV-11-001}.

\end{itemize}

Note that the latest release of \textsc{Delphes} can also simulate the presence of pile-up events and the corresponding degradation of the detector reconstruction performance. This functionality is however not used in this project due to the lack of validation results.

\section{MONTE CARLO EVENT GENERATION}

Using the SUSY spectra created by {\sc SPheno}~3,
we produce events at the parton-level for sbottom and stop pair production with up to two additional partons with {\sc MadGraph}~5~\cite{Alwall:2011uj}. Our simulation setup differs from the default loaded when starting {\sc MadGraph}~5. First,
we take into account initial $b$-quarks and antiquarks. We then work with the \texttt{mssm-full} model~\cite{Duhr:2011se}
of {\sc MadGraph}~5 and the CTEQ6L1 set of parton densities. We then apply the loose generator cuts given in Table~\ref{tab:Cuts_used_for_production}. 
\begin{table}[h!]
\caption{Loose cuts applied during the event generation.}
\label{tab:Cuts_used_for_production}
\begin{center}
\begin{tabular}{ l l }
\hline
Minimal distance between two (parton-level) jets & 0.001  \\
Minimal distance between two (parton-level) $b$-jets & 0.001 \\
Minimal distance between a (parton-level) $b$-jet and a (parton-level) light jet & 0.001\\
\hline
\end{tabular}
\end{center}
\end{table}
As we produce up to two extra partons in the final state, we will need to merge the different multiplicities 
and match them to the parton shower. This is performed with {\sc Pythia}~6~\cite{Sjostrand:2006za} and following
the MLM standard procedure~\cite{Mangano:2006rw}, although the exact values for the merging parameters 
need to be determined process by process.

\begin{figure}
\centering
\includegraphics[width=0.42\textwidth]{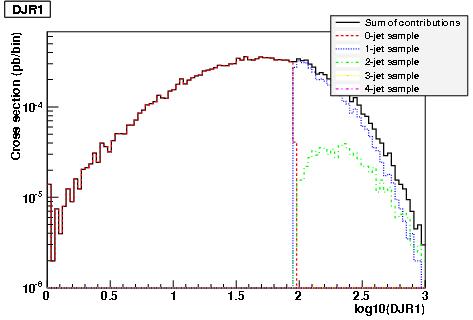}
\hspace{1eM}
\includegraphics[width=0.42\textwidth]{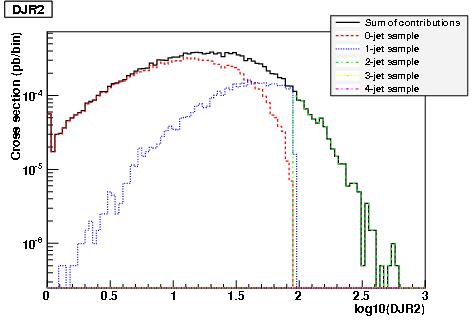}
\caption{Differential jet rate spectra for stop pair production with $m_{\tilde{t}_1} = 600$ GeV. 
On the left the DJR distribution for $1 \rightarrow 0$ transition is shown, while on the right we show
the $2 \rightarrow 1$ DJR spectra.}
\label{fig:nsusy-DJR_stop600}
\end{figure}

A correct choice for the merging parameters is crucial as it yields a consistent and smooth splitting of the phase 
space into regions dominated by matrix-element-based predictions and regions where parton showering correctly describes
QCD radiation. The smoothness of the transition between these regions, or equivalently a check of the merging setup, 
can be investigated via differential jet rate distributions, a class of variables consisting of the distributions of 
the scale at which a specific event switches from a $N$-jet configuration to a $N+1$-jet configuration.
In order to determine the setup for the processes under consideration, 
we performed a scan over the two main parameters to determine the \texttt{xqcut} and \texttt{qcut} parameters
of {\sc MadGraph}~5 and {\sc Pythia}~6, respectively. 
We scanned {\tt xqcut} from 10 to 70 GeV, and varied {\tt qcut} in the range [{\tt xqcut} + 5 GeV, 2$\cdot${\tt xqcut}]. 
For each combination we generated 60000 events, processed them up to the parton shower and computed the DJR spectra.
A wide range of {\tt xqcut} and {\tt qcut} combinations were observed to be acceptable, 
with only a slight mass dependence. 
We illustrate this validation for stop pair production with $N=0$ (left panel) and $N=1$ in 
Figure~\ref{fig:nsusy-DJR_stop600} (right panel). 
The final values are {\tt xqcut}=50~GeV and {\tt qcut}=90~GeV,
for pair production of fully left-handed stop quarks of mass  $m_{\tilde{t}_1} = 600$ GeV.
These values were found to be appropriate for stop and sbottom masses ranging up to 1 TeV, 
and were cross-checked by varying the gluino mass between 500 GeV and 2 TeV, 
and by changing the stop and sbottom mixing matrices. 

With this choice of parameters, we produced 50000 events for each process and parameter space point. 
The computational resources required for this were accessed via the GlideinWMS \cite{Sfiligoi:2009} on the Open Science Grid~\cite{OpenScienceGrid}. We decayed the sparticles in the events according to the branching ratios given in the SLHA file from {\sc SPheno} and applied parton showering and hadronization with {\sc Pythia}~6. 
The resulting event samples can then be parsed through the re-implemented analyses described in Section~\ref{sec:nsusy-recast}
after including a simulation of the detector response (see Section~\ref{sec:nsusy-moddelphes3}).

\begin{figure}
\centering
\includegraphics[scale=1]{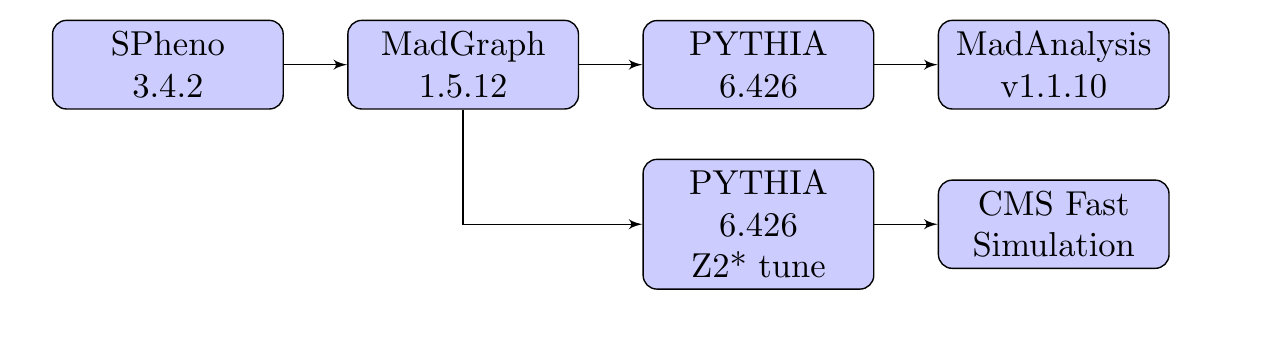}
\caption{Schematic depiction of the programs used in the production chain.}
\label{fig:nsusy-production_chain}
\end{figure}

The (hard-scattering-level) LHE files stemming from {\sc MadGraph}~5 will be also handed over to CMS for further processing. 
The sparticles therein will first be decayed using {\sc Pythia}~6, according the SLHA decay table 
for that particular parameter point. 
Once this is done, they will go through the usual CMS procedure for producing SUSY scans. 
The parton shower and hadronization will be done with {\sc Pythia}~6, using the Z2* tune, 
after which the events will be processed with the official CMS Fast Simulation software.  
Once completed, the scan will be made available to the SUSY analysis groups, 
who can then run their analysis on the sample, and provide efficiencies 
and upper limits for the considered parameter points. 
A flowchart of the production chain is depicted in Figure~\ref{fig:nsusy-production_chain}. 

\section{CONCLUSIONS}

Our project is targeted at
\begin{itemize}
\item[1)] assessing in detail the status of ``natural supersymmetry'' after the 8~TeV LHC run, and 
\item[2)] developing a general analysis framework, within {\sc MadAnalysis}\,5,
which can serve as a platform for re-implementing ATLAS and CMS search results
for new physics.
\end{itemize}
In the present contribution, we reported on the progress of
this project. We presented a scan over physical parameters, namely stop and
sbottom masses and mixing angles, and higgsino masses, which lays the basis of
any later re-interpretation of the experimental results. The SLHA files from
this scan are readily available and may also be used by ATLAS and/or CMS 
for interpreting their results.

Analyzing these SLHA spectra with {\sc SModelS} in the Simplified Models approach, we found that the exclusion 
limits indeed seem to depend quite sensitively on the precise scenario at hand -- even if it contains only 
a light $\tilde t_1$, light higgsinos ($\tilde\chi^0_1$, $\tilde\chi^0_2$ and $\tilde\chi^\pm_1$), and 
possibly a   light $\tilde b_1$.

The implementation of the relevant ATLAS and CMS searches for stops and sbottoms in the 
{\sc MadAnalysis}\,5 framework first required several extensions of {\sc MadAnalysis}\,5 as well as
some modifications of {\sc Delphes}\,3, in order to adapt them for this project. A large part of the work 
thus concerned developing the necessary tools. This part is now concluded. 

The process of implementation and validation is still ongoing. It is impeded by the fact that  
the information publicly available in the experimental papers and/or analyses notes and/or 
on Twiki pages is usually not sufficient for an unambiguous validation. It would be of immense 
advantage for efforts like the one which we are undertaking here, if the experimental collaborations 
followed more closely the 
{\it Les Houches Recommendations for the Presentation of LHC Results}~\cite{oai:arXiv.org:1203.2489}. 
In particular it would be extremely helpful if well-defined benchmark points, efficiencies and cut-flows were made 
available in a more systematic manner.  

Two other program packages for confronting new physics scenarios with LHC Data were published recently.
They are largely complementary to the ones we have been discussing in this contribution. 
The first one, {\sc CheckMATE}~\cite{Drees:2013wra}, is based on fast simulation and determines  
whether a model is excluded or not at 95\% confidence level by comparing to several recent experimental analyses; 
the analyses currently implemented in   {\sc CheckMATE} are mostly from ATLAS and have only little 
overlap with the ones we are interested in in this project. 
The second program is {\sc Fastlim}~\cite{Papucci:2014rja} and reconstructs the visible cross sections 
[of SUSY events] from pre-calculated efficiency tables and cross section tables for simplified event topologies. 
{\sc Fastlim} has only ATLAS analyses implemented, including however in particular those targeted 
at stop and bottom searches.
It will be interesting to compare the different approaches -- a comparison between {\sc SModelS} and {\sc Fastlim} is on the way.

\section*{ACKNOWLEDGEMENTS}

We gratefully acknowledge the use of the computational power provided by the Open Science Grid~\cite{OpenScienceGrid}
that were accessed via GlideinWMS~\cite{Sfiligoi:2009}.

This work was financed in part by the Theory-LHC-France initiative of the
CNRS/IN2P3, the PEPS-PTI project ``LHC-itools'' and by the EGIDE/DAAD project nr.\  PROCOPE 54366394. 
EC and BF acknowledge partial support from the French ANR 12 JS05 002 01 BATS@LHC.
LM and WP are supported by the BMBF, project nr.\ 05H12WWE. 
LM acknowledges moreover support from the Elitenetzwerk Bayern. 
NS is supported by a Ph.D.\ fellowship of the Research Foundation - Flanders (FWO).


%% file: lrmssm/lrmssm.tex

\def\bpm{\begin{pmatrix}} 
\def\epm{\end{pmatrix}} 
\def\bea{\begin{eqnarray}}
\def\eea{\end{eqnarray}}


\chapter{Reviving Minimal Left-Right Supersymmetry in the Light of LHC Data}

{\it A.~Alloul, L.~Basso, B.~Fuks, M. E.~Krauss, W.~Porod}


\begin{abstract}
In the context of left-right supersymmetric theories we present a reinterpretation
of three experimental searches dedicated to signatures
induced by a $W'$-boson, of which one is sensitive to the presence of
right-handed neutrinos. We emphasize that according to the way the experimental
results are provided, this task can be either easily feasible or could be
rather complex, if not impossible in the general way. We consequently provide
recommendations to improve the situation in the future.
\end{abstract}

\section{INTRODUCTION}
Large classes of theories have been proposed over the last decades to
extend the Standard Model (SM) and provide tentative answers to one or several of
its conceptual questions. Among those, weak scale
supersymmetry is one of the most studied
options, both at the theoretical and experimental levels. However, no hint for
any superpartner has been found so far, both through direct searches at
colliders and via indirect probes at low-energy experiments.
Most results have been
derived under either the framework of the so-called Minimal Supersymmetric
Standard Model (MSSM) or by assuming simplified models inspired by the 
latter.
Often these minimal assumptions are too limiting. There exist
alternative, non-minimal realizations of
supersymmetry that evade the current bounds and at the same time
have novel features possibly not covered by current searches.

In this report, we consider one of these non-minimal supersymmetric
theories that exhibit a left-right (LR) symmetry~\cite{Francis:1990pi,
Huitu:1993uv,Huitu:1993gf,Babu:2008ep}, and focus on the consequences of the
presence of an additional charged gauge boson, commonly denoted by $W'$ 
(or, as we will use from now on, $W_R$),
arising from the extended gauge structure. In contrast to the
non-supersymmetric case, the new boson can decay into lighter supersymmetric
particles, which implies that the branching ratios into commonly
searched for signatures, and thus the derived bounds on the $W_R$-boson mass
and couplings, could be reduced. To illustrate this statement, we recast three
recent CMS searches in a left-right supersymmetric framework.
We start by revisiting two classical $W_R$ analyses
where the new charged gauge boson is expected to decay either into a pair of
jets~\cite{CMS-PAS-EXO-12-059} or into a two-body system comprised of a top and of a bottom
quark~\cite{CMS-PAS-B2G-12-010}. In the first analysis, the experimental results have been
derived in the very specific theoretical framework constructed in
Refs.~\cite{Pati:1974yy,Mohapatra:1974hk,Mohapatra:1980yp,Mohapatra:1983aa}
while in the second analysis, an effective Lagrangian encompassing a $W'$-boson
coupling uniquely to quarks has been employed. Next, we focus on
regions of the parameter space where right-handed neutrinos can be copiously
produced at the Large Hadron Collider (LHC), and reinterpret the
results of a recent search for right-handed neutrinos based on a
simplified setup where they always decay into a lepton plus dijet final
state~\cite{CMS-PAS-EXO-12-017}.

The rest of this contribution is organized as follows: in
Section~\ref{sec:lrmssm_model}, we briefly describe our theoretical framework
and provide information on our choices of benchmark scenarios and on the
simulation chain that has been employed to generate signal events. Our results
related to the reinterpretation of the three above-mentioned analyses are then
shown in Section~\ref{sec:lrmssm_results}, where we also
emphasize that the way in which the
results are presented is not often appropriate for recasting. We therefore
take the opportunity to provide recommendations aiming to facilitate the
communication between theorists and experimentalists in the future.
Our conclusions are presented in Section~\ref{sec:lrmssm_conclusions}

\section{THE MODEL}\label{sec:lrmssm_model}
\subsection{Model description}
There exist many versions of left-right symmetric supersymmetric 
theories~\cite{Francis:1990pi,Huitu:1993uv, Huitu:1993gf,
Babu:2008ep},
each based on the $SU(3)_c \times SU(2)_L \times SU(2)_R \times
U(1)_{B-L}$ gauge group. Aside the gauge structure,
all these models include four vector superfields, directly
related to the gauge structure, and (s)quark and (s)lepton degrees of 
freedom organized
in left- and right-handed doublets of both $SU(2)$ symmetries. 
They differentiate in
the Higgs sector, that is generally very rich in order to allow for a successful
symmetry breaking scheme without spoiling current data. In the
scenario we have adopted, it comprises two $SU(2)_L \times SU(2)_R$
bidoublets $\Phi_1$ and $\Phi_2$ not sensitive to the $B-L$ symmetry, two
$SU(2)_L$ ($SU(2)_R$) triplets $\Delta_{1L}$ ($\Delta_{1R}$)
and $\Delta_{2L}$ ($\Delta_{2R}$) with  $B-L$ charges of  $-2$ and $+2$, respectively, and one gauge singlet $S$.

Focusing on the model-dependent part of the superspace action,
the superpotential is given by
\bea
   W &=&\nonumber
   Q_L y_Q^1 \Phi_1 Q_R +
   Q_L y_Q^2 \Phi_2 Q_R +
   L_L y_L^1 \Phi_1 L_R +
   L_L y_L^2 \Phi_2 L_R +
   L_L y_L^3 \Delta_{2L} L_L +
   L_R y_L^4 \Delta_{1R} L_R \\
&&
   + \Big( \mu_L + \lambda_L S \Big)\Delta_{1L} \cdot \Delta_{2L}
   + \Big( \mu_R + \lambda_R S \Big)\Delta_{1R} \cdot \Delta_{2R} 
   + \Big( \mu_1 + \lambda_1 S \Big)\Phi_1\cdot \Phi_1\\
&&\nonumber
   + \Big( \mu_2 + \lambda_2 S \Big)\Phi_2\cdot \Phi_2
   + \Big( \mu_{12} + \lambda_{12} S \Big)\Phi_1 \cdot \Phi_2
   + \frac13 \lambda_s S^3 + \mu_s S^2 + \xi_s S\ ,
\label{eq:Wtrip}\eea
where we refer to Ref.~\cite{Alloul:2013fra} for details on the underlying
(understood) index structure as well as for a
more extensive description of the model.
In this expression, $Q_L$ ($L_L$) and $Q_R$
($L_R$) are the left- and right-handed doublets of quark (lepton) superfields and
the interaction strengths have been embedded into
$3\times 3$ Yukawa matrices ($y^i_Q$ and $y^j_L$, $i=1,2$, $j=1,\dots,4$), a single linear $\xi$ term,
a set of supersymmetric mass ($\mu$) terms and trilinear Higgs(ino)
self-interactions ($\lambda$).
From 
the superpotential, we can derive the form of the
soft supersymmetry-breaking Lagrangian which 
contains, in addition to scalar and gaugino mass
terms, trilinear scalar interactions of sfermions and Higgs bosons ($T_Q$ and
$T_L$) and linear ($\xi$), bilinear ($B$) and trilinear ($T$) Higgs
self-interactions.

\subsection{Benchmark scenarios and technical setup}
In order to design theoretically motivated benchmark scenarios, we have
implemented the left-right supersymmetric model of the previous subsection into
{\sc Sarah}~\cite{Staub:2012pb,Staub:2013tta}, which allows one to
automatically calculate all the model mass matrices, vertices and tadpole
equations and pass the information to {\sc SPheno}~\cite{Porod:2011nf}
for spectrum calculation at the one-loop level. Several analytical cross
checks with the model implementation in
{\sc FeynRules}~\cite{Christensen:2008py,Alloul:2013bka} have been performed.
The model information has then been translated into the UFO
format~\cite{Degrande:2011ua} and linked to
{\sc MadGraph}~5~\cite{Alwall:2011uj} for simulating LHC collisions and
reinterpreting the three CMS analyses of Section~\ref{sec:lrmssm_results}.
The model implementations have been further validated via a confrontation of
numerical results as computed by both {\sc MadGraph} and
{\sc CalcHep}~\cite{Belyaev:2012qa}.

We have employed the tadpole
equations to calculate the soft masses of the Higgs bidoublets and triplets in terms of the vacuum expectation values (vevs).
We have then scanned over the remaining parameter space on the basis of a setup
analogous to the one of Ref.~\cite{Alloul:2013fra}, to which we refer for more
information. All bilinear superpotential
terms  $\mu_i$ and the corresponding soft breaking parameters ($B_i$) have hence been set to zero,
together with the $\lambda_1$, $\lambda_2$, $\xi_s$ and $\xi$ parameters. The
vev of the gauge singlet Higgs field has been
imposed to be $v_s = 10$~TeV\footnote{Variations from this choice can be
obtained
via the $\lambda$ parameters.}, while most of the other vevs have been taken to be
vanishing, with the exception of those of the neutral scalar component of the
$\Delta_R$ superfields, controlling the $W'$-boson mass, and those of the
scalar component of the $(\Phi_1)^1{}_1$ and $(\Phi_2)^2{}_2$ superfields
related to the weak boson masses.
Moreover, we have imposed that both $SU(2)_L$ and $SU(2)_R$ gauge coupling strengths
are equal at the weak scale, that the spectrum exhibits a Standard Model-like
Higgs boson with a mass in the $122-128~$GeV range, and that a lower bound of
220~GeV on the mass of the doubly-charged Higgs boson is
satisfied~\cite{CMS-PAS-HIG-12-005}\footnote{The lowest bounds on a doubly-charged Higgs boson is 198~GeV when it decays with a branching fraction of 100~\% into two tau leptons. 
Since in general all three lepton flavours are possible as a decay product and the bounds on  the other final states are more severe, we restrict
ourselves to the case where $m_{H^{\pm \pm}_1}>220~$GeV and BR$(H^{\pm \pm}_1 \to \tau \tau) > 0.8$.}.

\section{RECASTING LHC RESULTS}\label{sec:lrmssm_results}
\subsection{$\mathbf{W_R}$ decays into SM-particles}
The search for new heavy charged gauge bosons comprises several search channels. The seemingly most stringent bounds are set by investigating the signature of a charged lepton and missing energy, 
assuming that the new gauge boson decays into a lepton and a low-mass neutrino which escapes detection, see \textit{e.g.} Ref.~\cite{CMS-PAS-EXO-12-060}. 
In a left-right symmetric scenario the results associated with this
search however do not apply.
When neutrino data is explained by some sort of seesaw mechanism, the decay of the $W_R$ into a lepton and a low-mass neutrino is generally suppressed by the small neutrino mixing, while the naive $W_R \to \ell \nu_R$ mode does not lead to missing energy since the right-handed neutrino decays.

\begin{figure}[h]
\begin{center}
\includegraphics[width=0.49\linewidth]{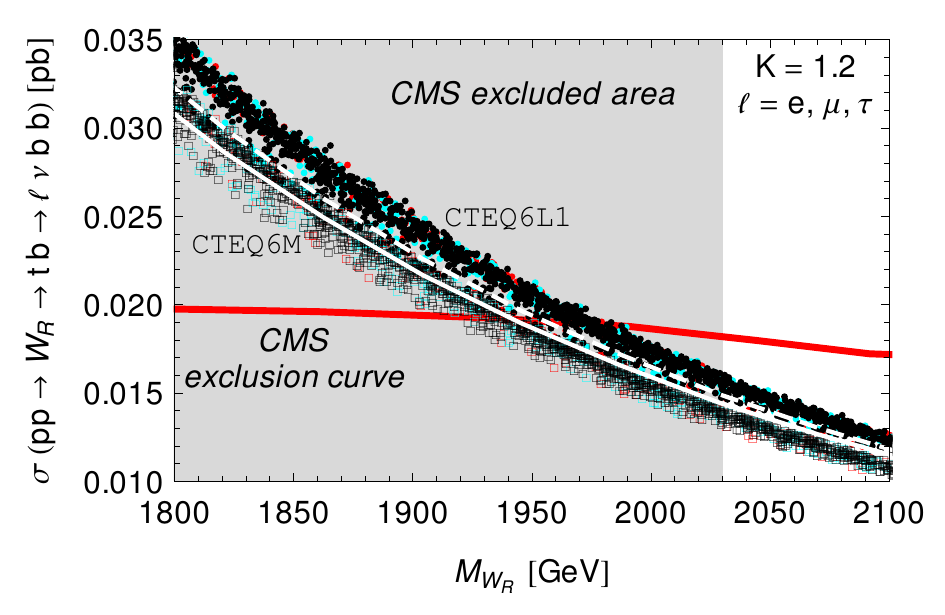}
\includegraphics[width=0.49\linewidth]{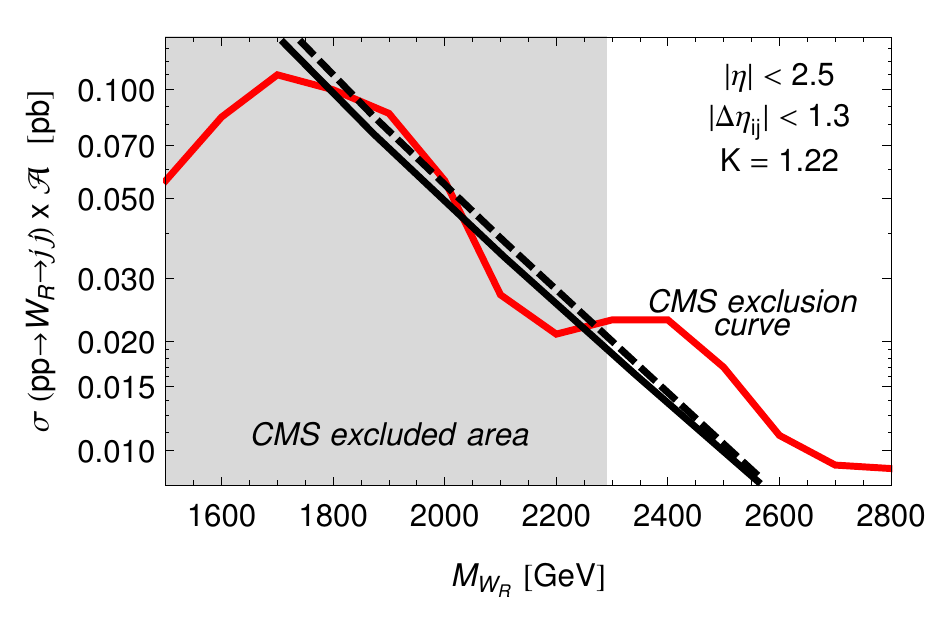}
\caption{(Left) Bounds from $W_R$-boson searches in the $tb$ mode: the red line shows the CMS upper limit on the cross section 
$pp \to W_R \to tb \to \ell \nu b b$, where $\ell = e,\mu,\tau$. Red benchmark points present
a valid spectrum but with $198 < m_{H^{\pm \pm}_1}/\mbox{GeV}<220$ or BR$(H^{\pm \pm}_1 \to \tau \tau) < 0.8$. Cyan points exhibit a more sensible doubly-charged Higgs boson
($m_{H^{\pm \pm}_1}>220~$GeV and BR$(H^{\pm \pm}_1 \to \tau \tau) > 0.8$). Black points are a subset of the former that is relevant for section~\ref{lrssm_subsect_RH}, \textit{i.e.}
when $m_{H^\pm_1} < m_{\nu_{R,i}}$.
The shown cross sections visualized by dots (squares) have been obtained using the PDF
set {\sc CTEQ6L1} ({\sc CTEQ6M}), as explained in the text. The parameters have been varied within the ranges $3.8\leq v_R/\text{TeV}\leq 4.6,~-1 \leq \lambda_{s/L/R} \leq 1,~-0.1 \leq \lambda_{12} \leq 0.1,~200 \leq M_{2L/2R}/\text{GeV} \leq 900,~0 < M_1/\text{GeV} \leq 500,~4 \leq \tan \beta \leq 15$. We also show in white the cross sections of two specific parameter points with reduced branching ratios of the $W_R$-boson to SM particles, evaluated with {\sc CTEQ6L1}. The values have been set to $\lambda_s = 0.8$, $\lambda_L=\lambda_R = -1$, $\lambda_{12}=0.025$, $M_1=M_{2R}=250~$GeV, $M_{2L}=500$~GeV, $\tan \beta=30$ (solid line) and $\lambda_s = 0.3$, $\lambda_L=\lambda_R = -0.4$, $\lambda_{12}=0.04$, $M_1=200$~GeV, $M_{2R}=M_{2L}=550$~GeV, $\tan \beta=10$ (dashed line). \newline
(Right) Bounds from $W_R$-boson searches in the dijet mode: shown is the cross section exclusion line from CMS (red) at 8~TeV center-of-mass energy and 19.6~fb$^{-1}$~\cite{CMS-PAS-EXO-12-059}. The black solid and dashed lines depict the calculated cross section for two parameter choices as in the left figure. 
  \label{lrmssm_fig_1bound}}
\end{center}
\end{figure}

On the other hand, searches with hadronic two-body final states do apply to left-right models since the coupling strength
of the $W_R$-boson to a pair of quarks is equal to that of the SM $W$-boson, given that $g_R=g_L$ holds. The 
tightest current bounds are provided by the CMS searches into a) a top-bottom pair of
quarks~\cite{CMS-PAS-B2G-12-010}, and b) a dijet final state~\cite{CMS-PAS-EXO-12-059}.
In order to have a meaningful comparison, we apply the same strategy of calculating the cross section 
(\textit{i.e.} parton distribution functions (PDFs), $K$-factors and selection requirements) as was used by the
experimental collaborations when deriving the bounds. However,
concerning the bounds for the $tb$ final state of Ref.~\cite{CMS-PAS-B2G-12-010}, 
the choice of the PDF set has been used inconsistently:  in  \cite{CMS-PAS-B2G-12-010} leading order (LO) matrix elements have been convoluted with next-to-leading order (NLO) PDFs and
 the resulting cross section has then been multiplied by an additional $K$-factor.
The correct procedure, however, is to use both (matrix elements and PDFs) at NLO,
or both at LO and to apply in the latter case  a 
$K$-factor such that $\sigma_{LO} \cdot K = \sigma_{NLO}$. In Fig.~\ref{lrmssm_fig_1bound}
we display both the cross section evaluated as it should be, 
using the LO PDF sets {\sc CTEQ6L1}~\cite{Pumplin:2002vw} and a $K$-factor $K=1.2$, 
and as done in Ref.~\cite{CMS-PAS-B2G-12-010} with the NLO set of parton densities
{\sc CTEQ6M} and a $K$-factor also set to $K=1.2$. 
One observes that the difference in cross section between both approaches is of about 10~\%. Moreover,
the bound on the $W_R$-boson mass in our model from $tb$ searches is around $M_{W_R}>1970~$GeV, which is
about 50~GeV lower than the bounds on a $W'$-boson decaying into standard modes only. For completeness we
remark that in  Ref.~\cite{CMS-PAS-B2G-12-010} the results are given for three generations of leptons.

The reason for the experimental choice is to have a better handle on the PDF systematics. Our investigation showed that this procedure, albeit theoretically inconsistent, however only mar\-gi\-nal\-ly affects the results. We suggest to experimentalists to use a consistent framework, either with NLO matrix elements when available, or with a more suitable definition of the $K$-factor as in their setup to avoid NLO over counting. Aside for this technical point, we confirm the suitability of employing simplified models in these simple searches, since they capture the essence of the models within few percent.

Moving to the dijet search, we give again the results obtained after
applying the same selection requirements as in Ref.~\cite{CMS-PAS-EXO-12-059}, namely
imposing that the pseudorapidity of the jets satisfies $|\eta|<2.5$, and that
the relative difference in pseudorapidity
between the jets $j_i$ and $j_j$ fullfils $|\Delta \eta_{ij}|<1.3$. The subsequent 
acceptance resulting from these selections is 
$\mathcal A \approx 0.5$. In Fig.~\ref{lrmssm_fig_1bound} we compare the cross sections for two parameter choices
with the CMS bounds in this dijet analysis. 
We chose the PDF set {\sc CTEQ6L1} and a $K$-factor of 1.22 as in Ref.~\cite{CMS-PAS-EXO-12-059}. 
We see that the $W_R$-boson mass is constrained to lie above $2250~$GeV, with the exception 
of a small window between 1800 and 2030~GeV. Combining this with the results on the $tb$-channel,
this window gets reduced to the range 1970-2030 GeV.

By virtue of several difficulties we encountered when intending to confront our results to the experimental bounds, we would like to encourage our experimental colleagues to present their results in a more transparent way when comparing measured cross sections to theory expectations. This encompasses in our understanding a profound explanation of the understood theory assumptions and/or simplifications, the details of modelling the
 signal as well as the provision of a table of the excluded cross sections as has been done by the authors of Ref.~\cite{CMS-PAS-EXO-12-059}. 
All details of how signals should be produced in order to be suitable for comparing to the exclusions should also be given (\textit{e.g.} PDF set, final state,  selection requirements,
$K$-factor, model assumptions, and so on) to allow for reproducibility and cross checks.

\subsection{Interplay with right-handed neutrinos}\label{lrssm_subsect_RH}
It is usually assumed
in left-right symmetric models that the right-handed neutrinos can only decay via 
the $W_R$-boson, which can be either on- or off-shell.  This is used by experimental 
collaborations to set strict bounds on a combination of masses of $\nu_R$ and $W_R$, see, \textit{e.g.}, Ref.~\cite{CMS-PAS-EXO-12-017}. 

\begin{figure}[t]
\begin{center}
\includegraphics[width=0.49\linewidth]{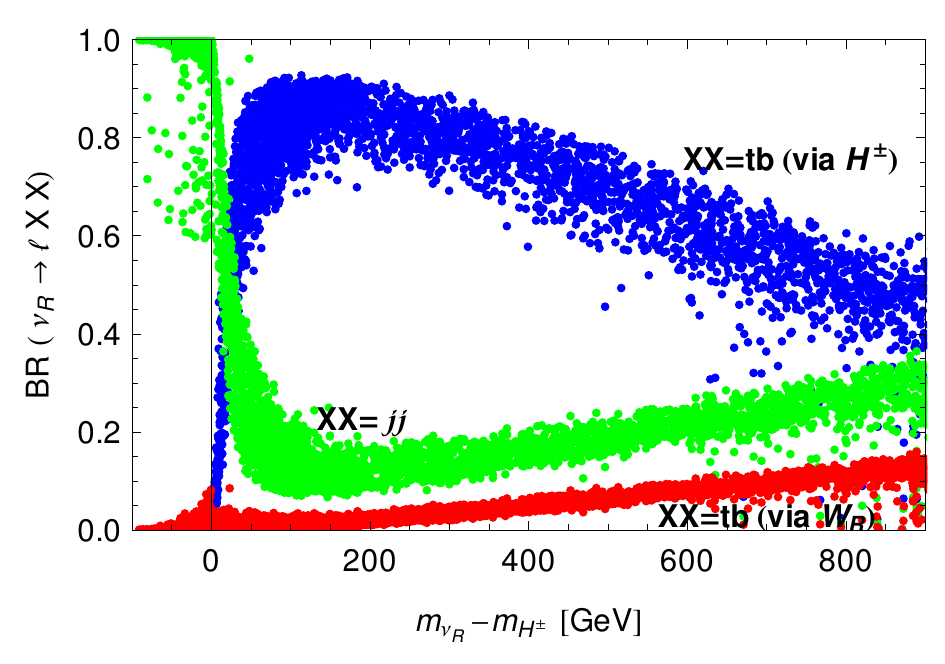}
\includegraphics[width=0.49\linewidth]{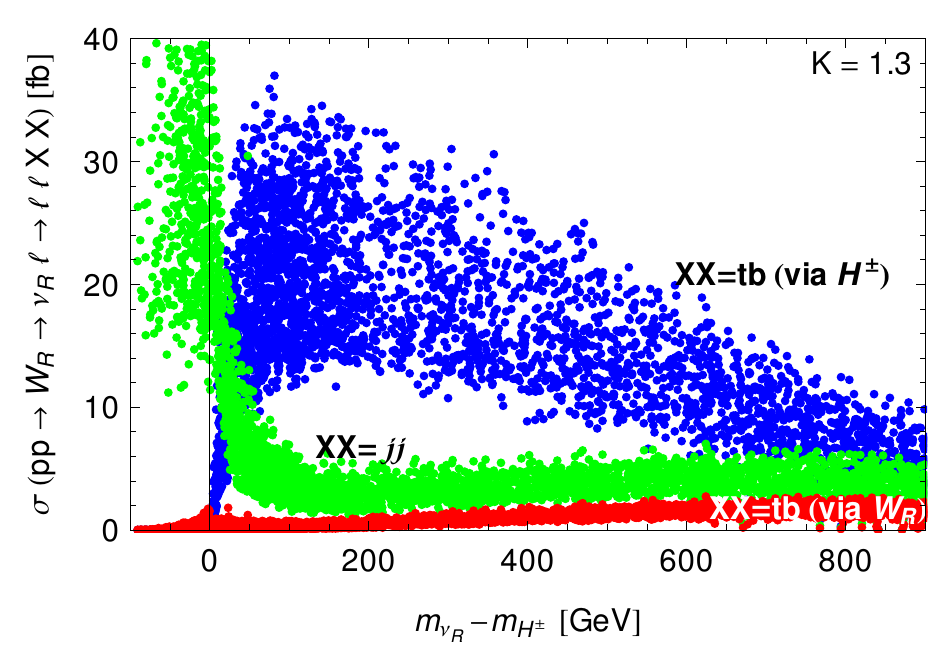}
\caption{Branching ratios of the right-handed neutrino as a function of the mass difference to the lightest charged Higgs state (left) and the corresponding cross section of the respective neutrino state produced by a $W_R$-resonance (right),
using the PDF set {\sc CTEQ6L1} as in Ref.~\cite{CMS-PAS-EXO-12-017} and a $K$-factor of 1.3.
Blue points depict cases where the right-handed neutrino $\nu_R$ decays to a
$\ell H^\pm_1$ state, with the charged Higgs boson $H^\pm_1$ further decaying to a $tb$
final state, whereas green (red) points represent cases where it decays into a
$\ell j j\, (tb)$ state mediated by an intermediate off-shell $W_R$-boson. The parameters have been varied within the ranges $3.8\leq v_R/\text{TeV}\leq 4.3,~-1 \leq \lambda_{s/L/R} \leq 1,~-0.1 \leq \lambda_{12} \leq 0.1,~200 \leq M_{2L/2R}/\text{GeV} \leq 900,~0 < M_1/\text{GeV} \leq 500,~4 \leq \tan \beta \leq 15$. The mass of the charged Higgs boson $H_1^\pm$ lies in the range of $180 < m_{H_1^\pm}/\mbox{GeV}<320$. Let us note that the branching fractions shown do not always add to 1 for some parameter points due to three-body decays into $\ell \tilde \chi^\pm \tilde \chi^0$ states that are kinematically accessible. 
\label{lrmssm_fig_2bound}}
\end{center}
\end{figure}

However, this simplified assumption can sometimes be too restrictive, in particular in our  model.
The right-handed neutrino acquires its mass via
the coupling $y_L^4$ to the $SU(2)_R$ triplet $\Delta_{1R}$. This implies
that $y_L^4 \sim \mathcal O(0.1)$
for a right-handed neutrino mass $m_{\nu_R}$ lying in the range 100~GeV$-$1~TeV and that
the vev $v_R$ is of ${\cal O}(1~\rm{TeV})$.
In addition, one of the six physical singly-charged Higgs bosons present in our model ($H^\pm_1$) can be rather light
with a mass of $\mathcal O(200~\text{GeV})$. The coupling of this state to the right-handed neutrino $\nu_R$ is $y_L^4$ times the
entry of the charged-Higgs mixing matrix connecting $H^\pm_1$ to $\Delta_{1R}$.
Hence, the two-body decay $\nu_R \to H^{\pm}_1 \ell$ will usually 
dominate over the three-body decays mediated by an off-shell $W_R$-boson. This can be seen in Fig.~\ref{lrmssm_fig_2bound} 
where we show the branching ratios $BR(\nu_R  \to H_1^{\pm} \ell)$, $BR(\nu_R  \to\ell j j)$ and $BR(\nu_R  \to \ell t b)$,
the latter two mediated by an off-shell $W_R$-boson.
As soon as $m_{\nu_R} > m_{H_1^\pm}$,  $\nu_R \to H^{\pm}_1 \ell$ becomes the dominant decay mode and the branching ratio for the 
$\nu_R \to W_R \ell \to \ell j j$ channel, which is considered by the experimental analyses,
gets significantly reduced. Finally, $H_1^\pm$ decays almost exclusively into a $t\bar{b}+\bar{t}b$ final state.

\begin{figure}[htb]
\begin{center}
\includegraphics[width=0.49\linewidth]{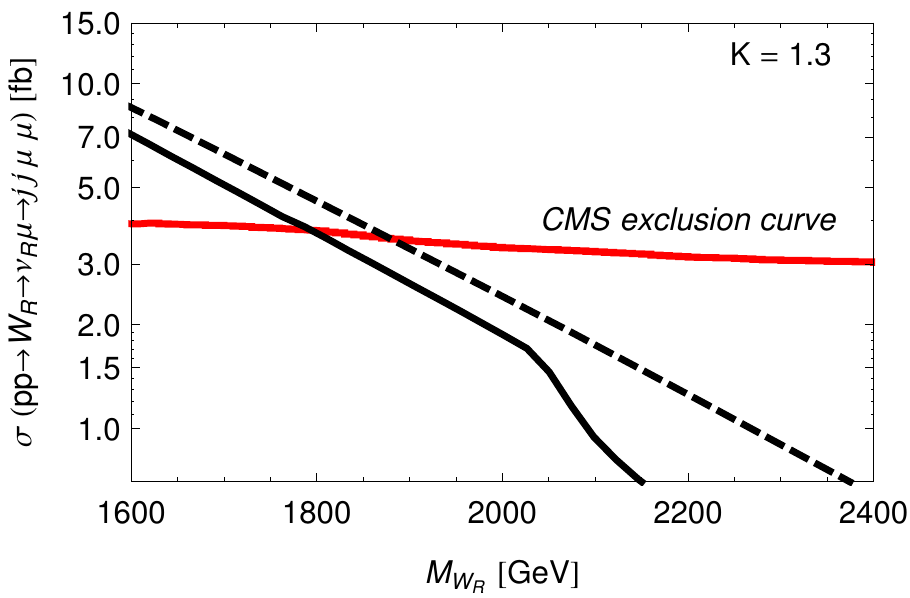}
\includegraphics[width=0.49\linewidth]{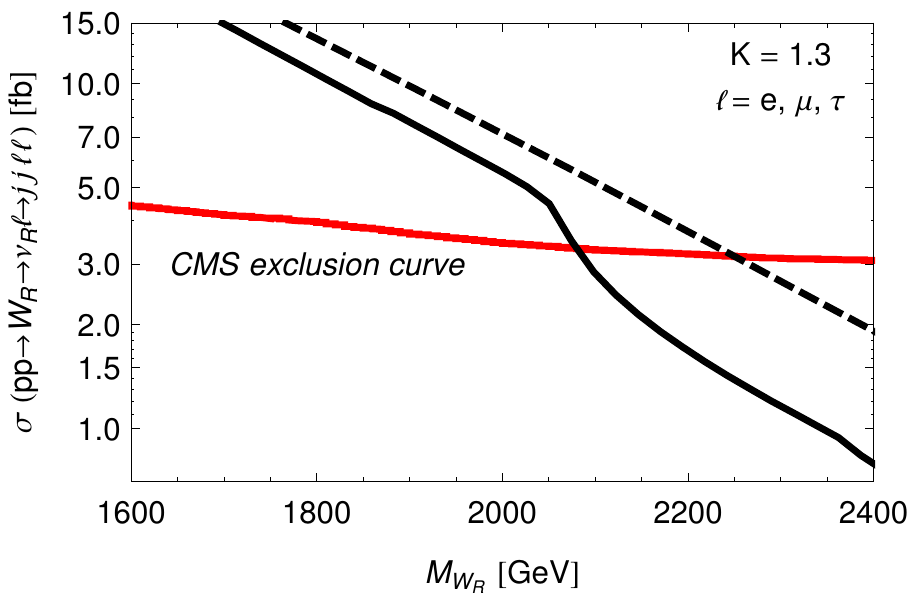}
\caption{Bounds on the cross section $\sigma (p p \to W_R \to \nu_R \ell \to jj \ell \ell)$ at $\sqrt{s}=8~$TeV for 
$m_{\nu_R} = \frac12 M_{W_R}$ considering $\sigma \cdot BR(j j\mu\mu)$ only (left) and for the case that all 
three right-handed neutrinos are degenerate in mass (right). Our curves have been
obtained from the results given in Ref.~\cite{CMS-PAS-EXO-12-017}. 
The same two parameter points as in Fig.~\ref{lrmssm_fig_2bound}, but with adjusted masses for the right-handed neutrinos, are shown, and the vev
$v_R$ has been varied from 3~TeV to 6~TeV. The black solid and dashed lines depict the calculated cross sections for the two parameter choices as in Fig.~\ref{lrmssm_fig_1bound}, 
and we note that for the parameter choice depicted by the solid black line, 
the two-body decay into a chargino and a slepton opens at $m_{\nu_R} \approx 1~$TeV. \label{lrmssm_fig_4bound}}
\end{center}
\end{figure}

Experimental results that are possible to recast are available only for 
$m_{\nu_R} = \frac12 M_{W_R}$ \cite{CMS-PAS-EXO-12-017}.
We depict in Fig.~\ref{lrmssm_fig_4bound} the results for our model and compare them with these bounds for two cases:
(a) the right-handed neutrino $\nu_{R,\mu}$ has half of the $W_R$-boson mass and (b)
all three heavy neutrinos are degenerate in mass. In the  setup
used in Ref.~\cite{CMS-PAS-EXO-12-017} the lower bounds on the mass of the $W_R$-boson
$M_{W_R}$ are about 2.45~TeV and 2.75~TeV, respectively.
Taking into account the additional decay modes we find that these 
bounds can be reduced by about 600~GeV each. 
Note that, although we can not compare to bounds for more general mass patterns, it is clear that in the case of even lighter right-handed neutrinos, the bounds are even more relaxed since the three-body decay
gets more suppressed with respect to the open two-body decay, as can be seen in Fig.~\ref{lrmssm_fig_2bound}.
Unfortunately currently there is no data available to study this case in detail.

As in the previous subsection, we would like to address  a message to experimentalists.
The simple exclusions as presented in Ref.~\cite{CMS-PAS-EXO-12-017} are not useful, because we cannot recast them for new models or modified assumptions (other than for $m_{\nu_R} = \frac12 M_{W_R}$) in the absence of expected and observed cross sections and relative efficiencies. In this case, a table of excluded cross sections in the $M_{W_R}-m_{\nu_R}$ plane would be of great help.

\section{CONCLUSIONS}\label{sec:lrmssm_conclusions}

We have studied a supersymmetric variant of a left-right-symmetric model focusing in particular
on the bounds on the mass of the additional charged vector-boson $W_R$. We have shown
that the bounds obtained by experimental collaborations get reduced as new decay channels are
open. Our findings are that this effect is strongest in scenarios where the $W_R$ can decay into
right-handed neutrinos as the latter have an additional decay channel into a
$\ell H^\pm$ state. In this
case, the bounds on the mass of the $W_R$-boson $M_{W_R}$ get reduced by about 600~GeV. Finally, we have remarked how in our view
the communication between experiment and theory could be improved by a more detailed way of
presenting the analysis results.

\section*{ACKNOWLEDGEMENTS}
We thank the organisers for setting up this nice event (including the
great food and amazing sceneries of the French alps) and for the nice working
atmosphere. MEK thanks Florian Staub for many useful discussions about the model implementation.
We further thank John Paul Chou, Bryan M. Dahmes, Robert M. Harris, Freya Blekman, and Maxim Perfilov for helpful clarifications concerning the experimental analyses. 
 The work of AA, LB and BF has received partial support
from the Theorie-LHC France initiative of the CNRS/IN2P3 and
by the French ANR 12 JS05 002 01 BATS@LHC.
LB is also supported by the
Deutsche Forschungsgemeinschaft through the Research Training Group grant
GRK\,1102 \textit{Physics at Hadron Accelerators} and by the
Bundesministerium f\"ur Bildung und Forschung within the F\"orderschwerpunkt
\textit{Elementary Particle Physics}. 
MEK and WP are supported by the DFG research training group
1147 and by the DFG project no. PO-1337/3-1. WP is also supported by DAAD 
project PROCOPE 54366394.



%% file: rho/draft_rho.tex
\chapter{Benchmark Models for Spin-1  Resonances in Composite Higgs Theories}

{\it R.~Contino, D.~Greco, C.~Grojean, D.~Liu, D.~Pappadopulo, A.~Thamm, R.~Torre,  A.~Wulzer}


\graphicspath{{rho/}}

\begin{abstract}
Heavy spin-1 resonances are a generic prediction of theories where electroweak symmetry breaking is triggered by 
new strongly-interacting dynamics at the TeV scale. In this contribution we study the phenomenology of spin-1 resonances
in  composite  Higgs theories based on the coset $SO(5)/SO(4)$. We introduce a simple model to describe their dynamics 
and give its implementation in the parton-level generator \texttt{MadGraph5}.
The model  captures the basic features of the resonances' phenomenology in terms of a minimal set of parameters, and can be
used as a benchmark in the search for heavy spin-1 states at the LHC and future colliders.  
\end{abstract}

\section{Introduction}

One of the robust predictions of theories 
with strong electroweak symmetry breaking (EWSB) is the existence of spin-1  resonances excited 
from the vacuum by the conserved currents of the strong dynamics.
They form multiplets of the unbroken global symmetry, which includes an $SO(4) \sim SU(2)_L \times SU(2)_R$
in the case of composite Higgs theories~\cite{Kaplan:1983fs}. The phenomenology of these resonances can be rather different from
the one of heavy $Z'$ states in weakly coupled extensions of the Standard Model (SM). They interact
strongly with longitudinally-polarized $W$ and $Z$ bosons and the Higgs boson, and thus tend to be broader than
weakly-coupled vectors. The strength of their couplings to SM fermions depends on whether these latter participate 
to the strong dynamics or are purely elementary states.  A simple possibility is that SM fermions couple in the EWSB 
dynamics according to their masses, so that the lightest ones are  the most weakly coupled.
This idea has an elegant implementation in the framework of partial compositeness~\cite{Kaplan:1991dc,Contino:2006nn} 
and can give a qualitative explanation of the hierarchies in the  Yukawa matrices of the SM fermions~\cite{Grossman:1999ra,Gherghetta:2000qt} 
in terms of RG flows~\cite{Contino:2004vy,Agashe:2004rs}.

In this work we study the phenomenology of spin-1 resonances in composite Higgs theories by means of a simplified
description based on an effective Lagrangian.  This is aimed at capturing the main features 
relevant for the production of the resonances at high-energy colliders and their effects in low-energy experiments,
avoiding the complication of a full model. Although simplified, our construction will be sophisticated enough
to properly include those aspects which are distinctive predictions of the class of theories under consideration,
such as for example the pseudo Nambu-Goldstone (NG) nature of the Higgs boson.
We will focus on  minimal $SO(5)/SO(4)$ composite Higgs theories and consider resonances transforming as
$(3,1)$ and $(1,3)$ of $SO(4)$, respectively denoted as $\rho_L$ and $\rho_R$ in the following. 
We assume that SM fermions are fully elementary and couple to the heavy resonances only 
through the mixing of the latter with elementary gauge fields. This implies small universal couplings of $\rho_L$, $\rho_R$ to fermions 
of order $\sim g_{SM}^2/g_\rho$, where $g_\rho$ sets the interaction strength of the resonances to other composite states,
including longitudinally-polarized $W$ and $Z$ bosons and the Higgs boson.
This construction can be generalized to include direct couplings of the heaviest SM fermions, in particular of the top and bottom quarks, 
to composite states as implied by partial compositeness.
Starting from the Lagrangian defining our model, we discuss the rotation to the mass-eigenstate basis and
derive the physical spectrum and  interactions.
We provide a calculator of physical quantities (masses and couplings) and an  implementation of the model
in the parton-level generator \texttt{MadGraph5}~\cite{Alwall:2011uj}
for the simulation of MonteCarlo events. These tools can be  downloaded from the HEPMDB website at the URL
\url{http://hepmdb.soton.ac.uk/hepmdb:0214.0154}
and are the main result of this work.

Aim of this work is to provide a benchmark model to be used in searches for heavy spin-1 states at the LHC and at future colliders. A simple kinematic model 
based on the width and the production cross section times the decay branching ratio ($\sigma \times BR$) is sufficient to guide searches for narrow resonances 
in individual channels and to set limits, see the recent discussion in Ref.~\cite{Pappadopulo:2014qza}. However, combining the results obtained in different final 
states as well as interpreting the limits on $\sigma \times BR$ in explicit models of physics beyond the SM requires an underlying dynamical description, such as 
the one given by a simplified Lagrangian. Here we provide such a dynamical description for spin-1 resonances appearing in a motivated and sufficiently large class 
of composite Higgs theories. Our construction fully takes into account the non-linear effects due to multiple insertions of the Higgs vev and does not rely on an 
expansion in $v/f$, where $v$ is the electroweak scale and $f$ is the decay constant of the NG boson Higgs. In the limit $v/f \ll 1$ our Lagrangian can be matched 
onto the more general one of Ref.~\cite{Pappadopulo:2014qza}, which covers a more ample spectrum of possibilities in terms of a larger number of free parameters. 
In this sense, the main virtue of our model is that of describing the phenomenology of spin-1 resonances in composite Higgs theories in terms of a minimal set of 
fundamental quantities: one mass and one coupling strength for each resonance. Expressing the experimental results in such restricted parameter space is thus 
extremely simple and gives an immediate understanding of the reach of current searches in the framework of strongly interacting models for EWSB.

In the next section we define our model and discuss the rotation to the mass eigenstate basis. The collider phenomenology of the  spin-1
resonances is briefly analyzed in Section~\ref{sec:pheno} by focussing on the LHC. Section~\ref{sec:program} explains in detail how to run the calculator and the 
\texttt{MadGraph5} model implementing the spin-1 resonances. We conclude in Section~\ref{sec:conclusions} and collect some useful analytic formulas in the Appendix.

\section{The Model}
\label{sec:model}

We describe our spin-1 resonances using vector fields
and write the effective Lagrangian  by adopting the CCWZ formalism~\cite{Coleman:1969sm,Callan:1969sn},  following the notation and conventions of 
Ref.~\cite{Contino:2011np}.
The Lagrangian reads:~\footnote{We normalize hypercharge as $Y = Q - T_{3L}$ and define the projector over  left-handed fermions 
as $P_L = (1-\gamma_5)/2.$}
\begin{equation}
\label{Lag}
\begin{split}
\mathcal{L} =  
& -\frac{1}{4 g_{el}^2} W_{\mu\nu}^a W^{a\, \mu\nu} -\frac{1}{4 g_{el}^{\prime\, 2}} B_{\mu\nu} B^{\mu\nu}
+ \bar{\psi} \gamma^\mu \!\left(i \partial_\mu- \frac{\sigma^a}{2} W_\mu^a  P_L-  Y B_\mu \right)\! \psi + \frac{f^2}{4} (d_\mu^{\hat{a}})^2 \\
&  - \frac{1}{4 g_{\rho_L}^2} \rho_{\mu\nu}^{a_L} \rho^{a_L \mu\nu} - \frac{1}{4 g_{\rho_R}^2} \rho_{\mu\nu}^{a_R} \rho^{a_R \mu\nu}  
+ \frac{m_{\rho_L}^2}{2g_{\rho_L}^2} (\rho_\mu^{a_L} - E_\mu^{a_L})^2 + \frac{m_{\rho_R}^2}{2g_{\rho_R}^2} (\rho_\mu^{a_R} - E_\mu^{a_R})^2\, .
\end{split}
\end{equation}
The CCWZ covariant variables $d^{\hat a}_\mu, E^{a_L, a_R}_\mu$ ($\hat a = 1,\dots 4$, $a_L, a_R = 1,2,3$) 
are functions of the four $SO(5)/SO(4)$ Nambu-Goldstone bosons $\pi^{\hat a}$ and are defined by
\begin{align}
\label{eq:dEdef}
- i U^\dagger D_\mu U = d_\mu + E_\mu\, ,
\end{align}
where $U =  \text{exp}(i\sqrt{2}\pi/f)$, $\pi = \pi^{\hat{a}} T^{\hat{a}}(\theta)$ and $T^{\hat a, a_L, a_R}$ are $SO(5)$ generators.
The Lagrangian~(\ref{Lag}) describes an elementary sector of $SU(2)_L \times U(1)_Y$ gauge fields~($W_\mu$, $B_\mu$) and fermions~($\psi$), 
as well as a composite sector comprising the NG bosons 
and the resonances $\rho_L$, $\rho_R$. While the elementary sector has a local $SU(2)_L \times U(1)_Y$  gauge invariance (with the fermions
falling into the SM representations), the composite sector has a global $SO(5)$ symmetry spontaneously broken to $SO(4)$.
The resonances $\rho_L$ and $\rho_R$ transform respectively like a $(3,1)$ and $(1,3)$ of $SO(4)\sim SU(2)_L \times SU(2)_R$.
The elementary $W_\mu$, $B_\mu$ fields weakly gauge a subgroup $SU(2)_L \times U(1)_Y$ of the global $SO(5)$ misaligned by an angle $\theta$
with respect to the unbroken  $SO(4)$.~\footnote{The derivative in Eq.~(\ref{eq:dEdef}) is thus covariant under the local 
$SU(2)_L \times U(1)_Y$, $D_\mu = \partial_\mu + i T^a W^a_\mu + i Y B_\mu$.} 
Such misalignment eventually implies the breaking of the low-energy electroweak symmetry, hence the angle $\theta$
can be seen as an EWSB order parameter.
More details about the symmetry construction and the CCWZ formalism for $SO(5)/SO(4)$ can be found in Ref.~\cite{Contino:2011np}.
The Lagrangian~(\ref{Lag}) provides a minimal description of the spin-1 resonances. Additional operators involving $\rho_{L,R}$ can be in general
included and play a relevant role at energies of order of the resonances' mass, see for example Ref.~\cite{Contino:2011np}. In the following we will omit them
for simplicity.

The only source of interactions in Eq.~(\ref{Lag}) among the composite $\rho_{L,R}$ and the elementary fields is the $\rho_L - W$ and $\rho_R - B$
mass mixings that follow from the last two terms in the second line of Eq.~(\ref{Lag}) ($\rho$ mass terms).
This can be seen explicitly by expanding  $d_\mu$ and $E_\mu$ at quadratic order in the fields ($i,j = 1,2,3$):
\begin{equation}
\label{dE}
\begin{split}
d_\mu^{\hat{a}} &= A_\mu^{\hat{a}} + \sqrt{2} \frac{\partial_\mu \pi^{\hat{a}}}{f} + \frac{\sqrt{2}}{2f}\delta^{\hat{a} j}\left(\epsilon^{i a j} \pi^{i}(A_\mu^{a_L}+A_\mu^{a_R})
     +\pi^{4}(A_\mu^{j_L} - A_\mu^{j_R})\right) - \frac{\sqrt{2}}{2f}\delta^{\hat{a} 4}\left(\pi^{i} A_\mu^{i_L} - \pi^{i} A_\mu^{i_R}\right), \\
E_\mu^{a_L} & = A_\mu^{a_L} + \frac{1}{2f^2} \left( \epsilon^{a_L i j} \pi^{i}\partial_\mu \pi^{j} + \delta^{a_L i}(\pi^{i}\partial_\mu \pi^{4}- \pi^{4}\partial_\mu \pi^{i})\right) 
     + \frac{\sqrt{2}}{2f}\left( \epsilon^{a_L i j}\pi^{i}A_\mu^{\hat \jmath} + \delta^{a_L i} (\pi^{i} A_\mu^{\hat{4}} - \pi^{4} A_\mu^{\hat{\imath}})\right), \\
E_\mu^{a_R} & = A_\mu^{a_R} + \frac{1}{2f^2} \left( \epsilon^{a_R i j} \pi^{i}\partial_\mu \pi^{j} - \delta^{a_R i}(\pi^{i}\partial_\mu \pi^{4}- \pi^{4}\partial_\mu \pi^{i})\right) 
     + \frac{\sqrt{2}}{2f}\left( \epsilon^{a_R i j}\pi^{i}A_\mu^{\hat \jmath} - \delta^{a_R i} (\pi^{i} A_\mu^{\hat{4}} - \pi^{4} A_\mu^{\hat{\imath}})\right), 
\end{split}
\end{equation}
where 
\begin{equation}
\label{Amu}
\begin{split}
A_\mu^{\hat{i}} &= \frac{\sin\theta}{\sqrt{2}}\left(W_\mu^i - \delta^{i3}B_\mu\right), \qquad A_\mu^{\hat{4}} = 0, \\[0.1cm]
A_\mu^{a_L} &= \left(\frac{1 + \cos \theta}{2}\right) W_\mu^a + \delta^{a3}\left(\frac{1 - \cos \theta}{2}\right) B_\mu, \\[0.1cm]
A_\mu^{a_R} &= \left(\frac{1 - \cos \theta}{2}\right) W_\mu^a + \delta^{a3}\left(\frac{1 + \cos \theta}{2}\right) B_\mu\, .
\end{split}
\end{equation}
Therefore, the global mass matrix of spin-1 fields ($W, B, \rho_L, \rho_R$) is non-diagonal and must be diagonalized by a proper field rotation.

Before discussing the rotation to the mass-eigenstate basis, let us first count how many parameters appear in our Lagrangian:
there are five couplings ($g_{el}, g_{el}^\prime, g_{\rho_L}, g_{\rho_R}, f$), two mass scales ($m_{\rho_L}, m_{\rho_R}$),  and the misalignment angle $\theta$,
for a total of 8 parameters. Notice that we have listed the NG decay constant $f$ as a coupling, since it controls
the strength of the NG boson interactions. The misalignment angle is determined by the radiatively-induced Higgs potential, and can be conveniently
traded for the variable $\xi \equiv \sin^2\!\theta$. 
All the Lagrangian (input) parameters can be re-expressed in terms of physical quantities in the mass eigenbasis. Three of them must be fixed 
to reproduce the basic electroweak observables, which we conveniently choose to be $G_F$, $\alpha_{em}$ and $m_Z$.
Of the remaining five input parameters, $\xi$  controls the modifications of the Higgs 
couplings from their SM values and is thus an observable, while the other
four can be traded for the following physical quantities: the masses of the neutral heavy resonances $m_{\rho_1^0}, 
m_{\rho_2^0}$ and their couplings to the charged leptons, $g_{\rho_1ll}$, $g_{\rho_2ll}$.

In order to fix three of the input parameters in terms of $G_F$, $\alpha_{em}$ and $m_Z$ we need the expressions of the latter in terms of the former.
It turns out that $G_F$ and $\alpha_{em}$ are very simple to compute and read:
\begin{align}
\label{eq:GF}
G_F & = \frac{1}{\sqrt{2} f^2 \xi}\, ,  \\[0.2cm]
\label{eq:alphaem}
\frac{1}{4\pi \alpha_{em}} & = \frac{1}{g_{el}^2} +\frac{1}{g_{\rho_L}^2} + \frac{1}{g_{el}^{\prime 2}} +\frac{1}{g_{\rho_R}^2} =\frac{1}{g^2} +\frac{1}{g^{\prime 2}} \, ,
\end{align}
where we have conveniently defined the intermediate parameters
\begin{equation}
\label{eq:ggprime}
\frac{1}{g^2} \equiv \frac{1}{g_{el}^2} +\frac{1}{g_{\rho_L}^2}, \qquad \frac{1}{g^{\prime 2}} \equiv \frac{1}{g_{el}^{\prime 2}} +\frac{1}{g_{\rho_R}^2} \, .
\end{equation}
Notice that $\alpha_{em}$ does not get corrections after EWSB at any order in $\xi$. 
The formula for $G_F$ can be most easily derived by integrating out first the composite $\rho$'s  using their equations of
motion at leading order in the derivative expansion, $\rho_\mu = E_\mu + O(p^3)$. From Eq.~(\ref{Lag}) one can then see that the low-energy 
Lagrangian for the elementary fields contains two extra operators, $(E_{\mu\nu}^L)^2$ and $(E_{\mu\nu}^R)^2$, which however do not contribute
to $G_F$. This means that the expression of $G_F$ in terms of the elementary parameters does not receive any tree-level contribution from the composite 
$\rho$'s, hence the simple formula~(\ref{eq:GF}). 
Finally, the formula of $m_Z$  is in general quite complicated and we do not report it here.
By making use of such expression and of Eqs.~(\ref{eq:GF}),~(\ref{eq:alphaem}), for given values of the other input parameters
($g_{\rho_L}$, $g_{\rho_R}$, $m_{\rho_L}$ and $m_{\rho_R}$) we can fix that of $g_{el}$, $g_{el}^\prime$ and $f$ 
so as to reproduce the experimental values of  $G_F$, $\alpha_{em}$ and $m_Z$.

We now discuss the rotation to the mass eigenstate basis. The mass matrix consists of a $3\times 3$ charged block and a $4\times 4$ neutral block,
their expressions are reported in Eqs.~(\ref{eq:mass1})-(\ref{eq:mass3}) of the Appendix.
Although it can be  diagonalized analytically,  the expressions of its eigenvectors and eigenstates are extremely complicated
in the general case. It is thus more convenient to perform a numerical diagonalization unless  specific limits are considered in which expressions simplify.
We provide a \texttt{Mathematica} code which makes such numerical diagonalization for given values of the input parameters 
$\{ m_{\rho_L}, m_{\rho_R}, g_{\rho_L}, g_{\rho_R},\xi\}$ and returns the values of all the relevant physical couplings and masses. The program is
illustrated in detail in section~\ref{sec:program}.
A limit in which it is useful to perform the diagonalization analytically is that of small $\xi$, which is also phenomenologically
favored by the constraints coming from  the electroweak precision tests.
The corresponding eigenvalues (physical masses) in this case are, at linear order in $\xi$:
\begin{equation}
\label{eq:spectrum}
\begin{split}
m_W^2 &= f^2 \xi  \, \frac{ g^2}{4} \, , \\[0.1cm]
m_Z^2 &= f^2 \xi\, \frac{  \left(g'^2 + g^2 \right)}{4} \, , \\[0.25cm]
m_{\rho_1^{\pm}}^2 &= m_{\rho_L}^2
   \frac{g_{\rho_L}^2}{g_{\rho_L}^2 - g^2} \left[ 1  - \frac{\xi}{2}\,  \frac{ g^2}{g_{\rho_L}^2}   \left( 1- \frac{g^2 f^2}{2m_{\rho_L}^2}  \right) \right]\, , \\
m_{\rho_2^{\pm}}^2 &= m_{\rho_R}^2 \, ,\\[0.25cm]
m_{\rho_1^0}^2 &=  m_{\rho_L}^2
   \frac{g_{\rho_L}^2}{g_{\rho_L}^2 - g^2} \left[ 1  - \frac{\xi}{2}\,  \frac{ g^2}{g_{\rho_L}^2}   \left( 1- \frac{g^2 f^2}{2m_{\rho_L}^2}  \right) \right]\, , \\[0.1cm]
m_{\rho_2^0}^2 &= m_{\rho_R}^2
   \frac{g_{\rho_R}^2}{g_{\rho_R}^2 - g'^2} \left[ 1  - \frac{\xi}{2}\,  \frac{ g'^2}{g_{\rho_R}^2}   \left( 1- \frac{g'^2 f^2}{2m_{\rho_R}^2}  \right) \right]\, ,
\end{split}
\end{equation}
We have defined the mass eigenstates so that for $\xi$ small $\rho_1$ and $\rho_2$ are mostly made of respectively $\rho_L$ and $\rho_R$.
If we use Eq.~(\ref{eq:GF}) and define (to all orders in $\xi$) the electroweak scale as $v = \sqrt{\xi} f$, then
$m_W$ and $m_Z$  in Eq.~(\ref{eq:spectrum}) have formally the same expression as in the 
SM.~\footnote{With this choice the $O(\xi^2)$ corrections appear in $m_W$ and $m_Z$ but not in $v$.
One could  equivalently define $v$ through the formula $m_W = g v/2$, so that $G_F$ in Eq.~(\ref{eq:GF}) deviates from its SM expression
at $O(\xi^2)$ once rewritten in terms of $v$.} The masses of the resonances instead arise at zeroth order in $\xi$ and get corrections after
EWSB.  In the case of~$\rho_1$, the $O(\xi)$ corrections to the charged and neutral masses are equal, since they do not depend on $g_{el}^\prime$,
which is the only parameter in the bosonic sector to break the custodial symmetry. Another accidental property of the charged sector is that if
$m_{\rho_L} = m_{\rho_R}$ then one of the  mass eigestates, i.e. $m_{\rho_2^\pm}$ in our notation, remains unperturbed at all orders in $\xi$.
That is, the formula for $m_{\rho_2^\pm}$ reported in Eq.~(\ref{eq:spectrum}) becomes exact in this limit.
This can be easily understood by noticing that for $m_{\rho_L} = m_{\rho_R} = m_\rho$ the charged mass matrix contains a $2\times 2$ sub-block proportional 
to the identity and, as a consequence, its characteristic polynomial $\text{det}(M^2 - \lambda \, 1)$ is proportional to $(\lambda - m_\rho)$.

Once the form of the rotation to the mass eigenbasis is derived, either numerically or analytically, it is straightforward to obtain the physical interactions 
between the heavy resonances and the SM fields. In the following we will focus on trilinear couplings, neglecting for simplicity quartic interactions. 
The terms in the Lagrangian with cubic interactions involving one heavy resonance are:~\footnote{All interaction terms between SM fermions and  
spin-1 resonances in this Lagrangian are flavor diagonal. This follows from assuming that fermions are fully elementary:
in absence of elementary-composite fermion mixings one can always make fields rotations to diagonalize 
the fermionic kinetic terms in flavor space.  Of course the assumption is very crude, since in this way fermions are massless.
By allowing for some degree of compositeness and non-vanishing elementary-composite couplings $\lambda$, the Lagrangian~(\ref{eq:cubic}) is valid at  
O($\lambda^0$) in the weak interaction eigenbasis for the fermions. In this basis the fermion masses are not diagonal in flavor space.
After rotating the fermion fields  to diagonalize their mass matrices, a $V_\text{CKM}$ matrix appears in the vertex $\rho^+ \bar\psi_u \psi_d$,
while the interactions of $\rho^0$ remain diagonal.
}
\begin{equation}
\label{eq:cubic}
\begin{split}
\mathcal{L}_\rho   =  \, 
& i  g_{\rho^+ W Z} \, \big[  (\partial_\mu \rho_{\nu}^+ - \partial_\nu \rho_{\mu}^+ ) W^{\mu -}Z^{\nu}    
                            -(\partial_\mu W_\nu^- - \partial_\nu W_\mu^- ) \rho^{\mu +} Z^{\nu }  \\[0.15cm]
& \phantom{ig_{\rho^+ W Z}\,\big[}           
                           +  (\partial_\mu Z_\nu - \partial_\nu Z_\mu ) \rho^{\mu +}W^{\nu -}  + h.c.\big]  \\[0.15cm]
& + i g_{\rho^0 W W}\, \big[  (\partial_\mu W_\nu^+ - \partial_\nu W_\mu^+ ) W^{\mu -}\rho^{0\nu}  
                                  + \frac 12 (\partial_\mu \rho_{\nu}^0 - \partial_\nu \rho_{\mu}^0 ) W^{\mu +}W^{\nu -}  + h.c.\big]  \\[0.15cm]
& + g_{\rho^+ W h}  \, ( h \rho_{\mu}^+ W^{\mu-} + h.c.)  +  g_{\rho^0 Z h}  \, h  \rho_{\mu}^0 Z^{\mu}   \\[0.15cm]
& + \frac{1}{\sqrt{2}} \, g_{\rho^+ ud} \left( \rho^+_{\mu}\bar{\psi}_{u} \gamma^\mu P_L \psi_{d}  + h.c. \right) \\[0.15cm]
& +  \rho^0_{\mu}\, \bar{\psi}_u \gamma^\mu \!\left[ \frac 12 (g_{\rho^0 ffL} - g_{\rho^0 ffY} ) P_L+ g_{\rho^0 ffY } Q[ \psi_u] \right]\! \psi_u  \\[0.15cm]
& + \rho^0_{\mu} \,\bar{\psi}_d \gamma^\mu  \!\left[ -\frac 12 (g_{\rho^0 ffL} - g_{\rho^0 ffY} )P_L + g_{\rho^0 ffY } Q[ \psi_d] \right] \!\psi_d \, ,
\end{split}
\end{equation}
where $\rho$ indicates either of $\rho_1$ and $\rho_2$, and $\psi_u$ ($\psi_d$) stands for any of the SM up-type quarks and neutrinos
(down-type quarks and charged leptons). The expressions of the couplings appearing in Eq.~(\ref{eq:cubic}) are reported in the Appendix at linear 
order in $\xi$.

The Lorentz structure of the vertices among three vector fields is the same as the one of triple gauge vertices in the SM.
This is because the kinetic terms for both composite and elementary fields in Eq.~(\ref{Lag}) imply interactions of the SM form,
and rotating to the mass eigenbasis does not obviously change the  Lorentz  structure.
The value of the $VV\rho$ and $Vh\rho$ couplings ($V= W,Z$) can  be extracted  by using the Equivalence Theorem for $m_\rho \gg m_V$.
In this limit 
the leading contribution to the interaction comes from the longitudinal polarizations of the SM vector fields, 
and the overall strength equals that of the 
coupling of one $\rho$ to two NG bosons, $\rho\pi\pi$, up to small corrections of $O(m_V^2/m_\rho^2)$. 
As it can be directly seen from  Eq.~(\ref{Lag}),
the $\rho\pi\pi$ coupling  is proportional to $g_\rho a_\rho^2$,  where $a_\rho \equiv m_\rho/(g_\rho f)$ is a quantity expected 
to be of order 1 according to  naive dimensional analysis (NDA).
Finally,  the interactions of the heavy resonances to the SM fermions follow entirely from the universal composite-elementary mixing,
that is, from the elementary component of the heavy spin-1 mass eigenstate (the fermions are  assumed to be fully elementary). 
As a consequence, the three couplings $g_{\rho^+ ud}$, $g_{\rho^0 ffL}$,  $g_{\rho^0ffY}$
are of order $\sim g^2/g_\rho$ and do not depend on the fermion species, i.e. they are universal. 
From the above discussion it follows that, in the limit $g_\rho \gg g$, the heavy resonances are most strongly coupled to  composite states,
i.e. the  longitudinal polarizations of $W$, $Z$ and the Higgs boson, while their coupling strength to the SM fermions is extremely weak.

\section{Production and decay of spin-1 resonances at the LHC}
\label{sec:pheno}

\begin{figure}[tb]
\begin{center}
\includegraphics[width=0.48\textwidth]{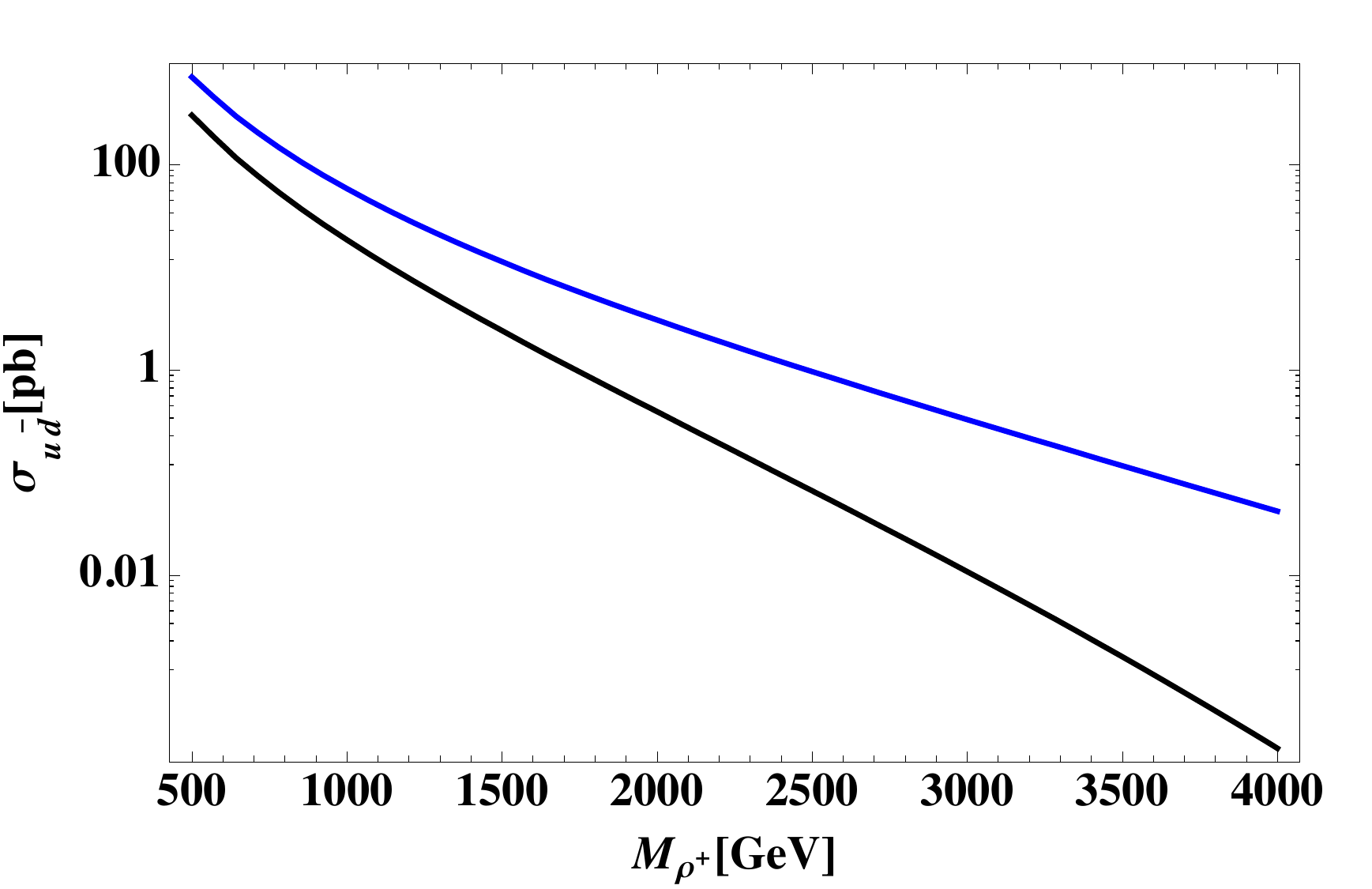}
\includegraphics[width=0.48\textwidth]{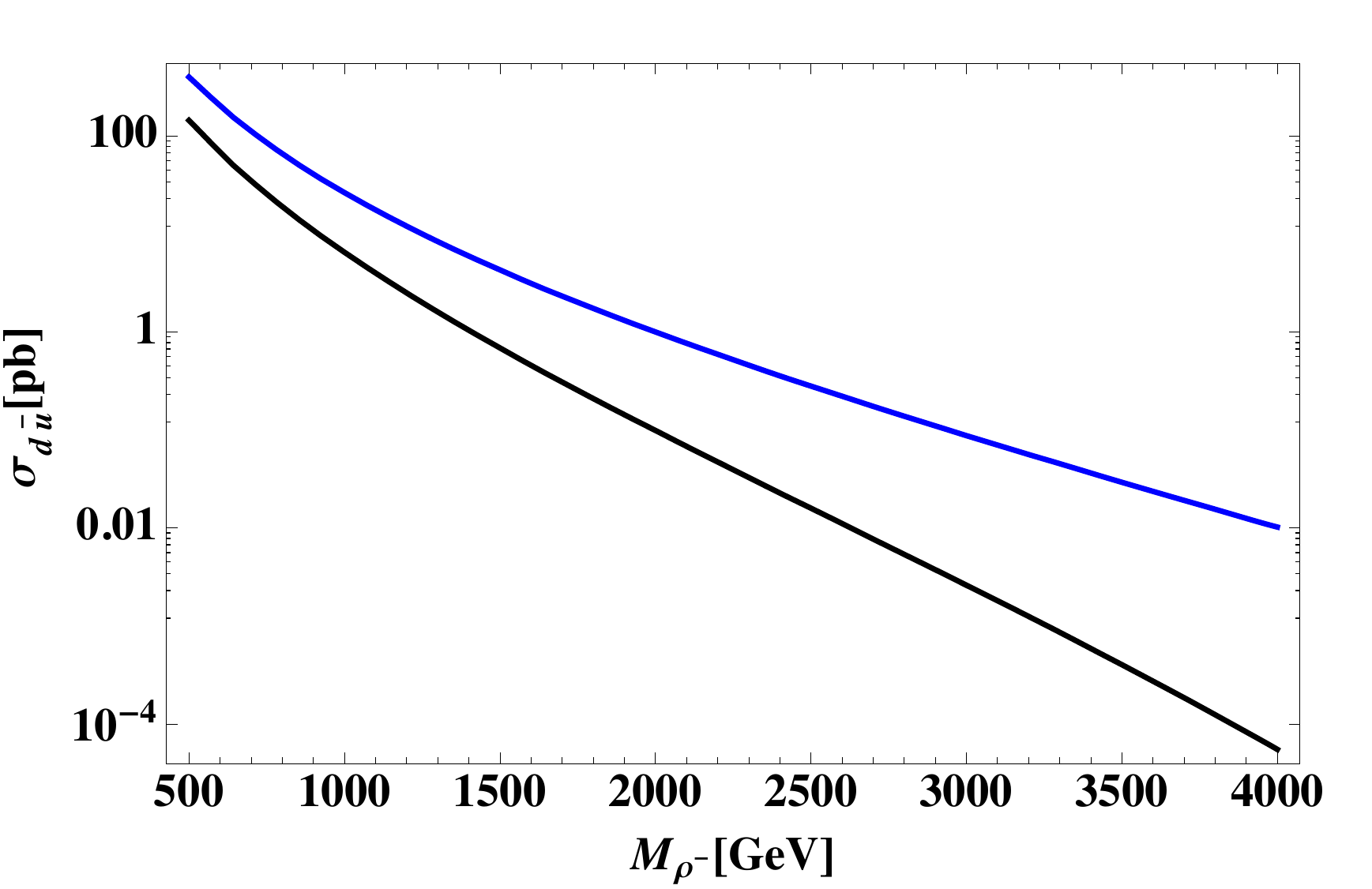} \\ 
\includegraphics[width=0.48\textwidth]{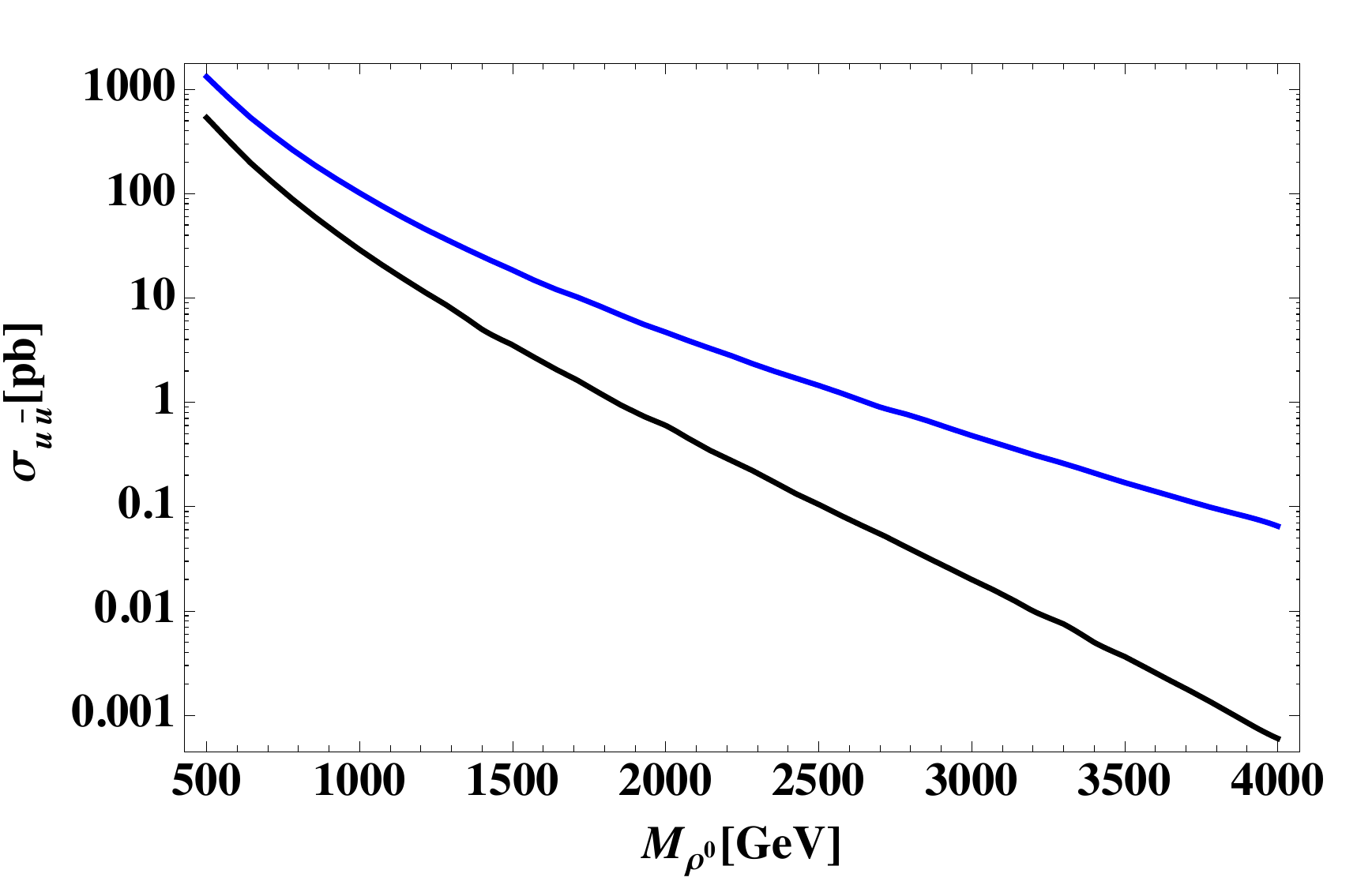} 
\includegraphics[width=0.48\textwidth]{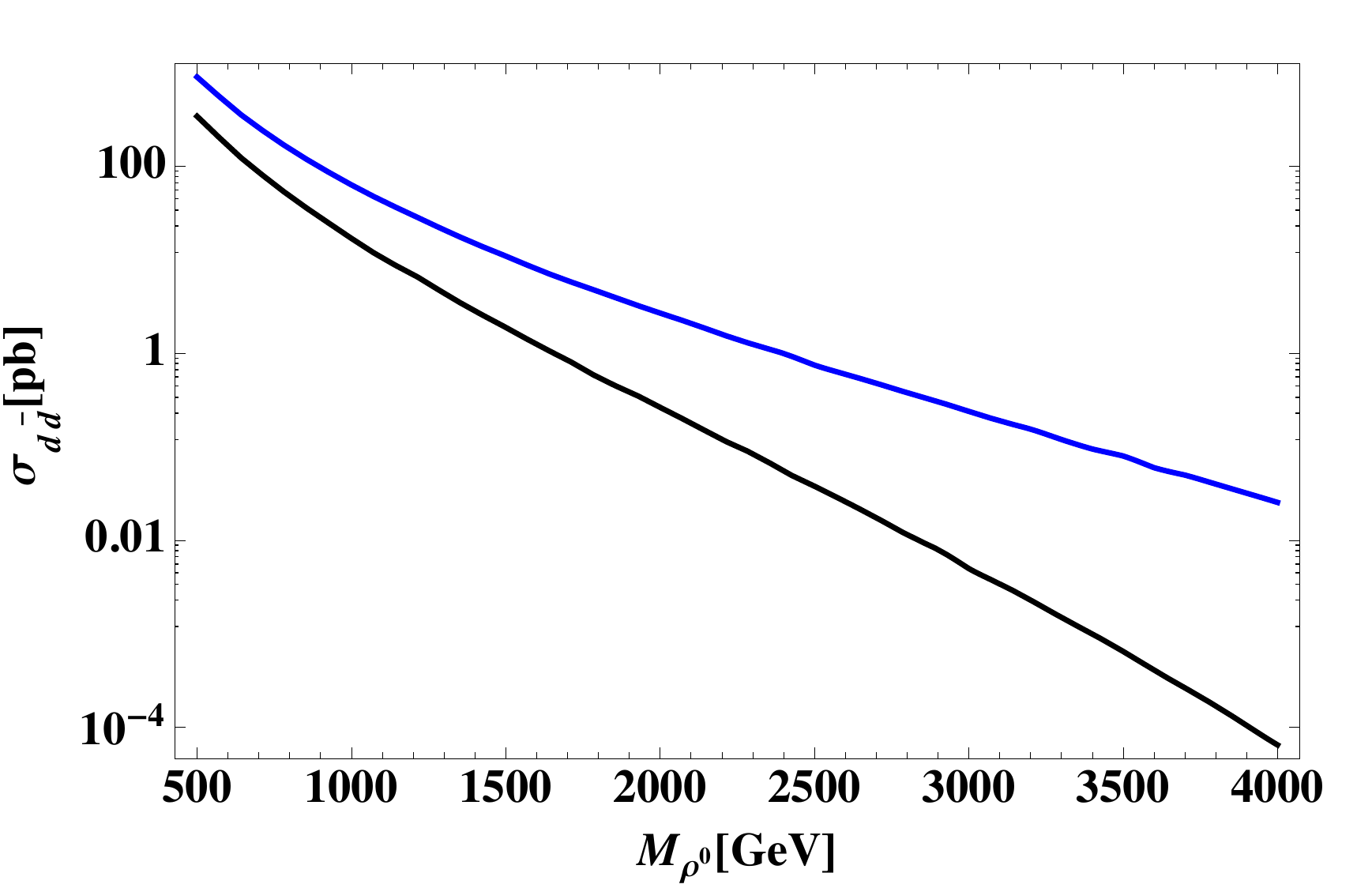} 
\end{center}
\caption{\small \label{fig:xsections}
Fundamental cross sections of Eq.~(\ref{eq:basicxsec}) as functions of the physical mass of the resonance.
Black (blue) curves are obtained for a collider center-of-mass energy $\sqrt{s} =8\,$TeV ($\sqrt{s} =14\,$TeV).
}
\end{figure}

Despite their suppressed couplings to the SM fermions, the main production mechanism of the heavy resonances at the LHC is Drell--Yan processes.
Under the validity of the Narrow Width Approximation (NWA), each production rate can be factorized into an on-shell cross section times
a decay branching fraction. The on-shell cross sections are controlled by the universal couplings 
$g_{\rho^+ ud}$, $g_{\rho^0 ffL}$,  $g_{\rho^0ffY}$ and can be written 
as~\footnote{Another convenient way to parametrize the production cross sections is in terms of partial widths and parton luminosities,
see Ref.~\cite{Pappadopulo:2014qza}.}
\begin{equation}
\label{eq:xsec}
\begin{split}
\sigma(pp \to \rho^+ + X) & = g^2_{\rho^+ ud} \cdot \sigma_{u \bar d}  \, ,\\[0.1cm]
\sigma(pp \to \rho^- + X) & = g^2_{\rho^+ ud} \cdot \sigma_{d \bar u} \, , \\[0.1cm]
\sigma(pp \to \rho^0 + X) & = g^2_{\rho^0 uu} \cdot \sigma_{u \bar u} + g^2_{\rho^0 dd} \cdot \sigma_{d \bar d} \, ,
\end{split}
\end{equation}
where $\rho$ stands for either $\rho_1$ or $\rho_2$, $g_{\rho^0 uu}$ and $g_{\rho^0 dd}$ are the coupling strengths of respectively 
up- and down-type fermions to the resonance,
\begin{equation}
\label{eq:couplings}
\begin{split}
g_{\rho^0 uu} & \equiv \left[ \left(\frac 12 \left(g_{\rho^0 ffL} - g_{\rho^0 ffY} \right) + \frac 23 g_{\rho^0 ffY}\right)^2+ \left( \frac 23 g_{\rho^0 ffY} \right)^2 \right]^{1/2} \, , \\
g_{\rho^0 dd} & \equiv \left[  \left(- \frac 12 \left(g_{\rho^0 ffL} - g_{\rho^0 ffY} \right) - \frac 13 g_{\rho^0 ffY}\right)^2+ \left( -\frac 13 g_{\rho^0 ffY} \right)^2 \right]^{1/2}\, ,
\end{split}
\end{equation}
and we defined ($\psi_u = u,c$, $\psi_{d} = d,s$)
\begin{equation}
\label{eq:basicxsec}
\begin{split}
\sigma_{u \bar d} & = \sum_{\psi_u, \psi_d} \sigma(pp \to \psi_u \bar \psi_d \to \rho^+ + X)\big|_{g_{\rho^+ ud} =1} \, ,\\[0.05cm]
\sigma_{d \bar u} & = \sum_{\psi_u, \psi_d} \sigma(pp \to \psi_d \bar \psi_u \to \rho^- + X)\big|_{g_{\rho^+ ud} =1} \, ,\\[0.05cm]
\sigma_{u \bar u} & = \sum_{\psi_u} \sigma(pp \to \psi_u \bar \psi_u \to \rho^0 + X)\big|_{g_{\rho^0 uu} =1} \, , \\[0.05cm]
\sigma_{d \bar d} & = \sum_{\psi_d} \sigma(pp \to \psi_d \bar \psi_d \to \rho^0 + X)\big|_{g_{\rho^0 dd} =1} \, .
\end{split}
\end{equation}
The total production rates~(\ref{eq:xsec}) are thus simply given in terms of the ``fundamental'' cross sections of Eq.(\ref{eq:basicxsec})
--which include the contributions of all the initial partons and can be computed once for all-- 
appropriately rescaled by $g_{\rho^+ud}$, $g_{\rho^0 uu}$ and $g_{\rho^0 dd}$.
Figure~\ref{fig:xsections} shows the fundamental cross sections  as functions of the physical  mass of the resonance 
for a collider center-of-mass energy $\sqrt{s} = 8\,$TeV and $\sqrt{s} = 14\,$TeV.
%
%

In order to illustrate the typical size of the production cross sections and decay rates, we will
show results as functions of one mass and one coupling strength, fixing the other
parameters.~\footnote{All the plots shown in this section have been obtained by making use of our \texttt{Madgraph5} model. 
Although several sanity checks have been performed to test it, a full validation of the model has not been done and is left for a future work.} 
We will consider the following two sets of benchmark values:
\begin{equation}
\label{eq:benchmarks}
\begin{split}
\text{(I)}  \quad m_{\rho_L} & = 0.5\, m_{\rho_R} \, , \quad g_{\rho_L} = g_{\rho_R} \equiv g_\rho \, , \quad \xi =0.1 \\[0.25cm]
\text{(II)} \quad m_{\rho_L} & = 2.0\, m_{\rho_R} \, , \quad g_{\rho_L} = g_{\rho_R} \equiv g_\rho \, , \quad \xi =0.1 \, .
\end{split}
\end{equation}
In case (I)  the lightest charged and neutral resonances are $\rho_1^\pm$ and $\rho_1^0$. They are mostly made of~$\rho_L$ and both
couple to SM fermions with  strength $\sim g_{el}^2/g_\rho$.
In fact, their couplings and masses are equal up to $O(\xi)$ terms, see Eq.~\ref{eq:rho1tofermion}, because the breaking of the custodial symmetry due to
the hypercharge coupling $g'_{el}$ enters only through EWSB effects.
In case (II), on the other hand, the lightest resonances are $\rho_2^\pm$ and $\rho_2^0$, mostly made of $\rho_R$.
While $\rho_2^0$ couples to SM fermions with  strength $\sim g'^2_{el}/g_\rho$, the coupling of $\rho_2^\pm$
to fermions is further suppressed by a factor $\xi$, see Eq.~\ref{eq:rho2tofermion}. Prior to EWSB, indeed, the lightest charged resonance is purely $\rho_R$
and thus does not couple to SM fermions as it does not mix with any elementary vector field. In fact, this holds true even after EWSB if
$m_{\rho_L} = m_{\rho_R}$. As  previously explained, in this case the mass eigenstate $\rho^\pm$ remains unperturbed at all orders in $\xi$, and does not couple to 
elementary fields.  In all cases in which its coupling to fermions is sufficiently suppressed, the  production of
this charged resonance at the LHC proceeds mainly
via vector boson fusion (VBF), $pp \to W^\pm Z jj \to \rho^\pm jj$, or through cascade decays of the heaviest ones.

\begin{figure}[tbp]
\begin{center}
\includegraphics[width=0.4\textwidth]{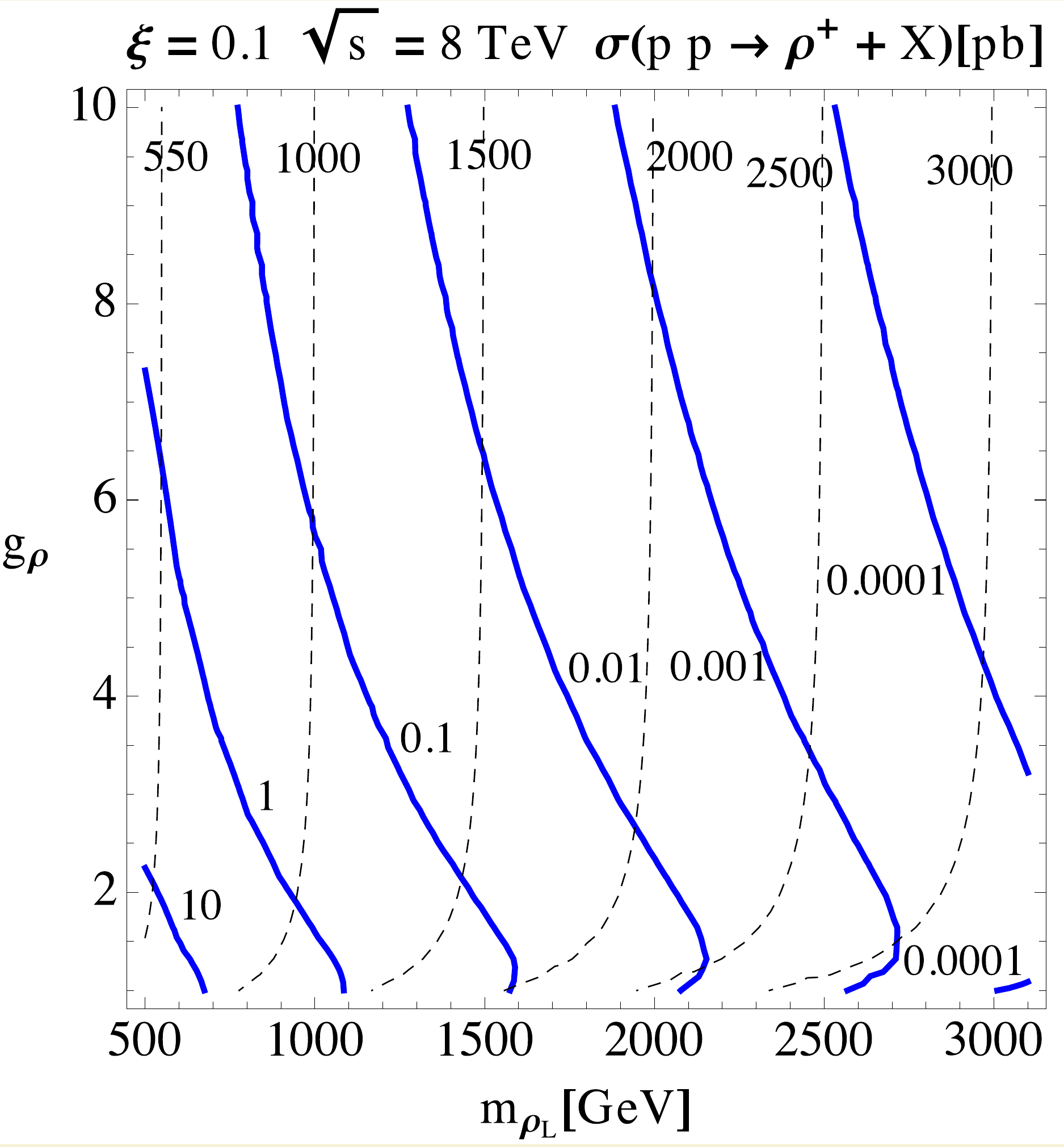}
\includegraphics[width=0.4\textwidth]{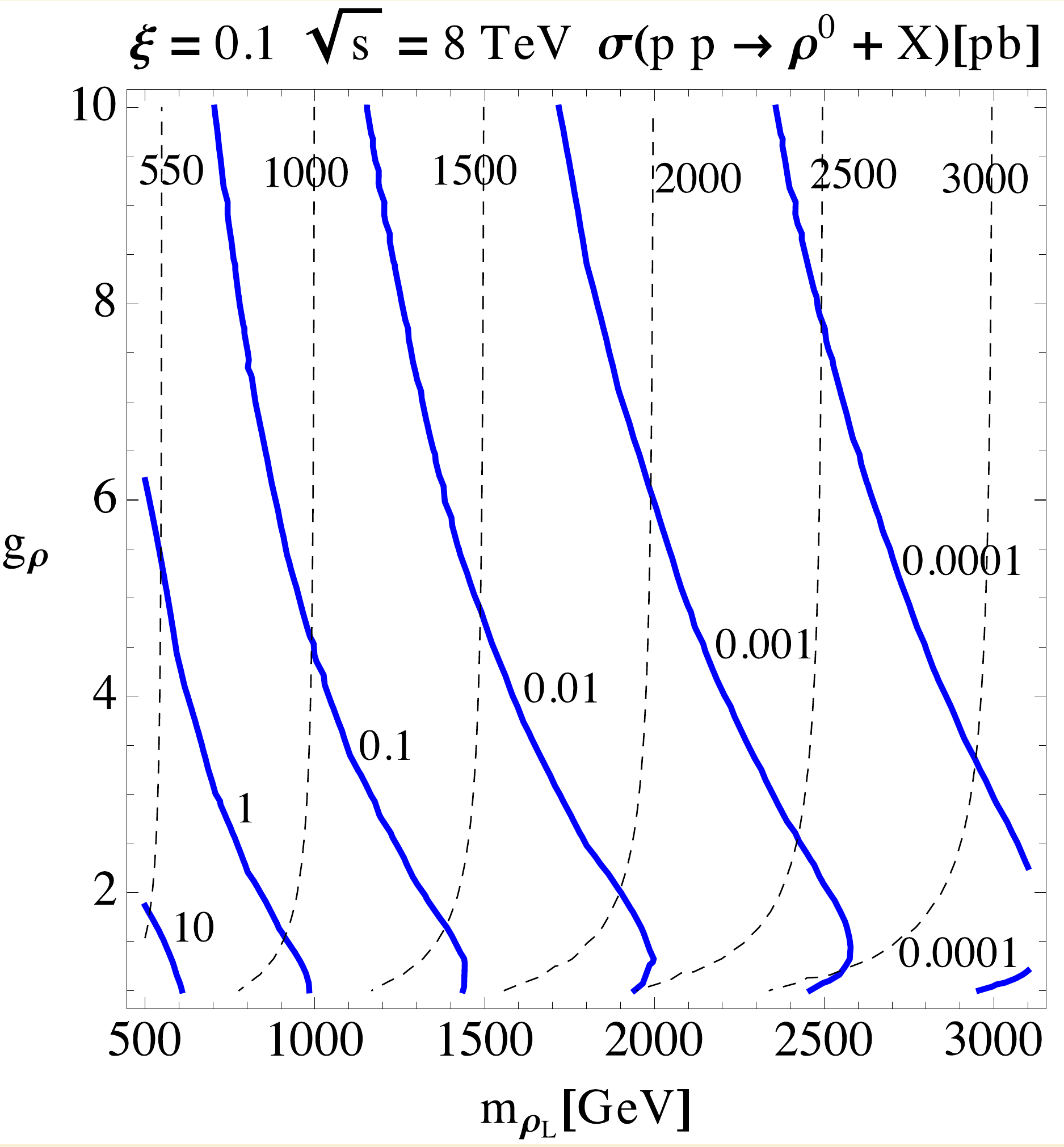} \\[0.5cm]
\includegraphics[width=0.4\textwidth]{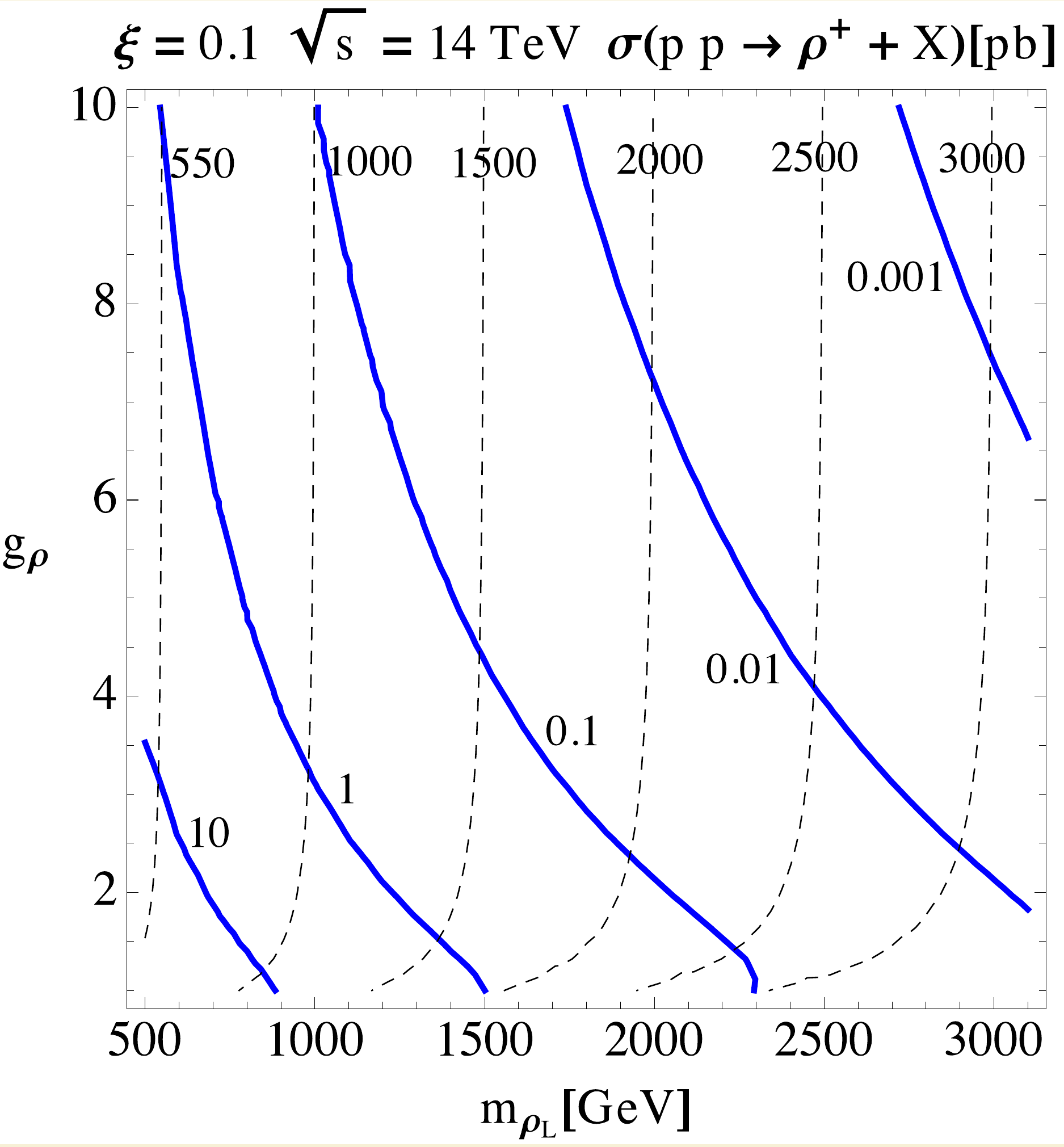} 
\includegraphics[width=0.4\textwidth]{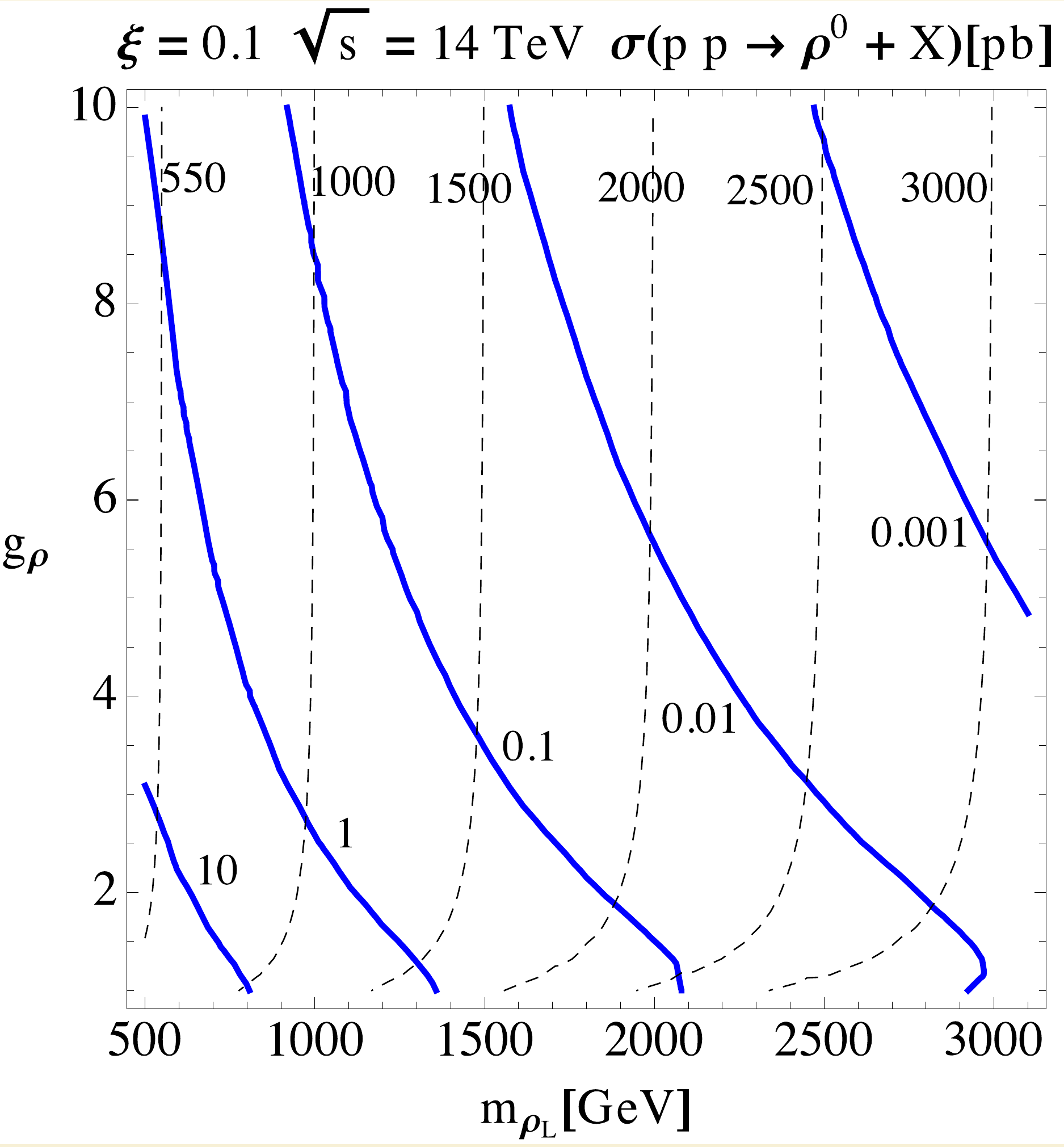} 
\end{center}
\caption{\small \label{fig:contours}
Contours of constant cross section  (blue lines, labels in picobarn) in the plane $(m_{\rho_L},g_{\rho})$  
for the production of the lightest charged and neutral resonances
at the LHC, in the benchmark case (I) of Eq.~(\ref{eq:benchmarks}). 
The dashed curves denote the contours of constant mass of the lightest  resonance (in GeV units). The plots in the upper row are done 
for $\sqrt{s} = 8\,$TeV, those in the lower row assume $\sqrt{s} = 14\,$TeV.
The contribution from vector boson fusion and possible cascade decays has been neglected for simplicity. 
}
\end{figure}

Figure~\ref{fig:contours} shows the contours of constant cross section  for the production of the lightest resonance in the plane $(m_{\rho_L}, g_{\rho})$
for the benchmark choice (I) of Eq.~(\ref{eq:benchmarks}). For simplicity we have neglected the contribution from VBF and possible cascade decays.
%
%
As expected, the cross section increases for smaller values of $g_{\rho}$, since in that limit the couplings to  SM fermions 
get larger as a consequence of the larger elementary-composite mixing. In case (II) the shapes of the contours are similar, but
the overall size of the cross section  is smaller by a factor $\sim (g'/g)^2$ and $\sim \xi\, (g'/g)^2$  respectively for the
neutral and the charged state.

\begin{figure}[tbp]
\begin{center}
\includegraphics[width=0.49\textwidth]{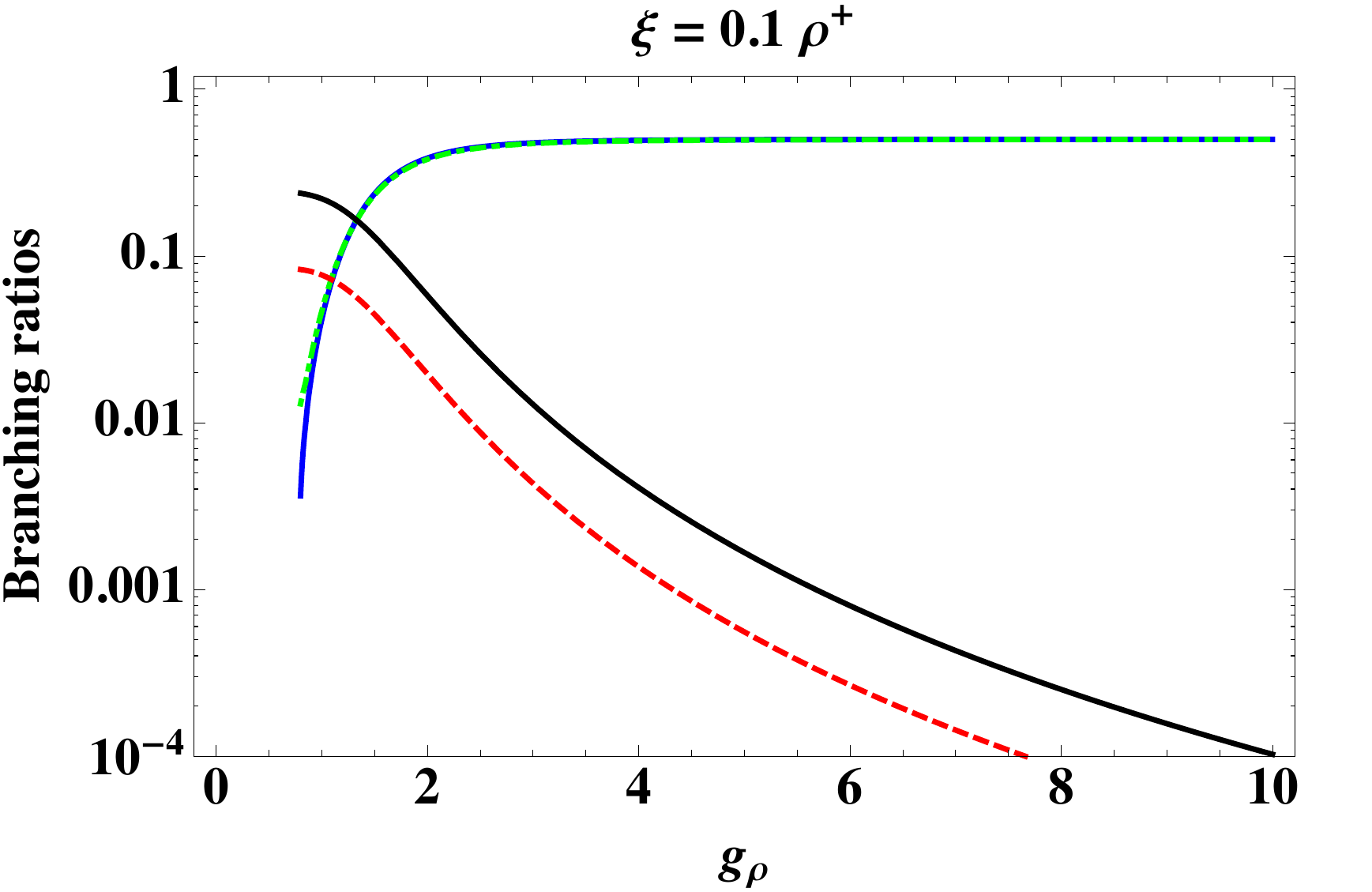}
\includegraphics[width=0.49\textwidth]{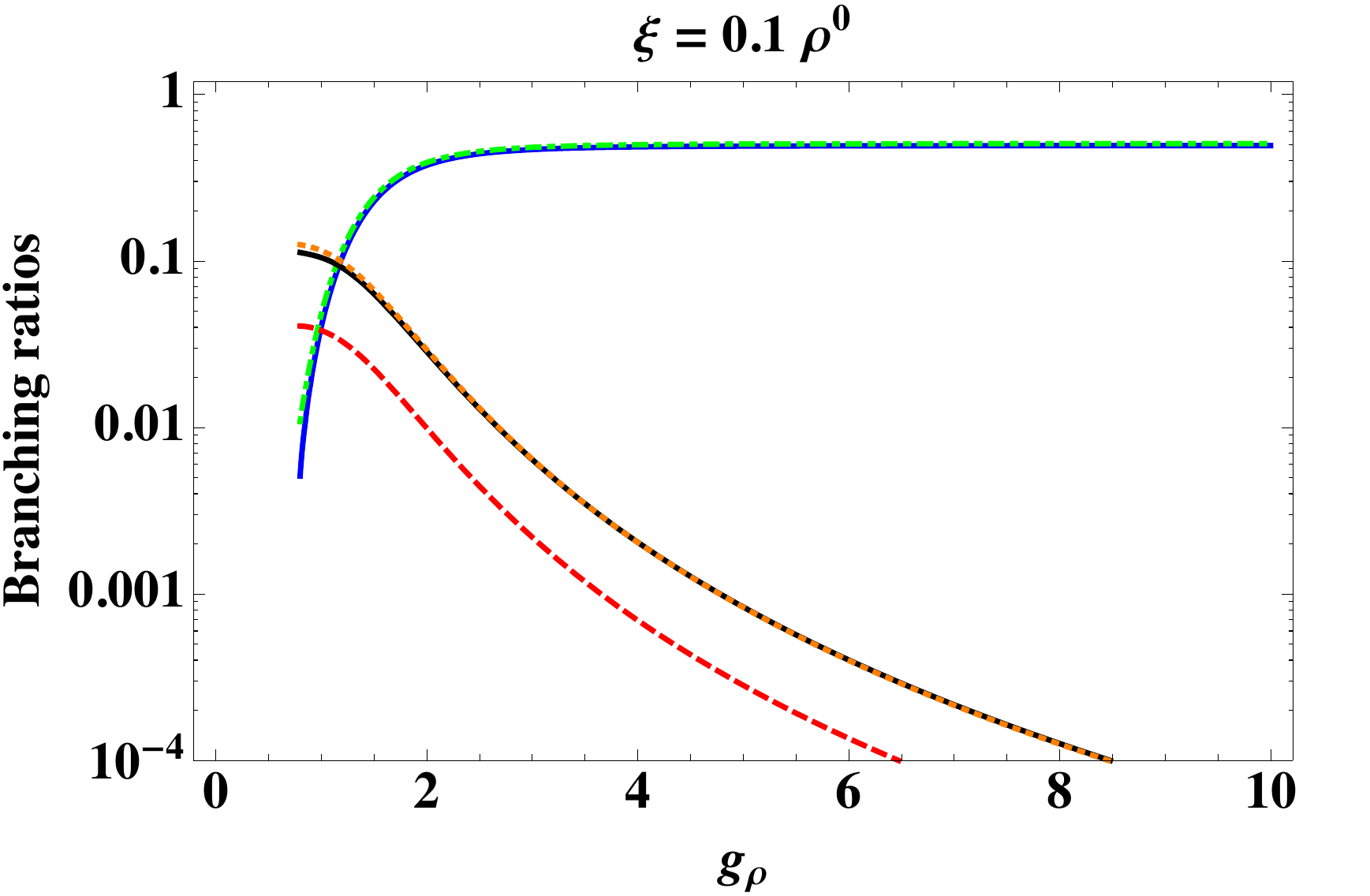} \\
\includegraphics[width=0.49\textwidth]{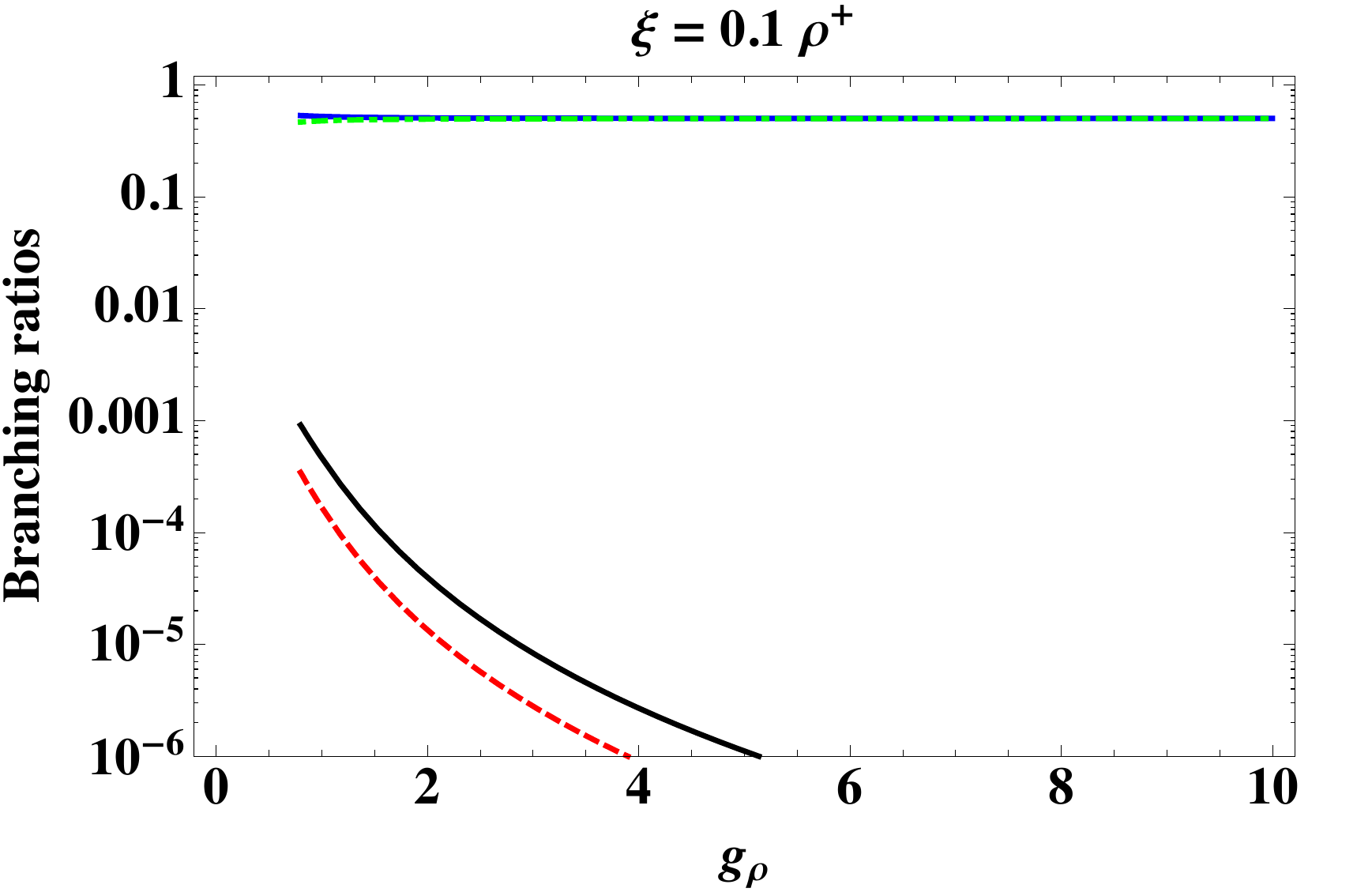}
\includegraphics[width=0.49\textwidth]{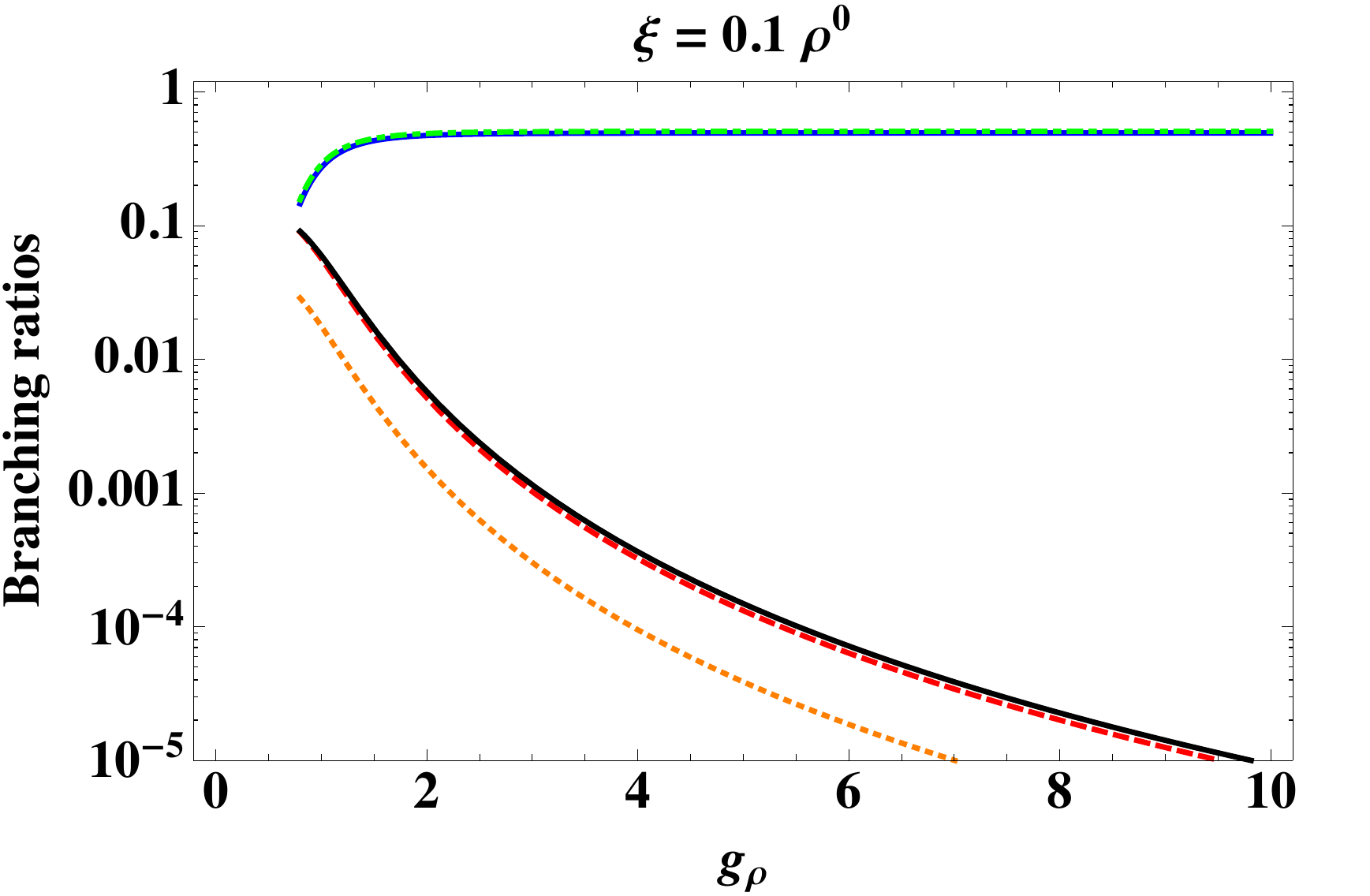}
\end{center}
\caption{\small \label{fig:BRs}
Decay branching ratios of the lightest charged (left plots) and neutral (right plots) resonance as functions of $g_{\rho}$.
Upper plots are done for the benchmark case~(I) of Eq.~(\ref{eq:benchmarks}), the lower ones refer to case~(II).
The various curves correspond to the following decay channels:
$WZ$ (solid blue), $Wh$ (dot-dashed green), $l\nu$ (dashed red),
$t\bar b$ (solid black) in left plots;  $WW$ (solid blue), $Zh$ (dot-dashed green), $l^+l^-$ (dashed red),
$t\bar t$ (solid black), $b\bar b$ (dotted orange) in right plots. 
}
\end{figure}

Concerning the decay of the resonances, for $g_{\rho} \gg g$ the dominant branching fractions are those to $VV$ and $Vh$ final states,
where $V = W,Z$. The decay rates to fermions, in the same limit, are strongly suppressed.
This is illustrated in Fig.~\ref{fig:BRs}, where we have plotted the branching ratios of the lightest neutral and charged resonance
as functions of $g_{\rho}$ for case (I) with $m_{\rho_L} = g_{\rho_L} f$ (hence $a_{\rho_L} \equiv m_{\rho_L}/(f g_{\rho_L})=1$), and 
for case (II) with $m_{\rho_R} = g_{\rho_R} f$ (hence $a_{\rho_R} \equiv m_{\rho_R}/(f g_{\rho_R})=1$).
Notice that the branching fractions to $WZ$ and $Wh$, as well as those to $WW$ and $Zh$, are equal to very good approximation. 
This is implied by the Equivalence Theorem, which works well since $m_{\rho_{L,R}} \gg m_{W,Z}$ for the chosen values of parameters.
%
%
As expected, the branching ratios of the charged resonance to fermions  are much smaller in case (II) than in case (I), as a consequence of the
suppressed couplings. In case (I) the approximate custodial symmetry implies that 
$BR(t\bar t)\simeq BR(b\bar b) \simeq 3\,  BR(l^+l^-)$. 
The equality of the $t\bar t$ and $l^+l^-$ decay fractions in case (II) is instead a numerical accident.

\begin{figure}[tbp]
\begin{center}
\includegraphics[width=0.49\textwidth]{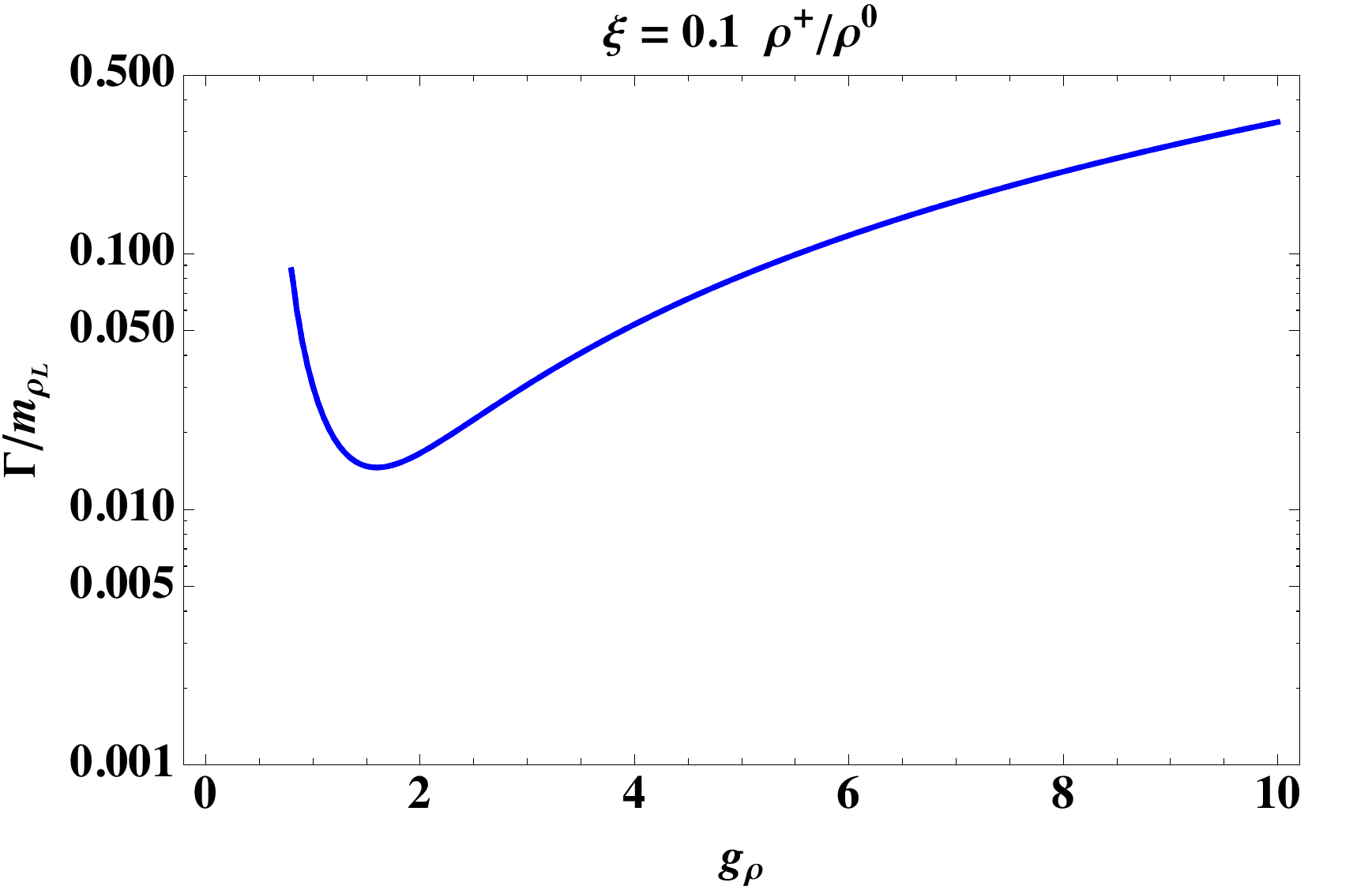}
\includegraphics[width=0.49\textwidth]{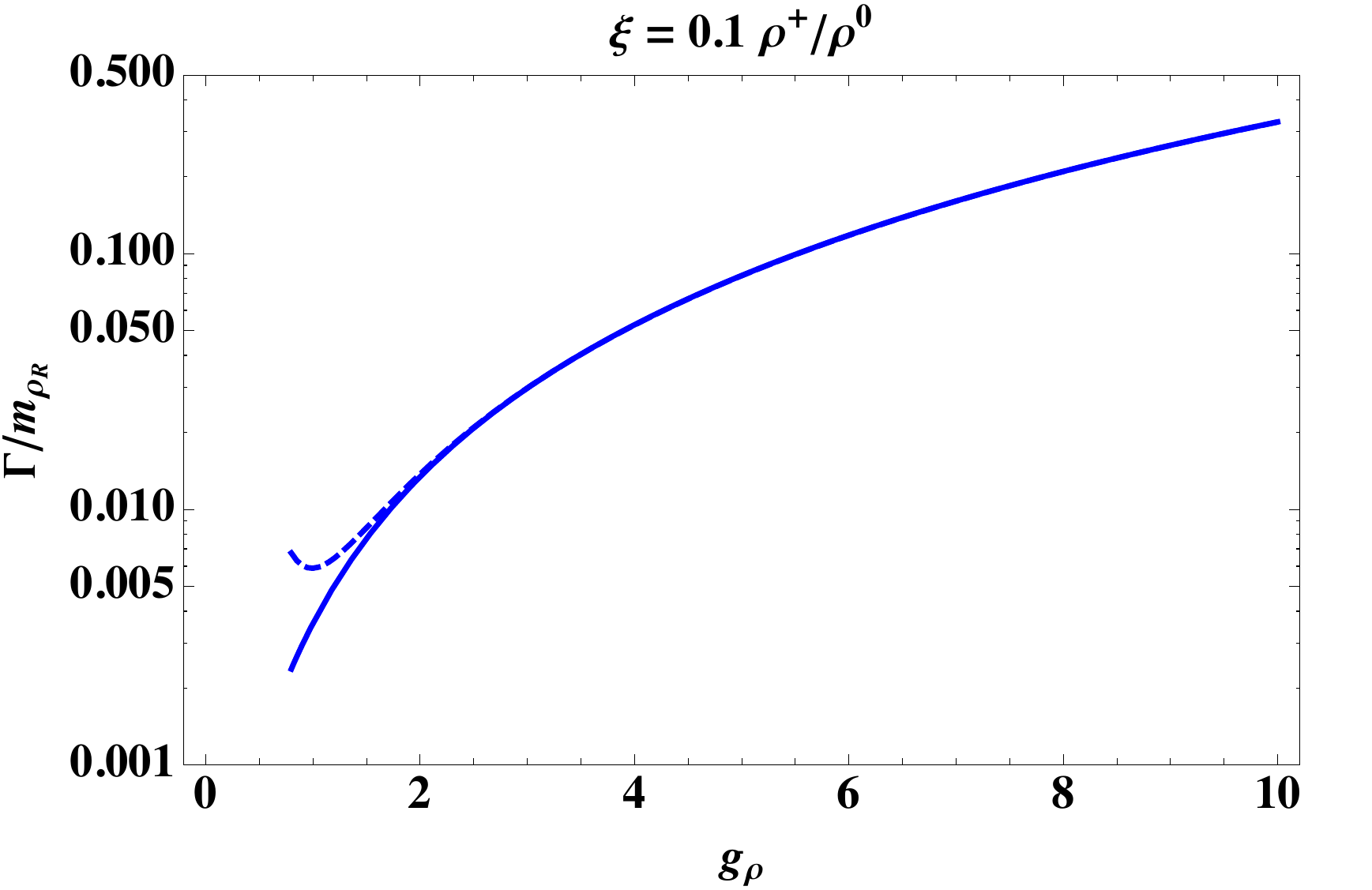} 
\end{center}
\caption{\small \label{fig:totalwidth}
Total decay width the lightest charged (solid curve) and neutral (dashed curve) resonance
for case~(I) (left plot) and case~(II) (right plot) of Eq.~({\ref{eq:benchmarks}}), respectively in units of $m_{\rho_L}$ and $m_{\rho_R}$.
In the left plot the solid and dashed curves lie on top of each other and cannot be distinguished. 
}
\end{figure}

Figure~\ref{fig:totalwidth} shows the total decay width of the lightest charged and neutral resonance 
for case (I) (left plot) and case (II) (right plot) of Eq.~({\ref{eq:benchmarks}}), respectively in units of $m_{\rho_L}$ and $m_{\rho_R}$.
%
%
In general, for large values of $g_{\rho}$ the total width is dominated by the decay modes $VV$, $Vh$ and scales as $g_{\rho}^2$.
Since the relative importance of these final states is fixed by the Equivalence Theorem, this explains why all total widths become equal
for large $g_\rho$. 
For very small $g_{\rho}$, on the other hand, the decay modes to fermions dominate and the width increases if $g_{\rho}$ decreases. 
This is in fact not true for the special case of the charged lightest resonance in case (II), where the couplings to fermions are extremely small
and the bosonic final states always dominate. Notice also that while the total decay widths of $\rho^0$ and $\rho^\pm$ are different at small $g_\rho$
in case (II), they are always equal to very good approximation in case (I). In fact, the two curves lie on top of each other in Fig.~\ref{fig:totalwidth} and cannot
be distinguished. This equality is again a consequence of the approximate custodial symmetry which is present for small $\xi$.
In all cases, the NWA is sufficiently accurate for $g_{\rho} \lesssim 4-5$.
This remains true also  for different values of  $\xi$ and $a_{\rho_{L}}$, $a_{\rho_{R}}$ provided the latter two parameters are of $O(1)$ as expected  from NDA.

All the cross sections and partial decay widths described in this section can be computed analytically
by using the \texttt{Mathematica} calculator provided with our package.
It is useful to stress that the results presented in this section are based on the validity of the Narrow Width Approximation.
This latter assumes that the production rate can be factorized into an on-shell cross section times a decay branching ratio, and 
neglects the interference with the SM background. 
Experimental analyses performed by following this approach must be carried out consistently  with its underlying assumptions,
e.g. the limits on the production rate of the new particles should be set by focusing on the on-shell signal region. 
When this is possible, it 
allows for a fast, analytical scan over the parameter space of the model. On the contrary, in the case of a broad resonance, or when the fast variation 
of the parton distribution functions makes the NWA no longer reliable due to a relevant off-shell tail at low-invariant masses, the factorized approach 
cannot be used. In this case, a robust shape analysis must rely on a full MonteCarlo simulation, thus becoming more model dependent and much more 
demanding (for a detailed discussion of these aspects see Ref.\cite{Pappadopulo:2014qza}).

\section{How to run the \texttt{Mathematica} calculator and the \texttt{MadGraph5} model}
\label{sec:program}

In this section we illustrate how to run the \texttt{Mathematica} calculator and the \texttt{MadGraph5} model
implementing the spin-1 resonances. All the software can be downloaded in a single package from the
HEPMDB website at the following URL:  \url{http://hepmdb.soton.ac.uk/hepmdb:0214.0154}.
De-compressing the downloaded archive creates a directory called \texttt{rho\_model} where all the main files are.
These include the file \texttt{frhoWSimp.nb}, which is the \texttt{Mathematica} calculator, the directory \texttt{rho}  containing
the \texttt{MadGraph5} model, and other scripts that can be used to set up the event generation.

\subsection{The \texttt{Mathematica} calculator}

The \texttt{Mathematica} calculator can be used to compute numerically the physical couplings and masses
by specifying the input parameters.
It also implements the analytical formulas for the computation of the cross sections
and partial decay rates discussed in the previous section.
The program is divided into four sections.

Section 1 is where all basic routines are defined and should be thus executed first. The user does not need to 
open it.

Section 2 is where the numerical calculation of the physical couplings and masses can be performed by means of the routine 
\vspace{-0.05cm}
\begin{flushleft}
\hspace{0.5cm} \texttt{ComputeOutputParameters}[ InputData, \texttt{fileinput.dat}, \texttt{fileoutput.dat}, mode ]
\end{flushleft}
\vspace{-0.05cm}
InputData is a list of input points, each point representing a list of input parameters. 
By appropriately setting the variable ``mode'', the user can choose among three operational modes:  in the first (``mode=LR''), 
both a $\rho_L$ and a $\rho_R$ are considered and the format of each input point must be $\{ \xi, m_{\rho_L}, g_{\rho_L}, m_{\rho_R},\ g_{\rho_R} \}$;
in the second (``mode=L''), only a $\rho_L$ is considered and the format of an input point is $\{ \xi, m_{\rho_L}, g_{\rho_L} \}$;
in the third (``mode=R''), only a $\rho_R$ is considered and the input point must be specified in the format $\{ \xi, m_{\rho_R}, g_{\rho_R} \}$.
For each of the input points the routine computes the physical masses and couplings and writes them in the file \texttt{fileoutput.dat}
in the following format: \\[0.2cm]
\hspace*{0.5cm} $n_p$, $m_W$, $\xi$, $m_{\rho_1^+}$, $m_{\rho_1^0}$, $g_{\rho_1^+ WZ}$, $g_{\rho_1^0 WW}$, $g_{\rho_1^+ Wh}$, $g_{\rho_1^0 Zh}$, 
$g_{\rho_1^+ ud}$, $g_{\rho_1^0 ffL}$, $g_{\rho_1^0 ffY}$,  \\[0.08cm]
\hspace*{0.5cm} $m_{\rho_2^+}$,  $m_{\rho_2^0}$, $g_{\rho_2^+ WZ}$, $g_{\rho_2^0 WW}$, $g_{\rho_2^+ Wh}$, $g_{\rho_2^0 Zh}$, $g_{\rho_2^+ ud}$, 
$g_{\rho_2^0 ffL}$, $g_{\rho_2^0 ffY}$ \\[0.4cm]
where $n_p$ is the number of the point. In the modes ``L'' and ``R", where only one $\rho$ is included, the couplings of the second
resonance $\rho_2$ are set to zero, and its mass is set to a default value.\footnote{In this way it never contributes to physical processes
generated by \texttt{Madgraph5}. Notice that $m_\rho \to \infty$ is a non-decoupling limit in our model, since  the
couplings to the NG bosons blow up. Therefore, one cannot  decouple one of the resonances by giving it a large mass. Rather, one should 
set all its physical couplings to zero.}
The value of the input parameters is recorded instead in the file \texttt{fileinput.dat} for convenience.

Section 3 contains the routines for the  calculation of the production cross sections and decay branching ratios.

Finally, Section 4 contains the \texttt{FeynRules}~\cite{Christensen:2008py,Alloul:2013bka} code which can be used to generate 
a UFO library~\cite{Degrande:2011ua} to be linked with the \texttt{MadGraph5} event generator.
The user does not need to execute this section unless he/she wants to modify the \texttt{MadGraph5} model, which is already 
provided in the directory \texttt{rho}.

\subsection{The \texttt{MadGraph5} model}

The main directory contains a few additional files which are useful to set up the automatic generation of events with \texttt{MadGraph5}.
The python script \texttt{wmg5.py}, in particular, should be executed first by using the syntax: 
\begin{flushleft}
\hspace{0.5cm} \texttt{python wmg5.py  --dataset=fileoutput.dat  --runmode=1  --runtimes=n} \\
\hspace{3.14cm} \texttt{ \   --scriptname=generateevents}
\end{flushleft}
It reads the physical parameters from the file \texttt{fileoutput.dat}  
and creates the new script \texttt{generateevents} which can be executed by \texttt{MadGraph5} to run the events for each of the input points.
The option \texttt{runmode} can be used to specify the run mode in \texttt{MadGraph5} (set \texttt{runmode}=1 to run at parton level), 
while \texttt{runtimes}  sets the  number of event generations to be run for each input point  (this is useful if one wants to execute multiple event 
generations by using the \texttt{multi\_run} command in \texttt{MadGraph5}).

Once the python script has been executed and the script \texttt{generateevents} has been created, there are a few more steps to follow
to generate the events. Let us indicate with \texttt{MG5dir} the main directory where  \texttt{MadGraph5} has been installed, and
with \texttt{Userdir} the user directory where events will be generated. One should then:
\vspace{0.15cm}
\begin{enumerate}
\item Copy the directory \texttt{rho} to \texttt{MG5dir/models}.
\item Edit the script \texttt{generateprocess} specifying: \textit{i)} the process to generate and \textit{ii)} the name of the directory \texttt{outdir} 
(a subdirectory of \texttt{Userdir}) to be used to store the files defining the process.
Then copy the script to \texttt{Userdir} and run: 
\vspace{0.15cm}
\begin{flushleft}
\hspace{0.5cm} \texttt{MG5dir/bin/mg5 \quad generateprocess}
\end{flushleft}
\vspace{0.15cm}
from a terminal.
\item Edit the cards \texttt{param\_card.dat}, \texttt{pythia\_card.dat} and \texttt{run\_card.dat} (a version of each card with default settings is provided 
in the main directory with the other scripts, for convenience of the user) and then copy them to \texttt{Userdir/outdir/Cards}.
\item Copy the script \texttt{generateevents} to \texttt{Userdir/outdir} and start the  generation of events by executing the command:
\vspace{0.15cm}
\begin{flushleft}
\hspace{0.5cm} \texttt{Userdir/outdir/bin/madevent \quad generateevents}
\end{flushleft}
\vspace{0.15cm}
from a terminal.
\end{enumerate}
Through this procedure the user can  generate events for each of the input points (i.e. for each choice of input masses and couplings)
in an automatic way. By creating suitable grids of points it is thus easy to scan over the model's parameter space.


\section{Conclusions}
\label{sec:conclusions}

In this work we have studied the phenomenology of spin-1 resonances arising in $SO(5)/SO(4)$ composite Higgs models.
We have focused on resonances transforming  as $(3,1)$ and $(1,3)$ of $SO(4)$ and introduced a simple Lagrangian
which contains  a minimal set of parameters: one mass and one coupling strength for each resonance. 
We have discussed the rotation to the mass eigenstate basis and provided a numerical calculator of physical masses and couplings.
We also derived approximate analytic formulas for the masses and couplings at linear order in $\xi = \sin^2\!\theta$, where~$\theta$ is the 
vacuum misalignment angle. The model has been implemented in \texttt{MadGraph5} and can be used as a benchmark in the search
for heavy spin-1 states at the LHC and future colliders.

\section*{Acknowledgments}
We are grateful to the Galileo Galilei Institute for Theoretical Physics in Florence for hospitality during part of this project. 
C.G. is supported by the Spanish Ministry MICINN under contract FPA2010-17747 and by the European Commission
 under the ERC Advanced Grant 226371 MassTeV and the contract PCIG14-GA-2013-631962. 
D.P. has been supported by the NSF Grant PHY-0855653. D.L. is supported by the China Schorlarship Foundation.  
A.T. and D.G. acknowledge support from the Swiss National Science Foundation under contract no. 200020-138131. 
The work of R.C, R.T. and A.W. was partly supported by the ERC Advanced Grant no. 267985 DaMeSyFla. 
R.T. also acknowledges support by the Research Executive Agency (REA) of the European Union under the Grant Agreement 
number PITN-GA- 2010-264564 LHCPhenoNet. A.W. acknowledges the MIUR-FIRB grant RBFR12H1MW.

\appendix

\section{Analytical formulas}

We collect here the analytical formulas discussed in the text.

The mass terms can be written as
\begin{equation}
\label{eq:mass1}
{\cal L}_{mass} = X^+ M_\pm^2 X^- + \frac{1}{2} X^0 M_0^2 X^0 \, ,
\end{equation}
where $X^\pm = (X^1 \pm i X^2)/\sqrt{2}$, $X^{1,2} = \{ W^{1,2}, \rho_L^{1,2}, \rho_R^{1,2} \}$ and $X^0 = \{ W^3, \rho_L^3 , B , \rho_R^3\}$.
The charged and neutral mass matrices are:
\begin{align}
M_\pm^2 & = 
\begin{pmatrix}
A & \displaystyle -\frac{(1 + c_\theta) g_{el} m_{\rho_L}^2}{2 g_{\rho_L}} & \displaystyle -\frac{( 1  - c_\theta) g_{el} m_{\rho_R}^2}{2 g_{\rho_R}} \\
 \displaystyle -\frac{(1 + c_\theta) g_{el} m_{\rho_L}^2}{2 g_{\rho_L}} & m_{\rho_L}^2 & 0 \\
 \displaystyle -\frac{(1 - c_\theta) g_{el} m_{\rho_R}^2}{2 g_{\rho_R}} & 0 &    m_{\rho_R}^2 
\end{pmatrix}
\\[0.3cm]
\label{eq:mass3}
M_0^2 & =
\begin{pmatrix}
A & \displaystyle -\frac{(1 + c_\theta) g_{el} m_{\rho_L}^2}{2 g_{\rho_L}} & B  & \displaystyle -\frac{(1 -c_\theta ) g_{el} m_{\rho_R}^2}{2 g_{\rho_R}} \\
 \displaystyle -\frac{(1 + c_\theta ) g_{el} m_{\rho_L}^2}{2 g_{\rho_L}} & m_{\rho_L}^2 & \displaystyle -\frac{(1 - c_\theta) m_{\rho_L}^2 g_{el}'}{2 g_{\rho_L}} & 0 \\
B & \displaystyle -\frac{(1 - c_\theta) m_{\rho_L}^2 g_{el}'}{2 g_{\rho_L}} & C & \displaystyle -\frac{(1 + c_\theta) m_{\rho_R}^2 g_{el}'}{2 g_{\rho_R}} \\
 \displaystyle -\frac{(1 - c_\theta) g_{el}   m_{\rho_R}^2}{2 g_{\rho_R}} & 0 & \displaystyle -\frac{(1 + c_\theta) m_{\rho_R}^2 g_{el}'}{2 g_{\rho_R}} & m_{\rho_R}^2
\end{pmatrix}\, , 
\end{align}
where $c_\theta \equiv \cos\theta$ and we have conveniently defined
\begin{equation}
\begin{split}
A = \, &  \frac{g_{el}^2}{4 g_{\rho_L}^2g_{\rho_R}^2} \Big[
               \left(g_{\rho_R}^2 m_{\rho_L}^2+g_{\rho_L}^2 m_{\rho_R}^2\right) \cos^2\!\theta+2\left(g_{\rho_R}^2
               m_{\rho_L}^2-g_{\rho_L}^2 m_{\rho_R}^2\right) \cos\theta \\
 & \phantom{\frac{g_{el}^2}{4 g_{\rho_L}^2g_{\rho_R}^2} \Big[} 
              +f^2 \sin^2\!\theta  \, g_{\rho_L}^2 g_{\rho_R}^2+g_{\rho_R}^2   m_{\rho_L}^2+g_{\rho_L}^2   m_{\rho_R}^2 \Big] \, ,  \\[0.2cm]
B = \, & \frac{g_{el}  g_{el}' \sin ^2\!\theta}{4g_{\rho_L}^2 g_{\rho_R}^2}
              \left(m_{\rho_L}^2 g_{\rho_R}^2+g_{\rho_L}^2m_{\rho_R}^2-f^2   g_{\rho_L}^2 g_{\rho_R}^2\right)\, , \\[0.2cm]
C = \, & \frac{g_{el}'^2}{4 g_{\rho_L}^2g_{\rho_R}^2} \Big[
              \left(g_{\rho_R}^2 m_{\rho_L}^2+g_{\rho_L}^2 m_{\rho_R}^2\right) \cos ^2\theta+\left(2g_{\rho_L}^2 m_{\rho_R}^2-2 g_{\rho_R}^2
              m_{\rho_L}^2\right) \cos\theta \\
& \phantom{\frac{g_{el}'^2}{4 g_{\rho_L}^2g_{\rho_R}^2} \Big[}
              +f^2 \sin^2\!\theta \, g_{\rho_L}^2 g_{\rho_R}^2+g_{\rho_R}^2 m_{\rho_L}^2+g_{\rho_L}^2 m_{\rho_R}^2 \Big]\, . 
\end{split}
\end{equation}

As explained in Section~\ref{sec:model}, it is useful to diagonalize the mass matrix analytically by expanding for $\xi$ small.
In the following we collect the expressions of the cubic vertices with  zero and one~$\rho$. Notice that our sign convention
for these vertices differs from that of Ref.~\cite{Contino:2011np} while it coincides with the one implemented in \texttt{Madgraph5}
for the SM vertices. One can easily pass from one convention to the other by redefining the sign of the input couplings: 
$g_{el} \to - g_{el}$, $g_{el}^\prime \to - g_{el}^\prime$, $g_{\rho_{L,R}} \to - g_{\rho_{L,R}}$. Obviously this redefinition does not
affect physical observables.

Let us start by discussing the vertices among three SM fields. By $U(1)_{em}$ gauge invariance, all the couplings appearing in vertices
with a photon ($WWA$ and $\bar\psi\psi A$) are equal (up to the particle's charge) to the electromagnetic coupling $\sqrt{4\pi \alpha_{em}}$
given in Eq.~(\ref{eq:alphaem}).  This holds to all orders in $\xi$.
There are no non-minimal couplings involving the photon, since there are no non-minimal operators which can generate them 
in the initial Lagrangian~(\ref{Lag}).
The triple gauge interaction $WWZ$ has the same structure as in the SM and its coupling reads, at linear
order in $\xi$:
\begin{equation}
g_{WWZ} = \frac{g^2}{\sqrt{g^2 + g^{\prime\, 2}}}\, ,
\end{equation}
where $g$, $g^\prime$ have been defined in Eq.~(\ref{eq:ggprime}).
Notice that the $O(\xi)$ terms are vanishing, and  EWSB corrections only arise at $O(\xi^2)$. 
We parametrize the couplings of the SM fermions to SM vector bosons as follows
\begin{equation}
\frac{1}{\sqrt{2}} g_{Wud} \, W^+_{\mu}\bar{\psi}_{u} \gamma^\mu P_L \psi_{d} + h.c. + Z_{\mu} \bar{\psi} \gamma^\mu \left( P_L (g_{Z ffL} - g_{Z ffY} ) T_{L}[\psi] + g_{Z ffY } Q[ \psi] \right) \psi \, ,
\end{equation}
where $T_L[\psi_u] = +1/2$, $T_L[\psi_d] = -1/2$ and $\psi$ can be any of the $\psi_u$ and $\psi_d$ fermions in the case of the $Z$ interaction.
We find, at linear order in $\xi$
\begin{equation}
\begin{split}
g_{Wud} & = g +\xi\,  \frac{{g}^3 \left(m_{\rho_L}^2-f^2 {g}^2\right)}{4 g_{\rho_L}^2 m_{\rho_L}^2}  \, ,\\[0.1cm]
g_{Z ffL} & = \frac{{g}^2}{\sqrt{{g}^2+{g'}^2}} -\xi\, \frac{  \sqrt{{g}^2+{g'}^2} \left(f^2 {g}^4-{g}^2 m_{\rho_L}^2\right)}{4 g_{\rho_L}^2 m_{\rho_L}^2}  \, ,\\[0.1cm]
g_{Z ffY} & =-\frac{{g'}^2}{\sqrt{{g}^2+{g'}^2}} + \xi\, \frac{ \sqrt{{g}^2+{g'}^2} \left(f^2 {g'}^4-{g'}^2 m_{\rho_R}^2\right)}{4 g_{\rho_R}^2 m_{\rho_R}^2}\, .
\end{split}
\end{equation}

The expressions for the $VV\rho$ and $Vh\rho$ couplings defined in Eq.~(\ref{eq:cubic}) are, at leading order in $\xi$:
\begin{equation}
\label{eq:VVrho}
\begin{split}
g_{\rho^0_1WW} &=  m_W^2 \left(-\frac{1}{f^2
   \left(g_{el}^2+g_{\rho_L}^2\right){}^{1/2
   }}  
+\frac{g_{el}^2
   g_{\rho_L}^2}{m_{\rho_L}^2
   \left(g_{el}^2+g_{\rho_L}^2\right){}^{3/2
   }}\right),  \\[0.1cm]
g_{\rho_1^{+}WZ} & = \frac{m_Z}{m_W} \, g_{\rho^0_1WW}\, , \\[0.1cm]
g_{\rho^0_2WW} &=  m_W^2 \left(-\frac{1}{f^2
   \left(g_{el}'^2+g_{\rho_R}
   ^2\right){}^{1/2}}
+\frac{g_{\rho_R}^2
   g_{el}'^2}{m_{\rho_R}^2
   \left(g_{el}'^2+g_{\rho_R}
   ^2\right){}^{3/2}}\right),  \\[0.1cm]
g_{\rho_2^{+}WZ} & =  -\frac{m_W m_Z}{f^2 g_{\rho_R}} \, ,
\end{split}
\end{equation}
and
\begin{equation}
\label{eq:Vhrho}
\begin{split}
g_{\rho_1^{+}Wh}& =   m_W \left(
\frac{m_{\rho_L}^2
   \sqrt{g_{el}^2+g_{\rho_L}^2}}{f^2
   g_{\rho_L}^2}-\frac{g_{el}^2}{\sqrt{g_{el}^2+g_{\rho_L}^2}} \right), \\[0.1cm]
g_{\rho_1^0Zh} &= \frac{m_Z}{m_W} \, g_{\rho_1^{+}Wh} \, , \\[0.1cm]
g_{\rho_2^{+}Wh}& = 
-\frac{m_{\rho_R}^2}{f^2 g_{\rho_R}} \, m_W \, , \\[0.1cm]
g_{\rho_2^0Zh} &=    m_Z  \left( \frac{g_{el}'^2}{\sqrt{g_{el}'^2+g_{\rho_R}^2}}-\frac{m_{\rho_R}^2
   \sqrt{g_{el}'^2+g_{\rho_R}
   ^2}}{f^2 g_{\rho_R}^2} \right)\, .  
\end{split}
\end{equation}
Here $m_W$ and $m_Z$ stand for the $O(\xi)$ expressions reported in Eq.~(\ref{eq:spectrum}).
Notice that each of the couplings in Eqs.~(\ref{eq:VVrho}),~(\ref{eq:Vhrho}), with the exception of $g_{\rho_2^{+}WZ} $ and $g_{\rho_2^{+}Wh}$, gets two 
$O(\xi)$ contributions, which 
have opposite relative sign: one comes from the $\rho$ mass terms, the other from the kinetic term of the NG bosons.
Some choices of input parameters lead to a cancellation between these two contributions resulting in an $O(\xi)$ term accidentally smaller
than $O(\xi^2)$ terms.  In these cases  the formulas of Eqs.~(\ref{eq:VVrho}),~(\ref{eq:Vhrho}) become numerically inaccurate.

Finally, at linear order in $\xi$, the couplings of the resonances to  SM fermions appearing in Eq.~(\ref{eq:cubic}) are given by:
\begin{equation}
\label{eq:rho1tofermion}
\begin{split}
g_{\rho_1^+ud} = \, & - \frac{g_{el}^2}{\sqrt{g_{el}^2+g_{\rho_L}^2}}  + \xi \, \frac{ g_{el}^2 g_{\rho_L}^2
   \left(m_{\rho_L}^2
   \left(g_{el}^2+g_{\rho_L}^2\right)-f^2
   g_{el}^2 g_{\rho_L}^2\right)}{4
   m_{\rho_L}^2
   \left(g_{el}^2+g_{\rho_L}^2\right){}^{5/
   2}}   ,  \\[0.3cm]
g_{\rho_1^0ffL}  = \, & g_{\rho_1^+ud} \, ,\\[0.3cm]
g_{\rho_1^0ffY} = \, &  \xi \, \frac{ g_{el}'^2}{4 m_{\rho_L}^2\left(g_{el}^2+g_{\rho_L}^2\right){}^{3/2} 
                                                    \left(g_{\rho_R}^2 m_{\rho_L}^2 \left(g_{el}^2+g_{\rho_L}^2\right)-g_{\rho_L}^2 m_{\rho_R}^2
                                                    \left(g_{el}'^2+g_{\rho_R}^2\right)\right) }  \\[0.15cm]
                            & \times \Big[  \, f^2 g_{el}^2 g_{\rho_L}^2 g_{\rho_R}^2 \left(g_{el}^2 m_{\rho_L}^2+g_{\rho_L}^2
                                                    \left(m_{\rho_L}^2-m^2_{\rho_R}\right) \right) \\ 
                           &  \hspace{0.65cm}         
                                                    + m_{\rho_L}^2 \left(g_{el}^2+g_{\rho_L}^2\right) \left(g_{\rho_L}^2 m_{\rho_R}^2
                                                    \left(g_{\rho_R}^2-g_{el}^2\right)-g_{\rho_R}^2 m_{\rho_L}^2
                                                    \left(g_{el}^2+g_{\rho_L}^2\right)\right ) \Big]\, ,
\end{split}
\end{equation}
and 
\begin{equation}
\label{eq:rho2tofermion}
\begin{split}
g_{\rho_2^+ud} = \, &   \xi \, \frac{g_{el}^2 g_{\rho_L}^2
   \left(m_{\rho_R}^2-m_{\rho_L}^2\right)}{4
   g_{\rho_R} m_{\rho_L}^2
   \left(g_{el}^2+g_{\rho_L}^2\right)-4
   g_{\rho_L}^2 g_{\rho_R} m_{\rho_R}^2}  ,     \\[0.3cm]
g_{\rho_2^0ffL}  = \, &  -\xi  \frac{ g_{el}^2}{4 m_{\rho_R}^2 \left(g_{el}'^2+g_{\rho_R}^2\right)^{3/2} \left(g_{\rho_L}^2
                                                m_{\rho_R}^2 \left(g_{el}'^2+g_{\rho_R}^2\right)-g_{\rho_R}^2 m_{\rho_L}^2
                                                \left(g_{el}^2+g_{\rho_L}^2\right)\right)}  \\[0.15cm]
                         & \times \Big[ \, f^2 g_{\rho_L}^2 g_{\rho_R}^4 m_{\rho_L}^2 g_{el}'^2 +g_{\rho_L}^2 m_{\rho_R}^4 \left(g_{el}'^2+g_{\rho_R }^2\right)^2 \\
                         & \phantom{\times \Big[ \,} 
                                                -g_{\rho_R}^2 m_{\rho_R}^2 \left(g_{el}'^2+g_{\rho_R}^2\right) \left(f^2 g_{\rho_L}^2 g_{el}'^2+m_{\rho_L}^2
                                                 \left(g_{\rho_L}^2- g_{el}^{\prime 2}\right)\right) \Big]  ,    \\[0.3cm]
g_{\rho_2^0ffY} = \, &    -\frac{g_{el}'^2}{\sqrt{g_{el}'^2+g_{\rho_R}^2}} + \xi \, \frac{g_{\rho_R}^2
   g_{el}'^2
   \left(m_{\rho_R}^2
   \left(g_{el}'^2+g_{\rho_R}^2\right)-f^2 g_{\rho_R}^2
   g_{el}'^2\right)}{4
   m_{\rho_R}^2
   \left(g_{el}'^2+g_{\rho_R }^2\right){}^{5/2}} \,  .
\end{split}
\end{equation}
%




%% file: nmfvsusy/nmfvsusy.tex

\newcommand{\sm}[2]{ \tilde{M}_{#1, \tilde{#2}}}
\newcommand{\smi}[3]{ \tilde{M}_{#1, \tilde{#2}_#3}}
\newcommand{\m}[3]{M_{#1,{#2_#3}}}


\chapter{Indirect Constraints on Non-Minimal Flavour Violating Supersymmetry}

{\it K.~De Causmaecker, B.~Fuks, B.~Herrmann, F.~Mahmoudi,
  B.~O'Leary, W.~Porod, S.~Sekmen, N.~Strobbe}


\begin{abstract}
We present an analysis of non-minimal flavour violating effects arising
in general versions of the Minimal Supersymmetric Standard Model. Considering
several flavour and electroweak observables and the recent discovery of a
Higgs boson, we perform a scan of the model parameter space and then
design a set of experimentally non-excluded reference scenarios for which
we study the dependence of the considered observables on squark flavour violation.
\end{abstract}

Among all candidate theories extending the Standard Model (SM), supersymmetry remains,
after more than 30 years, one of the most studied and popular choices.  As a 
consequence, the quest for the supersymmetric partners of the SM particles is one of
the key topics of the current high-energy physics experimental program.
As up to now no sign of supersymmetry has been found, in particular at the Large
Hadron Collider (LHC), either the superpartners are constrained to be heavy or the
spectrum must present very specific properties allowing the superpartners to
evade detection \cite{atlassusy,cmssusy}. However these statements are valid for interpretations within the context of simplified model spectra (SMSs), or the most constrained version of the Minimal Supersymmetric Standard Model (MSSM), where over a hundred free model parameters are reduced to a set of four parameters and a sign.
General realizations of the MSSM could therefore turn out to be less
constrained by data.  In this work we consider a version of the MSSM with general
flavour-violating squark mixings. This feature is expected to lead to interesting
phenomenological consequences, both with respect to indirect constraints on new
physics derived from low-energy, flavour and electroweak precision observables,
and in the context of the LHC~\cite{Bozzi:2007me,Nomura:2007ap,Hiller:2008wp,
Fuks:2008ab,Hurth:2009ke,Bartl:2009au,Hiller:2009ii,Bartl:2010du,Bruhnke:2010rh,
Muhlleitner:2011ww,Bartl:2011wq,Fuks:2011dg,Fuks:2014xpa}.
In the framework of a global effort to unveil flavour effects in supersymmetry in light of the present
high-energy physics data, we investigate non-minimal flavour-violation (NMFV) in the squark sector for
several flavour physics observables, for the anomalous magnetic moment of the
muon and for the mass of the recently observed Higgs boson.

In the following, we briefly detail the properties of the form of the NMFV
MSSM under consideration and present our strategy to design
theoretically motivated reference scenarios not yet experimentally excluded,
relevant to be considered in the context of LHC data. We then define two 
reference scenarios, for which we study in detail the dependence of the various 
considered observables on the NMFV parameters. Conclusions are given at the end
of this contribution.

In the super-CKM basis, the mass matrices of the up- and down-type squarks are represented by
\begin{align}
    \arraycolsep=1.4pt\def\arraystretch{1.2}
    M_{\tilde{q}}^2 =
    \left(
    \begin{array}{ccc|ccc}
        \m{L}{q}{1}^2 & \Delta^{q_1 q_2}_{LL} & \Delta^{q_1 q_3}_{LL} & m_{q_1} X_{q_1} & \Delta_{LR}^{q_1 q_2} & \Delta_{LR}^{q_1 q_3} \\
        \Delta^{q_1 q_2 *}_{LL} & \m{L}{q}{2}^2 & \Delta^{q_2 q_3}_{LL} & \Delta_{RL}^{q_1 q_2 *} & m_{q_2} X_{q_2} & \Delta_{LR}^{q_2 q_3} \\
        \Delta^{q_1 q_3 *}_{LL} &  \Delta^{q_2 q_3*}_{LL} & \m{L}{q}{3}^2 & \Delta_{RL}^{q_1 q_3 *} & \Delta_{RL}^{q_2 q_3*} & m_{q_3} X_{q_3} \\ \hline
        m_{q_1} X^*_{q_1} & \Delta_{RL}^{q_1 q_2} & \Delta_{RL}^{q_1 q_3} & \m{R}{q}{1}^2 & \Delta^{q_1 q_2}_{RR} & \Delta^{q_1 q_3}_{RR}\\
       \Delta_{LR}^{q_1 q_2 *} & m_{q_2} X^*_{q_2} & \Delta_{RL}^{q_2 q_3} & \Delta^{q_1 q_2 *}_{RR} & \m{R}{q}{2}^2 & \Delta^{q_2 q_3}_{RR}\\
       \Delta_{LR}^{q_1 q_3 *} & \Delta_{LR}^{q_2 q_3 *} & m_{q_3} X^*_{q_3} & \Delta^{q_1 q_3 *}_{RR} & \Delta^{q_2 q_3 *}_{RR} & \m{R}{q}{3}^2\\
    \end{array}
    \right) \qquad\text{with}\qquad (q = u,d) \ ,
\end{align}
where the diagonal entries of each block are related to the parameters of the
supersymmetry-breaking and SM sectors,
\begin{align}
	\m{L}{q}{i}^2 = \smi{L}{q}{i}^2 + m_{q_i}^2 + m_Z^2 \left( I_q - e_q s^2_w\right) \cos 2 \beta \, &, \quad
	\m{R}{q}{i}^2 = \smi{R}{q}{i}^2 + m_{q_i}^2 + e_q m_Z^2 s^2_w  \cos 2 \beta  \ , \nonumber \\
	 X_{u_i} = A^*_{u_i} - \mu \cot \beta\, &, \quad
	 X_{d_i} = A^*_{d_i} - \mu \tan \beta .
\end{align}
In these expressions, $m_Z$ and $s_w$ denote the $Z$-boson mass and the sine of
the electroweak mixing angle, respectively, and $e_q$ and $I_q$ are
the electric charge and isospin quantum numbers of the (s)quarks. The up-type
and down-type quark masses are represented by $m_{u_i}$ and $m_{d_i}$ with $i$
being a flavour index and the matrices $\sm{L}{q}^2$ and $\sm{R}{q}^2$ are
related to the usual soft squark mass matrices $\hat{m}_{\tilde{q}}^2$,
$\hat{m}_{\tilde{u}}^2$ and $\hat{m}_{\tilde{d}}^2$ through
\begin{align}
	\sm{L}{u}^2 = V_{CKM} \hat{m}^2_{\tilde{q}} V_{CKM}^{\dag} \, , \quad
	\sm{L}{d}^2 = \hat{m}^2_{\tilde{q}} \, , \quad
	\sm{R}{u}^2 = \hat{m}^2_{\tilde{u}}\, , \quad
	\sm{R}{d}^2 = \hat{m}^2_{\tilde{d}} \ .
\end{align}
Turning to the Higgs sector parameters, $\mu$ stands for
the off-diagonal Higgs mixing parameter,
$\tan\beta$ for the ratio of the vacuum expectation values of the
neutral components of the two Higgs doublets, and $A_{q_i}$ for the trilinear
couplings of the Higgs fields to squarks. As in earlier
works, the non-diagonal entries of the mass matrices are normalized relatively
to the diagonal ones,
\begin{equation}
   \Delta_{ab}^{q_iq_j} = \lambda^{q_iq_j}_{ab} \smi{a}{q}{i} \smi{b}{q}{j} \ ,
\end{equation}
such that non-minimal flavour violation is parameterized by the dimensionless
quantities $\lambda^{q_iq_j}_{ab}$. Since the matrices $\sm{L}{u}^2$ and
$\sm{L}{d}^2$ are related by $SU(2)_L$ gauge invariance, the
$\lambda^{u_iu_j}_{LL}$ parameters can be obtained from the knowledge of the
$\lambda^{d_id_j}_{LL}$ ones, so that we choose the latter as independent
parameters. Moreover, motivated by measurements in the kaon sector,
we ignore mixing of the first
generation squarks~\cite{Ciuchini:2007ha} and further assume, for simplicity,
\begin{align}
    \lambda_{L} \equiv  \lambda^{sb}_{LL}  \quad , \quad
    \lambda_{R} \equiv \lambda_{RR}^{sb} = \lambda_{RR}^{ct}    \quad , \quad
    \lambda_{LR} \equiv  \lambda^{ct}_{LR}  = \lambda^{ct}_{RL}  = \lambda^{sb}_{LR} =  \lambda^{sb}_{RL} \ .
\end{align}
This leaves us with three free parameters $\lambda_{L}$, $\lambda_{R}$, $\lambda_{LR}$ describing squark flavour violation.

In addition to these parameters, the gaugino sector is further
determined by the bino mass parameter $M_1$, which we relate to the wino and gluino tree-level masses $M_2$ and $M_3$ through the relation
$M_1 = M_2/2 = M_3/6$ inspired by Grand-Unified theories. Furthermore,
all diagonal sfermion soft-mass parameters are set to the common scale $M_{\rm SUSY}$, the trilinear couplings involving any third-generation sfermions are taken equal $A_t = A_b = A_{\tau}$, while all other trilinear scalar couplings are set to zero. Our model parametrization is finally
completed by including the pole mass of the pseudoscalar Higgs boson $m_{A^0}$.
We define all parameters at the electroweak scale and perform a grid scan over the following ranges in parameter space:
\begin{itemize}
  \item[-] $\tan \beta = [10,40]$,
  \item[-] $M_1 = [100, 350, 600, 850, 1100, 1350, 1600]~{\rm GeV}$,
  \item[-] $A_{t,b,\tau} = [-10000, \dots, 10000]$ GeV in steps of 1000 GeV,
  \item[-] $\mu = [100, 350, 600, 850]$ GeV,
  \item[-] $m_{A_0} = [100, 350, 600, 850, 1100, 1350, 1600]$ GeV,
  \item[-] $M_{\rm SUSY} = [100, 350, 600, 850, 1100, 1350, 1600, 2000, 2500,
     3000, 3500]$ GeV,
  \item[-] $\lambda_{L} = [-0.8, \dots, 0.8]$ in steps of $0.1$,
  \item[-] $\lambda_{LR} = [-0.0025, \dots, 0.0025]$ in steps of $0.0005$,
  \item[-] $\lambda_{R} = [-0.8, \dots, 0.8]$ in steps of $0.1$.
\end{itemize}

\renewcommand{\arraystretch}{1.2}
\begin{table}[t]
	\begin{center}
	\begin{tabular}{|c|c|c|} 
	\hline
	Observable    & Experimental result & Reference \\
	\hline\hline
 	${\rm BR}(B \rightarrow X_s\gamma) $ & $(3.43 \pm 0.21^{\rm stat} \pm 0.07^{\rm sys} \pm 0.24^{\rm th})\times 10^{-4}$ & \cite{Amhis:2012bh,Misiak:2006zs,Mahmoudi:2007gd} \\ 
	\hline
	${\rm BR}(B_s \rightarrow \mu \mu)$ & $(2.9 \pm 0.7^{\rm exp} \pm 0.29^{\rm th}) \times 10^{-9}$ & \cite{Aaij:2013aka,Chatrchyan:2013bka,CMS-PAS-BPH-13-007,Mahmoudi:2012un}  \\ 
	\hline
	${\rm BR}(B \rightarrow K^* \mu \mu)_{q^2 \in [1,6] ~\text{GeV}^2}$ & $1.7 \pm 0.3^{\rm exp} \pm 1.7^{\rm th} \times 10^{-7}$  & \cite{Aaij:2013qta,Descotes-Genon:2013vna,Hurth:2013ssa} \\ 
	\hline
	${\rm AFB}(B \rightarrow K^* \mu \mu)_{q^2 \in [1,6] ~\text{GeV}^2}$ & $(-0.17^{+ 0.06^{\rm exp} + 0.037^{\rm th}}_{- 0.06^{\rm exp} - 0.034^{\rm th}}) \times 10^{-7} $ & \cite{Aaij:2013qta,Descotes-Genon:2013vna,Hurth:2013ssa} \\ 
	\hline
	${\rm BR}(B \rightarrow X_s \mu \mu)_{q^2 \in [1,6] ~\text{GeV}^2}$ & $(1.60 \pm 0.68^{\rm exp} \pm 0.16^{\rm th} ) \times 10^{-6}$ & \cite{Aubert:2004it,Iwasaki:2005sy,Huber:2007vv,Hurth:2012jn} \\ 
	\hline
	${\rm BR}(B \rightarrow X_s \mu \mu)_{q^2 > 14.4 ~\text{GeV}^2}$ & $(4.18 \pm 1.35^{\rm exp} \pm 0.44^{\rm th}) \times 10^{-7} $ & \cite{Aubert:2004it,Iwasaki:2005sy,Huber:2007vv,Hurth:2012jn} \\ 
	\hline
	${\rm BR}(B_u \rightarrow \tau \nu)$ & $ (1.05\pm 0.25^{\rm exp} \pm 0.29^{\rm th} )\times 10^{-4} $ & \cite{Beringer:1900zz, Mahmoudi:2007vz, Mahmoudi:2008tp} \\ 
	\hline
	$\Delta M_{B_s}$ & ($17.719 \pm 0.043^{\rm exp} \pm 3.3^{\rm th}) {\rm ps}^{-1}$ & \cite{Amhis:2012bh,Ball:2006xx} \\ 
	\hline
	$\Delta a_\mu$ & $(26.1 \pm 12.8)\times 10^{-10}$ $[e^+e^-]$ & \cite{Hagiwara:2011af}  \\ 
	\hline
	$m_h$ & $125.5\pm2.5$ GeV & \cite{ATLAS-CONF-2013-014,CMS-PAS-HIG-13-002} \\ 
	\hline
	\end{tabular}
	\caption{Experimental constraints imposed in our parameter scan over the MSSM.}
	\label{tab:PLMs}
	\end{center}
\end{table}
\renewcommand{\arraystretch}{1.0}

For each point in the parameter space, we calculate the theory predictions for all the flavour observables listed in Table~\ref{tab:PLMs}.  We then compute a likelihood for each observable based on its measurement, and then an overall likelihood for each point given as the product of the likelihoods for each observable, which allows us to select reference scenarios.  
Concerning the theoretical predictions of the observables under consideration, the branching ratio of the rare $B$ decays
${\rm BR}(B \rightarrow X_s\gamma)$, ${\rm BR}(B \rightarrow K^* \mu \mu)$, 
${\rm BR}(B \rightarrow X_s \mu \mu)$, ${\rm BR}(B_u \rightarrow \tau \nu)$ and
the forward-backward asymmetry in the $B$ meson three-body decay
${\rm AFB}(B \rightarrow K^* \mu \mu)$
are calculated using the {\tt SuperISO}
package~\cite{Mahmoudi:2007vz,Mahmoudi:2008tp}, while
the computation of the SM-like Higgs boson mass
$m_h$, the supersymmetric contributions to the anomalous magnetic moment of
the muon $\Delta a_\mu$, the mass difference between the two neutral $B^0_s$
mesons and the branching ratio of $B_s \rightarrow \mu \mu$ rely on the {\tt SPheno} program~\cite{Porod:2003um, Porod:2011nf}\footnote{The Higgs boson mass 
shows a rather important variation with the $\lambda$-parameters, in particular in
the left-right sector for large values of $\tan\beta$. However, the associated
two-loop calculations implemented in {\tt SPheno} partly rely on
flavour-conserving expressions which would have  to be extended in case of flavour violation.
Therefore the Higgs masses obtained have to be taken with a grain of salt and are not shown.}.

In the $B$ physics sector, the observables that mostly constrain the choice
of experimentally allowed scenarios
are the branching ratio ${\rm BR}(B \rightarrow X_s\gamma)$, which is particularly sensitive to left-left mixing,
the $B$-meson oscillation parameter
$\Delta M_{B_s}$ involving
the products of left-left and right-right mixing parameters,
the branching fraction  ${\rm BR}(B_s \rightarrow \mu \mu)$, largely depending
on the parameters of the Higgs sector
$\tan\beta$ and $m_{A^0}$, as well as the branching ratio
${\rm BR}(B_u \rightarrow \tau \nu)$ which
is sensitive to $\tan\beta$ and the mass of the charged Higgs boson.
Additionally, the measurement of the anomalous magnetic moment of the muon
leads to a preference for light sleptons, whereas achieving
a correct mass for the Higgs boson requires
a large left-right mixing in the stop sector and/or heavy stop masses and is
in principle also sensitive
to flavour mixing in the squark sector \cite{Heinemeyer:2004by,Bruhnke:2010rh,Bartl:2012tx}.
Moreover, the complicated structure of the scalar potential possibly allows 
for charge- and
color-breaking minima~\cite{Camargo-Molina:2013sta}. We have verified with
the program $\tt Vevacious$ \cite{Camargo-Molina:2013qva}
that the selected benchmark scenarios were stable or sufficiently long-lived
against tunneling to vacua with combinations of non-zero sstrange, sbottom,
scharm and stop vacuum expectation values.

\begin{figure}
	\begin{center}
		\includegraphics[scale=0.35]{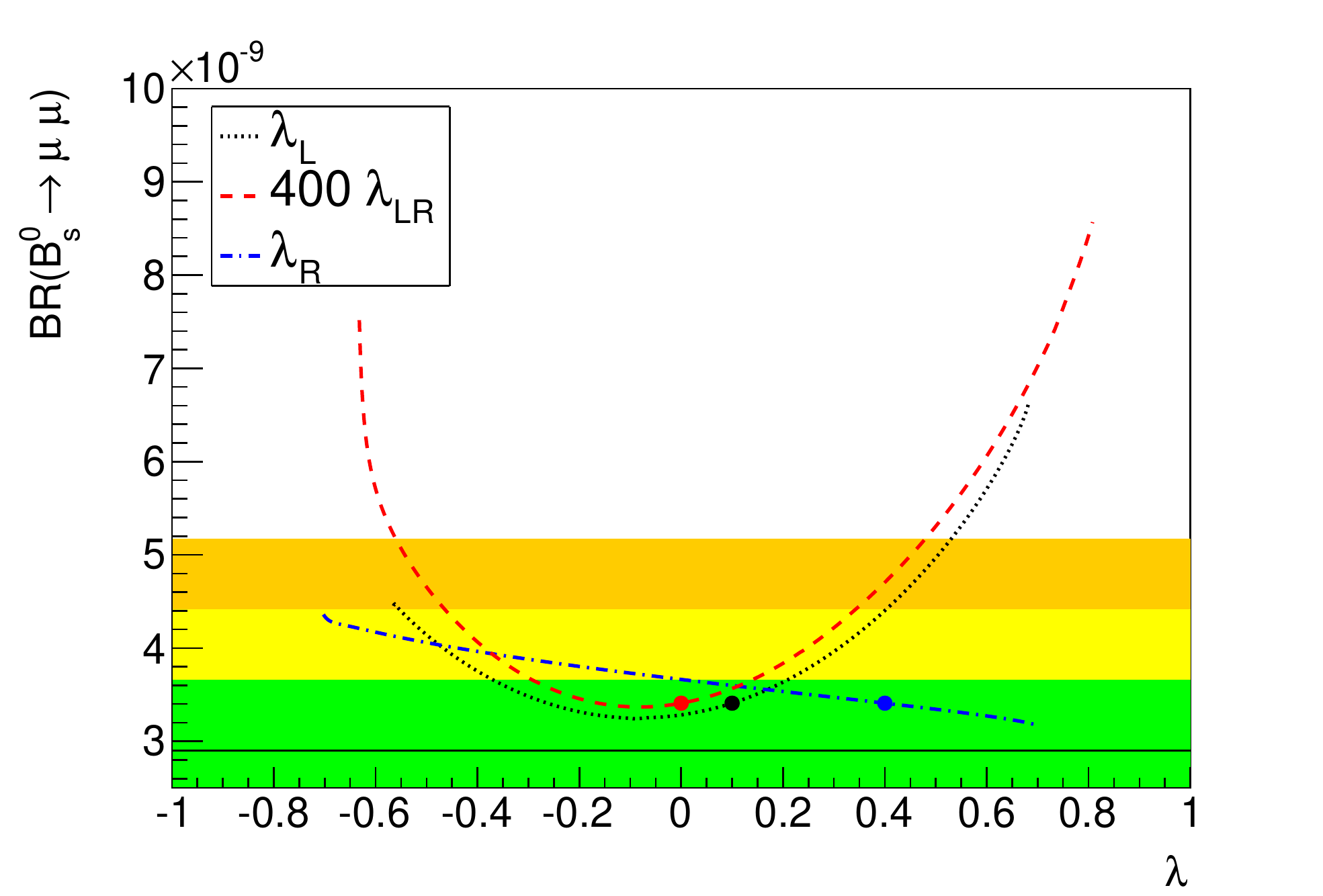}
		\includegraphics[scale=0.35]{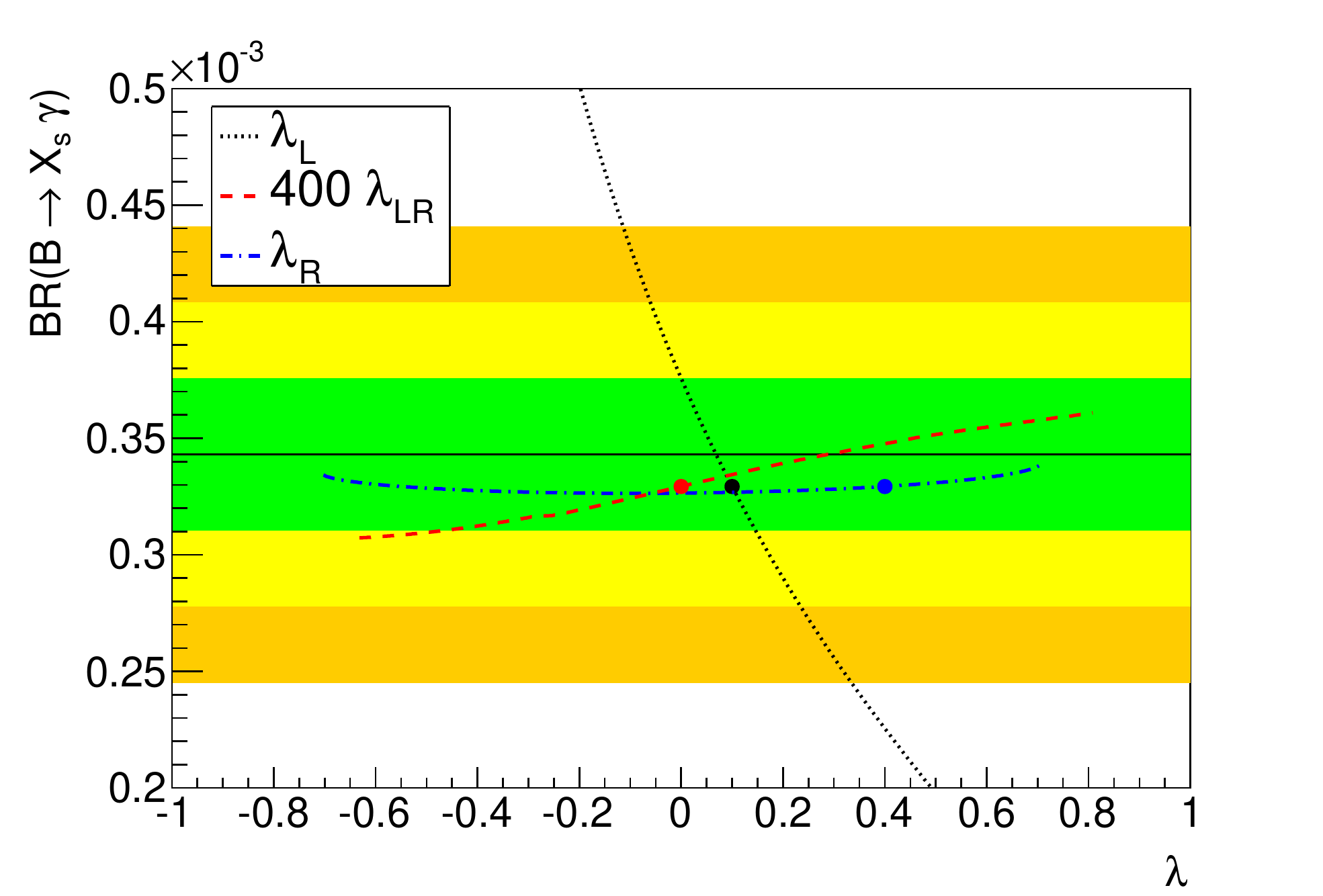}
		\includegraphics[scale=0.35]{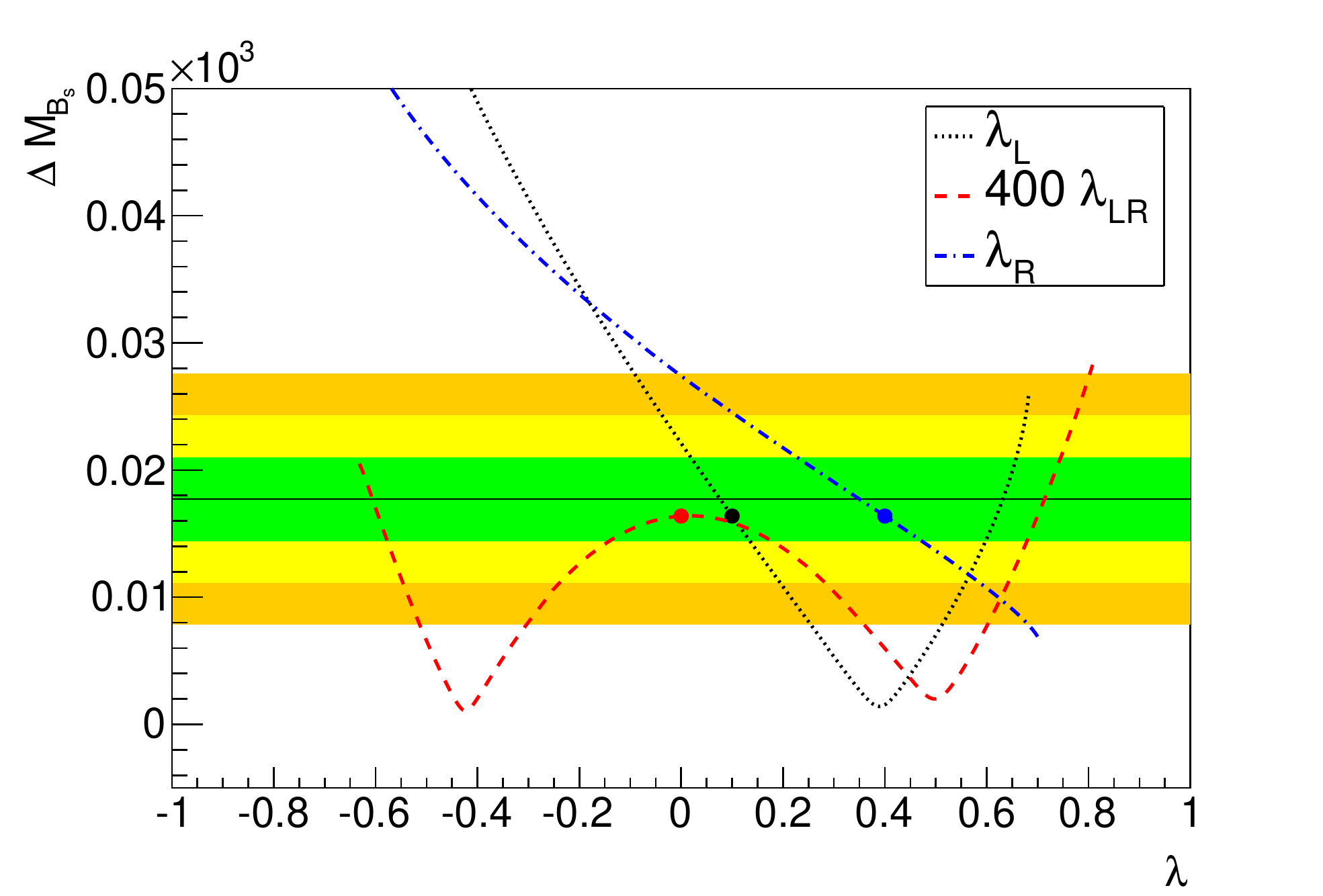}
		\includegraphics[scale=0.35]{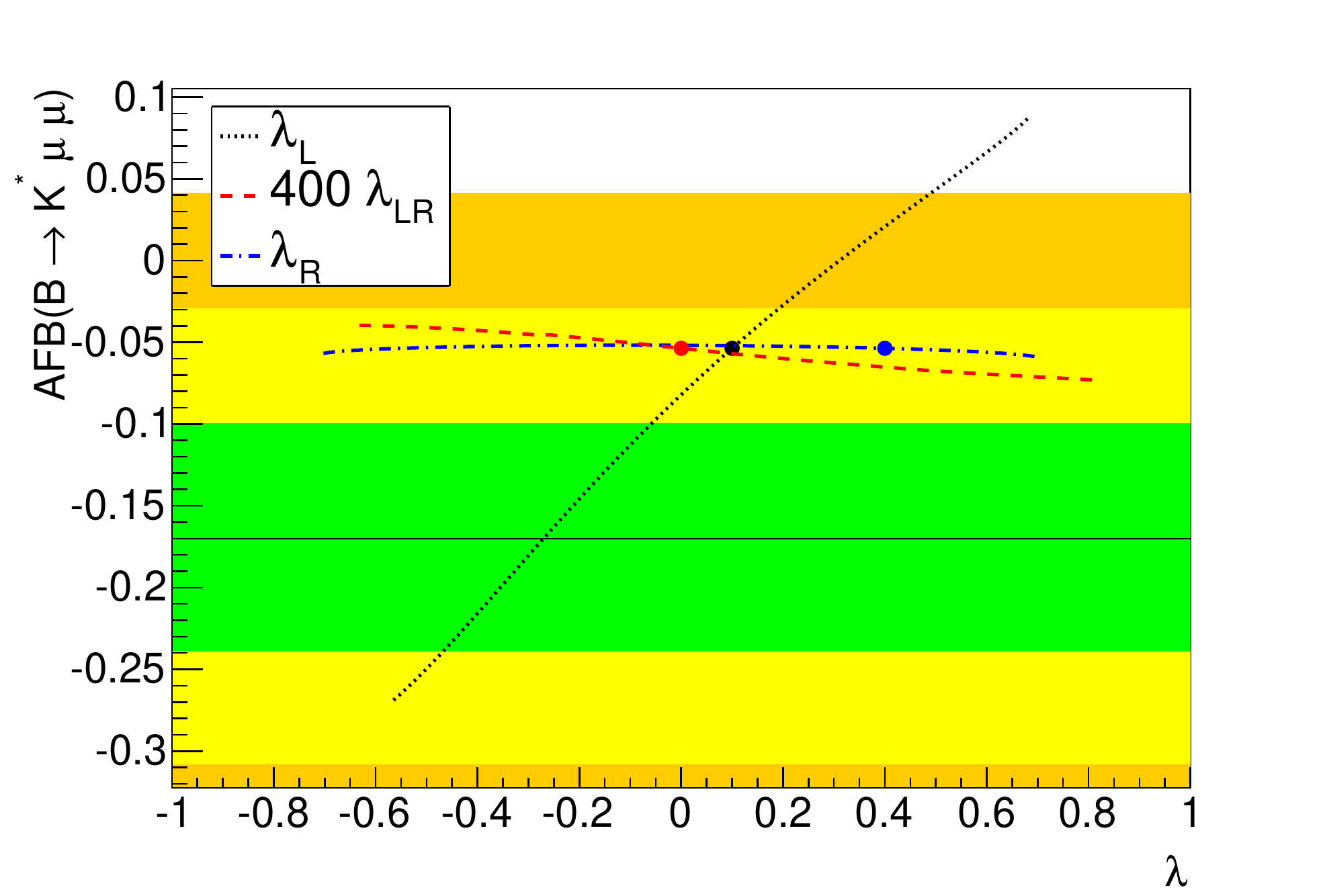}
		\vspace*{-3mm}
		\caption{Dependence of selected observables on the NMFV-parameters $\lambda_{L}$, $\lambda_{LR}$, and $\lambda_{R}$ around reference scenario I. The markers correspond to the scenario as defined in Table~\ref{tab:points}. The black horizontal lines indicate the experimental central value of Table~\ref{tab:PLMs}. The green, yellow, and orange bands correspond to the limits at $1\sigma$, $2\sigma$, and $3\sigma$ confidence level, respectively. 
		}
		\label{fig:point1}
		\vspace*{7mm}
		\includegraphics[scale=0.35]{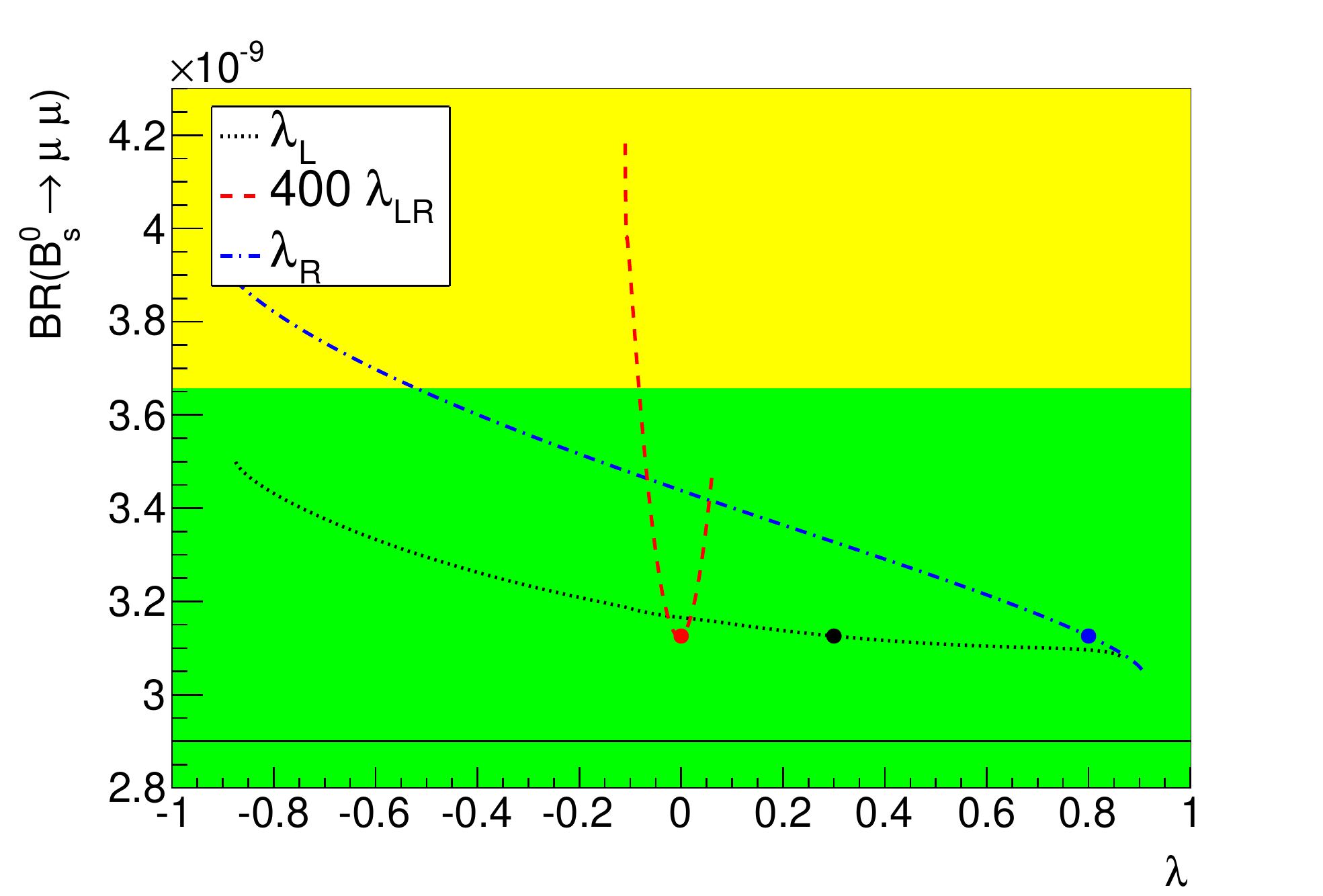}
		\includegraphics[scale=0.35]{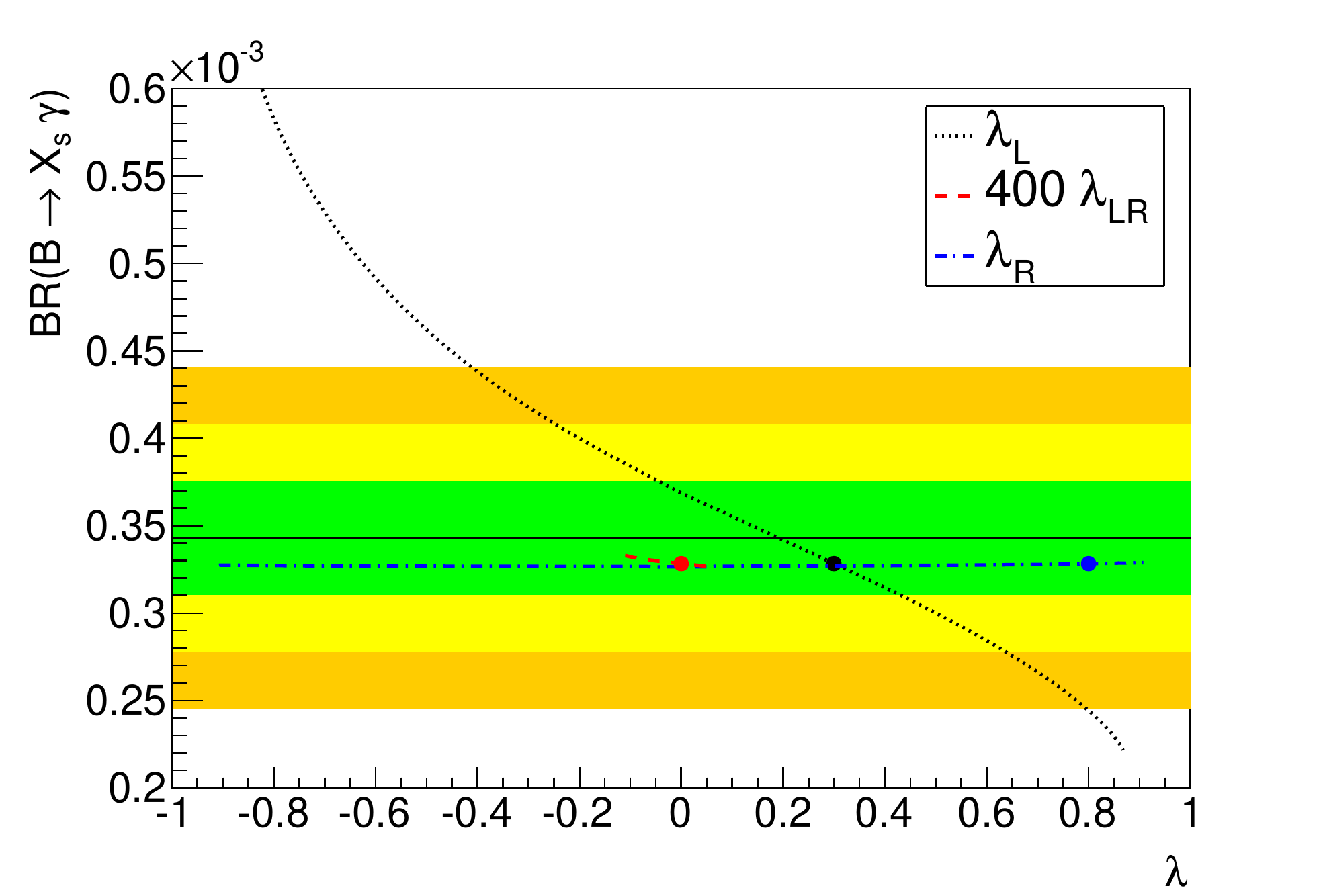}
		\includegraphics[scale=0.35]{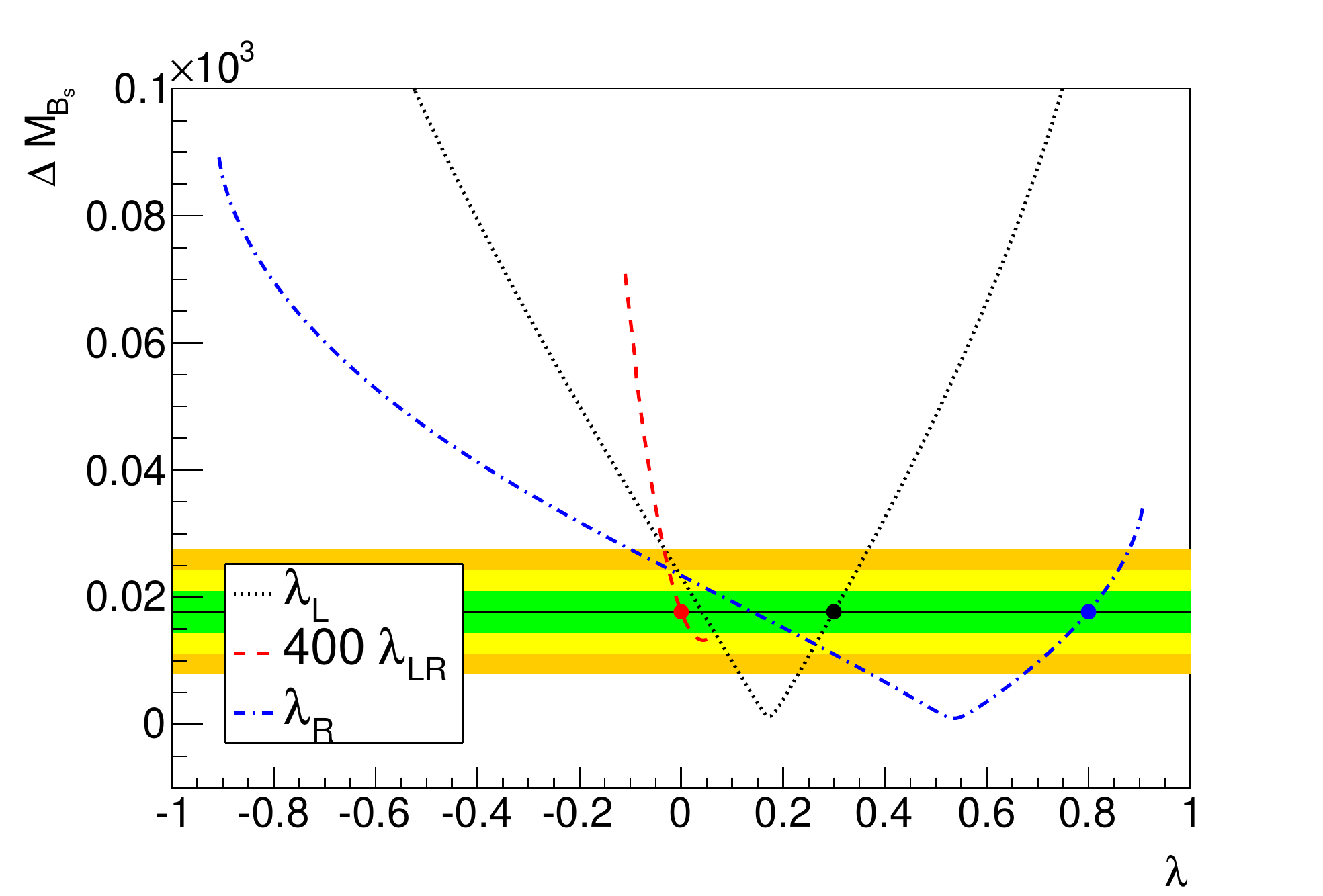}
		\includegraphics[scale=0.35]{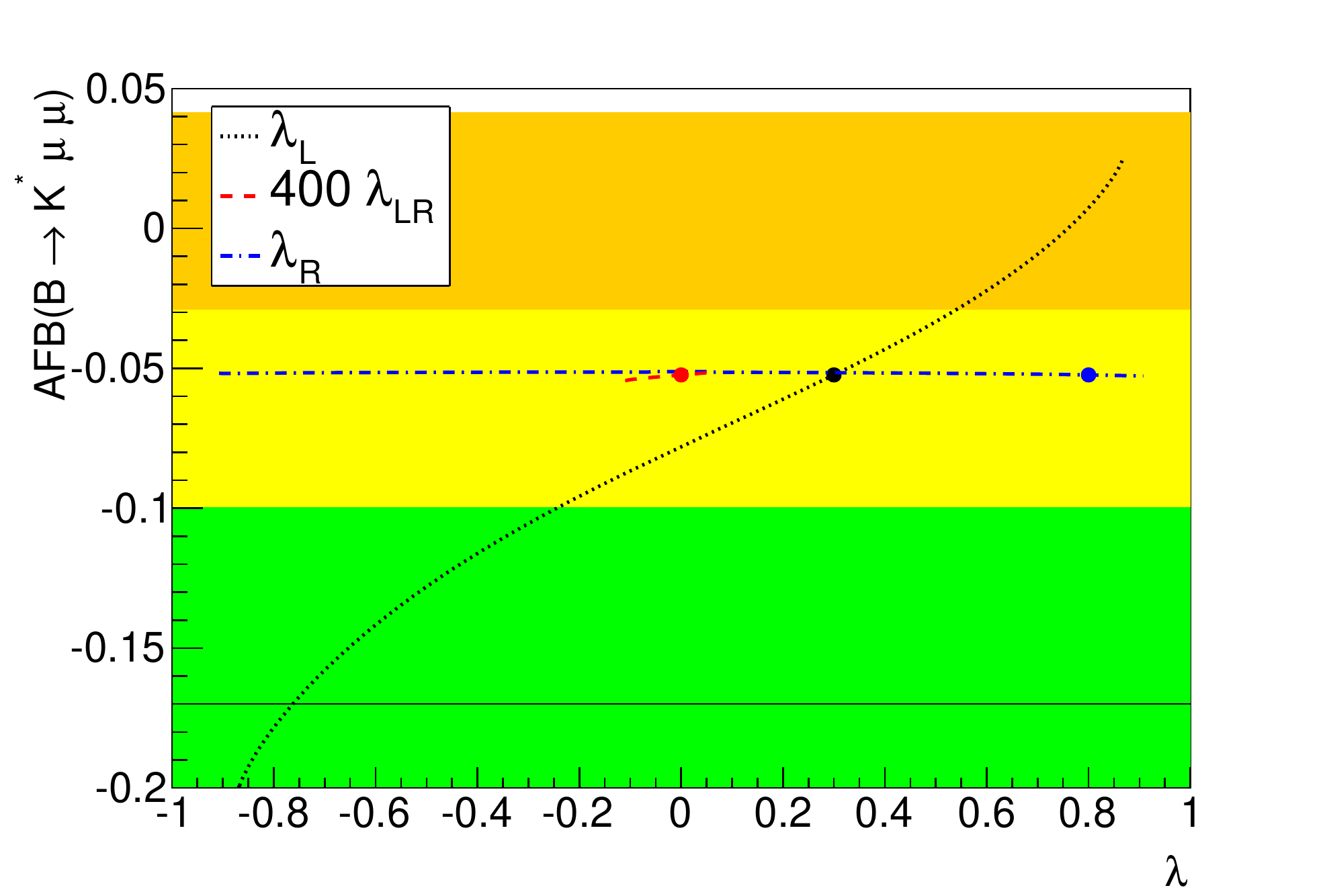}
		\vspace*{-3mm}
		\caption{Same as Fig.\ \ref{fig:point1} for variations of $\lambda_{L}$, $\lambda_{LR}$, and $\lambda_{R}$ around reference scenario II.}
		\label{fig:point2}
	\end{center}
\end{figure}

Based on the above-mentioned scan, we have identified two reference 
scenarios (given in Table~\ref{tab:points}) which are the most favoured by the 
current experimental constraints and which capture quite generic and interesting features 
of squark flavour mixing.  The main differences between the two points reside in the value of the scalar mass scale $M_{\rm SUSY}$ and the Higgs-sector parameters $\mu$, $m_{A^0}$ and $\tan\beta$. Moreover, scenario II exhibits stronger generation mixings in both the left-left and right-right sectors. We have studied the dependence of the observables summarized in
Table~\ref{tab:PLMs} on the three NMFV parameters and shown a selection of
the most relevant results in Figs.\ \ref{fig:point1} and \ref{fig:point2}.
\begin{table}
	\begin{center}
	\begin{tabular}{|c|ccccccccc|}
		\hline
		 & $M_1$ & $M_{\rm SUSY}$ & $A_{t,b,\tau}$ & $\mu$ & $\tan\beta$ & $m_{A^0}$ & $\lambda_{L}$ & $\lambda_{R}$ & $\lambda_{LR}$ \\
		\hline
		\hline
		 I & 350 & 1100 & 2000 & 600 & 40 & 1600 & 0.1 & 0.4 & 0 \\
		II & 350 & 1600 & 2000 & 100 & 10 &  600 & 0.3 & 0.8 & 0 \\
		\hline
	\end{tabular}
	\caption{Two reference scenarios which are favoured by the experimental constraints of Table~\ref{tab:PLMs}.}
	\label{tab:points}
	\end{center}
\end{table}
As expected, the $B \rightarrow X_s\gamma$ decay is in both cases very sensitive to mixing in the left-left sector, but almost 
independent of any mixing in the right-right and left-right sectors. For scenario I, the $B^0_s \to \mu^+ \mu^-$ decay shows a
strong dependence on both $\lambda_L$ (in contrast to scenario II for which
the influence of $\lambda_{L}$ is less pronounced) and
$\lambda_{LR}$, whilst the dependence
on $\lambda_R$ is found to be milder. Finally, any mixing induces large modification
of the $B$-meson oscillation parameter $\Delta M_{B_s}$, with interference effects involving $\lambda_{LR}$ being in particular observed for scenario~I.

In the first scenario, the $\Delta M_{B_s}$ observable constrains the parameters
$\lambda_L$ and $\lambda_R$ the most, reducing the allowed interval to 
$\lambda_L \in [ -0.09, 0.25]$ (and the $B \to X_s \gamma$ constraints is slightly
less stringent) and \mbox{$\lambda_R \in [- 0.01, 0.68]$}.
The combination of constraints from the $B \to X_s \gamma$ and
$B^0_s \to \mu^+ \mu^-$ branching ratios restricts the $\lambda_{LR}$ parameter
to lie in the ranges
$[-14.0, -12.8] \cdot 10^{-4}$ and $[-7.6, 9] \cdot 10^{-4}$. In the second
scenario, the $\Delta M_{B_s}$ observable is again the most constraining one,
the three parameters being bound to satisfy $\lambda_L \in [ -0.03, 0.11]$
and $[0.23, 0.37]$ (and the constraints derived from the $B \to X_s \gamma$ measurement
are less important here due to the lower value of $\tan\beta$ and the heavier
squark masses), \mbox{$\lambda_{R} \in [ -0.10, 0.37 ]$} and $[ 0.67, 0.88]$,
and $\lambda_{LR} \in [-0.9, 1.50 ] \cdot 10^{-4}$.
Finally, the forward-backward asymmetry observed in $B \to K^* \mu^+ \mu^-$
decays mostly constrains, for both scenarios, flavour mixing in the
left-left sector, but its influence is less stringent than that of
the $B \to X_s \gamma$ branching ratio or $\Delta M_{B_s}$.

In summary, we have shown that, despite all existing precise experimental
measurements at low-energy, there is still room in the MSSM for sizable flavour-violating entries in the squark mass matrices concerning the mixing of second and third generations of squarks. We have defined and studied two
scenarios favoured by current data that we aim to confront, in a future
publication, with the LHC results.
These scenarios are indeed characterized by new decay modes of squarks and
gluinos~\cite{Bartl:2009au, Hurth:2009ke, Bruhnke:2010rh, Bartl:2010du, Bartl:2011wq, Bartl:2012tx}, which modify the standard decay patterns
employed in the derivation of squark and gluino mass limits at the LHC.
As a consequence, the reinterpretation of the experimental limits after accounting for flavour mixing in the squark sector might significantly reduce the bounds on the superpartners.

\section*{ACKNOWLEDGEMENTS}

{\small B.O'L.\ and W.P.\ are supported by DFG research training group GRK 1147,
 DFG project PO 1337-2/1  and by DAAD, project PROCOPE 54366394.
 B.H.\ is supported by France Campus, project PHC PROCOPE no.\ 26794YC.
 The work of B.F.\ has been partially supported by the Theory-LHC
 France-initiative of the CNRS/IN2P3 and by the French
 ANR 12 JS05 002 01 BATS@LHC.
 K.D.C.\ and N.S.\ are supported by an ``FWO-Vlaanderen'' aspirant fellowship
 and K.D.C.\ also by the Belgian Federal Science Policy Office through the
 Interuniversity Attraction Pole P7/37.
 K.D.C.\ acknowledges the support of the Strategic Research Program
 ``High Energy Physics" and the Research Council of the Vrije Universiteit
 Brussel, the hospitality of the CERN TH group, and the support from the ERC
 grant 291377, ``LHCtheory: Theoretical predictions and analyses of LHC physics:
 advancing the precision frontier''. }



%% file: cpvmssm/cpvmssm.tex

\chapter{Effects of CP Violation in MSSM Scenarios}

{\it A.~Arbey, J.~Ellis, R.~M.~Godbole, F.~Mahmoudi}



\begin{abstract}
We consider the MSSM with CP violation in view of the LHC Higgs searches and the results of
searches for dark matter, in both constrained and unconstrained MSSM scenarios, taking into account the effects of
electric dipole moment constraints for the scenarios under consideration.
\end{abstract}

\section{FRAMEWORK}

Constrained MSSM scenarios such as the CMSSM or NUHM, with their limited number of parameters, offer simple frameworks to study the implications of the latest results from collider and dark matter searches (see, for example, \cite{Buchmueller:2013rsa}
for the most recent global analysis of these constraints in the CMSSM and NUHM). These scenarios assume 
R-parity conservation and CP invariance, but they can be easily extended to incorporate CP-violating phases. On the other hand, the phenomenological MSSM (pMSSM) does not rely on any universality assumptions, and with its 19 parameters 
(assuming CP invariance and R conservation) provide a more general set-up. Thorough studies have been performed to reinterpret the LHC results in this context (see, for example, \cite{Conley:2011nn,Sekmen:2011cz,Arbey:2011un,Arbey:2011aa,CahillRowley:2012kx}). A preliminary exploration of these issues including CP violating phases in the framework of a MCMC analysis, before the LHC data became available also exists~\cite{Brooijmans:2010tn}.

We consider here the implications of the latest collider data and dark matter searches, including the possibility of CP violation in both constrained MSSM scenarios and the pMSSM, 
which has six additional parameters corresponding to the CP-violating phases of the gaugino masses $M_{1,2,3}$ and of the trilinear couplings $A_{t,b,\tau}$. To study these scenarios, we generate spectra and compute the Higgs decay widths and Electric Dipole Moments (EDMs) using {\tt CPsuperH}~\cite{Lee:2003nta,Lee:2007gn,Lee:2012wa}. Flavour physics observables are also computed with {\tt CPsuperH} and cross-checked with a development version of {\tt SuperIso}~\cite{Mahmoudi:2007vz,Mahmoudi:2008tp}. The dark matter relic density is computed with {\tt SuperIso~Relic}~\cite{Arbey:2009gu} for the pMSSM, and 
{\tt MicrOMEGAs}~\cite{Belanger:2006is,Belanger:2008sj} is used for the relic density in constrained scenarios and for dark matter direct detection cross-sections.

For the constrained scenarios, we consider two CP-conserving benchmark scenarios that correspond to the CMSSM and NUHM1 best fit points obtained in the analysis of \cite{Buchmueller:2013rsa}:
\begin{center}
\begin{tabular}{|l||c|c|c|c|c|c|}
 \hline
 Description & $m_0$ (GeV) & $m_{1/2}$ (GeV) & $A_0$ (GeV) & $\tan\beta$ & sgn($\mu$) & $M_H^2$ (GeV$^2$)\\
 \hline\hline
 CMSSM & 5650 & 2100 & 780 & 51 & + & --\\
 \hline
 NUHM-1 & 1380 & 3420 & 3140 & 39 & + & $1.33\times10^7$\\
 \hline
\end{tabular}
\end{center}
These points satisfy simultaneously constraints from the LHC Higgs and SUSY searches, flavour physics and dark matter relic density.
Starting from these CP-conserving points, we perform a flat scan over the six CP-violating phases at the low-energy scale between -$180^\circ$ and $180^\circ$, in order to study the influence of CP violation on the different observables. 

For the pMSSM study, we have set up the scan machinery in two ways: a random flat scan over the $19+6$ parameters, as well as an optimised scan using the geometric approach presented in \cite{Ellis:2010xm}. Due to the large number of parameters, the flat scan is not very efficient since most of the points with large CP phases are excluded by EDM constraints, but it has the advantage of a flat distribution in the CP angles. The geometric scan, on the other hand, is much more efficient, and the neutron, thallium and mercury EDM are used in addition to the lightest Higgs mass to determine the optimised directions in the model parameter space.

We consider the set of constraints as described in \cite{Arbey:2013aba}, in addition to the EDM constraints. In particular, we require that the lightest neutralino is the lightest supersymmetric particle (LSP) and constitutes dark matter. We require that one of the three Higgs states lies in the mass range 121-129~GeV. If the $h_2$ or $h_3$ is at about 125~GeV, we impose LEP and LHC Higgs search constraints. We also consider the constraints on the signal strengths for $h_1$. The Higgs signal strengths $\mu_{XX}$ are defined for each channel as the ratio of the Higgs production cross-section times branching ratio relative to the Standard Model values. Furthermore, we impose flavour constraints and use the upper limit on the relic density constraint. We compare the results with the case without phases in order to investigate the impact of the CP phases. Table~\ref{tab:constraints} summarizes the Higgs and EDM constraints that are applied to the MSSM points.

\begin{table}[!t]
\begin{center}
\begin{tabular}{|c|c|}
 \hline
 Higgs observable & Constraint\\
 \hline\hline
  $M_H$  & $125\pm4$ GeV\\
 \hline
  $\mu_{\gamma\gamma}$ & $1.20\pm0.30$\\
 \hline
  $\mu_{ZZ}$ & $1.10\pm0.22$\\
 \hline
  $\mu_{WW}$ & $0.77\pm0.21$\\
 \hline
  $\mu_{\tau\tau}$ & $0.78\pm0.27$\\ 
 \hline
  $\mu_{bb}$ & $1.12\pm0.45$\\ 
 \hline
 \end{tabular}\qquad\qquad\qquad
 \begin{tabular}{|c|c|}
 \hline
 EDM & Upper limit (e.cm)\\
 \hline\hline
 Thallium & $9\times10^{-25}$ \\
 \hline
 Mercury & $3.1\times10^{-29}$ \\
 \hline
 Neutron & $3\times10^{-26}$  \\
 \hline
 Muon & $1.9\times10^{-19}$ \\
 \hline
\end{tabular}
\caption{Higgs and EDM constraints used in this study.\label{tab:constraints}}
\end{center}
\end{table}

\section{HIGGS SECTOR}

\begin{figure}[!ht]
\begin{center}
\includegraphics[width=5.cm]{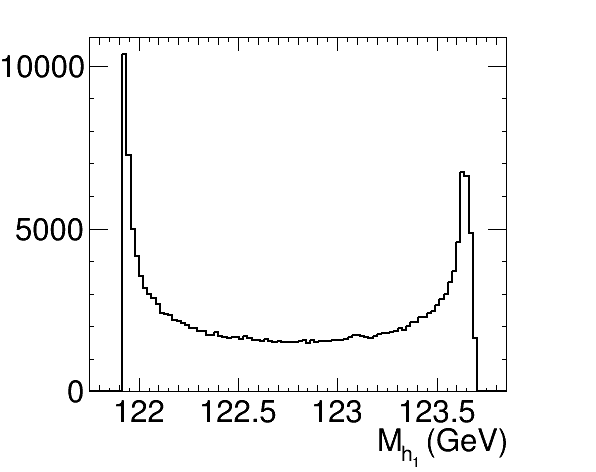}\includegraphics[width=5.cm]{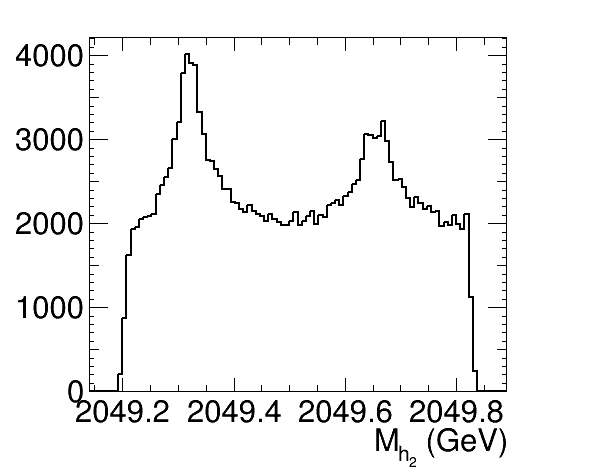}
\includegraphics[width=5.cm]{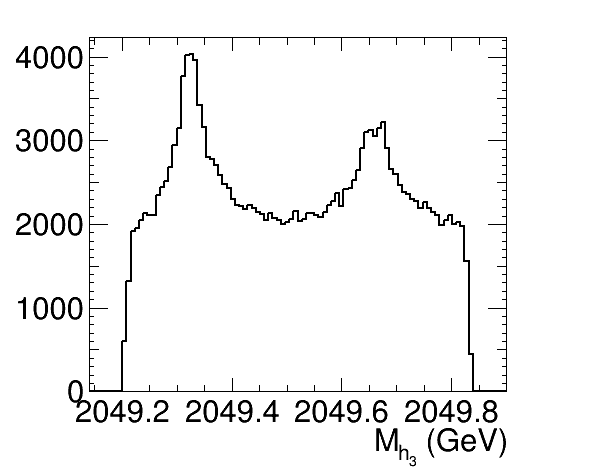}\\ 
\includegraphics[width=5.cm]{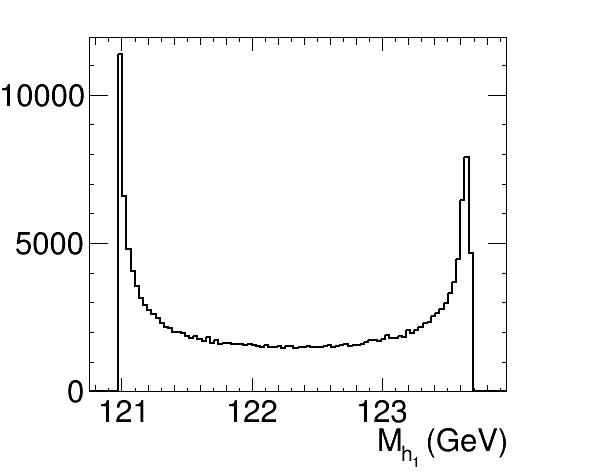}\includegraphics[width=5.cm]{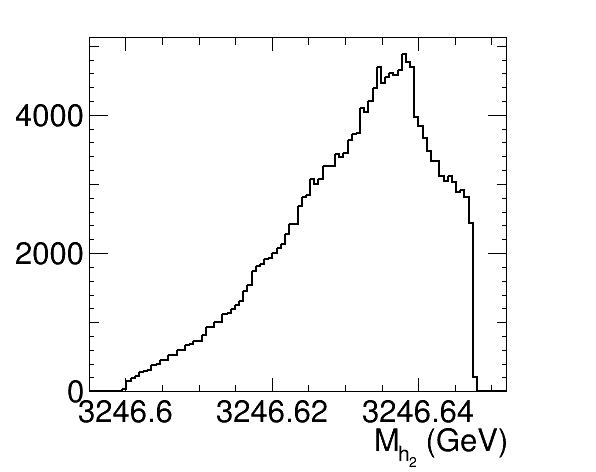}
\includegraphics[width=5.cm]{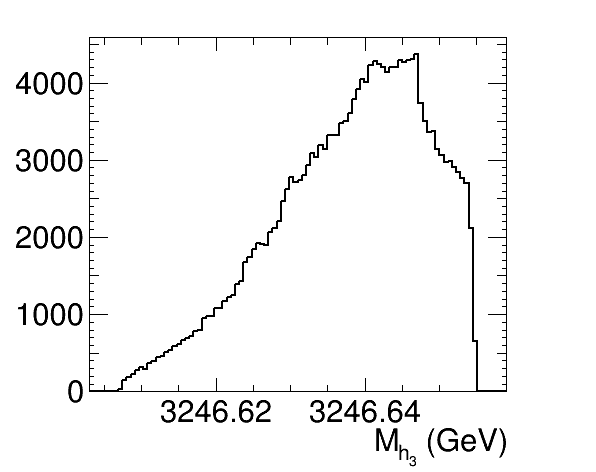}
 \caption{Distributions of the $h_1$ (left), $h_2$ (center) and $h_3$ (right) Higgs masses for the CMSSM (top), and NUHM1 (bottom) in the best-fit benchmark scenarios \emph{before} applying the EDM constraints.\label{cmssm-higgs}}
\end{center}
\end{figure}

\begin{figure}[!ht]
\begin{center}
\includegraphics[width=5.cm]{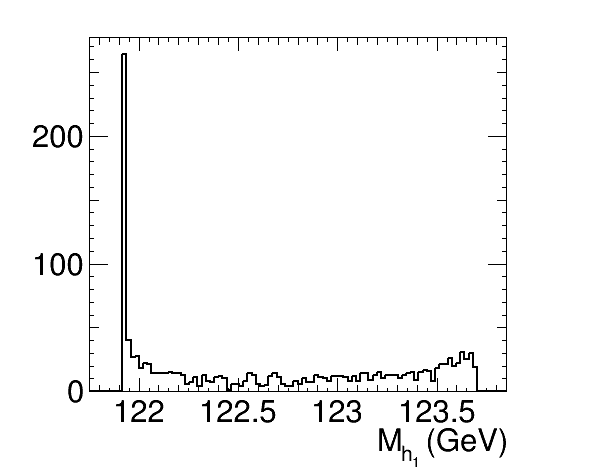}\includegraphics[width=5.cm]{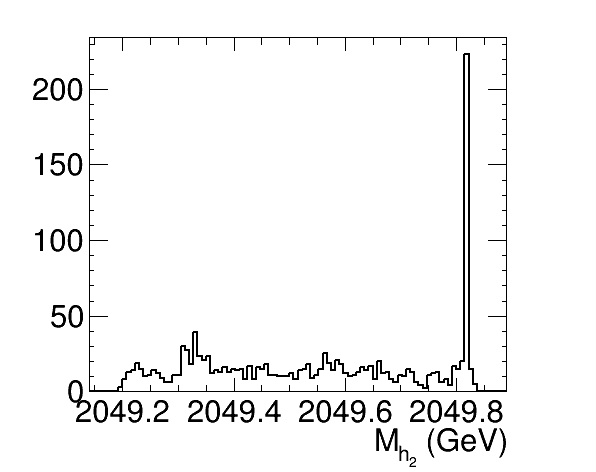}
\includegraphics[width=5.cm]{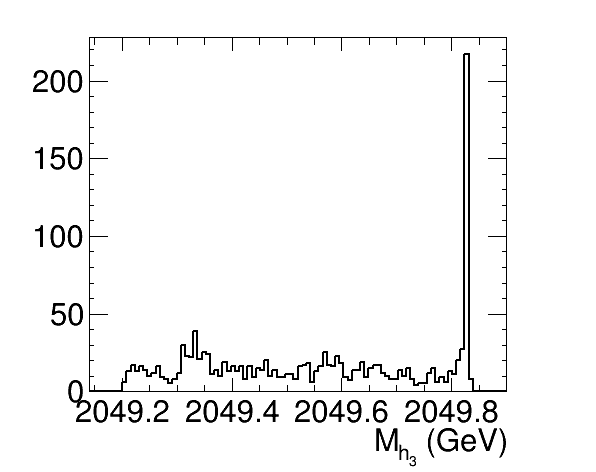}\\ 
\includegraphics[width=5.cm]{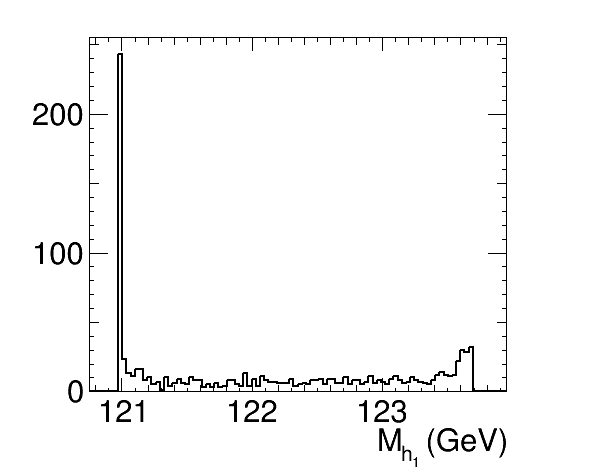}\includegraphics[width=5.cm]{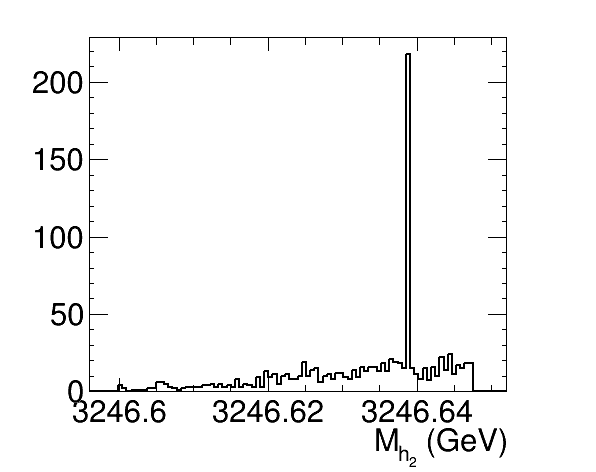}
\includegraphics[width=5.cm]{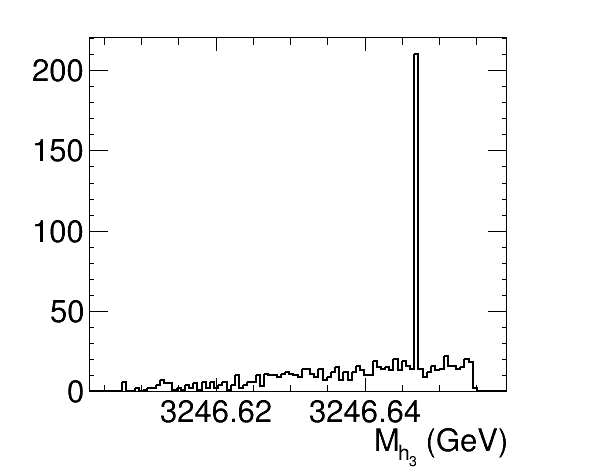}
 \caption{Distributions of the $h_1$ (left), $h_2$ (center) and $h_3$ (right) Higgs masses for the CMSSM (top) and NUHM1 (bottom) best-fit benchmark scenarios \emph{after} applying the EDM constraints.\label{cmssm-higgs-edms}} 
\end{center}
\end{figure}

In the CMSSM and NUHM1 scenarios, the Higgs discovered at the LHC is identified with the lightest MSSM Higgs boson,
and also in the presence of CP-violating phases. 
In Fig.~\ref{cmssm-higgs}, we present the distributions of the masses of the three Higgs bosons for the benchmark scenarios after varying the six phases, before applying the EDM constraints. For comparison, in Fig.~\ref{cmssm-higgs-edms} the distributions are shown after applying the EDM constraints.
The figures reveal that the heavy Higgs masses are nearly unaffected by CP violation, whereas the lightest Higgs mass can be modified by a few GeV. Once the EDM constraints are applied, only the scenarios with very small EDMs survive, resulting in distributions with peaks at the positions of the CP-conserving points. After applying the EDM constraints, the statistics reduces substantially, whilst still allowing for a modification of the Higgs mass by up to 2-3 GeV.
We have also studied the Higgs couplings for the benchmark points, and found that all the signal strengths of Table~\ref{tab:constraints} remain close to unity, and varying the CP phases leads to less than 2\% modifications.

The Higgs sector of the pMSSM with CP violation can differ more strongly from the CP-conserving case, with substantial mixing between the CP-even and CP-odd states. However, the recently-discovered Higgs boson has a large decay rate to two vector bosons, which implies that it is mainly CP-even. If this Higgs is identified with the lightest CPV-pMSSM Higgs boson $h_1$, in most of the cases $h_2$ is CP-even and $h_3$ CP-odd, which corresponds to a set-up similar to that of the CP-conserving pMSSM.

\begin{figure}[!ht]
\begin{center}
\includegraphics[width=0.35\textwidth]{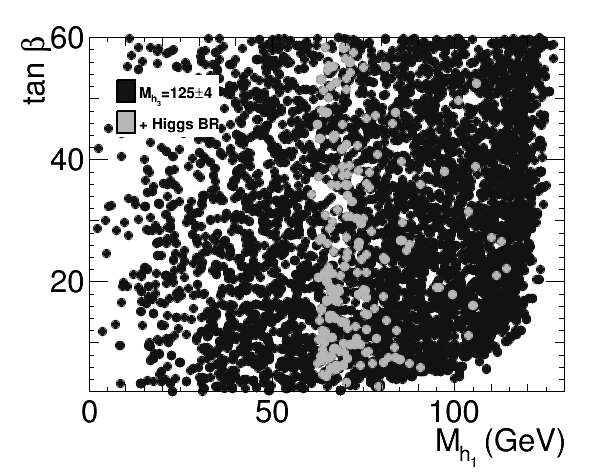}\qquad\qquad\includegraphics[width=0.35\textwidth]{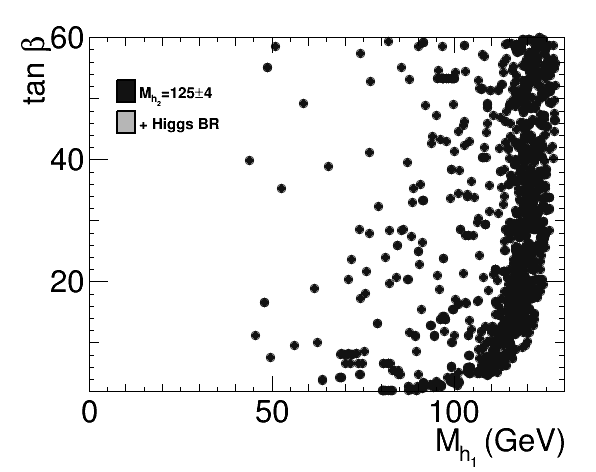}
\caption{CPV-pMSSM model points satisfying the EDM constraints in the plane ($M_{h_1},\tan\beta$), in the case where either the $h_2$ (left) or the $h_3$ (right) is identified with the Higgs state discovered at the LHC.\label{mh1_tanb}}
\end{center}
\end{figure}
In Fig.~\ref{mh1_tanb} we show all the points that satisfy the EDM constraints and for which either $h_2$ or $h_3$ has a mass close to 125~GeV. Once the constraints on the Higgs signal strengths are applied, only a small region remains in the case of $h_2$ at 125~GeV, while no possibility is found for the case where $h_3$ is at 125~GeV. However, if we consider in addition flavour physics constraints, no solution is found any more also for former case. This result is similar to that obtained in the 
CP-conserving pMSSM~\cite{Arbey:2012bp,Arbey:2013jla}. Hence, the Higgs boson observed at the LHC has to be identified with $h_1$ in the CPV-pMSSM, which we will assume in the following. 

\begin{figure}[!ht]
\begin{center}
\includegraphics[width=0.35\textwidth]{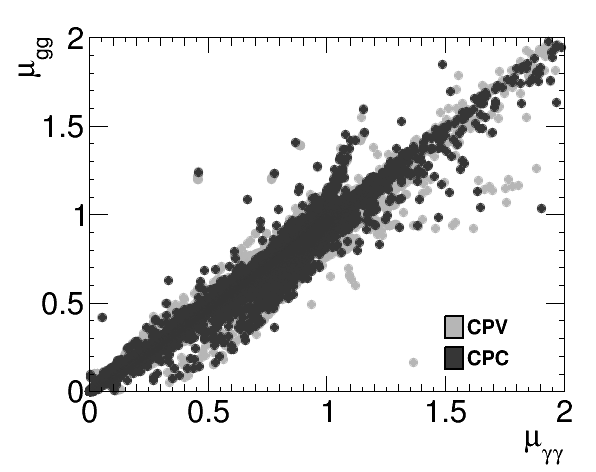}\qquad\qquad\includegraphics[width=0.35\textwidth]{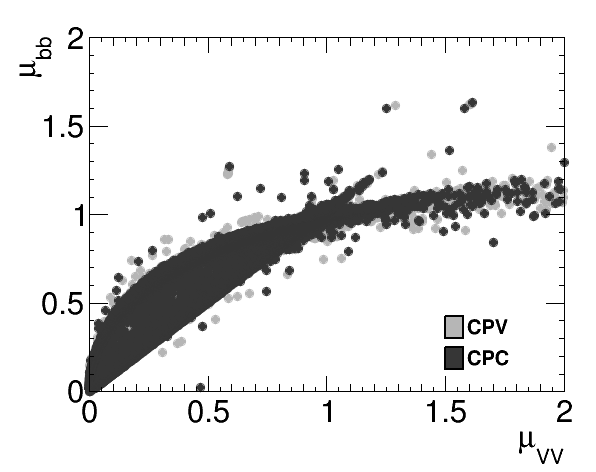}
\caption{The strength of the coupling of the lightest Higgs boson to two gluons as a function of the coupling strength to photon pairs (left), and the strength of the coupling to $b {\bar b}$ pairs as a function of the coupling strength to two vector bosons (right).\label{mu}}
\end{center}
\end{figure}
In Fig.~\ref{mu} we present the correlations between the signal strengths in the $\gamma\gamma$, $gg$ and $b\bar{b}$, $VV$ channels, comparing the CP-conserving and CP-violating cases. The results are very similar in both cases, with the CP-violating case offering a possibility of small deviations in comparison to the CP-conserving case. This shows that it will be difficult to discriminate between the two cases via more precise measurements of the Higgs signal strengths. 

However, we have seen that CP violation can modify the Higgs mass by a few GeV, as well as the signal strengths by a few percent. In addition to the Higgs observables, the flavour sector and EDM observables provide direct probes of the CP properties. The interplay between the different sectors can therefore help determining the MSSM parameters and the CP properties of the model. Another sector of interest is the neutralino dark matter sector, that we consider in the next Section.

\section{DARK MATTER DIRECT DETECTION}

Dark matter searches provide important constraints, even if they suffer in general from astrophysical and cosmological uncertainties. In the following, we apply different constraints on the MSSM scenarios into consideration. First, we impose the relic density constraint, but only as an upper bound, in order to account for the possibility of cosmological modifications of the properties of the early Universe \cite{Arbey:2008kv,Arbey:2009gt}.
Older analyses of the effects of CP violating phases in Supersymmetric theories, on the relic density of neutralino dark matter exist in literature~\cite{Belanger:2006qa,Belanger:2006pc}. Here we analyze the issue in light of the latest  data on  various fronts. We do not apply constraints from indirect dark matter detection searches, as they are subject to large astrophysical uncertainties. However, we consider the results of direct dark matter detection searches, which are sensitive primarily to the local density of dark matter.

\begin{figure}[!t]
\begin{center}
\includegraphics[width=0.5\textwidth]{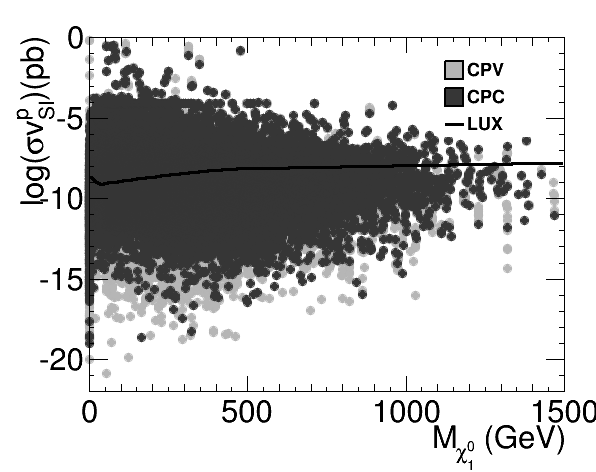}
\caption{Spin-independent dark matter scattering cross-section on a proton as a function of the lightest neutralino mass.\label{DDpSI}}
\end{center}
\end{figure}
The LUX collaboration has recently released strong upper limits on direct dark matter detection \cite{Akerib:2013tjd}, in particular on the spin-independent scattering cross-section of a WIMP with the proton, that we consider in the following. The impact of the direct dark matter detection results from the LUX experiment on the CP-conserving and CP-violating pMSSM is presented in 
Fig.~\ref{DDpSI}. The figure reveals that the inclusion of CP phases can lead to smaller scattering cross-sections, therefore allowing the neutralino to evade more easily the direct detection limits.

\begin{figure}[!t]
\begin{center}
\includegraphics[width=5.cm]{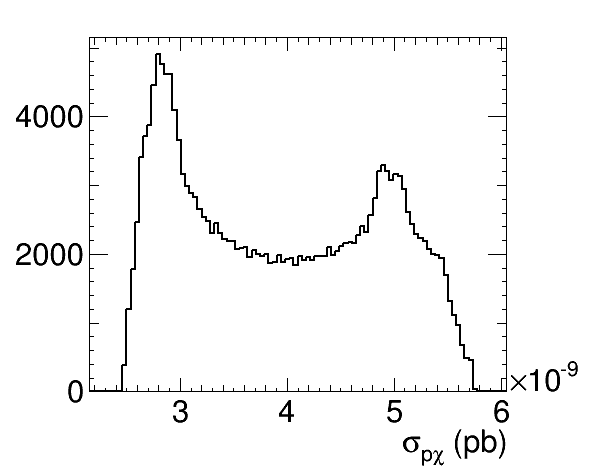}\qquad\qquad\includegraphics[width=5.cm]{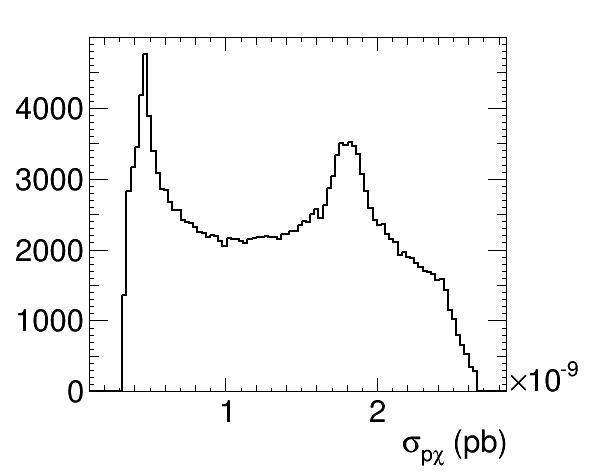}
 \caption{Distributions of the spin-independent dark matter scattering cross-section on the proton in the CMSSM (left) and NUHM1 (right) benchmark scenarios \emph{before} applying the EDM constraints.\label{cmssm-dd}}
\end{center}

\begin{center}
\includegraphics[width=5.cm]{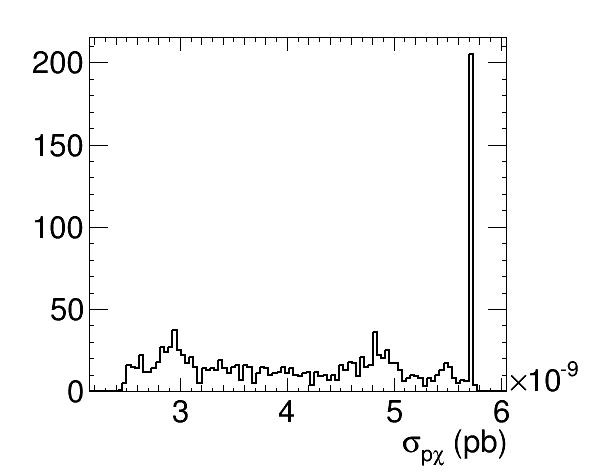}\qquad\qquad\includegraphics[width=5.cm]{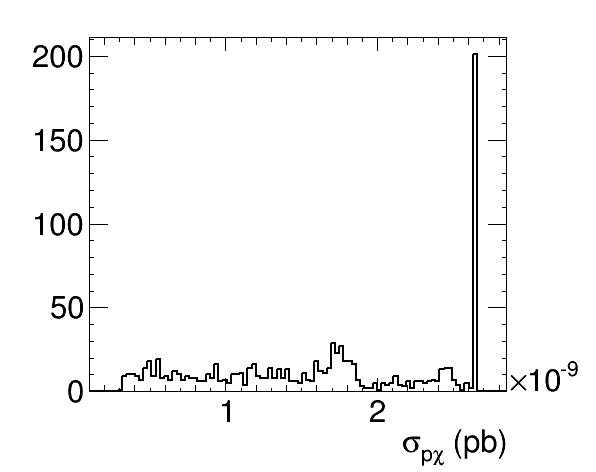} 
 \caption{Distributions of the spin-independent dark matter scattering cross-section on the proton in the CMSSM (left) and NUHM1 (right) benchmark scenarios \emph{after} applying the EDM constraints.\label{cmssm-dd-edms}} 
\end{center}
\end{figure}

The constrained MSSM benchmark scenarios, being simpler, give more possibilities to test CP violation with direct dark matter detection. In Figs.~\ref{cmssm-dd} and \ref{cmssm-dd-edms}, we show the distributions of the spin-independent neutralino-proton scattering cross-section in the constrained MSSM benchmark scenarios, before and after applying the EDM constraints. As can be seen, the scattering cross-sections are close to the LUX limit. After including the CP phases, the scattering cross-sections may decrease by a factor up to 5. Applying the EDM constraints, while strongly decreasing the statistics, does not change this feature.
Therefore, dark matter direct detection can probe CP violation. If neutralino dark matter is discovered in direct searches, a better determination of the local density of dark matter could help in understanding the CP properties of the MSSM.

\section*{CONCLUSIONS}

We have studied the effects of CP violation in the MSSM, in particular in view of the EDM constraints, Higgs observables and direct dark matter detection. The EDM constraints are well-known probes of CP violation, as well as flavour observables such as CP asymmetries in $B$ meson decays. We have shown that the observed Higgs boson has to be identified with the lightest Higgs boson state $h_1$, and that CP violation mostly changes the $h_1$ mass but does not affect much the Higgs signal strengths, rendering more difficult the possibility of discriminating between CP conserving and CP violating models in the Higgs sector. 

We have also considered direct dark matter detection, and shown that CP violation can modify the neutralino-proton scattering cross-section by a factor up to 5. However, direct detection currently suffers from cosmological uncertainties that can limit the interpretations.

Finally, once all the present constraints are applied, and since all present observables are compatible with the CP-conserving results, discovering CP violation in this type of new physics scenario will be a difficult task, as they are already strongly constrained. However, the direct dark matter detection situation may improve in the near future.

\section*{ACKNOWLEDGEMENTS}
The authors would like to thank the organizers of the Les Houches workshop where this work was initiated. The work
of A.A. was supported by the F\'ed\'eration de Recherche A.-M. Amp\`ere de Lyon. The work of J.E. is supported in part by the London Centre for Terauniverse Studies (LCTS), using
funding from the European Research Council
via the Advanced Investigator Grant 267352.
RMG wishes to thank the Department of Science and Technology, Government of India, for support under the J.C. Bose Fellowship under grant no. SR/S2/JCB-64/2007.



%% file: dmeft/DMeff.tex

\chapter{Dark Matter Effective Field Theory at Colliders}

{\it A. Arbey, M. Battaglia, G. B\'elanger, A. Goudelis, F. Mahmoudi, S. Pukhov}



\begin{abstract}
The absence of signals for physics beyond the Standard Model at the LHC and the strong evidence for dark matter motivate the use of a fairly 
model-independent effective field theory approach to the study of dark matter physics. We present some preliminary results on constraints 
of the effective scalar operators coupling dark matter to quarks and gluons derived by combining results from monojet searches at the LHC, 
the recent upper limits on spin-independent WIMP-nucleon scattering cross section and the determination of the relic dark matter abundance in 
the Universe. We also comment on the potential of the next LHC runs as well as that of a future $100$ TeV $pp$ collider to further constrain
the dark matter properties.
\end{abstract}

\section{INTRODUCTION}
The Large Hadron Collider has been extremely successful in pushing the mass scale of new physics towards the TeV
scale. The discovery of a particle compatible with the Standard Model (SM) Higgs boson has given no hint
of physics beyond the SM, while all direct searches for new physics have placed stringent bounds on
the masses and couplings of new particles. 

At the same time, there is strong evidence that physics beyond the SM (BSM) does exist, the most striking example 
being the existence of dark matter (DM). In the most commonly adopted freeze-out picture, the so-called ``WIMP miracle'' 
occurs: stable particle(s) with electroweak (EW) scale masses and couplings can actually account for the 
observed relic DM abundance in the universe. If the mass of the DM particle(s) lies indeed at the EW scale, it is 
highly likely that it can be produced at colliders, in particular at the LHC. This has motivated several experimental 
searches for final states with large missing transverse energy which would signal the production (and escape) of weakly 
interacting massive particles (WIMPs) from the detector.

A common problem of these searches stems from the fact that the interpretation of their results require rather strong 
assumptions on the nature of the underlying model, in particular when long decay chains are involved. In recent years, 
there has been an increasing interest in approaching DM physics using effective field theory (EFT) techniques, to overcome 
the limitations due to model-dependence. The basic underlying hypothesis of the EFT approach is that the sector mediating 
the interactions between dark matter particles and the SM is heavy and no light state is relevant for DM physics. 
Most analyses reported in the recent literature make the additional assumption that a single effective operator mediates 
the DM-SM interactions.

However, it should be kept in mind that the EFT approach is not completely free of assumptions. In fact, in many concrete models,  
the DM particles couple at tree-level to the Higgs and $Z$ bosons, whereas even if all DM-SM mediators are heavy enough for an 
EFT approach to be valid, upon integrating them out a \textit{set} of effective operators is obtained, and not a single operator. These may 
interfere destructively or constructively changing the resulting DM-SM interaction. Last but not least, the standard EFT approach 
fails to capture specific features common to several DM models, such as the existence of resonances or co-annihilation processes 
which may be crucial in determining the DM relic density. Keeping these limitations in mind, we should also note that the EFT 
approach to dark matter has the advantage of encompassing different theoretical frameworks and, to some extent, manages to provide 
us with relatively generic information on DM.

In this work, we perform an investigation of the constraints on the dark matter EFT coming from the LHC, direct detection and 
the DM relic abundance. We examine the potential of the increases in LHC energy and luminosity to constrain the DM EFT for the case of 
scalar operators. In view of the discussions currently taking place in the HEP community on future collider projects, we comment on 
the possible impact of a future hadron collider reaching a centre-of-mass energy of $100$~TeV. 


\section{THE DARK MATTER EFFECTIVE FIELD THEORY}

In this preliminary study, we assume that the dark matter particle is not charged under the SM gauge group and the DM stability is ensured 
by some discrete ${\cal{Z}}_2$-like symmetry under which the SM particles are even and the DM particles are odd, which amounts to the 
DM particles only appearing in pairs in the Lagrangian of the full (UV-complete) theory. This feature is assumed to be preserved 
upon integration of the heavy degrees of freedom, while now the DM interactions with the SM are suppressed by powers of some mass scale 
$M_*$, which should be understood as $M_* \sim M/\sqrt{g_1 g_2}$, with $M$ being the mass of the heavy mediator and $g_1, g_2$  
the mediator couplings to the SM and DM particle, respectively. 

We focus on effective operators linking two DM particles to two quarks or two QCD field strengths, which are those most relevant for hadron 
colliders. In general, the DM particle can be either a real or complex scalar, a Dirac or Majorana fermion or a vector, when restricting 
ourselves to spin up to $1$. Sets of effective operators coupling these classes of DM particles to the SM have been 
already defined under various assumptions~\cite{Cao:2009uw,Goodman:2010yf,Goodman:2010ku,Rajaraman:2011wf}. 
In this note, we focus our preliminary study on two of these operators assuming the DM particles are real scalar fields, namely
\begin{eqnarray}
R_1 & = & \frac{m_q}{2 M_*^2} \chi \chi \bar{q} q \\ \nonumber
R_3 & = & \frac{\alpha_s}{8 M_*^2} \chi \chi G_{\mu\nu} G^{\mu\nu} \\ \nonumber
\end{eqnarray}
where $m_q$ is the mass of the quark $q$, $\chi$ is the dark matter particle, $\alpha_s$ is the strong coupling and $G$ is the 
QCD field strength tensor \footnote{Our operator naming scheme follows the notations of Ref.~\cite{Goodman:2010ku}.}.

As mentioned above, the validity of the effective field theory approach must be verified. Whereas the description of DM 
interactions with the SM particles via effective operators can be justified in the case of direct detection  for DM particle masses 
as low as ${\cal{O}}$(few GeV) and mediator masses as low as ${\cal{O}}$($10^2$ MeV), this is certainly not the case for the 
searches at the LHC. The EFT approach is at best valid for $M > 2 M_{\mathrm{WIMP}}$ and embedding the EFT in a weakly coupled theory would
require $g_1 g_2 < (4\pi)^2$. This leads to the requirement $M_{\mathrm{WIMP}} < 2\pi M_*$ for the EFT description to be valid. 
In fact, this bound may be overly optimistic for the validity of the effective field theory, since the actual momentum transfer involved 
in each process must also be considered\cite{Busoni:2013lha,Busoni:2014sya}. For simplicity, in this note we will adopt to the 
limit $M_{\mathrm{WIMP}} < 2\pi M_*$ to define the region of validity of the EFT approach.
\section{ANALYSIS AND FIRST RESULTS}

We consider dark matter constraints in combination with monojet search results from ATLAS and CMS, for the two effective operators $R_1$ and 
$R_3$ defined above. For this study, we have set-up a software framework similar to that described in \cite{Arbey:2013aba}. 
In particular, we have implemented the effective operator Lagrangians in FeynRules \cite{Alloul:2013bka}, which provides us with an automatic export 
of the model to the {\tt CalcHEP} \cite{Pukhov:2004ca,Belyaev:2012qa} and Universal FeynRules Output (UFO) \cite{Degrande:2011ua} file formats. 
The model files have then been imported in {\tt micrOMEGAs} \cite{Belanger:2001fz,Belanger:2006is,Belanger:2008sj,Belanger:2014hqa} 
and {\tt Madgraph 5}~\cite{Alwall:2011uj}. The former program calculates the WIMP relic density and dark matter direct detection scattering cross-sections 
off matter, and the {\tt CalcHEP} model file has been specifically modified to compute the scattering cross-section of WIMPs with gluons. 
These scattering cross-sections allow us to re-cast the LHC monojet searches in terms of DM direct detection constraints, 
and the results are compared 
to the recent LUX~\cite{Akerib:2013tjd} and XENON~\cite{Angle:2011th,Aprile:2012nq} 90\% C.L. upper limits. We also consider the upper limit on the relic 
cold DM density $\Omega h^2 = 0.1199 \pm 0.0027$ \cite{Ade:2013zuv}, derived from latest PLANCK cosmic microwave background measurements.

\begin{figure}
\begin{center}
\begin{tabular}{cc}
\includegraphics[width=0.5\textwidth]{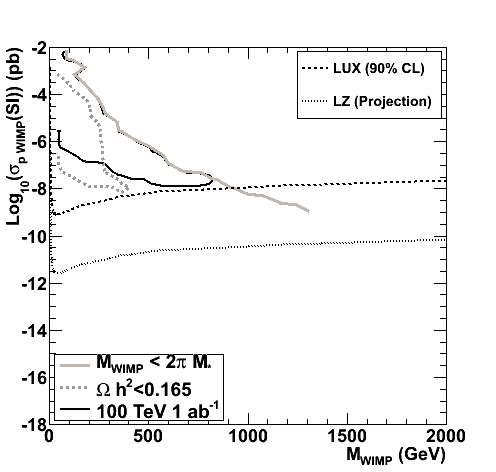} &
\includegraphics[width=0.5\textwidth]{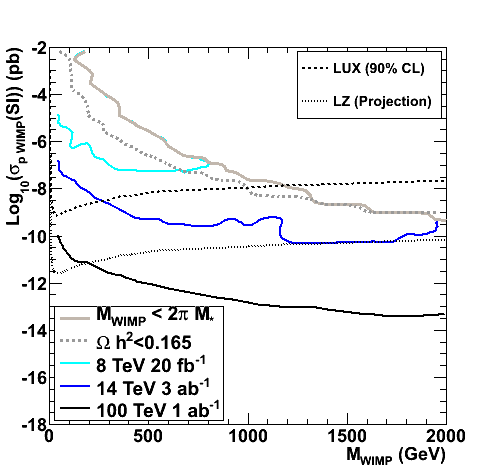} \\
\end{tabular}
\caption{Constraints obtained for the $R_1$ (left panel) and $R_3$ (right panel) operators in the spin-independent scattering cross section 
vs.\ WIMP mass plane. The preliminary results for monojet searches at 8, 14 and 100~TeV are compared to the current LUX 90\% C.L. upper limits and the 
expected reach of the future LZ experiment. The constraint derived from relic dark matter density and the limit of validity of the EFT approach are 
also shown.}
\label{fig:LHC}
\end{center}
\end{figure}

WIMP masses are probed up to 2~TeV, and $M_*$ is varied between 0 and 3 TeV. For each parameter point and each operator, 
the dark matter relic density and direct detection spin-independent scattering cross-sections are computed and the detectability through 
monojet searches at 8, 14 and 100 TeV centre-of-mass energies is evaluated as follows: We use the event reconstruction and selection criteria 
of the ATLAS~\cite{ATLAS-CONF-2013-068} and CMS~\cite{CMS-PAS-EXO-12-048} preliminary monojet analyses performed on $\sim$20~fb$^{-1}$ of 8~TeV data. 
This analysis follows the procedure discussed in Ref.~\cite{Arbey:2013iza}. Monojet signal events have been generated for each parameter point 
using {\tt Madgraph 5} and {\tt Pythia 6}~\cite{Sjostrand:2006za} with the CTEQ6L1 parton distribution functions (PDFs)~\cite{Pumplin:2002vw}.
Physics objects have been reconstructed using the {\tt Delphes 3.0.7}~\cite{Ovyn:2009tx,deFavereau:2013fsa} parametric detector simulation. 
The reconstruction and selection cuts of the original analyses have been applied, namely we required the events to 
have one jet with large transverse momentum, $p_t$, and  missing transverse energy, MET, no electrons or muons 
fulfilling the fiducial $p_t$ and $|\eta|$ experimental cuts and at most one additional jet with $p_t$ in excess to 30~GeV. 
In the ATLAS analysis the leading jet $p_t$ must exceed 280~GeV and the MET must be larger than 220~GeV. The CMS analysis requires 
instead a jet with $p_t >$ 110~GeV and defines seven MET signal regions (MET $>$ 250, 300, 350, 400, 450, 500, 
550~GeV). We use the background estimates obtained by the experiments rescaled by the ratio of the 
products of production cross section times cut acceptance at the different energies of interest compared 
to 8~TeV and the ratio of the assumed luminosity to that used in the original 8~TeV analyses.
The production cross section and acceptance of signal events have been computed using a similar procedure 
as that described above for the monojet signal. The 95\% confidence level (C.L.) exclusion of each generated 
model point in presence of background only is determined using the CLs method~\cite{Read:2002hq}.
These results are projected towards the reach of the HL-LHC project for 3~ab$^{-1}$ at 14~TeV and a more ``futuristic'' 
100~TeV $pp$ collider delivering an integrated luminosity of 1~ab$^{-1}$~\cite{Barletta:2013ooa}. In this preliminary analysis the 
experimental cuts are not optimised at the two energies and a further improvement of these exclusion contours can be therefore expected. 
The results from the monojet analyses are then interpreted in terms of exclusion contours in the spin-independent scattering cross section 
vs.\ WIMP mass plane as shown in Figure~\ref{fig:LHC} where they are also compared to the EFT validity region defined by $M_{\mathrm{WIMP}} < 2\pi M_*$ 
and the DM relic density constraint.
We observe that the elusive $R_1$ could be probed only with energies of the order of 100~TeV and then down to scattering DM-nucleon cross section of the 
order of the current LUX and XENON bounds. On the contrary, monojet searches at HL-LHC at 14~TeV will push the sensitivity to the operator $R_3$ well 
below the current LUX and XENON cross section bounds to match the projected accuracy of LZ~\cite{Malling:2011va} and to WIMP masses of about 2~TeV. 
The availability of 100~TeV $pp$ collision data will push this sensitivity further down into a completely un-chartered territory. We expect that the 
optimisation of the monojet event selection for 14 and 100~TeV will further extend this region of sensitivity.    

\section*{CONCLUSIONS}

In this preliminary investigation we have considered the impact of the latest results from the LHC monojet searches at 8 TeV as well as their projections 
at 14 and 100 TeV in the context of two effective models with real scalar WIMPs. We have seen that the $R_1$ operator is not constrained by the 
8 TeV LHC searches and will be hardly probed by the LHC future run. It will be necessary to perform searches at future higher energy colliders to set 
constraints on this elusive model. On the contrary, the $R_3$ operator is already strongly probed by the current LHC result, and the experimental 
prospects to tighten these bounds in the 14~TeV and high luminosity LHC runs are favourable.

The study presented here represents the first step of a more general program of studies of the effective approach supported by the development of 
a versatile software framework design which will make easy the extension of the study to other operators. Planned developments consist of the 
interpretation of the LHC monojet searches in a broader range of effective models, for real and complex scalar, Dirac and Majorana fermion, vector, 
spin 3/2 and tensor WIMPs. Moreover, in view of our introductory comments, we believe it is appropriate to move 
past the minimal picture where a single effective 
operator is assumed to be responsible for the DM couplings to the SM and examine the consequences of the inclusion of multiple operators 
and of direct couplings of dark matter particles to the Higgs and $Z$ bosons in the analysis. These issues will be addressed in 
details as part of a future study.

\section*{ACKNOWLEDGEMENTS}
The authors would like to thank the organisers of the Les Houches workshop where this work was initiated. The work
of A.A. was supported by the F\'ed\'eration de Recherche A.-M. Amp\`ere de Lyon. 



%% file: nsusydm/NSUSYDM.tex
\newcommand{\txt}[2]{{\color{#1}{#2}}}
\newcommand{\met}{\diagup\hskip -8ptE_T}


\chapter{The Interplay of the LHC and Direct Dark Matter Detection in
Unravelling Natural Supersymmetry at the Focus Point}

{\it D.~Barducci, S.~Belyaev, A.~Bharucha, W.~Porod and V.~Sanz}


 
\begin{abstract}
In this paper we present our results on the interplay of the LHC at 13 TeV (LHC13TeV) 
and direct dark matter (DDM) experiments in probing the far focus point (FFP) 
region of natural supersymmetry, within the MSSM framework.
This parameter space  is characterised by low values of $\mu$
and a compressed spectrum for $\chi^0_1$, $\chi^0_2$ and $\chi^\pm_1$
-- the lightest MSSM particles -- which can only be probed
via mono-jet signatures. We therefore study such signatures in our analysis.
The low signal-to-background ratio is a challenging but important characteristic
of this search which, as we show, never exceeds 6\%,
such that the control of the systematic error is crucial.
We take into account a) realistic systematic errors
and b) fast detector simulation which are both essential in 
estimating the correct LHC sensitivity to the FFP region.
We have have found a high degree of the complementarity
between the LHC13TeV and DDM search experiments:
LHC13TeV@ 1.5 ab$^{-1}$ would be able to exclude FFP parameter space for $m_{\tilde{\chi}^0_1}$
 with a mass {\it below} 120 (200) GeV for a 5\% (3\%) systematic error on the background
  while XENON1T would be able to probe FFP space  with $m_{\tilde{\chi}^0_1}$ {\it above} 320 GeV.
The sensitivity to the mass gap between 120 (200) GeV and 320 GeV is problematic even for the combination 
of the LHC13TeV and XENON1T experiment and requires further attention.
Our findings on the collider searches are also applicable to a more general SUSY framework
 beyond the MSSM.

\end{abstract}

\section{Introduction}
The naturalness of supersymmetry (SUSY), 
which has already been discussed for close to two decades has became even more relevant today, 
at the time of the LHC running, when the scale of SUSY is finally being tested in the TeV region.
Indeed, the lack of evidence for superparticles at the CERN LHC, along with the rather high value
of the Higgs boson mass for SUSY, raised the questions of whether the remaining allowed parameter space 
suffers from a high degree of fine-tuning, and if there is any parameter space of Natural SUSY (NSUSY) left.
We discuss this problem in the framework of the well-motivated minimal supersymmetric standard model
(MSSM), however note that our findings on collider searches are applicable to a general SUSY framework.

In the first papers on this subject, the NSUSY space was connected to light gluino and 
stop masses~\cite{Dimopoulos:1995mi}, on which the limits have already reached the 1 TeV scale
in the case where the gluino and stops are not degenerate with the LSP (see for example Refs.~\cite{Chatrchyan:2013wxa,Chatrchyan:2013lya}).
Note, though, that experimental limits relying on certain dominant decay channels
(e.g. $\tilde t\to t \chi^0_1$) can be significantly relaxed in the scenario we consider below 
with Higgsino-like dark matter, where the branching ratios would depend on the left-right admixture of the lightest stop.
At the same time it has been shown that fine-tuning can be low even if the masses of the supersymmetric scalars and 
gluino are large. This happens in so called "hyperbolic branch"(HB)~\cite{Chan:1997bi} or "focus point" 
(FP)~\cite{Feng:1999mn,Feng:1999zg,Feng:2011aa} region of parameter space, where the value of 
the $\mu$-parameter is low.

The states directly related to naturalness (primarily the stop and higgsinos) are 
especially challenging, and model independent collider bounds are weak or 
non-existent. Light stops can be searched directly via missing energy signatures, or 
indirectly making use of the Higgs data~\cite{Espinosa:2012in,D'Agnolo:2012mj,Carena:2013iba,Kribs:2013lua,Hardy:2013ywa,Fan:2014txa}. 

This study is devoted to NSUSY in the HB/FP region.
In the constrained MSSM, for example, $\mu$  is driven to low values when
$m_0$ parameter is being increased. 
In this region the magnitude of the $\mu$ parameter falls, 
and the higgsino components of the lighter neutralinos increase.
It was recently argued~\cite{Baer:2013gva} that electroweak fine tuning in SUSY 
can be grossly overestimated by neglecting additional non-independent terms which lead
to large cancellations favouring  HB/FP for NSUSY.
In the case of large  $M_1$ and $M_2$ gaugino masses, MSSM particles, namely $\chi^0_1$, $\chi^0_2$ and $\chi^\pm_1$ 
become quasi-degenerate and acquire a significant higgsino component.
This scenario also provides a naturally low dark matter (DM) relic density
via gaugino annihilation and co-annihilation processes into Standard Model gauge and Higgs
bosons. We therefore have relatively light higgsinos-electroweakinos compared to the other SUSY particles. 
This scenario is not just motivated by its simplicity, but also by the lack of evidence for SUSY to date, 
indicating that a weak scale SUSY spectrum is likely non-universal.
Already one decade ago it was shown that  HB/FP parameter space is challenging to probe at the LHC~\cite{Baer:2004qq} 
even if the mass gap between  gauginos is large enough to provide leptonic signatures.
The most challenging case takes place when only $\chi^0_1$, $\chi^0_2$ and $\chi^\pm_1$
are accessible at the LHC, and the mass gap between them is not enough to produce any leptonic signatures.
We call this scenario Far Focus Point region (FFP). The  only way to probe FFP is via a mono-jet signature, as 
suggested in~\cite{Alves:2011sq}. This has been applied to studies on compressed SUSY spectra~\cite{Dreiner:2012gx,Han:2013usa,Han:2014kaa}.

In this contribution we present our results on the interplay of LHC13TeV
 with direct dark matter search experiments in order
to probe the FFP region of NSUSY. We focus on scenarios
where the lightest states are nearly pure higgsinos, and plan to 
consider scenarios with higgsino-bino mixing in future work.
We also take into account realistic systematic 
error and perform a fast detector simulation analysis for FFP,
crucial in correctly estimating the correct LHC sensitivity to FFP.

\section{Spectrum and decays}~\label{spect}

We consider scenarios where the lightest neutralinos and charginos
are higgsino-like and where all sfermions have masses in the multi-TeV
range. 

In the limit $|\mu|\ll|M_1|,|M_2|$ one finds
\begin{eqnarray}
m_{ \tilde{\chi}^0_{1,2} } &\simeq& \mp \left[ |\mu|  \mp \frac{m_Z^2}{2} (1\pm s_{2\beta}) \left(\frac{s_W^2}{M_1} +\frac{c_W^2}{M_2}\right)\right]\\
m_{\tilde{\chi}^\pm_{1} } &\simeq& 
 |\mu| \left(1+\frac{\alpha(m_Z)}{\pi}\left(2+\ln\frac{m^2_Z}{\mu^2}\right)\right)
- s_{2\beta}  \frac{m^2_W}{M_2} 
\end{eqnarray}
 where we have defined $s_{2\beta} = \sin(2 \beta) sign(\mu)$. We have included the EM corrections in case of $
 m_{\tilde{\chi}^\pm_{1} }$. In the case of $\mu>0$, the eigenstates are
 
 \begin{eqnarray}
 \tilde{\chi}^0_{1,2} & \simeq & \frac{1}{\sqrt{2}} (\tilde H_d^0 \mp \tilde H_u^0) \\
  \tilde{\chi}^\pm_{1} & \simeq & \tilde H_{u,d}^\pm 
 \end{eqnarray}
The mass separation is given by
\begin{eqnarray}
\Delta m_o & = &  m_{\tilde \chi_2^0} -  m_{\tilde \chi_1^0} \simeq m_Z^2 \left(\frac{s_W^2}{M_1} +\frac{c_W^2}{M_2}\right) \\
\Delta m_\pm & = & m_{\tilde \chi_1^\pm} -  m_{\tilde \chi_1^0} \simeq \frac{\Delta m}{2}  + \mu \frac{\alpha(m_Z)}{\pi}\left(2+\ln\frac{m^2_Z}{\mu^2}\right)
\end{eqnarray}
where we have neglected corrections of the order $1/\tan\beta$ and $\mu/M_i^2$.

In the case of pure higgsinos, the three body decays are dominated by 
virtual vector bosons. However, due to the small mass differences
the decays into third generation fermions are suppressed. 
Note that in the scenario where  $M_1$ is close to
$|\mu|$ also the off-shell light Higgs boson $h^0$ can give sizeable
contributions \cite{Bartl:1999iw}, in particular if $\tan\beta$ is large.
Therefore the essential parameters for the scenario under study are
$\mu$ and  $M_1$.

 Three body decays in the limit of small mass separation are discussed in~\cite{DeSimone:2010tf}, where an effective theory study of the pseudo-Dirac Dark Matter scenario~\cite{Nelson:2002ca,Hsieh:2007wq,Belanger:2009wf} such as the higgsino-like was done. In this limit, the decay width does not depend on the overall neutralino mass, just on the mass difference, 
 \begin{eqnarray}
 \Gamma( \tilde{\chi}^\pm_{1}, \tilde{\chi}^0_{2} \to f\, f' \,  \tilde{\chi}^0_{1}) = \frac{C^4}{120 \pi^3} \frac{\Delta m^5}{\Lambda^4}
 \end{eqnarray}
where $\Lambda \simeq m_{W,Z,h^0}$ and $\Delta m$ is either
$m_{\tilde \chi^0_2}-m_{\tilde \chi^0_1}$ or 
$m_{\tilde \chi^+_1}-m_{\tilde \chi^0_1}$. 

For example, for off-shell $Z$ exchange and decay into leptons, the coefficient $C$ is as follows 
\begin{eqnarray}
C^4 = \frac{1}{4}  \frac{g^4}{c^4_W} (s^2_w-1/2)^2 (N_{13} N_{23}-N_{14} N_{24})^2 \simeq \frac{1}{4}  \frac{g^4}{c^4_W} (s^2_w-1/2)^2 
\end{eqnarray}
and similarly for the off-shell $W$-decay.

The proper decay length is very sensitive to the value of $\Delta m$, and values below the GeV  could lead to a displaced vertex, or a collider-stable situation. Indeed, for the decay $\tilde{\chi}^0_{2} \to f\, \bar{f} \,  \tilde{\chi}^0_{1}$ with a $Z$-exchange, the proper decay length is given by
\begin{eqnarray}
L = c \tau \simeq 0.01 \textrm{ cm } \left(\frac{\Delta m}{1 \textrm{ GeV}}\right)^{-5}  \textrm{ (Z-exchange)}
\end{eqnarray}
which implies that for $\Delta m \lesssim 0.1$ GeV, $\tilde \chi_2^0$ would be collider stable. Similarly, for $\Delta m \lesssim 1$ GeV one could look for displaced vertices of order 100 $\mu$m. Note that the measured decay length would depend on the boost factor of the decaying neutralino, and in Ref.~\cite{DeSimone:2010tf} a detailed discussion on how to introduce it is presented.

\begin{figure}[htb]
\centering
\includegraphics[width=0.5\textwidth]{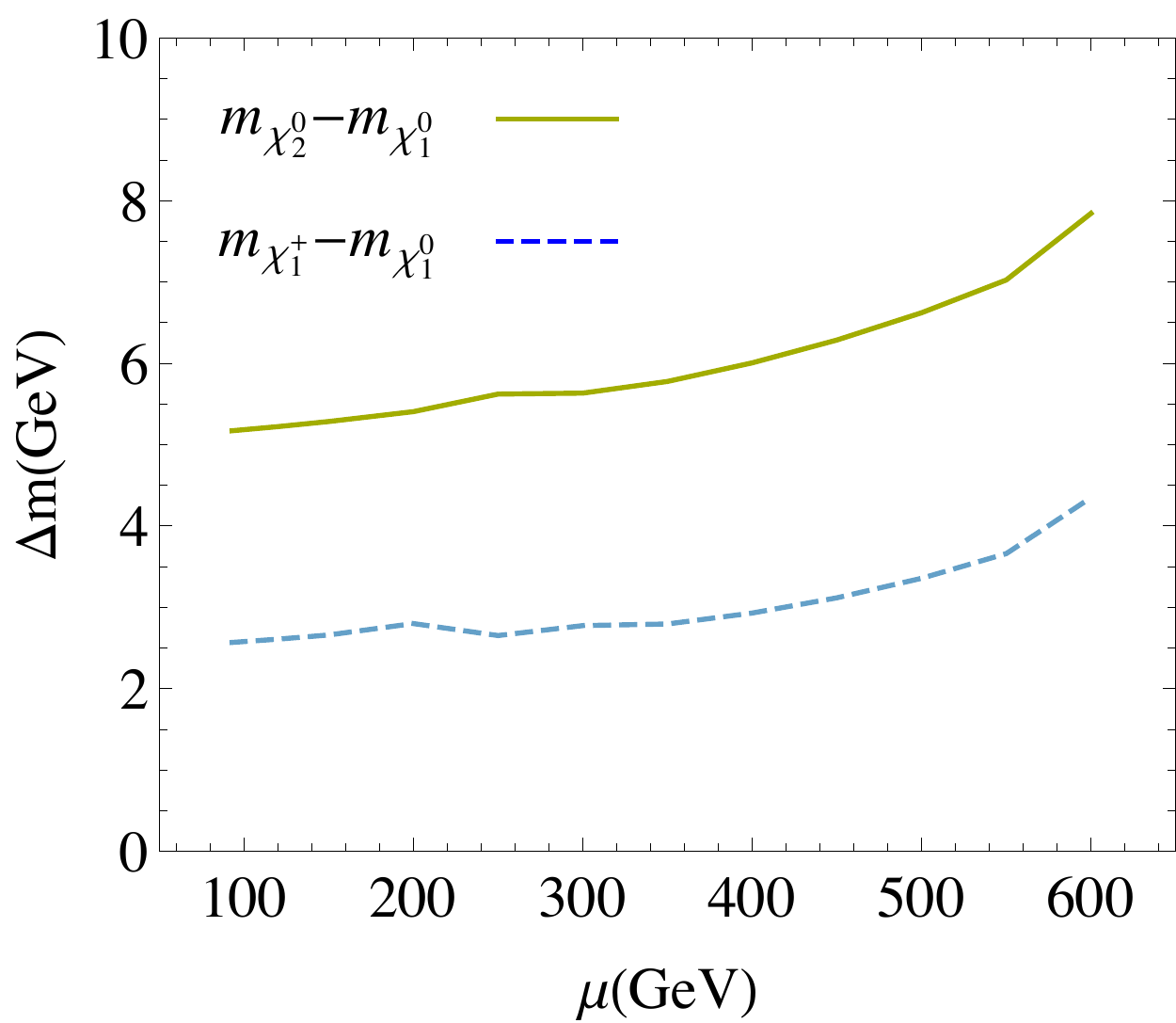}
\caption{\label{fig:mass-spectrum} The mass splitting
 between $\chi^0_2$/$\chi^\pm_1$  and $\chi^0_1$ for the case $M_1$= 1 TeV and $\tan{\beta}$= 5 versus~$\mu$.}
\end{figure}
For $W$-exchange the situation is very similar,
\begin{eqnarray}
\tilde{\chi}^\pm_{1} \to f\, f' \,  \tilde{\chi}^0_{1}
\end{eqnarray}
with $W$-exchange leading to
\begin{eqnarray}
L = c \tau \simeq 0.006 \textrm{ cm } \left(\frac{\Delta m}{1 \textrm{ GeV}}\right)^{-5}  \textrm{ (W-exchange)}
\end{eqnarray}
\begin{table}[htb]
\centering
\begin{tabular}{c||c|c|c|c}
\hline
\hline
$\mu$ (GeV)     & $m_{\chi^0_1}$ (GeV) & $m_{\chi^0_2}$ (GeV) & $m_{\chi^0_3}$ (GeV) & $m_{\chi^\pm_1}$ (GeV) \\
\hline
93           & 98.4                 & 103.6                     & 994.2                     &  101.0                      \\
\hline
200          & 201.9                & 207.2                     & 994.4                     &  204.6                      \\
\hline
300          & 289.8                & 295.4                     & 994.5                     &  292.6                      \\
\hline
400          & 400.0                & 406.0                     & 994.8                     &  402.9                      \\
\hline
500          & 502.7                & 509.3                     & 995.1                     &  506.1                      \\
\hline
\hline
\end{tabular}
\caption{\label{tab:benchmarks}Masses of the higgsino-like 
 lightest gauginos as a function of $\mu$ for the scenario with $M_1=1$ TeV and $\tan{\beta}=5$}
\end{table}

In Table~\ref{tab:benchmarks} and Fig.~\ref{fig:mass-spectrum} we show the mass spectrum and the mass splitting 
for the higgsino-like lightest electroweakinos as a function of $\mu$.
These results are presented for $M_1=1$ TeV and $\tan{\beta}=5$, however for such large values
of  $M_1$ and $M_2$ the mass spectrum,  as well as mass splitting pattern varies slightly (below 1\% level) for the whole range
of $\tan\beta$ (5-50) under consideration.  

In the following sections \ref{sec:DM} and \ref{sec:LHC13}, where we have studied the dark matter and collider phenomenology respectively,
$M_2$ and $M_3$ are kept fixed to  2 TeV and 1.5 TeV respectively, while we decouple the effect of squarks 
and sleptons in the electroweakino production and decay by keeping their mass at 
$\simeq$ 2 TeV. 
The dark matter results are presented for the cases $M_1=\mu$, $M_1=(\mu+1~\mathrm{TeV})/2$ and $M_1=$1 TeV and for $\tan\beta=5$, 15, 25, 50, 
whereas the collider results are presented for $M_1=1$ TeV and $\tan{\beta}=5$.

\section{Dark Matter}\label{sec:DM}
The newly released results from Planck ~\cite{Baltz:2006fm} (see also WMAP~\cite{Hinshaw:2012aka}) mean that the 
uncertainty on the already very precise measurement of the dark matter (DM) relic density ($\Omega_{\rm DM} h^2$), 
assuming $\Lambda_{\rm CDM}$ cosmology, has become even smaller, $\Omega_{\rm DM}^{\rm Planck} h^2=0.1198\pm0.0026$.

\begin{figure}[htb]
\includegraphics[width=0.99\textwidth]{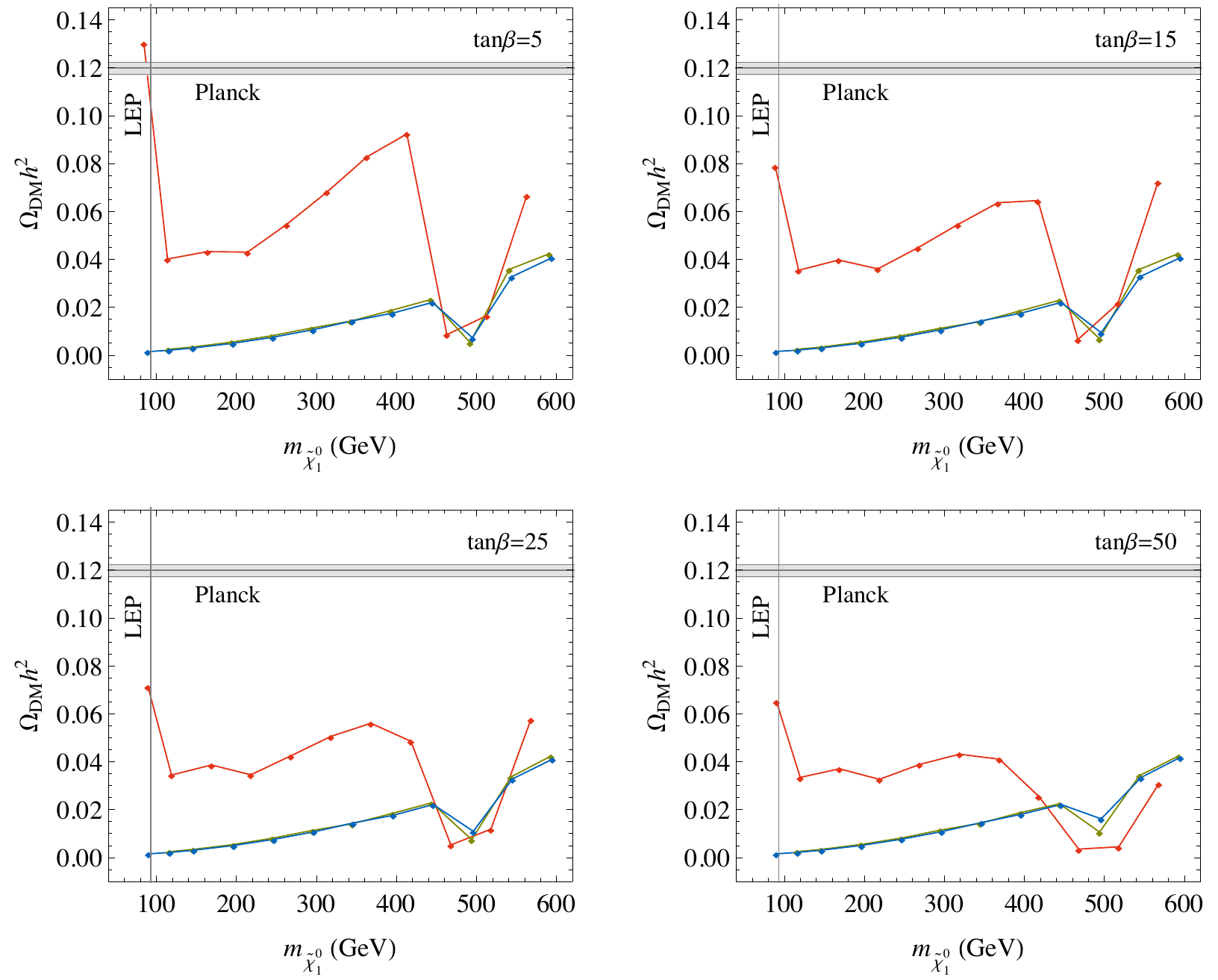}
\caption{The predicted value of the dark matter relic density $\Omega_{\rm DM} h^2$ is shown as a function 
of the mass of the lightest neutralino $m_{\tilde{\chi}^0}$ for four different values of $\tan\beta$ as indicated. 
The red, yellow and blue lines indicate $M_1=\mu$, $M_1=(\mu+1~\mathrm{TeV})/2$ and $M_1=$1 TeV respectively. The 
relic density measured by the Planck satellite, $\Omega_{\rm DM}^{\rm Planck} h^2$, is also shown for comparison. The vertical line shows
the LEP limit on $m_{\tilde{\chi}^+}$ \cite{lep:susy}.} 
\label{fig:Omega}
\end{figure}

The lightest supersymmetric particle (LSP), if stable, will contribute to this relic density. 
In the scenarios considered here, the LSP is the lightest neutralino, composed
predominantly of the higgsino and, to varying degrees, of the bino.
It is well known that a higgsino-like LSP produces an under-density of dark matter, i.e.~the annihilation cross-section
is too high to obtain the relic abundance observed by Planck. On the other hand, if the LSP 
is bino-like, annihilation is suppressed, and an over-density is predicted. 
Therefore a mixed LSP can, at low LSP masses $\sim$$100$ GeV, lead to a correct prediction of the relic abundance
and this is known as the focus point(FP) region.
Therefore we chose this  $\mu\lesssim 
M_1$ scenario which  results in the value of $\Omega_{\rm DM}h^2$ being pausably
not above the measured value from Planck
which solves the typical problem of the over-closure of the universe for generic SUSY parameter space.
 
\begin{figure}[t]
\includegraphics[width=0.99\textwidth]{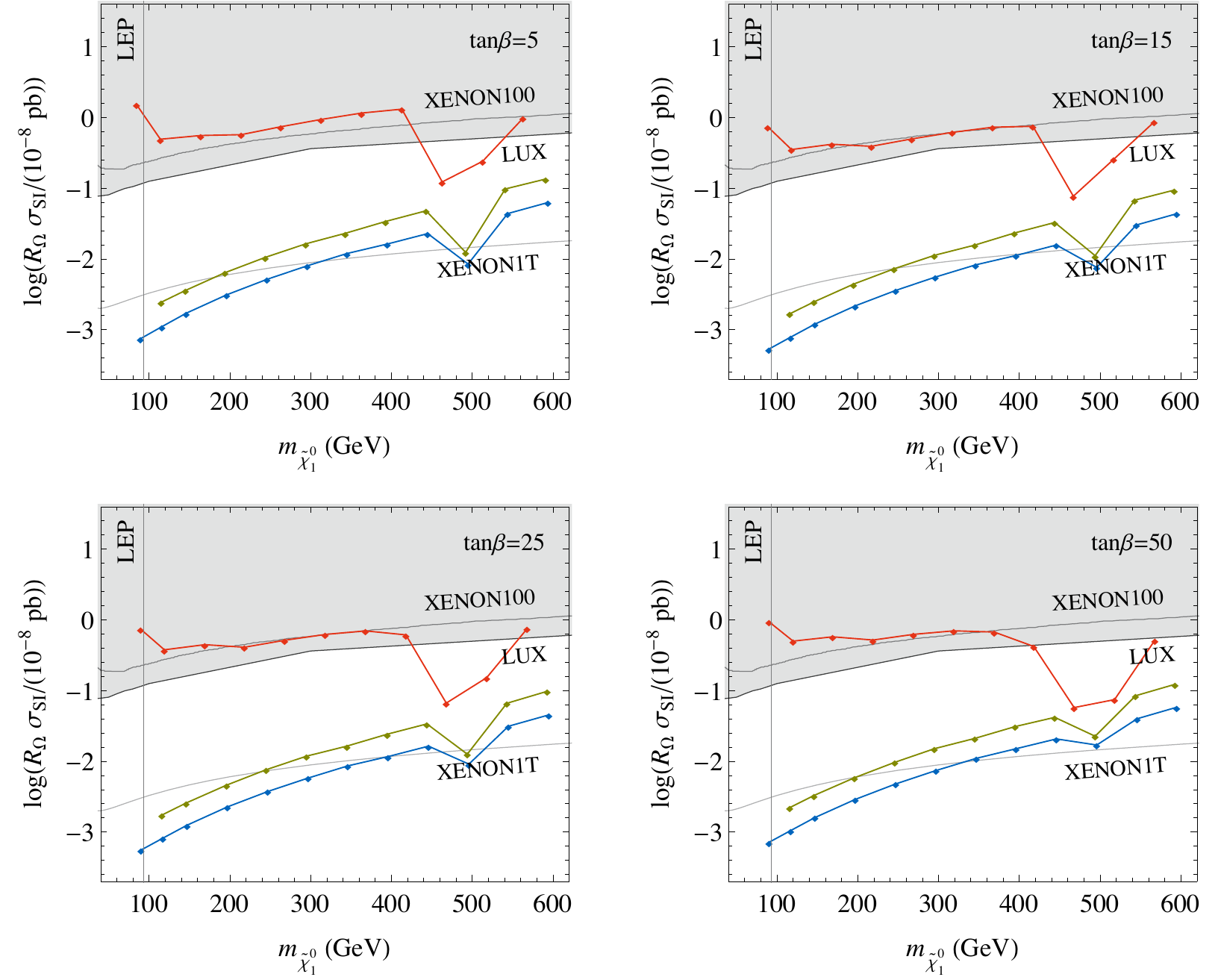}
\caption{The predicted value of the spin-independent annihilation cross section for direct detection $R_\Omega\,\sigma_{\rm SI}$ (pb), 
rescaled by $R_\Omega=\Omega_{\rm DM}/\Omega^{\rm Planck}_{\rm DM}$, is shown as a function of the mass of the lightest neutralino $m_{\tilde{\chi}^0}$, where colours are as in Fig.~\ref{fig:Omega}. The 
exclusion limits from XENON100 and LUX, as well as the projected limit from XENON1T are also shown for comparison.} 
\label{fig:sigmaSI}
\end{figure}
We then assume that the 
remaining relic abundance is accounted for by other means, for example:
\begin{itemize}
 \item Adding a light multi-TeV moduli field where the higgsino LSP is non-thermally produced (e.g.~Ref.~\cite{Allahverdi:2012wb})
 \item Mixed axion-higgsino dark matter (e.g.~Ref.~\cite{Baer:2012uy})
\end{itemize}

In order to assess the compatibility of the scenarios studied with existing results from 
dark matter experiments, we have calculated $\Omega_{\rm DM} h^2$ and the spin-independent cross section
for direct detection using {\texttt{micrOmegas 2.4.1}}~\cite{Belanger:2006is, Belanger:2010gh}.
In Fig.~\ref{fig:Omega} we show the results for $\Omega_{\rm DM} h^2$ as a function of
the mass of the LSP, i.e. the lightest neutralino, for four different values of $\tan\beta$ as indicated.
The red, blue and yellow lines indicate $M_1=\mu$, $M_1=(\mu+1~\mathrm{TeV})/2$ and $M_1=$1 TeV respectively.
From this plot we see, as expected, that in general $\Omega_{\rm DM} h^2$ lies below $\Omega_{\rm DM}^{\rm Planck} h^2$,
and decreases as the neutralino becomes increasingly higgsino-like.
The reason causing the difference between the mixed gaugino/higgsino case and the nearly pure higgsino case is
the Higgs contribution. The Higgs bosons can couple maximally to neutralinos if they are a nearly equal admixture
of gauginos and higgsinos as is the case for  $M_1=\mu$ (red line). Moreover, the dip close to
$m_{\tilde \chi^0_1} \simeq 500$~GeV is due to the pseudo-scalar Higgs boson $A^0$ which has a mass close to
1 TeV in our case. Again the effect is more pronounced in case of $M_1=\mu$.
In Fig.~\ref{fig:sigmaSI} we further show the spin-independent annihilation cross section for direct detection, 
again for four different values of $\tan\beta$ as indicated, and with colour-coding analogous to Fig.~\ref{fig:Omega}.
For convenience, on these plots we additionally indicate the most recent 
limits from XENON100~\cite{Aprile:2013doa} and LUX~\cite{Akerib:2013tjd}, as well as the projected limits from 
XENON1T after 2 years live-time and 1 ton fiducial mass (see e.g.~Ref.~\cite{Aprile:2012zx}).
Instead of $\sigma_{\rm SI}$, we plot the rescaled $R_\Omega\,\sigma_{\rm SI}$ (pb), where the scaling factor
$R_\Omega=\Omega_{\rm DM}/\Omega^{\rm Planck}_{\rm DM}$ allows easy comparison with these bounds,
which in general assume the relic density to be the value measured by Planck.
Fig.~\ref{fig:sigmaSI} illustrates that in the focus point region for low LSP masses, 
where the correct relic density is predicted and which is also easiest to see at
 colliders, is in fact excluded as the 
the spin-independent cross section for the direct detection experiments is too high. We would further like to highlight the interesting complementarity between the reach of the collider searches and the direct detection searches,
particularly interesting for low dark matter masses.

\section{Projections for the LHC run at 13 TeV}\label{sec:LHC13}

In this section we perform a study to obtain the projected sensitivity of the LHC run at 13 TeV to the electroweakino sector in Natural SUSY. 
Conservatively, we assume that the sfermions are heavy, and consider a scenario where 
the mass separations $\Delta m$ and $\Delta m_\pm$ is not large enough as to produce visible decay products. 

\begin{figure}{}
\centering
\includegraphics[width=0.9\textwidth]{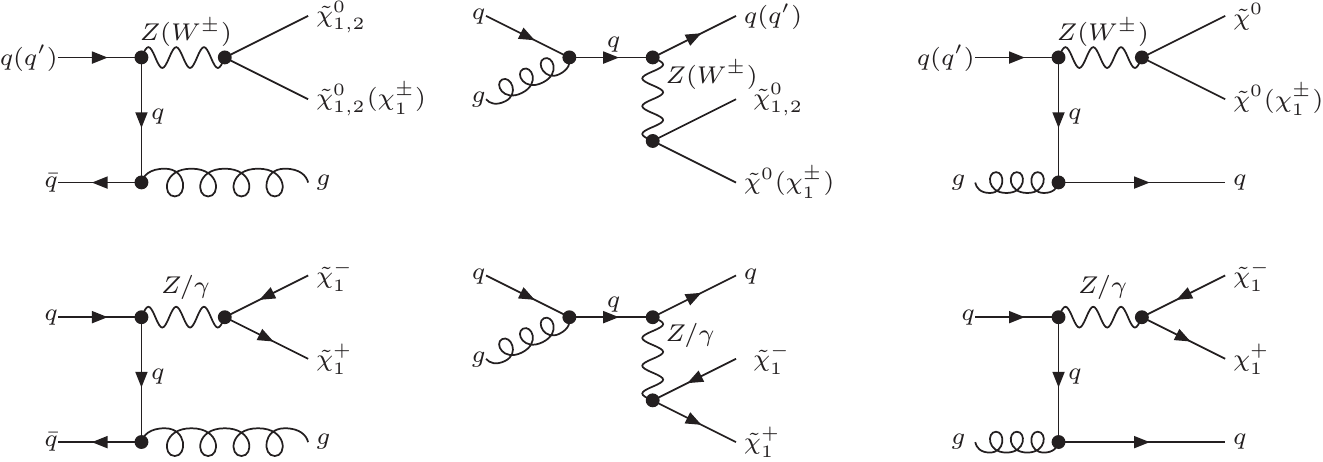}
\caption{\label{fig:diagrams} Representative diagrams for
pair neutralino-chargino production in association with quark/gluon 
leading to mono-jet signature.}
\end{figure}

This scenario with degenerate higgsinos, i.e.~the NSUSY scenario, is the most difficult one to test since pair 
electroweakino production {\it per se} will be not detectable. One requires the help of  initial- or final-state radiation (ISR, FSR), 
which could produce a high $p_T$
jet~\cite{Dreiner:2012gx} or photon~\cite{Belanger:2012mk,Lee:2012ph} recoiling against the neutralino system.~\footnote{ 
Note that mono-Z~\cite{Bell:2012rg} or mono-W~\cite{Bai:2012xg} signatures could also be used to constrain this scenario, 
although monojet is the most promising channel.} The diagrams for the mono-jet + $\met$ signature are shown in Fig.~\ref{fig:diagrams} 
where we have omitted contribution from heavy squarks. Therefore the signal subject of our study will be
\begin{equation}
p p \to \chi \chi j \qquad \chi=\chi^0_{1,2}, \chi^\pm_1
\end{equation}

In the following, the results are presented for $M_1=1$ TeV and $\tan{\beta}=5$, and other parameters are as mentioned in section~\ref{spect}.

\subsection{Analysis Setup}

We performed a parton-level simulation using MadGraph~\cite{Alwall:2011uj} 
with the MSSM model from FeynRules~\cite{Alloul:2013bka}\footnote{\tt{http://feynrules.irmp.ucl.ac.be/wiki/MSSM}}, and cross-checked with
 CalcHEP~\cite{Belyaev:2012qa} with the MSSM model from the HEPMDB website\footnote{\tt{http://hepmdb.soton.ac.uk/hepmdb:0611.0028}}.
Parton-level Standard Model background simulations have been also cross checked between two packages. Our choice of PDF sets is CTEQ6L1~\cite{Pumplin:2002vw} and we used the MadGraph dynamical choice of renormalization scale. 
Parton-level events went through hadronization and parton-showering using PYTHIA~\cite{Sjostrand:2006za},
followed by the the Delphes3 \cite{deFavereau:2013fsa} package for fast detector simulation.

\subsection{Signal versus background analysis and LHC prospects}

It should be stressed that soft leptons and quarks coming from $\chi^0_2$ and $\chi^\pm_1$ decays will not be visibly boosted by the ISR since 
the momenta of the boosted particles is proportional to their mass. The boost will mostly be taken by the neutralino, whose 
mass is already limited by LEP to be above 90 GeV in this scenario~\cite{Eidelman:2004wy}.

\begin{figure}[h!]
\includegraphics[width=0.49\textwidth]{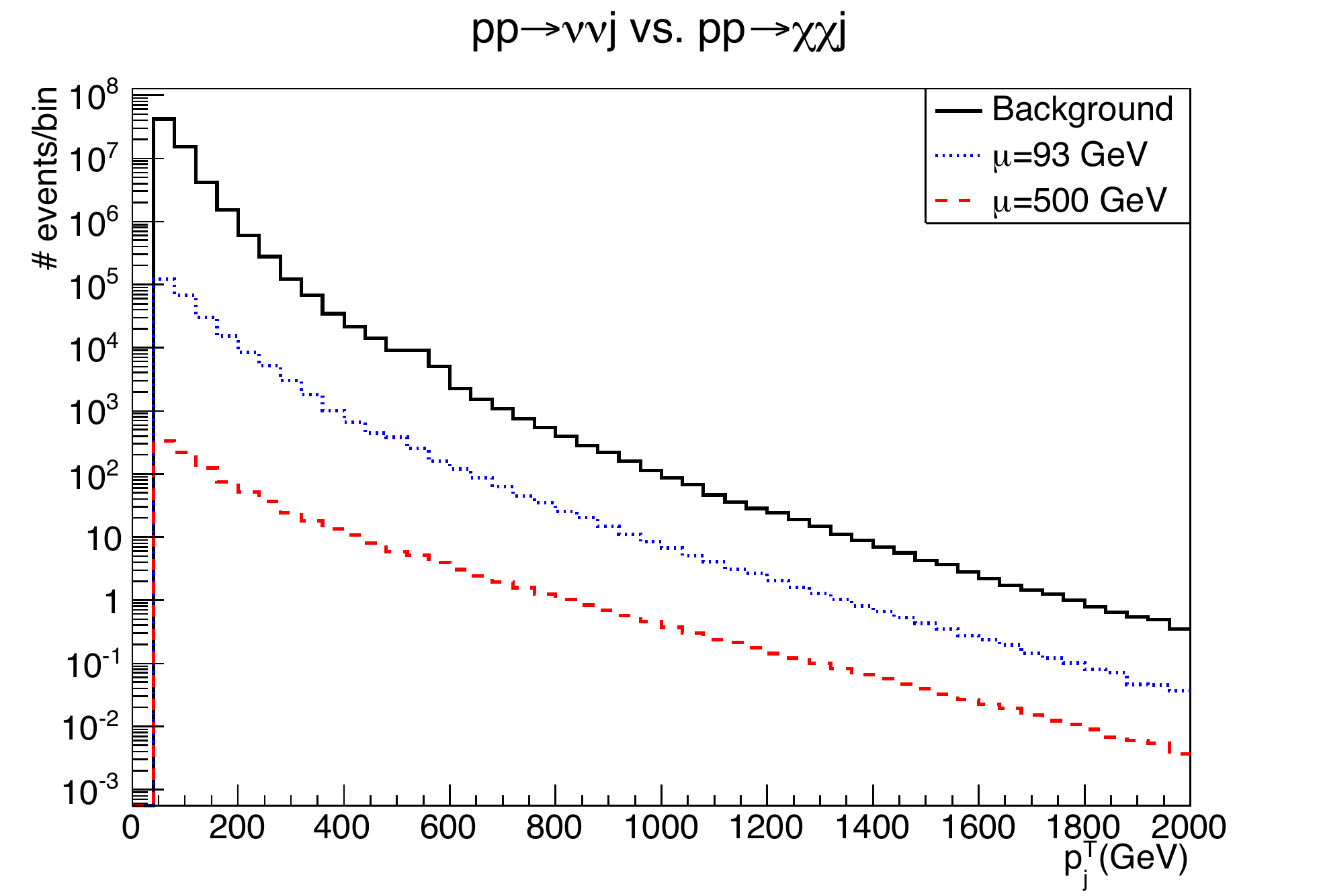}\hfill
\includegraphics[width=0.49\textwidth]{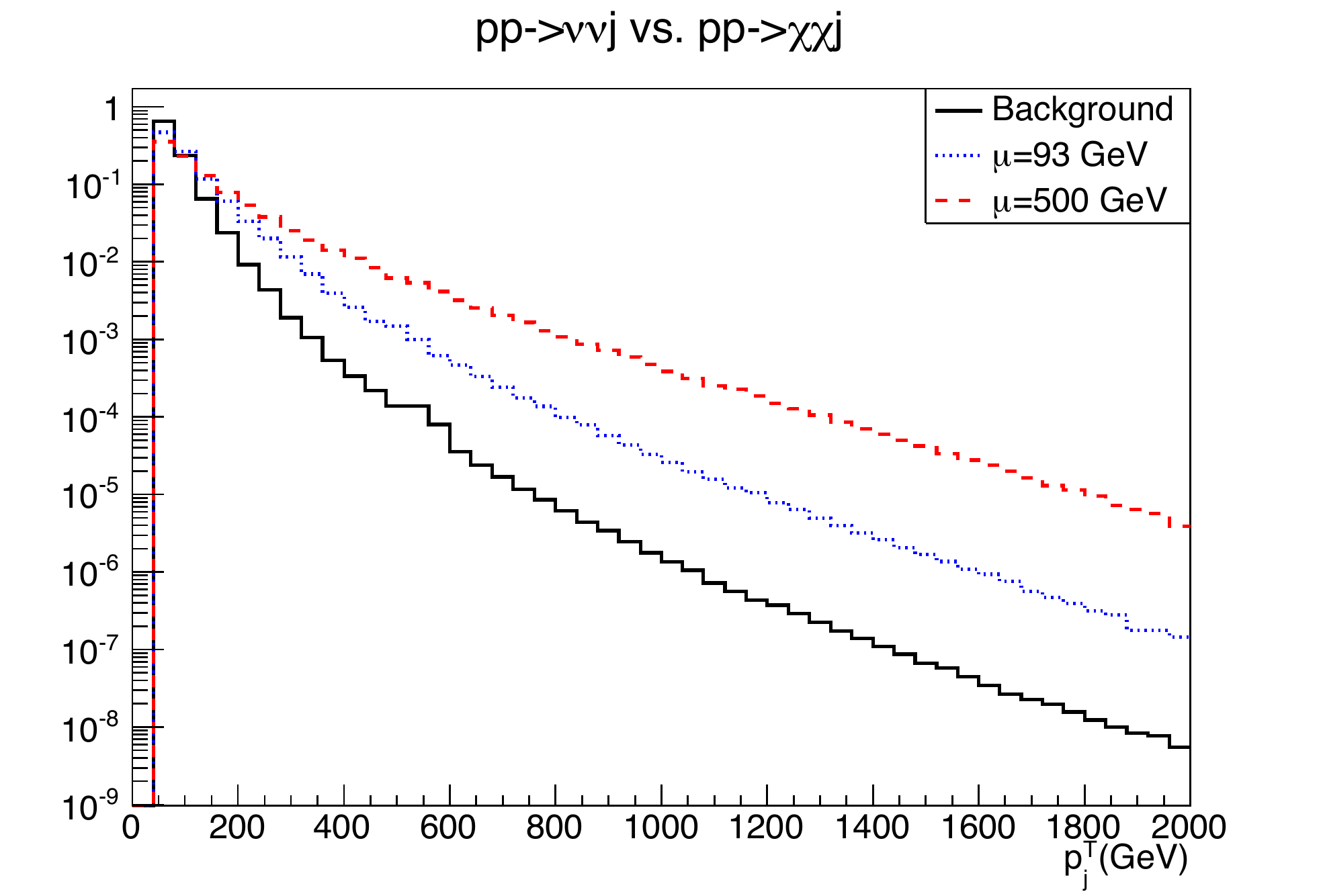}
\caption{Signal (dotted blue and dashed red) and $Zj$ background (solid black) parton-level
$p^T_j$ distributions for the 13 TeV LHC for the DHS
scenario. Left: $p^T_j$ distributions for  100 $fb^{-1}$ integrated luminosity.
Right: normalised signal and  $Zj$ background distributions.}
\label{fig:signal-vs-bg}
\end{figure}

The main SM background to our high $p_T$ jet + high  $\met$ (monojet) signature is the irreducible  $Z+jet \to \nu\bar{\nu}+jet$ ($Zj$). 
The relative size and shape difference of the signal versus the $Zj$ background is presented in the left and right frames of 
Fig.~\ref{fig:signal-vs-bg} respectively for $p_T^j$ distribution. In this figure we see that a sizeable cut on $p_T$ needs to be applied. 
With a basic $P_T^j>$ 20 GeV cut the $Zj$ background is about 3 orders of magnitude higher than the signal for the lowest allowed mass of $m_{\chi^0_1}\simeq 100$~GeV. 

An important feature of the signal versus background is that the $shape$ of the background distribution is quite different from the signal: the background falls more rapidly with $p^T_j$, and the difference of slope with respect to the signal is bigger for higher neutralino masses. The different slope is  mainly due to the mass difference  between the neutralino from signal and neutrino from the background. One should also notice that the difference between shapes of signal and background $p^T_j$ distributions  vanishes for very large 
values of $p^T_j \gg m_{\chi^0_1}$, as one would expect.

After inspecting these distributions,  one expects that  the best sensitivity  will be achieved for a high enough values of $p^T_j$ cut and eventually correlated values of $\met$ cut. Ultimately, though, the sensitivity will be limited by the systematic uncertainty on the background prediction. From Fig.~\ref{fig:signal-vs-bg} (left) one can see that even for very large values of $p^T_j$ cut 
the highest signal to background ratio (S/B) will be about 1/10, hence one needs to control the systematic error at the few percent level. 
Therefore, in our projections for 13 TeV LHC we carefully take  the systematic error into account~\footnote{Note that in a similar study~\cite{Han:2013usa}, the authors relied on 1\% systematic uncertainty which we believe is unrealistic.}. 

\begin{figure}[htb]
\centering
\includegraphics[width=0.50\textwidth]{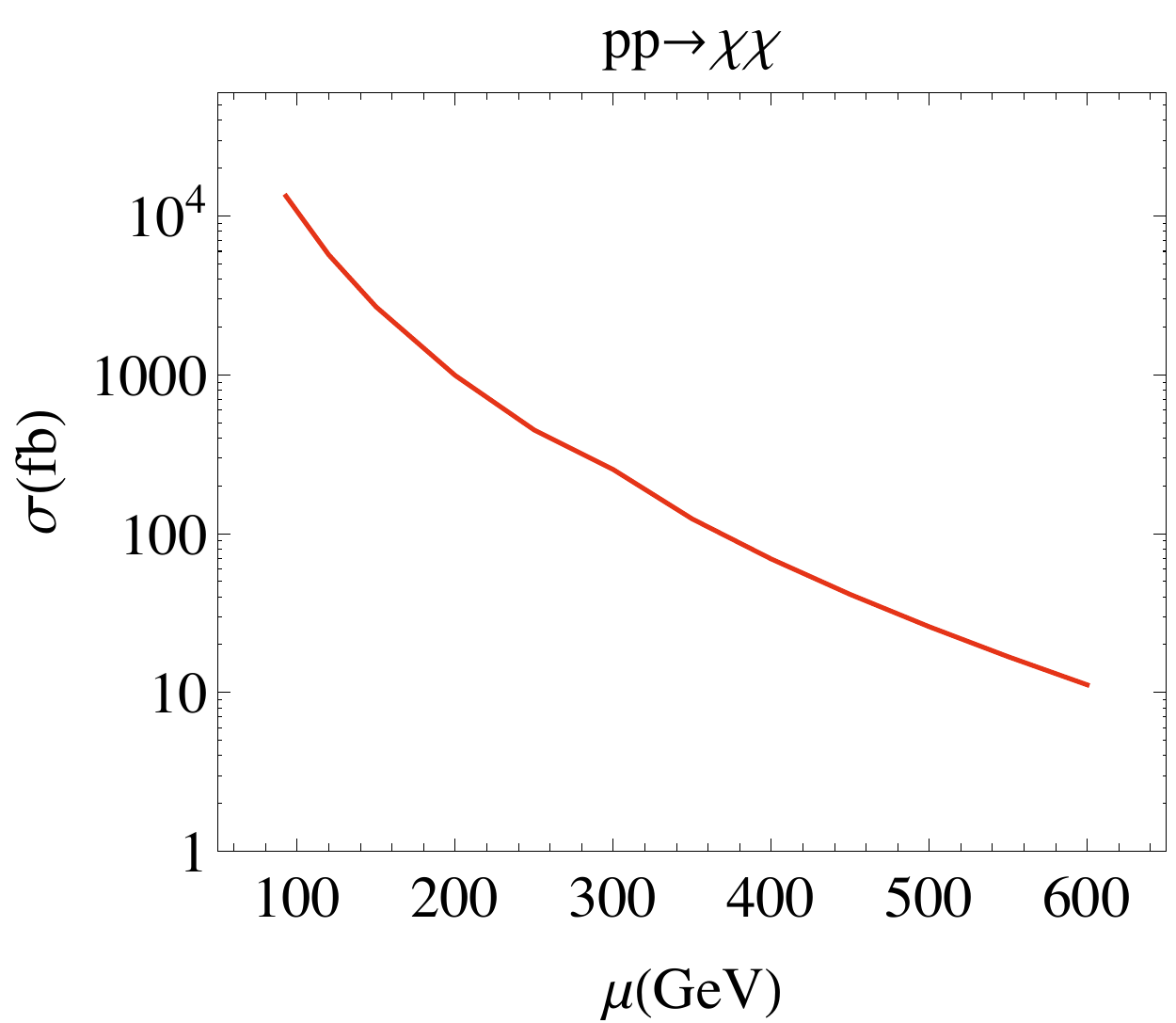}%
\includegraphics[width=0.50\textwidth]{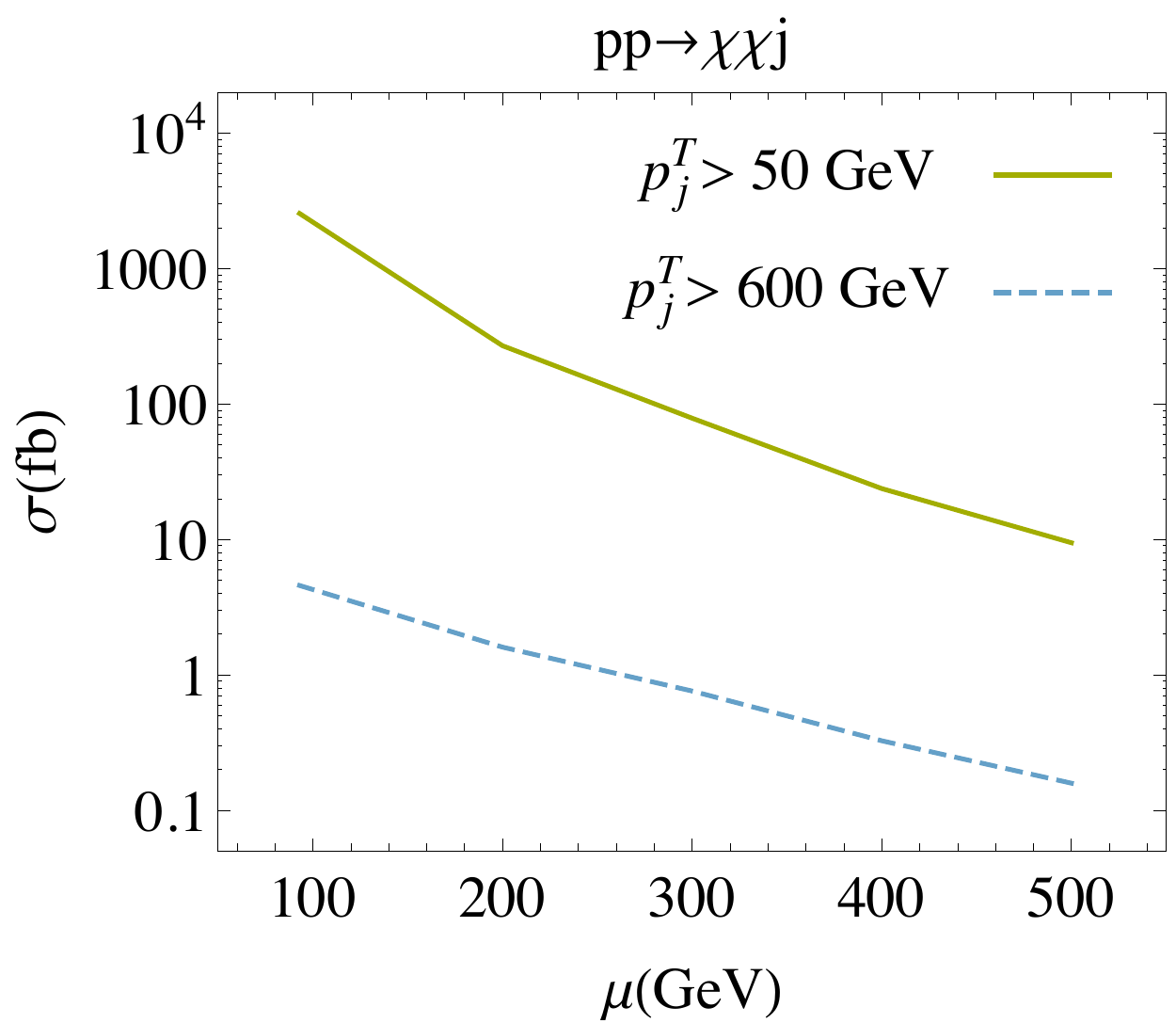}
\caption{Cross section of the  $pp\to\chi\chi$ (Left) and  $pp\to\chi\chi j$ (Right) processes ($\chi=\chi^0_{1,2},\chi^\pm_1$)
at the 13 TeV LHC for two choices of cuts at parton level on the leading jet $p^T$.}
\label{fig:cs_vs_mu}
\end{figure}

To give the reader an idea on expected signal rates we present the total cross section of the  $pp\to\chi\chi j$ process as a function of $\mu$ in Fig.~\ref{fig:cs_vs_mu} for two different values of $p^T_j$ cut: 50 GeV (green curve)  and 600 GeV (blue dashed curve).
One can see that a high $p^T_j$ cut such as 600 GeV or even higher would be required, which provides high enough S/B to comply with control over systematic uncertainties.

Besides taking into account crucial aspect of controlling on systematic uncertainties, the steps of going beyond parton-level simulation is essential. By performing a fast simulation, one can take into account multiple ISR effects, realistic detector energy resolution and particle acceptances.
In particular, the perfect correlations between $p^T_j$ and $\met$ which take place at parton level can be considerably spoilt
at the level of the fast detector simulation, as we can see from Fig.~\ref{fig:pt-met-corr}. As we will see below, the lack of the perfect $p^T_j$ vs  $\met$ correlations leads to a visible difference of the S/B ratio and significance, and should be taken into account.  Note that in Ref.~\cite{Han:2014kaa} degenerate higgsinos were studied at parton-level, hence this analysis misses the effects we just mentioned-- as the authors already pointed out.

\begin{figure}[htb]
\includegraphics[width=0.45\textwidth]{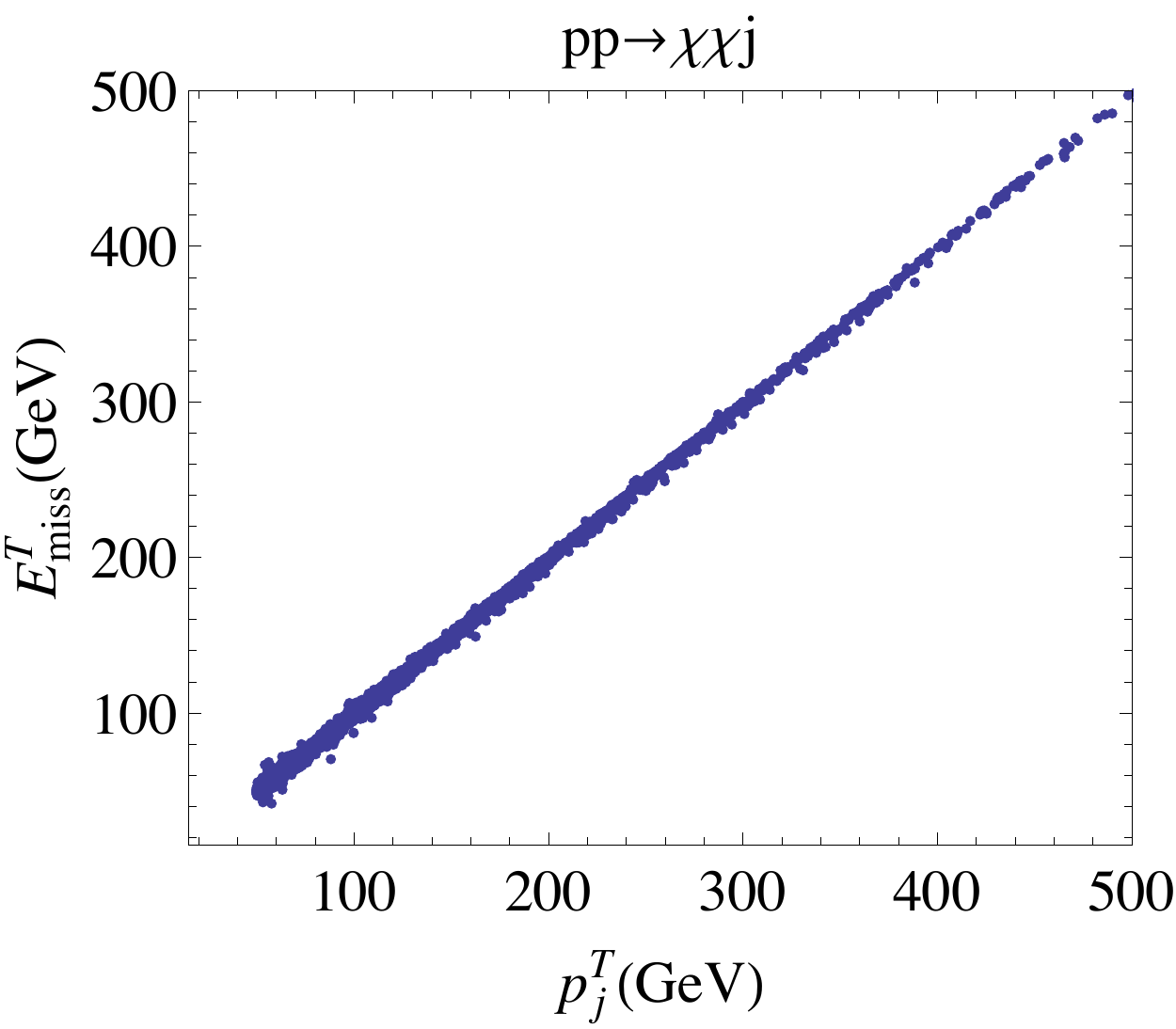}\hfill
\includegraphics[width=0.45\textwidth]{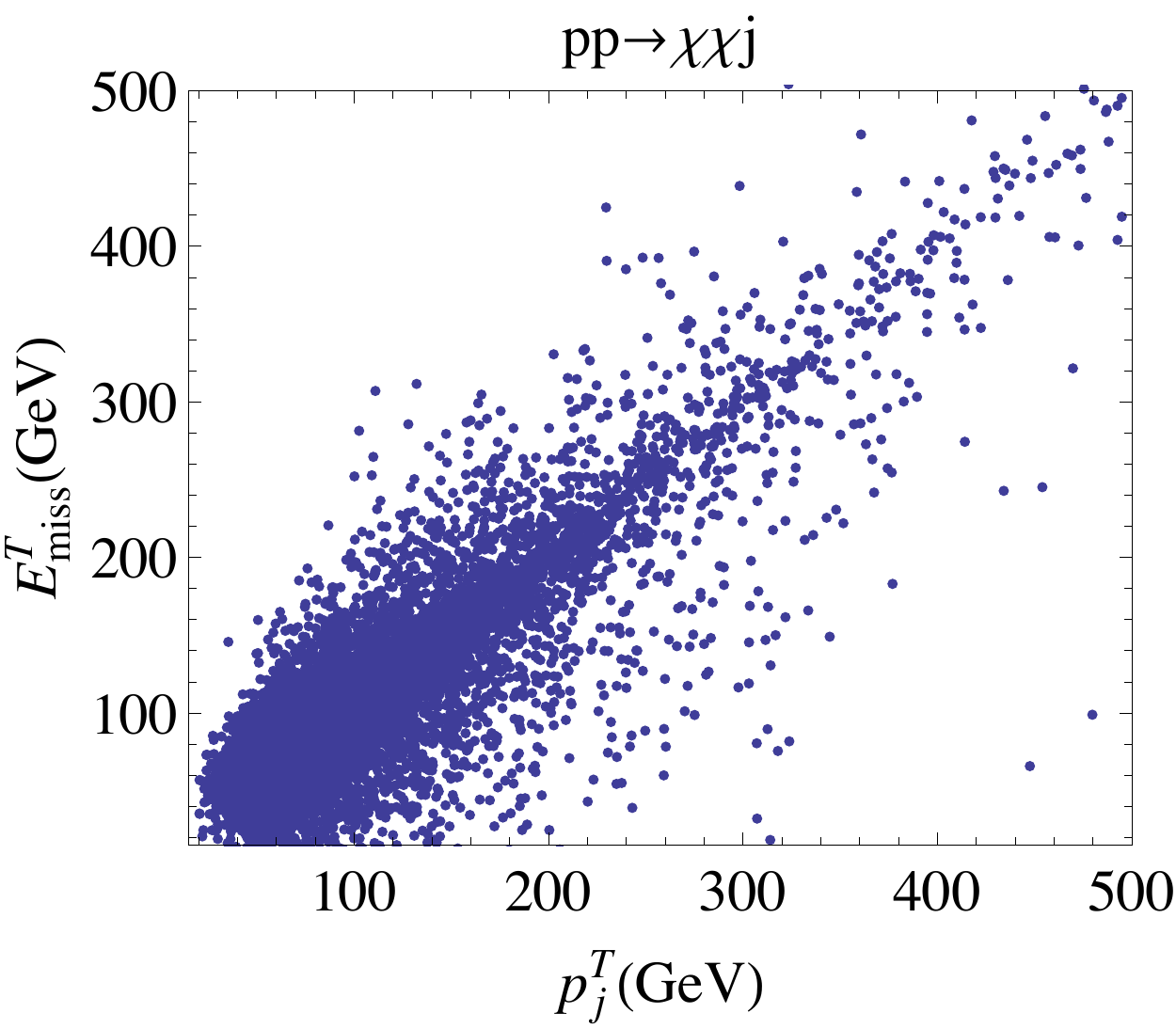}
\caption{$p^T$ of the leading jet versus $\met$ for the  $pp\to\chi\chi j$ process ($\chi=\chi^0_{1,2},\chi^\pm_1$) process in the NSUSY scenario at parton level (left)
and detector simulation level (right).}
\label{fig:pt-met-corr}
\end{figure}

Besides the leading $Zj$ background, we also considered the irreducible $W+jet\to l\nu+jet~(Wj)$ background
which mimics the signal when the charged lepton goes below the
lepton acceptance cuts. To suppress this background we have applied veto
on the charged leptons with the following $p^T$ cuts: 
\begin{equation}
 p^T_{e^\pm,\mu^\pm}> 10~\textrm{GeV}\qquad p^T_{\tau^\pm}> 20~\textrm{GeV}
\end{equation}
and standard acceptance pseudorapidity cuts
\begin{equation}
|\eta_{e^\pm,\mu^\pm}|<2.5.
\end{equation}
 We also checked that the $t\bar{t}$ QCD background is subleading to the electroweak SM background, once final cuts
on the $p^T_j$ an $\met$  bigger than $\simeq$ 1 TeV are applied.

\begin{table}[htb]
\centering
\begin{tabular}{l|c|c|c|c}
\hline
\hline
						& $Z(\nu\bar\nu)j$ 		 & $W(\ell\nu)j$  	& $\mu=93$ GeV 	  & $\mu=500$ GeV\\
\hline
\hline
$p^T_{jet}>50$~GeV, $|\eta_{jet}|<4.5$		& 6.4 E+7   	 & 2.9 E+8     		 & 2.6 E+5		  & 948    \\
\hline
Veto $p^T_{e^\pm,\mu^\pm/\tau^\pm}>$10/20 GeV   & 6.2 E+7     	 & 1.2 E+8     		 & 2.5 E+5	 	  & 921   \\
\hline
$p^T_j>$500 GeV 				& 2.5 E+4        & 2.0 E+4                & 1051                   & 32     \\
\hline
$p^T_j=\met>$500 GeV 				& 1.5 E+4        & 4.1 E+3               & 747                    & 27      \\
\hline
$p^T_j=\met>$1000 GeV 				& 315  (375)     & 65  (32)  		 & 21 (31)	 	  & 2	 (2)	\\
\hline
$p^T_j=\met>$1500 GeV				& 18   (20)  	 & 2  (1)    		 & 1  (2)		  & 0	 (0)	\\
\hline
$p^T_j=\met>$2000 GeV				& 1    (1)   	 & 0  (0)     		 & 0  (1)		  & 0	 (0)	\\
\hline
\hline
\end{tabular}
\caption{
Number of events at the 13 TeV LHC with 100 $fb^{-1}$ integrated luminosity for the backgrounds and
for the signal, $\mu$=93 GeV and $\mu$=500 GeV after the respective cuts indicated in the left column.
 Numbers in round brackets correspond to the parton-level predictions.
Numbers are rounded to the integer. 
}
\label{tab:nevents_delphes}
\end{table}

In Table~\ref{tab:nevents_delphes}
we present our results for the cut-flow for signal and background
events for a centre-of-mass energy of
13 TeV and an integrated luminosity of 100 fb$^{-1}$.

\begin{figure}[htb]
\includegraphics[width=0.49\textwidth]{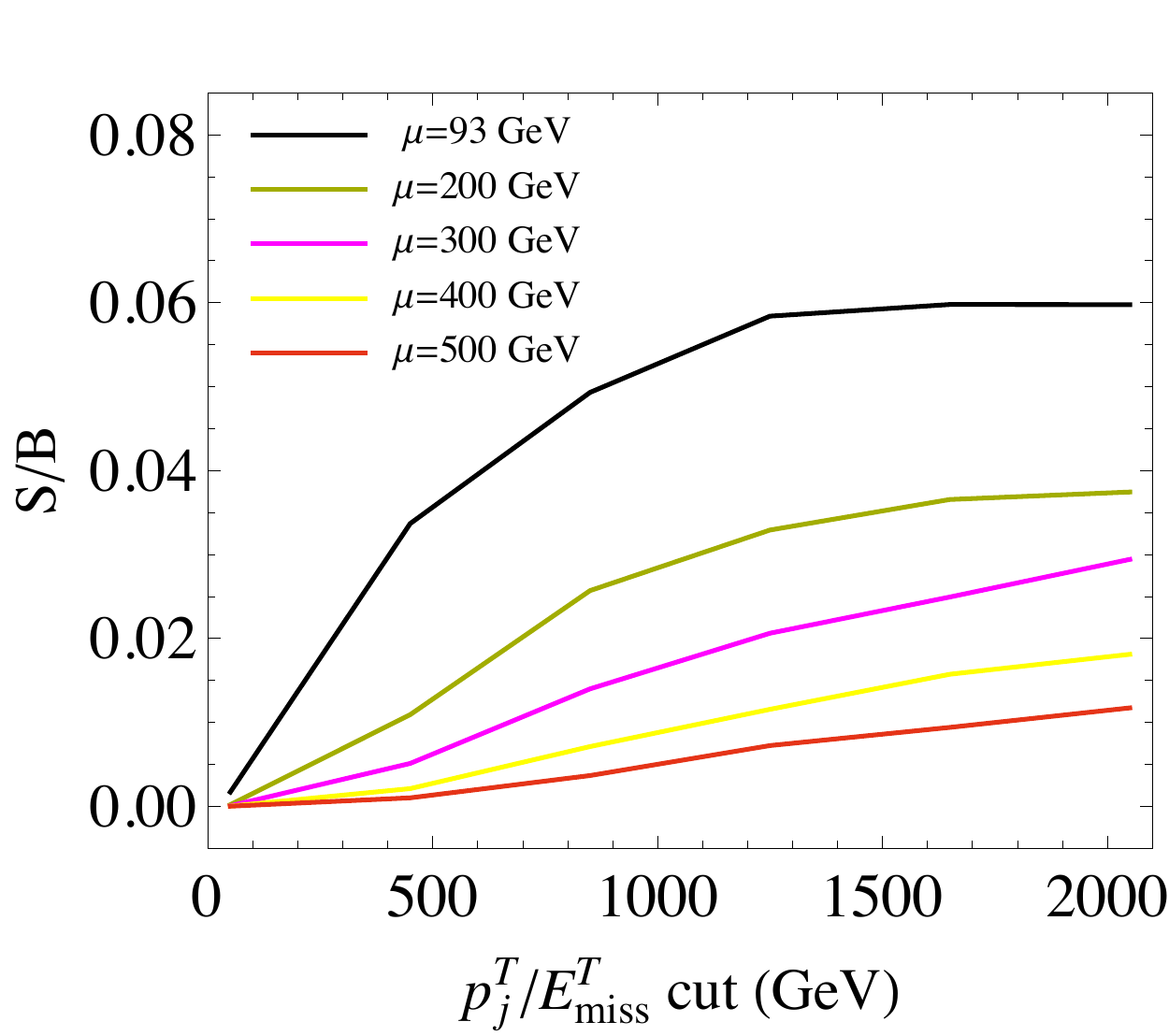}\hfill
\includegraphics[width=0.49\textwidth]{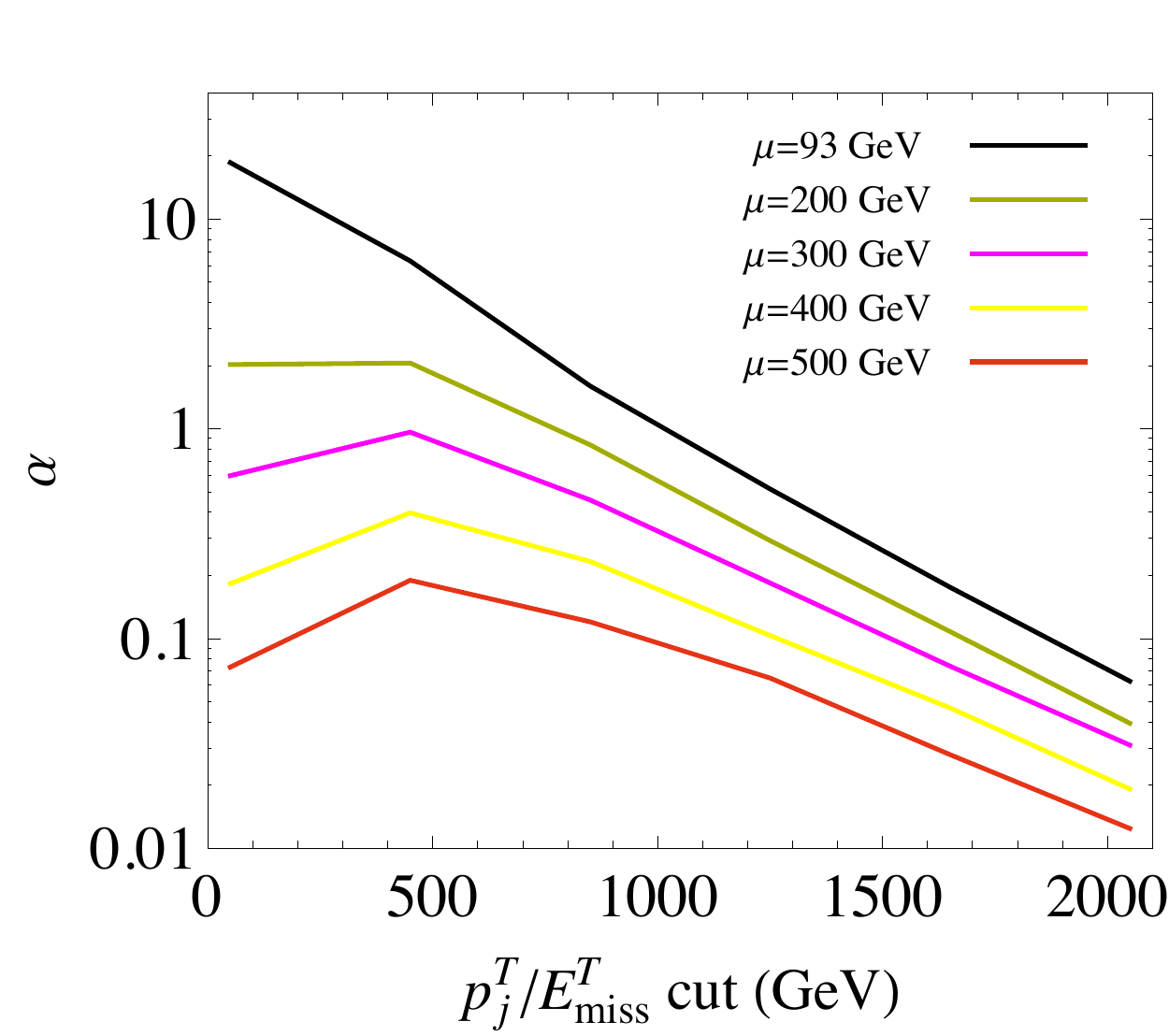}  
\caption{Signal over background ratio (left) and  statistical significance (right)  for the 13 TeV LHC with
100 $fb^{-1}$ of integrated
luminosity as a  function of the $p^T_j=\met$ cut at detector simulation level after the applications of the lepton veto. Colour code in the legend.}
\label{fig:signif_delphes}
\end{figure}

The effect of the $p^T_j$ and $\met$ on the S/B and significance is presented in Fig.~\ref{fig:signif_delphes}.
The significance $\alpha$ is based only on statistical error and is calculated as 
\begin{eqnarray}
\alpha=2(\sqrt{S+B}-\sqrt{B})
\end{eqnarray}

\begin{figure}[htb]
\includegraphics[width=0.49\textwidth]{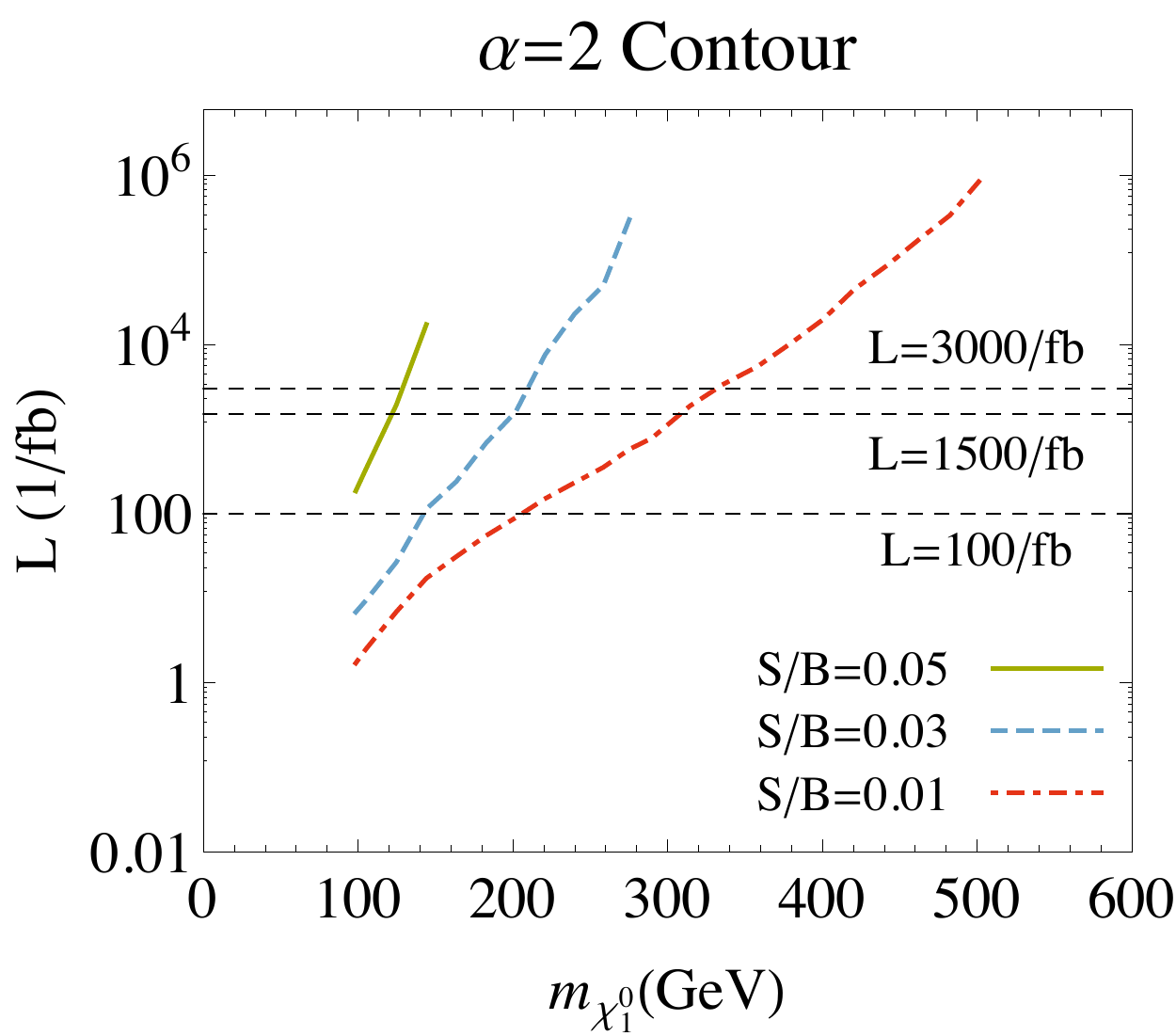}\hfill
\includegraphics[width=0.49\textwidth]{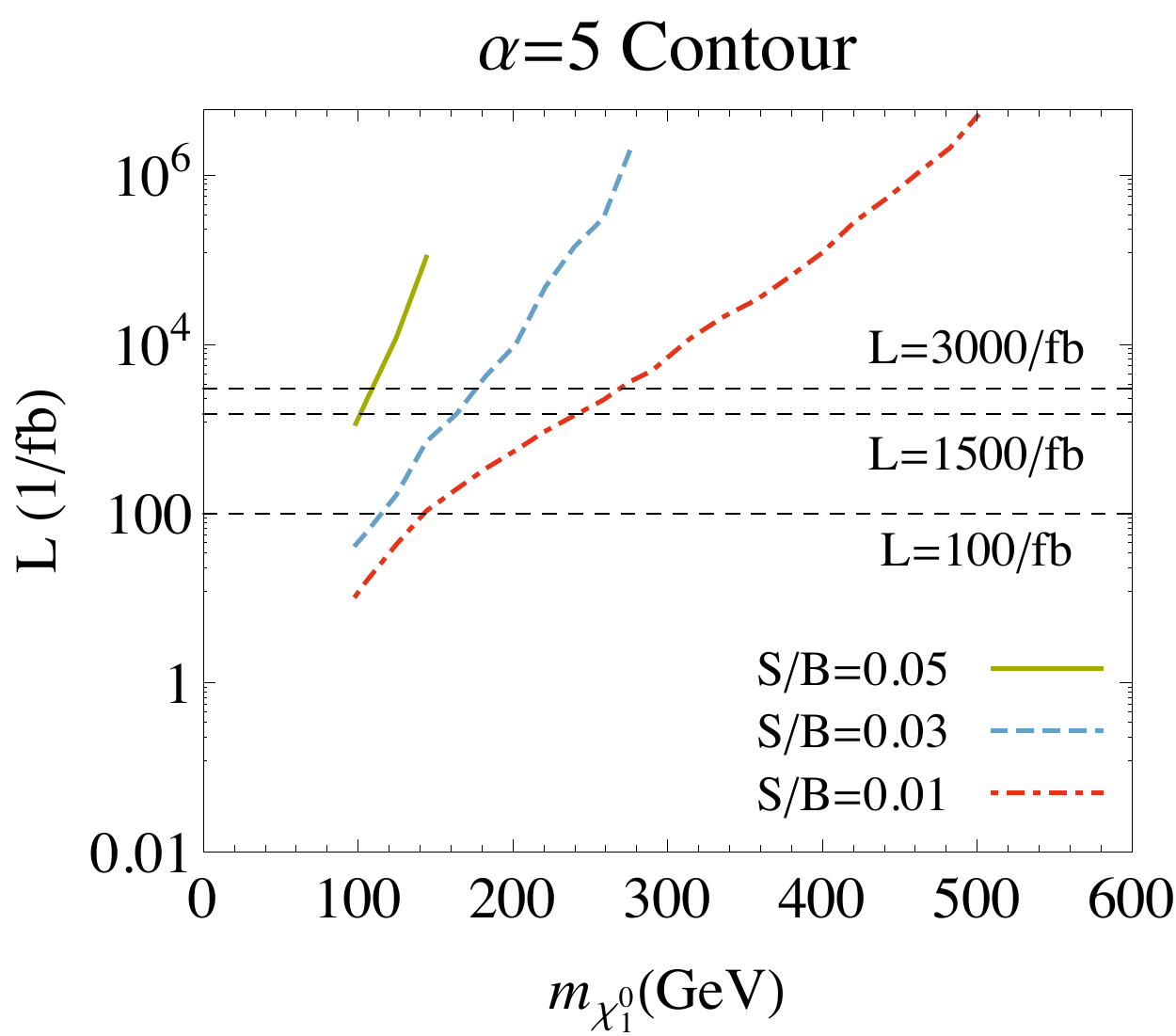}
\caption{$\alpha$=2 and $\alpha$=5 contour plot in the $m_{\chi^0_1},L$ plane for the 13 TeV LHC.}
\label{fig:lumivsn1mass_delphes}
\end{figure}

From Fig.~\ref{fig:signif_delphes} one can clearly see the tension between 
the increase of S/B and decrease of statistical significance as a function of $p^T_j$.
One can also note that S/B ratio is never better than 6\%, which demands the respective systematic error to be well under control.
In Fig.~\ref{fig:lumivsn1mass_delphes} we present our final results for the 13 TeV LHC sensitivity in Luminosity-$m_{\chi^0_1}$ plane for different S/B assumptions. Under each assumption, for each point in the parameter space, the cut on $p^T_j/\met$ has been chosen in order to have at least the chosen S/B ratio.
Even for optimistic S/B=5\% ratio, corresponding to the respective systematic error which is consistent with the recent CMS and ATLAS analyses for LHC8\cite{CMS-PAS-EXO-12-048,ATLAS-CONF-2012-147}, 
the LSP exclusion is limited to about 120 (130) GeV at 95\%CL at 1.5 (3) ab$^{-1}$ (left figure).

If there is a chance for systematics to  go below the LHC8 mark the situation would be drastically better.  For example, we show the results for S/B=3\%, a case where the
LHC would be able to probe the NSUSY scenario considerably better: $m_{\chi^0_1}>200$~GeV  would be 
excluded with 1.5 ab$^{-1}$ integrated luminosity while even for L=100 fb$^{-1}$  LHC would be able to exclude $m_{\chi^0_1}>140$~GeV.
Unfortunately, control of S/B=1\% assumed in previous works  is not realistic.

Finally in the right figure we show the prospects of discovering such a scenario. In case  of S/B ratio at
the 5\% level, we could be able to claim a discovery up to 110 GeV LSP with 3 ab$^{-1}$.

\section{Complementarity of the LHC and Dark Matter Direct Detection search experiments}

\begin{figure}[htb]
\includegraphics[width=\textwidth]{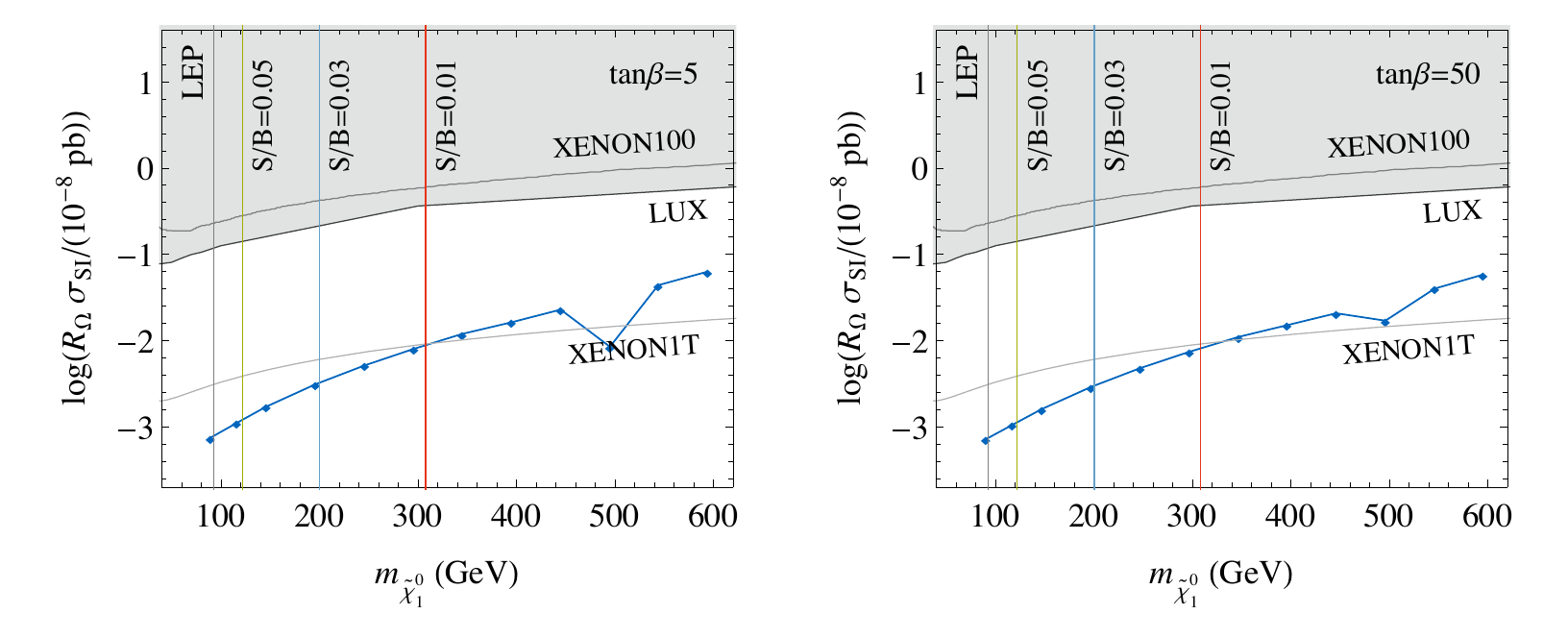}
\caption{Current (XENON100, LUX) and future (XENON1T) cross section limits on SI DDM scattering off the nuclei as a function of the $m_{\chi^0}$. 
The cross-section is rescaled with $R_\Omega=\Omega_{\rm DM}/\Omega^{\rm Planck}_{\rm DM}$ factor
discussed in the text. Vertical lines represent the projected sensitivity of LHC13TeV to the FFP
parameter space, for different assumptions on the control of the systematic error (S/B).}
\label{fig:combined}
\end{figure}

Let us now take a look at the combined sensitivity of the LHC and DDM experiments to the FFP parameter space presented in Fig.~\ref{fig:combined}.
One can see that sensitivity of the present DDM experiments such as XENON100 and LUX is 
clearly not enough to probe FFP parameter space with the naturally low DM relic density.
Another point to stress is that cross section of LSP-nuclei scattering grows {\it faster} 
with the DM mass than the sensitivity of XENON1T does, which make XENON1T actually 
able to access the region of FFP parameter space starting from $m_{\tilde\chi_0}\gtrsim 320$~GeV.
One should again note that in order to ease comparison with the DDM search experiment bounds, 
the cross-section calculated should be rescaled with the respective $R_\Omega$ factor as in section~\ref{sec:DM} 
to take into account the lower DM relic density in our scenario.
At the same time, collider sensitivity is eventually independent of $R_\Omega$.
One can see that the LHC reach in probing the FFP parameter space via monojet signature from production of higgsinos 
is highly complementary to DDM search experiments. The vertical lines presents LHC reach for 1.5 ab$^{-1}$
integrated luminosity and different assumptions on the systematic error (S/B).
One can see that for S/B control at 5\% level, $m_{\tilde\chi_0}$ up to 120 GeV
is covered, while making the very optimistic assumption that S/B $\simeq$ 3\%, 
the sensitivity of the LHC could extend up to  $m_{\tilde\chi_0}\simeq 200$~GeV.
Just for reference, in order to check against results 
of previous studies, we present S/B=1\% case (which is not realistic)
for which LHC would cover FFP parameter space up to  $m_{\tilde\chi_0}\simeq 300$~GeV.
One should note that  LHC sensitivity to FFP region can be inferred from 
ATLAS\cite{ATLAS:2012ky} and CMS\cite{Chatrchyan:2012me} results on monojet searches 
which can be translated into limits on to the effective $q\bar{q}\tilde\chi_0\tilde\chi_0$
operators and then compared with the spin-dependent(SD) and spin-independent(SI) limits from DDM search experiments.
The best current limit from DDM search experiments is the SI one from  LUX experiment which is slightly better than the XENON100 one.
It turns out that even rescaled XENON100 limit on the SI scattering of the neutralino off the nuclei
is more than two orders of magnitude better than  the analogous limits from ATLAS and CMS searches
mentioned above.

\section{Conclusions}
In this study we explore the
 LHC13TEV sensitivity to the focus point region of the MSSM parameter space 
which is characterised by low values of $\mu$
parameter, a necessary condition for natural SUSY
with low fine-tuning.
In the case of the $M_1$ and $M_2$ parameters being large, 
three MSSM  particles, namely $\chi^0_1$, $\chi^0_2$ and $\chi^\pm_1$
become quasi-degenerate and acquire a significant higgsino component.
This scenario also provides a naturally low dark matter relic density
via gaugino annihilation and co-annihilation processes into Standard Model gauge and Higgs
bosons.
In the case where coloured SUSY particles are heavy and out-of-reach of the LHC,
as in the FFP region, the only way to probe SUSY is via monojet signature.

We present here the first realistic results on the LHC13TeV projections
to probe FFP parameter space taking into account 
a) realistic systematic errors and b) fast detector simulation. 
Both components are crucial in order to
estimate  the  correct LHC sensitivity to FFP.
It turns out that for a very optimistic estimate of the control on S/B at the 3\% level, the 
LHC will be able to exclude (at 95\%CL) FFP parameter space with neutralino masses below 140 GeV with 
with  100 fb$^{-1}$ and below 200 GeV with 1.5 ab$^{-1}$ integrated luminosity.
The LHC sensitivity depends drastically on the systematic uncertainty level: for
a 5\% S/B ratio the limits drop down to $m_{\tilde{\chi}^0_1}=120$~GeV 
(compared to 200 GeV at S/B=3\%) with 1.5 ab$^{-1}$ integrated luminosity.

We have found that while the LHC is sensitive to the low end of $\tilde{\chi}^0_1$ mass range,
the DDM search experiments, especially the future XENON1T, become sensitive 
to the upper end of FFP $\tilde{\chi}^0_1$ mass range, starting from about 320 GeV.
This high degree of complementarity of the LHC and DDM search experiments
is crucial for pinning down the whole range of FFP parameter space.
One should note that coverage of the mass gap between 200 GeV and 320 GeV is problematic even for the combination 
of the LHC13TeV and XENON1T experiment and requires further attention.

\section*{Acknowledgments}
The authors acknowledge the use of the IRIDIS High Performance Computing Facility, and associated support services at the University of Southampton, in the completion of this work.
DB and AB are financed in part through the NExT Institute.
WP acknowledges partial support from German Ministry of Education and Research 
(BMBF) under contract no.\ 05H12WWE.

